%% file: main.tex
\documentclass[a4paper, 12pt, openright, twoside, table]{report}
\usepackage[utf8]{inputenc}
\usepackage{graphicx}
\usepackage{subcaption}
\graphicspath{ {images/} }
\usepackage{hyperref}
\usepackage[table]{xcolor}
\usepackage{multirow}
\usepackage{amsmath}
\usepackage{amssymb}
\usepackage{minitoc}
\usepackage{blkarray}
\usepackage{url}
\usepackage{wrapfig}
\usepackage{comment}
\usepackage{setspace}
\usepackage[spanish, british]{babel}
\usepackage{titlesec}
\usepackage{tikz}
\usepackage{emptypage}
\usepackage{booktabs}
\usepackage{siunitx}
\usepackage{makecell}
\usepackage{enumitem}
\usepackage[official]{eurosym}
\usepackage{upgreek}
\usepackage{pdfpages}

\definecolor{headerYellow}{RGB}{255, 242, 204}
\definecolor{headerGreen}{RGB}{198, 224, 180}

\titleformat{\chapter}[display]
    {\Huge}
    {\scalebox{4}{\fontfamily{pbk}\selectfont \thechapter}} % Scales the chapter number
    {10pt}
    {\Huge\bfseries}

\newcommand{\chapquote}[3]{
    \begin{flushright}
    \begin{minipage}{0.6\textwidth}
        
        \begin{quotation} 
            \footnotesize
            \textit{#1} 
        \end{quotation}
        
        \begin{flushright}
            \footnotesize
            - #2, \textit{#3}
        \end{flushright}
    \end{minipage}
    \end{flushright}
}

\hypersetup{
    colorlinks=true,
    citecolor=purple,
    linkcolor=black,      
    urlcolor=blue,
    }

\usepackage[style=ieee,
            backref=true,
            backrefstyle=none 
            ]{biblatex}

\usepackage[a4paper, inner=35mm, outer=25mm, top=25mm, bottom=25mm]{geometry}

\usepackage{fancyhdr}
\usepackage{adjustbox}
\usepackage{array, booktabs}

\newcolumntype{R}[2]{%
    >{\adjustbox{angle=#1,lap=\width-(#2)}\bgroup}%
    l%
    <{\egroup}%
}
\begin{document}
\pagenumbering{roman} %Use lowercase Roman numerals for page numbers
\dominitoc% Initialization, [n] for not Contents word, [c] for centering Contents word
\csname @openlefttrue\endcsname
\onehalfspace

%%%% PORTADA NORMALIZADA %%%% 
\begin{comment}
\includepdf[pages=1]{Chapters/Portada_normalizada_tesis.pdf}
% Insert a completely blank white page
\newpage
\thispagestyle{empty}
\null
\newpage  
\end{comment}  

%%%% CUBIERTA %%%% 

\includepdf[pages=1]{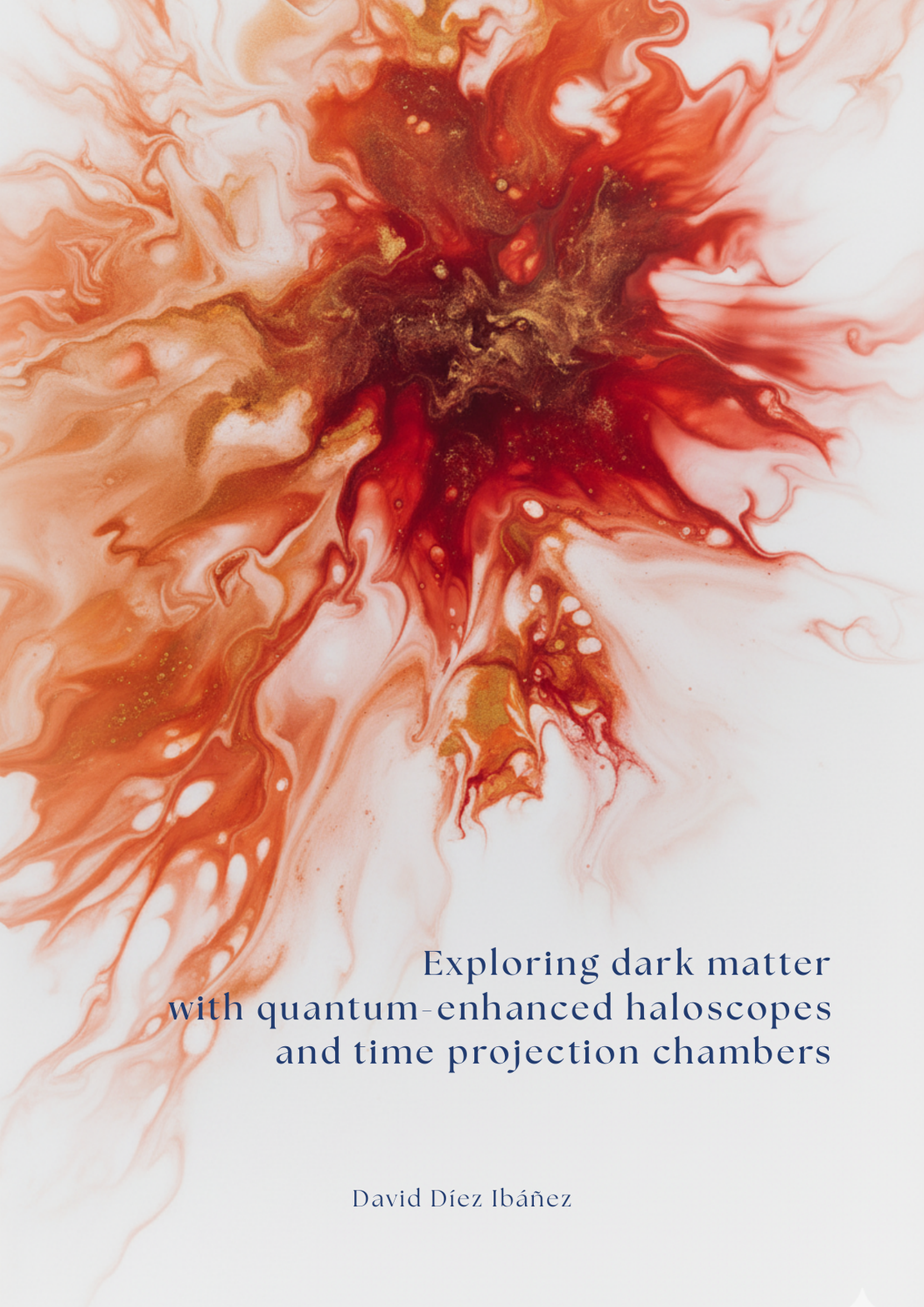}
% Insert a completely blank white page
\newpage
\thispagestyle{empty}
\null
\newpage

%%%% PORTADA %%% 
%\maketitle

\begin{titlepage}
   \begin{tikzpicture}[remember picture,overlay]
   % Background image at bottom-right corner
   \node[anchor=south east, inner sep=0pt] at (current page.south east) {
   \includegraphics[width=12cm]{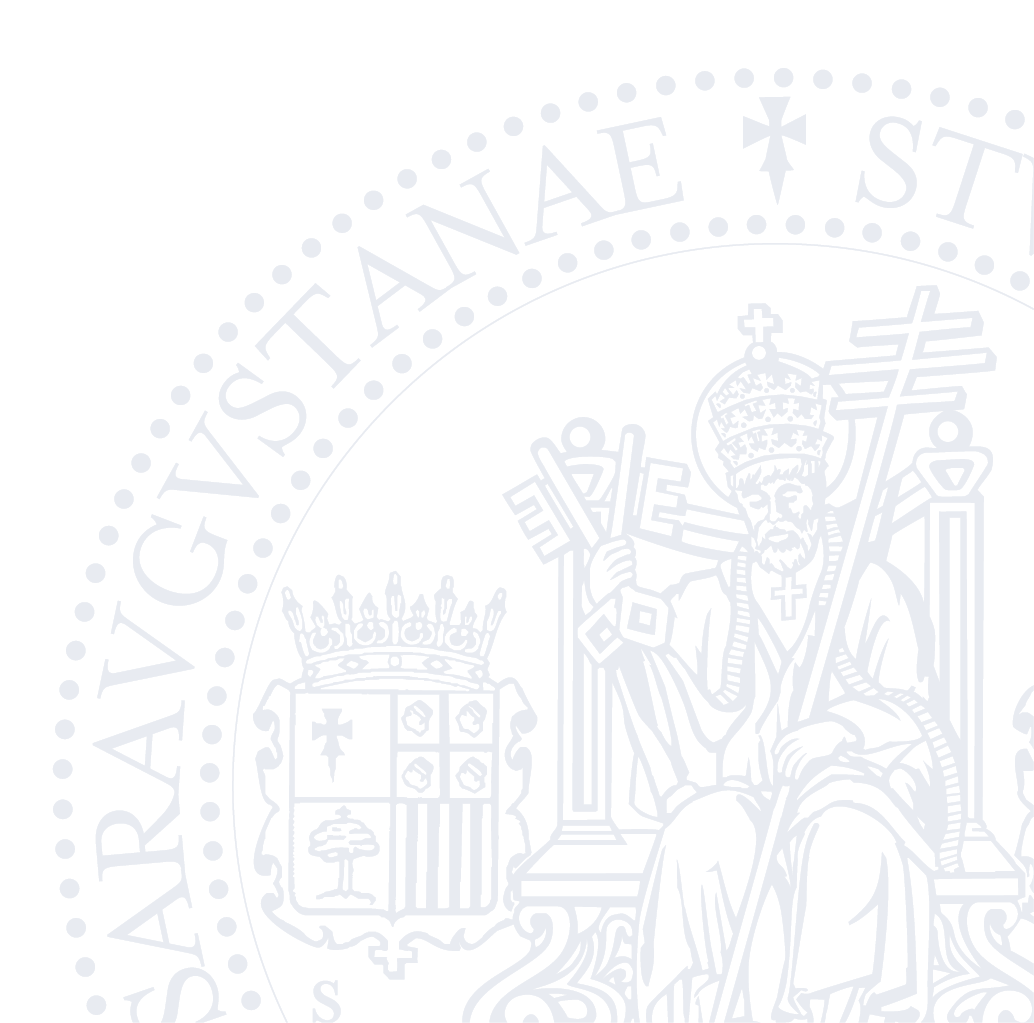}
   };
  \end{tikzpicture}
  
  \begin{center}
  
    \vspace*{\fill} % pushes the title to vertical center
    
    \includegraphics[width=5cm]{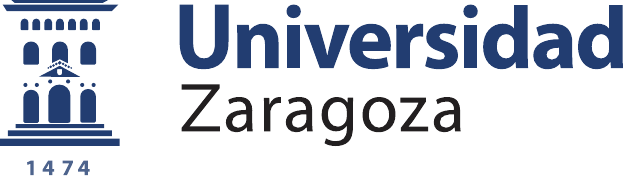}\\[2cm]

    {\large PhD. Thesis }\\[2cm]
    \setstretch{2.5} % Adjust line spacing (1.3 = ~1.5 spacing)
    \parbox{0.9\textwidth}{\centering
      {\Huge \textbf{Exploring dark matter with quantum-enhanced haloscopes and time projection chambers}}
      
      %Dark matter searches: \\[0.3cm] TREX-DM experiment and quantum technologies for axion searches
    }
    \\
    \vspace{3cm}
    \setstretch{1.3}
    {   {\sc\Large David Díez Ibáñez} \\[1.5 cm]
        \large Supervised by \\[0.2 cm]
        {\sc\large Prof. Theopisti Dafní}
    }

    \vspace{2cm}
    \setstretch{1.2}
    {\large Área de Física Atómica, Molecular y Nuclear \\ Departamento de Física Teórica \\ \textbf{Universidad de Zaragoza}} \\[2em]
    {\sc Julio 2025}

  \end{center}
\end{titlepage}

% Insert a completely blank white page
\newpage
\thispagestyle{empty}
\null
\newpage

%%%% PROLOGO %%% 
\pagenumbering{roman} %Use lowercase Roman numerals for page numbers
\chapter*{\vspace{3ex}\Huge Preface }

\vspace{-37ex}
\chapquote{``Los labios apretados en el tubo de ámbar de la pipa, la barba aplastada contra el gorjal de amatistas, los dedos de los pies curvados nerviosamente en las pantuflas de seda, Kublai Kan escuchaba los relatos de Marco Polo sin alzar la vista.''}{Italo Calvino}{\textit{Las ciudades invisibles}}
\vspace{9ex}

\input{Chapters/prologo}

%%%% AGRADECIMIENTOS %%% 
\chapter*{El \\ lugar exacto \\ entre las olas del mar}

\vspace{-45ex}
\chapquote{``Quien ha entrevisto el Universo, quien ha entrevisto los ardientes designios del Universo, no puede pensar en un hombre, en sus triviales dichas o desventuras, aunque ese hombre sea él.''}{Jorge Luis Borges}{\textit{La escritura del Dios}}
\vspace{20ex}

\input{Chapters/agradecimientos}

%%%% TABLE OF CONTENTS %%% 
\tableofcontents

%%%% CHAPTER 1 %%% 
\chapter{Dark Matter}
\pagenumbering{arabic}  % Now Use Arabic numerals for page numbers
\vspace{-40ex} 
\chapquote{``Fue el verano en que el hombre pisó por primera vez la luna. Yo era muy joven entonces, pero no creía que hubiera futuro. Quería vivir peligrosamente, ir lo más lejos posible y luego ver qué me sucedía cuando llegara allí."}{Paul Auster}{El Palacio de la Luna}
\vspace{12ex}

\minitoc
\vspace{5ex}

\input{Chapters/1_DarkMatter}

%%%% CHAPTER 2 %%% 
\chapter{Working principles of a Micromegas gaseous detector}\label{Ch:GasDect}

\vspace{-40ex}
\chapquote{``El especialista 'sabe' muy bien su mínimo rincón de universo; pero ignora de raíz todo el resto.''}{José Ortega y Gasset}{\textit{La rebelión de las masas}}
\vspace{18ex}

\minitoc
\vspace{5ex}
\input{Chapters/2_GaseousDetectors}

%%%% CHAPTER 3 %%% 
\chapter{TREX-DM experiment}\label{Ch:TREX}

\vspace{-40ex}
\chapquote{``A vosotros, los que vengáis a hacer lo que nosotros no hemos hecho (...), os confío mi fracaso, y os deseo la victoria.''}{Pedro Salinas}{\textit{A vosotros}}
\vspace{18ex}

\minitoc
\vspace{5ex}
\input{Chapters/3_TREXDM}

%%%% CHAPTER 4 %%% 
\chapter{New developments with gaseous detectors}

\vspace{-45ex}
\chapquote{``A lonely impulse of delight \\
drove to this tumult in the clouds''}{William Butler Yeats}{\\ \textit{An Irish Airman foresees his Death}}
\vspace{23ex}

\minitoc
\vspace{5ex}
\input{Chapters/4_NewGaseousDetectors}

%%%% CHAPTER 5 %%% 
\chapter{Dark photons and axions}\label{ch:DarkSector}

\vspace{-40ex}
\chapquote{``We astronomers are nomads, \\ merchants, circus people, \\ all the Earth our tent.''}{Rebecca Elson}{\textit{We Astronomers}}
\vspace{18ex}

\minitoc
\vspace{5ex}
\input{Chapters/5_DarkPhotons_Axions}

%%%% CHAPTER 6 %%% 
\chapter{Quantum sensors for haloscopes}\label{QuantumTech}

\vspace{-40ex}
\chapquote{``Dame luz, Señor, para ver mis defectos —pero, por favor, no todos a la vez.''}{Ignacio Peyró}{\textit{Ya sentarás cabeza} }
\vspace{21ex}

\minitoc
\vspace{5ex}
\input{Chapters/6_QuantmSensinghaloscopes}

%%%% CHAPTER 7 %%% 
\chapter{Single microwave photon counter: DarkQuantum} \label{Ch:DarkQuantum}

\vspace{-50ex}
\chapquote{``No one knows the reason for all this, but it is probably quantum."}{Terry Pratchett}{Pyramids}
\vspace{32ex}

\minitoc
\vspace{5ex}
\input{Chapters/7_DarkQuantum}

%%%% CONCLUSIONS  %%% 
\addcontentsline{toc}{chapter}{Summary and conclusions}
\chapter*{\vspace{5ex}Summary and conclusions}

\vspace{-35ex}
\chapquote{``And what is the use of a book, thought Alice, without pictures or conversations?''}{Lewis Carroll}{\textit{Alice’s adventures in Wonderland}}
\vspace{21ex}

\input{Chapters/8_Conclusiones}

%%%% CONCLUSIONES %%% 
\addcontentsline{toc}{chapter}{Resumen y conclusiones}
\chapter*{\vspace{5ex}Resumen y conclusiones}

\vspace{-40ex}
\chapquote{``Ojalá que haya encontrado a alguien que cuente por ella las partículas subatómicas de un grano de arena de una playa infinita.''}{Javier Aznar}{\textit{¿A donde vamos a bailar esta noche?} }
\vspace{25ex}

\input{Chapters/8_Conclusiones_ES}

%%%% BIBLIOGRAFÍA %%% 

%\addcontentsline{toc}{chapter}{Bibliography}

% Define custom heading for bibliography
\defbibheading{bibliography}[\bibname]{%
  \chapter*{\vspace{2ex}#1}
  \addcontentsline{toc}{chapter}{#1}
  \vspace{-35ex}
  \chapquote{``Libros. Libros, libros, libros. Libros viejos, libros nuevos, libros caros, libros baratos, libros en escaparates, estanterías, carretillas, bolsas, arrojados en montón sin orden ni concierto o cuidadosamente alineados tras un cristal.''}{Walter Moers}{\textit{La ciudad de los libros soñadores}}
  \vspace{7ex}
}

%En hondas, frías, huecas estancias donde se juntan sombras con sombras, donde los libros sueñan distancias y al contemplarlos siempre te asombras.

\printbibliography[heading=bibliography]

%%%% CONTRAPORTADA %%% 

\newpage 
\thispagestyle{empty}
\null
\newpage 
\includepdf[pages=1]{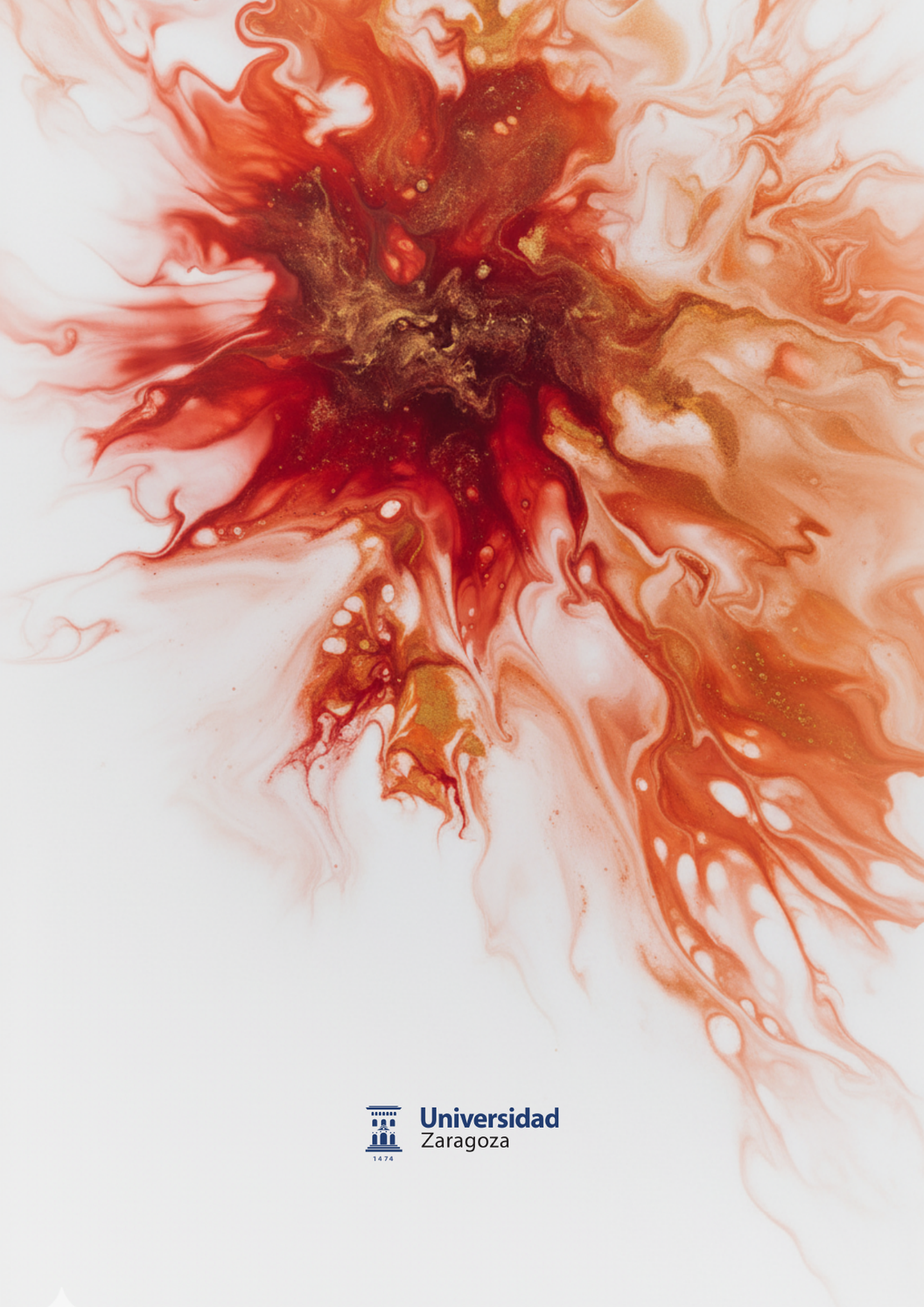}

\end{document}

%% file: Chapters/prologo.tex
This work collects the most significant results obtained during my years working in the Grupo de Investigación de Física Atómica, Nuclear y Astropartículas (GIFNA) of the Universidad de Zaragoza. We are devoted to dark matter searches, ranging from WIMPs to axions. Our group has specialized in utilizing various detection technologies, including scintillators such as the ANAIS experiment and gaseous detectors like CAST. My contributions are primarily centred on the use of time projection chambers equipped with Micromegas readout planes. During my initial year with the group, I participated in the final data collection of CAST, an axion helioscope at CERN, and the initial tasks with the detectors for the next generation helioscope BabyIAXO. However, a significant portion of my research efforts focused on TREX-DM, a low-mass WIMP detector located in the Laboratorio Subterráneo de Canfranc. The main challenge was to reduce the background events, which involved exploring various solutions, testing different gases, and developing novel analysis techniques to identify their sources. Within this context, we conducted experiments with new gaseous detectors aimed at enhancing the performance of Micromegas detectors, including new calibration techniques and operating them at high pressures with innovative gas mixtures. 

Chapter 1 motivates the dark matter problem and the proposed solutions for it. In chapter 2, gaseous detectors are presented, with the basic properties of particle interaction in gases and the general behaviour of Micromegas readout planes. Chapter 3 is devoted to TREX-DM experiment, with the description of the setup, the treatment of radon background and the identification of alpha particles, and the last improvement with a mixed readout plane composed of a GEM amplification stage on top of the Micromegas plane. The threshold measurement for this new configuration shows an enhanced behaviour, crucial to achieve the expected sensitivity. In chapter 4 two new developments for gaseous detectors are shown: a UV LED calibration source and high pressure gain studies with argon + 10\% isobutene mixture. The UV LED allows a cheap source for variable energy calibrations. This could be ideal for Micromegas performance validation, threshold measurements and gas properties testing like diffusion or drift velocity.

Two years ago, I became involved in a promising new project: the development of a single-photon counter for axion haloscopes as part of the efforts of the RADES collaboration. This endeavour required the adoption of a novel technology —microwave resonators for detecting dark matter axions— and introduced us to an entirely new field: quantum optics. As part of this initiative, I delved into the study of superconducting qubits and the stringent requirements associated with microwave single-photon detection. To further this work, I had the opportunity to visit the Low Temperature Laboratory at Aalto University in Helsinki, where I acquired much of my current knowledge in quantum optics. This fruitful collaboration has culminated in the experimental results presented in the second half of this work. 

Chapter 5 motivates the axion and other dark sector candidates as the dark photon, which shares detection signs. Chapter 6 deals with axion haloscopes and the general context of superconducting qubits and their manipulation. And in the last one, chapter 7, the design of the double cavity developed as a single photon counter and the associated transmon is  described, together with its first measurements for dark photons, showing the readout protocol, the analysis and interpretation of the  data, and finally the estimated dark photon sensitivity extracted from these measurements.

All figures and tables in this work clearly indicate their sources through citations in the captions, unless they are original and created by the author. Most of the code developed for data analysis and plotting can be found in \cite{ThesisCodes}.

This work explores two complementary approaches to address a fundamental question in modern physics: the nature of dark matter. It showcases a representative sample of the diverse experimental techniques currently under investigation, as well as the imagination required to devise new strategies for probing increasingly elusive particle models. Inevitably, this effort will be continued...

%% file: Chapters/agradecimientos.tex
\begin{otherlanguage}{spanish}
He querido escribir este último pedazo al sol, casi todo me sale mejor bajo su augurio. Sospecho que se convertirá en uno de los fragmentos más leídos entre mi gente, mi tribu, así que me he propuesto afrontar una particular tarea autoimpuesta y, por tanto, placentera: haceros ver que este trabajo no es una isla, que no he aparecido en ella por sorpresa, abandonando a todos atrás. Si estas páginas se han dejado escribir se debe a los incontables puentes, transbordos y singladuras por los que vosotros me habéis guiado.

Yo, que soy hombre de secano, no puedo evitar escribir pensando en el mar que es infinito y está fuera de mi alcance. Este anhelo lo heredo de mi madre, Ana, que fue la primera en lanzarme al agua, la que me enseñó a flotar. Desde entonces, y a pesar de seguir echando de menos el velero que nunca hemos tenido, todos mis caminos sobre la mar han partido de ella. Es la que suelta mi última amarra en cada viaje, su inspiración la que fija mi rumbo. Junto a ella mi padre, Ricardo. Él es distinto, se marea en el mar. Y sin embargo me lo encuentro allí a donde vaya, cada rincón que he visitado había sido anticipado, cada nuevo atolón presagiado en sus historias. Tiene algo de Cheshire, se lo noto en la sonrisa. Y entre ambos Adrián, que esté donde esté siempre custodia uno de mis ojos. Y yo el suyo aquí mientras escribo.

Los rumbos están siempre trazados, lo que no quiere decir que se sigan a la perfección. En uno de esos «voyage autour» a los que nada obliga me encontré con Theopisti en esta isla, tan mediterránea ella aunque nada de su aspecto lo anticipe. Ha sido curioso tener su guía en en este nuevo mundo, del que nada sabemos. Me vio y me dijo: ven, aquí hay dragones, te gustará. Sorprendentemente ha tenido razón. Y no piensen que me atrapó como la magia de Circe, ella tiene la rara habilidad de ofrecer lo que puede a quien lo quiere. En mis paseos por esta isla he descubierto algunas cosas, entre otras, a sus pobladores: Igor, Gloria, Juanan, Laura y Héctor, Juan, Alfonso, Javier y Javier, Jorge y Ángel y María, Susana, Iván. Y de otras tribus: Akash, Enrique, Pepa y Jorge. 
Con Héctor y Óscar nos sumergimos en sus profundidades, probamos el sol de sus montañas y admiramos el mar en el horizonte, sabiendo en ocasiones que nunca lo alcanzaríamos. Luis parecía estar por aquí algo más perdido que el resto, aunque sospecho que siempre tuvo más claro el camino. A mi desde luego me desbrozó infinitos senderos y tuve la suerte de recorrer algunos a su lado. Además, el resto de náufragos: Cristina, Elisa, David, Álvaro, María, Jorge, Ana, Konrad, Tamara, David y el tinglao que montamos, Yikun, Sophia, Mathieu, Fran, Carmen, Jaime, Itxaso. Algunos ya se fueron, otros aún seguirán un poco más por aquí cuando yo me vaya. Sin ellos todo esto habría sido más aburrido.

Navegar hasta aquí es a todas luces intrascendente. Y sin embargo la gente sigue llegando. Esto, como casi todo, tiene una explicación, yo puedo dar la mía. El primer puente fue tendido gracias a tres personas: Guillermo, César y Óscar. Yo aun andaba en tierra firme cuando ellos ya surcaban la mar océana. Siguiendo sus estelas atraqué en muchas costas, la más antigua Manu y Esteban, calas gemelas donde descansar. En María y Blanca y María y Elisa y Mónica y Alicia y por supuesto Lorena he encontrado incontables frutos, una inspiración especialmente satisfactoria y el extraño placer que produce siempre reconocer una costa en el horizonte. Fernando es un peñón que se desgajó de esos farallones, por eso ahora es el mejor sitio para nadar. Existe una furia protectora que he visto alguna vez en el mar. Roberto, Carmen, Sandra, Ángel y Jaime, que durante tanto tiempo han navegado a mi lado, lo son para mí en tierra. También he encontrado singulares gemas en las marejadas nocturnas: Alejandro, Pablo y Almudena, Cristina, Bárbara, Dani y Salva y Fiorella, Juanma, Lucía y Andrés, José Ramón, Ilya y Marko, Alicia y Carla y Lorenzo, cada una presagio inevitable de la siguiente, ¿cómo me iba a detener?

Entre tanta jarcia y tanta vela uno a veces puede llegar a pensar que el mundo es eso, un arriba y un abajo y el viento empujando de través. Pero se nos olvida que la mar y sus corrientes nos mecen sin que podamos opinar. En esas profundidades encontré a Rebeca que me llevó a reinos olvidados y me hizo escuchar lo que nunca antes había visto. Clara fue la primera que me contó algunas leyendas de esos pueblos ignotos y fui yo el que acompañó a otra María a sus profundidades particulares. Todos seguimos batiéndonos en estas aguas, como si lo que dictan las velas tuviera la menor importancia. Esther sabe que no.

Hablar de lo que he encontrado en esta isla tiene una trascendencia relativa, mayor para los que la habitamos, escasa para los que jamás habéis oído hablar de ella. Por eso no me importa demasiado propagar su fama, solo trato de hacer ver que por lejos que parezca, hay una linea que os une con ella, que mis pasos están señalados en alguna carta náutica. Porque al fin y al cabo los hombres no somos islas, somos parte de un continente y también del océano.
\end{otherlanguage}

%% file: Chapters/1_DarkMatter.tex
This first chapter is devoted to outlining the theoretical framework that underpins the dark matter problem and the proposed solutions to explain it. More complete reviews about dark matter can be found easily, below are the most useful ones for this work. My colleague Iván Coarasa made a clear overview that set its roots in the knowledge of the history and evolution of the Universe in \cite{Coarasa2021anais}. The Particle Data Group annual publication \cite{PDG2024review} is a reference always updated with concise explanations and emphasis in the experimental efforts. And, of course, there are rigorous and extensive works devoted to analysing the dark matter problem, like \cite{bertone2005particle}. Also, from a historical point of view, many interesting articles have focused on the scientific developments that lead to the understanding of the problem and the efforts made to unravel its mysteries \cite{bertone2017dark, bertone2005particle}.

\section{Experimental evidence that points towards an unseen component of the Universe}

Dark matter is a fundamental component of modern astrophysics and cosmology, playing a crucial role in the formation and evolution of cosmic structures. Although it does not interact electromagnetically and remains undetectable via direct optical observations, its existence is strongly inferred from multiple independent astrophysical and cosmological observations. This section explores evidences supporting the existence of dark matter: galaxy rotation curves, galaxy clusters, and cosmological measurements from the cosmic microwave background (CMB).

Each kind of observation in galaxies, clusters, or cosmology requires that there be at least a certain minimum amount of dark matter. The true abundance must be at least as high as the highest of these minimum estimates, and remarkably, all the major lines of evidence point to a consistent, substantial amount of dark matter in the universe. This abundance is parametrized through the mass density, $\Omega_{DM}$, where $\Omega_{i} = \rho_i/\rho_c$, being $\rho_c$ the mass density required for a flat Universe. The total energy density of the Universe is thus expressed as $\Omega = \sum \Omega_i = \sum \rho_i/\rho_c = 1$. Recent measurements point to $\Omega_{DM} = 0.25$ and total mass content of the Universe of $\Omega_{M} =0.31$~\cite{aghanim2020planck}.

\subsection{Galaxies}
One of the first compelling pieces of evidence for dark matter emerged from the study of galaxy rotation curves. Vera Rubin and her collaborators in the 1970s conducted spectroscopic studies of spiral galaxies, measuring the velocity of stars and gas as a function of their radial distance from the galactic center \cite{rubin1970rotation}.

In a galaxy dominated by visible matter, Keplerian dynamics predict that the orbital velocity of stars should decrease when increasing radial distance, following the relation:
\begin{equation}
v(r) = \sqrt{\frac{GM(r)}{r}}
\end{equation}
with $M(r) = 4\pi \int \rho(r) r^2 dr$ the total mass inside the orbit and $\rho(r)$ the matter density as a function of the distance to the centre of the galaxy, $r$.

\begin{figure}[h]
    \centering
    \includegraphics[width=0.5\linewidth]{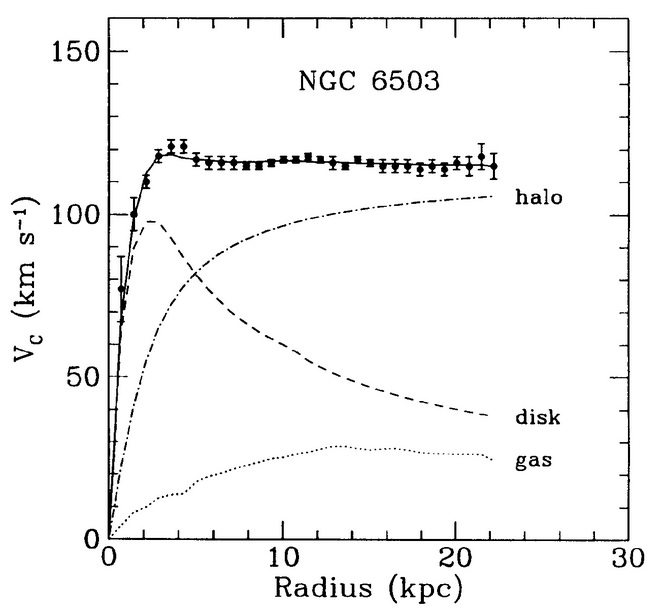}
    \caption{Rotation curve from NGC 6503 galaxy: when moving away from the galactic centre, the speed of the stars remains constant instead of falling. The halo pattern can be explained by dark matter \cite{bertone2005particle}.}
    \label{fig:GalaxyRotationCurve}
\end{figure}

However, observations consistently show that rotation curves remain flat at large distances \cite{rubin1980rotational}, an example can be seen in figure \ref{fig:GalaxyRotationCurve}. This discrepancy suggests the presence of an unseen mass component that extends beyond the visible disk of the galaxy, resulting in an additional gravitational pull. This distribution of invisible mass is called dark matter halo and it should have $M(r)\propto r$ and therefore $\rho(r) \propto 1/r^2$, which at some point should decay faster to keep the galaxy size finite.

From these rotation curves it is possible to deduce the density and speed that dark matter particles should have in these halos that surround the visible matter of galaxies. This knowledge is especially interesting for direct detection experiments of dark matter, because both these parameters affect its interaction cross section, the probability of interaction with the baryonic particles of a detector. The estimated values for the solar environment are a density of 0.2-0.6 GeV/cm$^3$ and a velocity of 220-240~km/s~\cite{PDG2024review}.

\subsection{Galaxy clusters}
Galaxies gather in the Universe forming greater structures that we called galaxy clusters. The movement of these structures is governed by gravity and therefore they are excellent objects to examine the influence of dark matter. Galaxy clusters provide a crucial piece of evidence for dark matter through three independent observational methods: the virial theorem, X-ray measurements of intracluster gas, and gravitational lensing.

\subsubsection{The virial theorem and mass discrepancy}
The first hint of the presence of an unknown matter component in the Universe is due to Fritz Zwicky, in 1933 \cite{zwicky1933rotverschiebung}. He applied the virial theorem to the Coma Cluster, finding that the galaxies' high velocity dispersions implied a much larger gravitational mass than what was inferred from visible stars alone. He introduced the term "dunkle Materie" (dark matter) to describe this missing mass that glues together the different galaxies in the clusters. 

\subsubsection{X-ray observations of intracluster gas}
Clusters of galaxies contain hot gas that emits X-rays due to bremsstrahlung radiation. The temperature and spatial distribution of this gas, determined through X-ray observations from space borne telescopes like Chandra or XMM-Newton, indicate the presence of a deep gravitational potential well that cannot be explained solely by visible matter.

\begin{figure}
    \centering
    \includegraphics[width=0.8\linewidth]{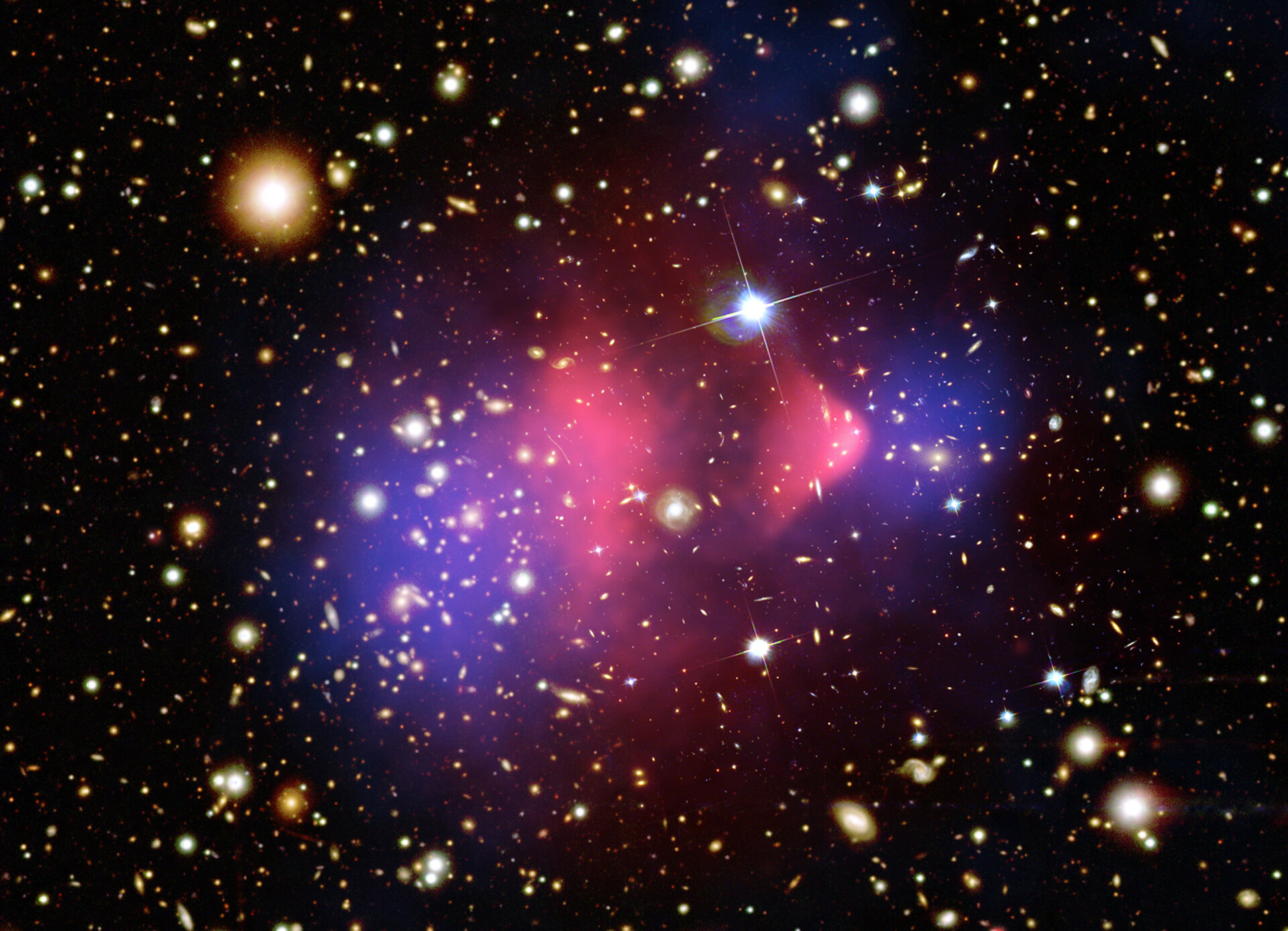}
    \caption{Bullet cluster composite image combining optical, X-ray and gravitational lensing. The optical image from the Magellan and the Hubble Space Telescope shows galaxies in orange and white in the background. Hot gas, which contains the bulk of the normal matter in the cluster, is shown by the Chandra X-ray image in pink. Gravitational lensing reveals the mass of the cluster, dominated by dark matter, in blue. Original study and similar pictures in \cite{clowe2006direct}.}
    \label{fig:BulletCluster}
\end{figure}

\subsubsection{Gravitational Lensing}
Einstein’s theory of general relativity predicts that mass bends spacetime. This means that light trajectories are distorted around massive objects like galaxy clusters, allowing us to receive light from background galaxies. Studying the patterns of distorted light, the mass distribution of the gravitational lens can be inferred. Therefore, strong and weak gravitational lensing studies provide direct maps of mass distribution in clusters. The Bullet Cluster (1E 0657-56) is a striking example where the separation between baryonic matter and total mass provides strong evidence for dark matter. In the Bullet cluster, two clusters have collided and we can see the effects of this event. Observations at these two regimes, X-ray and gravitational lens, allow to discriminate between baryonic and dark matter because the electromagnetic interactions in the hot gas induce a distortion in baryonic matter distribution whereas dark matter only interacts gravitationally so they pass through each other easily. This can be seen in the composite image in Figure~\ref{fig:BulletCluster}.
\\

\subsection{Evidence from Cosmology}
The most precise measurement of dark matter's contribution to the contents of the Universe comes from cosmological observations. This is the largest scale we can probe because it is the scale of the Universe. All observations at this scale are based on the cosmic microwave background (CMB), an isotropic radiation with the spectrum of a blackbody with a temperature of T = 2.725~K. This signal, predicted by Gamow in 1948 and detected by Penzias and Wilson in 1965, is the strongest argument for the Big Bang model. It comes from the recombination period in which electrons joined with protons to form the first hydrogen nuclei, allowing photons to start travelling freely without being absorbed by the proton-electron plasma. This radiation has been travelling unchanged since that moment, and due to the expansion of the Universe, its wavelength has been shifted.

\begin{figure}[h]
    \centering
    \includegraphics[width=0.9\linewidth]{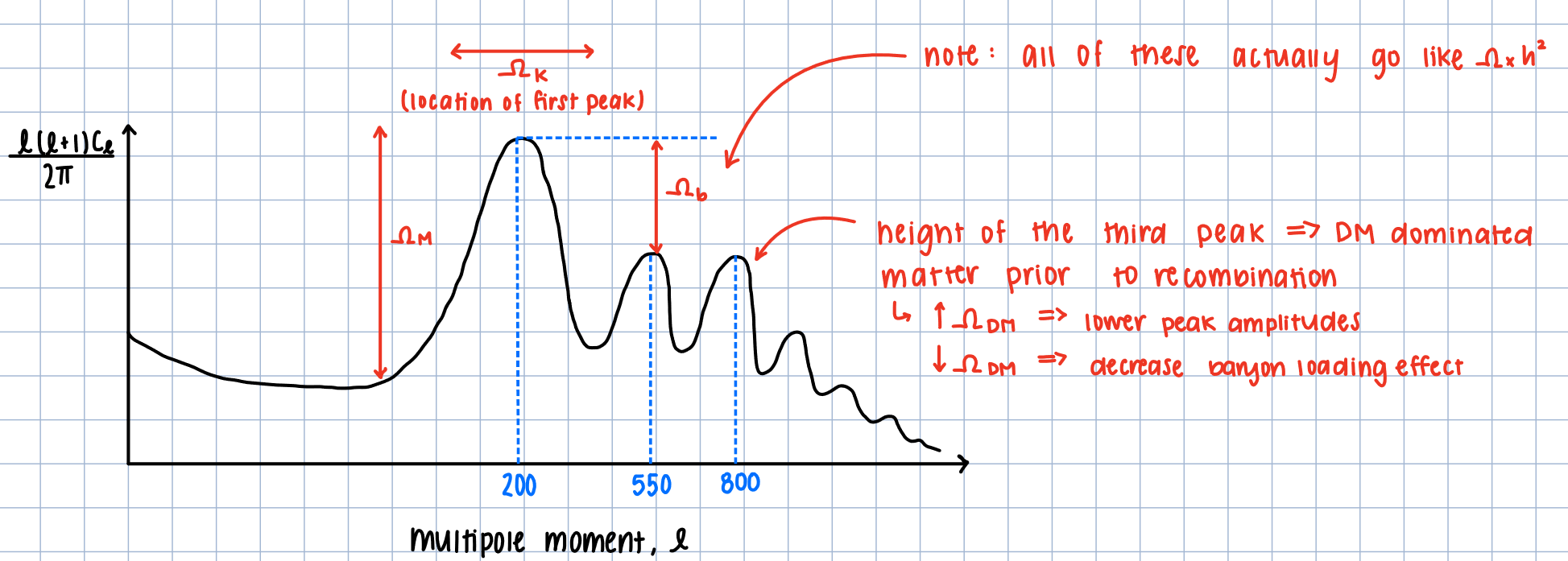}
    \caption{The dark matter content in the Universe can be extracted from the power spectrum of anisotropies of the CMB. A deeper and pedagogical approach can be found in the CMB section of \cite{AstroWiki}, from where this image was extracted. }
    \label{fig:CMBfitHand}
\end{figure}

Since the 1990s, three space observatories have been launched to accurately measure the cosmic microwave background: Cobe in 1992, WMAP in 2003, and Planck in 2013 \cite{durrer2015cosmic}. These missions have determined with increasing precision the tiny anisotropies in the microwave background, which, among other things, allow us to study the distribution of matter in the Universe. The basis for this is the Standard Cosmological Model, whose parameters include the abundances of baryonic and non-baryonic matter. Figure \ref{fig:CMBfitHand} shows how the abundances of each type of matter can be derived from the power spectrum of anisotropies of the CMB.

The Planck 2018 results \cite{aghanim2020planck} report a dark matter density parameter of $ \Omega_{DM} \approx 0.26$, meaning that dark matter constitutes about 26\% of the total energy density of the Universe.

Another hint comes from Big-Bang nucleosynthesis (BBN). It is an essential ingredient in the study of the current composition of the Universe, since it provides an estimate of the density of baryons ($\Omega_b$) independent of that supplied by the analysis of the CMB anisotropies.

At the beginning, the Universe was so hot and dense that everything was in its most elementary constituents, the elementary particles. While the Universe was cooling down, compound objects were formed, like neutrons and protons. Around a temperature of $k_BT \sim 80$ keV, the combination of protons and
neutrons to form deuterons stopped the decay of free neutrons ($\tau_n \simeq 880$ s), when the neutron to proton ratio was 1/7, launching primordial nucleosynthesis.

According to BBN, the fusion of protons and neutrons generated a large quantity of helium nuclei and smaller amounts of other light nuclei, such as deuterium, helium-3, and lithium-7. Present cosmological models and our knowledge of nuclear reaction rates make possible the calculation of the percentage of these nuclei formed in those first minutes.

From the agreement between the abundance of light elements calculated from the BBN and the observations, it follows that the amount of ordinary matter is of the order of 5\% , in good agreement with CMB anisotropies. From this number follows that most of the matter content of the Universe is not baryonic and, therefore, we call it dark matter.

\section{Proposed explanations for the dark matter\index{Dark matter} problem}

Considering all the observations suggesting the existence of some form of unseen matter, humanity has proposed several plausible explanations for these phenomena. Some of these hypothesis are more appealing because they are simpler, they fulfill the expectations better or they help to explain other open problems in particle physics. Nevertheless, in the absence of an experimental discovery that definitively rules out the alternatives, all of the following explanations should remain under consideration.

\subsection{Standard Model dark matter}
The Standard Model of particle physics is our most precise explanation of the fundamental constituents of the Universe. All particles known and their interactions gathered in this model allow to explain most of the natural phenomena that humanity has seen. Despite its success it is not fully satisfactory because it is known that there are experimental observations that cannot be explained within the Standard Model, one of these open problems being  dark matter.

Some of the particles of the Standard Model can satisfy the conditions of being part of an invisible component of the Universe or can form structures that behave like dark matter seen at great distances.

\subsubsection{Neutrinos}
Neutrinos are the only particle in the Standard Model that can satisfy the properties described for dark matter. However, ordinary neutrinos were relativistic particles at the time of decoupling, and therefore could not produce as many small scale structures as currently observed in the Universe. Cosmological observations and simulations are incompatible with neutrinos being the only particle composing dark matter, although they can be a fraction of it.

\subsubsection{MaCHOs}\index{MaCHOs} 
Massive compact halo objects, MaCHOs, are baryonic objects, such as brown dwarfs, neutron stars, and black holes, that reside in the halos of galaxies and could contribute to dark matter. These objects do not emit sufficient light to be easily detected but can be observed through gravitational microlensing events. Studies performed by MACHO and EROS collaborations have placed stringent limits on the fraction of dark matter that could be composed of MaCHOs, suggesting that they are not the dominant form of dark matter \cite{alcock2000macho}.

\subsubsection{Primordial black holes (PBH)}\index{Primordial black holes}
Some theories suggest that dark matter could consist of primordial black holes (PBHs) formed in the early Universe. Unlike regular black holes formed from stellar collapse, PBHs would have originated from high-density fluctuations shortly after the Big Bang. Since they formed in the radiation-dominated era, before BBN, PBHs are effectively non-baryonic. Although the calculation of their lifetime depends on the details of the gravitational collapse, if their mass is greater than $5 \times10^{11}$ kg, their lifetime is longer than the age of the Universe. Although observational constraints, such as gravitational lensing surveys and cosmic microwave background (CMB) distortions, place limits on the abundance of PBHs, they are still viable candidates to form all dark matter or at least a relevant portion of it, despite not being elementary particles \cite{PDG2024review}.

\subsection{Dark matter particles}
Most hypotheses include new particles to explain dark matter. The motivation varies, and although there are models driven by the dark matter problem that include ad hoc candidates, most of them are proposed for other open issues in particle physics, but fit well as dark matter.

\subsubsection{WIMPs}
Weakly Interacting Massive Particles have been the preferred candidate to form dark matter since they were proposed by Steigman and Turner in 1985 \cite{steigman1985cosmological}. They appear naturally in models that contain electroweak-scale new physics that addresses the hierarchy problem, being the most paradigmatic example the minimal supersymmetric extension to the Standard Model.
They typically interact with Standard Model particles via weak nuclear force and have a mass range between 10 GeV and 1 TeV, although this depends strongly on the model. These particles with weak interaction cross section and mass in the range of hundreds of GeV naturally provide a candidate for cold dark matter whose primordial abundance would be in agreement with the present observed abundance of dark matter. This is called the “WIMP miracle”. But as most of the parameter space for simple supersymmetric models is ruled out, other theories have came up with particles with similar properties. Other non-supersymmetric WIMP models include models with a Higgs or Z portal, universal extra dimensions, and other models with extra (warped or flat) dimensions, little Higgs theories, technicolor and composite Higgs theories, among others \cite{PDG2024review}. 

\subsubsection{Axions and axion-like particles}

Axions were first introduced by Peccei and Quinn to address the strong CP problem in quantum chromodynamics (QCD). They are very light scalar or pseudoscalar fields, viable dark matter candidates arise for masses up to $\sim 1$ eV and at present are one of the most compelling candidates for dark matter. Most searches are based on the inverse Primakoff effect, axion-to-photon conversion in the presence of very strong electromagnetic field. Direct axion detection experiments vary considerably depending on the source of axions. For cold dark matter axions, called also relic axions, or axion-like particles (ALPs) the main detection scheme is based on resonant cavities called \textit{haloscopes} in which the axion is converted into a photon on resonance with the cavity in the presence of a magnetic field. And \textit{helioscopes} are the instruments devoted to detect solar axions, axions originated in the Sun; in this case, the converted photon is more energetic, typically at the range up to 10~keV, and can be focused through X-ray lenses and detected with ultra low background detectors. Further details about axions can be found in chapter \ref{ch:DarkSector}.

\subsubsection{Sterile neutrinos}

Sterile neutrinos arise in extensions of the Standard Model that include a right-handed neutrino. They are called sterile because they do not interact weakly with any other particle, apart for the mixing with the ordinary neutrinos. Although the lifetime of sterile neutrinos must be greater than the age of the Universe for them to be a candidate for dark matter, some of them may have decayed through various channels that may allow its indirect detection. Nowadays, radiative decay, one neutrino plus one photon with half the energy of the sterile neutrino mass, is under strong review because an unidentified emission line at $3.5$ keV in the X-ray spectra of galaxy clusters has been reported both by XMM and ChandraX-ray observatories. Its origin is not clear and it is being investigated further \cite{PDG2024review}.

\subsubsection{Dark photons}

Dark photons appear in extensions of the Standard Model where an additional U(1) gauge symmetry introduces a new force carrier similar to the regular photon that may have a small mass. These particles could interact with Standard Model particles through kinetic mixing with standard photons. Light vector bosons like this dark photon can be cosmologically stable, depending on its mass and kinetic mixing coupling with the visible photon. This makes them a viable dark matter candidate. A broad assortment of experiments look for these particles in the range of masses between $10^{-22}$ eV and $10^{-2}$ eV. Depending on the mass range sensitivity torsion balance experiments, atom interferometry, comagnetometers, gravitational wave detectors, broadband axion experiments, LC resonators, lumped-element LC resonators and cavity resonators have been employed for dark photon searches \cite{PDG2024review}. Most of the axion experiments can look for dark photons if they turn off the magnet. A more detailed description of dark photons will be presented in chapter \ref{ch:DarkSector}.

\subsection{Modified gravity}\index{Modified gravity}

The third possibility imagined by humanity makes use of another argument: there is no missing matter in our observations, it is the gravity that behaves differently at different scales and our knowledge of it is not complete.
This direction was first explored in the pioneering work on modified Newtonian dynamics (MOND) published in 1983 \cite{milgrom1983modification}.  Many attempts have been made since then to solve the dark matter problem by modifying Einstein’s theory of general relativity. The success of these efforts, however, remained limited to the rotation curves of galaxies and current understanding suggests that the only way that these theories can be reconciled with large scale observations is by mimicking the behaviour of cold dark matter on cosmological scales effectively and very precisely. 

The coincident observation of gravitational waves and electromagnetic radiation from GW170817~\cite{abbott2017gw170817} has allowed to set very stringent constraints on the propagation velocity of gravitational waves. The fact that this velocity does not differ from the speed of light by more than one part in $10^{-15}$ severely constrains all modified-gravity theories in which gravitational waves travel on different geodesics with respect to photons and neutrinos. This has in particular allowed us to rule out tensor–vector–scalar theories \cite{amendola2018fate}.

\begin{figure}\label{DMCandidates}
    \centering
    \includegraphics[width=0.8\linewidth]{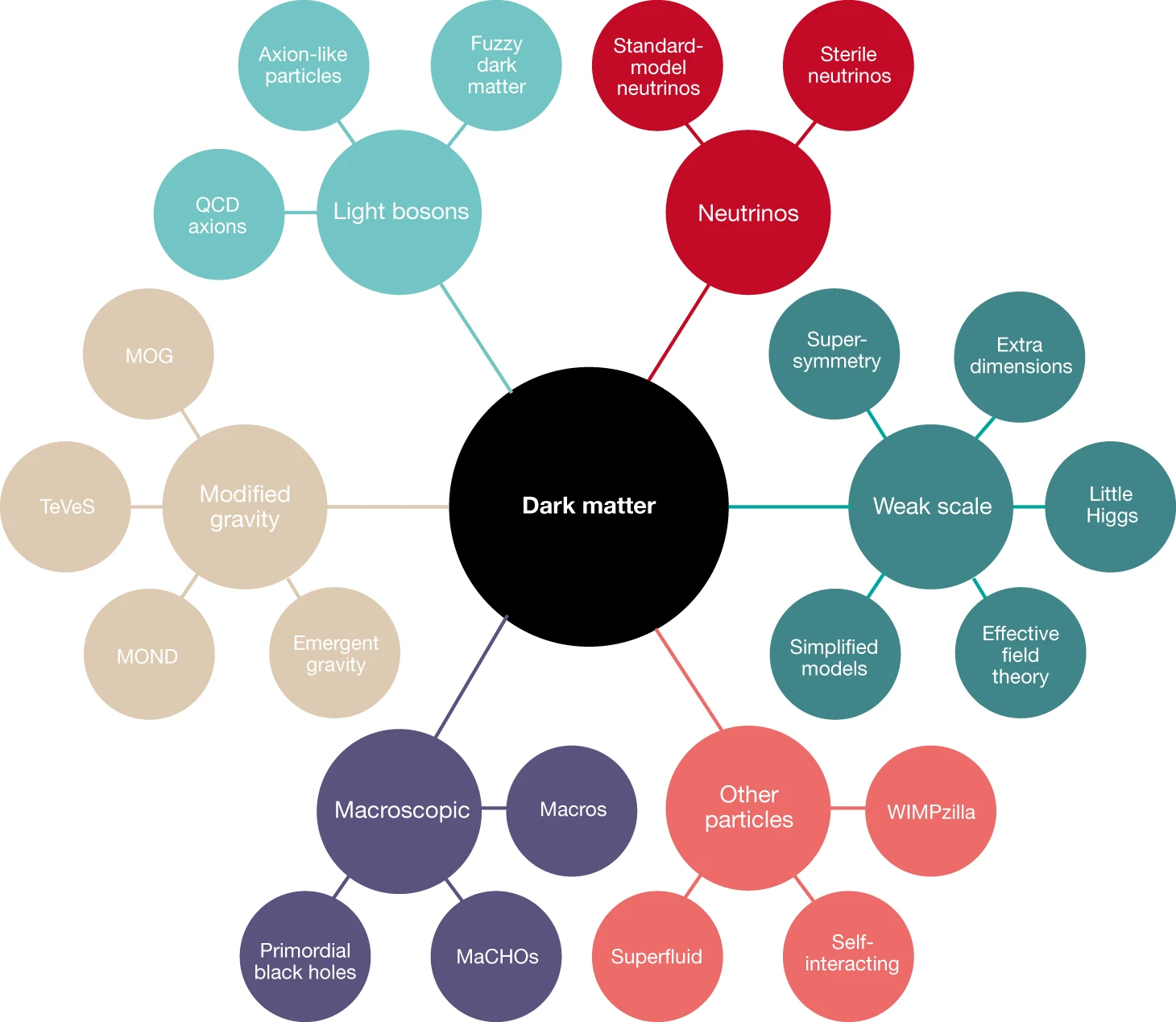}
    \caption{Tree of proposed solutions for dark matter problem. Beautiful image from \cite{bertone2018new}.}
    \label{fig:DarkMatterCandidates}
\end{figure}

``No stone left unturned" \cite{bertone2018new}.
There is a plethora of other possible explanations for the nature of dark matter, see figure \ref{fig:DarkMatterCandidates}, including fuzzy dark matter, gravitationally produced WIMPzillas, superfluid dark matter, macroscopic objects...
Therefore, searches are no longer constrained by what theoretical predictions dictate, instead experiments should explore all accessible avenues. A wider theoretical net offers the possibility of discovering new classes of dark matter candidates and opens up novel experimental opportunities to detect them.

\section{Hypothesis of dark matter as a new particle beyond the standard model}

Candidates for dark matter particles need to show properties compatible with astrophysical observations and production mechanisms that adjust to the current estimated abundances for dark matter content in the Universe.

\subsection{Dark matter properties}
Dark matter particles have to fulfill some conditions: 

\begin{itemize}
    \item Electric charge
    
    The ``darkness" of this matter comes from its non-radiative behaviour. Neither does it emit light nor does it interact electromagnetically with Standard Model particles. The most stringent limits come from the baryon acoustic peak structure due to the requirement of a dark matter completely decoupled from the baryon-photon plasma at recombination epoch. 

    \item Cold
    
    Structure formation in the Universe requires that dark matter particles be cold, meaning non-relativistic at the time of galaxy formation.

    \item Stable
    
    The dark matter lifetime must be long compared to cosmological timescales, otherwise it would have decayed by now.

    \item Mass
    
    Lower limits in the mass for dark matter particles come from quantum effects. They differ for fermionic or bosonic dark matter, as do the sources of these limits. For fermionic dark matter, these limits are $m_F > 70$ eV and for bosonic dark matter $m_B >10^{-22}$~eV~\cite{PDG2024review}. Upper limits come from stability against tidal disruption of structures immersed in dark matter halos, like galactic discs and globular clusters and they are around 5 $M_\odot$.

    In all cases, they must provide the correct amount of relic density to explain current observations.

    \item Weakly interacting
    
    They must have very weak interaction with other Standard Model particles, otherwise their effect could have been seen in collisions of clusters, for example.

    \item Non-baryonic
    
    Primordial nucleosynthesis and CMB data constrain the amounts of baryonic matter in the Universe, which is insufficient to explain present observations; therefore, it is deduced that dark matter consists of non-baryonic particles.
\end{itemize}

\subsection{Cosmological evolution}

Several explanations have been put forward to answer the question of where the dark matter comes from. 
In the early Universe, dark matter could have been generated via thermal or non-thermal production or both, or it may result from a particle-antiparticle asymmetry.

\subsubsection{Freeze out}
The main mechanism for particle production in the early Universe can be applied for dark matter production as well. During freeze-out, a particle species chemically decouples from the high-temperature, high-density thermal bath when the rate of its number-changing interactions drops below the Hubble expansion rate $H$.
The idea relies on the fact that when the temperature falls below a certain limit, the probability of the process that allowed to mix the dark matter candidate and the rest of species decays abruptly. There is not enough energy to continue mixing the dark matter with other particles and it decouples, fixing its abundance at that time, from which the present abundance can be estimated according to the expansion of the Universe. This calculation strongly depends on the assumed background cosmology.

\subsubsection{Freeze in}
In this case, dark matter was never in thermal equilibrium with the hot plasma of the early Universe. It is assumed to interact very weakly with Standard Model particles, with interaction rates much smaller than the Hubble expansion rate $H$. In this scenario, dark matter particles are produced via rare interactions or decays of Standard Model particles. This out-of-equilibrium production makes dark matter particles accumulate progressively over cosmic time, a process called freeze-in. As the Universe expands and cools, the production of dark matter effectively freezes in, leaving behind a relic abundance.

\subsubsection{Other mechanisms}
Other mechanisms are conceivable for dark matter production: cannibalization and other dark-sector number-changing processes, non-thermal production, asymmetric dark matter, Primordial Black Holes production. They are mechanisms postulated to account for dark matter production in different scenarios. Further information and references can be found in \cite{PDG2024review}.

\section{Dark matter searches}

Since the first claims of presence of an invisible matter component in the Universe, different strategies have been developed to pierce inside this unknown constituent of our reality.
Few properties are known for dark matter, so efforts have extended all across known technology, using every possible detector at hand to see possible effects of dark matter. Systematic searches have been performed in the last decades ranging from space-borne detectors to accelerators. It is common to classify dark matter searches in three groups: direct searches, indirect searches and laboratory production.

The direct searches group includes all experiments aiming for direct interaction between dark matter particles and baryonic matter. They attempt to detect nuclear recoils produced by the elastic scattering of dark matter particles in the target nuclei of the detector. Indirect searches are based on the detection of annihilation products of dark matter particles, such as gamma rays, neutrinos, antiprotons and positrons. Finally, laboratory production is based on particle colliders where events with missing energy may hint the creation of dark matter particles in high energy particle collisions. These experimental strategies are complementary and test different dark matter models.

\subsection{Laboratory production}

Production of dark matter in laboratories relies on particle accelerators. The CMS and ATLAS collaborations at the LHC, at CERN,  have set limits to dark matter production in proton-proton collisions. They have dedicated  experimental programs for dark matter searches that include searches for invisible particle production mediated by a Standard Model boson, generic searches for invisible particles produced via new particle mediators and specific searches for precise models.

A dark matter signal may have several characteristic features: the imbalance in the transverse momentum in an event due to the presence of dark matter particles, produced together with one Standard Model particle; a bump in the two-jet or two-lepton invariant mass distributions; or an excess of events in the two-jet angular distribution produced by a dark matter mediator. Signals with any of these features are searched for in favoured channels in which the signal-to-background ratio is high enough. No signal for dark matter has been observed in the LHC experiments so far \cite{universe4110131}. Fixed target experiments are also useful to prove some dark matter models, especially with sub-GeV particles \cite{PDG2024review}. Accelerator searches provide unique ways to test light dark matter models with lower dependence on the particle nature. Deeper searches and better characterization can be achieved in these experiments, very difficult to achieve in direct searches experiments, in which the dark matter particles have particular velocity distributions, cross sections, kinetic thresholds, etc...
No signal has been detected up to date but all colliders in the world still have research programs and dedicated detectors aiming for dark matter production \cite{PDG2024review}.

\subsection{Indirect searches}

Indirect searches refers to the detection of the annihilation or decay products from dark matter particles resulting in detectable particles, including especially gamma rays, neutrinos, and antimatter particles. Since the fluxes of annihilation and decay products are proportional to dark matter density -quadratically for annihilation and linearly for decay processes-, the largest number of particles is expected to come from astronomical objects. Following this reasoning, potential signatures are gamma rays from the Galactic Center, neutrinos from the Sun's or Earth's cores, and positrons and  antiprotons from the galactic halo \cite{baudis2016dark}.

\subsubsection{Gamma rays}

Almost any possible dark matter decay or annihilation may produce gamma rays. The rates may be very different for each process, but the sky is not so full of gamma sources: looking for excesses is one of the preferred indirect ways of searching for dark matter. Gamma rays propagate almost unperturbed from the source where they were produced. They are relatively easy to detect, but can be affected by absorption in the interstellar medium and known sources produce not negligible background. They are detected with high angular resolution by space-borne telescopes, like Fermi-LAT, and Cherenkov observatories on ground, like MAGIC, HESS or VERITAS, for higher energy gamma rays.
Searches for gamma ray emission from dark matter annihilation have focused on targets chosen primarily to maximize signal to noise ratios. Nearby dwarf spheroidal galaxies contain very small amounts of gas, and are dark matter dominated, and do not host any significant astrophysical background at gamma ray or X-ray frequencies, so they are the best objects for dark matter searches. Constraints from the non-observation of emission from these objects are dominated by systematics that may be reduced with observations from the future Cherenkov Telescope Array (CTA). Another promising source is the inner region of the Milky Way. While nearby and potentially hosting a large density of dark matter, the Galactic center region is very bright at almost any wavelength, making the extraction of a signal highly problematic.
Some intriguing excesses of gamma rays over the expected astrophysical flux have been reported by several experiments, for instance, the Fermi-LAT excess from the Galactic Center \cite{hooper2011dark}, but they remain unconfirmed to date.

\subsubsection{Neutrinos}
Like gamma rays, neutrinos travel across the Universe almost unaffected, so they point to the original source. Nevertheless, due to their extremely low cross section with ordinary matter, their detection is a real challenge, requiring large detectors and exposure times. Moreover, the angular resolution of neutrino detectors, around 1\% at energies near 100 GeV, is limited, making it challenging to accurately point back to sources in searches for relatively low-mass dark matter candidates. Despite that, neutrinos remain an attractive signature for dark matter searches not only because they can reach us from very long distances but also because there is not a significant neutrino background from astrophysical objects.

A very distinctive feature of neutrino signals originated in dark matter decays would be their increased flux coming from massive objects. Dark matter can be captured in celestial bodies in considerable amounts, depending on the scattering cross section off of nucleons, the dark matter mass, and the flux incident on the celestial body of interest. If enough dark matter accumulates, its annihilation inside the celestial body can then lead to the production of Standard Model particles. Such particles can heat up the body, if they lose most of their energy before escaping. This mechanism allows to establish constraints on dark matter models based on heat production in planets or stars. Alternatively, dark matter annihilation in celestial bodies can result in the production of particles that can escape the body, like neutrinos. Typical neutrino energies for these processes usually exceed the energy of neutrinos from the Sun, the best target for this type of searches, making this a virtually background-free dark matter search. Large neutrino telescopes like IceCube \cite{aartsen2018search}, ANTARES \cite{albert2020search} or Super-Kamiokande \cite{abe2020indirect} are able to search for dark matter annihilation into neutrinos, with no confirmed evidence yet.

\subsubsection{Cosmic ray charged particles}
Stable charged particles can also be produced by the annihilation or decay of dark matter. They interact strongly with the interstellar magnetic fields until they reach the Earth and can also lose energy during propagation through inverse Compton scattering or synchrotron radiation which makes it almost impossible to track them back to their source. Also energy spectra are affected. Nevertheless, it is possible to extract information from them and competitive limits for dark matter decays and annihilation have been set \cite{PDG2024review}.

To maximize signal-to-noise ratio, these searches focus on relatively rare species, such as positrons, antiprotons and anti-nuclei. These antiparticles would come from nearby sources such as the Galactic Center or the halo due to their high probability of annihilation with ordinary matter. Satellites like PAMELA and AMS-02, this one on board of the ISS, have been measuring the antimatter content of cosmic rays. Certain excess of positrons can be explained by dark matter annihilations, however, conventional explanations based on positron production by astrophysical sources like pulsars or supernova remnants are also possible.

\subsection{Direct searches}
Direct searches aim to detect direct interactions between dark matter and Standard Model particles. Since all experiments to date are located on Earth, they rely on the hypothesis of dark matter being present here, among us. Nothing points to a different direction, but our current evidence for dark matter comes from much bigger scales therefore it is possible that its distribution is not as uniform as we expect.

Direct detection experiments aim to observe dark matter particles interacting with ordinary matter through non-gravitational forces. Typically this means that they are scattering off nuclei in the active volume of the detector. The recoil of these nuclei leaves a trace that can be detected through different channels: ionization, light production or phonons. Every experiment is optimized according to the expected form of energy and the precise range of energies to which it is aiming. There are detectors sensitive to electron scattering as well as to absorption, although this will not be dealt here because our main interest is focused on nuclear recoils. 

These experiments are typically conducted in underground laboratories to shield them from cosmic rays. Specific additional shielding strategies are developed to prevent external background to reach the detector, in addition to a thorough screening of materials to construct a radiopure detector that avoids any internal contamination that can degrade the sensitivity of the experiment.

\subsubsection{Expected WIMP signal}\label{sec:WIMPsignal}
\begin{figure}[h]
    \centering
    \includegraphics[width=0.8\linewidth]{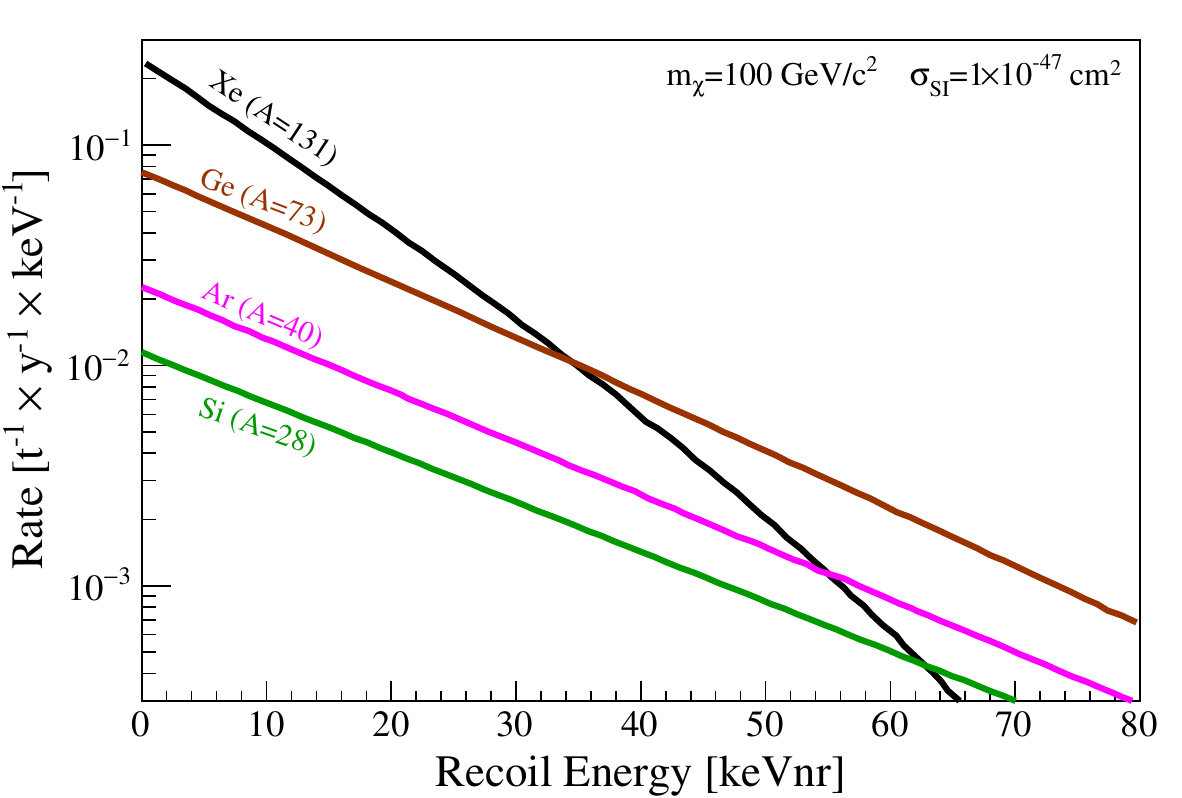}
    \caption{WIMP differential rates from nuclear recoils induced by a $m_\chi = 100$ GeV WIMP for several target materials considering $\sigma_{SI} = 10^{-47} \text{cm}^2$. Equation \ref{eq:DiffRate2} is used to obtain these spectra. Image from \cite{schumann2019direct}.}
    \label{fig:WIMPrates}
\end{figure}

The signal dark matter particles may leave in the detector depends on their number abundance, their energy, the type of interaction with atoms in the active volume of the detector and the design of the detector itself. Some of these possibilities have already been discussed: the interaction would involve scattering off nuclei, occurring with an extremely low probability due to its very small cross section. The effect of the design of the detector in maximising the sensitivity to certain models of dark matter particles will be made clear in chapter \ref{Ch:TREX}, when the TREX-DM experiment, object of this thesis, will be discussed. Here, several parameters of interest will be addressed, including the local distribution of dark matter and its associated dependencies. The outcomes of these discussions are rate distributions for target materials like the ones shown in figure \ref{fig:WIMPrates}.
\\

\begin{itemize}
    \item \textbf{Event rate}
\end{itemize}

The rate of interactions of the dark matter in the detector is proportional to the number of target nuclei in the active volume $N= M_{\text{det}}/m_N$ and the incoming flux of WIMPs $\phi$. The proportional constant is called cross section $\sigma_{\chi N}$. Therefore the rate is expressed as: 

\begin{equation}
    R = \sigma_{\chi N} N \phi 
\end{equation}

Usually, the quantity of interest when comparing the various experiments is the differential rate, given in number of counts per kilo, keV and day: 

\begin{equation}\label{eq:DiffRate1}
    \frac{dR}{dE_R} = \frac{d\sigma_{\chi N}}{dE_R} \frac{N}{M_{\text{det}}} \phi .
\end{equation}

The differential rate is given in terms of the recoil energy $E_R$ of the nucleus and the mass of the sensitive volume $M_{\text{det}}$. Immediately, one design consideration can be extracted from here: increasing the number of target nuclei increases the rate in any experiment. This means bigger detectors are favourable.

The incoming flux of dark matter particles is defined as the WIMP density times their mean velocity. Typically, the density of dark matter is given as energy density $\rho_0$ in units of $GeV/cm^3$, so dividing this by WIMP mass $m_\chi$ one gets the density of dark matter particles. The mean velocity is extracted from the velocity distribution $f(v)$; this depends on the model but the most widely used is the Standard Halo Model (SHM) that will be described below. Therefore, the flux of dark matter particles can be expressed as: 

\begin{equation}
    \phi  = \frac{\rho_0}{m_{\chi}} \int_{v_{\min}}^{v_{\max}} v f(v) dv
\end{equation}

Substituting these expressions in equation \ref{eq:DiffRate1} the differential rate per energy, time and mass is written:

\begin{equation}\label{eq:DiffRate2}
    \frac{dR}{dE_R} = \frac{\rho_0}{m_\chi} \frac{1}{m_N} \int_{v_{\min}}^{v_{\max}} v f(\vec{v}) \frac{d\sigma_{\chi N}}{dE_R} (\vec{v}, E_R) d\vec{v}
\end{equation}

The integration limits for the velocity distribution depend on the recoil energy $E_R$ due to the kinematics of the collision: 

\begin{equation}
    E_R = \frac{\mu^2 v^2}{m_N} \left( 1 - \cos(\theta) \right) \Rightarrow v_{\min} = \sqrt{\frac{E_{R} m_N}{2 \mu^2}}
\end{equation}

\noindent where $\theta$ is the scattering angle in the WIMP-nucleus center of mass frame, and $\mu = m_\chi \cdot m_N/(m_\chi+m_N)$ is the WIMP-nucleus reduced mass. The minimum velocity is determined by the recoil energy $E_{R}$. Incoming particles with smaller velocity will not be able to leave a recoil of certain  $E_R$ energy so the integral is bounded for each value. On the contrary, $v_{max}$ may be, in principle, infinite, but it is bounded by the escape velocity $v_{esc}$ of particles in the dark matter halo. 

To calculate the total rate of events in an experiment, the differential rate needs to be integrated over energy and mass. In this step, the detection efficiency can be included in the integration as sometimes it is energy dependent, for example when selection criteria are applied during the analysis.

\begin{equation}
    R = \int_{E_{\text{th}}}^{E_{\max}} \varepsilon(E_R) \frac{\rho_0}{m_\chi} \frac{M_{\text{det}}}{m_N} \int_{v_{\min}}^{v_{\max}} v f(\vec{v}) \frac{d\sigma_{\chi N}}{dE_R} (\vec{v}, E_R) d\vec{v} \, \,  dE_R
\end{equation}

Although the total rate may not be useful when comparing different experiments, for daily analysis tasks it is important: in practice what one measures is the rate of events in the detector. The total rate allows  to directly compare the expected signal with the measured experimental background, given usually in counts per day in the energy range of interest. 

\vspace{3ex}

\begin{itemize}
    \item \textbf{Cross section}
\end{itemize}
Several pieces in the differential rate formula \ref{eq:DiffRate2} need further explanation. Some of them need inputs from particle and nuclear physics, like the cross section factor, and others will profit from astrophysical data.

The WIMP-nucleus differential cross section can be generally separated into a spin-independent (SI) contribution and a spin-dependent (SD) one. These two contributions depend on the interaction model chosen for the dark matter particle -does the WIMP have spin?- and the relative weights may be very different. In most cases, the spin-independent cross section is used as a tool to compare sensitivities among experiments.

Expressions for both contributions are derived using Fermi's Golden Rule. The WIMP-nucleus differential cross section is divided into two terms: one independent of the momentum transfer $\sigma_0$ and the other containing the entire dependence on the momentum transfer in a function called \textit{form factor} $F(q)$, which arises from the finite size of the nucleus. 

\begin{equation}
    \frac{d\sigma_{\chi N}}{dE_R} = \left( \frac{d\sigma_{\chi N}}{dE_R} \right)_{SI} + \left( \frac{d\sigma_{\chi N}}{dE_R} \right)_{SD} = \frac{m_N}{2\mu^2 v^2} \left[ \sigma_0^{SI} F_{SI}^2 (E_R) + \sigma_0^{SD} F_{SD}^2 (E_R) \right]
\end{equation}
\\

\textbf{Spin independent}

The Spin independent WIMP-nucleus cross section is defined as a point-like interaction and can be expressed as:
\begin{equation}
    \sigma_{0}^{SI} = \frac{4\mu^2}{\pi} \left[ Z f_p + (A - Z) f_n \right]^2 ,
\end{equation}

\noindent with $Z$ the atomic number, $A$ the number of nucleons per nucleus of the precise isotope and $f_p$, $f_n$ the coupling to protons and neutrons of the WIMP particle. If $f_p=f_n$ is assumed, the expression is simplified to:

\begin{equation}
    \sigma_0^{SI}=\left(\frac{\mu}{\mu_{n}}\right)^2A^2\sigma_{SI}
\end{equation}

Here appears the reduced WIMP-nucleon mass $\mu_n = m_\chi \cdot m_n/(m_\chi+m_n)$, as well as the previously defined WIMP-nucleus reduced mass $\mu$; and $\sigma_{SI} = \dfrac{4\mu_n^2f_n^2}{\pi}$ the spin-independent WIMP-nucleon cross section.

\vspace{10ex}

\textbf{Spin dependent}

In this case, the point-like interaction can be written as.
\begin{equation}
    \sigma_0^{SD}=\frac{32\mu^2}{\pi}G_F^2\frac{J+1}{J}\left(a_p\left\langle S_p\right\rangle+a_n\left\langle S_n\right\rangle\right)^2
\end{equation}

Where $G_F$ is the Fermi coupling constant, $J$ is the total nuclear spin, $a_p$ and $a_n$ are the effective WIMP couplings to protons and neutrons, respectively, and $\left\langle S_{p,n}\right\rangle = \left\langle N | S_{p,n} | N\right\rangle$ are the expectation values of total proton and neutron spin operators in the limit of zero momentum transfer to the target nucleus, and it has to be estimated using detailed nuclear model calculations.
\\

\begin{itemize}
    \item \textbf{Form factor}
\end{itemize}

This function accounts for the loss of coherence in momentum transfer due to the fact that the nucleus is not point-like. For the spin-independent contribution, this function is approximated by the Fourier transform of the nuclear density, which can be obtained from experimental data. In the case of analytical calculations, different parametrizations are often used, one of the most common being the Helm form factor \cite{lewin1996review}.

\begin{equation}
    F_{SI}^2(q) = \left( \frac{3 j_1(qR_1)}{qR_1} \right)^2 e^{-q^2 s^2}
\end{equation}

With $j_1(x)$ being the first Bessel function $j_1(x) = \frac{\sin(x)}{x^2} - \frac{\cos(x)}{x}$. And the parameters  $s = 0.9$ fm  representing the thickness of the nuclear surface layer and $R_1 = \sqrt{(1.23 A^{1/3} - 0.6)^2 + \frac{7}{3} \pi^2 0.52^2 - 5s^2}$ being a parametrization of the effective nuclear radius in fm.  

This form factor is expressed in terms of the transferred momentum  $q$ , but it is used in terms of the deposited energy  $E_r$ . Since the WIMPs considered are non-relativistic, this variable change is straightforward, as  $E_r = \frac{q^2}{2m_N}$. An explicit expression in terms of $E_R$ and plots for different target elements can be found in \cite{DiezIbanez2019deteccion}. 
\\

\begin{itemize}
    \item \textbf{Velocity distribution}
\end{itemize}

The exact distribution and structure of dark matter halo has not yet been measured but it can be modelled taking into account the gravity-based observational evidence, mainly due to the observation of the galactic rotational curves.
With considerations extracted from this observational evidence, standard halo model (SHM) has been assumed as the base model to interpret dark matter searches. It predicts an isothermal, spherical and isotropic halo, with density profile $\rho(r) \propto r^{-2}$ and Maxwellian velocity distribution:

\begin{equation}
    f(\nu)=\frac{1}{\sigma_{\nu}\sqrt{2\pi}}\cdot e^{-\frac{\nu^{2}}{2\sigma_{\nu}^{2}}}
\end{equation}

where the dispersion in the velocities is related with the local circular velocity by $\sigma_v = \sqrt{3/2} v_c$, where in the SHM $v_c = (220 \pm 20)$ km/s, being $\sigma_v \sim 270$ km/s. The WIMP distribution is truncated at the local galactic escape velocity, $v_{esc} \sim 500-650$ km/s. The velocity of the Earth with respect to the local system, ignoring the motion of the Earth around the Sun, is around 12 km/s, so our velocity with respect to the halo is $v_r = v_c + 12$ km/s $\simeq 232$ km/s. The local dark matter density $\rho_0$ can be estimated from the measured rotation curve of the Milky Way, resulting in $\rho_0 =~0.3$ $\text{GeV/cm}^3$, the common value adopted by direct detection experiments to facilitate the comparison of their results.
\\

\subsubsection{Detector technologies}
Nuclear recoils are the main interaction channel with dark matter particles for most direct detection experiments. The energy transferred to the nucleus can be observed through three different signals: by measuring the phonon excitations produced by the conversion of the kinetic energy of the scattering particles to lattice vibrations in solid targets (heat), by measuring the scintillation photons released after de-excitation of the target nuclei in scintillating materials (light) and by direct measurement of the ionization produced in the target atoms (charge).

\begin{figure}[h]
    \centering
    \includegraphics[width=0.8\linewidth]{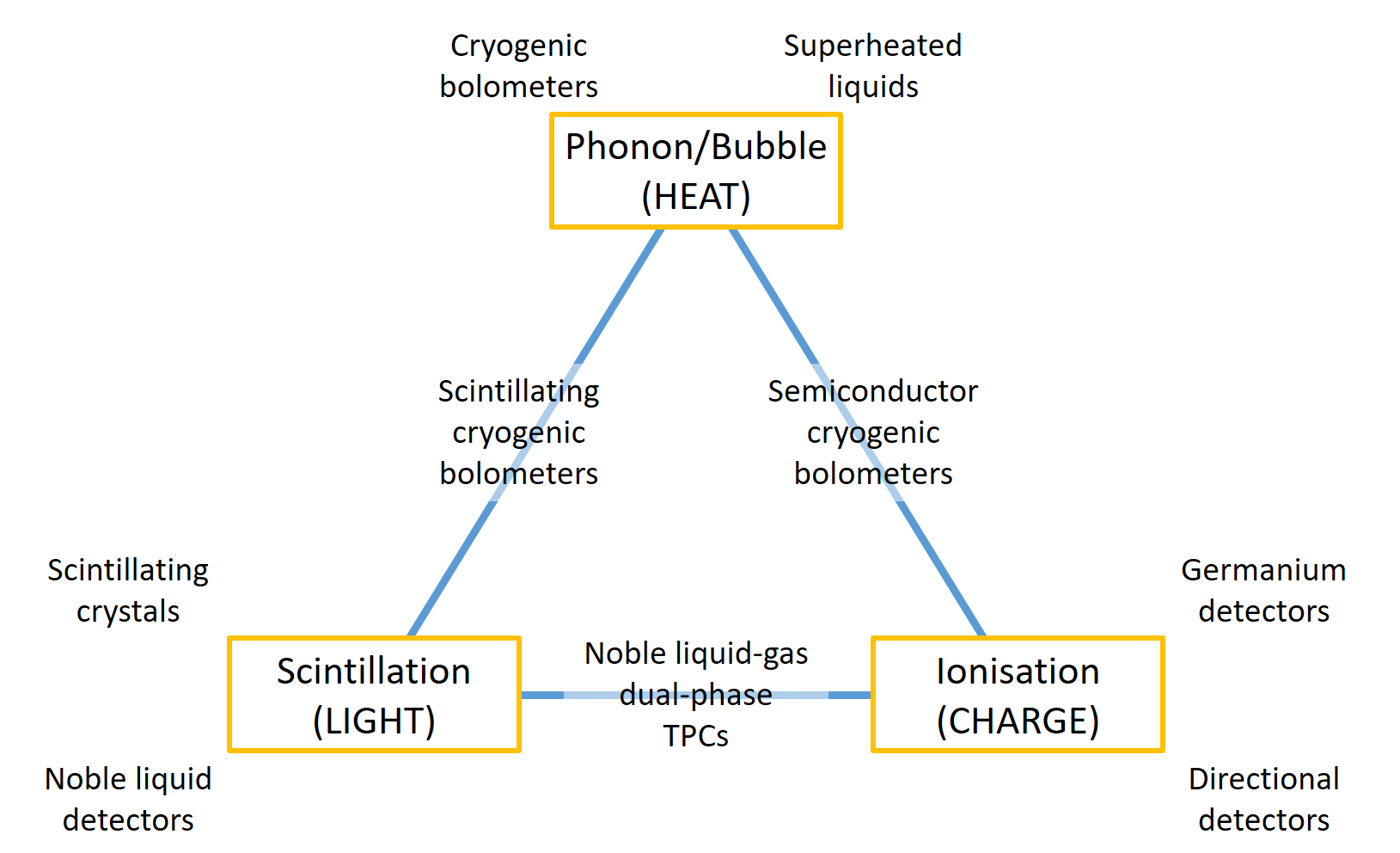}
    \caption{Most common detection techniques attending to the signal channel they exploit. Diagram from \cite{Coarasa2021anais}.}
    \label{fig:DetectionTechniques}
\end{figure}

Some experiments make use of two of the signals to achieve better event rejection rates (see figure \ref{fig:DetectionTechniques}). This strategy allows to distinguish between electronic and nuclear recoils for example. The cost of this extra information for each event is a more complex detector design able to read two very different signals. Up to date, no experiment is making use of all three channels due to this complexity.

Each of these signal channels has a leading readout technology associated, however, for double signal detectors some concessions are needed. Bolometers are used for phonons, scintillators for photons and gaseous detectors for charges. They are not the only possible technologies available, for example, superheated liquids and semiconductor detectors are also used in certain experiments.

\begin{itemize}
    \item \textbf{Room temperature ionization detectors}
\end{itemize}

Silicon charged-coupled devices (CCDs) are employed for low-mass dark matter searches. The charge generated by the interaction is drifted towards the pixel gates, until it reaches the readout. The 3D position of any interaction can be reconstructed due to the correlation between the interaction depth and the transversal charge diffusion, which allows to distinguish the particle type (electron,neutron, muon,alpha particles, etc) based on the recorded track pattern. Ionisation events are observed with charge resolutions around 1-2 $e^-$ and extremely low leakage currents. These type of detectors are very well suited for low-mass WIMP searches and they are especially competitive proving the dark matter-electron interaction.

The DAMIC (DArk Matter In CCDs) experiment at SNOLAB is the main example of this type of detectors \cite{aguilar2020results}. It consists on 7 CCDs with a total mass of 40~g, and it has been operating at SNOLAB since 2017, reaching a leakage current of 2~$e^-/\text{mm}^2/\text{day}$ and 50 eV energy threshold.  Skipper CCDs are the last development for these detectors, increasing the sensitivity by reducing its threshold. DAMIC-M (DAMIC at Modane) aims for a kg-size target mass using these skipper CCDs. First prototypes have been measuring recently and achieved constraints on sub-GeV dark matter particles interacting with electrons in the mass range (0.53-1000) MeV \cite{arnquist2023damic}.

Another collaboration using similar technology is SENSEI. It employs the skipper technology as well to achieve single-electron sensitivity. A run with a 2 g detector in the MINOS cavern at Fermilab yielded constraints on dark matter-electron scattering for a large range of sub-GeV dark matter masses.

\begin{figure}[h]
    \centering
    \includegraphics[width=0.8\linewidth]{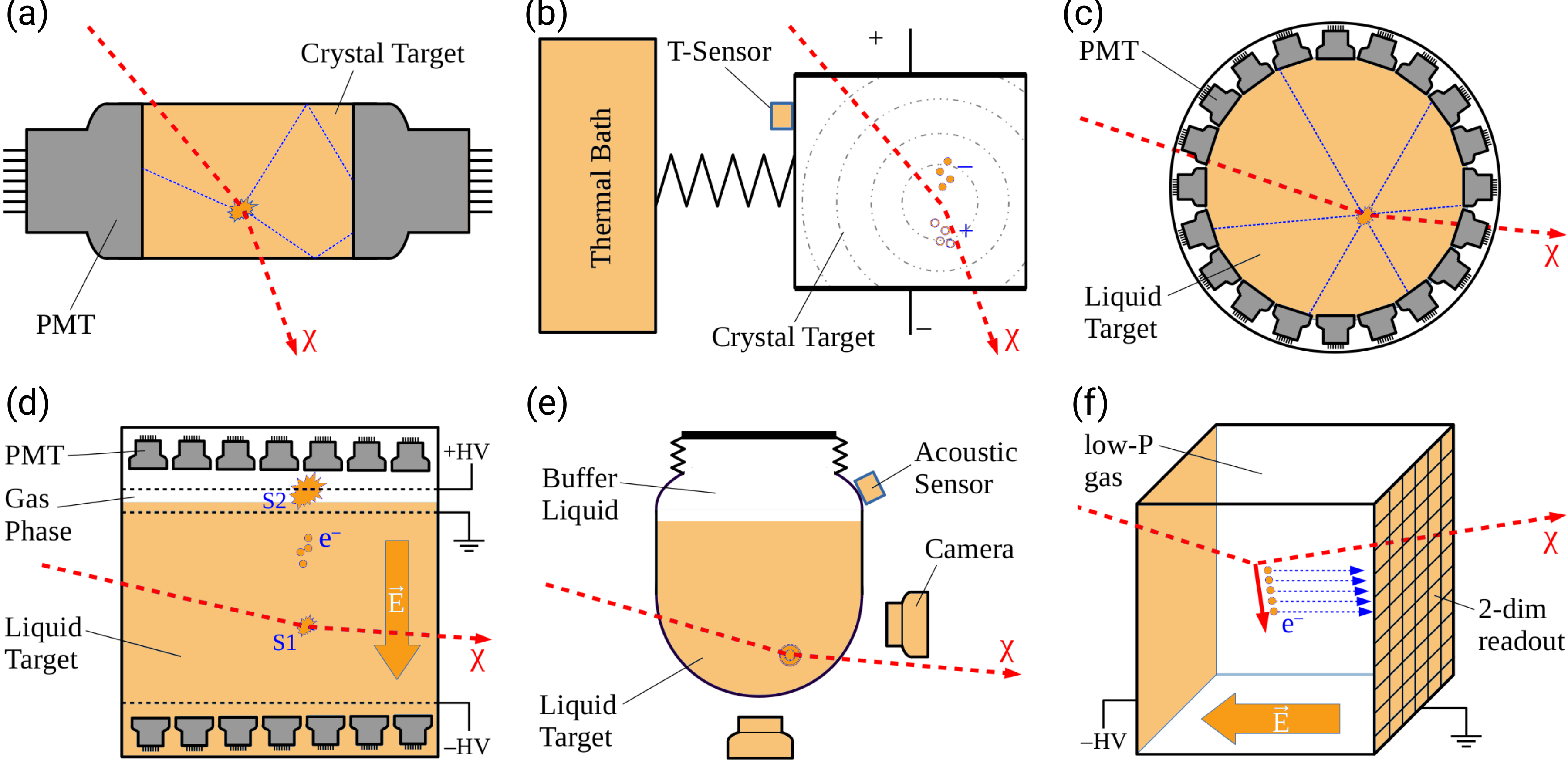}
    \caption{Working principle of common detector types for the direct WIMP search: (a) scintillating crystal, (b) bolometer, here with additional charge-readout, (c) single-phase and (d) dual-phase liquid noble gas detectors, (e) bubble chamber, (e) directional detector. From \cite{billard2022direct}, and there inherited from \cite{schumann2019direct}. }
    \label{fig:detectors}
\end{figure}

\begin{itemize}
    \item \textbf{Room temperature scintillators}
\end{itemize}

Several large dark matter experiments are using high-purity NaI(Tl) crystals placed in underground laboratories. When a particle interacts within these crystals, atoms in the lattice get excited and the subsequent de-excitation releases scintillation light, which can be easily collected by photomultipliers (PMTs). The simplicity of this technology allows the detectors to operate stably for long periods of time and accumulate large amounts of exposure. These detectors cannot distinguish nuclear recoils from electronic recoils at low energy, so achieving an ultra-low background is needed. 

Experiments using scintillation crystals make use of the expected annual modulation of the WIMP signal due to the inherit Earth motion around the Sun. This modulated signal over a much larger constant background is the leading analysis strategy for this type of detectors. Using ultrapure NaI(Tl) scintillators, the DAMA/LIBRA experiment at the LNGS in Italy has reported a long-standing positive result compatible with the dark matter annual modulation expected from the Standard Halo Model \cite{bernabei2023annual}. This result has been controversial since the WIMP mass and cross section derived from it have been widely excluded by many other experiments, although using different technologies. For this reason several initiatives appeared to confirm or refute the DAMA/LIBRA result in a model independent way using the same detection technology. The ANAIS experiment at the Laboratorio Subterráneo de Canfranc operates 112.5 kg of NaI(Tl) scintillators with an energy threshold of 1 keV and a background rate of 3.6 events/(kg d keV) in the (1-6) keV region. A blind analysis of 6 years of data is incompatible with the DAMA/LIBRA modulation signal at a $4\sigma $ confidence level \cite{amare2025towards}. The COSINE-100 experiment, located at the Yangyang Underground Laboratory in Korea, operates 106 kg of NaI(Tl) crystals in a liquid scintillator, with an energy threshold of 1 keV and a background rate of 3.6 events/(kg d keV). Results from 3 years data in the (1-6) keV energy range are consistent, at 68.3\% confidence level, with both the null hypothesis and DAMA/LIBRA’s best fit value in the same energy range \cite{adhikari2022three}. Other initiatives and future upgrades are foreseen: COSINE-100U, SABRE, COSINUS and PICOLON.

\begin{itemize}
    \item \textbf{Solid-state cryogenic detectors}
\end{itemize}

Refrigerated ionization detectors and bolometers are highly sensitive to low-mass WIMPs. The CDEX-10 experiment at the China Jinping Underground Laboratory (CJPL) uses p-type, point-contact germanium ionisation detectors operated in liquid nitrogen to probe dark matter masses down to 3 GeV \cite{jiang2018limits}. These detectors operated at 77 K can reach sub-keV energy thresholds and very low backgrounds, but lack the ability to distinguish electronic from nuclear recoils.

Bolometers are ideal to read two signals, the thermal signal and light or charge, as presented in figure \ref{fig:detectors}. These detectors must operate at extremely low temperatures, typically tens of mK  in order to minimize thermal fluctuations, allowing to measure temperature increases of the order of $\upmu$K. Leading experiments of this type are SuperCDMS at Soudan and SNOLAB, EDELWEISS at the Laboratoire Souterrain de Modane (LSM) and CRESST at the Laboratori Nazionali del Gran Sasso (LNGS). 
\\

    \textbf{Semiconductor cryogenic detectors}
    
They measure simultaneously ionisation and heat signals, making the discrimination between nuclear and electronic recoils possible. EDELWEISS operated cryogenic germanium detectors with NTD (Neutron Transmutation Doped) germanium sensors for heat signal and aluminium electrodes for charge collection \cite{arnaud2020first}. SuperCDMS (Cryogenic Dark Matter Search) operates silicium and germanium bolometers \cite{agnese2018results}. It is optimized for the range of 1.5 to 250 GeV mass WIMPs but the spin-off experiment  CDMSlite has achieved exclusion limits for masses down to 30 MeV.
\\

    \textbf{Scintillating cryogenic detectors}
    
Light from scintillation crystals and heat is measured in this type of experiments. A transparent bolometer is needed for the wavelength of the scintillation photons and a double readout system that allows to distinguish nuclear from electronic recoils should be implemented. The CRESST (Cryogenic Rare Event Search with Superconducting Thermometers) experiment has operated bolometers made of different materials (CaWO4, Al2O3, Si, LiAlO2) and has reached competitive limits for spin independent WIMP-nucleon interaction up to $m_\chi = 160$  MeV \cite{abdelhameed2019first}.

\begin{itemize}
    \item \textbf{Noble element detectors}
\end{itemize}

Noble elements like argon and xenon can be used as target nuclei. Some of the most competitive experiments use dual-phase detectors combining liquefied gases with a small portion of gas volume. They record the signals, scintillation from the liquid and ionization from the gas phase. Also single phase detectors are in operation, both liquid and gas. In figure \ref{fig:detectors} the schematics of these detectors can be seen.
\\

    \textbf{Single-phase detectors}
    
The NEWS-G collaboration operates spherical proportional counters filled with a noble gas. The advantages of this technology are the low intrinsic electronic noise and a high amplification gain, allowing for low energy thresholds down to single-electron detection, and the possibility to use different light targets (He, Ne, etc) \cite{balogh2023news}. A 60 cm diameter chamber operated at LSM with a gas mixture of Ne + CH4(0.7\%) at 3.1 bar, excluded spin independent WIMP-nucleon cross sections at 0.5 GeV with an energy threshold 100 eV \cite{arnaud2018first}. TREX-DM uses the same principle of detection as well.

Liquid detectors measure scintillation signal form the target volume with a $4\pi$ covering of PMTs. The DEAP-3600 (Dark matter Experiment using Argon Pulse-shape discrimination) experiment at SNOLAB is a single-phase liquid argon detector with a total mass of 3.3 tons of argon \cite{ajaj2019search}. In these experiments background discrimination is handled by pulse shape analysis that allows to distinguish between electronic and nuclear recoils.
\\

    \textbf{Dual-phase detectors}

Dual-phase -liquid and gas- TPCs  detect simultaneously the scintillation and ionisation signals generated by an energy deposition. After the particle interaction, the primary scintillation signal, called S1, is recorded by two arrays of PMTs at the top and bottom of the detector. Ionisation electrons released in the interaction are drifted upwards to the gas phase where they generate a secondary light signal, S2, proportional to the charge. The ratio between charge and light (S2/S1) can be used to distinguish electronic from nuclear recoils, and 3D event reconstruction is possible with spatial resolution at the order of mm. These detectors lead the efforts to constraint the parameter space for a wide range of WIMP masses. 

The XENON1T experiment at the LNGS has the best sensitivity to spin independent WIMP-nucleon interaction for masses above 6 GeV using a dual-phase LXe TPC with an active mass of 2 tons of xenon \cite{aprile2017xenon1t}. Close to this, the PandaX-II (Particle and Astrophysical Xenon Experiments) experiment, operating at CJPL using a dual-phase LXe TPC with an active mass of 580 kg of xenon \cite{cui2017dark}, and the already finished LUX experiment in the Sanford Underground Research Facility (SURF) in USA \cite{akerib2017results}, are the most sensitive liquid xenon detectors up to date. 

The DarkSide-50 at LNGS, is the best dual-phase liquid argon TPC, established the best limit in the WIMP mass range from 1.8 to 3.5 GeV \cite{agnes2018low}. It uses 46 kg of argon depleted in the radioactive isotope $^{39}$Ar, produced by cosmogenic activation. DarkSide-20k plans to produce 20 tonnes of this radiopure argon extracting it from an old mine facility in Colorado, USA \cite{aalseth2018darkside}.

\begin{itemize}
    \item \textbf{Superheated liquid detectors}
\end{itemize}

Bubble chambers are an old technology in particle physics, they were used in the experimental discovery of weak currents, for example. Nowadays, they are mainly used for dark matter searches. These bubble chambers consist of a metastable superheated liquid kept at a temperature just above their boiling point so that when an energy deposit occurs, above some threshold, a bubble is created. Figure \ref{fig:detectors} shows this detection principle. Bubbles can be counted and localized by cameras, allowing mm resolution in the original interaction. These detectors can be tuned in order that only nuclear recoil events will create bubbles. Also, different target fluids with different composition can be operated inside. This is of great interest for the spin-dependent interaction channel because it needs a target nucleus with uneven total angular momentum. A particularly favourable candidate is $^{19}F$, where the spin is carried mostly by the unpaired proton, yielding a cross-section which is almost ten times higher than of other employed nuclei with spin \cite{PDG2024review}. The PICO collaboration is the main project working with bubble chambers. Several of them have been operating at SNOLAB, like PICO-60 \cite{amole2019dark} or the recent PICO-500 which is in construction, yielding to stringent constraints on the dark matter-proton spin-dependent interaction.

\begin{itemize}
    \item \textbf{Directional detectors}
\end{itemize}

These detectors aim to detect the incoming direction of dark matter particles. This would unequivocally confirm the galactic origin of a signal and could probe the region below the neutrino floor. These two advantages, though, face technological challenges unsolved at the moment. The nuclear recoil track of WIMP-induced interactions is below 100 nm for energies less than 200 keV in liquid and solid materials.
The solution proposed to this is the use of low pressure gases, at around 0.1~atm, where tracks can be resolved, as presented in figure \ref{fig:detectors}. The first to try this technique was the DRIFT (Directional Recoil Identification From Tracks) experiment \cite{battat2017low}. DRIFT operated at the Boulby Underground Laboratory in UK over more than a decade using 0.140~kg of CS$_2$+CF$_4$+O$_2$ mixture at a pressure of 55~mbar. Several directional detectors are presently in operation, with volumes around 1~m$^3$ and masses of $\sim 100$ g, that can measure the sense of an incoming nuclear recoil above a few tens of keV. Most of them are gathered in the proto-collaboration Cygnus \cite{vahsen2020cygnus}, which coordinates the R\&D efforts for gas based TPCs with 1~keV threshold.

\begin{figure}[h!]
    \centering
    \includegraphics[width=0.9\linewidth]{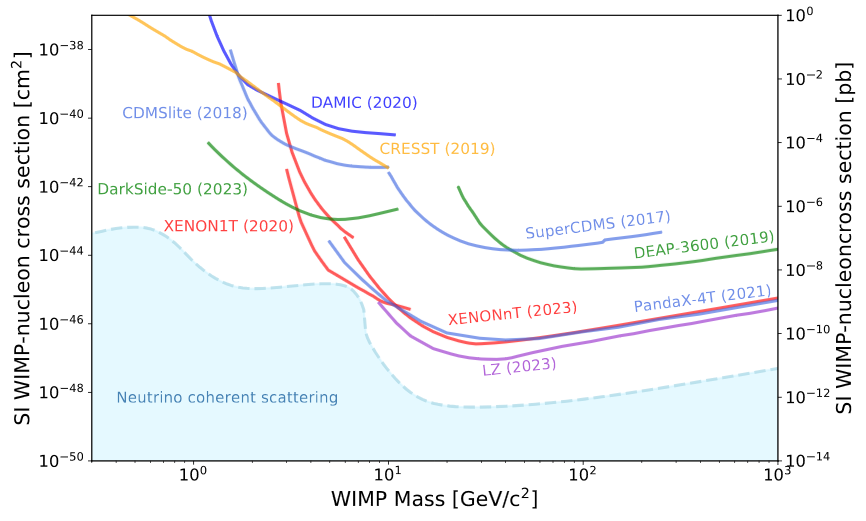}
    \caption{Current exclusion limits for WIMP searches in the spin-independent parameter space ($m_\chi$ and $\sigma_{SI}$). Elastic WIMP–nucleus scattering computed assuming the standard parameters for an isothermal WIMP halo: $\rho_0 = 0.3$ GeV cm$^{-3}$ , $v_0 = 220$km s$^{-1}$ , $v_{esc}$ = 544 km s$^{-1}$. Plot from \cite{PDG2024review}.}
    \label{fig:ExclusionLimitsWIMP}
\end{figure}

\begin{comment}
    \begin{figure}[h!]
    \centering
    \includegraphics[width=0.9\linewidth]{images/ExclusionLimitsWIMP2.png}
    \caption{Current exclusion limits for WIMP searches in the spin-independent parameter space ($m_\chi$ and $\sigma_{SI}$). Elastic WIMP–nucleus scattering computed assuming the standard parameters for an isothermal WIMP halo: $\rho_0 = 0.3$ GeV cm$^{-3}$ , $v_0 = 220$km s$^{-1}$ , $v_{esc}$ = 544 km s$^{-1}$. Results labelled ‘M’ were obtained assuming the Migdal effect. Results labelled ‘Surf’ are from experiments not operated underground. Plot from \cite{billard2022direct}.}
    \label{fig:ExclusionLimitsWIMP}
\end{figure}
\end{comment}

A new technique, based on fine-grained nuclear emulsions, has been proposed for directional searches \cite{alexandrov2021directionality}. It makes use of solid-state detectors with silver halide crystals uniformly dispersed in a gelatine film, where each crystal works as a sensor for charged particles. They are able to trace tracks smaller than 1 $\upmu$m in size, achieving a superior spatial resolution compared to gaseous detectors. Current emulsions allow for 100 nm tracking and target masses are around 1 kg, which is still bellow the requirements for setting competitive limits.

\subsection{Present status}
Many experiments have been working hard to unravel the mysteries of dark matter for decades. To date, no conclusive dark matter signal has been detected (despite the long-standing claims of DAMA/LIBRA) but this does not diminish the value of these experimental efforts. Dark matter masses and interaction cross sections have been explored across many orders of magnitude, and experimental technology continues to be refined to extend the sensitivity reach and probe increasingly deeper into the WIMP parameter space. The results of past and ongoing efforts are gathered in exclusion plots like \ref{fig:ExclusionLimitsWIMP}. The Standard Halo Model is used to compute the sensitivities of all experiments, allowing to compare different technologies and results. 

Exclusion limits are computed when no signal is present in the results, so when the detection rate is compatible with the only background hypothesis. A likelihood test is performed to address the maximum intensity of the dark matter signal that can be hidden in the fluctuations of these backgrounds. For every possible dark matter mass, the smallest cross-section that would have been discovered is computed, so higher values are ruled out. In plots such as Figure \ref{fig:ExclusionLimitsWIMP}, the white regions represent areas of WIMP parameter space that remain untested. Masses above 10 GeV are quite constrained and experiments are about to reach the hypothesized neutrino floor, a background for dark matter searches never measured yet, while below 10 GeV there remains significant unexplored parameter space, and many experiments are increasingly focusing their efforts in this region. One of them is TREX-DM, the main focus of this work, a low-mass WIMP high-pressure gaseous detector with Micromegas readout planes. It will be described in detail in chapter \ref{Ch:TREX}. Before, in chapter \ref{Ch:GasDect} particle interactions with matter in gaseous detectors and the behaviour of Micromegas readout planes will be presented.

%% file: Chapters/2_GaseousDetectors.tex
Dark matter searches draw on over a century of technological knowledge from experimental particle physics. Since the development of the Geiger-Müller counter at the dawn of the 20th century, gaseous detectors have proved to be one of the most useful technologies, helping in the discovery of many new particles and present nowadays in some of the biggest experiments like CMS or ATLAS at the LHC in CERN or AMS-02 on board of the International Space Station. These detectors offer remarkable versatility, they can be finely tuned in terms of pressure, gas composition, geometry, and other parameters, making them highly adaptable to specific experimental goals. The technology has reached a high level of maturity and now extends beyond fundamental research into industrial and applied domains like medical physics, archaeology material science, geology... For dark matter searches, gaseous detectors have reached the ton-scale in dual-phase detectors (liquid-gas). These experiments, the most sensitive in the field, are not the only ones. Directional detectors, optimized to infer the incoming direction of dark matter particles, or low-mass WIMP detectors like TREX-DM, one of the main interests of this work, are other examples of gaseous detectors for dark matter searches. 

This chapter discusses the physics of particle interactions in gaseous media, providing the necessary background to understand the processes of ionization and electron collection, which form the basis of gaseous detector operation. The focus is placed on one of the most recent advancements in this field: Micromegas readout planes. This technology represents a core area of expertise for the Grupo de Investigación en Física Nuclear y Astropartículas (GIFNA) of the Universidad de Zaragoza, where I have developed my work. Its specific implementation in the TREX-DM experiment will be detailed in chapter \ref{Ch:TREX}.

\section{Particle interactions in gaseous detectors}

Particle detectors adopt many different forms depending on their purpose. Specialized devices for a determined type of particle, for high detection efficiency, wide range of energies, large volume, pixelized detectors, ultra fast timing setups... Many different possibilities arise when the challenge is there; and no technology is well suited for everything. 

Gaseous detectors are devices that use a gaseous medium as sensitive volume. Different pressures and gas mixtures allow to adapt these detectors for a wide range of the above applications. The energy deposited in the gas can be collected in different ways. The working principle of our gaseous detectors is based on ionization, but there also exist gaseous scintillators where energy is collected as photons. 

\begin{figure}[h]
\centering
\includegraphics[width=0.80\textwidth]{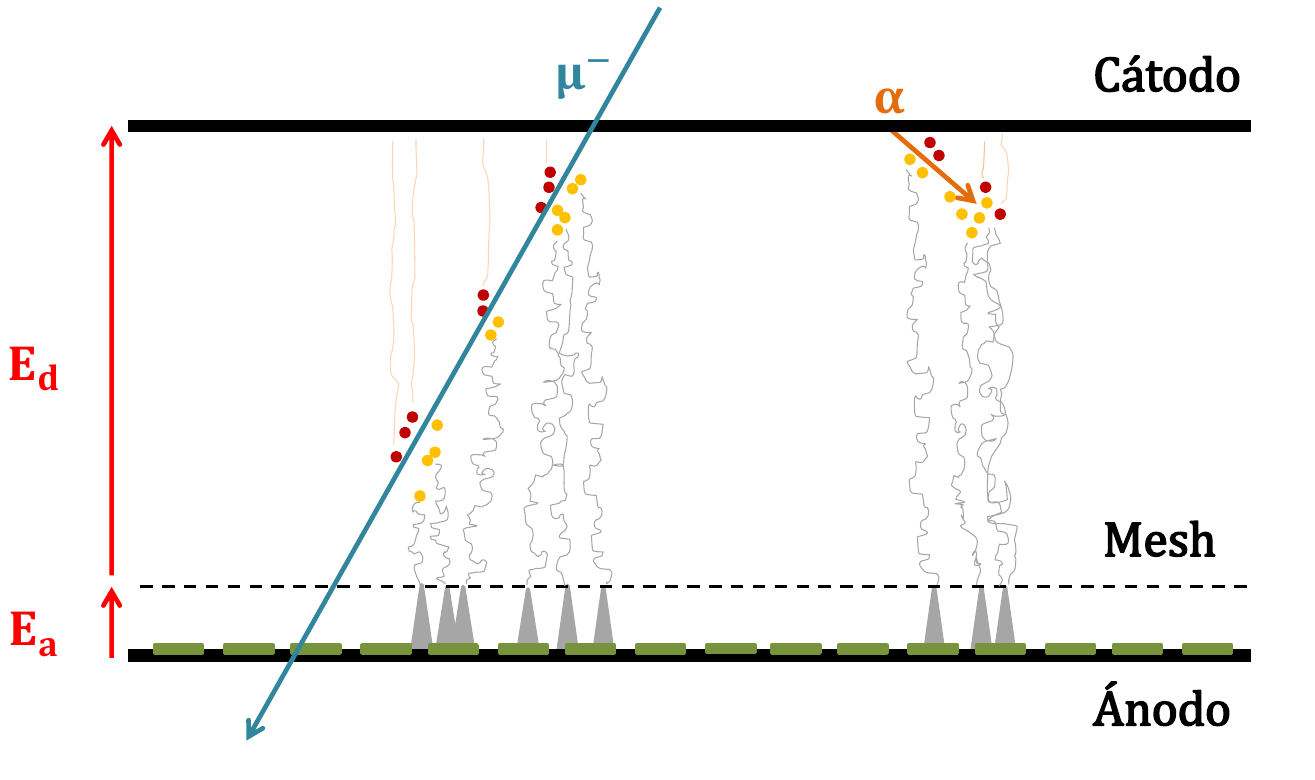}
\caption{Schematics of a Time Projection Chamber (TPC) with a Micromegas readout. Ions, in red, drift towards the cathode (Cátodo) and electrons, in yellow, towards the anode (Ánodo). The cathode  has the highest voltage in the system and the anode is usually grounded. The avalanche region is defined by the mesh, a grid close to the anode with an intermediate voltage that creates an intense electric field ($E_a$) to amplify the electron signal. 
}\label{fig:TPC}
\end{figure}

In an ionizing gaseous detector, like the Time Projection Chamber (TPC) shown schematically in figure \ref{fig:TPC},  when a particle interacts with the atoms of the gas, some electrons may be extracted. In an non-instrumented volume, these electrons quickly recombine with the atoms from which they were originally extracted. But when an electric field is applied this is prevented, electrons and ions drift in opposite directions and they are collected in the electrodes. If the signal induced in the electrodes by the moving charges is read, one can detect the passing of an ionizing~particle.

The primary ionization, formed by the electrons extracted directly by the incoming particle, is very different depending on the nature of the particle and its energy. In general, higher-energy particles deposit more energy, however, this depends significantly on the particle type. The primary distinction is between charged and neutral particles. Charged particles, such as muons, can ionize atoms along their path if they carry sufficient energy. While neutral particles interact very differently, scattering through the atoms and depositing energy only when they interact, like photons and neutrons. 

In this section interaction methods for photons, electrons, alpha particles and other heavy charged particles, as well as muons are described. A dedicated subsection is devoted to neutrons as they are neutral particles that interact as dark matter is expected to interact. More extensive reviews can be found in \cite{gracia2016micromegas}, \cite{ruiz2019ultra} and \cite{PDG2024review}.

\subsection{Photons}\label{sec:photons}

Photons are neutral particles without mass. They can travel through matter meaningful distances without interacting. And when they do, they transfer totally or partially their energy to the medium in a point-like interaction. The parameter that measures how far a photon may travel through a certain material is called attenuation length. It reflects the amount of photons of certain energy that remains in a parallel beam without interactions after a certain length. The attenuation coefficient is denoted by $\mu$ and it can be extracted from measurements of the intensity of a photon beam traversing certain thickness $x$ of material, being $I_0$ the incident intensity of the beam: $$ I(x) = I_0 e^{-\mu x}$$
This parameter $\mu$ sets the distance in which the intensity of the beam is reduced by a factor $e$: the mean free path $\lambda = 1/\mu$. Photons that do not cross the thickness have interacted with the material. Depending on the energy of the incident photon and the material, different interaction mechanisms are possible. 

\begin{comment}
    
\begin{figure}
    \centering
    \includegraphics[width=0.8\linewidth]{images/MeanFreePaths.png}
    \caption{Mean free paths at different pressures in argon, in the left. And for different noble gases in the right.}
    \label{fig:MeanFreePath}
\end{figure}

\end{comment}

\begin{figure}[h]
    \centering
    \includegraphics[width=0.8\linewidth]{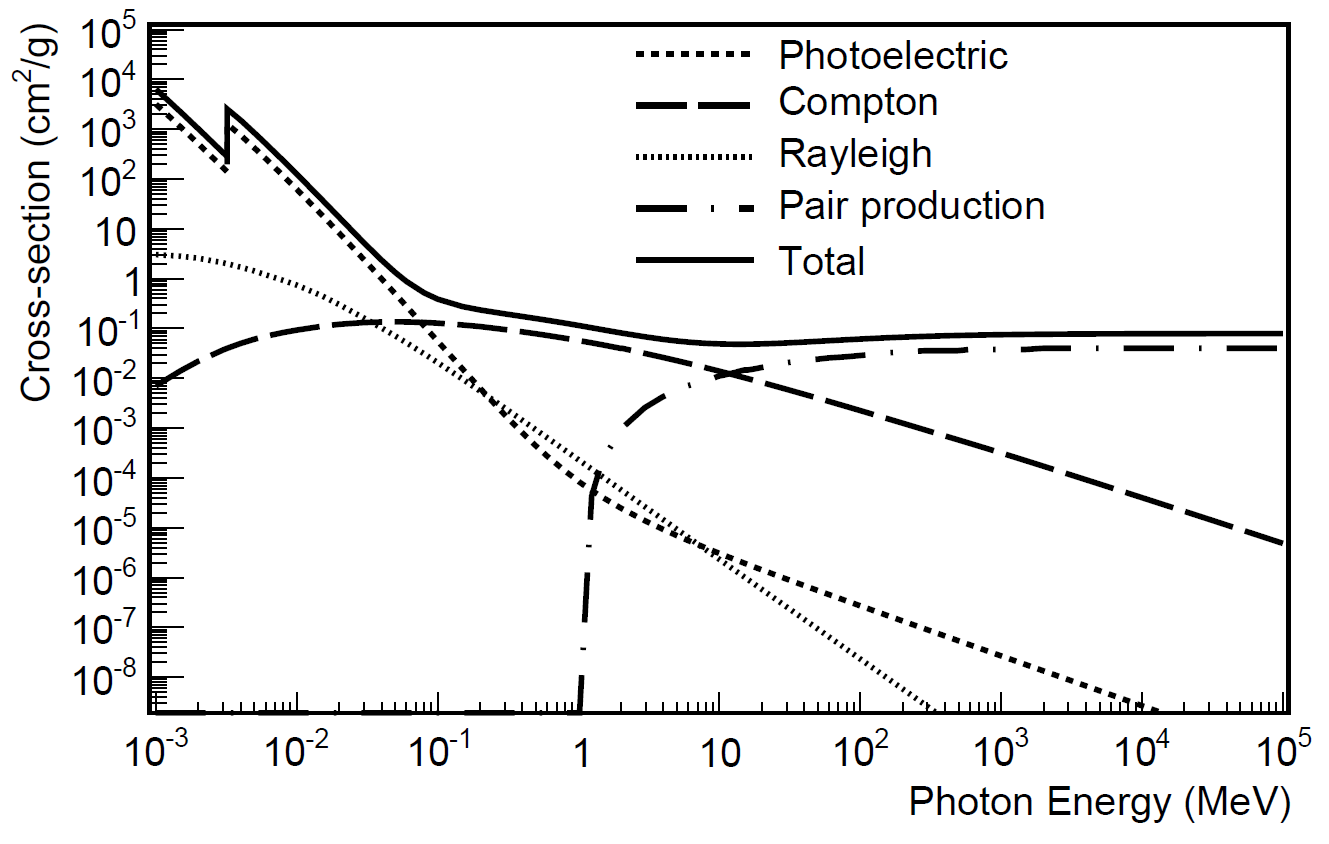}
    \caption{Interaction cross section of photons in argon, decomposed into contributions from the relevant processes. Extracted from \cite{garcia2015solar}. Data available in XCOM \cite{berger2010xcom}.}
    \label{fig:InteractionsCrossSections}
\end{figure}

Photons can deposit energy when they interact with matter in three ways: Photoelectric effect \index{photoelectric effect}, Compton scattering \index{Compton scattering} and pair production \index{pair production}. Other types of scattering like Thomson or Rayleigh scatterings, are not taken into account here. 

\begin{comment}
    Thomson scattering is an elastic process that happens for energies lower than X-rays and the amount of energy transferred to electrons in the atom is negligible. Rayleigh scattering \index{Rayleigh scattering}, that also appears in figure \ref{fig:InteractionsCrossSections}, is a non-ionizing photon interaction in which light is scattered in a coherent way (it gets polarized) by the atom as a whole. It appears when the wavelength of the incoming photon is bigger than the atomic radius. But in this process no energy is transferred to the atom, the out-coming photon has the same energy as the incident one,  so it is of no interest for a gaseous detector. This effect is responsible for the blue sky.
\end{comment}

The cross section  $\sigma$ measures the probability of interaction for each method and energy. It is a rescaling of the attenuation coefficient $\mu$ taking into account the density $\rho$ of the material: $\mu = \sigma \rho$. It varies with the incident photon energy and the atomic number of the target material. As a rule of thumb, in gaseous materials, the photoelectric effect dominates up to 100 keV, Compton scattering is the most abundant interaction between 100 keV and 1 MeV, and above 1.022 MeV, minimum energy for pair production, this method appears, being the dominant from 10 MeV onwards.

\begin{comment}
\begin{figure}
    \centering
    \includegraphics[width=0.5\linewidth]{images/ArXCrossSection.png}
    \caption{Interaction cross section of photons in Argon (same as figure \ref{fig:InteractionsCrossSections})}
    \label{fig:ArXCrossSection}
\end{figure}
\end{comment}

In order to compute the attenuation length from the cross section $\sigma$ tabulated values, like shown in figure  \ref{fig:InteractionsCrossSections}, one has to apply the ideal gas law:
$$
PV = nRT \Rightarrow \dfrac{n}{V} = \dfrac{P}{RT}
$$

Using units of $bar$, $K$ and $m^3$ the value for the ideal gas constant is 
$$R = 8.2 \cdot 10^{-5} \dfrac{m^3 \cdot bar}{K \cdot mol}$$

As an example of how the values from table \ref{tab:AttLengthAr} are obtained, let us explicitly compute the attenuation length for a 6 keV photon at 1 bar. 
$$
\mu = \sigma \rho \dfrac{n}{V} = 2.6 \cdot 10^2 \, \dfrac{cm^2}{g} \cdot 39.95 \, \dfrac{g}{mol} \cdot 4.16 \cdot 10^{-5} \, \dfrac{mol}{cm^3} = 0.43 \, cm^{-1}
$$

So the attenuation length is $ \lambda = 1 / \mu = 2.31$ cm. In table \ref{tab:AttLengthAr} values are approximated, since their utility for detector design purposes does not need further precision. What is more interesting is to highlight the inverse proportionality with pressure.  

\begin{table}[h!]
    \centering
    \begin{tabular}{c | c c}
         \textbf{Argon}&  6 keV& 20 keV\\
         \hline
         1 bar&  25 mm& 75 mm\\
         5 bar&  5 mm& 15 mm\\
         10 bar&  2.5 mm& 7.5 mm\\
    \end{tabular}
    \caption{Approximated mean free path of photons in argon gas at different pressures for different energies. These energies are close to photons produced in typical calibration sources $^{55}Fe$ and $^{109}Cd$. Values for cross sections extracted from XCOM~\cite{berger2010xcom}.}
    \label{tab:AttLengthAr}
\end{table}

\subsubsection{Photoelectric effect}
The photoelectric effect consists in the absorption of an incident photon with energy $E_0$ by an atom and the emission of an electron from the atomic shells with energy $E= E_0-E_{shell}$. The explanation of this effect granted Albert Einstein the Nobel prize in physics in 1921 and in 1923 to Robert A. Millikan for his measurements of the photoelectric effect among other discoveries (he is known for the measurement of the charge of the electron). 

The absorption of a photon excites the atom and the energy excess is dissipated with the ejection of an electron from the bounded shells. Only atomic shells with binding energy below the energy of the incoming photon are available for this interaction so the cross section grows with sharp steps when the energy of a new shell is reached. This effect can be seen in figure \ref{fig:InteractionsCrossSections}. Above the K-shell binding energy, the photoelectric cross-section with this shell represents more than about 80\% of the total photoelectric cross section. It is also very dependent on the atomic number of the element, being easier to extract electrons by photoelectric effect from bigger atoms, due to the higher electron density in the material. This means that heavier atoms are more effective in stopping X-rays. The energy of the incoming photon is also critical, hard X-rays are more penetrating than soft ones \cite{Littlejohn}. 

\begin{comment}
The expression for the photoelectric effect cross section can be derived for non relativistic photons in an hydrogen-like atom: 

$$
\sigma_{\text{tot}} = \dfrac{16 \pi \sqrt{2}}{3}  \alpha^8 Z^5 a_0^2\left( \dfrac{m_ec^2}{E_0} \right)^{7/2}
$$
Being $\alpha$ the fine structure constant, $Z$ the atomic number of the atom, $a_0^2$ is the cross-sectional area of a hydrogen atom, $m_e$ the electron mass and $E_0=\hbar w_0$ the energy of the incoming photon.
This formula is approximately valid for the ejection of K-shell electrons by X-rays in any atom. The strong dependence in atomic number Z indicates that heavier atoms are more effective in stopping X-rays, and the strong inverse dependence with energy indicates that hard X-rays are more penetrating than soft ones \cite{Littlejohn}. 
\end{comment}

When the energy available is enough to eject K-shell electrons, these are the most probable electrons to be emitted. Therefore very often vacancies appear in inner shells. This makes other electrons in outer shells jump to fill the vacancy and the energy excess of these outer electrons is emitted through two competing channels: fluorescence photons and Auger electrons. 
In fluorescence the energy is emitted as an X-ray photon with the energy difference between both shells. This photon can ionize again the gas or escape the sensitive volume. This missing energy lead to ``escape peaks" in the energy spectrum.
In the emission of Auger electron the energy is transferred to another electron of the atom. If this energy is higher than the binding energy the electron is ejected and can ionize again surrounding atoms. In this case, all the energy of the original photon is recovered.

\begin{comment}

\begin{itemize}
    \item Fluorescence: Energy is emitted as an X-ray photon with the energy difference between both shells. This photon can ionize again the gas or escape the sensitive volume. This missing energy lead to ``escape peaks" in the energy spectrum.  
    
    \item Auger electron: The energy of the descending electron is transferred to another electron of the atom. If this energy is higher than the binding energy the electron is ejected and can ionize again surrounding atoms. In this case, all energy of the original photon is recovered.
\end{itemize}

In both cases the first vacancy is filled with another vacancy, so this can involve several steps. The fraction of the deexcitation through fluorescence is called the fluorescence yield. It is negligible in helium where a photoelectron is always followed by an Auger electron. In argon, the K-shell fluorescence yield is equal to 13.5 \% while the one of the L-shell is negligible \cite{garcia2015solar}.
\end{comment}

\subsubsection{Compton scattering}
Compton scattering is the inelastic scattering of photons in the electrons bound to an atom. In this process, the photon transfers energy to an electron in the atomic shells. The binding energy for this electron is several orders of magnitude smaller than the photon energy when this process dominate (binding energies of few eV and photons in the X-ray range, tens of keV), so they can be considered free electrons.

The energy shift in the bounced photon can be calculated using energy and momentum conservation principles. Calling $\Delta \lambda$ the shift in wavelenght ($\Delta \lambda = \lambda - \lambda_0$):
$$\Delta \lambda = \frac{h}{m_e c} (1 - \cos \theta)$$

$\theta$ is the deviation angle with respect to the incoming photon. So the shift in energy is related with dispersion angle. 

The maximum energy transfer occurs when the photon bounces back, $\theta = 180º$, and it can be expressed in terms of the energy of the incoming photon $E$ and de electron mass: $$\Delta E_{\text{max}} = E \cdot\dfrac{ \frac{2E}{ m_e c^2}}{1 + \frac{2E}{m_e c^2}}
$$

When the transferred energy to the electron is higher than the binding energy it is ejected. The spectrum of these ejected electrons is a continuum, due to the variable transferred energy, but it has a characteristic end point at $\Delta E_{max}$ called \textit{Compton edge}.

The cross section for this process was one of the first to be derived using quantum electrodynamics, known as the Klein–Nishina formula \cite{yazaki2017klein}, which can be consulted in \cite{garcia2015solar}.

\begin{comment}

$$
\frac{d\sigma_c}{d\Omega} = \dfrac{r_e^2 }{2}\left[ \frac{1 + \cos^2\theta}{\left(1 + \gamma (1 - \cos\theta)\right)^2} \right] 
\left[1 + \frac{ \gamma^2 (1-\cos \theta )^2}{(1 + \cos^2\theta)(1 + \gamma (1 - \cos\theta))} \right]
$$

Where $r_e = 2.82 \times 10^{-13}$ cm is the classical electron radius, $\theta$ is the photon scattering angle, $\nu = c/\lambda$ is the photon frequency and $\gamma = h\nu/mc^2$.

\begin{figure}
    \centering
    \includegraphics[width=0.5\linewidth]{images/KleinNishina.png}
    \caption{Klein-Nishina angular cross section for Compton scattering. Energies of incoming photon grow towards brighter colors.}
    \label{fig:enter-label}
\end{figure}

\end{comment}

\subsubsection{Pair production}
Pair production is the conversion of a photon into a pair electron-positron in the presence of the electromagnetic field of an atomic nucleus. This third body is needed to ensure momentum conservation. The pair production cross-section increases proportional to $Z^2$.

In order to create these two particles, the minimum energy at which this process can happen is the sum of the two rest masses: $2 \times 511$ keV $= 1022$ keV, and it becomes dominant at slightly higher energies. For example, in argon pair production is the most probable interaction for photons with energy higher than 10 MeV, as can be seen in \ref{fig:InteractionsCrossSections}. Therefore, for photons in the range of tens of keV, this process can be neglected.

\subsection{Charged particles}

When a particle traverses matter, compared to its relative movement, it finds static nuclei and electrons bouncing around. This can be seen as a lattice that holds an electronic gas. An abstraction like that is closer to our present view of a metal but, in general, from the perspective of a subatomic particle, most of the space is empty with electrons floating around heavy charged nucleus. In this landscape, electromagnetic interactions govern the behaviour of charged particle interactions with matter. 

When one of these particles traverses matter, the Coulomb force enters into play. They ionize the medium extracting electrons while the energy deposited by the Coulomb interaction is higher than the binding energy: for a energetic particle, electrons will be extracted all along its path.

The shape of the path and the distribution of deposited energy along the path depend on the nature of the particle. Alpha particles, for example, are very heavy and their charge is $2e$ so they interact strongly with the medium. Their paths are short and straight and while they slow down they deposit more energy per unit length, therefore most of the charge is found at the end of the track. Muons are light but very energetic. They are difficult to stop, very often they cross all the sensitive volume leaving behind long straight tracks. Electron paths are tortuous, they are very influenced by Coulomb forces, bouncing from one nucleus to another. 

There are two main ways of interactions: Inelastic scattering (Coulomb) with electrons in the atoms and Bremsstrahlung radiation. Other types of interactions are possible, like elastic scattering with the nucleus, nuclear reactions or Cherenkov radiation, but they are very rare (nuclear reactions) or deposit negligible amounts of energy (elastic interactions).

\begin{comment}
\begin{figure}
    \centering
    \includegraphics[width=0.8\linewidth]{images/ElectronStoppingPower.png}
    \caption{Fractional energy loss per radiation length in lead as a function of electron or positron energy \cite{PDG2024review}.}
    \label{fig:ElectronStoppingPower}
\end{figure}
\end{comment}

\begin{figure}[h]
    \centering
    \includegraphics[width=0.8\linewidth]{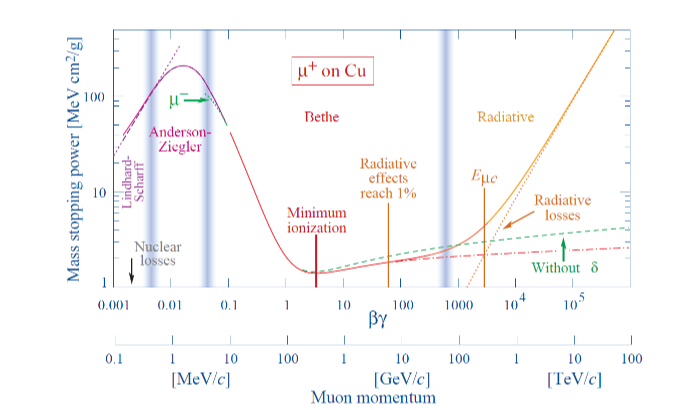}
    \caption{Stopping power of positive muons in coper as function of the muon momentum \cite{PDG2024review}.}
    \label{fig:StoppingPower}
\end{figure}

The rate at which a charged particle losses its energy while traversing through matter is known as the stopping power. In figure \ref{fig:StoppingPower} it is presented for muons in copper. The first calculation of energy loss due to Coulomb interactions was due to Bethe in the 30's of the XX century and the expression he derived is now known as Bethe-Bloch formula, that can be reviewed in \cite{ruiz2019ultra} and \cite{PDG2024review}. This expression and its corrections are valid in the range from tens of MeV to hundreds of GeV, depending on the target material. There are other analytical models for different energy ranges and tabulated experimental values for when they fail \cite{EstarPestarAstar}.

\begin{comment}

$$-\frac{dE}{dx} = 4 \pi N_A r_e^2 m_e c^2 \rho \frac{Z}{A} \frac{z^2}{\beta^2} 
\left[
\frac{1}{2} \ln\left(\frac{2 m_e c^2 \beta^2 \gamma^2 W_{\text{max}}}{I^2}\right) 
- \beta^2 - \frac{\delta(\beta \gamma)}{2} - \frac{C}{Z}
\right]
$$

\begin{center}
\begin{tabular}{clcl}

$r_e$ & Classical electron radius & $z$ & Charge of the incident particle \\

$m_e$ & Electron mass & $\beta$ & $v/c$ of the incident particle \\ 

$N_A$ & Avogadro’s number & $\gamma$ & $\frac{1}{\sqrt{1 - \beta^2}}$ \\ 

$I$ & Mean excitation potential & $\delta$ & Density correction \\

$Z$ & Atomic number of the material & $C$ & Shell correction \\ 

$A$ & Atomic weight of the material & $W_{\text{max}}$ & Maximum energy transfer in a single collision \\ 

$\rho$ & Density of the gas & & \\ 

\end{tabular}
\end{center}

$$W_{\text{max}} = \frac{2 m_e c^2 \beta^2 \gamma^2}{1 + \left( \frac{2\gamma m_e}{M} \right) + \left( \frac{m_e}{M} \right)^2}
$$

This expression includes some corrections added to take into account smaller effects. An extensive interpretation of this formula is done for example in \cite{PDG2024review}, where the range of validity of the different terms is discussed. In figure \ref{fig:StoppingPower} the stopping power for muons in cooper is shown. Bethe-Bloch formula only accounts for intermediate range of energies presented there. 

\end{comment}

\subsubsection{Electrons}
Due to the low mass of electrons and given that they interact with identical particles, other electrons around the nuclei, relativistic effects need to be taken into account. For example radiative energy losses by Bremsstrahlung become significant at relatively low energies. 

\begin{figure}
    \centering
    \includegraphics[width=0.7\linewidth]{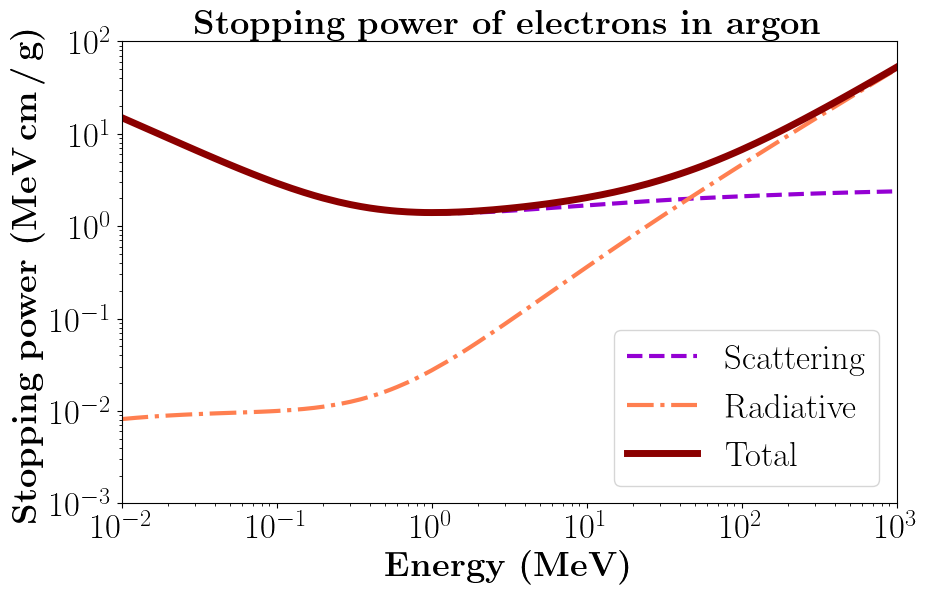}
    \caption{Stopping power for electrons in argon due to scattering and Bremstrahlung. Data from ESTAR \cite{EstarPestarAstar}.}
    \label{fig:ElectronStpPowerArgon}
\end{figure}

The energy at which ionizing losses and radiative losses become equal is called critical energy. Figure \ref{fig:ElectronStpPowerArgon} shows that for electrons in argon, the critical energy is slightly below 10 MeV. For practical uses, values for stopping powers are tabulated in \cite{EstarPestarAstar} and ranges for electrons in different elements can be computed from there.

\subsubsection{Alpha particles}  

Their bigger mass and double charge makes them quite unique because their deposition of energy is highly non-homogeneous. They interact strongly with matter, they are stopped easily with a thin sheet of material, but they can be seen if they are generated inside the active volume of the detector. Due to the rapid loss of energy their velocity decays very fast, typically in the range of few centimetres in 1 bar of any noble gas. At the end of their track, they go slower and, therefore, they deposit more energy per unit length. This is called Bragg peak and it is clearly visible in alpha events. In the next chapter this feature will be used to identify alpha particles in the TREX-DM experiment, see figure \ref{fig:AlphaParticle}.

Tabulated data for stopping power in different elements can be extracted from ASTAR database \cite{EstarPestarAstar}.

\subsubsection{Muons}

Detectors used in rare event searches are typically located in underground facilities to significantly reduce the flux of cosmic muons. 

Muons are highly energetic and can traverse the entire detector without appreciable energy loss, producing ionization uniformly along their track. Rare event searches detectors minimize their impact when installed in underground facilities and through detailed simulations \cite{diez2025efficient}. In pixelated gaseous detectors, the signature of muons is particularly distinctive, as they pass through the entire detection chamber, leaving a continuous, easily recognizable track.

\subsection{Neutrons and dark matter}

Neutral particles cannot interact electromagnetically with matter, so they are neither affected by the electric fields of nuclei nor capable of ionizing atoms. This is the case of neutrons and possibly WIMPs. They can interact via different processes, such as elastic and inelastic scattering, spallation, transmutation or radiative capture. As a secondary product of the interaction, charged particles may appear, which ionize the surrounding atoms and thus can be detected.

The main neutron interactions with matter is elastic scattering, in which the neutron bounces off the nucleus in a perfect elastic collision.  Kinetic energy is conserved and no extra particles are emitted. A nuclear recoil due to this interaction may produce secondary ionization that can be detected. This is the expected signal for a WIMP recoil.
Inelastic scattering is also possible. The incoming neutron is absorbed by the nucleus forming an unstable compound, that quickly emits another neutron with a lower kinetic energy. The nucleus is left in an excited state, and it returns to the ground state through $\gamma$ emission.
In transmutation, a neutron is absorbed by a nucleus, causing it to transform into a different isotope. Usually it decays very fast emitting $\alpha$ or $\beta$ particles in the process.
Radiative capture is similar, again the nucleus absorbs a neutron and becomes excited, but in this case it doesn't decay into another element. To relax, the nucleus emits $\gamma$-rays and becomes a different isotope.
In both spallation and fission, the outcome is similar: the nucleus breaks apart. Spallation involves the fragmentation of a nucleus into several smaller components as a result of a collision with a high-energy neutron. In contrast, fission occurs when a heavy nucleus captures a slow neutron and subsequently splits into two or more fragments, releasing additional neutrons and photons in the process.

\begin{comment}
    
\begin{itemize}
    \item Elastic scattering: it is main neutron’s interaction with matter, in which the neutron bounces off the nucleus in a perfect elastic collision.  Kinetic energy is conserved and no extra particles are emitted. Nuclear recoil due to this interaction may produce secondary ionization that can be detected. This is the expected signal for a WIMP recoil.
    
    \item Inelastic scattering: in this process, the incoming neutron is absorbed by the nucleus forming an unstable compound, that quickly emits another neutron with a lower kinetic energy. The nucleus is left in an excited state, and some $\gamma$-decays can happen until it returns to the ground state.
    
    \item Transmutation: a neutron is absorbed by a nucleus, causing it to transform into a different isotope. Usually it decays very fast emitting $\alpha$ or $\beta$ particles in the process.
    
    \item Radiative capture: again a nucleus absorbs a neutron and becomes excited, but in this case it doesn't decay into another element. To relax, the nucleus emits $\gamma$-rays, so no transmutation occurs, the nucleus becomes a different isotope.
    
    \item Spallation: this process consist of the fragmentation of a nucleus into several parts due to the collision with a high energy neutron.
    
    \item Fission: a slow neutron is captured by a heavy nucleus which splits in fragments emitting several neutrons and photons.  
\end{itemize}

\end{comment}

\section{Electron behaviour in a gaseous detector}

\subsection{Charge generation}

Incoming particles interact with gaseous media through different interactions as explained before. Most of the times, the result of this interaction is not only a single electron extracted from one atom, but also many secondary decays that follow the original interaction. Photons, for example, give rise to electrons ejected from the gas molecules, but they are usually accompanied by Auger electrons and fluorescence X-rays. Also, heavy charged particles and electrons lose their energy through inelastic collisions with the gas molecules that result in the production of electron-ion pairs and excited states of the gas atoms, which relax to their fundamental state through fluorescence emission and Auger electrons. Collisions with complex molecules can also activate rotational and vibrational modes which do not contribute to the ionization yield.

The different contributions can be classified in two groups: 
\begin{itemize}
    \item Primary ionization: electron-ion pairs generated directly by the collision of the incident particle with the atoms of the gas.
    \item Secondary ionization: some of the electrons generated in the primary ionization may have enough energy to ionize further atoms. They are called $\delta$-rays. 
\end{itemize}

These subsequent ionizations continue until the energy of the ejected electrons is lower than the ionization energy of the gas atoms. 
\\

The ionizing collisions occur randomly along the track of the charged particle. The distribution of the number of ionizing collisions $k$ in a segment $s$ of the track follows Poisson statistics

\begin{equation}
    P(k) = \frac{(s/\lambda)^k}{k!} e^{-s/k}
\end{equation}

with $\lambda=1/(N_e\sigma_I)$ the mean free path, the mean distance between charge clusters. $N_e$ is the electron density of the gas and $\sigma_I$ the ionization cross-section. The number of primary electrons per unit length is $n_p = 1/\lambda$, which depends on the type of particle, its energy and
the gas mixture. 

\subsubsection{Work function W}
Not all energy involved in the interaction is converted into electron-ion pairs. Part of the incoming energy is lost in other excitation channels, such as scintillation or vibration and rotation of the gas molecules. Experimentally, above few tens of eV the mean number of extracted electrons, $N_e$ is proportional to the absorbed energy $E$:

\begin{equation}
    N_e=E /W
\end{equation}

$W$ is the mean energy needed to create an electron-ion pair in a certain gas 
These parameters have been measured for many gases and with different ionizing particles. For electrons and alphas in noble gases, similar values for $W$ are measured, with small deviations for complex hydrocarbons gases. The $W$ dependence on the gas composition is complex. In general, it increases with the ionization potential and with the probability of the gas molecules to undergo non-ionizing mechanisms, such as the promotion to excited, vibrational, and rotational modes.

For mixtures, the final $W$ value can be computed as the weighted average:

\begin{equation}
    \frac{1}{W_{AB}} = \frac{C_A \sigma_A}{W_A} + \frac{C_B \sigma_B}{W_B}
\end{equation}
where $C_i$ and $\sigma_i$ are the concentrations and ionizing cross-sections of the individual gases.

\begin{table}[h]
\centering
\renewcommand{\arraystretch}{1.3} % Slightly increase row spacing
\begin{tabular}{c|c|c|c|c}
\hline

\textbf{Gas} & $I_{\text{exc}}$ \textbf{(eV)} & $I_{\text{ion}}$ \textbf{(eV)} & $W$ \textbf{(eV)} & $F$ \\
\hline
\hline

He        & 19.8 & 24.5 & 45 & 0.17 \\
\hline
Ne        & 16.7 & 21.6 & 30 & 0.17 \\
\hline
Ar        & 11.6 & 15.7 & 26 & 0.23 \\
\hline
Xe        & 8.4  & 12.1 & 22 & 0.17 \\
\hline
CH$_4$    & 8.8  & 12.6 & 30 & 0.26 \\
\hline
iC$_4$H$_{10}$ & 6.5  & 10.6 & 26 & 0.26 \\
\hline

\end{tabular}
\caption{Lowest excitation energy ($I_{\text{exc}}$), lowest ionisation energy ($I_{\text{ion}}$), mean energy for electron-ion pair production ($W$) and measured Fano factor ($F$) for different gases. Table from \cite{Oscar2025development}, data from \cite{gracia2016micromegas}. }
\end{table}

\subsubsection{Fano factor}
Ionization processes are statistical by nature. Therefore two identical charged particles depositing the same energy $E$ in a gas will not necessarily produce the same number of electron-ion pairs. If the ionizing collisions would be completely independent, the number of electron-ion pairs $N_e$ woudl fluctuate following a Poisson distribution with variance $\sigma_{N_e}=N_e$. But the distribution of ionizing collisions is not fully independent, as the number of ways an atom can be ionized is limited by the discrete electronic levels. This shows as a reduced variance in the generation of electron-ion pairs, accounted by the Fano factor: 

\begin{equation}
    \sigma_{N_e}=F N_e
\end{equation}

The Fano factor depends on the gas composition and the nature and energy of the ionizing particle. The fluctuation in the number of primary electron-ion pairs is the first contribution to the energy resolution of a gaseous detector. The fact that an atom can only become ionized in a certain number of ways, results in a better energy resolution than predicted by purely statistical considerations. 

The lower limit to the relative energy resolution, $R = \sigma_E /E$, imposed by this fluctuation is called the Fano limit, or intrinsic energy resolution, and it is given by:

\begin{equation}\label{PrimaryElectrons}
    R(\% \text{ FWHM}) = 2.35 \sqrt{\frac{W}{E}} F
\end{equation}

Note that $R$ increases $\propto 1/\sqrt{N_e}$ since $\sigma_E \propto \sqrt{N_e}$ and $E \propto N^{-1}_e$. The intrinsic energy resolution of any gaseous detector can be estimated at any energy using the tabulated values. For example, in argon at 5.9 keV, the energy of the $^{55}Fe$ isotope main X-rays, the intrinsic energy resolution is 7.5$\%$ FWHM, equivalent to 0.44 keV.

\subsection{Drift velocity}

Electrons generated by incoming particles are drifted towards the anode, while resulting ions move towards the cathode. Their masses are very different and if the electric field is homogeneous in all the conversion volume, electrons drift faster. Ion drift will not be discussed further as only electrons are amplified in the readout plane and induce a signal.

The movement of electrons under the presence of an electric field tends to follow the general pattern of field lines but due to their small mass, thermal energy also affects their paths. The overall result is a convoluted path with the general tendency to follow field lines but far from being straight. Electrons scatter continuously in surrounding atoms and molecules of the gas. Precisely this effect makes the drift velocity stabilize. Electrons are accelerated by the electric field and, their velocity would grow unlimited, for example in vacuum. But in a gas, electrons lose energy through inelastic collisions with atoms. These two processes, one fostering their velocity and the other stopping them, balance to reach the final drift velocity. It is different for every gas mixture because the frequency of scatterings and the mean energy loss per collision depends on the scattering sections of every atom and molecule. For example, gas molecules have rotational and vibrational modes, what favours inelastic interactions; while noble gas atoms tend to interact through elastic scattering, which produces almost no decrease in electron velocity, only changes their direction.

\begin{figure}[h]
    \centering
    \includegraphics[width=0.8\linewidth]{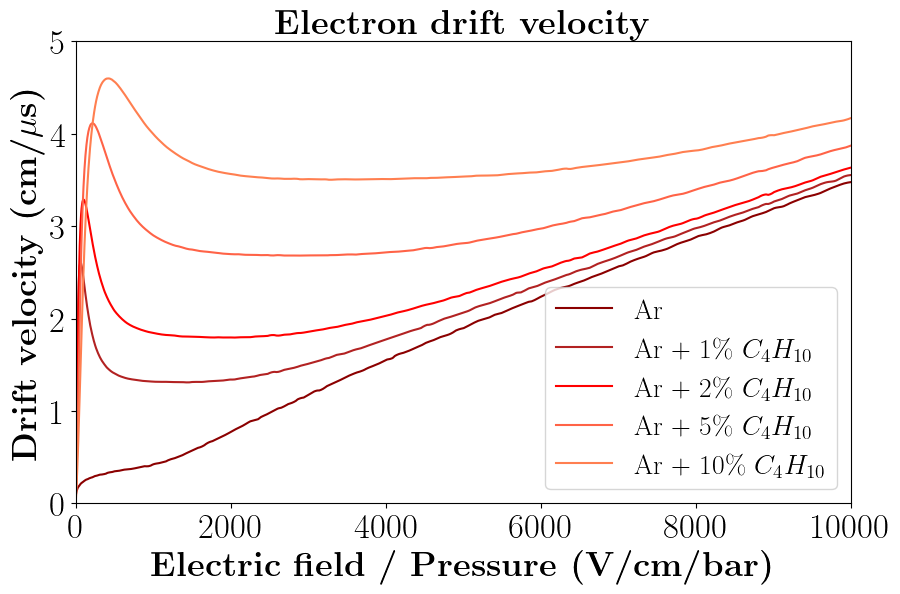}
    \caption{Drift velocity of electrons with the reduced drift field for different argon-isobutane mixtures, obtained from Magboltz \cite{biagi1999monte} simulations and easily available in \cite{ObisGas}.}
    \label{fig:DriftVelocityArgon}
\end{figure}

The results of these effects are not always obvious. Pure noble gases tend to have lower drift velocities because electrons suffer a lot of collisions before reaching the anode, therefore, the total length of the path may be very long. On the contrary, efficient inelastic scattering helps to reduce their velocity, so they follow straighter lines. However, if the velocity decreases significantly, electrons may get captured by surrounding atoms and molecules in the gas, an effect known as attachment (see section \ref{sec:attach}). Typical gases used in particle detectors are a mixture of noble gases with organic molecular quenchers to reach a balance between different competing properties, one of them the drift velocity. Also energy resolution, sparking limit, work function $W$, etc., have to be taken into account.

The mean drift velocity can be parametrized as \cite{blum2008particle}: 

\begin{equation}
    v_d = \dfrac{e \, \tau}{m_e}E
\end{equation}

where $e$ and $m_e$ are the charge and mass of the electron, $E$ is the value of the electric field and $\tau(E) = 1/(N \sigma_s u)$ is the mean time between collisions. This time is determined by the molecular density of the gas, $N$, the scattering cross-section, $\sigma_s$, and the mean
instantaneous velocity, $u$. Since the drift velocity depends on the molecular density of the gas it will change with pressure. Figure \ref{fig:DriftVelocityArgon} shows simulated electron drift velocities in different argon mixtures with Magboltz, the code used to extract and display the values can be found here \cite{ObisGas}.

\begin{comment}
\begin{figure}[h]
    \centering
    \includegraphics[width=0.6\linewidth]{images/DriftSimulation.png}
    \caption{Drift simulation in 2D with Garfield++ with vertical electric field. All dimensions in cm. In yellow electron paths, in brown ionizations in the avalanche region, in green excitations.}
    \label{fig:DriftSimulation}
\end{figure}
\end{comment}

\subsection{Diffusion}

Electrons generated after the primary interaction drift towards the anode. During this movement, the electron cloud expands due to the collisions with gas atoms. This is called diffusion and it introduces one of the main uncertainties in topological reconstruction. The electron cloud expands in all three dimensions but the extent of these elongations is different depending on the electric field applied. So, one can distinguish between longitudinal diffusion, in the direction of the field lines, and transversal diffusion, in the two orthogonal dimensions to the electric field lines. 

Charge distribution along the drift path is modelled with a Gaussian shape whose variance is determined by the diffusion coefficients, $D_L$ and $D_T$:

\begin{equation}
    \sigma = \sqrt{2D_it} = \sqrt{\frac{2D_il}{v}}
\end{equation}

The diffusion coefficients, $D_i$, depend on the electric field applied and properties of the gas, as the longer the drifting time $t$, the more the spread in the cloud, and this time can be obtained from the drifted distance $l$ and the drift velocity $v$.

\begin{figure}[h]
   \begin{minipage}{0.5\textwidth}
     \centering
     \includegraphics[width=0.999\linewidth]{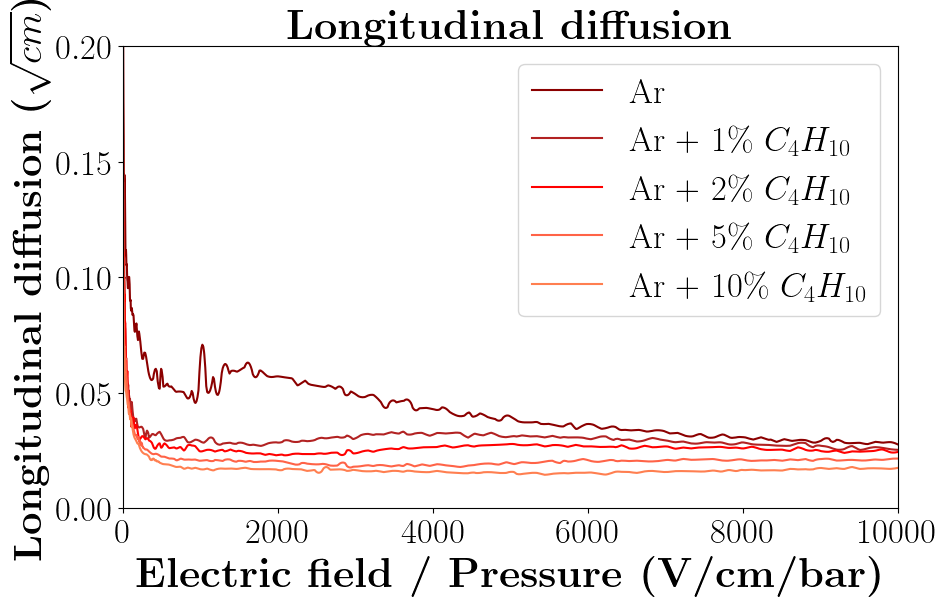}
     \label{fig:LongDiffArgon}
   \end{minipage}\hfill
   \begin{minipage}{0.5\textwidth}
     \centering
     \includegraphics[width=0.999\linewidth]{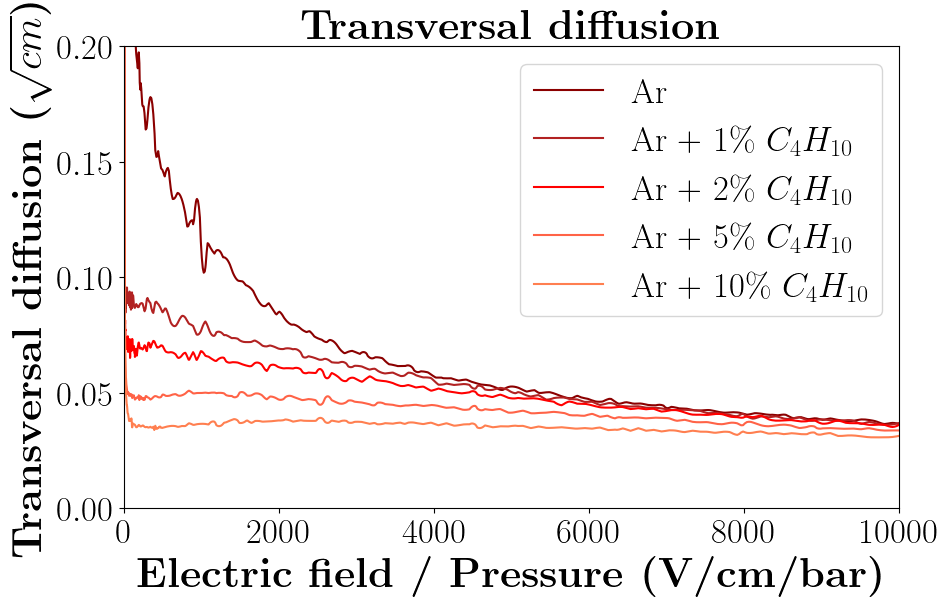}
     \label{fig:TransDiffArgon}
   \end{minipage}
   \caption{Longitudinal and transversal coefficients for different argon-isobutane mixtures from Magboltz \cite{biagi1999monte}. }
\end{figure}

\subsection{Recombination and attachment}\label{sec:attach}

Not all electrons generated in the primary interaction reach the anode. There are processes that reduce the number of collected electrons, the most important of which are recombination and attachment.

Recombination refers to the encounter of an electron and an ion to form a neutral atom. This effect can be highly suppressed by applying an electric field, as it is the case. But it is still relevant in environments with high electron-ion density, for example along the development of an avalanche. 

Attachment appears when a gas atom absorbs an electron, becoming an ion. Noble gases have the lowest cross section for this process, their electron affinity is very low, but organic molecules and above all impurities -oxygen and water mainly, with high electron affinity- tend to absorb electrons drifting to the anode. If they are present, the effect is very clear because less charges are collected per primary energy deposition, shifting the energy spectrum towards lower gain.

\section{Gaseous detectors and Micromegas readout planes}

\subsection{Evolution of gaseous detectors}

In early early XX century, the physicist Hans Geiger developed a particle detector for alpha particles, later refined by Walther Müller in the 1920s, the so called Geiger-Müller counter. This device is composed of a cylindrical cathode and a wire anode passing through its centre. It detects ionizing radiation through gas ionization and avalanche multiplication but provides only binary information, whether a particle passed through or not. 

In the 1940s and 1950s, proportional counters improved the performance by operating at lower voltages, allowing the output signal to be proportional to the energy of the incoming particle. This enabled basic energy measurements and expanded the detector’s applications in nuclear and X-ray physics.

A major breakthrough came in the 1960s with Georges Charpak’s invention of the multiwire proportional chamber (MWPC). By arranging many wires in a plane and reading them out individually, MWPCs provided position-sensitive detection with high efficiency. They were widely adopted in high-energy experiments and medical X-ray equipment.

The need for full 3D tracking led to the development of the time projection chamber (TPC) in the 1970s. TPCs allow for three-dimensional reconstruction of charged particle tracks by measuring both the position and drift time of ionization electrons. Figure \ref{fig:TPC} shows a schematic version of a TPC. These chambers became critical in collider experiments, handling high track densities with precision, and they are also widely used in astrophysics and medical physics.

The Micromegas detector (MICRO-MEsh GAseous Structure) belongs to a wider class of particle detectors commonly known as MPGD (Micropattern Gaseous Detectors) \cite{giomataris1996micromegas}, together with GEM (Gas Electron Multiplier) \cite{sauli1997gem} devices, that have in common the fabrication techniques of microelectronics and photolithography to achieve micrometric structures. Their main advantages are the high granularity and fast response, so they become standard readout planes for gaseous detectors in high energy physics, but also in nuclear physics, biology, medical applications, etc...

\subsection{Micromegas readout planes}

\begin{figure}
    \centering
    \includegraphics[width=0.5\linewidth]{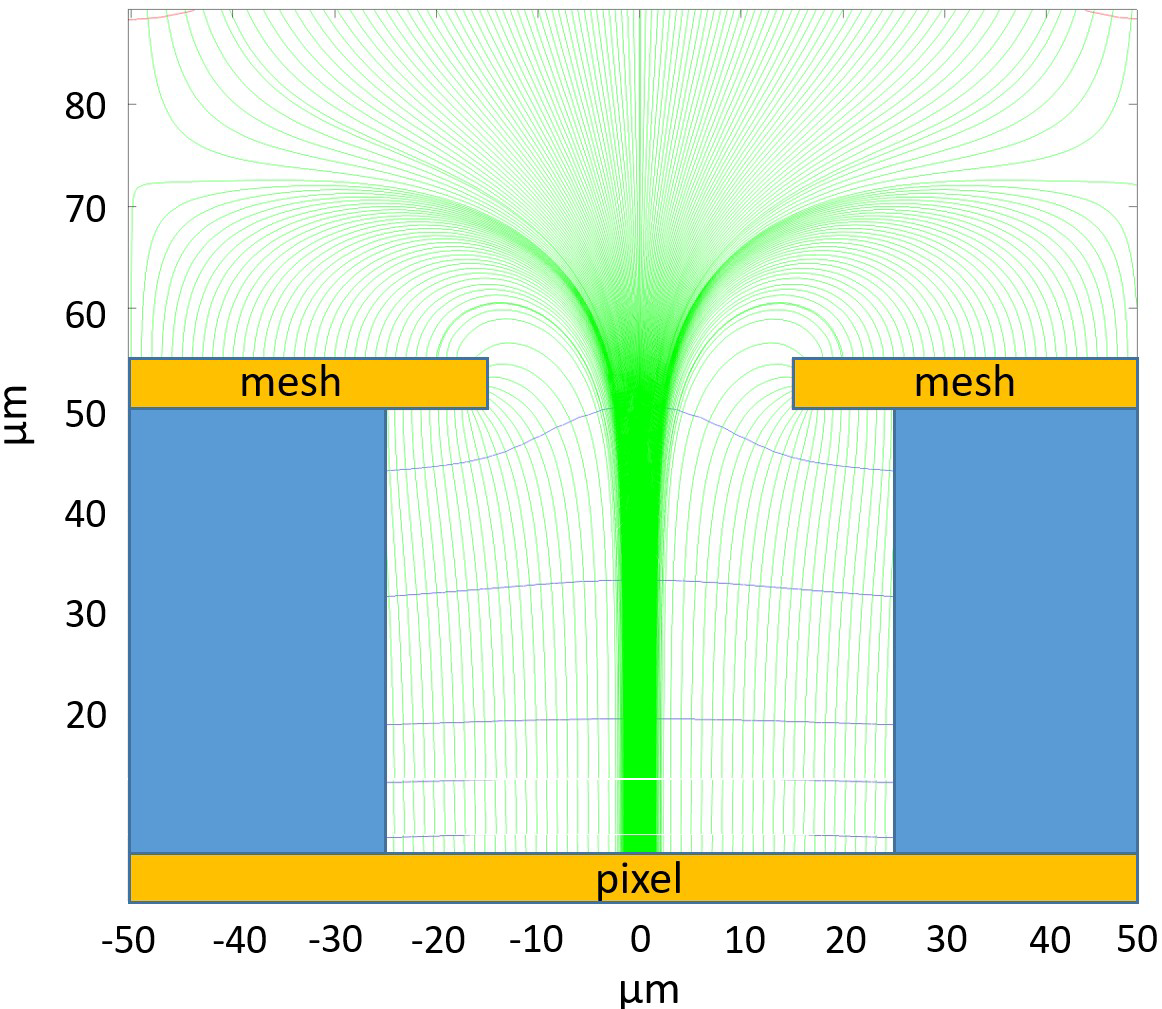}
    \caption{Simulation of electric field lines in the amplification gap for 30 $\upmu$m hole diameter and 50 $\upmu$m gap. Figure from \cite{mirallas2024planos}.}
    \label{fig:AvalancheGapField}
\end{figure}

The main purpose of the Micromegas readout planes is to collect and multiply the charges generated in the conversion volume, as depicted in figure \ref{fig:TPC}. The number of primary electrons generated in an ionization event is not sufficient to be directly registered by the acquisition chain, so gaseous detectors make use of the avalanche effect to multiply the number of charges by successive ionizations under the effect of a very intense electric field. In a Micromegas detector, this behaviour is achieved in a small gap, of the order of tens of $\upmu$m, between two metallic planes. The bottom one is the anode where the charge of the avalanche is collected, and the upper one is a mesh, an electrode with many holes. With an adequate application of  electric fields above and below this mesh, all the electric field lines that guide the electrons go through the holes, as can be seen in a simple simulation in figure~\ref{fig:AvalancheGapField}.

The geometrical configuration of the structure (distance between the electrodes, their thickness, the hole disposition...) affects the performance of the detector, and can, up to a point determined by the fabrication limitations, be chosen to fit the application at hand.  The latest generations of Micromegas rely on two different fabrication technologies: Bulk \cite{giomataris2006micromegas} and Microbulk \cite{andriamonje2010development}. Both have the advantage, compared with previous iterations, that both planes, anode and mesh, form a single structure. The difference is in the mesh. Bulk technology uses woven wire mesh (tenths of $\upmu$m thick), which are very robust to stretching and handling and are industrially produced. Microbulk Micromegas are even more refined: the two electrodes are produced together from a copper-kapton-copper foil, where the anode pattern is engraved at the lower copper foil, the mesh holes are etched at the upper copper foil and finally the kapton is removed only below these holes. This technology offers better homogeneity in the height of the gap, makes use of intrinsic radiopure materials for its fabrication (kapton and copper) and allows gaps as small as 25~$\upmu$m. All Micromegas used along this work are microbulk Micromegas. They will be further detailed in chapter \ref{Ch:TREX}. Their radiopurity, thanks to the techniques pioneered in close collaboration with Rui de Oliveira, from the Micro Pattern Technologies workshop at CERN under the collaborations RD51 and DRD1, have allowed to our group to tailor them for rare event searches like axion searches in CAST and BabyIAXO, dark matter searches in TREX-DM or neutrinoless double beta decay in PandaX-III.

\subsection{Amplification region in a Micromegas}

As explained previously, when electrons reach the Micromegas plane, they enter in a region called ``amplification gap". A high electric field is applied in that region, typically around 50 kV/cm/bar, sufficient to produce many secondary ionizations. Electrons are accelerated so that when they collide with an atom they are able to ionize it. New electrons generated this way can ionize further and an avalanche of continuous ionizations happens. This is the first amplification stage of the detector. 
The number of charges is modelled with:

\begin{equation}\label{Townsend}
    \dfrac{dN}{dx}=N\alpha
\end{equation}

\begin{figure}[h]
    \centering
    \includegraphics[width=0.8\linewidth]{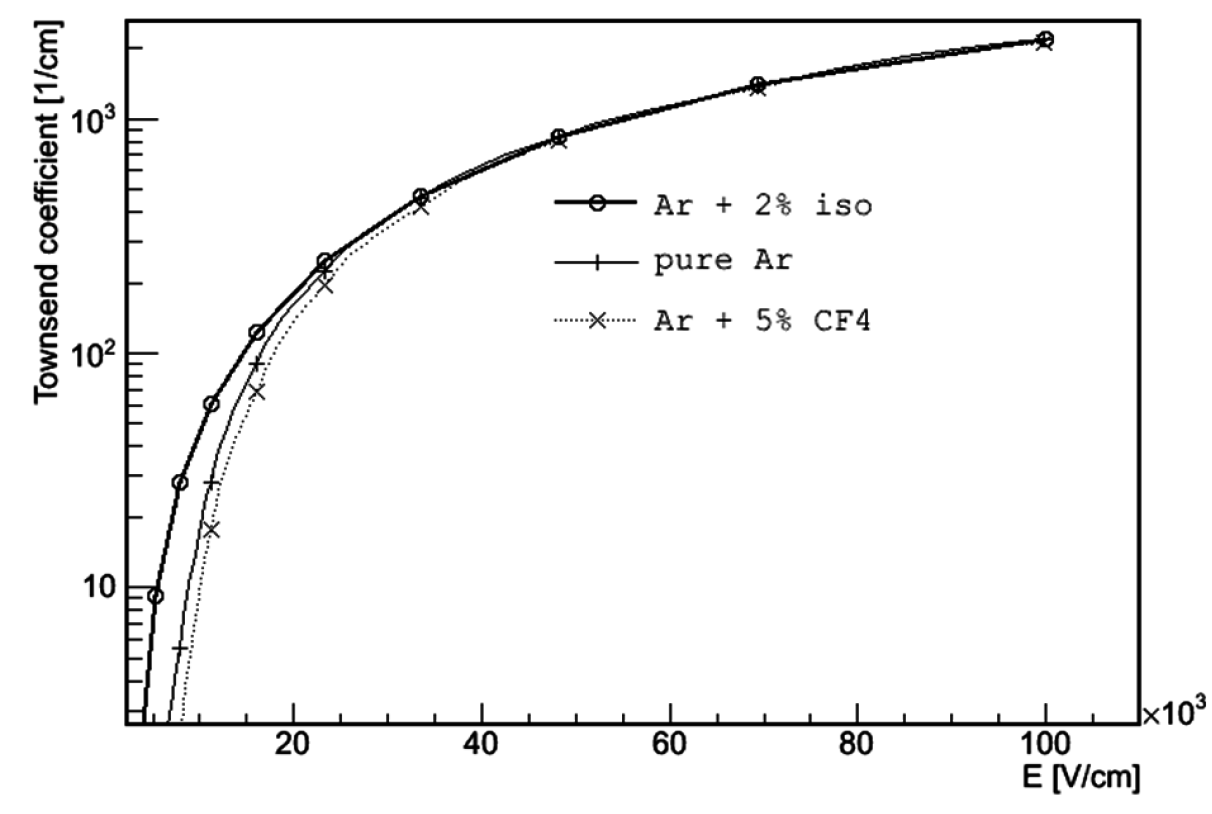}
    \caption{Townsend coefficient as a function of the electric field for argon and argon mixtures. Data from Magboltz \cite{biagi1999monte}, image from \cite{ruiz2019ultra}}
    \label{fig:Townsend}
\end{figure}

$\alpha$ is the Townsend coefficient that represents the number of collisions that will create an electron-ion pair per unit length. It shows a strong dependence on the electric field applied, as shown in figure \ref{fig:Townsend}. It can be expressed as $\alpha=\lambda^{-1}$, where $\lambda$ is the mean free path of the electron.

Integrating equation \ref{Townsend} along the amplification gap, typically of the order of 50~$\upmu$m, the gain factor, $G$, is obtained:

\begin{equation}
    N(x) = N_0 e^{\alpha x} \qquad G = \frac{N}{N_0} = e^{\alpha x}
\end{equation}

This gain factor $G$ represents the number of ionizations created by a single electron in an avalanche. This number cannot be increased at will, maximum values are around $G\simeq 10^8$ due to the Raether limit of the gas in which sparks appear and the detector is no longer stable.

The number of final electrons after the avalanche can fluctuate and this affects the energy resolution of the detector. These fluctuations can be modelled with a Polya distribution \cite{Derre2000}:

\begin{equation}
    p(x, G, m)=\frac{m^m}{\Gamma(m)} \frac{1}{G}\left(\frac{x}{G}\right)^{m-1} e^{-m \frac{x}{G}}
\end{equation}

\begin{comment}
\begin{figure}[h]
    \centering
    \includegraphics[width=0.7\linewidth]{images/Polya.png}
    \caption{Polya distributions for several values of the relative variance $b = m^{-1}$. Plot adapted from \cite{gracia2016micromegas}.}
    \label{fig:Polya}
\end{figure}
\end{comment}

It is this gain fluctuation effect, together with the variance in the generation of primary electrons of equation \ref{PrimaryElectrons}, that give rise to the energy resolution for a Micromegas readout:

\begin{equation}
R(\% FWHM) = 2.35 \sqrt{\frac{W}{E} (F + \dfrac{1}{m})}
\end{equation}
As an example, the limit to the intrinsic energy resolution at 5.9 keV in Ar + 2$\%$ isobutane (i$C_4H_{10}$)  of a 50 $\upmu$m gap Micromegas detector operated at 60 kV/cm at 1 bar is about 11$\%$ FWHM \cite{Cebrian_2010} \cite{IGUAZ2012448}.

\section{Some useful gas values for experimental measurements}

In the frenzy of the laboratory it is very useful to have at hand some of these gas parameters. Do photons of certain energy interact enough in the sensitive volume to see a peak in the spectrum? How big will the electron cloud reaching the detector be? Which time window is needed to see long events?

All these questions involve tabulated gas properties and some basic relations to estimate properties of the signal. Gas parameters are extracted from Garfield++, which at the end makes use of Magboltz data. My colleague, Luis Obis, developed a very nice tool to generate and visualize online this data, it can be found here \cite{ObisGas}. Ranges and attenuation values for electrons and photons are tabulated in XCOM \cite{berger2010xcom} and ESTAR \cite{EstarPestarAstar} databases. Also data for alpha particles and protons can be found in the same ESTAR webpage, in the databases called ASTAR and PSTAR.

For photon mean free path values, the ideal gas law is used, as explained in subsection \ref{sec:photons}, to obtain the density for each pressure and gas, and this allows to translate the attenuation tabulated values to mean free path, $\lambda$, following the relation $$\lambda=\frac{1}{Attenuation \cdot Density}$$

%\vspace{-1em}

Table \ref{tab:AttenuationLength} presents photon attenuation length and table \ref{tab:AlphaLength} the alpha track length for argon and neon at different pressures. And table \ref{tab:DriftVelocty} show values for electron drift velocity in mixtures of these gases with isobutane. These numbers are useful to determine the expected features of the detected events according to the dimensions of the TPC.

%%%%%%%% Attenuation lenght %%%%%%%%%%%

\begin{table}[h]
\centering
\small % Reduce font size to fit the table
\renewcommand{\arraystretch}{1.3} % Increase vertical spacing
\rowcolors{3}{white}{gray!10} % Apply zebra striping from row 3 onward
\begin{tabular}{
    >{\columncolor{white}}c
    >{\columncolor{headerYellow}}c
    >{\columncolor{headerYellow}}c
    >{\columncolor{headerYellow}}c
    >{\columncolor{headerYellow}}c
    >{\columncolor{headerGreen}}c
    >{\columncolor{headerGreen}}c
    >{\columncolor{headerGreen}}c
    >{\columncolor{headerGreen}}c}
\multicolumn{9}{c}{\cellcolor{white}\textbf{Photon attenuation length (cm)}} \\
\hline
\cellcolor{white} & 
\multicolumn{4}{c}{\textbf{Argon}} & 
\multicolumn{4}{c}{\textbf{Neon}} \\
\hline
\textbf{Energy (keV)} & {\cellcolor{headerYellow}\textbf{1 bar}} & {\cellcolor{headerYellow}\textbf{2 bar}} & {\cellcolor{headerYellow}\textbf{4 bar}} & {\cellcolor{headerYellow}\textbf{10 bar}} & 
{\cellcolor{headerGreen}\textbf{1 bar}} & {\cellcolor{headerGreen}\textbf{2 bar}} & {\cellcolor{headerGreen}\textbf{4 bar}} & {\cellcolor{headerGreen}\textbf{10 bar}} \\
\hline\hline
3  & 3.58    & 1.79    & 0.89    & 0.36    & 2.98    & 1.49    & 0.74    & 0.30 \\
6  & 2.35    & 1.18    & 0.59    & 0.24    & 22.07   & 11.04   & 5.52    & 2.21 \\
8  & 5.17    & 2.58    & 1.29    & 0.52    & 51.86   & 25.93   & 12.96   & 5.19 \\
22 & 93.12   & 46.56   & 23.28   & 9.31    & 968.88  & 484.44  & 242.22  & 96.89 \\
24 & 119.69  & 59.85   & 29.92   & 11.97   & 1211.10 & 605.55  & 302.77  & 121.11 \\
88 & 2539.89 & 1269.94 & 634.97  & 253.99  & 7118.06 & 3559.03 & 1779.51 & 711.81 \\
\bottomrule
\bottomrule
\makecell{\textbf{Density} \\ ($\times 10^{-3}$ g/cm\textsuperscript{3})} & 1.64 & 3.28 & 6.56 & 16.4 & 0.83 & 1.66 & 3.31 & 8.28 \\
\end{tabular}
\smallskip
\caption{Photon attenuation length in cm for argon (yellow) and neon (green) at different pressures for different energies.}\label{tab:AttenuationLength}
\end{table}

%%%%%%%% Alpha range %%%%%%%%%%%

\begin{table}[h]
\centering
\small % Reduce font size to fit the table
\renewcommand{\arraystretch}{1.3} % Increase vertical spacing
\rowcolors{3}{white}{gray!10} % Apply zebra striping from row 3 onward
\begin{tabular}{
    >{\columncolor{white}}c
    >{\columncolor{headerYellow}}c
    >{\columncolor{headerYellow}}c
    >{\columncolor{headerYellow}}c
    >{\columncolor{headerYellow}}c
    >{\columncolor{headerGreen}}c
    >{\columncolor{headerGreen}}c
    >{\columncolor{headerGreen}}c
    >{\columncolor{headerGreen}}c}
\multicolumn{9}{c}{\cellcolor{white}\textbf{Alpha track length (cm)}} \\
\hline
\cellcolor{white} & 
\multicolumn{4}{c}{\textbf{Argon}} & 
\multicolumn{4}{c}{\textbf{Neon}} \\
\hline
\textbf{Energy (MeV)} & {\cellcolor{headerYellow}\textbf{1 bar}} & {\cellcolor{headerYellow}\textbf{2 bar}} & {\cellcolor{headerYellow}\textbf{4 bar}} & {\cellcolor{headerYellow}\textbf{10 bar}} & 
{\cellcolor{headerGreen}\textbf{1 bar}} & {\cellcolor{headerGreen}\textbf{2 bar}} & {\cellcolor{headerGreen}\textbf{4 bar}} & {\cellcolor{headerGreen}\textbf{10 bar}} \\
\hline\hline
5.3  & 4.34 & 2.17 & 1.08 & 0.43    & 7.25 & 3.62 & 1.81 & 0.72 \\
5.5  & 4.58 & 2.29 & 1.14 & 0.46    & 7.64 & 3.82 & 1.91 & 0.76\\
6  & 5.20 & 2.60 & 1.30 & 0.52   & 8.65 & 4.33 & 2.16 & 0.87 \\
7.7 & 7.59 & 3.80 & 1.90 & 0.76   & 12.54 & 6.27 & 3.14 & 1.25 \\

\bottomrule
\end{tabular}
\smallskip
\caption{Alpha track length (CSDA range) in cm for argon (yellow) and neon (green) at different pressures for different energies.}\label{tab:AlphaLength}
\end{table}

%%%%%%% Drift velocity %%%%%%%%%%%%
\begin{table}[h]
\centering
\small % Reduce font size to fit the table
\renewcommand{\arraystretch}{1.3} % Increase vertical spacing
\rowcolors{3}{white}{gray!10} % Apply zebra striping from row 3 onward
\begin{tabular}{
    >{\columncolor{white}}c
    >{\columncolor{headerYellow}}c
    >{\columncolor{headerYellow}}c
    >{\columncolor{headerYellow}}c
    >{\columncolor{headerYellow}}c
    >{\columncolor{headerGreen}}c
    >{\columncolor{headerGreen}}c
    >{\columncolor{headerGreen}}c
    >{\columncolor{headerGreen}}c}
\multicolumn{9}{c}{\cellcolor{white}\textbf{Electron drift velocity (cm/$\upmu$s)}} \\
\hline
\cellcolor{white} & 
\multicolumn{4}{c}{\textbf{Argon}} & 
\multicolumn{4}{c}{\textbf{Neon}} \\
\hline
\makecell{\textbf{E field} \\ (V/cm/bar)} & {\cellcolor{headerYellow}\textbf{Pure}} & {\cellcolor{headerYellow}\textbf{1\% Iso}} & {\cellcolor{headerYellow}\textbf{2\% Iso}} & {\cellcolor{headerYellow}\textbf{10\% Iso}} & 
{\cellcolor{headerGreen}\textbf{Pure}} & {\cellcolor{headerGreen}\textbf{1\% Iso}} & {\cellcolor{headerGreen}\textbf{2\% Iso}} & {\cellcolor{headerGreen}\textbf{10\% Iso}} \\
\hline\hline
10  & 0.13    & 0.63    & 0.45    & 0.18    & 0.15   & 0.48   & 0.46   & 0.20 \\
100  & 0.24   & 2.43    & 3.29    & 2.51    & 0.4    & 1.55   & 1.88   & 1.88 \\
500  & 0.35   & 1.41    & 2.08    & 4.56    & 0.85   & 1.86   & 2.31   & 3.84 \\
1000 & 0.42   & 1.32    & 1.84    & 4.0     & 1.67   & 2.39   & 2.72   & 3.93 \\
\bottomrule
\end{tabular}
\smallskip
\caption{Electron drift velocity in cm/$\upmu$s for argon (yellow) and neon (green) mixtures with different percentages of isobutane and different electric fields.}\label{tab:DriftVelocty}
\end{table}

%% file: Chapters/3_TREXDM.tex
\section{The experiment}

TREX-DM (TPC for Rare Event eXperiments - Dark Matter) \cite{iguaz2016trex} is a time projection chamber with ultra low background and Micromegas readout planes, developed at the University of Zaragoza for low mass WIMP searches, below 10  GeV c$^{-2}$. It hosts two twin TPCs sharing the same cathode at high pressure, up to 10 bar. It is installed in the Laboratorio Subterráneo de Canfranc (LSC) since late 2018, and during these years, continuous improvements in background level and threshold have been achieved. 

The project T-REX (Time projection chambers for Rare Event eXperiments) got a ERC Starting Grant between 2009 and 2015 to develop low background gaseous detectors for rare event searches. Two main developments came out of this R\&D project. For neutrino-less double beta decay a partnership first with NEXT \cite{cebrian2010micromegas} and after with PANDA \cite{chen2017pandax} collaborations explored the usefulness of Micromegas readout planes for this rare decay searches. And regarding dark matter searches, TREX-DM emerged \cite{iguaz2016trex} as a competitive experiment due to the intrinsic low background of these detectors.
This experiment profits from the experience of the group working with gaseous detectors and Micromegas readout planes as X-ray detectors for the solar helioscope CAST \cite{cast2017new} which has been producing physics results until very recently \cite{altenmuller2024new}. For the next generation helioscope, BabyIAXO \cite{abeln2021conceptual}, similar detectors are being developed pushing background levels to even lower values \cite{altenmuller2024background}. As mentioned, this technology has been applied in the past for other rare event searches like neutrino-less double beta decay in the PandaX-III experiment, and also for dedicated detectors for radiopurity measurements like AlphaCAMM \cite{altenmuller2022alphacamm}, a surface contamination detector. 

With this background, all ingredients were available to develop a dark matter detector based on a gaseous TPC. The main challenge was the limited mass available in a gaseous detector, that can be partially mitigated by using a big chamber able to operate at high pressure. This implies big readout planes and a lot of material near the active volume that need to be carefully selected to minimize radioactivity-related background. Pressure was not a problem, as a dedicated chamber was designed to withstand pressures exceeding 10 bar. And a thorough campaign of radiopurity measurements were performed to select the material of most inner components. For readout planes, two large radiopure microbulk Micromegas, 25x25 cm$^2$, were produced, the biggest and more radiopure up to that date. 

The target gas should maximize the sensitivity at low WIMP masses, what explains why neon was initially selected -smaller atomic mass means more energetic nuclear recoils for same WIMP mass, as already explained in section~\ref{sec:WIMPsignal}, see figure \ref{fig:WIMPrates} -. However, due to the scarcity of this gas provoked by the Russian war on Ukraine (the latter being one of the biggest producers of noble gases worldwide), and because it is a mixture we are very familiar with, argon-isobutane ($C_4H_{10}$) is currently employed. Isobutane is a quencher gas, called like this because it helps to prevent UV photons escaping from electron avalanches and ionizing further in the gas volume, so it mitigates the sparks, making the detector more stable and increasing the maximum gain -at least at low pressures-. The low atomic mass of carbon and hydrogen also helps to increase sensitivity to low mass WIMPs.

The detector was tested in University of Zaragoza before being installed in the LSC. There, together with the TPC, a shielding made of lead and polyethylene for environmental gammas and neutrons was installed partially surrounding the chamber. The main effort during these years has been reducing the background. As it will be further detailed later, this involved changing some of the critical elements of the detector, like the readout planes. Efforts to further reduce the energy threshold are ongoing and a hybrid structure of a GEM plus Micromegas readout plane is in use nowadays.

\begin{figure}
    \centering
    \includegraphics[width=0.8\linewidth]{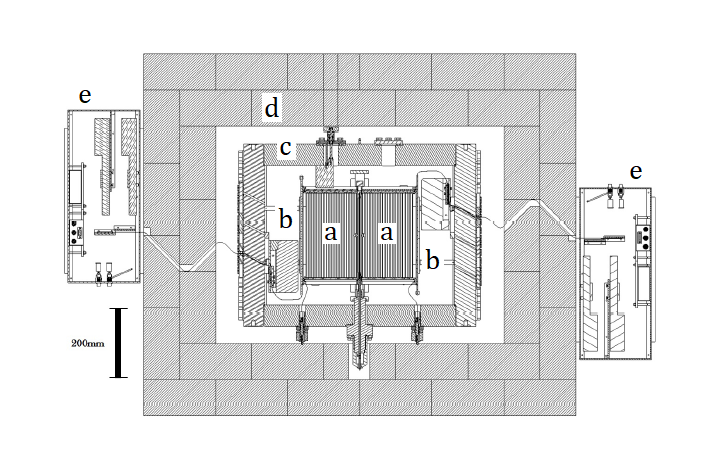}
    \caption{Experimental setup of TREX-DM: a) Twin time projection chambers. b) Micromegas readout planes. c) Copper vessel. d) Lead shielding. e) DAQ electronics. It is outside of the shielding to prevent radioactivity from electronic components.}
    \label{fig:TREXschema}
\end{figure}

\subsection{TPC design}

The TPC of TREX-DM is made of several elements: a vessel to contain the gas, a field cage to apply the drift field and the readout planes. Several ancillary systems are also crucial for the routine operation of the experiment like the calibration system, the gas panel, the electronics and data acquisition and the slow control to monitor everything.
\\

\textbf{Copper vessel}

\begin{figure}[h]
    \centering
    \includegraphics[width=0.9\linewidth]{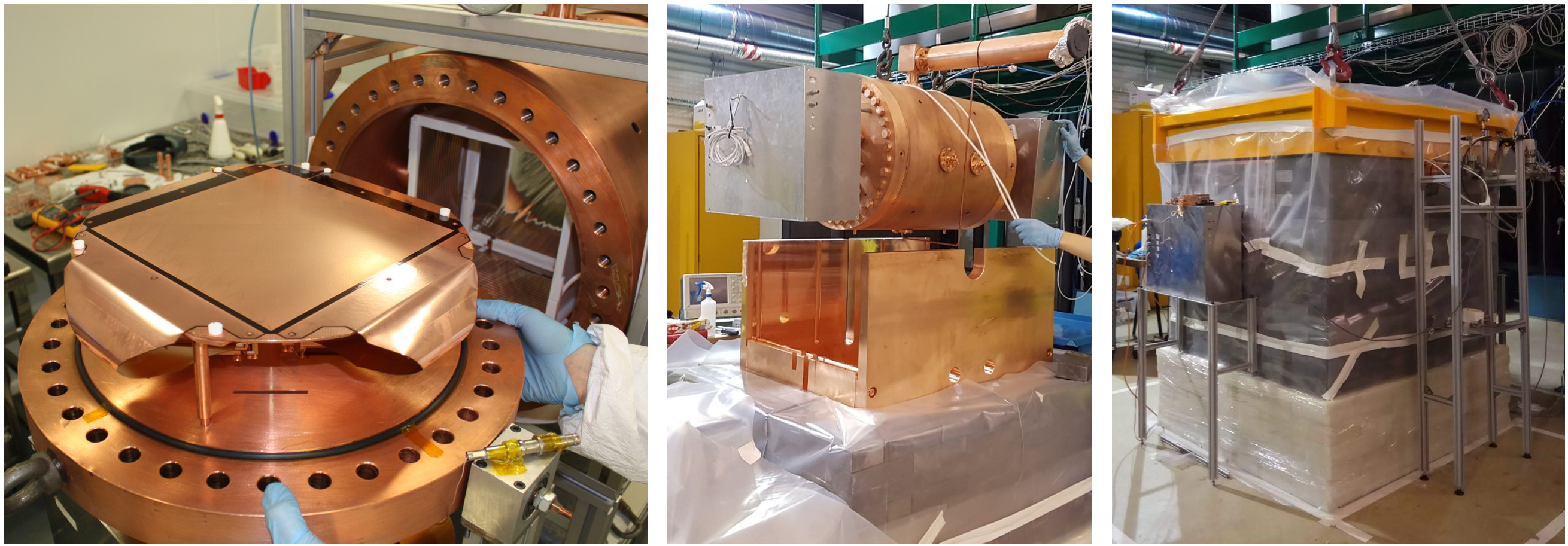}
    \caption{Left: Micromegas readout plane attached to one of the end cups. Middle: Closed vessel about to be placed inside the shielding. Right: TREX-DM is installed and operating in Hall A of the LSC. The lead shielding surrounding the chamber and the lower part of the polyethylene neutron shielding are visible \cite{mirallas2024planos}. }
    \label{fig:Vessel}
\end{figure}

\begin{figure}[h]
    \centering
    \begin{subfigure}{.6\textwidth}
        \centering
        \includegraphics[width=.95\linewidth]{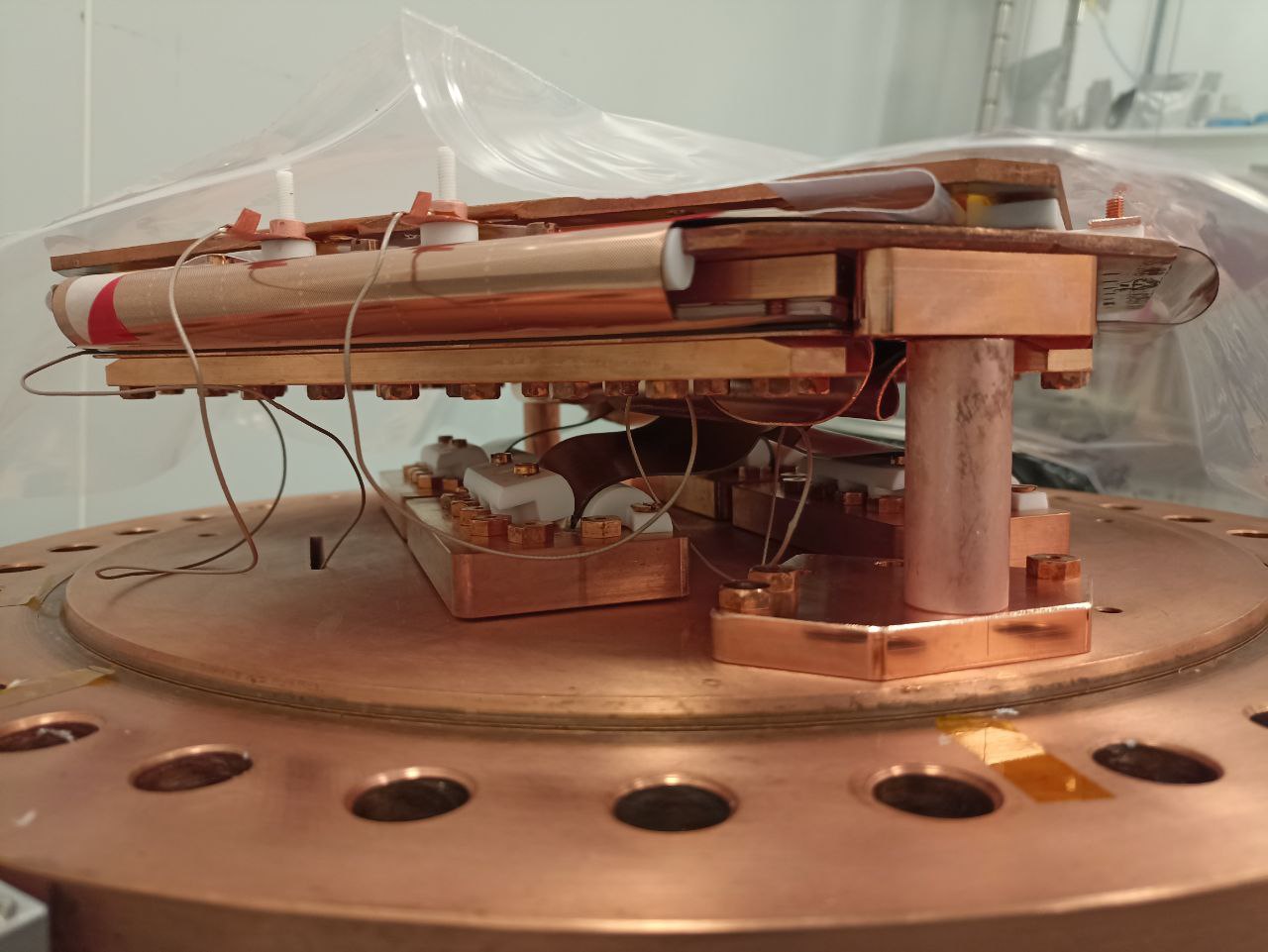}
    \end{subfigure}%
    \begin{subfigure}{.4\textwidth}
        \centering
        \includegraphics[width=.95\linewidth]{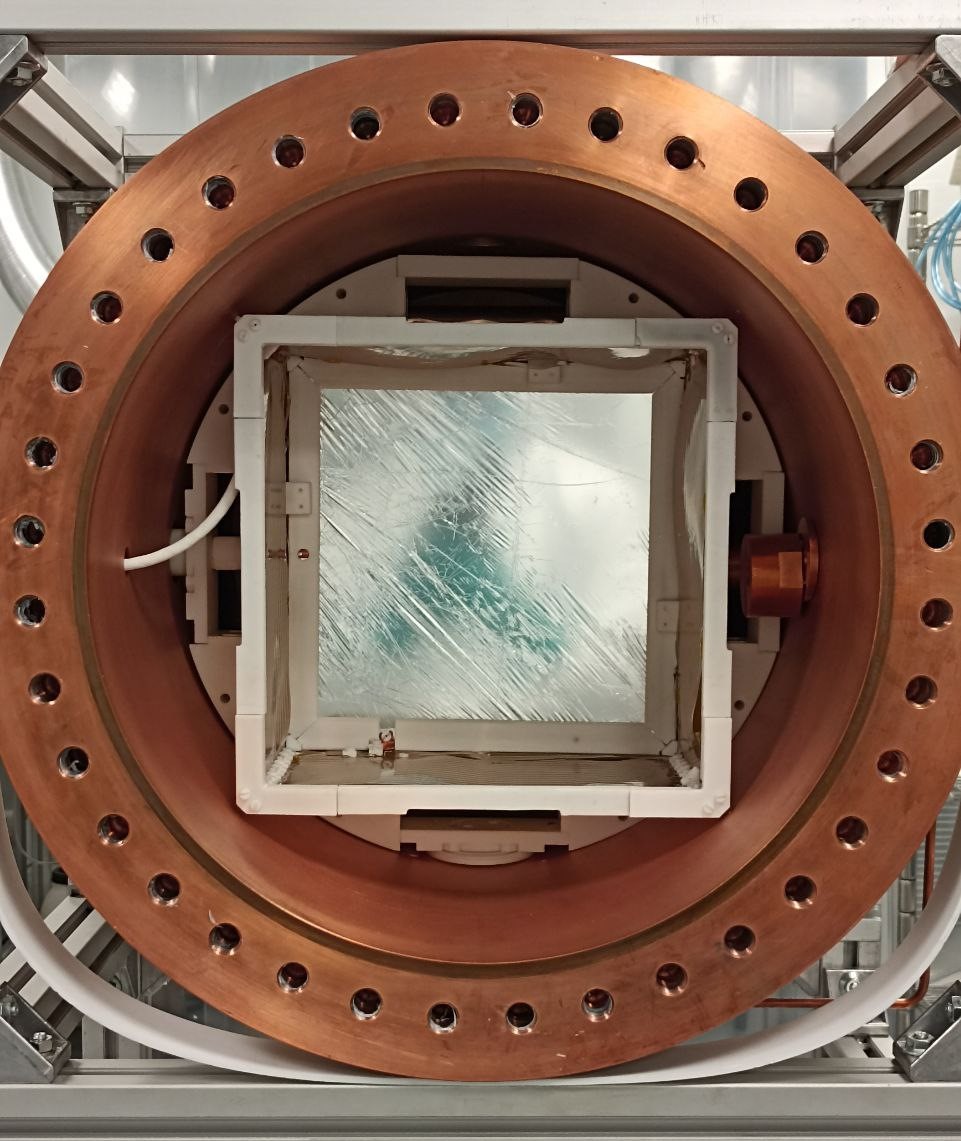}
    \end{subfigure}
    \caption{Left: Feedthroughs and support structure for a Micromegas plane in the end cap of the vessel. Right: View of one of the two halves of the field cage: the aluminized mylar cathode, reflecting in the middle; the teflon frame of the support structure of the field cage holding the field shaper. The copper block to shield the calibration source can be seen in the right. On the left, one of the high voltage cables to power the drift cage.} \label{fig:TREXinner}
\end{figure}

A 0.5 m length, 0.5 m diameter cylinder is the central body of the chamber, closed in both ends by two end caps. The body is made of Electrolytic Tough Pitch Copper (ETP Cu) and the end caps of  Oxygen Free Electronic Copper (OFE Cu). Their thickness is 6 cm, withstanding up to 12 bar; it also serves as the first component of the passive shielding to block external gamma radiation, see figure \ref{fig:Vessel}. The end caps have two additional features: on one hand, they act as the holder for the readout planes, which are attached to them through a copper structure; on the other, they guide the feedthrough cables with which the recorded signals are extracted, figure \ref{fig:TREXinner} left. The feedthroughs slits have been machined in diagonal with respect to the perpendicular of the endcap to impede any straight travelling particle reaching the active volume.
\\

\begin{comment}
\begin{figure}
    \centering
    \includegraphics[width=0.8\linewidth]{images/MMsupports.jpg}
    \caption{Feedthroughs and support structure for Micromegas plane in the end cup of the vessel.}
    \label{fig:MMsupports}
\end{figure}
\end{comment}

\textbf{Field cage}

Inside the vessel, the active volume is delimited by the field cage. It is a structure designed to generate an electric field capable to move electrons produced in ionizations towards the readout plane. TREX-DM is composed by two twin volumes sharing the same central cathode, that can operate as two independent detectors, each with half of the total volume and its own field cage. Each drift volume is 25x25x16 cm$^3$.

The field cages are composed by a cathode and the field shapers. In this case, they are copper rings printed on a thin sheet of kapton that follow the squared shape of the Micromegas readout plane. They are connected through 100~M$\Omega$ resistors with the purpose of maintaining appropriate voltages on each ring so that the electric field stays as homogeneous as possible. A detailed electronic diagram of the field cage can be seen in figure 5 of \cite{iguaz2016trex}.
\\
\begin{comment}
\begin{figure}
    \centering
    \includegraphics[width=0.5\linewidth]{images/MylarFieldCage.jpg}
    \caption{View of one of the two halves of the field cage: the auminized mylar cathode, reflecting in the middle; the teflon frame of the support structure of the field cage holding the field shaper. The copper block to shield the calibration source can be seen in the right. On the left, one of the high voltage cables to feed the drift cage.}
    \label{fig:MylarFieldCage}
\end{figure}
\end{comment}

\textbf{Calibration ports}

Regular energy calibrations are done between background runs. With this in mind, two special feedthroughs were designed to introduce radioactive sources close to the active volume. Each one consists of a $^{109}$Cd placed at the end of a rod in a small hole that crosses the vessel. It can be placed slightly inside the inner chamber but usually it is sufficient to leave as is. Each calibration port has a movable copper block that shields the activity when the source is not needed. The ports are denoted in the layout of the experiment, figure \ref{fig:TREXschema}, while the copper block corresponding to the nearest active volume can be spotted on the right hand part of figure \ref{fig:TREXinner} right picture.
\\

\textbf{Gases and recirculation system}

TREX-DM has been conceived to operate with neon and argon noble gases up to 10 bar with certain amount of isobutane as quencher gas \cite{castel2019background}. This means 0.3 kg of argon or 0.16 kg of neon as target mass. The main two mixtures,  $Ar + 1\% \, iC_4 H_{10}$ and $Ne+ 2\% \, iC_4 H_{10}$ were explored at different pressures in \cite{iguaz2022microbulk}. Transmission fields and amplification voltages for these were studied in a small time projection chamber with a non-segmented microbulk Micromegas.

\begin{figure}
    \centering
    \includegraphics[width=0.8\linewidth]{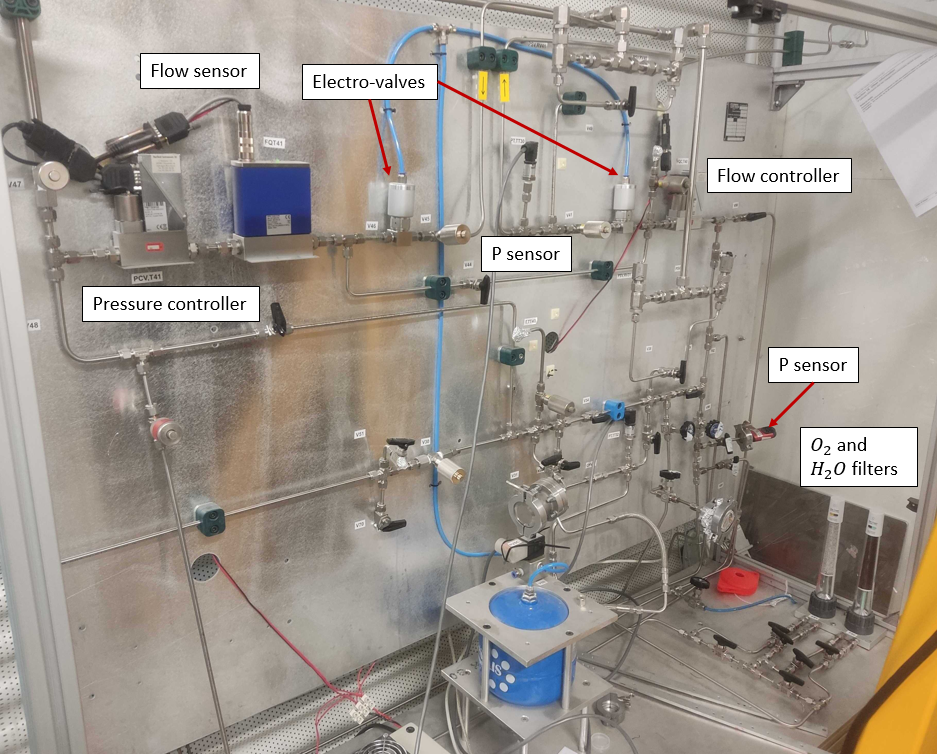}
    \caption{Gas panel of TREX-DM in LSC. It allows to fill the vessel make vacuum if needed, operate in open or closed loop and control flows and pressures.}
    \label{fig:GasPanel}
\end{figure}

Handling gases requires some effort and good logistics. In the LSC, gases come in bottles of 50 liters at 100 bar that have to be stored outside the lab or in safety cabinets. Dedicated pipes carry gas to the experiment passing through a gas panel that allows to control the flow and the pressure, see figure \ref{fig:GasPanel}. It also contains a binary gas analyzer in case gas purity checks are needed, oxygen and moisture filters and a pump. TREX-DM has operated in closed and open loop. In the first case, once the chamber is filled, gas circulates through the gas panel to filter it and avoid degradation of the mixture. This allows to use more expensive gases because the consumption is reduced. In the open loop schema, new gas is flowing inside the vessel all the time and it is continuously evacuated through the exhaust. In this case, the flux has to be tuned to minimize the expenditure of gas while preserving the performance of the experiment. 

The system can be coupled to a pumping station, that allows to pump it down to values of the order of $10^{-5}$ mbar after it has been opened to air and before injecting gas anew.

\begin{comment}
The chamber itself, has one inlet in the upper part of the vessel and the outlet in the bottom part, both in the middle of the cylindrical body. Emptying the chamber requires open the exhaust and sometimes flux with nitrogen. Injecting gas is more cumbersome in order to assure the good quality of the gas. The vessel has to be pumped for several ours to remove traces of oxygen and humidity, typically levels of $10^{-5}$ mbar are reached. This includes making vacuum in the pipes through with gas is going to flow. And then, new gas from the bottle is injected. If the filling needs to be done faster, both inlet and outlet pipes can be used. Then, a high gas flow is set for several hours, maybe between 20 and 50 liters per hour to make sure several refillings of the chamber are done, before going back to the usual flow, around 1 l/h for 1 bar. The total internal volume of the vessel is 60 liters.
\end{comment}

\subsection{Micromegas detectors} \label{MicromegasTREX}

\begin{figure}[h]
    \centering
    \begin{subfigure}{.45\textwidth}
        \centering
        \includegraphics[width=.9\linewidth]{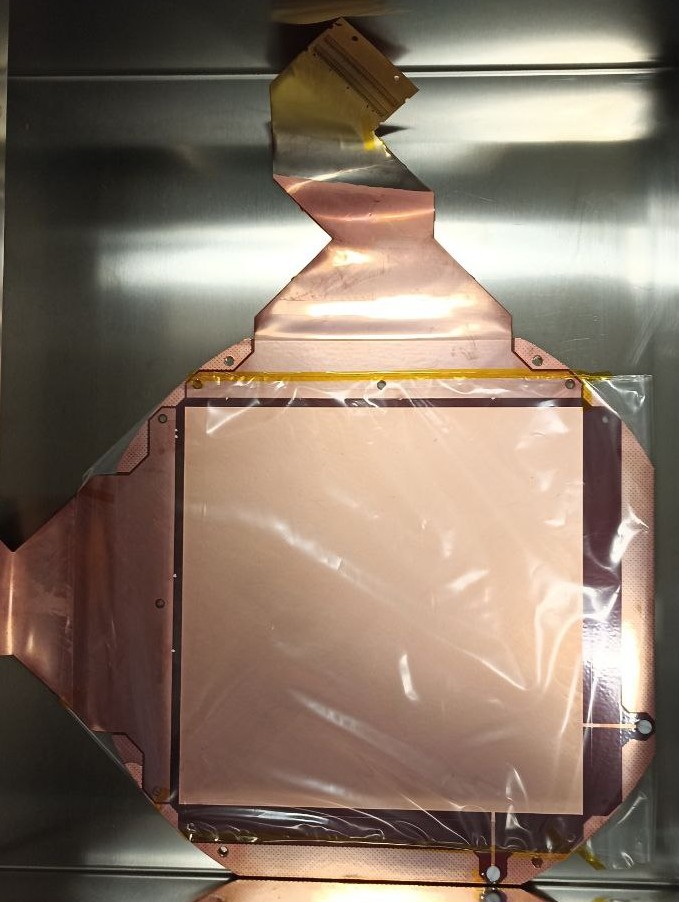}
    \end{subfigure}%
    \begin{subfigure}{.55\textwidth}
        \centering
        \includegraphics[width=.9\linewidth]{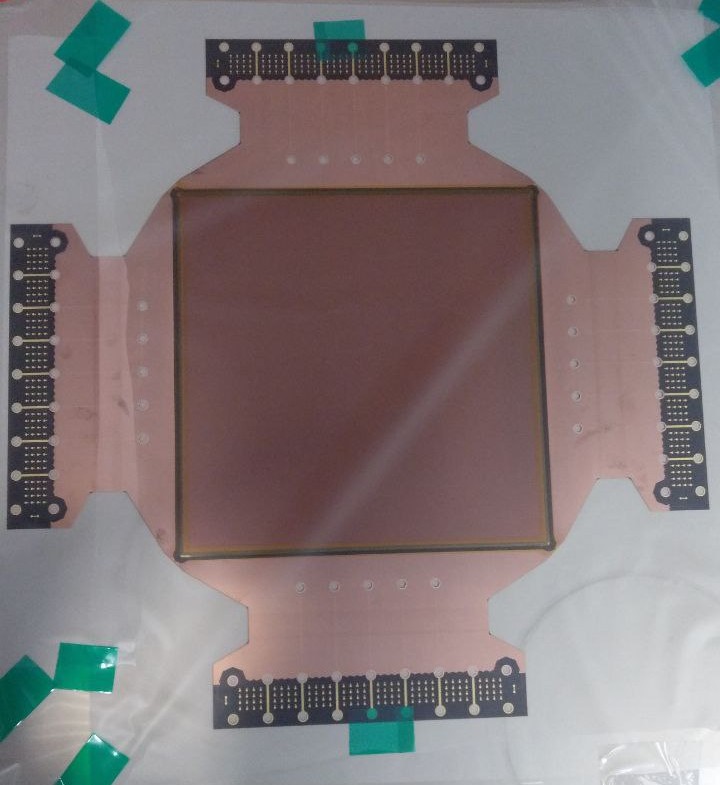}
    \end{subfigure}
    \caption{Left: First Microbulk Micromegas installed in TREX-DM in late 2018 (V1). Right: Second design of Microbulk Micromegas detector installed in TREX-DM in 2022 (V2).} \label{fig:TREXmicromegas}
\end{figure}

\begin{figure}
    \centering
    \includegraphics[width=0.8\linewidth]{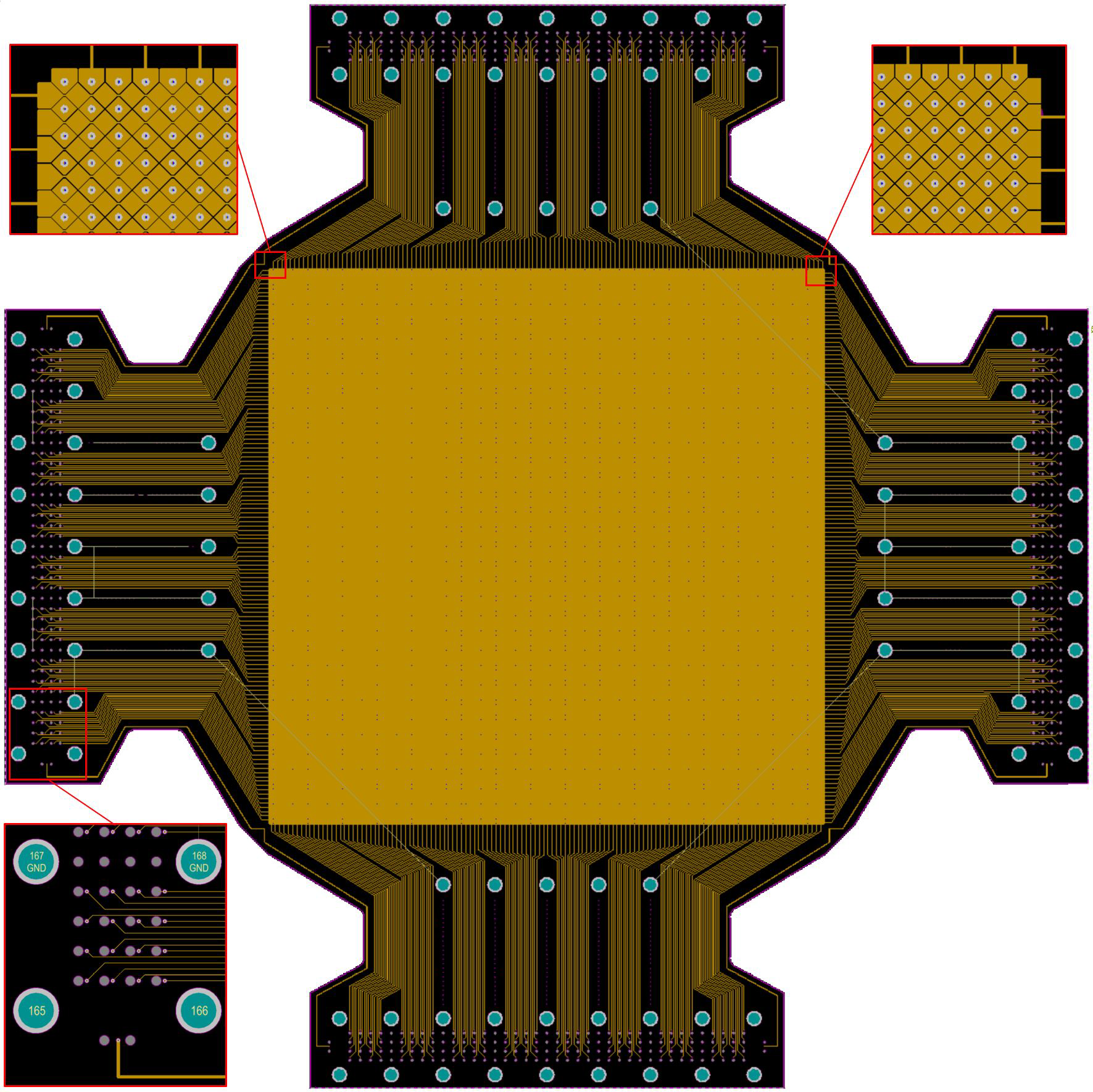}
    \caption{Drawing of the second layer  of the Microbulk Micromegas detectors version 2. In this layer, 512 channels extracting signals from 25x25 cm$^2$ active area can be seen. Upper inlets with close ups of the pixels of each strip and the extraction in both directions. Lower inlet, small portion of the face-to-face connector. Image from \cite{mirallas2024planos}. }
    \label{fig:MM_v2design}
\end{figure}

\begin{figure}
    \centering
    \includegraphics[width=0.99\linewidth]{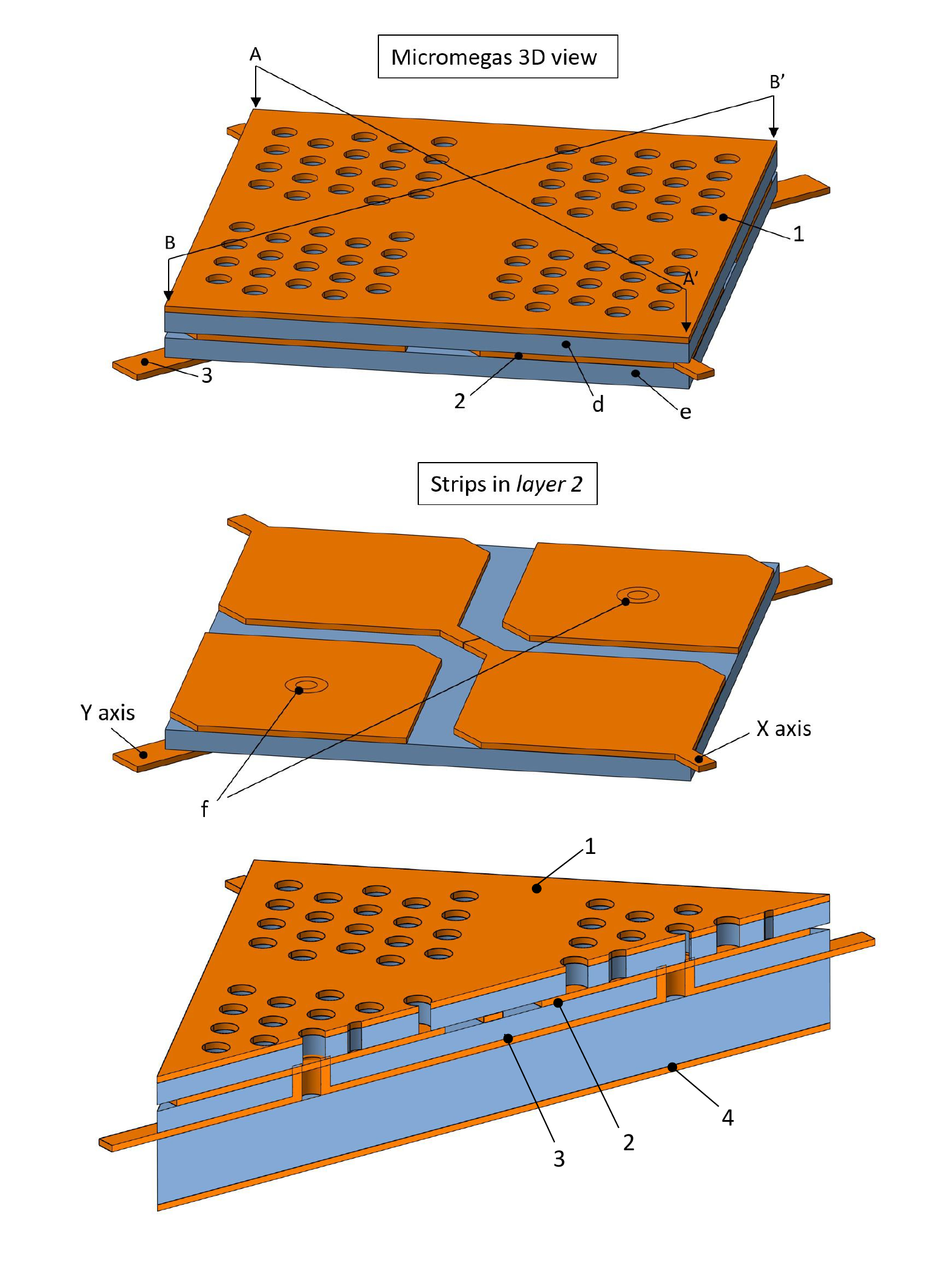}
    \caption{Close look of the 3D model of the microbulk Micromegas V2 design. In blue kapton, in orange copper. \textit{Upper}: Top layer (1) showing the mesh. Hole pattern over each pixel can be seen, with strips in X direction in layer 2, and Y direction in layer 3. \textit{Middle}: View of layer 2. Charge collected in X pixels is read through strips in this layer while Y pixels have connections, denoted by (f), with layer 3 where its signal is extracted. \textit{Bottom}: Slice of the general view of the Micromegas. Another layer is included, number 4, 150 $\upmu$m below for grounding. Distance between layers 1, 2 and 3 is 50 $\upmu$m.  Copper layers have 17 $\upmu$m thickness. Adapted from \cite{mirallas2024planos}.}
    \label{fig:MM_3Dmodel}
\end{figure}

The Micromegas of TREX-DM have an active area of 25x25 cm$^2$ and 512 channels in total, half in the X direction and the other half in the Y. Table \ref{tab:micromegas_comparison} shows the differences of the two designs, V1 and V2, that have been used along the years (figure \ref{fig:TREXmicromegas}). The main changes concern improvements towards avoiding crosstalk and leak currents, by increasing the distance in the routing of the channels (see figure \ref{fig:MM_v2design} for a detailed view of one of the routing layers), and reducing the density of pads in the connections, as will be commented in section \ref{sec:MMTREX}. This lead to the signals being extracted in four flaps instead of two, and the necessity to implement a new decoding (correspondence between detector position and DAQ channel). For a close look of the layers and pixel disposition in the Micromegas V2 see figure \ref{fig:MM_3Dmodel}. 

However, a big difference was related to the radiopurity of the new planes. The main source of radioctivity had been identified as $^{40}$K coming from the etching of the holes in the mesh carried out using Potassium Hydroxide (KOH). This chemical compound deposits as residue on the walls of the holes, thereby contributing to contamination through the isotope $^{40}$K. Consequently, the MPT group at CERN, manufacturers of these devices, proposed a cleaning process with deionized water that could be applied in the final step of the production process to reduce this contamination. To track the source of radioisotopes, samples were measured in germanium detectors in the LSC to trace their evolution along the different steps of the fabrication. In \cite{mirallas2024planos} the whole process is described in detail. 

\begin{comment}
In TREX-DM two different microbulk Micromegas designs have been used. First design can be seen in figure \ref{fig:MM_v1} and second in \ref{fig:MM_v2}. Both have an active area of 25x25 cm$^2$ and 512 channels in total, half to read strips in X direction and the other half for Y direction. Two of these readout planes are installed inside the vessel. In the second Micromegas version a number of changes were introduced. The idea behind most of them was to increase the distances between channels in all the device to prevent possible crosstalk and leak currents. In table \ref{tab:micromegas_comparison} can be seen the parameters that were changed. A detailed explanation of the modifications can be found in \cite{mirallas2024planos}. From the physics point of view, maybe the slightly different hole pattern in the mesh is the most significant change. Form a practical side, in version 2, Micromegas has four flaps from which signals are extracted, so a new electronic decoding had to be implemented in order to assign a position in the detector to each electronic channel.     
\end{comment}

\begin{table}[h]
    \centering
    \renewcommand{\arraystretch}{1.2}
    \begin{tabular}{|l|c|c|}
        \hline
        \textbf{Parameter} & \textbf{MM V1} & \textbf{MM V2} \\
        \hline
        Minimum distance between channels (\textmu m) & 75 & 500 \\
        Minimum distance between channels and ground (\textmu m) & 200 & 4000 \\
        Number of flaps (units) & 2 & 4 \\
        Distance between pads in the connector bellows (\textmu m) & 150 (fujipoly) & 4000 (FtF) \\
        Amplification gap (\textmu m) & 50 & 50 \\
        Distance between \textit{strips} in the active area (\textmu m) & 50 & 100 \\
        Hole pattern (Diameter-Pitch) (\textmu m) & 50 - 100 & 60 - 110 \\
        \hline
    \end{tabular}
    \caption{Comparison of design parameters between Micromegas V1 (installed in TREXDM in 2018) and Micromegas V2 (installed in TREXDM in 2022). Table extracted from \cite{mirallas2024planos}.}
    \label{tab:micromegas_comparison}
\end{table}

\begin{comment}
\begin{figure}
   \begin{minipage}{0.48\textwidth}
     \centering
     \includegraphics[width=0.85\linewidth]{images/MM_v1.jpg}
     \caption{First Microbulk Micromegas installed in TREX-DM in late 2018.}\label{fig:MM_v1}
   \end{minipage}\hfill
   \begin{minipage}{0.48\textwidth}
     \centering
     \includegraphics[width=0.99\linewidth]{images/MM_v2.jpg}
     \caption{Second design of Microbulk Micromegas detector installed in TREX-DM in 2022.}\label{fig:MM_v2}
   \end{minipage}
\end{figure}   
\end{comment}

\subsection{Electronics and DAQ system}

\begin{figure}
    \centering
    \includegraphics[width=0.8\linewidth]{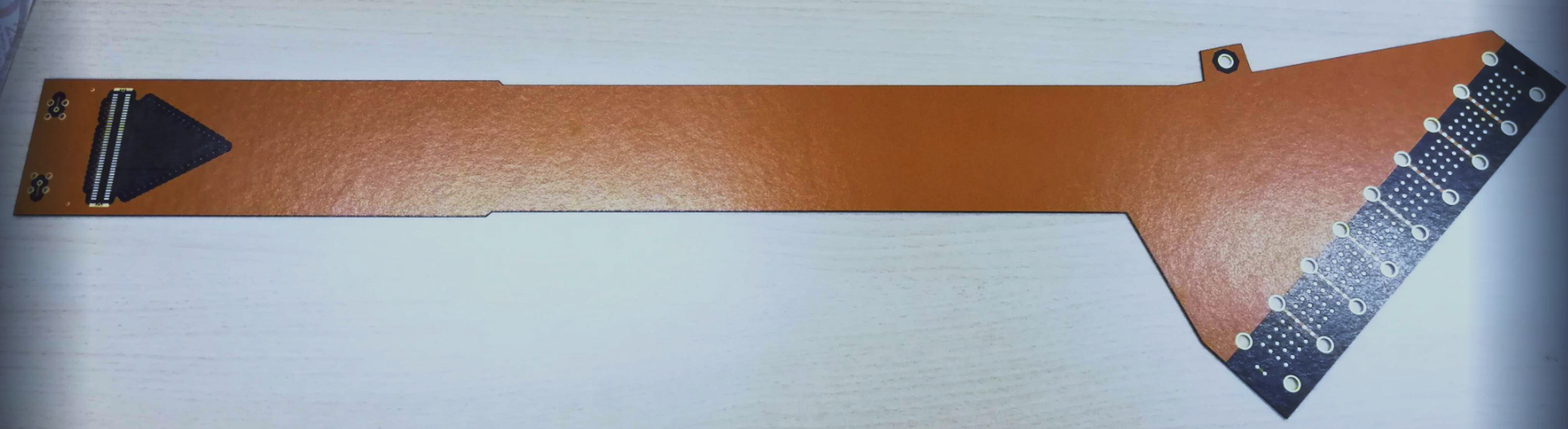}
    \caption{Flat cable for Micromegas V2 made of copper and kapton.}
    \label{fig:FlatCable}
\end{figure}

\begin{figure}
    \centering
    \includegraphics[width=0.8\linewidth]{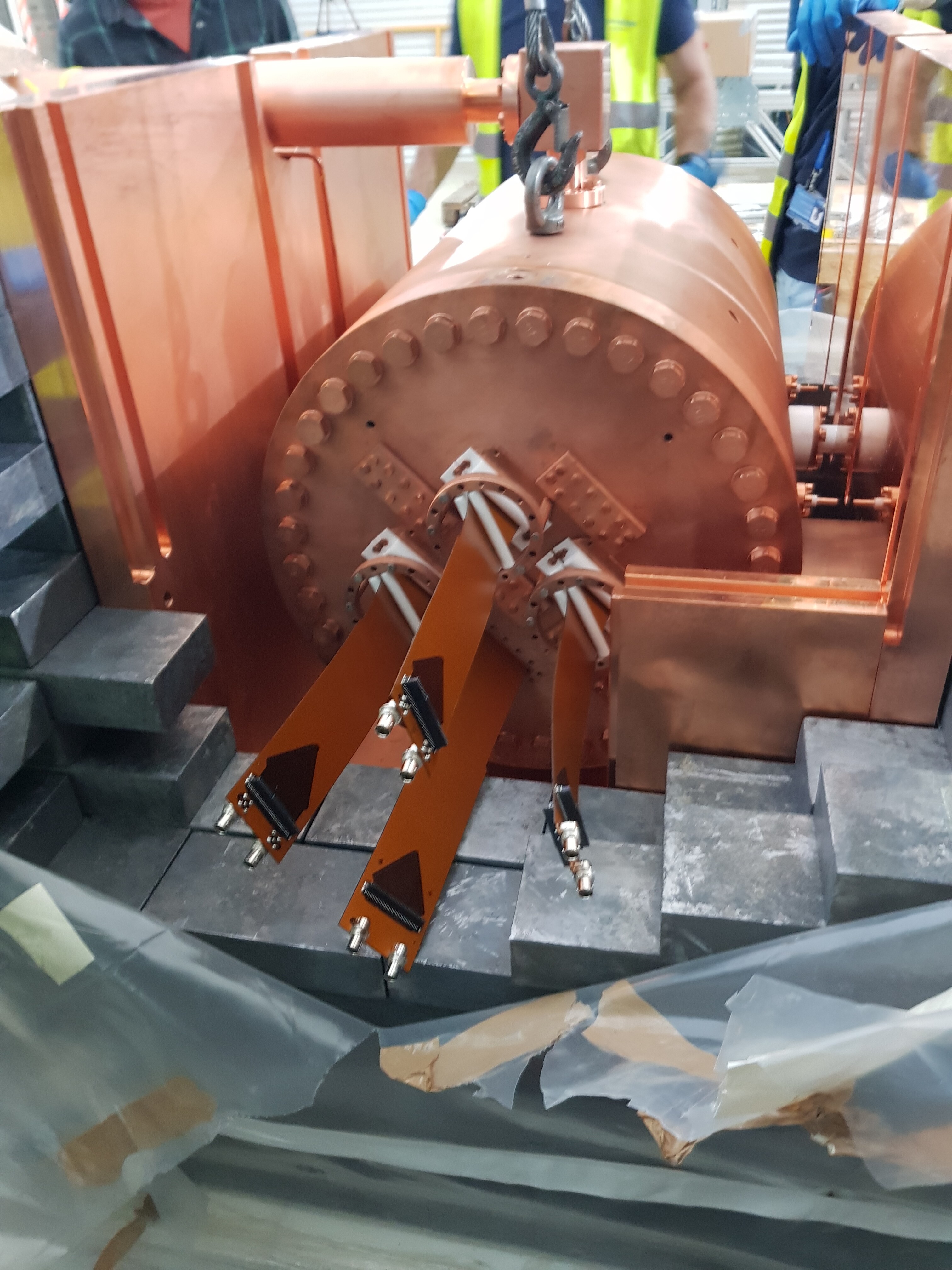}
    \caption{Installing the vessel inside the shielding. The four flat cables of one of the Micromegas can be seen coming out from one of the end caps.}
    \label{fig:VesselCables}
\end{figure}

Signals from the Micromegas are extracted outside the vessel through flat radiopure cables. They are fabricated following the same procedure and with the same materials as the Micromegas. Two of them are needed for Micromegas V1 and four for V2. In figure \ref{fig:FlatCable} one of the flat cables for V2 can be seen. These flat cables are long enough to traverse the first layer of lead shielding, as seen in figure~\ref{fig:VesselCables}, and leave the Front-End electronics, as a source of background, outside. Specific connectors were installed at each end of the flat cables. Both versions make use of ERNI connectors towards the Front-End electronics and BNC ports for grounding. In figure \ref{fig:VesselCables} the four flat cables from one side can be seen, each carrying 128 channels, coming out of the chamber. But towards the Micromegas an upgrade was implemented between V1 and V2. Micromegas V1 had Fujipoly connectors whose tight footprint lead to leak currents. This is prevented in Micromegas V2 with a new design of Face-to-Face connections.

\begin{comment}
For Micromegas V1, Fujipoly connectors were used to couple Micromegas with flat cables but due to leak currents discovered in the connectors, for V2 they were substituted by Face-to-Face connections. In the other end, outside the vessel, flat cables have ERNI connectors for the signals and BNC for grounding. In figure \ref{fig:VesselCables} the four flat cables from one side can be seen, each carrying 128 channels, coming out of the chamber.
\end{comment}

\begin{figure}
    \centering
    \includegraphics[width=0.6\linewidth]{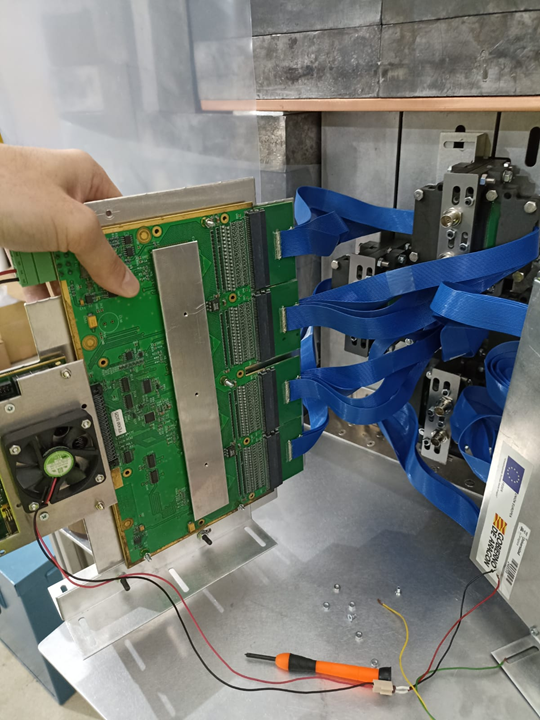}
    \caption{Connection between the flat cables and the FEC-Feminos through the blue cables. }
    \label{fig:BlueCables}
\end{figure}

\begin{figure}
    \centering
    \includegraphics[width=0.8\linewidth]{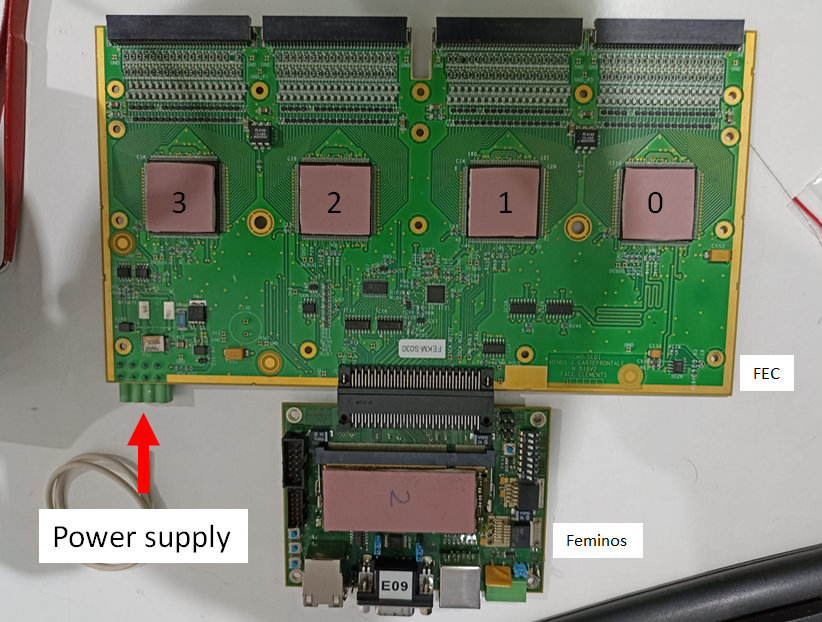}
    \caption{FEC-Feminos with four AGET chips. Two of these boards are needed for each Micromegas, four in total for the TREX-DM experiment. }
    \label{fig:FecFeminos}
\end{figure}

\begin{figure}
    \centering
    \includegraphics[width=0.9\linewidth]{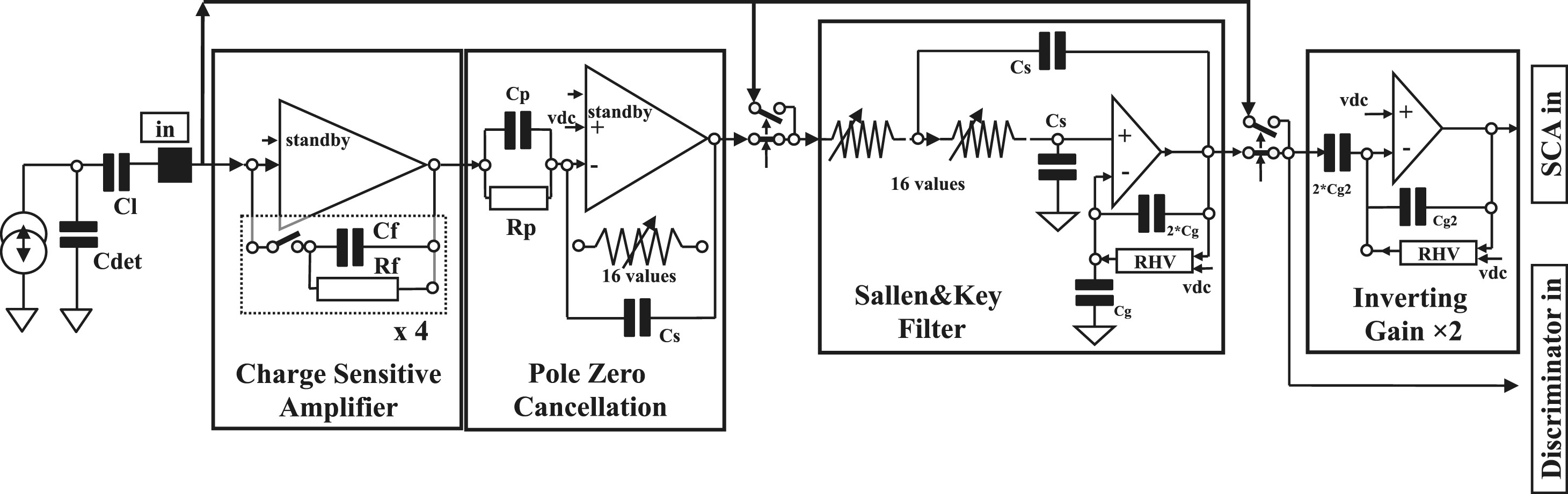}
    \caption{Schematic overview of the steps inside a channel of the AGET chip that consists of the charge sensitive amplifier (CSA), the pole zero cancellation (PZC), the Sallen and Key Filter (SK) and the inverting 2 gain (Gain-2) stages. Arrows at the top of the figure indicate which of these stages can be bypassed. From \cite{pollacco2018get}.}
    \label{fig:AGETSchema}
\end{figure}

\begin{figure}
    \centering
    \includegraphics[width=0.8\linewidth]{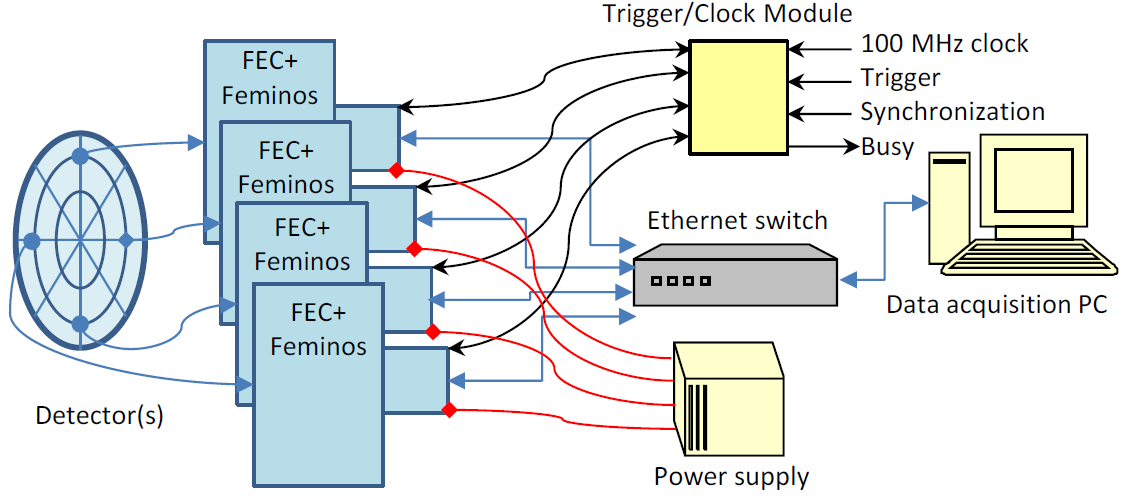}
    \caption{Diagram of the acquisition system. Four FEC-Feminos boards synchronized through a TCM board and connected to the control computer with a switch. Image from \cite{calvet2014versatile}.}
    \label{fig:AcquisitionSchema}
\end{figure}

\begin{figure}
\centering
\begin{subfigure}{.5\textwidth}
  \centering
  \includegraphics[width=.99\linewidth]{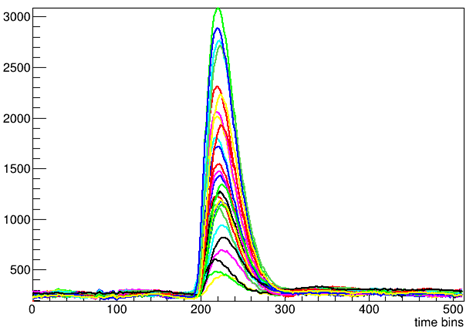}
  \caption{Only hit channels.} \label{fig:PulsesHit}
\end{subfigure}%
\begin{subfigure}{.5\textwidth}
  \centering
  \includegraphics[width=.99\linewidth]{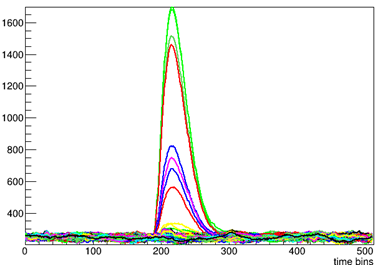}
  \caption{All channels of the Micromegas.} \label{fig:PulsesAllChannels}
\end{subfigure}
\caption{Examples of recorded pulses with AGET chips.}\label{fig:Pulses}
\end{figure}

Signals reach the acquisition system with the help of blue flat cables that connect ERNI connectors in the flat cables with those in the boards, as shown in  figure \ref{fig:BlueCables}. FEC-Feminos boards are used to digitalize and save the signals that are sent to the computer, see figure \ref{fig:FecFeminos}. They contain the AGET chip, the core element of the acquisition system, whose internal electronic chain is described \ref{fig:AGETSchema}. Each board has 4 chips, recording 64 channels each. This means that two boards are needed for every Micromegas of 512 channels, so 4 in total for both sides of TREX-DM experiment. These chips have the capability of recording signals with different sampling rates, sampling times, gain, position in the time window... Auto trigger is the main feature compared with older versions (AFTER) and a TCM board (Trigger Clock Module) is used to distribute trigger signals sent by any of the chips to other boards. This enables both sides to function together as a single detector for coincidence event detection. In figure \ref{fig:AcquisitionSchema} the schema of the acquisition system is shown. Also, several acquisition modes are possible: save only hit channels when trigger is launched or save all channels, whether there is a pulse or not. These capabilities are extremely useful to better extract the topology of the events registered in the TPC because some have features that are better seen in different acquisition configurations (gain, time length, sampling rate...). Some examples of pulses recorded with FEC-Feminos boards and AGET chips can be seen in \ref{fig:Pulses}.
 
\subsection{Shielding}

The LSC is located in a tunnel in the Spanish Pyrenees, under 850 m of rock of the ``Tobazo" mountain. This amount and type of rock blocks very efficiently the cosmic radiation, a factor of five orders of magnitude. This characteristic makes it the 3$^{rd}$ deeper one in Europe only after the Gran Sasso Laboratory in Italy and Laboratoire Souterrain de Modane in France. The mountain constitutes the first shielding layer of the experiment, as the cosmic rate flux in Hall A, where TREX-DM was located between 2018 and 2023, is $(5.26 \pm 0.21)\times 10^{-3}$ m$^{-2}$s$^{-1}$ \cite{trzaska2019cosmic}.

However, the natural radioactivity around the detector, from the rock itself and all the materials surrounding the experiment, is important. For this reason, a specific shielding for environmental gammas and neutrons has been designed for TREX-DM. It consists in 5~cm copper cradle for the vessel, 20~cm thick layer of lead and 40~cm of polyethylene at top and bottom plus water tanks surrounding the lead castle \cite{castel2019background}. The only neutron shielding in place until early 2025 was the layers below the chamber; in early 2025 the upper part was completed and a number of water tanks were placed around the detector covering almost 40\% of the perimeter. This was a decision motivated by the continuous interventions to the detector needed for the improvements. The full neutron shielding is expected to be mounted early 2026. 
In figure \ref{fig:Vessel} the copper cradle and the complete lead castle with the polyethylene bottom layer can be seen. Picture \ref{fig:VesselCables} is also illustrative of the thickness of the lead shielding.

\section{Data management and analysis}

\subsection{Data storage}
\begin{comment}
Data recorded with the FEC-Feminos boards is sent to the acquisition computer. This computer is synchronized via cron job with a data server for long term storage. Periodically, data from acquisition computer is removed to prevent problems with storage. 
\end{comment}

FEC-Feminos boards send the data to the acquisition computer and from there to a data server for long term storage. Raw files are text files (.aqs) that store for every event the ID number for each recorded channel and a string of 512 numbers with the amplitude level of the signal for each bin. Plotting these raw events is what is shown in figures \ref{fig:Pulses}. A raw file also keeps some metadata of the run, like the settings of the AGET chip, time duration of the run, etc. For every event, an ID is assigned and the time stamp is also stored. Since late 2024, data is also stored directly in ROOT files thanks to our colleague Luis Obis.

\subsection{REST-for-Physics}
The REST-for-Physics (Rare Event Searches Toolkit for Physics) framework \cite{altenmuller2022rest} is a software mainly developed by the Grupo de Física Nuclear y Astropartículas (GIFNA) of the University of Zaragoza. It is a modular software created to analyse gaseous detectors data both from experiments and simulation. Having everything in the same framework simplifies the comparison between them.

With experiential data, like the .aqs or .root recorded with FEC-Feminos boards in TREX-DM, the file is conveniently interpreted by REST and recast into ROOT files, if needed. ROOT \cite{brun1997root} is an analysis software developed by CERN and widely used in particle physics community. ROOT files are particularly well suited to optimize the data storage of huge amounts of data, at the end, it was created to store and handle data from large particle accelerators like the LHC.

\begin{figure}[h]
    \centering
    \includegraphics[width=0.99\linewidth]{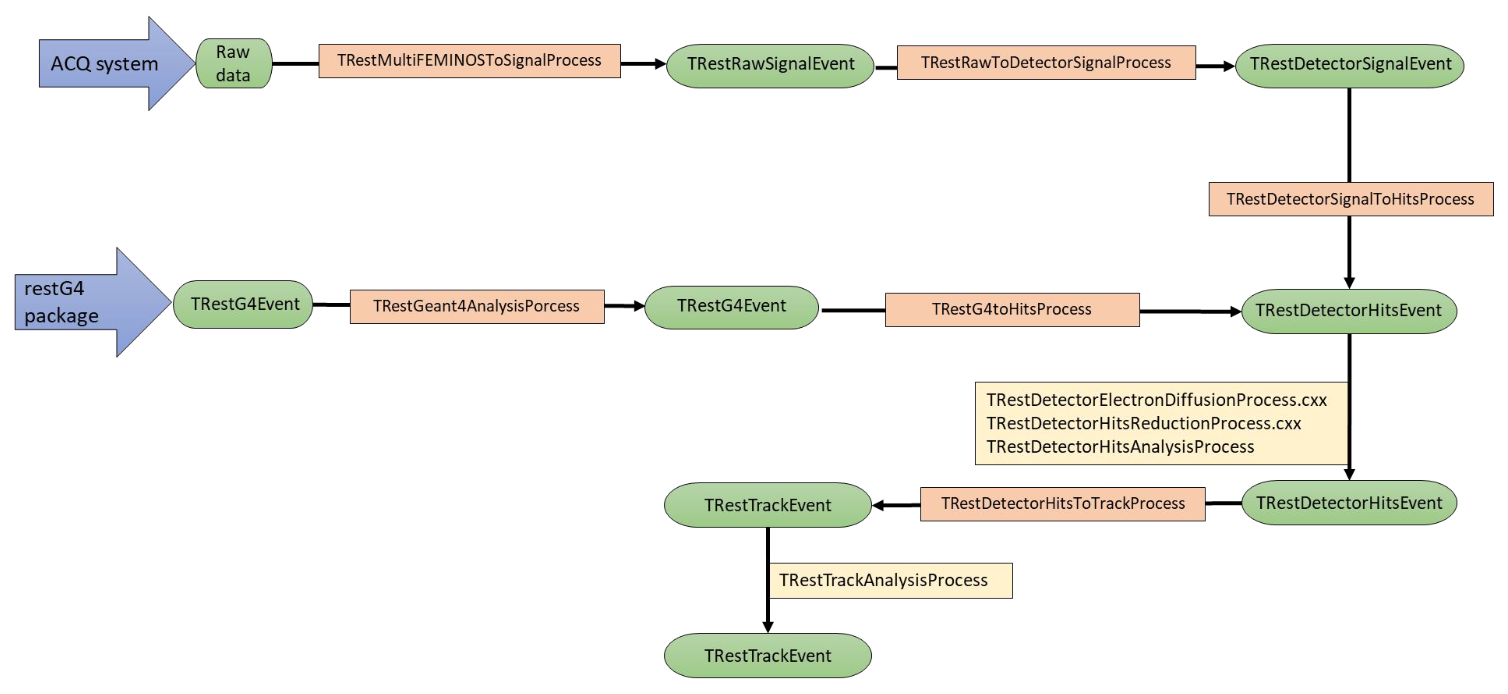}
    \caption{An example of a REST-for-Physics processing chain. In green different data types, in orange processes that transform between data types and in yellow analysis processes. Experimental and simulation branches merge at a certain point in the processing chain, facilitating the comparison between both.}
    \label{fig:RESTprocessingChain}
\end{figure}

The basic unit for data storage is the ``\texttt{TRestEvent}" \cite{REST}, but it can have different event types. The pulses shown if figure \ref{fig:Pulses} are ``\texttt{TRestRawSignalEvent}". At higher levels of analysis, pulses can be converted into ``hits", energy deposited in certain coordinates and time, this is a ``\texttt{TRestDetectorHitEvent}". And this can be further reinterpreted as a ``track", a path along which the particle has deposited energy, this is called ``\texttt{TRestDetectorTrackEvent}". The transformation between types of events is made through the ''\texttt{TRestEventProcess}" class. Many processes have been created, not only to transform data but also to analyse it. Many characteristics of the events are measured and stored in variables called ``observables". And the third main element of REST is the metadata class ``\texttt{TRestMetadata}" which stores the info of the run and the processes applied to the data. This is one of the main features of the framework as it allows to keep track of all the details of the analysts inside the data file. An example of a data processing chain with Rest-for-Physics is given in figure~\ref{fig:RESTprocessingChain}.

REST-for-Physics was developed knowing how crucial simulations are for an experiment devoted to rare event searches. Therefore, since its birth, GEANT4 \cite{agostinelli2003geant4} simulations have been implemented inside the framework. An extra layer was adopted to launch GEANT4 simulations from \textit{.rml} files using \textit{.gdml} to define the geometries. This was developed in the library ''\texttt{restG4}". Simulations are then treated in a way that result comparable with experimental data, this is shown in figure \ref{fig:RESTprocessingChain}. Electron drift in gaseous media is simulated, also the shaping to the collection of charges and noise can be added, and the spacial distribution that allow to identify in which channel charge would be collected... All processes that are needed to transform simulated data into the shape of real data are available. When simulated data arrive at ``\texttt{TRestRawSignalEvent}" satisfactorily, the same analysis chain can be applied to both types of data, simulated and experimental.

\subsection{Quick Analysis}
\begin{figure}[h]
    \centering
    \includegraphics[width=0.99\linewidth]{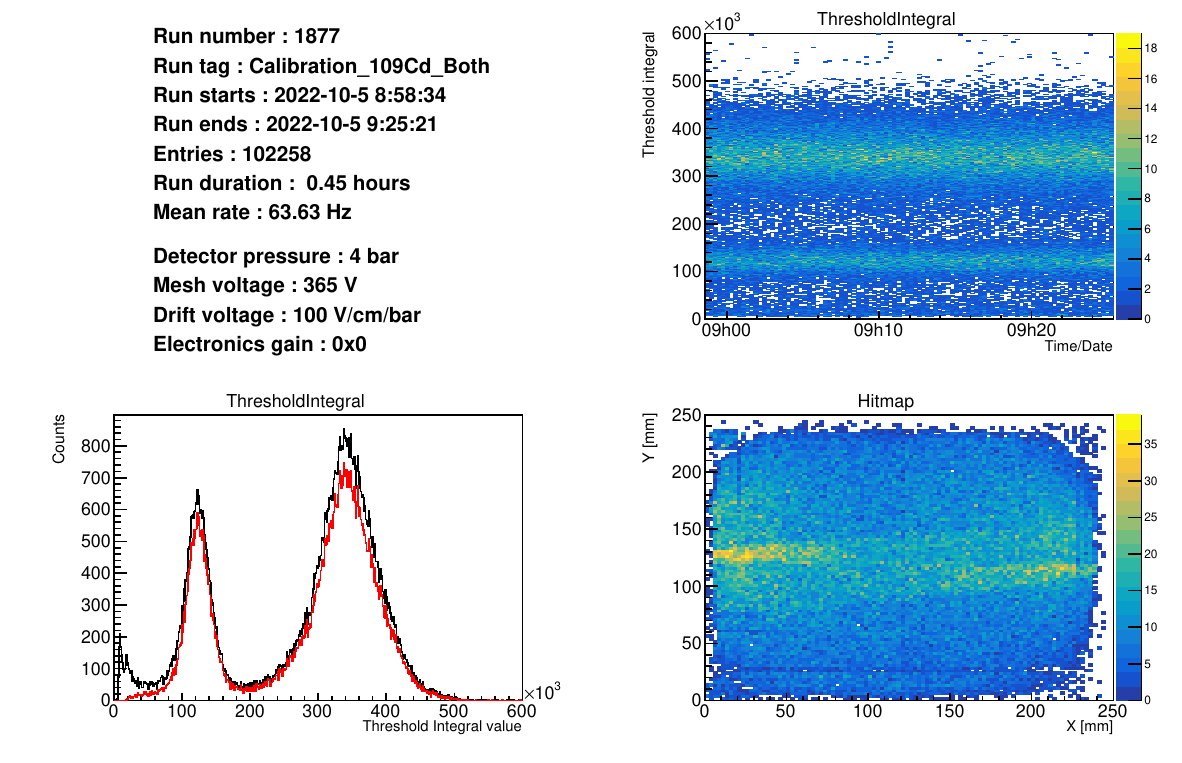}
    \caption{Quick Analysis report from a $^{109}$Cd calibration run. \textit{Upper left:} Metadata. \textit{Upper right:} Evolution plot of the energy spectrum. \textit{Lower left:} Energy spectrum. \textit{Lower right:} Hitmap. Keep in mind that events from both sides of the experiment are plotted together, this explain the two rays in the hitmap, one from each side.}
    \label{fig:SummaryQuickAna}
\end{figure}

Data is stored in the ``Sultan" server of GIFNA. Once a new file arrives, a cron job takes care of analysing it and producing several files with different event types and observables associated to them. To keep track of the evolution of the performance of the experiment, several reports are automatically generated with the metadata of the run and plots of key observables. ``Raw" contains maximum peak time delay, baseline sigma, average rise time, average peak time, number of signals and energy observables.``Hitmaps" plots the mean position of events applying different selection rules. ``Fiducial" plots energy spectra selecting events for smaller areas of the readout plane. ``Evolution" plots the same observables as ``Raw" but in 2D plots versus time. And ``Summary" is the shortest with just the metadata and three plots: energy spectrum, hitmap and evolution plot of the energy (energy vs time), see \ref{fig:SummaryQuickAna}. Metadata is the same in all reports. It gathers information of the conditions of the experiment for this run: gas pressure, mesh and cathode voltages, electronic gain, date, run time, number of events...

To facilitate the tracking of the data taking, the ``Summary" report is automatically sent through a dedicated Slack channel when it is produced.

\begin{figure}[h]
    \centering
    \includegraphics[width=0.9\linewidth]{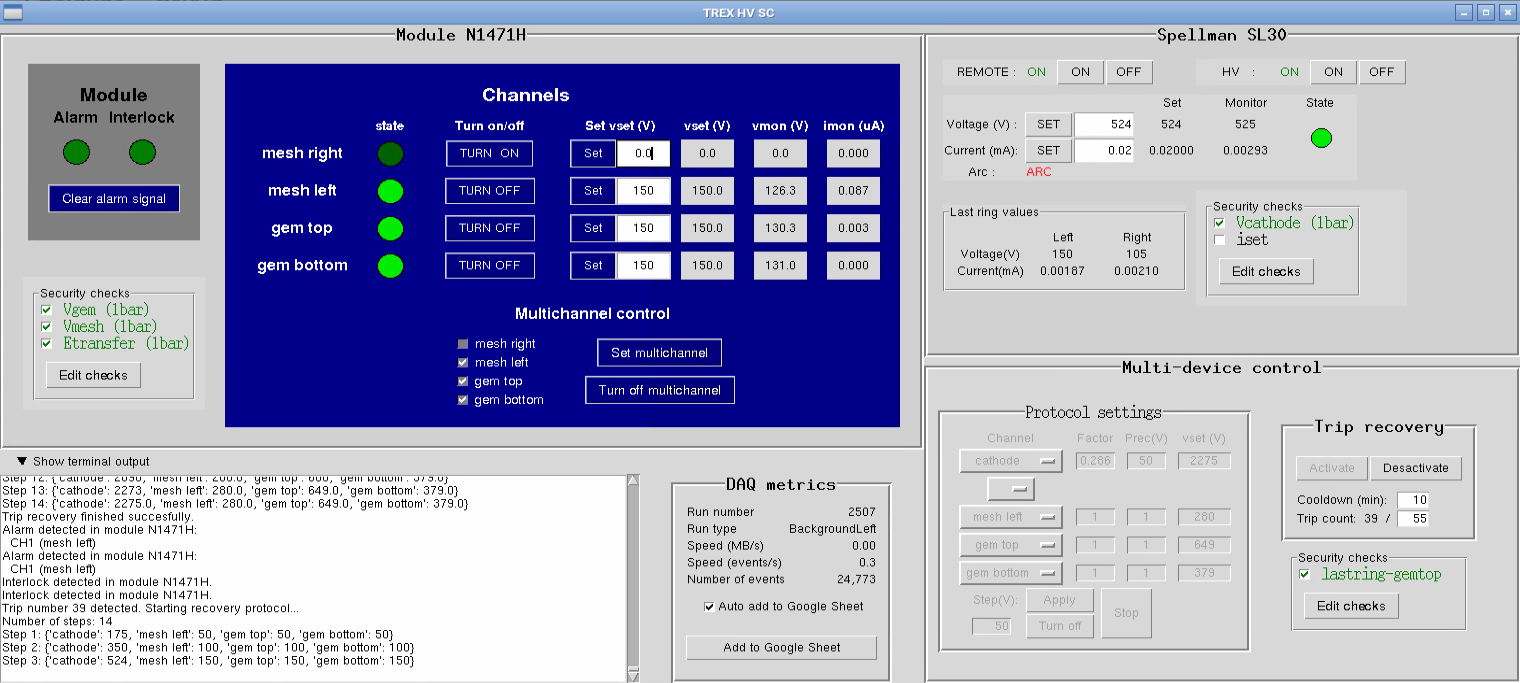}
    \caption{Window of the slow control for HV power supplies. Dark green channel available and OFF, vivid green channel ON, red TRIP. It allows automatic trip recovery and gradual ramp up of the voltages.}
    \label{fig:SlowControlHV}
\end{figure}

\subsection{Slow control}

The experiment has a long list of components and devices that are crucial for the behaviour of the detector. The most obvious ones are the power supplies used for the drift and amplification fields; but also pressure gauges, power supplies for the acquisition boards, gas sensors... Everything is monitored and controlled through a parallel system called ``slow control".

\begin{figure}
    \centering
    \includegraphics[width=0.9\linewidth]{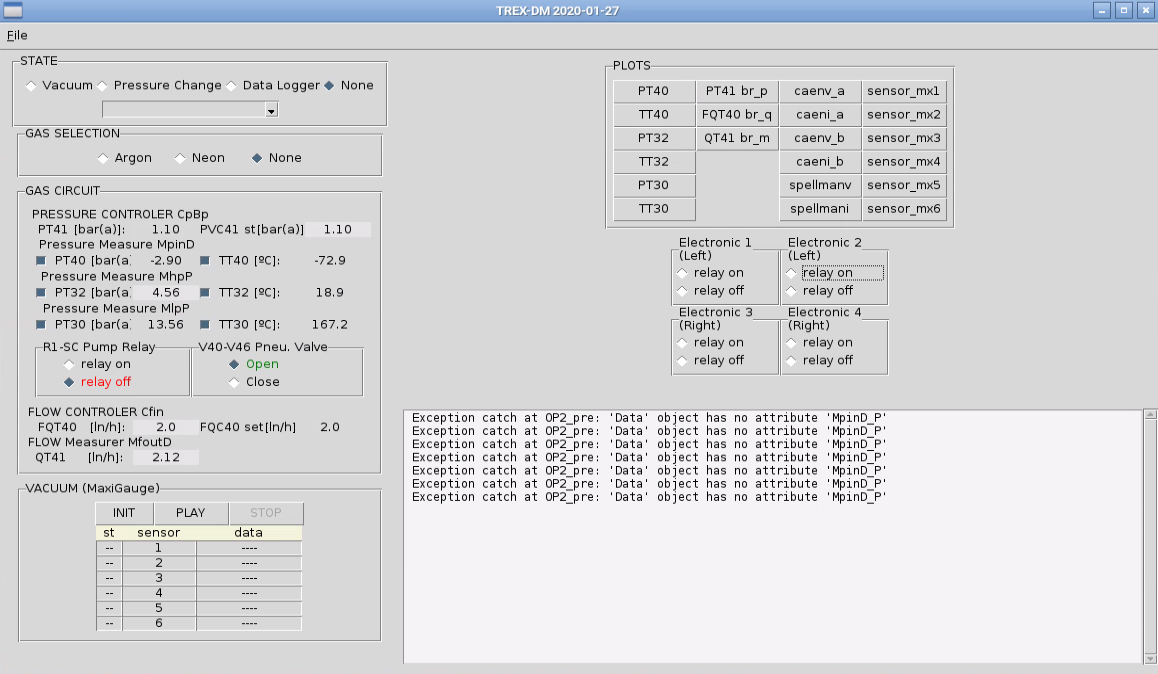}
    \caption{Gas panel. Pressures, flows and temperatures in different points of the gas system.}
    \label{fig:SlowControlGas}
\end{figure}

For developing reasons, the overall system is split in several panels. The slow control has been evolving with time -new sensors, GEM + Micromegas, different gases, etc.- but in figures \ref{fig:SlowControlHV} and \ref{fig:SlowControlGas} are shown the two main windows in March 2025. The first one controls the HV power supplies: Caen N1471H and Spellman SL30. The first module has four channels and is used for Micromegas and GEM voltages. The system allows to detect possible trips -sparks that produce important peaks in current that may damage the detector-, and react to this by switching off all the surrounding channels, including the cathode, if necessary. This slow control, developed by my colleague Álvaro Ezquerro \cite{EzquerroSlowControl}, allows to program ramp up routines, the logic for trip recovery and shut downs, monitor voltage and current in all channels, switch on and off channels and the equipments itself.

The second window \ref{fig:SlowControlGas} is the gas panel. It displays pressure, temperature and flow in several points of the gas system, as well as the position of several electro-valves. The ``PLOTS" section is still in the window but is not used any more.

\subsection{On-the-fly event viewer}

\begin{figure}
    \centering
    \includegraphics[width=0.8\linewidth]{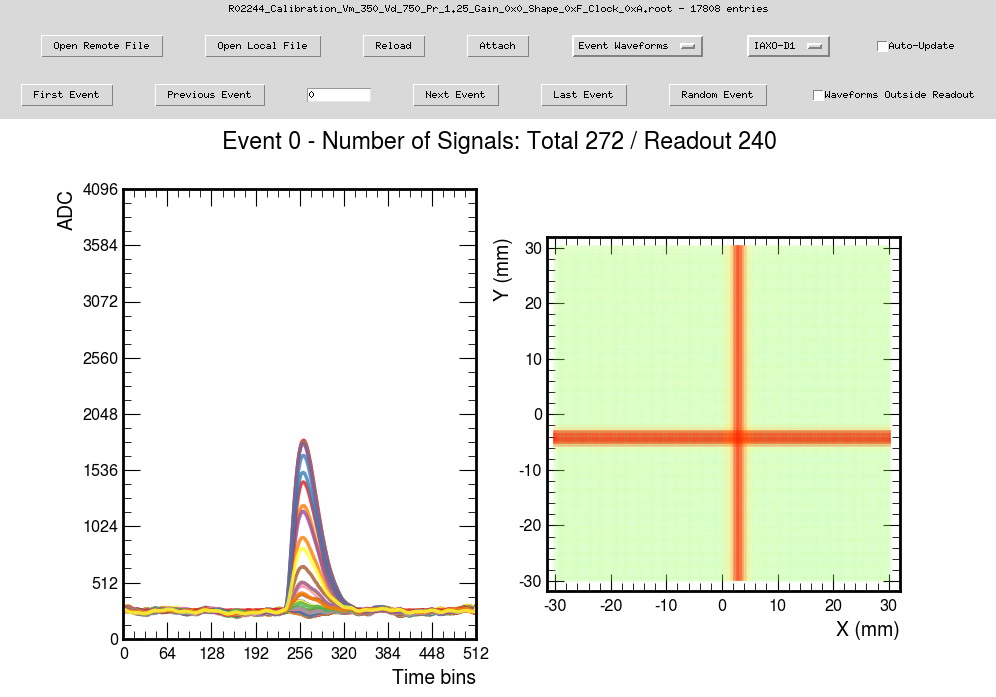}
    \caption{Window for plotting events. On the left event pulses, on the right readout channels hit. It allows past files or the one for the current data taking, it was developed with real time plotting in mind.}
    \label{fig:Viwer}
\end{figure}

Last year a nice upgrade of the data taking software by Luis Obis allowed to see on real time the recorded signals and plot with a small delay some basic observables. This viewer allows a look at the data in real time, simplifying and shortening several tasks and tests especially at commissioning times. An example of the display window is shown in figure \ref{fig:Viwer}.

\section{Background levels}\label{sec:BkgLevels}

\begin{figure}[h!]
    \centering
    \includegraphics[width=0.8\linewidth]{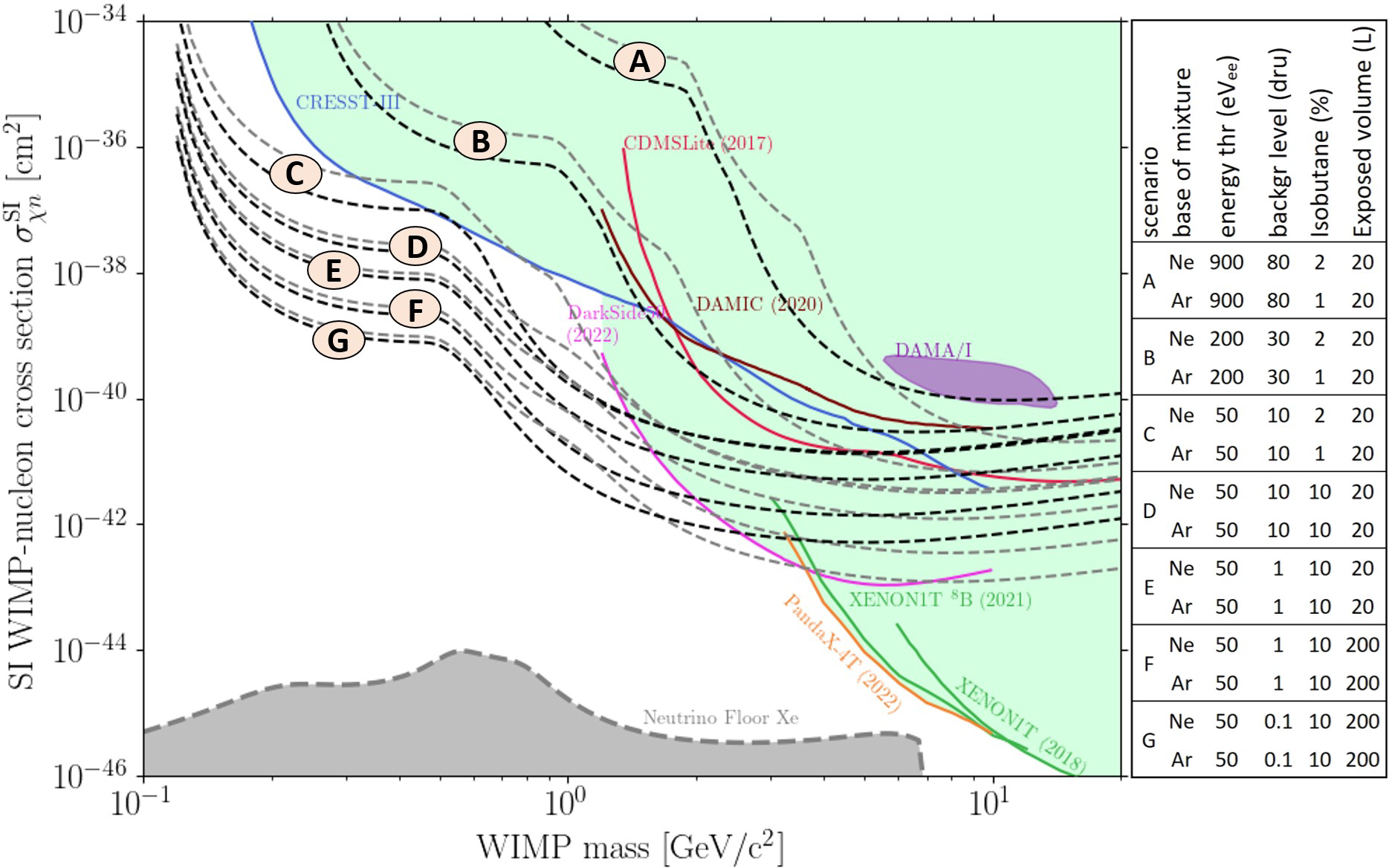}
    \caption{WIMP parameter space (WIMP mass and WIMP-nucleon cross section) with current bounds from experimental searches. Claimed discovery from DAMA/LIBRA and solar neutrino floor for a Xe target are also shown. TREX-DM different scenarios labelled from A to G (all of them 1 year of exposure time) are plotted, Ne-based mixtures in black and Ar-based mixtures in grey. Figure from \cite{pablo2025micromegas}.}
    \label{TREXsensitivityScenarios}
\end{figure}

TREX-DM background levels were assessed in the design phase. A long campaign of radiopurity measurements with high-purity germanium detectors was performed in the Laboratorio Subterráneo de Canfranc in order to quantify the radioactivity of all materials susceptible of being included in the detector. The impact of these measured activities were simulated in the Geant4 model of TREX-DM and summarized here \cite{castel2019background}. Intrinsic contributions, as well as cosmogenic activity in the copper vessel and in the gas, environmental gammas, neutrons and muons. With these contributions, the expected background level was 10 counts/keV/day/kg (dru). Recently, several scenarios were reviewed, and sensitivity prospects were calculated. In figure \ref{TREXsensitivityScenarios}, scenario C reflects this estimated background level with 10 dru.

Unfortunately, measurements from the initial data-taking period revealed that one background contribution had been underestimated: radon-induced activity. The atmospheric $^{222}$Rn decay chain can be seen in \ref{fig:Radon222Chain}: the contribution of electrons from $\beta$ decays was already counted in the 10 dru level estimation; and alpha particles emitted along the chain have energies much higher than the region of interest for dark matter searches, below 7 keV: nonetheless, secondary emissions due to energy transitions between excited atomic levels of the product of the decays were not taken into account. This included a non-negligible production of low energy photons precisely in the range of interest of TREX-DM. 

Simulating the full $^{222}$Rn chain for volumetric activity in the target gas, almost 8 events per disintegration were recorded, which are exactly the 4 alpha and 4 $\beta$ decays in the full chain. The energy deposition of these events is very different, and some are seen by the detector as low energy events. For example, particles appearing close to the border may leave only a fraction of their energy in the sensitive volume. Others generate secondary particles of lower energy. This leads to a contribution of events below 50 keV of the order of 1.3 per $^{222}$Rn disintegration. 

\begin{figure}[h]
    \centering
    \includegraphics[width=0.8\linewidth]{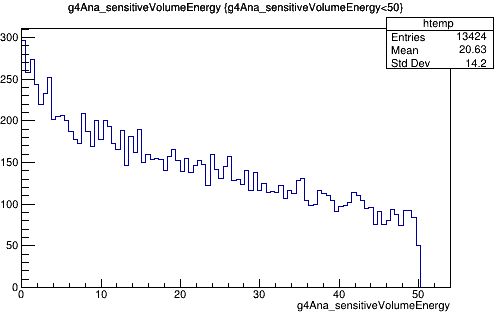}
    \caption{Low energy spectrum from 10000 simulated $^{222}$Rn decays in the gas volume of TREX-DM. }
    \label{fig:Rn222below50keV}
\end{figure}

This estimation of the contribution at low energies from the $^{222}$Rn decay chain admits a refinement. Due to the electrostatic field present in the gas volume, many isotopes in the chain end up attached to the cathode, not floating in the gas as in the simulation shown above. This is particularly relevant for $^{210}$Pb because, with a half-life of 22 years, it accumulates in the cathode. Therefore, decays of this isotope from the cathode were simulated to assess the impact at low energies. A total of 29487 simulated decays produced 52222 events below 50 keV, roughly 1.8 low-energy events per disintegration. This shows that surface contamination from the cathode probably is more dangerous than the volumetric $^{222}$Rn. Volumetric $^{222}$Rn can be prevented, as it will be described in the section \ref{sec:alphas}. The solution is to prevent radon exposition to minimize the amount of surface deposition. This implies fast and careful manipulation from production until installation in the detector of the most critical pieces, the readout planes and the cathode. This last piece can be changed regularly to prevent accumulation if exceptionally low levels are needed.

\begin{figure}[h]
    \centering
    \includegraphics[width=0.8\linewidth]{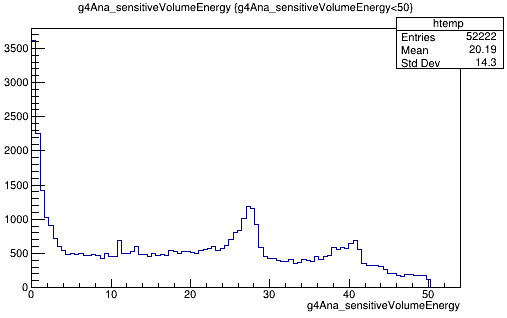}
    \caption{Low energy spectrum, below 50 keV, from 29487 simulated $^{210}$Pb decays from the cathode of TREX-DM.}
    \label{fig:Pb210below50keV}
\end{figure}

A detailed explanation of the impact of the alpha particle background to the low energy range is given in section \ref{sec:alphas}, along with the methodology developed to reduce it. Nowadays, the levels of volumetric contribution have been significantly reduced and the remaining contribution probably comes from surface contamination. During the last year, alpha levels have been around 1 count per hour, being the low-energy counterpart the dominant contribution to the low-energy background. With the present background levels, the sensitivity of the experiment is in scenario A in \ref{TREXsensitivityScenarios}. A reduction of background of one order of magnitude is needed to achieve the first scientifically relevant scenario, C, after a reduction of alphas of a factor 5 to 10.

\section{Characterization of Micromegas detectors}\label{sec:MMTREX}
\begin{comment}
Fabrication of Micromegas V1 was a great technological challenge due to its dimensions. It was the biggest Microbulk Micromegas plane fabricated up to that day \cite{mirallas2024planos}.  But their performance was not perfect, since their installation in TREX-DM in mid 2018, leak currents appeared in the fujiPoly connectors between the Micromegas and the flat cables and between the flat cables and electronic board. These leak currents were associated to short circuit channels, that even being disconnected due to the proximity of the channels in the fujipoly connectors reached the electronics. This was a limitation in the maximum voltage reachable in the detector, limiting as well the maximum pressure of the chamber. These fujipoly connectors were chosen by is high radiopurity but for the Micromegas V2 werw replaced by face-to-face connectors with wider separation between channels. 
\end{comment}

Despite the problems with leak currents mentioned in \ref{MicromegasTREX}, the first generation of TREX-DM Micromegas allowed to test all the setup while they were characterized with different gases and at different pressures. Their performance was good enough to identify setup improvements and assess the background reduction achieved after these changes, despite the large number of inoperative channels. Figure \ref{fig:HitmapFiducialOld} shows the pattern of ``dead" strips at the end of the life of one on those Micromegas. 

As already mentioned this motivated a new design of the microbulk planes and connectors, detailed in~\cite{mirallas2024planos}. The new design was ready by spring 2020 and the fabrication was completed by the end of 2021. During the first months of 2022, the four Micromegas V2 produced were tested and characterized in Zaragoza, with Argon +1\% Isobutane at 1.1 bar.

Specimens labelled MMv2-2 and MMv2-3 were chosen for installation in TREX-DM in spring 2022 based on the low number of possibly malfunctioning channels, as indicated by their unexpected capacitance (2 and 6 respectively) and voltage stability (below 295 V and 305 V).

\subsection{Comparison between the two designs}

\subsubsection{Gain studies}
\begin{figure}[h]
   \begin{minipage}{0.48\textwidth}
     \centering
     \includegraphics[width=0.99\linewidth]{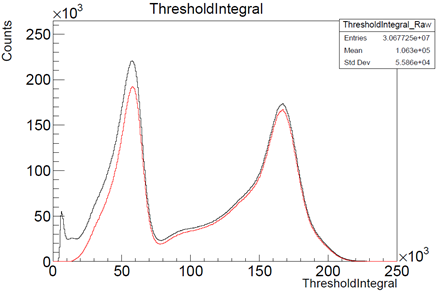}
    
   \end{minipage}\hfill
   \begin{minipage}{0.48\textwidth}
     \centering
     \includegraphics[width=0.99\linewidth]{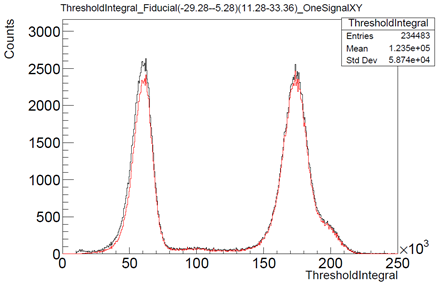}
     
   \end{minipage}
   \caption{\textit{Left}: $^{109}$Cd calibration spectrum (24h) from full readout plane, Micromegas V1, in neon + 2\% isobutane at 4 bar (run number 1344). The resolution is quite limited, pointing towards different gain along the Micromegas surface. In black all recorded events, in red those that fulfil the quality checks. \textit{Right}: Spectrum from the same run selecting events in a small fiducial area without dead strips. It is the same region as shown in \ref{fig:HitmapFiducialOld}. The copper fluorescence peak at 8 keV, and photopeaks from $^{109}$Cd calibration source at 22 and 24.9 keV can be identified with increased resolution. In black all recorded events, in red those that fulfil the quality checks.} \label{fig:SpectrumFullAndFiducial}
\end{figure}

The TPC detector gain is highly dependent on the gas properties. When this is under control, the variations in the multiplication of the charges come from the Micromegas itself. These gain variations are due to the manufacturing process of each specimen and in such big planes typically vary along the surface. The most obvious effect of this behaviour is a widening of the calibration peaks, a decrease in resolution. This effect can be seen in figure \ref{fig:SpectrumFullAndFiducial}, where peaks are broader in left spectrum due to the gain inhomogeneity in different areas of the detector. A gain map of the detector plane can be obtained if the active area is divided in several parts and the individual energy spectra are plotted, provided all the active area is well populated with calibration events, like shown in figure~\ref{fig:GainMaps}. On each small region the spectrum with events falling in this precise region has better resolution, as can be seen in the right plot of figure \ref{fig:SpectrumFullAndFiducial} (red spectra are made of events that satisfy certain quality checks, black ones with all events). They may vary slightly depending on the characteristics of the run, but in general they assure that events are physical events by checking the shape and distribution of signals. The four observables used for this are \textit{MaxPeakTimeDelay}, \textit{NumberOfGoodSignals}, \textit{AveragePeakTime} and \textit{RiseTimeAverage}.

\begin{figure}[h]
    \centering
    \includegraphics[width=0.9\linewidth]{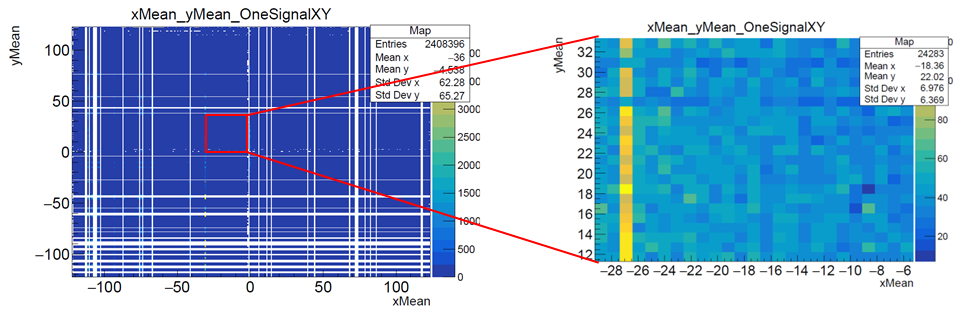}
    \caption{Hitmap from Micromegas V1: on the left, after a 24h-long $^{109}$Cd calibration. It is made with events with one signal in X and one in Y, low energy events in which the mean position is well defined. In white can be seen the dead strips, 27 in X direction, 35 in Y. Right inlet: Selecting a fiducial area without dead strips homogeneity improves. A high activity channel can be seen in yellow with roughly double the number of events than the neighbouring ones. }
    \label{fig:HitmapFiducialOld}
\end{figure}

\begin{figure}[h]
\centering
\begin{subfigure}{.5\textwidth}
  \centering
  \includegraphics[width=.99\linewidth]{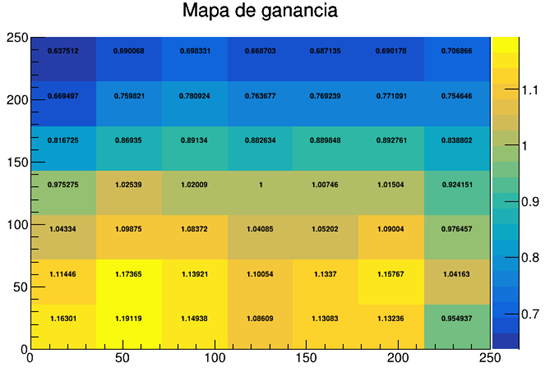}
  \caption{}\label{fig:GainMapSouthv2}
\end{subfigure}%
\begin{subfigure}{.5\textwidth}
  \centering
  \includegraphics[width=.99\linewidth]{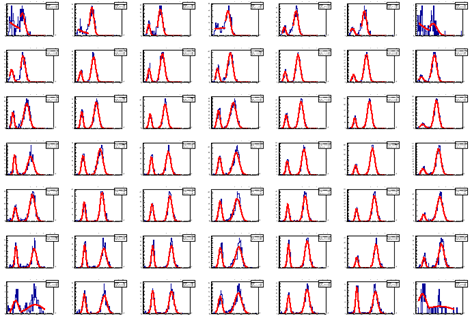}
  \caption{}\label{fig:GanMapSpectrumsSouthv2}
\end{subfigure}
\caption{Gain map of the right Micromegas V2 detector (MMv2-2). The area of the readout plane is divided in several regions and the spectra of each one are plotted (b). Fitting the peaks of these spectra the relative gain of every region is obtained, taking as a reference the central region (a). This allows energy correction of every event according to its position in the readout plane. }\label{fig:GainMaps}
\end{figure}

In the case of Micromegas V1 they had an additional problem: the ``dead" channels. They affect the charge collection with a combination of charge loss in these strips and increased activity of neighbouring ones that collect extra charge due to the distorted electric field. Figure \ref{fig:HitmapFiducialOld} shows the pattern of ``dead" strips in one of the Micromegas V1 before its substitution. Almost all the plane is affected by disconnected channels. In figure \ref{fig:HitmapFiducialOld} a homogeneous region is selected from which the spectrum in figure \ref{fig:SpectrumFullAndFiducial} is taken.

\begin{comment}
\begin{figure}[h]
    \centering
    \includegraphics[width=0.7\linewidth]{images/HitmapLongOld.png}
    \caption{Hitmap from Micromegas V1 north side after a 24h-long $^{109}$Cd calibration in order to achieve high statistics in all the readout plane. Here, only events with one signal in X and one in Y, so low energy events in which the mean position is well defined. 27 dead channels identified in X direction, 35 in Y. Few more have high activity, in some cases related to a contiguous dead strip.  }
    \label{fig:HitmapLongOld}
\end{figure}
\end{comment}

A gain map of the detector plane can be obtained if the active area is divided in several parts and the individual energy spectra are plotted, provided all the active area is well populated with calibration events. Figure~\ref{fig:GainMaps} contains the gain map for Micromegas V2 and the calibration spectra for each region. Afterwards, background spectra can be corrected according to the position of each event and the gain of the corresponding region.

\begin{comment}
In Micromegas V1 the ``dead" strips prevent from further gain corrections but a protocol was already developed developed to account for different gains in different areas. Plots in figure \ref{fig:GainMaps} show the result of this protocol applied to a Micromegas V2. The strategy is based on a tesselation of the readout plane \ref{fig:GainMapSouthv2}. With events falling in each piece an spectrum is obtained \ref{fig:GanMapSpectrumsSouthv2}, from whose peaks a relative gain is computed taking as reference the central area. In \ref{fig:GainMapSouthv2} brighter colors show areas with higher gains, so peaks slightly displaced to the right. Dedicated calibration runs with enough events to achieve reasonable tessellations are performed (see \ref{fig:BestCalibRun}) and then background spectra can be corrected according to the position of each event.
\end{comment}

\begin{figure}[h]
\centering
\begin{subfigure}{.6\textwidth}
  \centering
  \includegraphics[width=.99\linewidth]{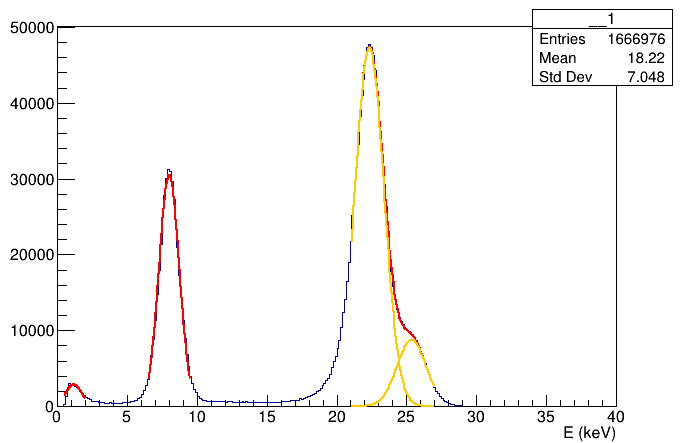}
  \caption{}\label{fig:BestSpectraV2}
\end{subfigure}%
\begin{subfigure}{.45\textwidth}
  \centering
  \includegraphics[width=.9\linewidth]{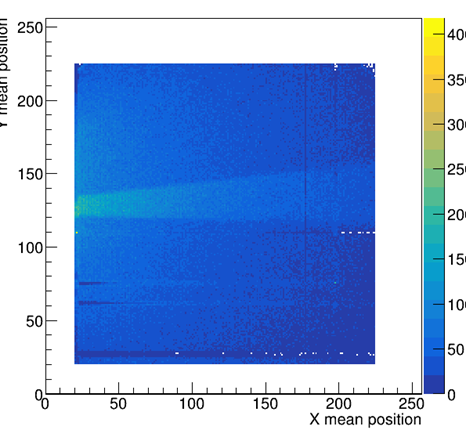}
  \caption{}\label{fig:BestHitmapV2}
\end{subfigure}
\caption{ (a) Best $^{109}$Cd calibration spectra for the left detector (specimen MMv2-3), with an energy resolution of 21\% FWHM at 8 keV and 11\% FWHM at 22 keV. (Run number 1876, neon + 2\% isobutane at 4 bar and Vmesh = 365 V, with duration of 16 hours, gain corrected with 13x13 tessellation and removing events from 2 cm frame in the border of the readout plane). (b) Hitmap of selected events from the same calibration. An area of 2 cm around the outer part of the plane has been removed to avoid partial events and populations associated to surface contamination. The calibration source emission can be seen as a horizontal ray.} \label{fig:BestCalibRun}
\end{figure}

\subsubsection{Decoding checks}

Crucial in the event reconstruction step is the proper correlation between electronic channel and position on the readout plane. This is done with a map called ``decoding" that associates the ID number of each channel in the electronics with a specific X or Y position in the Micromegas. With 512 channels per detector, split in several flat cables that are read with 8 AGET chips in two different boards, it is easy to lose the track. The version used for Micromegas V1 when the writer arrived to the experiment was shown to have errors that were difficult to be solved due to the large number of dead strips. 

The new Micromegas V2 came with several differences in the routing of the strips and the number of connections involved. The process to produce a new decoding implies four steps through which every channel has to be tracked: position in the detector, pad in both ends of the flat cables, interface between the flat cables and the FEC-Feminos and AGET ID channel. These four steps, in which channels change notation and also position along the path, have to be written in a binary file to be given to REST-For-Physics \cite{REST} software that will generate a readout file which our processing chain can interpret and finally assign positions to every recorded signal. This task is not light and requires focus and attention to the details. If someone reading this ever faces such a challenge, I advice to take the effort very calmly and being humble.

\begin{comment}
, it is quite probable that you will need to do it several times until you solve all mistakes that surely will appear.  
\end{comment}

\begin{figure}
    \centering
    \includegraphics[width=0.99\linewidth]{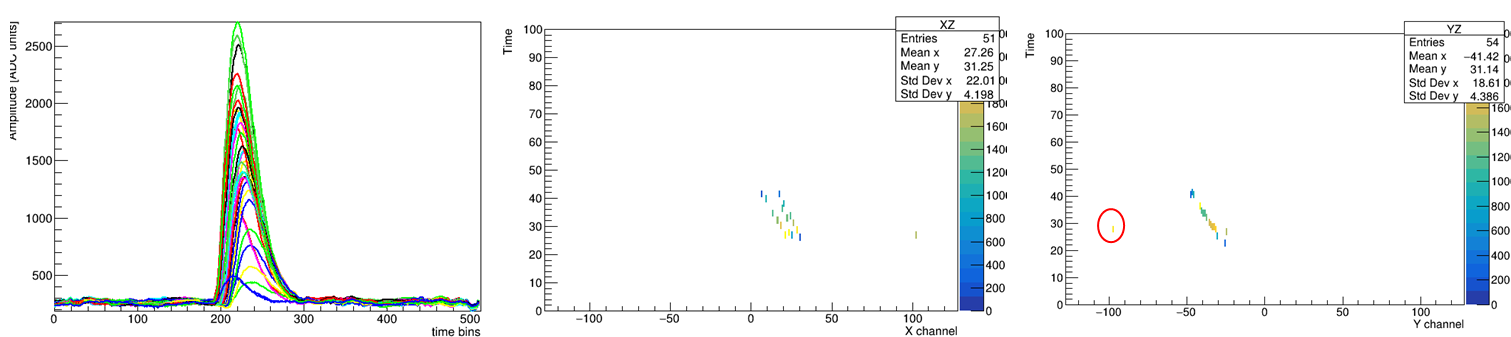}
    \caption{Many events show unusual reconstructions in Micromegas V1. Right: pulses of a calibration event. Center: Hits in X channels versus time. Colour code represents energy of each pulse. Strange two-line pattern points to bad decoding. Right: Hits in Y channels versus time. Isolated channel with high energy marked with a red circle.}
    \label{fig:EventBadDecoding}
\end{figure}

Identifying defects in the decoding is not always obvious, even with very few dead strips around. Energy spectra and mean positions may not be affected enough to sound the alarm, and due to the high symmetry of many of the possible errors in the decoding, the spread of the signals is not significantly affected either. The best place to identify them is the reconstruction of single events.  Figure \ref{fig:EventBadDecoding} presents an event from a calibration run. The signals (on the left) look correct (and they are), yet when plotted according to their position and peak time (at centre and on the right) the pattern is not as expected. In a typical interaction, the electron cloud is continuous and the same continuity in arrival times and energy distribution is expected for the signals. However, the X signals (centre plot) appear with a double linear pattern, a V shape, which might point to a permutation of contiguous channels in the readout. In the Y channels (right plot), all of the signals follow the linear distribution but one, in the red circle, off of the area of the interaction; this could be due to a decoding error or a cross-talk within the electronic readout chain.

\begin{figure}
    \centering
    \includegraphics[width=0.8\linewidth]{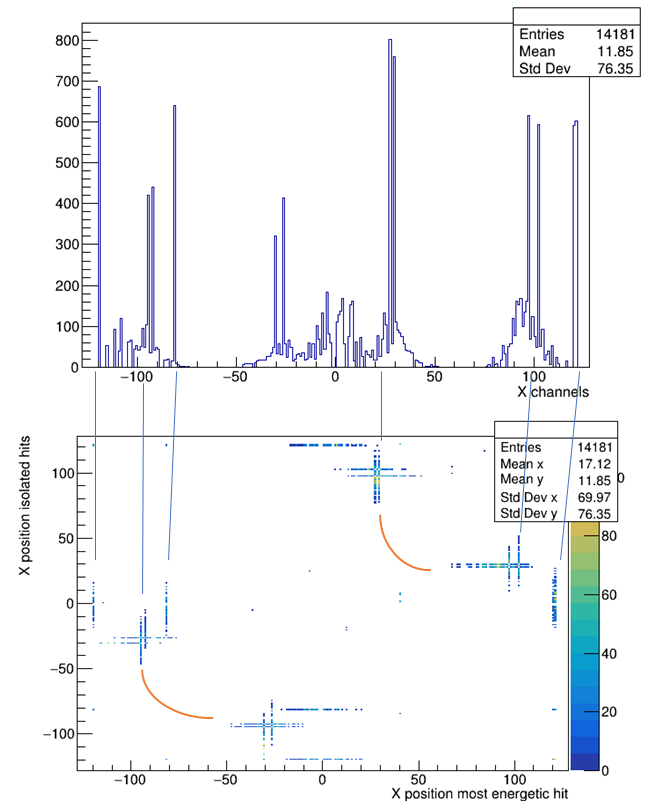}
    \caption{Channel activity of identified signals far from the most energetic signal recorded in the event. Some channels have unusual behaviour, pointing to bad decoding or cross-talk. The top plot represents the number of times a X channel has been activated in these events. The bottom reflects the position of the most energetic hit in the even versus the position of the isolated pulse. There are pairs of channels clearly correlated.}
    \label{fig:IsolatedSignalsV1}
\end{figure}

Regarding isolated signals, a broader search revealed frequent activation of certain channels far from the main cluster of signals, as illustrated in Figure \ref{fig:IsolatedSignalsV1}. The top graph shows channel activity for signals distant from the main deposition, with around ten to twelve channels identified as consistently active. These channels are correlated, as demonstrated in the second plot, where isolated signals consistently appear in specific regions when the most energetic pulse occurs in certain areas. These interconnected regions suggest a relationship that may involve cross-talk, impacting both high and low energy events, or decoding issues.

\begin{comment}
Regarding this isolated signals, a wider search was performed over many events and some channels appear to be activated very often far from the main cluster of signals. This is shown in figure \ref{fig:IsolatedSignalsV1}. In top graph, channels activity for signals far from the main deposition is depicted. Ten, twelve channels are identified there as highly active. And they seem to be correlated because in second plot, when the most energetic pulse is in certain area, the isolated signal always appear in a precise region. These pairs of regions are interrelated, it doesn't rule out the cross-talk but being in both directions and affecting also low energy events, decoding issues seem probable. 
\end{comment}

\begin{figure}[h!]
    \centering
    \includegraphics[width=0.8\linewidth]{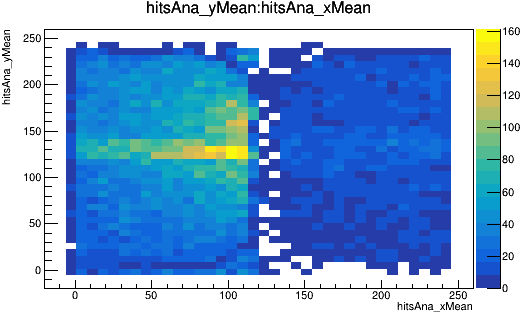}
    \caption{The first attempt for the decoding of Micromegas V2 had several problems. The process building a decoding is always cumbersome and needs several iterations until it is correctly done.}
    \label{fig:HitmapBadDecoding}
\end{figure}

\begin{figure}[h!]
    \centering
    \includegraphics[width=0.8\linewidth]{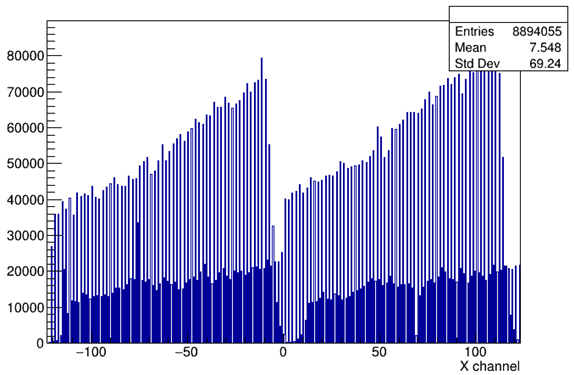}
    \caption{Channel activity in X channels in Micromegas V2: number of activations of every channel during a calibration run. it shows the effects of an incorrect association between readout channels and their position in the detector. A smooth pattern is expected if the decoding is correct.}
    \label{fig:ChannelActivityBadDecoding}
\end{figure}

These issues were solved when Micromegas V2 arrived. The new decoding required significant effort and multiple corrections during the first months of operation. 
The first iteration had obvious problems in X direction, intermediately identified in the hitmap, figure \ref{fig:HitmapBadDecoding}. Clearly, what looked like a symmetric distortion in X channels was affecting the decoding. Further analysis showed that one every two channels was swapped, as can be seen in figure \ref{fig:ChannelActivityBadDecoding}.

\begin{figure}[h!]
\centering
\begin{subfigure}{.4\textwidth}
  \centering
  \includegraphics[width=.99\linewidth]{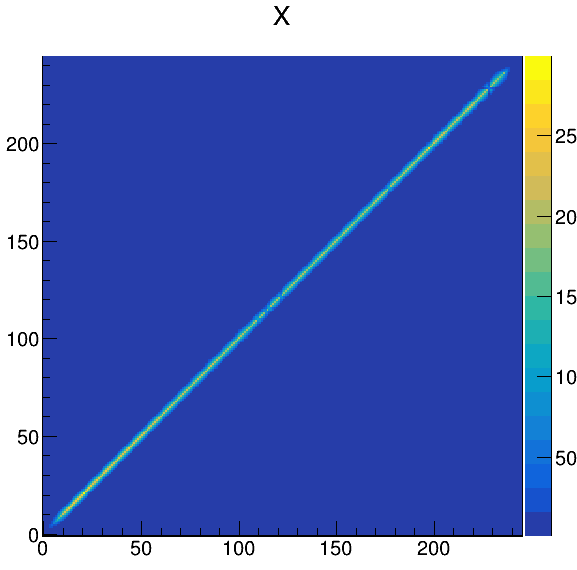}
  \label{fig:ChannelCorrelationXv2}
\end{subfigure}%
\begin{subfigure}{.4\textwidth}
  \centering
  \includegraphics[width=.99\linewidth]{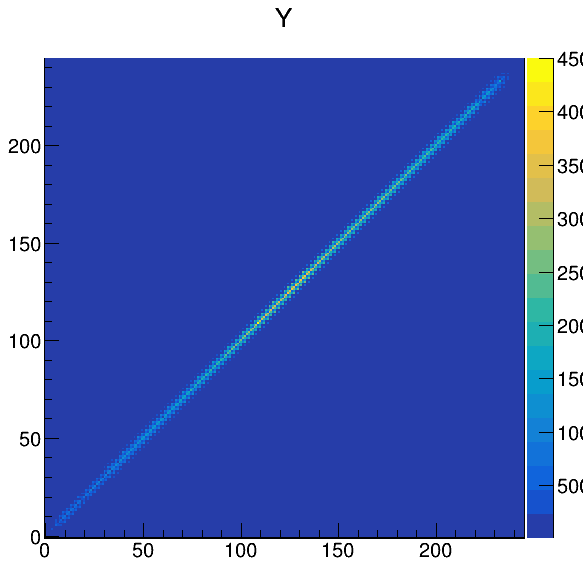}
  \label{fig:ChannelCorrelationYv2}
\end{subfigure}
\caption{Correlation matrices of simultaneous activated channels. X channels on the left, Y channels on the right. one.}\label{fig:ChannelCorrelation}
\end{figure}

\begin{figure}[h!]
\centering
\begin{subfigure}{.4\textwidth}
  \centering
  \includegraphics[width=.99\linewidth]{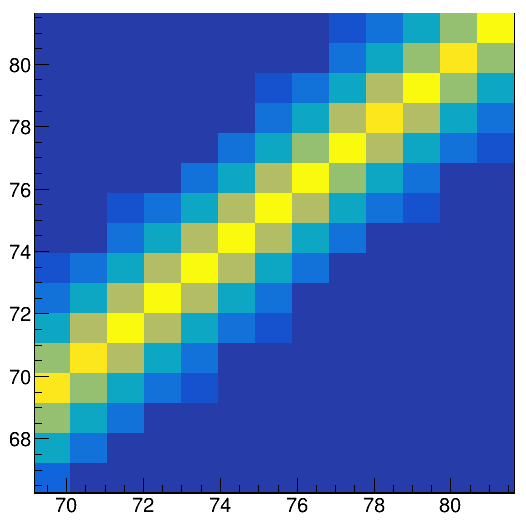}
  \label{fig:ChannelCorrelationZoomXv2}
\end{subfigure}%
\begin{subfigure}{.4\textwidth}
  \centering
  \includegraphics[width=.99\linewidth]{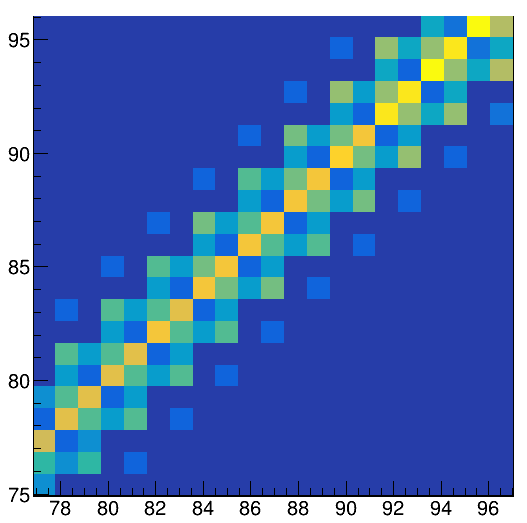}
  \label{fig:ChannelCorrelationZoomYv2}
\end{subfigure}
\caption{Zoom in the channel correlation matrix. The pattern in X channels (on the left) is as expected, neighbouring channels are more often activated together. Y channels (right) do not show the same pattern, pointing to errors in the decoding.} \label{fig:ChannelCorrelationZoom}
\end{figure}

After solving this first problem, another issue was hinted in the reconstruction of single events, with problems in the Y channels reconstruction. Correlation matrices were obtained, figures \ref{fig:ChannelCorrelation} and \ref{fig:ChannelCorrelationZoom} show the correlation of simultaneously activated channels. At first sight, both X and Y look the same, but upon a closer inspection, the pattern is very different. In Y channels a permutation every two channels was identified.

\begin{figure}[h!]
\centering
\begin{subfigure}{.5\textwidth}
  \centering
  \includegraphics[width=.99\linewidth]{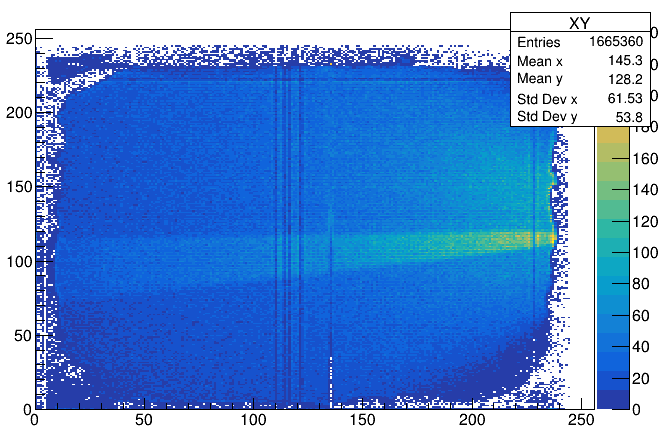}
  \label{fig:HitmapSouthv2}
\end{subfigure}%
\begin{subfigure}{.5\textwidth}
  \centering
  \includegraphics[width=.99\linewidth]{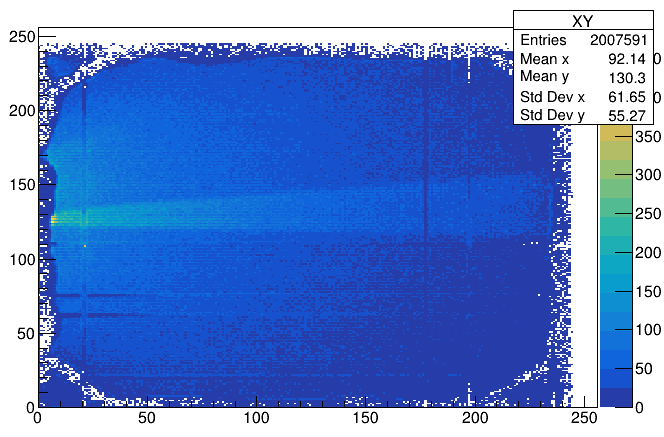}
  \label{fig:HitmapNorthv2}
\end{subfigure}
\caption{Hitmaps in both Micromegas V2 of TREX-DM with the correct decoding, of a long $^{109}$Cd calibration run. The distinctive feature of an opening in the teflon foil in the inner part of the field cage, which forms a ray can be clearly seen in both active volumes. It  allows to identify the sides, because the  position of calibration sources is known. \textit{Left}: Right side of TREX-DM (MMv2-2). \textit{Right}: Left side of TREX-DM (MMv2-3). }\label{fig:Hitmaps}
\end{figure}

The third version of the decoding gives the correct hitmap as seen in figure \ref{fig:Hitmaps}.

\subsubsection{Threshold}

\begin{figure}[h!]
\centering
\begin{subfigure}{.5\textwidth}
  \centering
  \includegraphics[width=.99\linewidth]{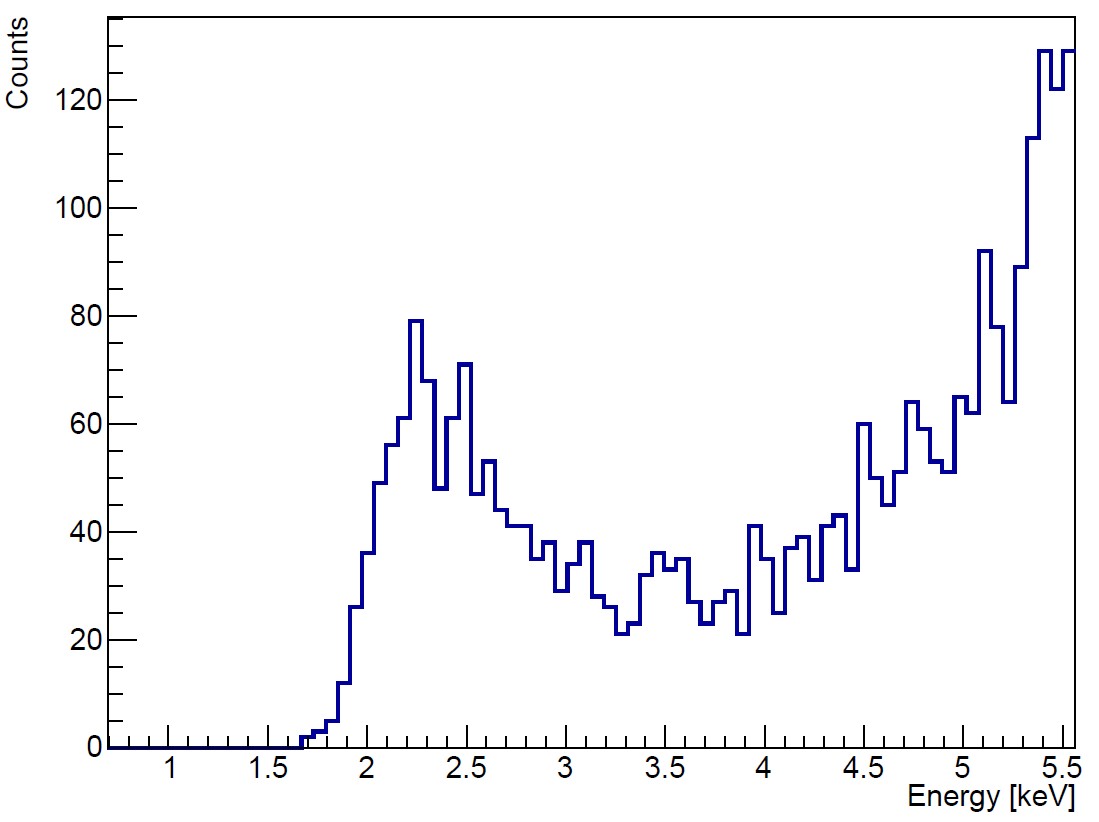}
  \caption{}
  \label{fig:Threshold1344CdCalibrationMMv1Neon}
\end{subfigure}%
\begin{subfigure}{.5\textwidth}
  \centering
  \includegraphics[width=.9\linewidth]{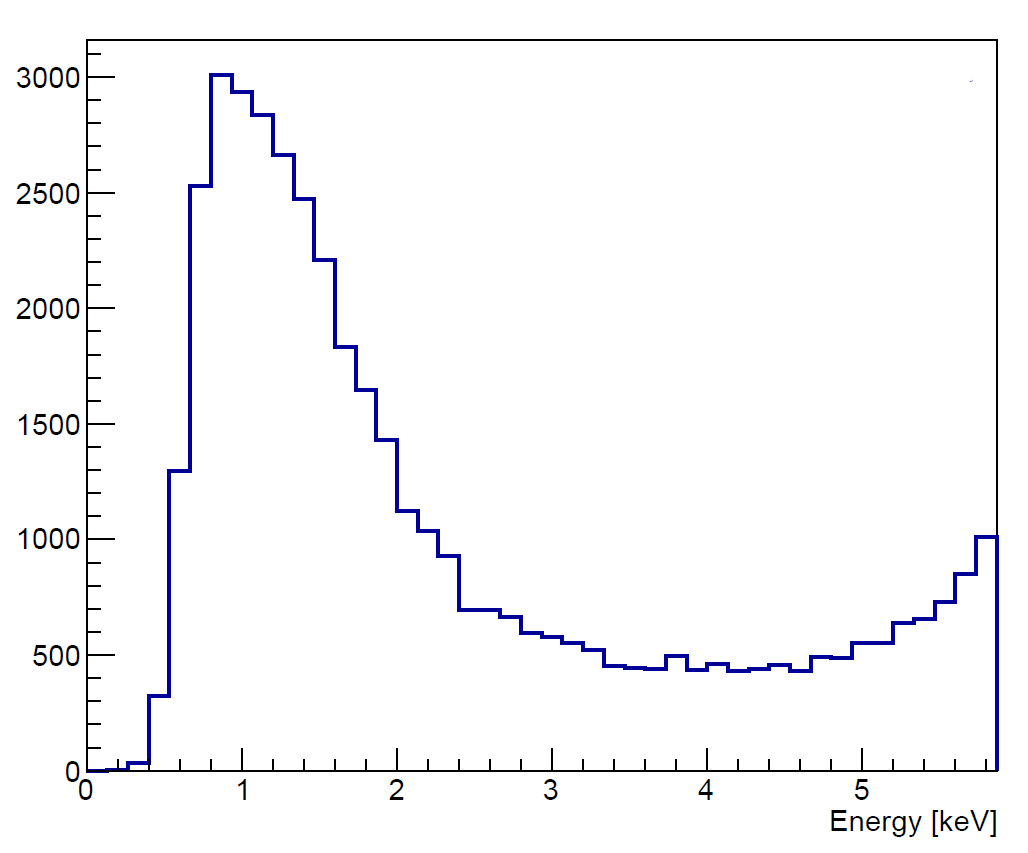}
  \caption{}
  \label{fig:Threshold1876CdCalibrationMMv2Neon}
\end{subfigure}
\caption{Low energy region of the spectra of two $^{109}$Cd calibrations performed in same conditions, neon + 2\% isobutane at 4 bar and $V_{mesh} = 365$ V: (a) Micromegas V1 (run number 1344 see figure \ref{fig:SpectrumFullAndFiducial}). Energy threshold is around 2 keV. (b) Micromegas V2 (specimen MMv2-3, run number 1876 with spectrum presented in \ref{fig:BestSpectraV2}). This points towards a threshold of 0.8 keV.
}\label{fig:MicromegasThresholdSpectrums}
\end{figure}

\begin{figure}[h]
    \centering
    \includegraphics[width=0.6\linewidth]{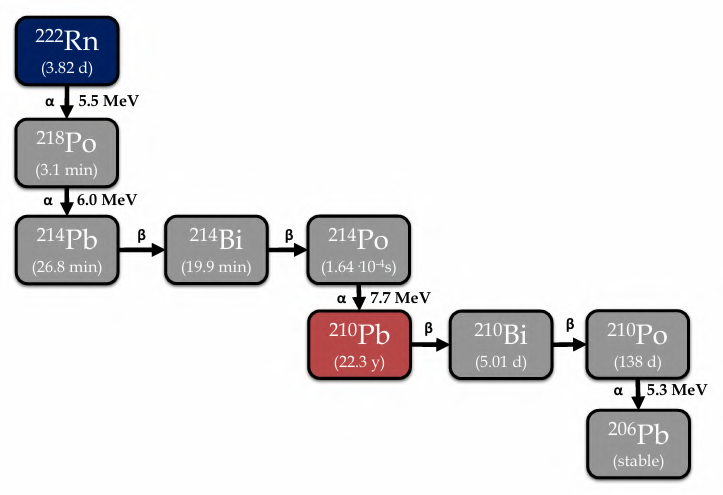}
    \caption{Decay chain of $^{222}$Rn. Highlighted are the two isotopes with longest half-lives, $^{222}$Rn with 3.8 days and $^{210}$Pb with 22 years. Alpha particles are emitted along the chain with energies between 5.3 and 7.7 MeV.}
    \label{fig:Radon222Chain}
\end{figure}

For low mass WIMP searches the energy threshold is critical, since the lower the threshold, the higher the sensitivity for lower masses. The poor connectivity and the large number of dead strips in V1 impeded triggering to low energy; the threshold at typical operation values was around 2 keV, and the best values were obtained when pushing the detector gain to high values, at the cost of instabilities and a higher risk of sparks and therefore damaging more strips.
In figure \ref{fig:Threshold1344CdCalibrationMMv1Neon}, a close up of the low energy region of the calibrated spectrum shown in figure \ref{fig:SpectrumFullAndFiducial} can be seen, with a threshold around 2 keV.

Right after the installation of Micromegas V2 in TREX-DM, several runs were taken with the same conditions applied previously to Micromegas V1. Figure \ref{fig:Threshold1876CdCalibrationMMv2Neon} shows the low energy region of the $^{109}$Cd calibration presented \ref{fig:BestSpectraV2}. The threshold was reduced significantly, down to 0.8 keV. In both cases, calibration runs were taken with neon + 2\% isobutane at 4 bar and $V_{mesh} = 365$ V.

\subsection{Alpha background}\label{sec:alphas}

The first version of the background model for the TREX-DM experiment underestimated the contribution from alpha particles to the energy region of interest \cite{castel2019background}, as explained in section \ref{sec:BkgLevels}. The commissioning phase in early 2020 showed an unexpected amount of events in the low energy region, approximately 1000 dru, that, although not alpha particles themselves,  were associated to the decay chain of $^{222}$Rn.

\begin{comment}
And this still holds, alpha particles from atomic decays are energetic particles in the range of MeV, far from the few keV region of interest for TREX-DM. The few months of background measuring during the commisioning phase in early 2020, however, showed an unexpected amount of events in the low energy region, approximately 1000 dru (1 dru = 1 c/keV/kg/d), several orders of magnitude higher than expected. Soon they were associated to the decay chain of $^{222}$Rn. The hypothesis was that along the decay chain low energy events are produced, mainly electrons and photons. 
\end{comment}

\begin{figure}[h!]
\centering
\begin{subfigure}{.5\textwidth}
  \centering
  \includegraphics[width=.99\linewidth]{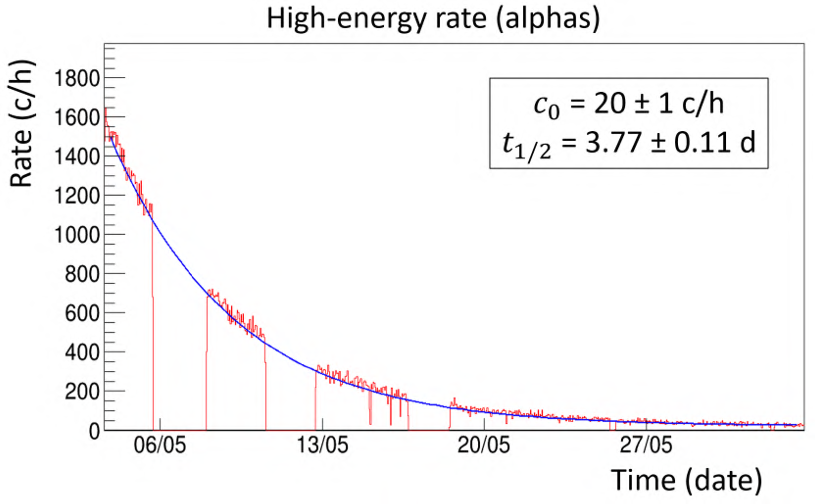}
  \caption{}
  \label{fig:HighRateAlphas}
\end{subfigure}%
\begin{subfigure}{.5\textwidth}
  \centering
  \includegraphics[width=.99\linewidth]{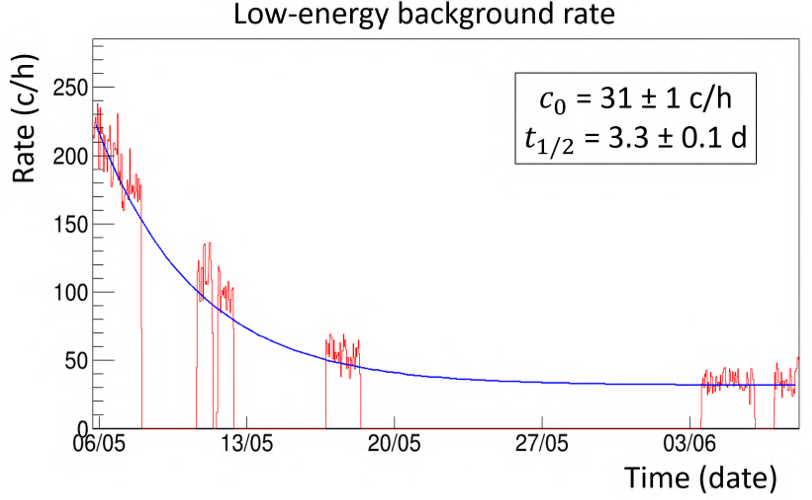}
  \caption{}
  \label{fig:LowEnergyAssAlphas}
\end{subfigure}
\caption{Evolution of high and low energy events with the chamber isolated from the gas system. Alternate runs were taken during one month with the same gas content, neon + 2\% isobutane at 4 bar. 
(a) Evolution of alpha rate, which is compatible with the $^{222}$Rn half-life of 3.82~d. There is a remaining content of 20~c/h that probably comes from other sources.
(b) Evolution of the number of low energy events. It decays at a rate similar to the alpha content, pointing clearly to its dependence. 
}\label{fig:SealedModeRuns}
\end{figure}

Radon 222 is a gaseous element that appears in the $^{238}$U chain. Uranium is present in the rocks that surround the Underground laboratory of Canfranc and, in general, it is  always present in small quantities in construction materials. Its gaseous condition allows it to escape and therefore, its presence is ubiquitous in the atmosphere. Discovering how detectable amounts of this gas reach the inside of the TREX-DM vessel took months of careful screening, isolating different regions of the system and measuring. The definitive proof came from the sealed mode runs, in which the chamber was isolated from the gas system for weeks. Low gain runs for alpha background measurements were alternated with high gain runs to monitor the evolution of the background in the low energy region of interest. Figure \ref{fig:SealedModeRuns} shows the evolution of both populations with time. The dependence of low energy events with alphas is shown here by the similar half-life value for both populations. 

In \cite{Oscar2025development} gives a detailed description of the efforts carried out to identify background sources, to develop radon filters and the solution applied to deal with this issue. Resulting of these works, the moisture filter in the recirculation system was identified as the main emanator, with a minor contribution, of the order of 10-20\%, coming from the recirculation pump and the oxygen filter. Several gas filters were examined, both commercial and custom made, but none  fulfilled the emanation levels required. At the end, a low-flow open-loop mode of operation without any filter was identified as the best option. Flow values below 1 l/h were proven sufficient to maintain the gas quality, with a complete gas renovation every two or three weeks.

The remaining component amounting to around  20 counts per hour is suspected to come from surface contamination associated to previous exposure to radon. As previously commented, the long lived $^{210}$Pb isotope may attach to materials inside the detector prior to their installation and therefore still affect the background. In addition, all $^{222}$Rn atoms that decay inside the chamber leave there their progeny. It has been shown \cite{nikezic2005radon} that subsequent isotopes are collected in the cathode and other surfaces electrostatically charged, so they may turn into a source of $^{210}$Pb.

\begin{figure}[h!]
\centering
\begin{subfigure}{.5\textwidth}
  \centering
  \includegraphics[width=.99\linewidth]{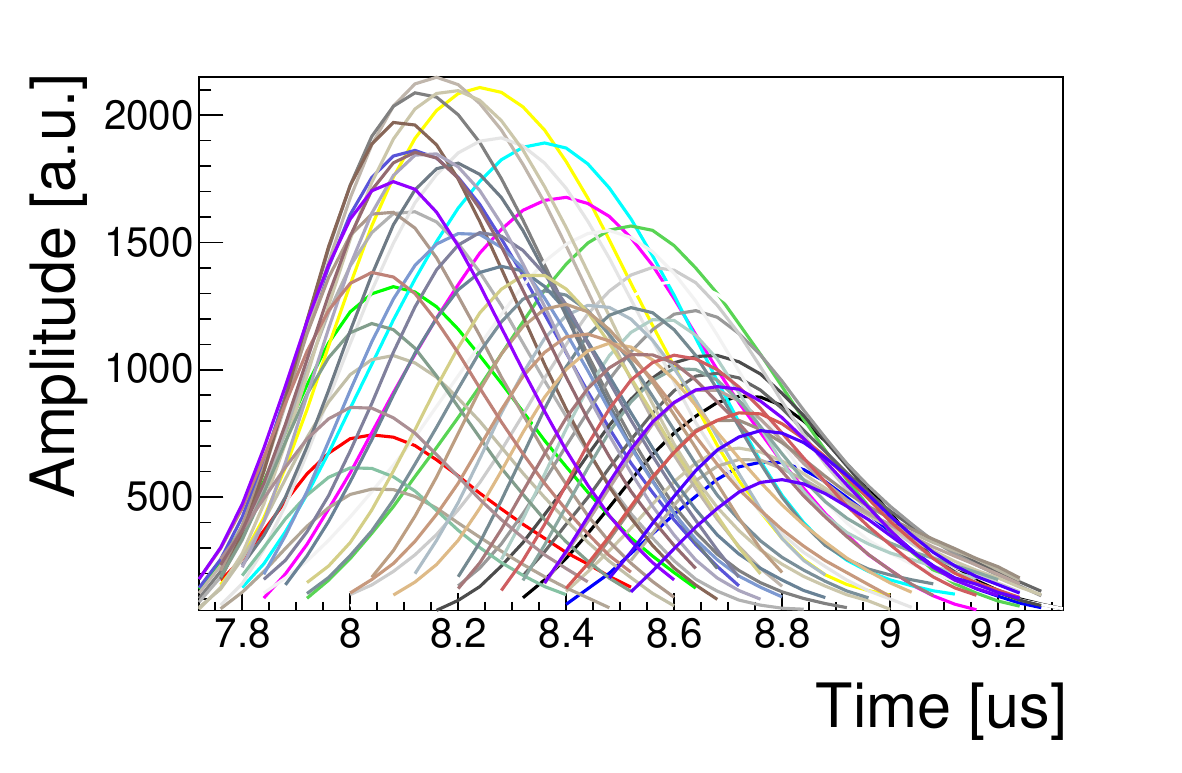}
 
\end{subfigure}%
\begin{subfigure}{.5\textwidth}
  \centering
  \includegraphics[width=.9\linewidth]{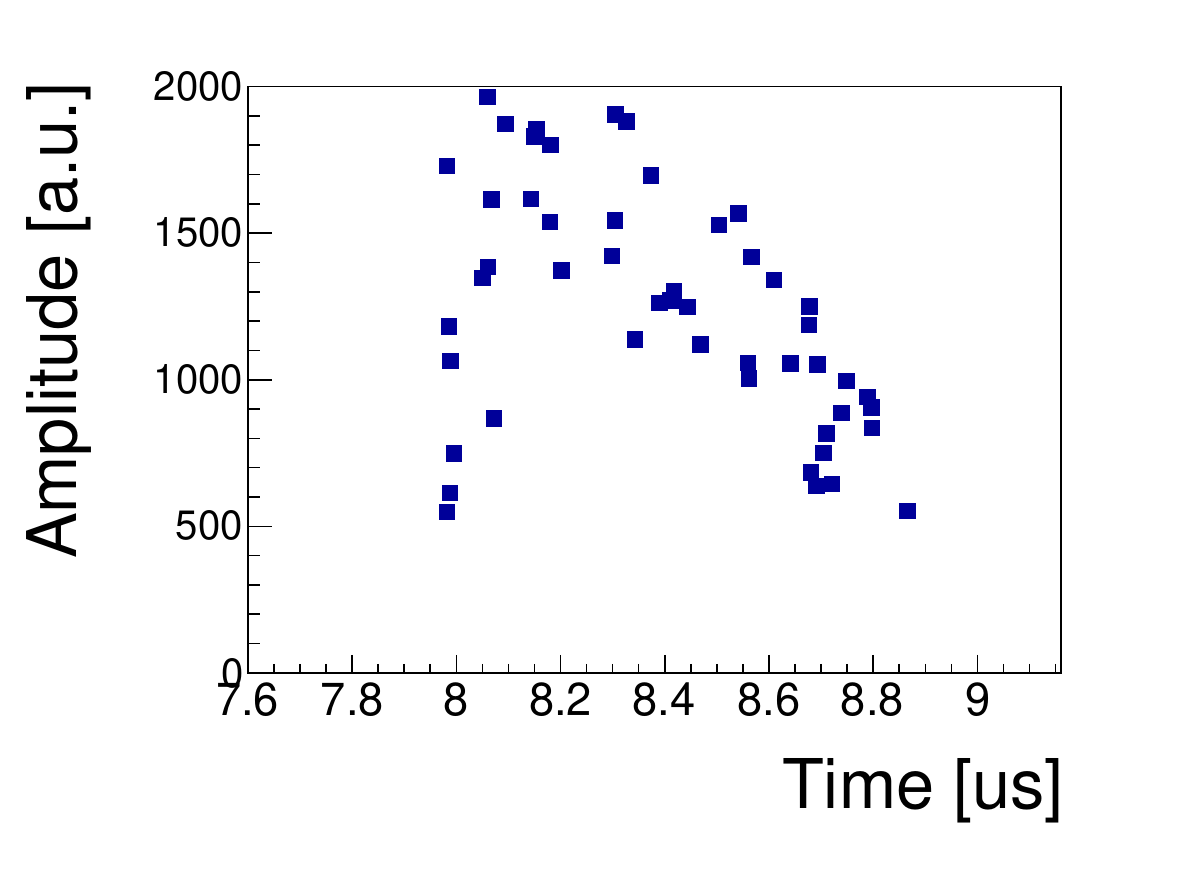}
  
\end{subfigure}
\caption{\textit{Left}: Pulses from an alpha particle registered in a low gain run in TREX-DM. The most energetic pulses arrive earlier, meaning the Bragg peak of the energy deposition is closer to the readout, and therefore the trajectory of the alpha particle points downwards:  it is a candidate of an event coming from the cathode.
\textit{Right}: Amplitude as function of time of arrival. Most of the energy of the event arrives in the first half of the time window. Measuring the energy balance with time for X and Y signals (here mixed) an angle for the trajectory can be extracted.} \label{fig:AlphaParticle}
\end{figure}

The interaction of alpha particles in matter has a distinctive energy deposition pattern that allows to distinguish their direction, the Bragg peak. This is produced by the slowing down of the particle when it loses energy, as explained in chapter \ref{Ch:GasDect}, depositing its last energy in a reduced spot. In figure \ref{fig:AlphaParticle} an alpha particle event with this characteristic pattern is shown, in (a) the registered sequence of pulses and in (b) the Bragg pattern showing the early arrival of the Bragg peak and therefore hinting a particle going towards the readout plane. In the analysis step, for every alpha particle the track is reconstructed and the origin and the end of the track is determined using this feature.

\begin{figure}[h]
    \centering
    \includegraphics[width=0.8\linewidth]{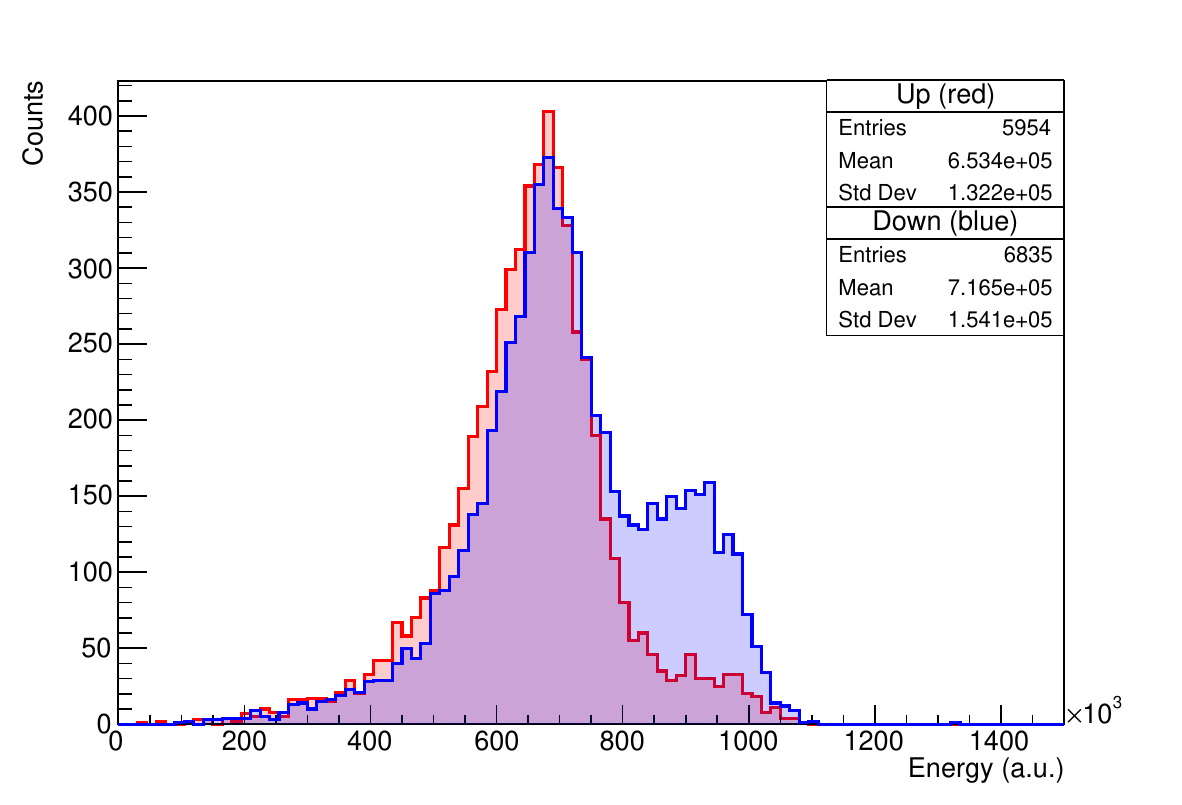}
    \caption{Alpha particle spectra. In red, particles with reconstructed angles compatible with the upward direction, in blue particles with downward direction. Almost all events come from the $^{222}$Rn chain. The first peak is due to 5.5 and 6 MeV events from $^{222}$Rn and $^{218}$Po, the second one from $^{214}$Po events from the cathode, explaining their appearance only in the blue spectrum.}
    \label{fig:R01739_AlphasUpDown_v4}
\end{figure}

With the direction of the track, the angle can be reconstructed from the total length and the temporal distribution that sets the Z component, permitting that upwards and downwards particles can be identified. In figure \ref{fig:R01739_AlphasUpDown_v4} two spectra with alpha particles from both directions are plotted. These are from early runs in TREX-DM, when the filters were introducing significant quantities of radon, therefore, most of the events are 5.5 MeV alphas that occur in the gas so they have random direction. However, there is a population of the down going events (originating from the cathode) that appear in a higher energy, associated to $^{214}$Po. They are alpha particles coming from $^{214}$Po decays which have had time to reach the cathode. The proportion between the peaks can be understood as follows: the first one comprises of  5.5 and 6 MeV events from $^{222}$Rn and $^{218}$Po in the gas volume, so equal proportions are expected to point towards going up and down;  the second peak at 7.7 MeV from $^{214}$Po, has only half of the events from the cathode facing the sensitive volume. This second peak has been continuously decreasing following the radon content reduction, until disappearing when operating at the low-flow open-loop mode. 

\begin{figure}[h!]
    \centering
    \includegraphics[width=0.8\linewidth]{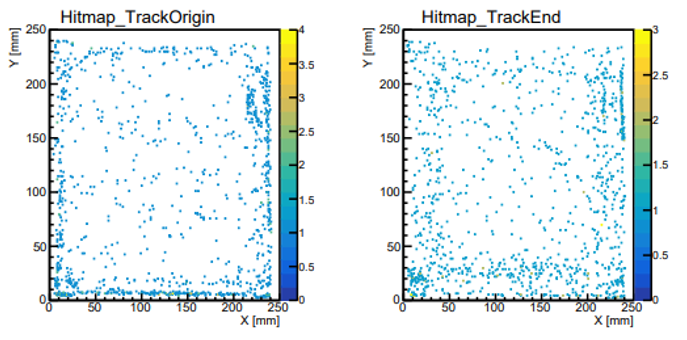}
    \caption{Hitmap of alpha events after 152 hours of low-gain data-taking in the ``right" detector (MMv2-2, run number 1997). The left plot shows the origin of the alpha track and the right one the end of the track. The effect of the surface contamination is clear in the origin hitmap in which most of the events are located close to the borders.  }
    \label{fig:AlphaTrackHitmaps}
\end{figure}

The remaining contribution of alpha contamination is due to superficial contamination in the walls of the field cage and, to some extent, radon content from the gas itself, maybe even emanation from the bottle. This last contribution, if finally it becomes dominant, may be addressed with a buffer volume to store the gas before entering the vessel and prevent emanation from the bottle. In figure \ref{fig:AlphaTrackHitmaps} the origin and the end of alpha tracks are plotted as hitmaps. Clearly, most events are concentrated in a frame along the borders of the readout plane. Selecting a fiducial area of 15x15 cm$^2$ in the centre of the chamber, only volumetric events and from the cathode remain. With this selection criterium, rates go from 20 counts per hour to below 1 count per hour. Although the statistics is quite low in this case, directional studies point towards around 70\% of the events coming downwards, indicating a non negligible contribution from the cathode. To further reduce this, a cathode made from radiopure copper wires has been designed, hoping that reducing the surface material in the cathode the amount of $^{210}$Pb will also decrease. To assure the low levels of this radioisotope in the grid cathode and the correct performance of the detector, a prototype is being tested in the AlphaCAMM detector \cite{altenmuller2022alphacamm}, developed by the group to measure surface contamination of samples foreseen to be placed in critical locations for low background detectors.

\section{GEM foils in TREX-DM\index{GEM foils}}

Micromegas V2 allowed to reduce the energy threshold below 1 keV. This improvement is not enough to reach Scenario C, which requires a threshold of 50 eV, as showed in \ref{TREXsensitivityScenarios}.
Therefore, in order to face this challenge, a Gas Electron Multiplier (GEM) was installed on top of the Micromegas detectors. It is an amplification device made of two copper foils and a kapton layer in between micro-patterned with holes. The GEM has a thickness of 60 $\mu$m (50 $\mu$m the kapton, 5 $\mu$m each copper layer), hole pitch of 140 $\mu$m, diameter of holes in copper of 70 $\mu$m and diameter of holes in kapton of 60 $\mu$m. Applying a voltage difference between the copper foils, electrons from the drift volume are collected in the holes and amplified by avalanche effect. We denote as $V_{GEM} = V_{top}-V_{bottom}$ the difference between voltages applied to the electrodes. The addition of the GEM foil acts as a preamplification stage that allows to further reduce the energy threshold, reaching values comparable to the single-electron ionization energy.

\begin{figure}[h!]
    \centering
    \includegraphics[width=0.6\linewidth]{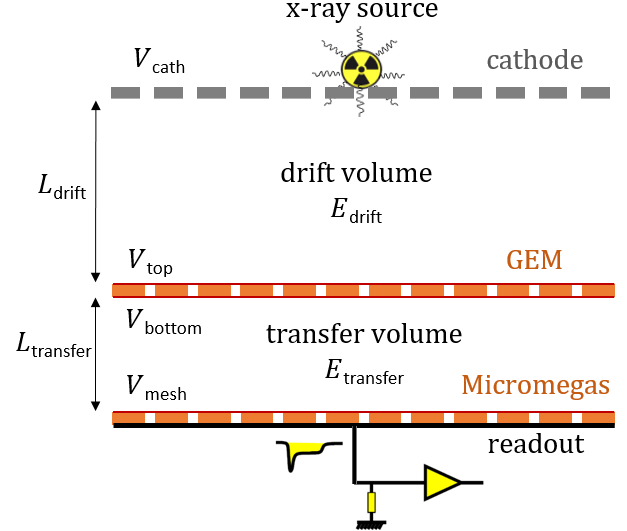}
    \caption{Micromegas + GEM schema.  In TREX-DM $L_{transfer}=1$ cm and $L_{drift}=16$ cm. For tests prior the GEM installation in TREX-DM, a X-ray source was placed in the cathode of the test bench chamber, this configuration can be seen in \ref{fig:GEMfullScaleSetup} and in \cite{pablo2025micromegas}. Figure from \cite{Oscar2025development}.}
    \label{fig:GEMschema}
\end{figure}

\begin{figure}[h!]
    \centering
    \includegraphics[width=0.8\linewidth]{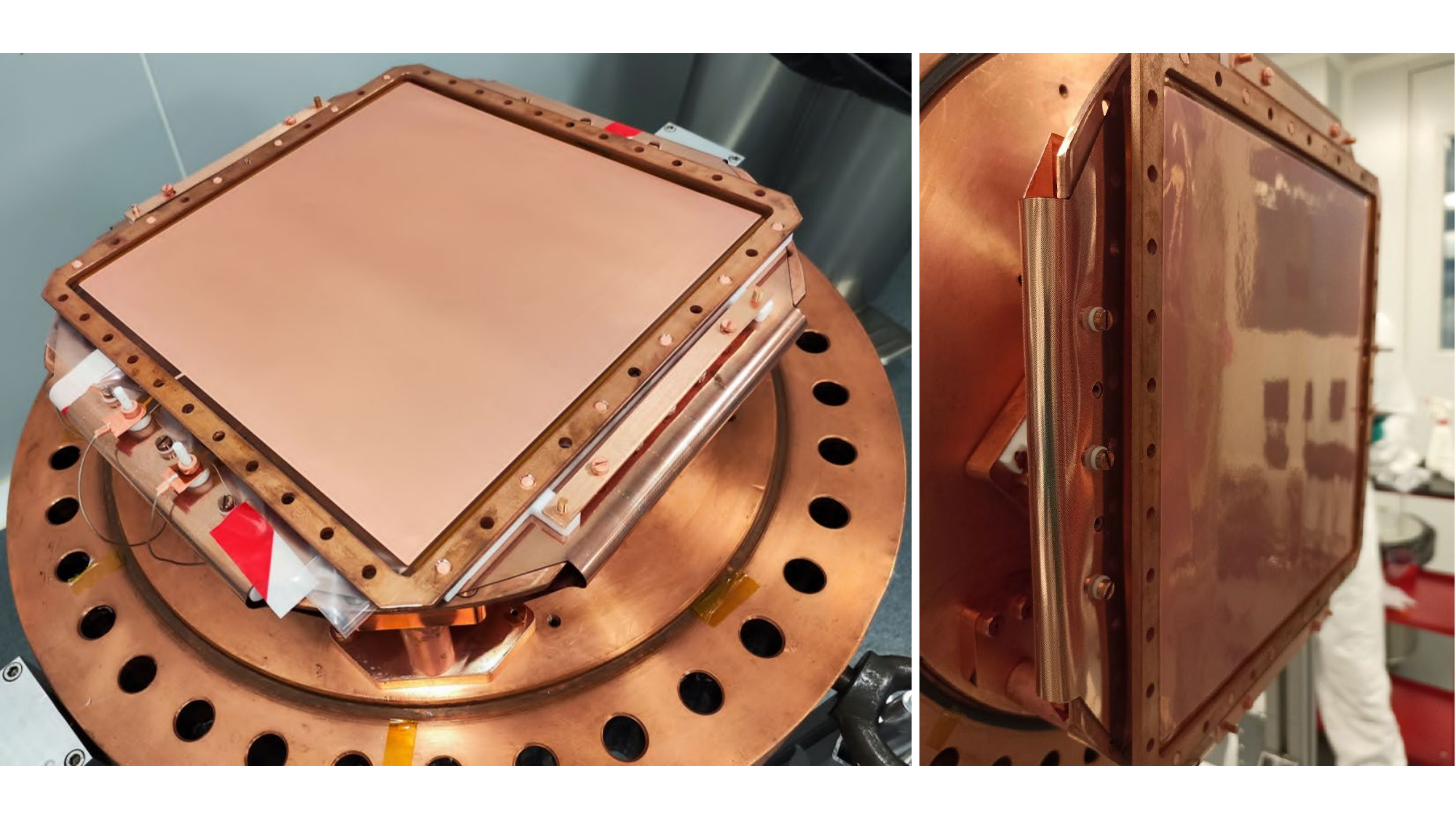}
    \caption{GEM foil installed on top of the Micromegas in TREX-DM. Pictures form first installation in the clean room of Lab2400 in LSC.}
    \label{fig:GEMinstallation}
\end{figure}

The first installation took place in March 2024 in the clean room of Lab2400 in LSC, see figure \ref{fig:GEMinstallation}, where one GEM foil was installed in the ``left" side of TREX-DM. Despite the proper tests after closing the chamber in the clean room, the transport to the Lab2500, where TREX-DM is currently installed, affected the GEM producing a short-circuit and leaving that half of the detector inoperative. In June 2024, a new clean room was finally installed in Lab2500 (in collaboration with the LSC) that enabled the first in-situ intervention, during which the damaged GEM was replaced with a new unit.

\begin{figure}[h!]
    \centering
    \includegraphics[width=0.8\linewidth]{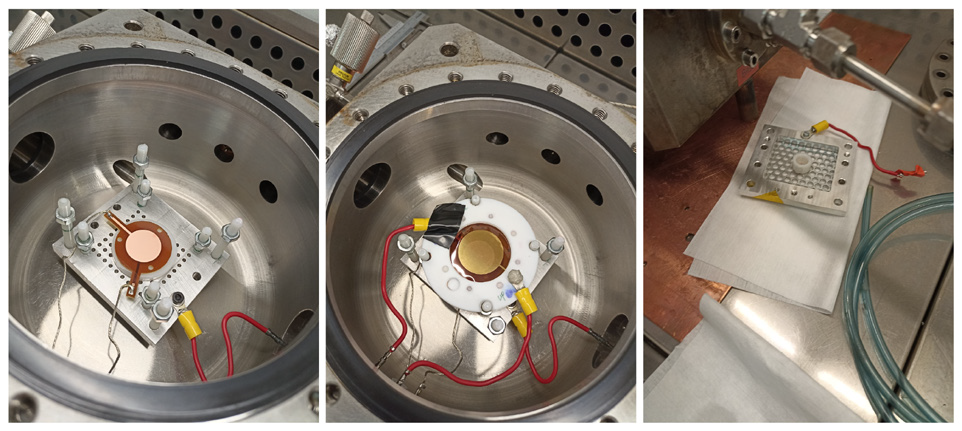}
    \caption{Small set-up to test the combination of Microbulk Micromegas with GEM foil on top. \textit{Left}: the 2 cm diameter Micromegas on the support plate. PTFE pillars isolate components at different voltages. \textit{Center}: GEM foil on top of the Micromegas. \textit{Right}: Cathode grid with $^{55}$Fe source attached facing down. Images published in \cite{pablo2025micromegas}.}
    \label{fig:GEMsmall}
\end{figure}

\begin{figure}[h!]
    \centering
    \includegraphics[width=0.8\linewidth]{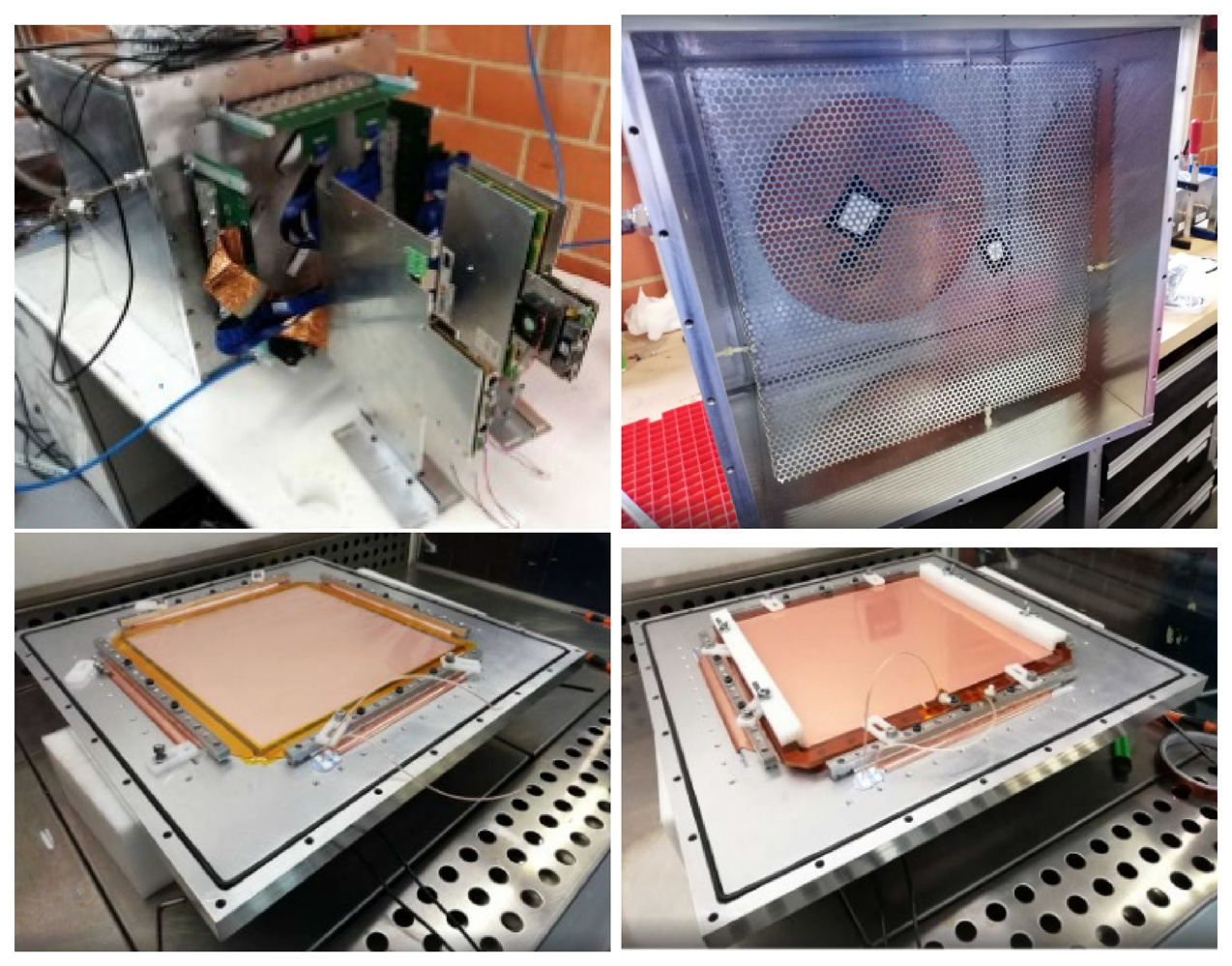}
    \caption{Full size test bench, named by myself \textit{ MicromegasBox}. \textit{Top left}: closed vessel with DAQ. \textit{Top right}: cathode with two $^{109}$Cd radioactive sources, attached with black tape. \textit{Bottom left}: Micromegas detector secured on the endcap of the chamber. \textit{Bottom right}: GEM foil placed on top of the Micromegas. \cite{pablo2025micromegas}.}
    \label{fig:GEMfullScaleSetup}
\end{figure}

The performance of the Micromegas plus GEM structure was  tested in the IAXOlab, prior to the installation in TREX-DM. Two set-ups were prepared, a scaled version and a full size setup. Starting with a small chamber hosting a single-channel Micromegas and a small GEM (figure \ref{fig:GEMsmall}), maintaining the same hole pattern of both devices as in the final set-up, allowed to test voltages with different gases and pressures. In parallel, a full size prototype with an Micromegas and GEM foil (figure \ref{fig:GEMinstallation}), identical to those of the TREX-DM experiment was tested with the same gas, argon plus 1 \% isobutane at 1 bar in the so-called \textit{MicromegasBox}, shown in figure \ref{fig:GEMfullScaleSetup}. 

These two test setups, their measurements and the results are the main object of~\cite{pablo2025micromegas}. Also in \cite{Oscar2025development} there is a chapter devoted to these measurements, including gain and electron transmission curves. Here, the main results are reproduced in table \ref{tab:GEMtests}, where the gain increase for both setups is shown. It is presented in two ways, that are called preamplification factor and GEM extra factor. Preamplification factors are defined as the gain ratio between GEM + Micromegas runs ($V_{GEM} \neq 0 $ V) and only Micromegas runs ($V_{GEM}  = 0$ V). Whereas GEM extra factor is obtained comparing the gain of Micromegas + GEM with the highest voltage that can be applied to the Micromegas only, $V_\text{mesh}^\text{opt}$, which is always higher than the one applied when the GEM is installed, $V_\text{mesh}^\text{ref}$. This is the real advantage obtained in the combined set-up. These results showed the feasibility of increasing the gain between one and two orders of magnitude depending on the configuration and, in particular, for a full scale set-up the increase in gain can be as high as a factor 80. 

With these works, values for drift and transfer fields were determined. As a reference, values of $E_{drift} = 100$ V/cm/bar and $E_{transfer}=150$ V/cm/bar were established. In TREX-DM data taking these values have been employed in a number of runs, although recently they have both been decreased around 90 V/cm/bar due to stability issues.

\begin{table}[ht!]
\centering
\begin{tabular}{|c|c|c|c|c|c|c|}
\hline
\textbf{Set-up} & \textbf{Pressure} & $V_\text{mesh}^\text{ref}$ & $V_\text{GEM}$ & \textbf{Preamp.} & $V_\text{mesh}^\text{opt}$ & \textbf{GEM} \\
 & (bar) & (V) & (V) & \textbf{factor} & (V) & \textbf{extra factor} \\
\hline
\multirow{3}{*}{Small} 
& 1  & 305 & 310 & -  & 315 & 90 \\
& 4  & 390 & 410 & 70 & 400 & 50 \\
& 10 & 535 & 550 & 21 & 540 & 19 \\
\hline
\multirow{1}{*}{Full scale}
& 1  & 290 & 285 & 85 & 293 & 80 \\
\hline
\end{tabular}
\caption{Gain improvements with the Micromegas + GEM combined setup for the small and full scale prototypes. Measured preamplification and GEM extra gain factors for both at different pressures with argon + 1\% isobutane. For the full scale setup the gain increase could be up to a factor 80, as stated by the GEM extra factor \cite{pablo2025micromegas}.}
\label{tab:GEMtests}
\end{table}

Since June 2024, TREX-DM has been operating with a Micromegas + GEM readout plane in its left side. This has increased the complexity of the experiment: new structure designed and machined to hold the GEM aligned on top of the Micromegas, two new feedthroughs needed to apply voltages to GEM electrodes, new power supply channels included in the slow control, new ramping-up and ramping-down protocols to prevent damages and sparks... From all these challenges, one stands above all that until today remains an open issue: stability. Since the installation of the GEM, the trip rate has significantly increased, what limits the live time of the experiment. Every time a trip happens, all voltages are down to 0~V and a recovery protocol ramp them up slowly, the whole process lasting almost 10 minutes. Two are the main suspects for these sparks: high energy events that are now amplified greatly, triggering discharges during the avalanche process; or spontaneous discharges between nearby electrodes. The first case also happened in Micromegas only operation, but it was limited to a small percentage of alpha events. Now, events from calibration sources saturate the DAQ and therefore the threshold to produce sparks is well lower than before. Reducing the operation voltage of the field cage's last ring, almost touching the GEM foil, tests the second suspicion. Although no clear conclusions could be drawn for the moment, the experiment has taken data the last months, albeit with a large dead time. A new field cage has been designed and is under fabrication that is expected to allow recover live times close to 100\%. Its installation is foreseen in autumn 2025. Stability issues haven't affected much the characterization of the detector. The biggest milestone achieved with the Micromegas + GEM set-up has been the reduction of the threshold until tens of eV. A dedicated calibration with $^{37}$Ar was performed to measure this and will be explained in the next subsection.

\begin{figure}[h!]
    \centering
    \includegraphics[width=0.8\linewidth]{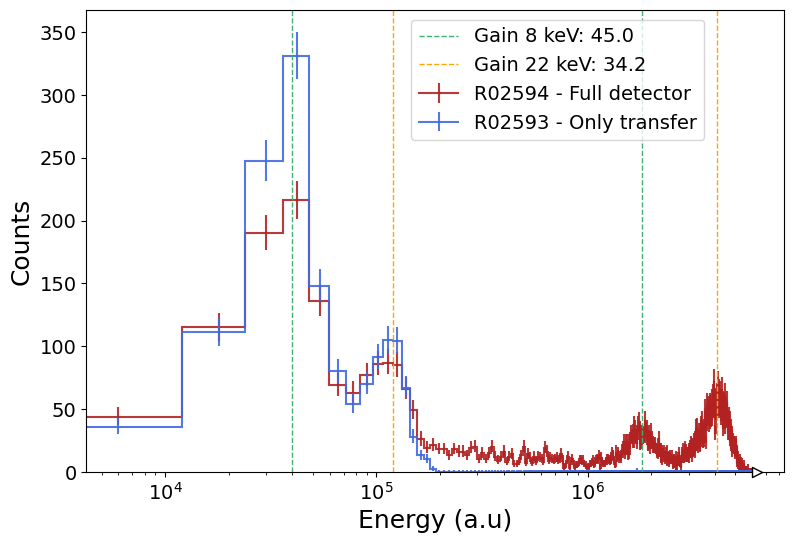}
    \caption{Micromegas and GEM $^{109}$Cd calibration spectra. In red, full detector spectrum, in blue GEM with no voltage difference so that only events in the transfer region are collected. This shows that in the red spectrum, both types of events are present, these from the drift volume that are preamplified by the GEM and those form the transfer region with no amplification. This allows to measure the intrinsic preamplification factor obtained by the GEM, which is computed from the 8 and 22~keV peaks identified by vertical lines. The preamplification value is 45 for the 8~keV peak and 34 for the 22 keV peak. (Run numbers 2593 and 2594 with 270 V in the mesh and in the GEM. In the case of 2593 GEM voltage was set to -10 V, so electric field in the reverse direction to prevent events from drift volume to trespass.)}
    \label{fig:MM_GEM_Cd109spectra}
\end{figure}

\begin{comment}
Regarding the increase in gain, it can be measured from routine $^{109}$Cd calibrations. Figure \ref{fig:MM_GEM_Cd109spectra} presents two spectra. The red one makes use of the full detector, with $V_{mesh} = 270$ V and $V_{GEM} = 270$ V. Peaks at 8 and 22 keV amplified both by GEM and Micromegas can be seen on the right. But also events in the transfer region, amplified only by the Micromegas and not the GEM, form both peaks in the low energy region of the spectrum, at he left. The origin of these two peaks in the low energy region is confirmed with a dedicated run in which the cathode is switched off and the voltage in the GEM is set to $V_{GEM} = -10$ V, so the electric field inside the GEM is in the reverse direction to prevent any event from the upper volume to reach the transfer region.  This is plotted in blue and both peaks coincide perfectly with the full detector spectrum. The preamplification factor due to the GEM can be extracted comparing positions of GEM amplified peaks and Micromegas-only amplified ones. Values presented there in a TREX-DM calibration are roughly half of the maximum values obtained in the MicromegasBox test bench, showed in table \ref{tab:GEMtests}. Computed values in TREX-DM calibration are 34 from the 8 keV peak and 45 from the 22 keV peak. They compare with preamplification factor measured in the IAXOLab of 85. This is due to the voltages applied, that were 15-20 V higher in the test set-up. Sparks prevented to go further in TREX-DM. 
\end{comment}

The gain increase can be measured using routine $^{109}$Cd calibrations. Figure \ref{fig:MM_GEM_Cd109spectra} shows two spectra. The red one corresponds to the full detector operating with $V_{mesh} = 270$ V and $V_{GEM} = 270$ V. Peaks at 8 and 22 keV, amplified by both the GEM and Micromegas, appear on the right. On the left, two lower-energy peaks originate from events in the transfer region, which are only amplified by the Micromegas.
To confirm their origin, a dedicated run was performed with the cathode off and $V_{GEM} = -10$ V, reversing the electric field in the GEM to block events from the upper volume. The resulting spectrum, shown in blue, overlaps perfectly with the Micromegas-only peaks from the full detector, confirming their source.
The GEM preamplification factor is estimated by comparing the positions of the GEM-amplified peaks to those amplified only by the Micromegas. In this TREX-DM calibration, values of 34 (8 keV peak) and 45 (22 keV peak) are obtained, about half the values from the MicromegasBox test bench (see Table \ref{tab:GEMtests}), where a factor of 85 was reached. The difference is due to the higher voltages, 15–20 V more, used in the test setup. In TREX-DM, sparks prevented increasing the voltage further.

\subsection{$^{37}$Ar low energy calibration}

The gain increase obtained with the GEM encouraged us to perform dedicated low energy calibrations to measure the energy threshold. With this purpose the isotope $^{37}$Ar was selected. It has a half-life of 35.04 days and it decays through electron capture ($Q=813.9$ keV) to $^{37}$Cl, emitting photons of 2.82 keV from K shell with 90\% probability and 0.27 keV form L shell with 9\% probability \cite{be2013table}. Also, as it is gaseous, it allows uniform calibrations along the readout plane. This isotope is not present in nature in significant quantities so it has to be produced. Three main production channels have been employed depending on the needs: Irradiation of $^{36}$Ar with thermal neutrons via the reaction $^{36}$Ar($n, \gamma$)$^{37}$Ar, proton irradiation via the reaction $^{37}$Cl($p, n$)$^{37}$Ar, and neutron activation of calcium oxide (CaO) targets via the reaction $^{40}$Ca($n, \alpha$)$^{37}$Ar. This last method has been used in other TPCs experiments like NEWS-G \cite{gerbier2014news} and it has been selected for TREX-DM too. 

\begin{figure}[h!]
    \centering
    \includegraphics[width=0.8\linewidth]{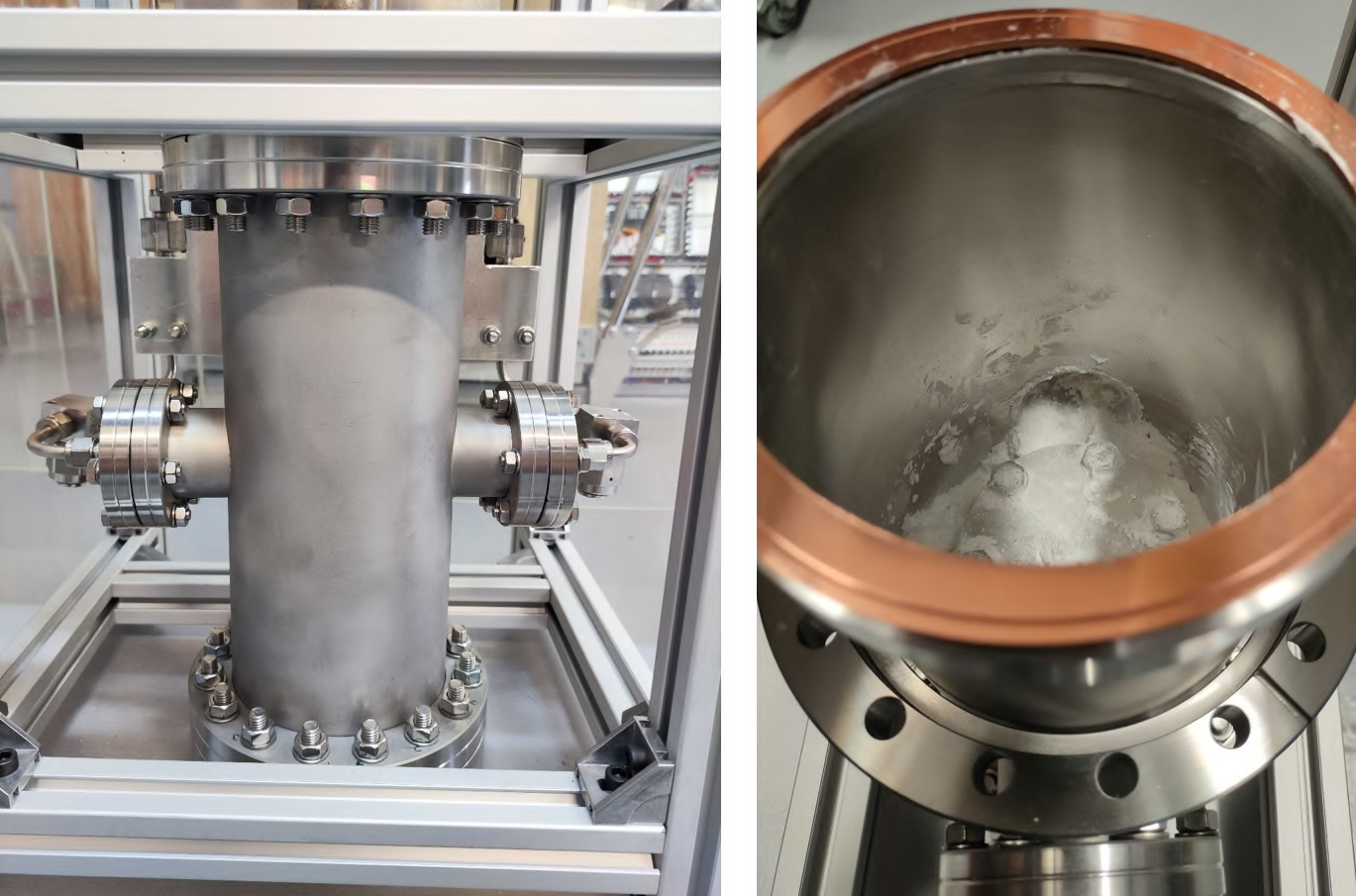}
    \caption{\textit{Left}: Vessel for CaO powder. \textit{Right}:  0.5 kg of CaO powder inside the vessel.}
    \label{fig:Ar37vessel}
\end{figure}

The Sigma Aldrich CaO powder was selected with a certified purity of 99.9\%. The Certificate of Analysis confirmed an actual purity of 99.985\%, with magnesium identified as the primary impurity, a minimal concern for our application. A quantity of 0.5 kg of CaO powder was deposited inside an iron vessel, leak and pressure-tested up to 10 bar. It has a cross shape, with two in-out ports protected from powder with Swagelok 0.5 $\upmu$m filters and an Ultra
High Vacuum (UHV) valve mounted on the top section of the vessel to achieve and maintain the required vacuum levels and pressure specifications during operation;  another 4 $\upmu$m filter was incorporated to prevent the risk of powder entering in the gas system. A picture of the vessel can be seen in figure \ref{fig:Ar37vessel}.

Powder activation was performed in Centro Nacional de Aceleradores (CNA) in Sevilla, specifically in the HiSPANoS irradiation facility. The neutron flux is produced bombarding with accelerated deuterons a thick beryllium target, which produces neutrons through the reaction $^9$Be($d, n$)$^{10}$B. The resulting neutron spectrum is a continuum with a mean energy of approximately 5 MeV. The irradiation was planned for approximately 6 hours. 

In order to assess the activity level of $^{37}$Ar due to neutron irradiation of $^{40}$Ca other by-product is used: $^{42}$K from $^{42}$Ca with half-life 12.4 hours. Argon-37 decays cannot be seen from outside the vessel as its low-energy emissions are absorbed by the stainless-steel wall but the $^{42}$K has a 1525 keV gamma emission that can escape. Activity of these isotope was monitored with a sodium iodide (NaI) and a lanthanum bromide (LaBr$_3$) scintillators. The relation between activities of both decays is estimated as $A(^{37}Ar)/A(^{42}K) \sim 20$ \cite{Oscar2025development}.

Following the irradiation process, the vessel remained in the irradiation hall until the dose rate decreased below the environmental background radiation level ($<$~0.2~$\upmu$Sv/h). Identifying the $^{42}$K gamma proved challenging due to significant material activation. The region of interest at 1525 keV contained numerous unidentified spectral components, presumably arising from stainless-steel vessel activation. An overall fast decaying component with half-life of 2.35~h was identified and fitted together with the decay of $^{42}$K. The initial activity for $^{42}$K was established in the range of 100-1000 Bq, which implies an approximate $^{37}$Ar activity of the order of 1-10 kBq.

\begin{figure}[h]
    \centering
    \includegraphics[width=0.8\linewidth]{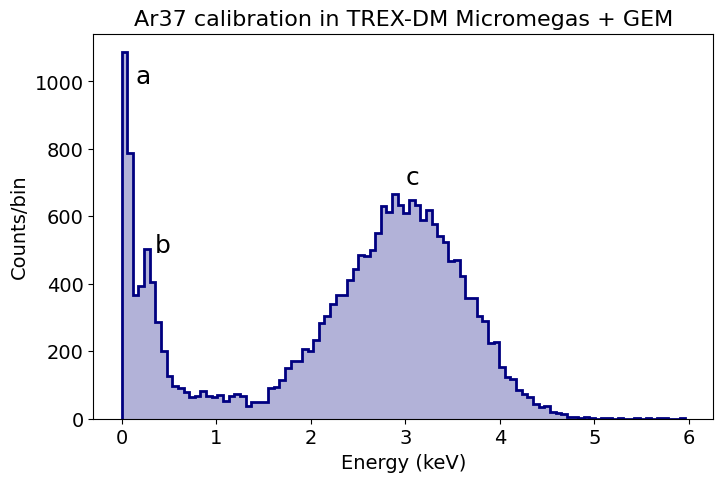}
    \caption{Low energy spectrum of an $^{37}$Ar calibration. Three peaks can be seen: a) 2.8 keV peak from transfer region, b) 0.27 keV and c) 2.8 keV.}
    \label{fig:Ar37SpectrumFilled}
\end{figure}

The first $^{37}$Ar calibration with the Migromegas + GEM readout plane took place in October 2024. Fifteen days after this second irradiation, $^{37}$Ar was injected into TREX-DM by a procedure in which the powder vessel is pressurized above the desired pressure for the detector chamber. In this case, 2 bar was used to reach 1.1 bar in the TREX-DM vessel. Repeated injection cycles allow to increase the amount of $^{37}$Ar pushed to the sensitive volume. In this case, two cycles were performed resulting in the transfer of an estimated 75\% of the container’s activity.

The development of these procedures to generate and inject the $^{37}$Ar and its validation as a low energy calibration source is further detailed in \cite{Oscar2025development}.

Activity was more than enough to see the expected peaks from $^{37}$Ar in seconds. In figure \ref{fig:Ar37SpectrumFilled} a 5 min-long calibration is presented. Three peaks are clearly seen in the spectrum, from low to high energy, 2.8 keV peak from transfer region, that is, without GEM amplification, and 0.27 keV and 2.8 keV from drift region, events amplified by the GEM. Voltages applied in this run are the highest achieved in TREX-DM, $V_{mesh}=295$ V and $V_{GEM}=280$ V, but due to that, stability was very short.

\begin{figure}[h]
    \centering
    \includegraphics[width=0.9\linewidth]{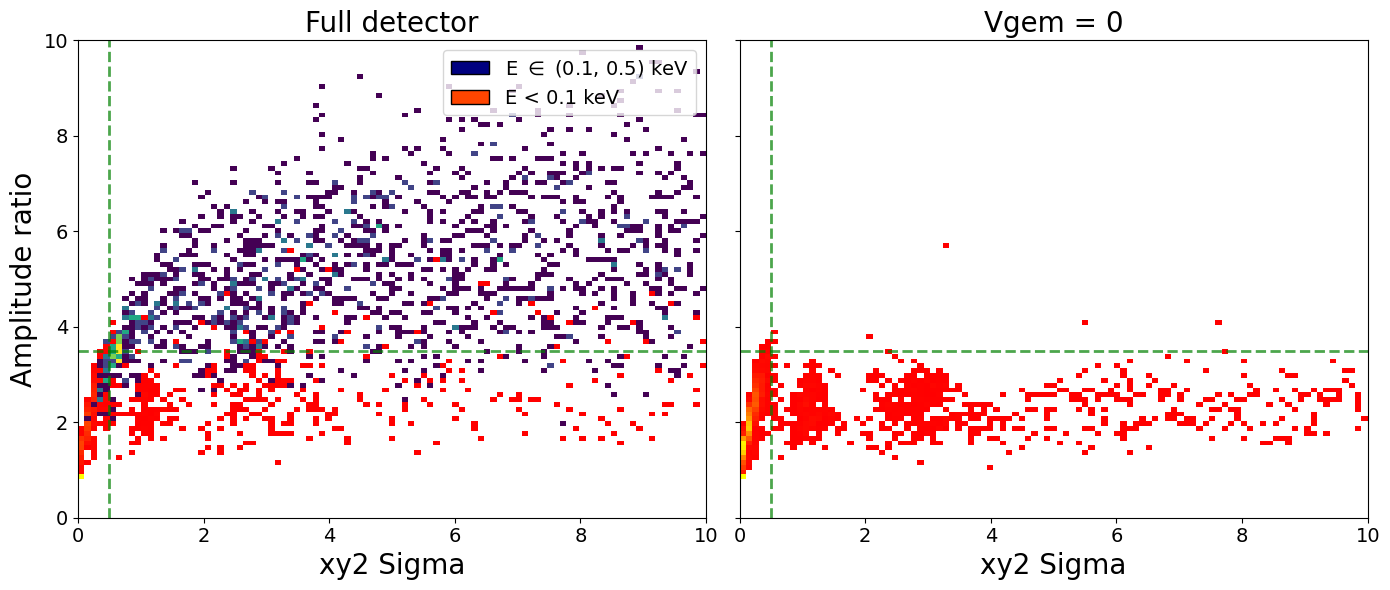}
    \caption{2D histograms of ``Amplitude ratio" vs ``xy2 Sigma". In red colour-scale events below 0.1 keV, in blue colour-scale events with energies between 0.1 and 0.5 keV. \textit{Left}: Events from a full detector run. Its spectrum is shown in figure \ref{fig:Ar37SpectrumFilled}. \textit{Right}: Events from a run with $V_{GEM} = 0$ V. This implies almost no events from drift volume reach the Micromegas and in any case there is no amplification in the GEM. This is the population we would like to remove from full detector runs to properly assess the threshold of the Micromegas + GEM readout plane.}
    \label{fig:xySigmaAmplitudeRatio_Full_GEM0}
\end{figure}

\begin{figure}[h!]
    \centering
    \includegraphics[width=0.9\linewidth]{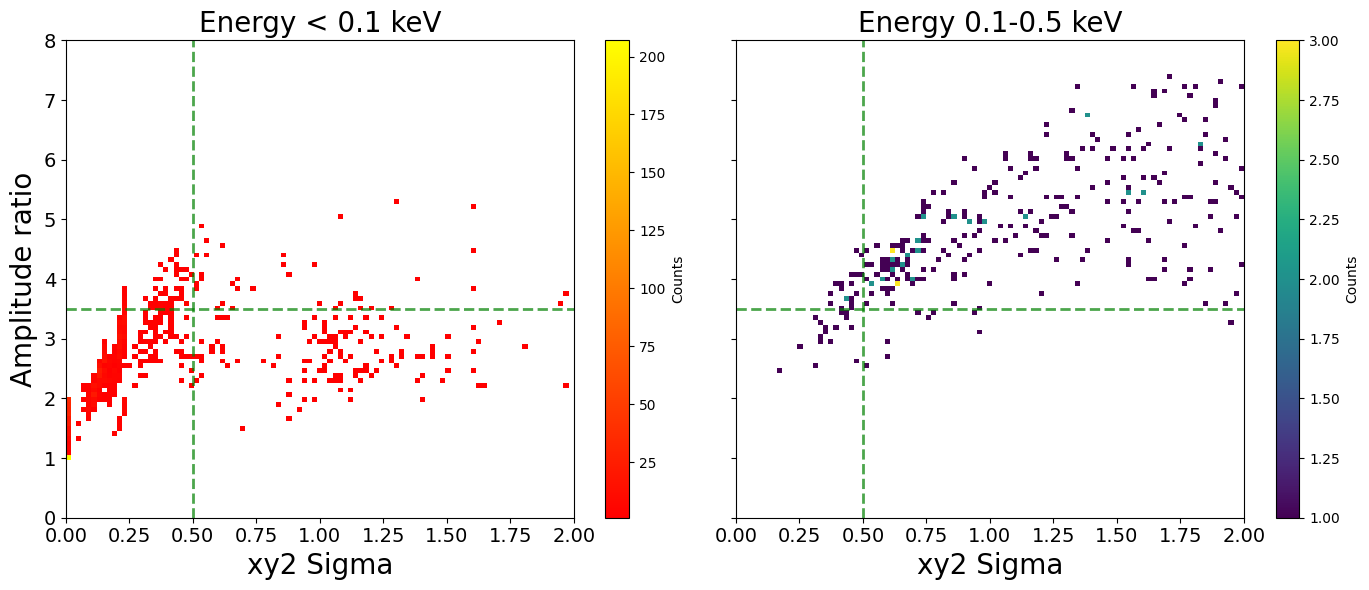}
    \caption{Close up of left plot in figure \ref{fig:xySigmaAmplitudeRatio_Full_GEM0}. Selecting appropriate values in both variables, \textit{Amplitude ratio} and \textit{xy2 Sigma},  GEM amplified events can be identified. \textit{Left}: events with energy below 0.1 keV. \textit{Right}: events with energy between 0.1 and 0.5 keV. Conditions applied to identify GEM amplified events are \textit{Amplitude ratio} $>$ 3.5 and \textit{xy2 Sigma} $>$ 0.5, so all that fall inside upper right region. }
    \label{fig:xy2SigmaAmpRatio_Cuts}
\end{figure}

\begin{figure}[h!]
    \centering
    \includegraphics[width=0.9\linewidth]{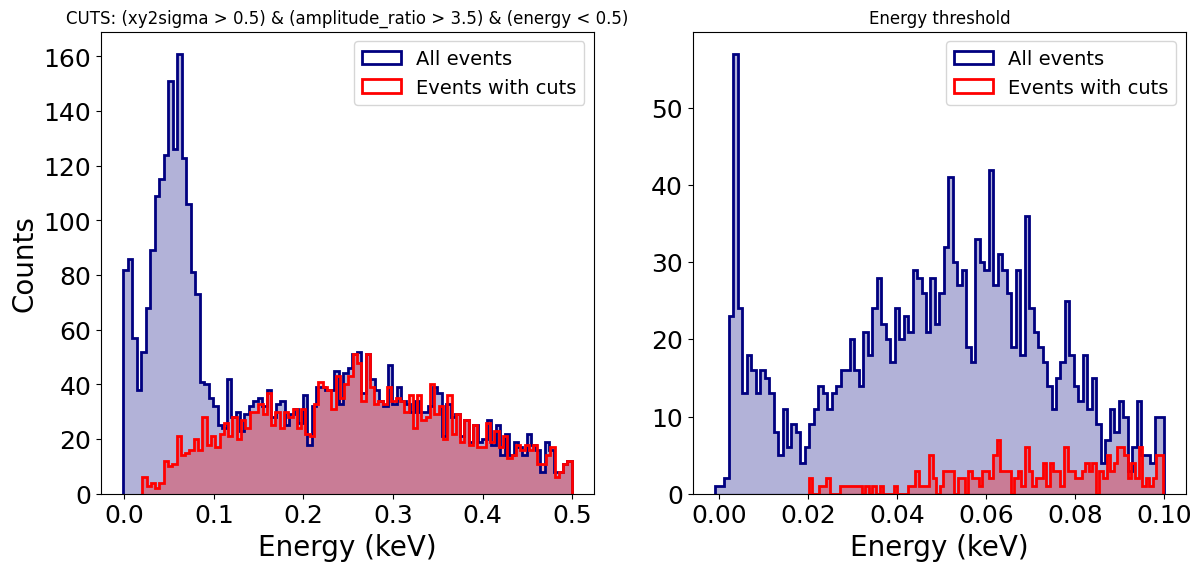}
    \caption{Low energy range of $^{37}$Ar spectra. The right plot is the augmented version of left plot. In both cases, the highest 2.8 keV peak from transfer region buries the smallest events amplified by the GEM. The application of several selection criteria allows to distinguish two different populations, in blue, full spectrum, in red only selecting events amplified by the GEM. The right plot focuses on the low energy part of the left plot and shows that events as weak as 20 eV can be detected with the combination of GEM and Micromegas. }
    \label{fig:Threshold}
\end{figure}

The first peak from the transfer region hides the smallest events coming from the drift region. In figure \ref{fig:Threshold} two close ups of the calibration spectrum can be seen in blue in different energy ranges. Events from the transfer region are piled up below 0.1~keV, in Micromegas +  GEM energy scale, and they are on top of the limit of the 0.27~keV peak from $^{37}$Ar in the drift region. It would be desirable to distinguish between these two populations to recover the lowest energy events that have been amplified by the GEM. To do so, a population free of transfer region events is identified: the peak of 0.27 keV. All these events have suffered the amplification in the GEM, therefore the strategy consists on identify a characteristic pattern that these events have and those of the transfer region do not. Between tens of observables, two were identified as promising for this task: ``Amplitude ratio" and ``xy2 Sigma". ``Amplitude ratio" is obtained  adding the amplitudes of all the pulses of the event and dividing them by the maximum amplitude among them. In some way it measures the energy density of the electron cloud when it reaches the Micromegas. The second is ``xy2 Sigma", obtained as the energy-weighted variance of the position of the hit channels in X and Y. This is again a way to measure the spread of the charges, in this case in XY plane. Plotting a 2D histogram with the values of these two observables, figure \ref{fig:xySigmaAmplitudeRatio_Full_GEM0}, both populations of interest can be identified in the ``xy2 Sigma"-``Amplitude ratio" plane. In the left plot the histogram is plotted with two colour scales, in red for events below 0.1 keV and in blue for events between 0.1 and 0.5 keV. In the first group, events from the transfer region dominate; in the second, all are from drift region. Two lines delimiting the separation between populations are presented, this would be the values looked for to identify GEM amplified events: ``xy2 Sigma"~$> 0.5$ and ``Amplitude ratio"~$> 3.5$. In figure \ref{fig:xy2SigmaAmpRatio_Cuts} close ups for both populations are presented together with the proposed criteria.

A dedicated run without voltage in the GEM was taken to test this discrimination schema. In figure \ref{fig:xySigmaAmplitudeRatio_Full_GEM0} right plot, the behaviour of these two observables for these data can be seen.  As in the full detector run, the selected area eliminates almost all these events that are from the transfer region. This supports the selected cuts for this purpose. 

Applying these selection criteria, events from the transfer region are discarded. In figure \ref{fig:Threshold} in red, the spectrum of the remaining events can be seen. The main feature contrasted here is that almost no events from the 0.27~keV peak are removed, while some remain below 0.1~keV. In the right plot, a zoom to low energies is presented. From there, an optimistic energy threshold can be established around 20~eV, and in any case 40~eV would be a conservative value. In argon, the mean energy to extract an electron is around $W=26$ eV, so this proves that the Micormegas + GEM configuration is sensitive to energy depositions compatible with single electron ionizations.

%% file: Chapters/4_NewGaseousDetectors.tex
This chapter presents two small test setups that were designed and developed to measure specific properties. The first one is a new calibration source with variable energy made with a UV LED and a photocathode. And the second one allowed a detailed study of a new gas mixture for us, argon + 10\% isobutane, at different pressures. Both provide useful information for future upgrades of TREX-DM that may include these techniques. They are just two stand alone examples of small gaseous detectors that we operated in IAXOLab. Small setups with GEM + Micromegas at different pressures or the ``MicromegasBox" test bench to characterize TREX-DM Micromegas V2 are other examples of these low size experimental setups that we can operate easily in our laboratory.

\section{UV calibrations}

\subsubsection{Objective}

Gaseous detectors like Micromegas are sensitive devices whose performance is affected by the properties of the gas and the precise manufacturing of the detector. Slight changes in gas composition or pressure affect the electron mobility and recollection, meaning different energy gain, resolution or threshold. Also, micro-pattern devices as Micromegas are susceptible to irregularities produced during fabrication that may result in varying performance along its surface, border effects, blocked holes or bad strip connections.

Working with big readout planes like in CAST ($6 \times 6$ cm$^2$), the axion helioscope, in its successors BabyIAXO and IAXO, described in chapter \ref{ch:DarkSector},  or in TREX-DM ($25 \times 25$ cm$^2$) implies that some of these non-ideal behaviour will appear. Therefore, a test bench where to examine some of these properties and to develop techniques, both in hardware and in post-analysis, to mitigate their effects was envisaged. A source to generate electrons in a controlled and precise way would be needed, so that the drift distance, the time of the interaction and its position would be known. Previous works have shown that ultraviolet light is extremely useful for this purpose. It can extract electrons by photoelectric effect in a number of metals and, generated at intense enough fluxes, can ionize impurities and some organic gases. Noble gas pulsed lamps \cite{colas2002electron, mcdonald2019electron} or UV lasers \cite{pellecchia2020uv} have been used together with micro-pattern detectors to study gas properties.  

\subsubsection{Methods and materials}

Gaseous detectors are routinely calibrated with radioactive sources emitting in the X-ray range. $^{55}$Fe with photons of 5.9 keV and $^{109}$Cd with several photons around 22 keV and the associated stimulated fluorescence of copper at 8 keV are the most common radioisotopes used for Micromegas calibrations. Their emissions tend to be isotropic, depending on the encapsulation, and random, as any atomic decay is. 

\begin{comment}
  Collimation is possible but then the rate decrease. Their intensity is usually limited in origin by their categorization as ``exempt source" by the ``Consejo de Seguridad Nuclear". Therefore, the use of radioactive sources for precise electron generation is not viable.  
\end{comment}

For precise characterizations of the readout planes, another type of source can be used: UV photons. They can be generated in a controlled way and allow electron extraction from metals by photoelectric effect. Common UV photon generators are noble gas lamps and UV lasers. Both are capable of generating high intensity fluxes of photons but with different spatial characteristics: UV lamps are not collimated and lasers, by definition, produce light in a preferred direction. 
\begin{comment}
X-ray generators, in which electrons are accelerated towards a target where UV photons are generated by Bremsstrahlung have been also used to characterize detectors but they require larger equipments so these tests are performed in dedicated laboratories like PANTER in Munich or lab 162/S-065 at CERN.    
\end{comment}
This makes lasers the perfect device for our purpose, the can focus the light in a small spot where electrons are extracted. Recently, another source has been commercially available: UV LED. They present the advantage of being more affordable compared to other UV sources (less than 100\euro{}) and easy to use. For the ``proof-of-concept" set-up that will be explained here, a UV LED was selected.

There are two ways that UV photons can be used to generate electrons in a gaseous chamber. The first technique is based on the interaction with small-concentration impurity molecules, which are naturally present due to
outgassing of the detector walls or the gas pipe. Such molecules have low ionization potential, less than 9~eV, and can then be ionized in multi-photon processes by ultraviolet laser beams. As these are multi-photon processes, high intensity fluxes are needed to achieve relevant electron production. A common wavelength for UV photons produced by these sources is 266~nm, or 4.66 eV, which explains why at least two photons have to interact together in order to ionize a molecule. 

\begin{figure}
    \centering
    \includegraphics[width=0.8\linewidth]{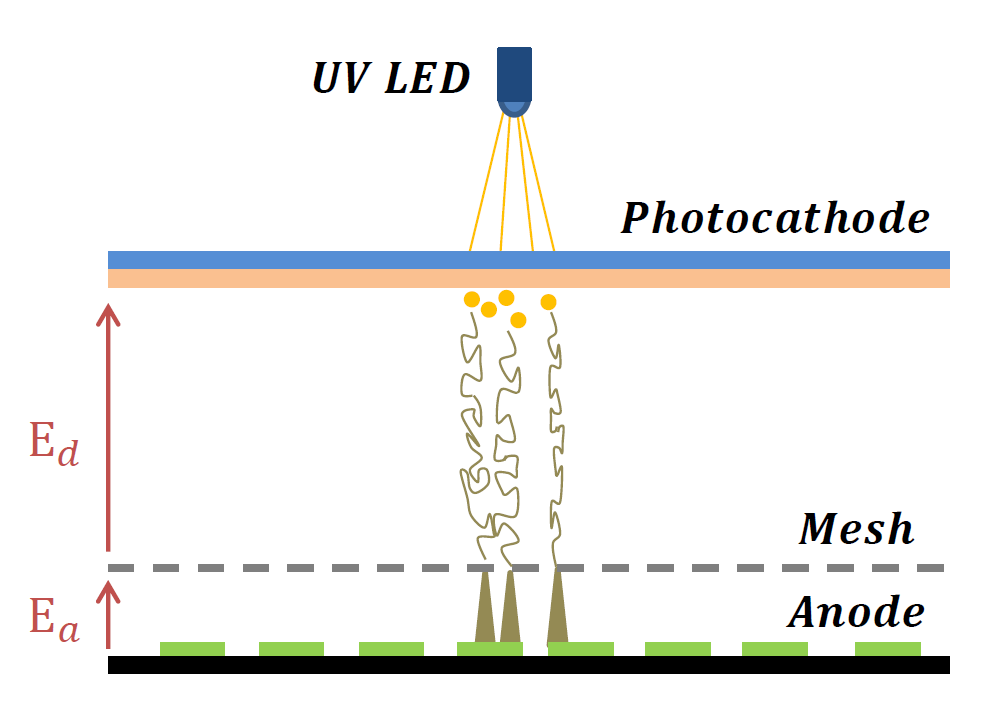}
    \caption{Schematic view of a UV LED calibration source with a photocathode. The LED shines through the quartz substrate, in blue, where the metallic layer (aluminium or copper) was deposited, in light orange. Electrons extracted, yellow dots, drift towards the Micromegas, where the charge is amplified. }
    \label{fig:UVphotocathodeSchema}
\end{figure}

The second technique is based on the photoelectric effect. Many materials, in particular many metals, have work functions with energies lower than 4.66 eV \cite{lang1971theory}. The work function is the minimum work needed to remove an electron from a solid to a point in the vacuum immediately outside the solid surface. It may vary depending on the crystallographic disposition of the atoms in the lattice of the crystal and the electric field applied. The Schottky effect indicates that the work function decreases as the electric field increases. The photoelectric effect is used to extract electrons from metals, typically placed at the top of the detector and used as cathode. Metal grids or thin layers deposited in quartz lenses may act as photocathode. Depending on the purpose, it could be convenient that some photons cross the cathode and reach the anode as well: this would generate a very characteristic double signal, useful for drift velocity measurements.

\begin{figure}[h!]
    \begin{minipage}{0.48\textwidth}
         \centering
         \includegraphics[width=0.99\linewidth]{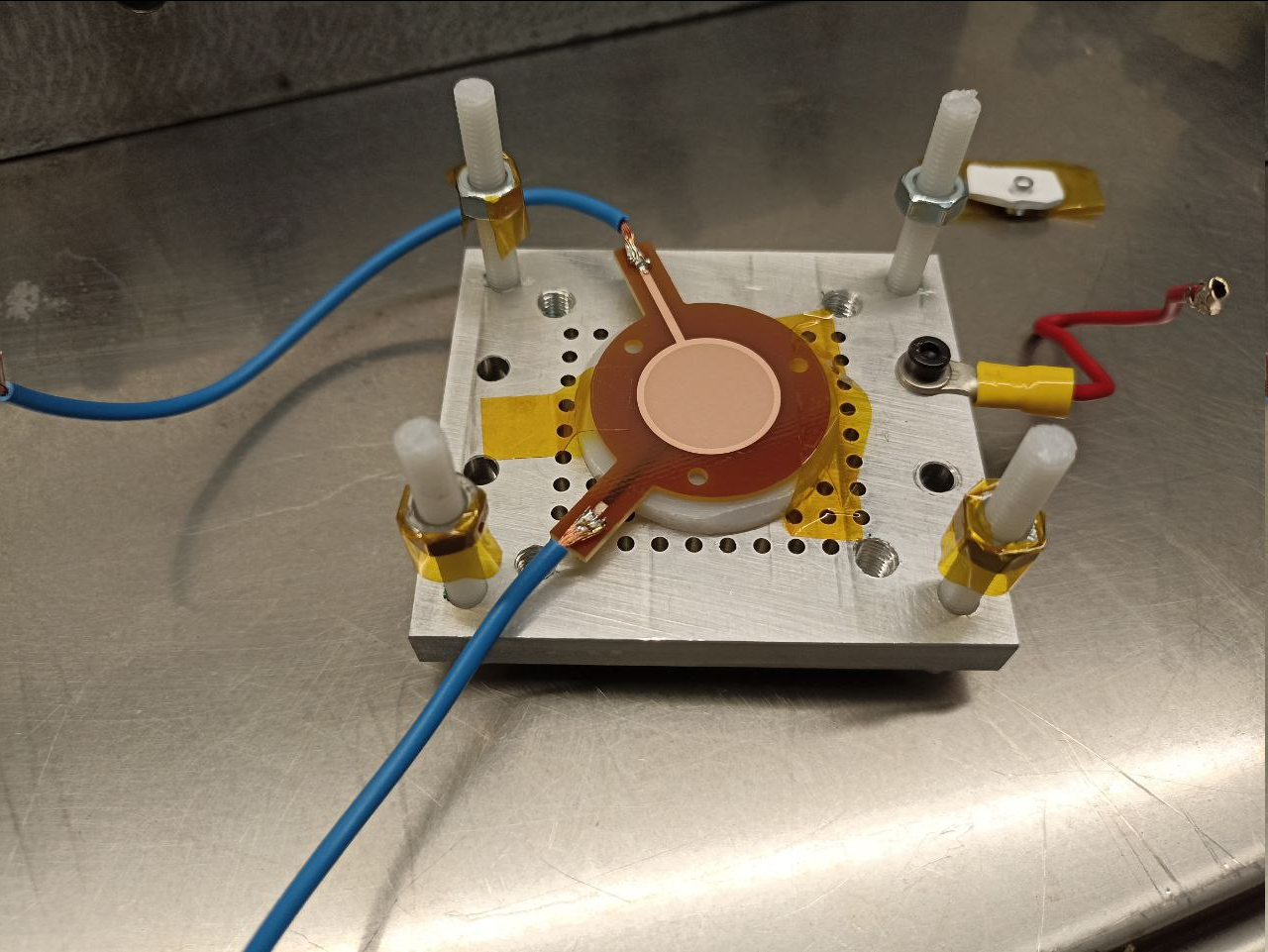}
   \end{minipage}
   \begin{minipage}{0.48\textwidth}
     \centering
     \includegraphics[width=0.99\linewidth]{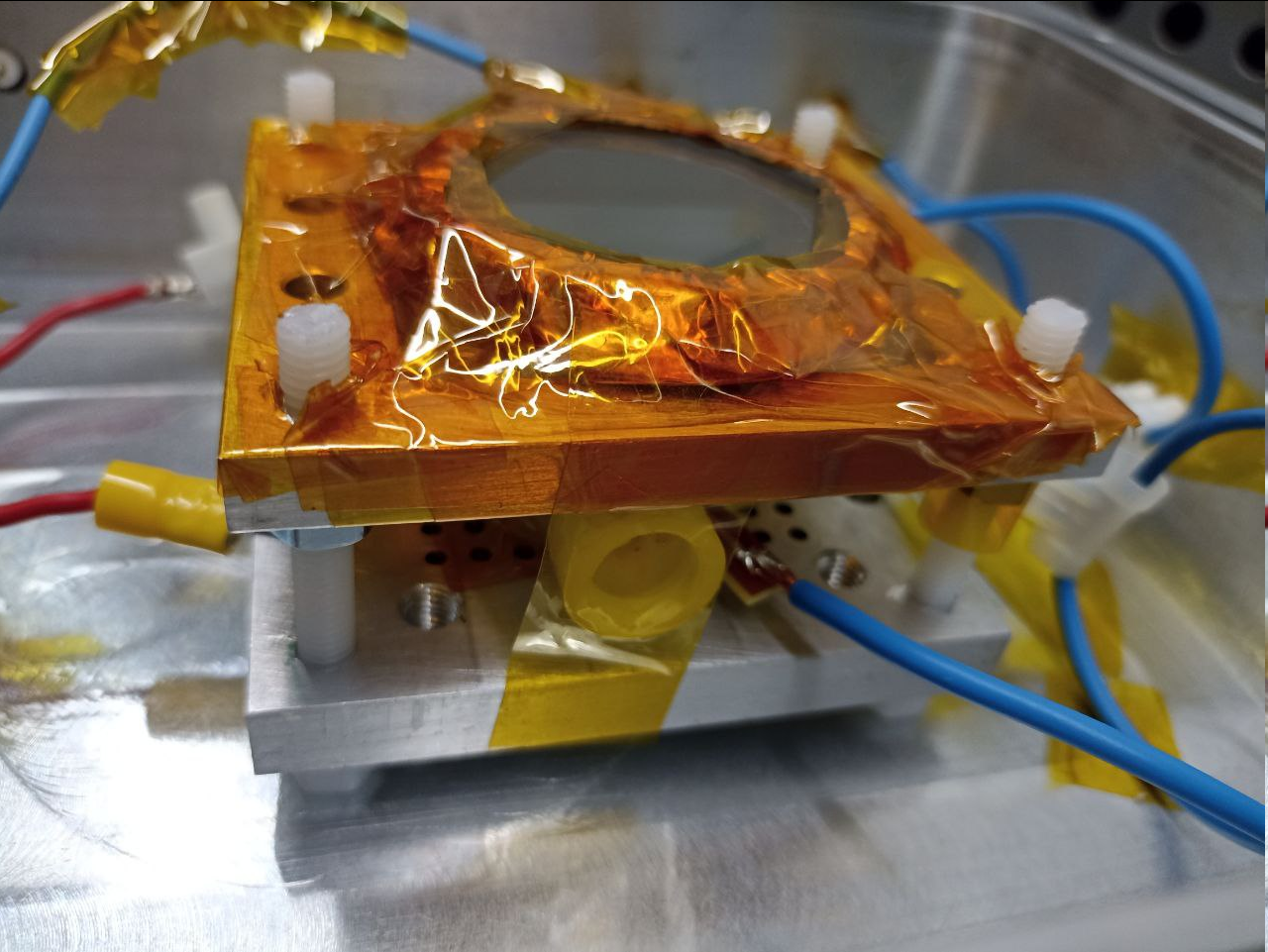}
   \end{minipage}\hfill
    \caption{\textit{Left}: Small Micromegas installed at the bottom of the TPC. \textit{Right}: Close up of the TPC: photocathode on top, $^{55}$Fe source in white piece taped on the side.}
    \label{fig:SmallMM}
\end{figure}

\begin{figure}[h!]
    \begin{minipage}{0.48\textwidth}
         \centering
         \includegraphics[width=0.99\linewidth]{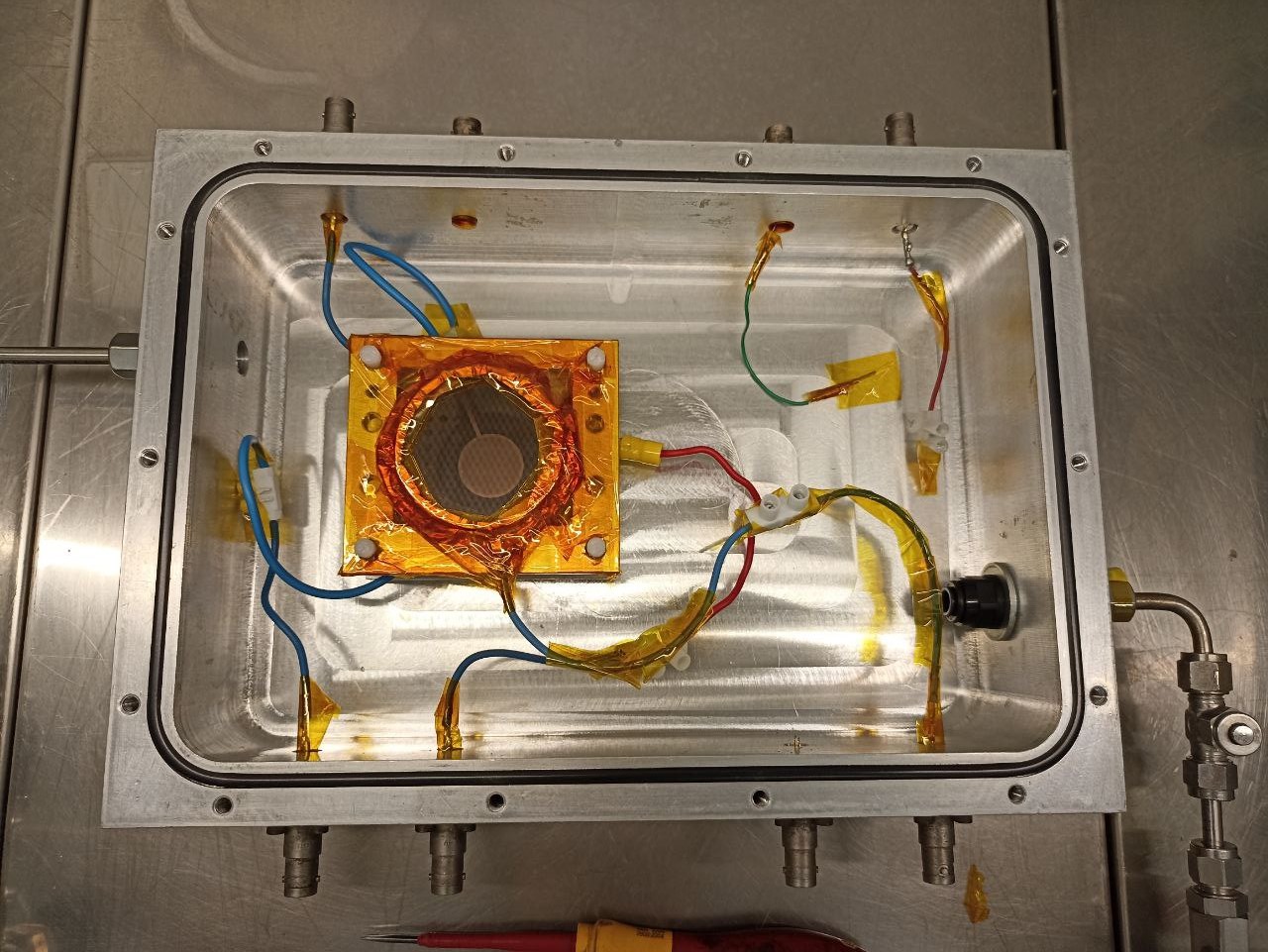}
   \end{minipage}
   \begin{minipage}{0.48\textwidth}
     \centering
     \includegraphics[width=0.99\linewidth]{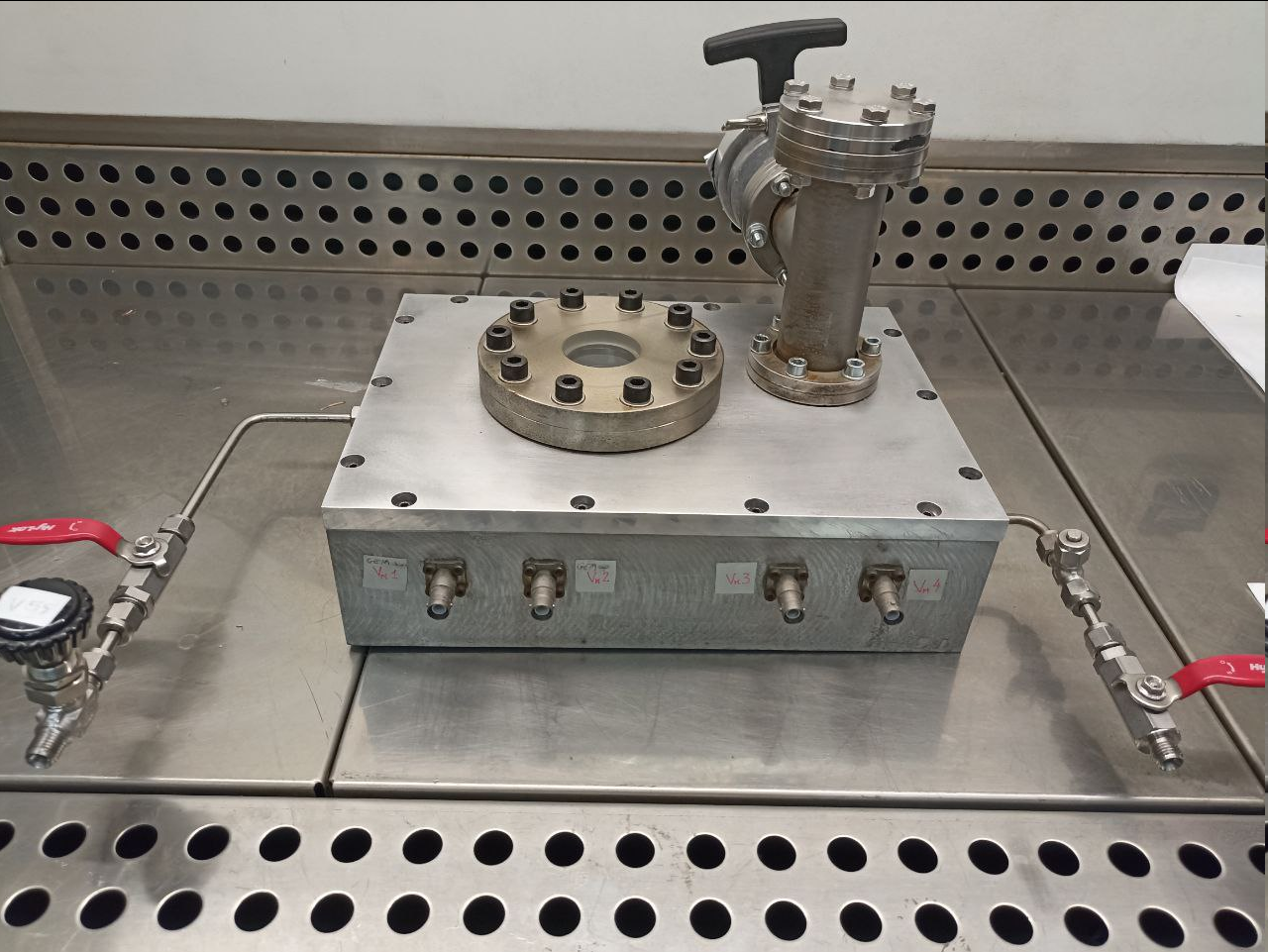}
   \end{minipage}\hfill
    \caption{\textit{Left}: Inner view of the chamber, where the  Micromegas can be seen through the photocathode. \textit{Right}: Outer view of the chamber, showing the quartz window and the port for vacuum pump. In both sides gas inlets, in the middle SHV feedthroughs.}
    \label{fig:ChamberFredy}
\end{figure}

This second technique was selected for these measurements. The setup, installed in our laboratory at University of Zaragoza, IAXOLab, consisted in a small aluminium chamber with a quartz window that allows to shine with a UV LED from outside and the TPC itself is built inside with the photocathode and the Micromegas. A schema of the experiment can be seen in figure \ref{fig:UVphotocathodeSchema}. For pictures of the experimental setup, figures \ref{fig:SmallMM} and \ref{fig:ChamberFredy} are interesting.

In our case, two LED diodes from QPhotonics were used: UVLED255TO46L and UVLED265TO46L. They both have 1mW of power, but their wavelength is slightly different, 255 and 265~nm with a spectral width of 11~nm, so they overlap. The first one has some collimation, a 6º angle cone thanks to a spherical lens. Both were tested with photocathodes and proved to be useful, but for the measurements presented here the 255~m one was used. From the specifications of both LED, the recommended current for pulsed operation is 100~mA, therefore the LED is coupled in series with a resistor of 50~$\Omega$. The typical voltage of operation is between 5 and 6 V,  although higher voltages have been applied without problems. Reverse bias could be fatal for the LED, so typically it is operated with a 1~V offset. A 3D printed holder fixed the LED on top of the window pointing towards the Micromegas detector. A signal generator was used to feed the LED, with duty cycles of around 1 \% and in no case surpassing the 10 \%.

\begin{figure}[b!]
    \centering
    \includegraphics[width=0.5\linewidth]{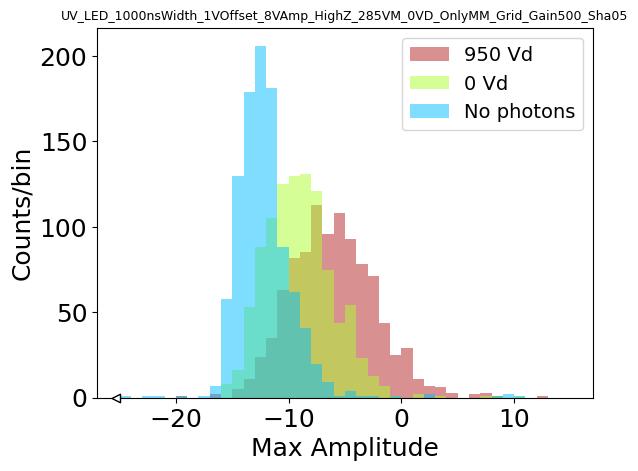}
    \caption{Spectra for LED pulses towards an aluminium grid cathode. In red with drift field on, in green with drift field off and in blue with LED not pointing to the window.}
    \label{fig:SpectrumsLed}
\end{figure}

In the first steps, an aluminium grid was used as a cathode, but soon a photocathode was installed. With the grid, the general schema was tested, the quartz window transparency was assured and electron extraction from the mesh was also observed. In figure \ref{fig:SpectrumsLed} the measurements that prove this: in red the amplitude of pulses recorded with the full detector operative, in green switching off the drift voltage and therefore only electrons generated in the mesh, and in blue random triggers due to noise when the LED was not pointing towards the window.

\begin{figure}[h!]

    \begin{minipage}{0.48\textwidth}
         \centering
         \includegraphics[width=0.8\linewidth]{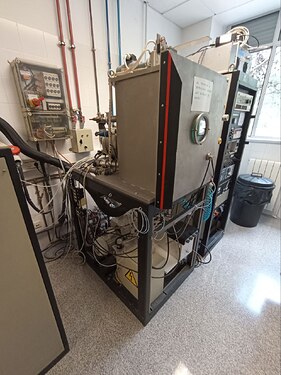}
   \end{minipage}
   \begin{minipage}{0.48\textwidth}
     \centering
     \includegraphics[width=0.8\linewidth]{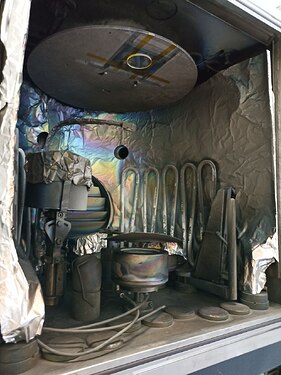}
   \end{minipage}\hfill
    \caption{\textit{Left}: General view of the thin layer deposition equipment. \textit{Right}: interior of the sputtering oven. The sample can be seen hanging from the ceiling. }
    \label{fig:SputteringOven}
\end{figure}

\begin{figure}[h!]
    \begin{minipage}{0.48\textwidth}
         \centering
         \includegraphics[width=0.99\linewidth]{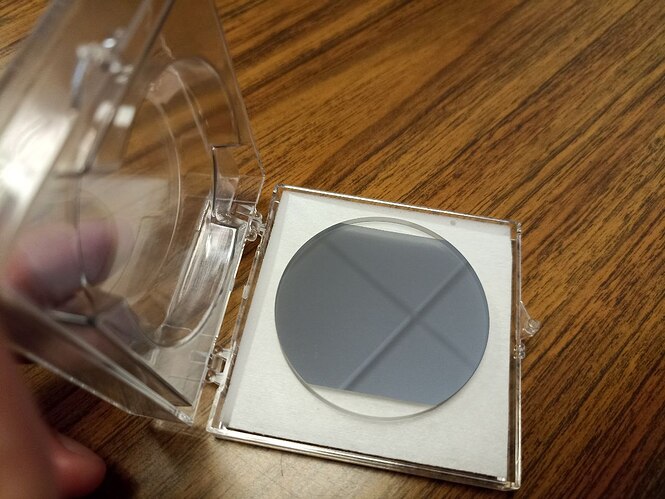}
   \end{minipage}
   \begin{minipage}{0.48\textwidth}
     \centering
     \includegraphics[width=0.99\linewidth]{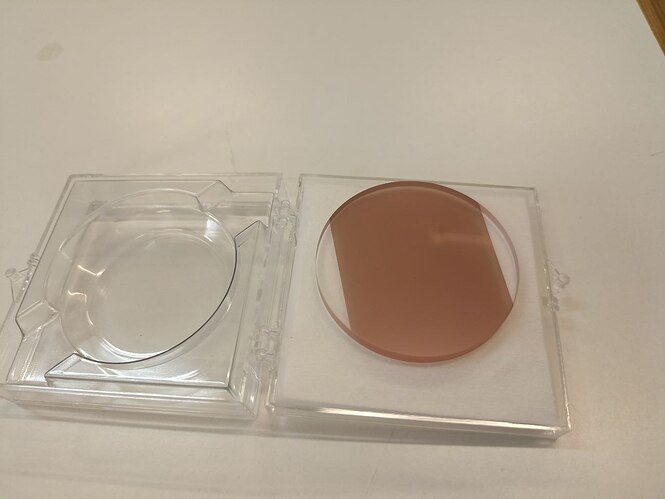}
   \end{minipage}\hfill
    \caption{Photocathodes, made of thin metallic layers over a quartz substrate. \textit{Left}: 30~nm thickness aluminium layer. \textit{Right}: 20~nm thickness copper layer. }
    \label{fig:Photocathode}
\end{figure}

\begin{figure}[h!]
    \centering
    \includegraphics[width=0.5\linewidth]{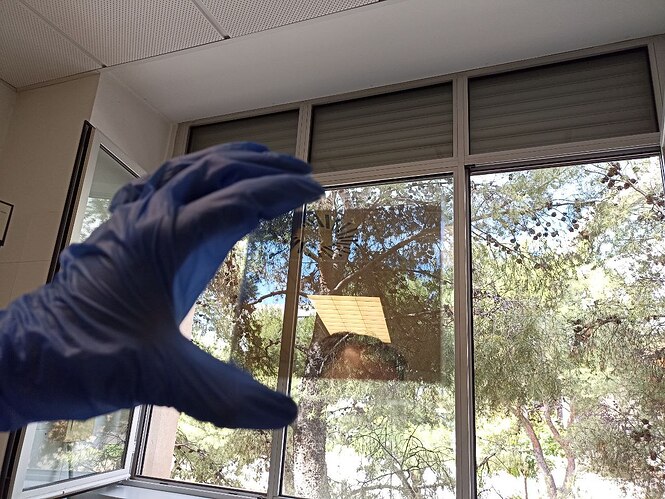}
    \caption{Glass sample used as a test for the deposition of the thin aluminium layer. Here transmission and reflection images for visible photons can be seen in the sample.}
    \label{fig:AlLayerSample}
\end{figure}

After testing the setup with the grid, the cathode was substituted by a phototocathode. Two were produced with the collaboration of professor Enrique Carretero, from the Applied Physics department at the University of Zaragoza, on top of quartz (also called fused silica) discs of 48.5 mm diameter and 4 mm thick. A sputtering oven was used for this, a general view and the interior of which can be seen in figure \ref{fig:SputteringOven} . The samples can be identified in the ceiling of the oven, held on top of the target. In a low-pressure argon atmosphere, high voltage is applied between the anode and the target, acting as cathode. Ionized argon atoms are accelerated towards the target reaching high energies, smashing the target and knocking atoms off of the surface that attach everywhere in the chamber, in particular in the substrate hanging from the ceiling. In two evaporation runs, thin layers of copper, 20~nm thickness, and aluminium, 30~nm thickness, were deposited on top of the discs (figure \ref{fig:Photocathode}). Prior to the deposition, the recipe was calibrated on glass samples like the one seen in Figure \ref{fig:AlLayerSample}. 

The Micromegas detector with the photocathode is placed inside a 2 litres aluminium chamber that can be seen in \ref{fig:ChamberFredy}. It has two inlets for gas flux, SHV ports for voltage supply and signal readout and the crucial quartz window installed in the end-cap. All measurements are performed at 1 bar with argon + 1\% isobutane. A CAEN HV power supply (NIM N1471 or stand alone DT5521E)  powers the photocathode and the mesh. The signal is extracted through one of the SHV feedthroughs, preamplified with a CANBERRA 2004, amplified further and then digitalised, whether with the oscilloscope, when pulses need to be recorded, or with an Multi-Channel Analyser (MCA), if histograms are needed. Gas injection was preceded by some hours of pumping the air of the chamber. Values of the order of  $10^{-4}$~mbar are reached before injecting argon + 1\% isobutane. Then, it is flushed in open loop at 10 l/hour during half an hour in order to assure the quality of the gas , and finally the chamber is sealed. Measurements are fast, usually less than 4 hours long, so no degradation of the gas is observed in this time scale.

Finally, the last item of the detector is the Micromegas itself. A small diameter single-channel Micromegas with avalanche gap of 50~$\upmu$m, distance between mesh hole centres (pitch) 110~$\upmu$m and diameter of mesh holes 60~$\upmu$m was used for all measurements. In figures \ref{fig:SmallMM} and \ref{fig:ChamberFredy} can be seen all the items of the detector: Micromegas, photocathode, calibration source, connections inside the chamber and the chamber itself.

\subsubsection{Results}

The first relevant result is the experimental confirmation of the working principle. To our knowledge, it is the first time a UV LED is used to extract electrons from a photocathode as a way to calibrate a gaseous detector. This is a significant improvement in some aspects compared with lasers and noble gas lamps, apart from the price. A LED diode is small and requires a minimal set-up to operate, so it offers the possibility to install it in narrow spaces, including the interior of the gas chambers, allowing routine calibrations that avoid the use of radioactive sources that cannot be switched off.

The work functions of the two metals used as photocathode, can be found in the literature: between 4.06 and 4.26~eV for the aluminium and 4.48 to 5.1~eV for copper depending on the crystallographic plane \cite {lide1995crc}, making it easier to extract electrons from aluminium than copper. However, as copper is everywhere in radiopure experiments (TREX-DM chamber, Micromegas readout planes, field cage wires...),  it is interesting to check the feasibility of using copper as target material. 

\begin{figure}[h!]
    \centering
    \includegraphics[width=0.5\linewidth]{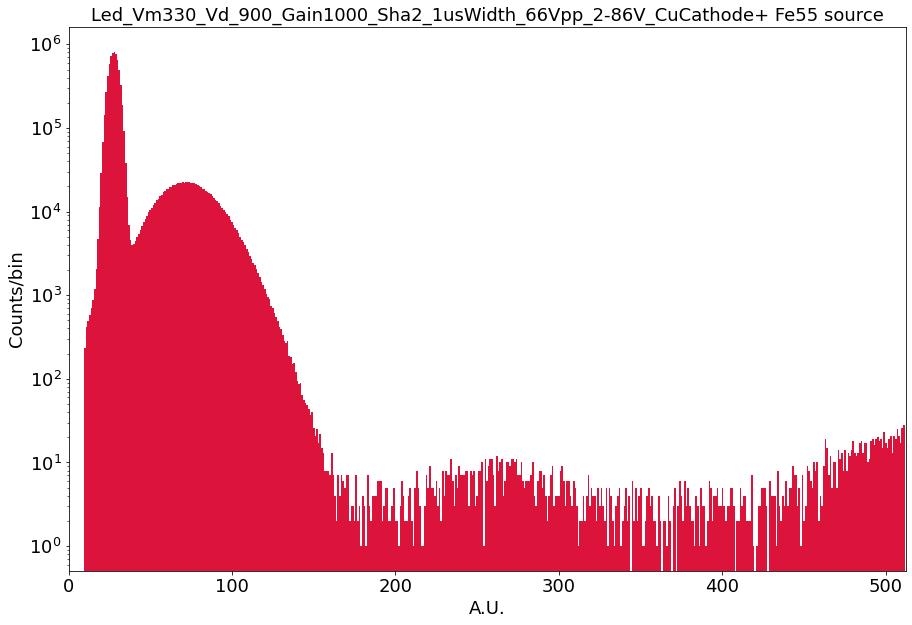}
    \caption{Energy spectrum with the copper photocathode using the 255~nm UV LED with 1~$\upmu$s width pulse and 9.6 V amplitude with 3 V offset, when a $^{55}$Fe source is also present. The UV peak is situated below the Ar escape 3 keV peak seen in the middle, while the 5.9 keV peak is only partially visible at the right of the spectrum. The sharp peak at the beginning are noise events.}
    \label{fig:CuLedSpectrum}
\end{figure}

\begin{figure}
    \centering
    \includegraphics[width=1\linewidth]{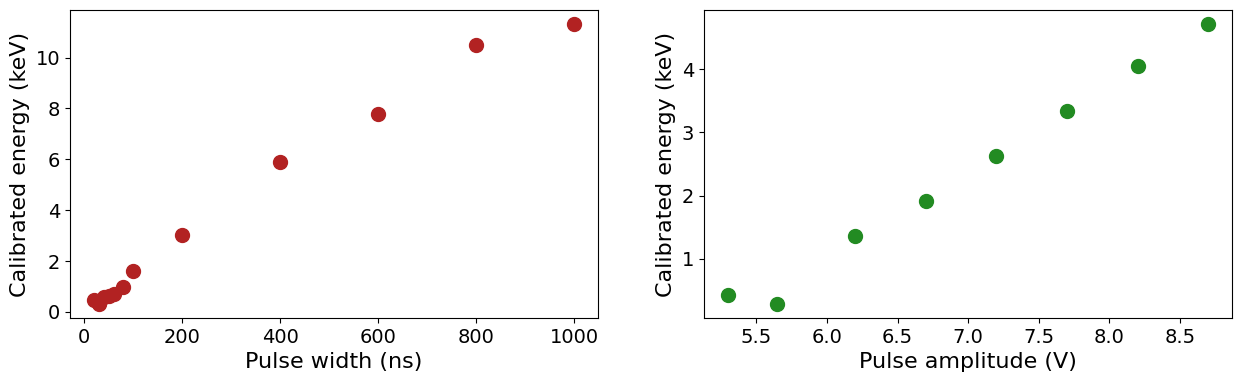}
    \caption{\textit{Left}: Energy vs pulse width, when the pulse amplitude fixed at 9.6 V counting 3 V offset. \textit{Right}: Energy vs pulse amplitude, with pulse width fixed at 200 ns, including 2 V offset. In both cases, threshold energy is 0.3 keV.}
    \label{fig:AlVaryingPulses}
\end{figure}

The copper photocathode showed less efficiency generating electrons for our purpose. In figure \ref{fig:CuLedSpectrum} a spectrum generated with it can be seen. The UV peak is the most visible feature of the spectrum, placed slightly below bin 100. A thinner noise peak is on top of it around bin 20. A $^{55}$Fe source was used together with the UV LED, and the characteristic 5.9 keV peak starts to be visible at the end of the acquisition range, above bin 500; however, the Ar escape peak at 3 keV can be seen in the centre, between bins 200 and 300. The  UV events were generated with 1 $\upmu$s width pulses and 9.6 V amplitude with 3 V offset. The equivalent energy of these events is well below 3~keV. In comparison, the fourth plot in \ref{fig:AlVaryingPulses} was taken with the aluminium photocathode, 800 ns width pulse and similar amplitudes. There, the UV peak is well above the 5.9 keV peak from the calibration source. This implies that more electrons are generated with the aluminium photocathode.

The main reason for this difference lies in the work function of each material. Lower work function for aluminium means higher efficiency extracting electrons with UV photons. But also the penetration depth matters. To estimate the fraction of 266~nm ultraviolet (UV) light transmitted through a material, we apply Beer–Lambert’s law, which models intensity attenuation in absorbing media as $T=e^{-\alpha d}$, where $T$ is the transmission, $d$ is the film thickness, and $\alpha$ is the absorption coefficient. For metals, $\alpha$ is given by $\alpha=\frac{4\pi k}{\lambda}$ , with $k$ the extinction coefficient and $\lambda$ the wavelength of incident light. Using tabulated optical constants for copper at 266~nm, where $k\approx 2.8$ \cite{polyanskiy2024}, we compute $\alpha\approx0.132$~nm$^{-1}$, yielding a transmission of $T\approx e^{-0.132 \times 20} \approx 7.1\%$. 
For aluminium with $k\approx 6.3$ and then $\alpha\approx 0.297$~nm$^{-1}$, it yields a transmission of $T\approx e^{-0.297 \times 30} \approx 0.62\%$. 
Therefore more photons interact in the aluminium than in the copper layer. This simplified model neglects surface reflections, which would further reduce the transmitted flux. 
And then, from these interactions only a fraction will generate electrons that can reach the sensitive volume of the TPC. Here is where the highest efficiency for aluminium due to the lower work function enters into play, reinforcing the suitability of this material.

\begin{figure}[h!]
    \centering
    \includegraphics[width=0.9\linewidth]{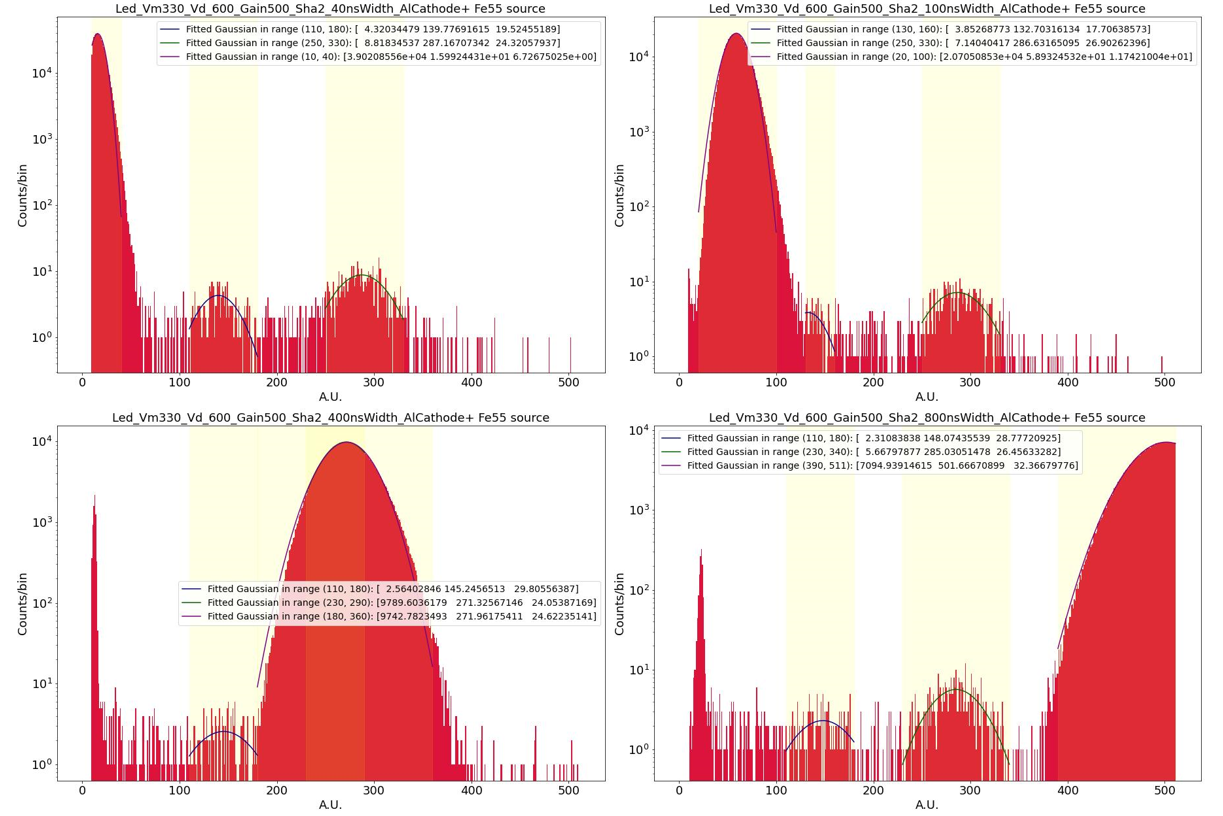}
    \caption{Some spectra where the $^{55}$Fe 3 and 5.9 keV peaks can be seen. UV peak, the biggest, moves towards increasingly higher energies when the pulse duration increases.}
    \label{fig:AlLedSpectrums}
\end{figure}

The main objective of this set-up was to develop a variable energy calibration source. The aluminium photocathode was selected and different pulses were fed to the 255~nm UV LED to measure the response of the detector to the varying UV flashes. The two parameters that were modified were amplitude and width of the voltage pulses feed to the the diode. In figure \ref{fig:AlLedSpectrums} four spectra for fixed amplitude and varying durations can be seen: 40, 100, 400, 800 ns width. The longer the pulse, the higher the energy extracted. In all cases the $^{55}$Fe spectrum can be seen to compare, with the peaks at 3 and 5.9 keV.

Fitting the UV LED peaks to gaussian functions and calibrating their equivalent energy thanks to the $^{55}$Fe calibration source their energy is related with the pulse width applied. This is shown in the left plot of figure \ref{fig:AlVaryingPulses}. If the width is fixed and amplitude is what varies, a similar relation is obtained, figure \ref{fig:AlVaryingPulses} right.

\begin{comment}
\begin{figure}[h!]
    \begin{minipage}{0.48\textwidth}
         \centering
         \includegraphics[width=0.99\linewidth]{images/AlVaryingPulseWidth.png}
   \end{minipage}
   \begin{minipage}{0.48\textwidth}
     \centering
     \includegraphics[width=0.99\linewidth]{images/AlVaryingPulseAmplitude.png}
   \end{minipage}\hfill
    \caption{\textit{Left}: Energy vs pulse width, when the pulse amplitude fixed at 9.6 V counting 3 V offset. \textit{Right}: Energy vs pulse amplitude, with pulse width fixed at 200 ns.}
    \label{fig:AlVaryingPulses}
\end{figure}
\end{comment}

\begin{figure}[h!]
    \centering
    \includegraphics[width=0.7\linewidth]{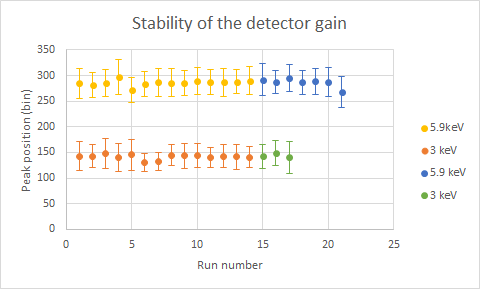}
    \caption{UV LED peak position (representing gain) along time: in orange and yellow, series from runs with varying pulse width; in green and blue, runs with varying pulse amplitude. The error bars are the standard deviation of the gaussian fit, errors in fitted positions are negligible.}
    \label{fig:AlStabilityDetector}
\end{figure}

During these runs, acquired in a single day during several hours with the chamber in sealed mode, the stability of the detector was monitored. In figure \ref{fig:AlStabilityDetector} the fitted position of the 3 and 5.9 keV peaks is represented. Most of them come from the same UV LED runs; however,  in the cases where the LED peak hides them, calibration runs without the UV LED were recorded to check the position of both peaks.
With this experiment it was proved that pulses with energies between 0.5 and 12 keV can be generated thanks to a UV LED diode of 1 mW. As a comparison, in \cite{pellecchia2020uv} a laser test bench was described, that makes use of a laser of 51 $\upmu$J with pulses of 1 ns, so 5 kW of maximum power, six orders of magnitude higher. 

\begin{comment}
\begin{figure}[h]
   \begin{minipage}{0.48\textwidth}
     \centering
     \includegraphics[width=0.99\linewidth]{images/MMGrid.jpeg}
     \caption{General view of the setup. }\label{fig:MMGrid}
   \end{minipage}\hfill
   \begin{minipage}{0.48\textwidth}
     \centering
     \includegraphics[width=0.99\linewidth]{images/MMWindow.jpeg}
     \caption{Micromegas and cathode seen through the quartz window.}\label{fig:MMWindow}
   \end{minipage}
\end{figure}
\end{comment}

\section{Argon + 10\% Isobutane}

\subsubsection{Objective}

As already discussed, TREX-DM aims for low-mass WIMP detection. The energy transfer in elastic collisions is more efficient with light nuclei, and that is the reason why neon was selected over more traditional gases for TPCs like argon or xenon. In this quest towards lighter masses, organic gases like isobutane, always present in our mixtures as a quencher to improve the energy collection and prevent sparks, could play a role, because both carbon and hydrogen are very light elements. The scenarios considered in figure \ref{TREXsensitivityScenarios} with an increased content of isobutane show higher sensitivity at lower masses. 

Isobutane is a flammable gas so only small amounts are allowed in the gas mixture for traditional installations. For higher percentages, like 10\%, safety measures have to be considered: pipes prepared for flammable gases, ventilation and big air volumes to dilute the isobutane in case of leaks, etc. The LSC has approved the use of argon + 10~\% isobutane in TREX-DM with the corresponding upgrade of the gas system and a safety protocol that is being defined right now. The information in the literature about this mixture is scarce and old, and has not been tested with these microbulk Micromegas before. 
\begin{comment}
But this is a new mixture that has not been tested before. Even in the literature there are few examples of the employ of this gas.    
\end{comment}
 Therefore, a campaign was designed to test this mixture for varying pressures between 1 and 10~bar. In \cite{iguaz2022microbulk} argon + 1~\% isobutane is tested at different pressures in the range we are interested in, and in \cite{iguaz2012characterization} several mixtures of argon and isobutane are examined, including 10~\% isobutane, at 1 bar. A small setup was prepared in IAXOLab laboratory  to test this mixture. The objective was to establish maximum voltage values to achieve a stable operation point.

\subsubsection{Methods and materials}

\begin{figure}[h!]
    \centering
    \includegraphics[width=0.8\linewidth]{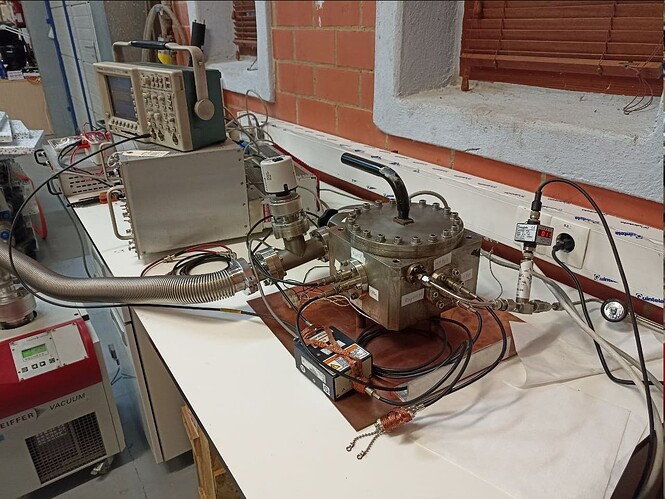}
    \caption{General view of the setup for argon + 10\% isobutane measurements. The chamber with preamplifier and pressure sensors in foreground, oscilloscope and power supplies in the back.}
    \label{fig:GeneralViewSetup}
\end{figure}

A very similar configuration to the one in the previous subsection was used, with the same type of small Micromegas and a grid cathode (figure \ref{fig:GeneralViewSetup}). The main difference was the chamber, we moved to a stainless steel one that withstands up to 10 bar. The same readout chain of preamplifier, amplifier and MCA or oscilloscope for acquisition was used. Analogous power supplies and SHV connections, and the same $^{55}$Fe calibration sources were used. Gas injection was handled following the same protocol as explained in the previous section: pumping during several hours before injecting gas and closing the chamber. Due to the flammable condition of this gas a fan was installed for extra venting around the chamber to prevent isobutane accumulations in case of unexpected leaks.

\subsubsection{Results}

\begin{figure}[h]
    \centering
    \includegraphics[width=0.8\linewidth]{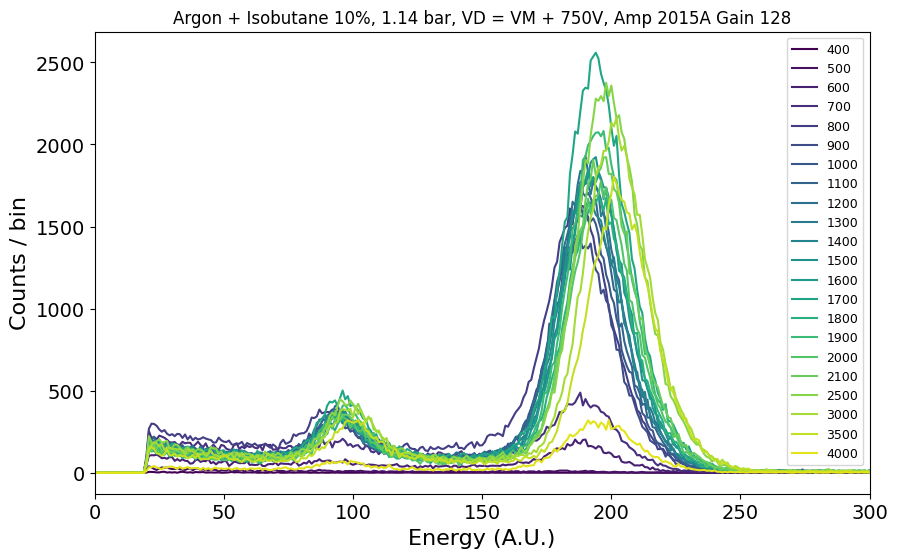}
    \caption{Spectra of $^{55}$Fe in argon + 10\% isobutane at 1 bar used for transparency curve. $V_{mesh}=330$ V and drift distance $25\pm3$ mm. $V_{drift}$ from 400 to 4000 V. The change in the main peak position indicates a loss of primary charge and therefore a lower electron transmission.}
    \label{fig:TransparencySpectrums}
\end{figure}

The first test consisted in establishing the available operation range. This translates in characterizing the detector in terms of gain and electron transmission. Too high mesh voltages can damage the Micromegas through spontaneous sparks and too low values do not provide enough amplification to record events above the noise threshold. In addition, a balance between the drift field and the amplification field has to be reached, to ensure maximum recollection of primary electrons generated in the drift volume. For these studies, the mesh voltage is kept fixed and the cathode voltage (and therefore the drift field) is varied. A $^{55}$Fe calibration source was used and the spectra can be seen in figure \ref{fig:TransparencySpectrums}: the mesh voltage was 330 V, and the drift voltage varied from 400 to 4000~V. The drift distance was 2.5~cm with an estimated uncertainty of 3 mm, the pressure was slightly above 1~bar, at 1.1~bar, and the drift field values tested go from 100 to 1300~V/cm/bar.

\begin{figure}[h]
    \centering
    \includegraphics[width=0.6\linewidth]{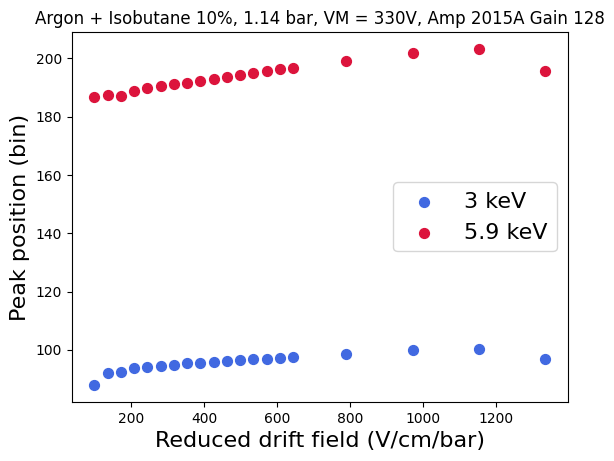}
    \caption{Electron transmission curve for argon + 10\% isobutane at 1.1 bar and $V_{mesh} = 330$ V. The two peaks of $^{55}$Fe, the escape peak of argon, 3 keV, and the characteristic 5.9 keV K$\alpha$ X-ray emission from iron were used. }
    \label{fig:TransparencyCurve}
\end{figure}

The electron transmission curve was extracted fitting the main two peaks of the spectra, at 3~keV and 5.9~keV, with gaussian functions to determine their position in bin units: the results can be seen in figure \ref{fig:TransparencyCurve}; both curves follow the same pattern, as expected, but there is not a clear plateau, they present a small slope until the beginning of the decrease.

\begin{figure}[h!]
    \centering
    \includegraphics[width=0.6\linewidth]{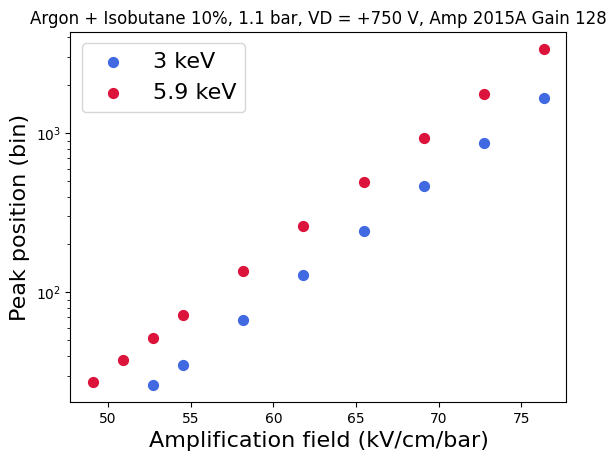}
    \caption{Relative gain curve for argon + 10\% isobutane at 1.1 bar. The drift field was set in all runs to 270 V/cm/bar.}
    \label{fig:GainCurve}
\end{figure}

For the gain curve, the peak position is registered for a fixed reduced drift field, in this case at 270 V/cm/bar, when the mesh voltage is varied. For the 50 $\upmu$m gap Micromegas we are using, values from 270 to 420 V were tested, corresponding to amplification fields from 50 to 75 kV/cm/bar. The measured gain curve can be seen in figure \ref{fig:GainCurve} and it is in good accordance with the one presented in \cite{iguaz2012characterization}.

Limitations appear in both ends of the transparency and gain curves. The lowest field values are determined by the lowest signals that can be distinguished from the electronic noise.  Therefore, lowest values for $V_{mesh}$ are the first for which calibration events can be detected. Measured transparency curves have more room downwards, probably events can be detected up to 20-50 V/cm/bar drift fields with this configuration. But in these cases, the resolution is so low that the peak broadening makes it impossible to determine its position. The useful drift field range is limited by attachment and recombination of electrons in the gas, if the field is too low the efficiency in the collection of charges decreases and no signals can be detected despite the amplification in the avalanche gap. For electric fields, maximum values limitations have to be handled more carefully. The voltage in the cathode, and therefore in the drift field, is limited only by spontaneous sparks. The cathode is a metallic grid, therefore small discharges are not problematic if not continuous. Basically, connections between the feedthroughs, cables in contact with chamber walls and the connection with the cathode grid are the sensitive areas where sparks may happen. Improved electrical insulation may help to reach higher voltages. High energy alpha particles in sensitive regions may also produce discharges. The Micromegas is a completely different case. Sparks may damage the mesh or the anode and continuous discharges can destroy them. In most of the cases during the tests with argon + 10\% isobutane, sparks at high voltages lead to the substitution of the Micromegas specimen.  Therefore, a conservative approach was followed and the measurements shown here, especially in the gain curves, are well below the sparking limit in most of the cases. Table \ref{tab:MaxAmpField} shows maximum values applied for different mixtures and  pressures. 

\begin{table}[h]
    \centering
    \caption*{\textbf{Maximum amplification fields}}
    \begin{tabular}{|c|c|c|c|c|c|c|c|c|c|}
        \hline
        Pressure & 1 & 2 & 3 & 5 & 6 & 8 & 10 & bar \\
        \hline
        Ar+1\%Iso & 60 & 68 & 75 & 88 & 94 & 104 & 114 & kV/cm \\
        \hline
        Ar+10\%Iso & 84 & 92 & 108 & 136 & \textcolor{gray}{150} & \textcolor{gray}{178} & \textcolor{gray}{207} & kV/cm \\
        \hline
        Spark (Ar+10\%Iso): & 92 & 94 & - & 140 & - & - & - & kV/cm \\
        \hline
    \end{tabular}
    \bigskip
    \caption{Maximum amplification fields applied to a Microbulk Micromegas with a 50 $\upmu$m amplification gap for two argon mixtures. Values for argon + 1\% isobutane extracted from \cite{iguaz2022microbulk}. Values for argon + 10\% isobutane measured in this work. During measurements for pressures 1, 2, 3, 5 bar several Micromegas were damaged due to sparks, at the values presented in the forth row. Maximum amplification values for 6, 8 and 10 bar, in gray,  are extrapolations from the linear fit obtained from values for lower pressures. This prevented damages in other specimens.}
    \label{tab:MaxAmpField}
\end{table}

Up to now, the mixture argon + 10\% isobutane has been tested in the context of optimizing the percentage of isobutane for TPC with Micromegas, or even in wider searches for gas mixture selection. These tests were performed at 1 bar with the aim at assessing the characteristics of different mixtures in the same set-up. Here, we are interested in the behaviour of the selected mixture at different pressures that have never been explored.

The reason to increase the percentage of isobutane was motivated by the increase in sensitivity for low mass WIMPs compared with argon + 1\% isobutane. But due to its role as quencher, it is also expected to affect the gain: an increase in isobutane allows higher voltages of operation due the effect of the quencher, it traps UV photons more efficiently preventing secondary ionizations and wide-spread avalanches in the amplification gap. On the other hand, the increase of pressure implies a reduction in the reduced electric field (V/cm/bar) and therefore in the gain, that can be recovered applying higher voltages. Therefore, due to the combination of these two effects, it was not clear beforehand how the gain will evolve with pressure for this mixture.

\begin{figure}[h]
    \centering
    \includegraphics[width=0.6\linewidth]{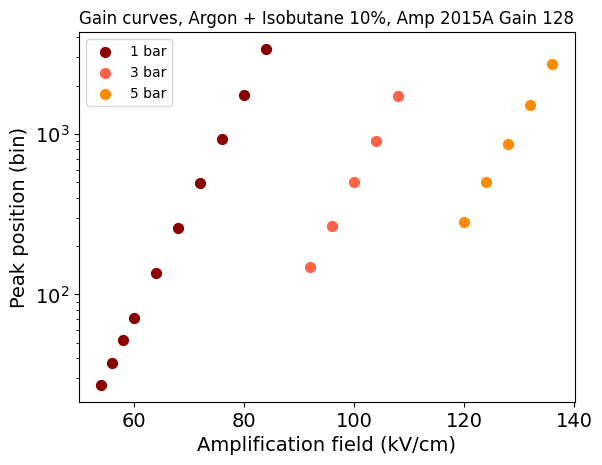}
    \caption{Gain curves for argon + 10\% isobutane at different pressures, using the 5.9 keV peak.}
    \label{fig:GainCurvesPressures}
\end{figure}

As mentioned before, unexplored territory in terms of pressures and voltages had the drawback of multiple unexpected sparks. At first, the maximum amplification voltage limit was checked without any reference and many Micromegas were lost. From these measurements, figure \ref{fig:GainCurvesPressures} was extracted. For example, the last point at 5 bar, slightly below 140 kV/cm, was extracted from the last spectrum obtained before a spark at 140 kV/cm shortcircuited that Micromegas. To prevent these destructive tests we keep voltages between 4 and 8 kV/cm below the estimated maximum voltages shown in table \ref{tab:MaxAmpField}. Measured maximum fields are presented in table \ref{tab:MeasuredMaxAmpField}.

The systematic measurement of the gain curves for different pressures involved several technical refinements learned in previous data takings. First, use a single Micromegas specimen. Different Micromegas show small performance differences even though they have the same hole pattern. This implies avoiding sparks during all the measurement. And second, the gas quality. The chamber was pumped for several hours before injecting gas to eliminate emanated contaminants, humidity and oxygen. The measurements were performed in sealed mode as fast as possible. To speed up the measurements, the most active $^{55}$Fe source available was used. Therefore, in the following measurements all peaks used correspond to its 5.9 keV emission.

\begin{figure}[h!]
    \centering
    \includegraphics[width=0.8\linewidth]{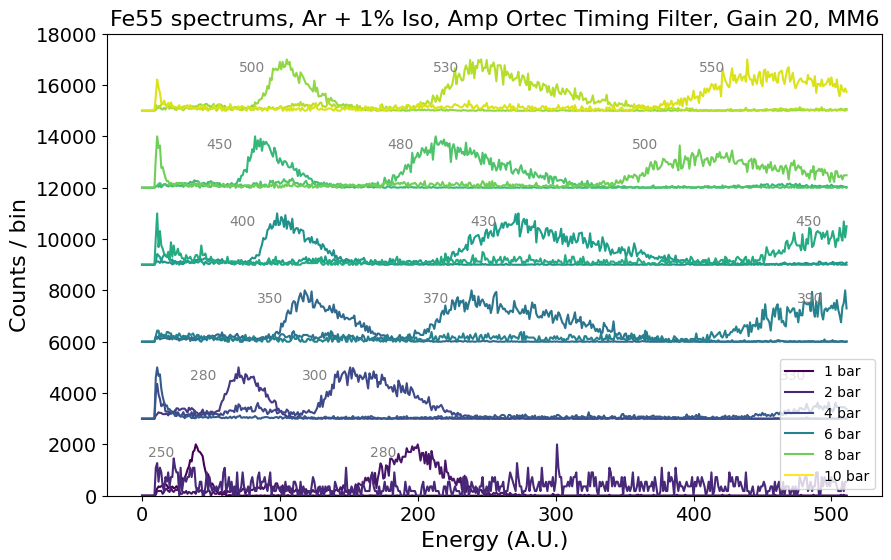}
    \caption{Spectra for different mesh voltages and pressures for argon + 1\% isobutane. The peaks correspond to the 5.9~keV from a $^{55}$Fe radioactive source. For every pressure, several spectra taken at different $V_{mesh}$, values in V written next to each peak, and plotted with a vertical offset for better visualization. From the peak positions, gain curves of \ref{fig:Ar1GainCurves} were obtained.}
    \label{fig:Ar1Spectrums}
\end{figure}

\begin{figure}[h!]
    \centering
    \includegraphics[width=0.8\linewidth]{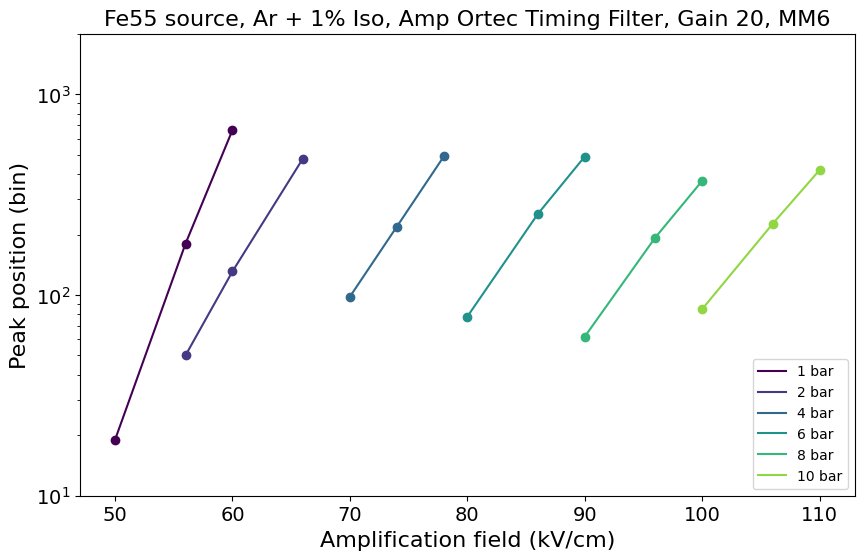}
    \caption{Gain curves for different pressures for argon + 1\% isobutane. 5.9 keV peak position extracted from spectra in \ref{fig:Ar1Spectrums}. Curve at 1 bar presents a unexpected slope compared with the rest of the series, raising suspicions about the quality of the gas during these measurements.}
    \label{fig:Ar1GainCurves}
\end{figure}

\begin{figure}[h!]
    \centering
    \includegraphics[width=0.8\linewidth]{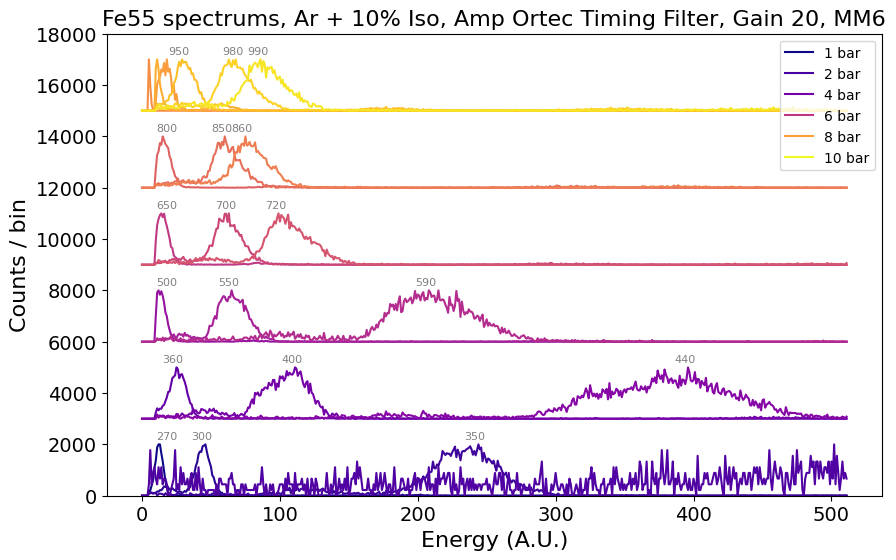}
    \caption{Spectra for different mesh voltages and pressures for argon + 10\% isobutane. Peaks are 5.9 keV from a $^{55}$Fe radioactive source. For every pressure, several spectra taken at different $V_{mesh}$, values in volts written next to each peak, and plotted with a vertical offset for better visualization.}
    \label{fig:Ar10Spectrums}
\end{figure}

In the same day, two data takings were recorded: first a series of calibrations for argon + 10\% isobutane at different pressures, figure \ref{fig:Ar10Spectrums}, and then another series with argon + 1\% isobutane, figure \ref{fig:Ar1Spectrums}. All runs were acquired in the span of less than five hours, with between 5000 and 10000 events per calibration.

\begin{figure}[h!]
    \centering
    \includegraphics[width=0.8\linewidth]{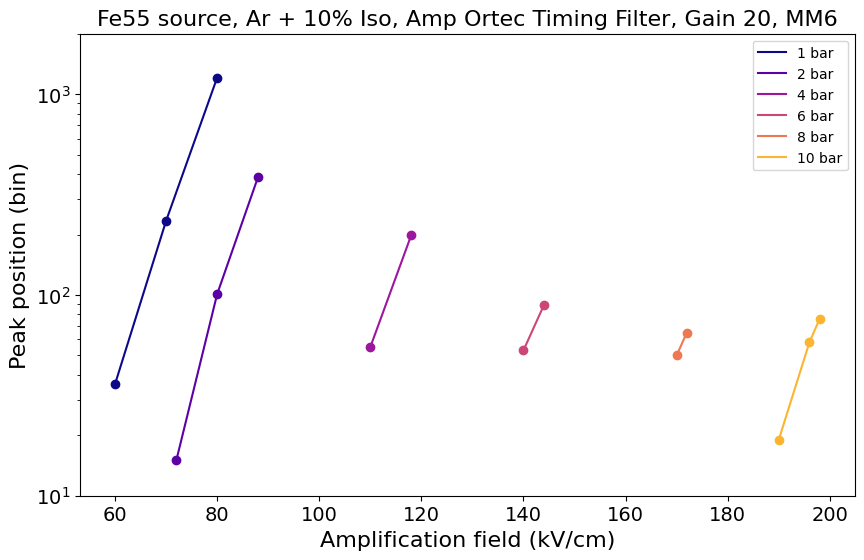}
    \caption{Gain curves for different pressures for argon + 10\% isobutane. 5.9 keV peak position extracted from spectra in \ref{fig:Ar10Spectrums}.}
    \label{fig:Ar10GainCurves}
\end{figure}

Figures \ref{fig:Ar1Spectrums} and \ref{fig:Ar10Spectrums} present the calibration spectra employed for the gain curves in \ref{fig:Ar1GainCurves}
and \ref{fig:Ar10GainCurves}. An offset on the Y axis has been applied with the increasing pressure of each run to facilitate the visualization. On each vertical level several spectra with different gains are plotted, with slight variations in colour. Each one has the peak coming from the 5.9~keV emission from the $^{55}$Fe source clearly visible and its position grows with the increase of $V_{mesh}$. In some of them, an extra noise peak can be seen next to the threshold. In some cases, pulses are so big that the peaks go beyond the acquisition range in ADC: in these cases, the amplifier gain was decreased and the peak position rescaled afterwards extracting the amplification factors from repeated measurements with different amplifier settings for lower $V_{mesh}$ values. This happened in the highest point for 1~bar in both figures \ref{fig:Ar1GainCurves} and \ref{fig:Ar10GainCurves}. This calibration procedure could introduce some distortion in the curves, possibly explaining the different slope of the curve at 1~bar in figure \ref{fig:Ar1GainCurves}. Also, the first point of this curve was the first of the series, rising possible suspicions about the gas quality.

\begin{table}[h]
    \centering
    \caption*{\textbf{Measured maximum amplification fields}}
    \begin{tabular}{|c|c|c|c|c|c|c|c|c|c|}
        \hline
        Pressure & 1 & 2 & 4 & 6 & 8 & 10 & bar \\
        \hline
        Ar+1\%Iso & 60 & 66 & 78 & 90 & 100 & 110 & kV/cm \\
        \hline
        Ar+10\%Iso & 80 & 88 & 118 & 144 & 172 & 198 & kV/cm \\
        \hline
    \end{tabular}
    \bigskip
    \caption{Maximum amplification fields applied to a Microbulk Micromegas with a 50 $\upmu$m amplification gap for two argon mixtures measured in the runs shown in figures \ref{fig:Ar1GainCurves} and \ref{fig:Ar10GainCurves}.}
    \label{tab:MeasuredMaxAmpField}
\end{table}

From these gain curves with conservative maximum amplification fields, shown in table \ref{tab:MeasuredMaxAmpField}, two conclusions are obtained: at lower pressures argon + 10\% isobutane reaches higher gain, almost double for 1 bar; but at higher pressures, argon + 1\% isobutane reaches further, being the gain  several times higher than for 10 bar. Probably the turning point is around 3 bar. This effect of decaying gain with pressure has been observed previously in other mixtures like argon + 2\% isobutane \cite{iguaz2016trex}, argon + 1\% isobutane and neon + 2\% isobutane in \cite{iguaz2022microbulk}, and xenon-trimethylamine \cite{cebrian2013micromegas}. This last reference is interesting because they examine different percentages of additive (trimethylamine) and they observe that the optimum amount changes with pressure. This could happen as well in our case , the optimum amount of isobutane for maximum gain at 10 bar could be smaller.

Our hypothesis is that at higher pressures, the isobutane concentration enhances the energy transfer in the collisions, and with electrons scattering constantly in isobutane molecules they lose too much energy to develop bigger avalanches, leading to less amplification.
Also, at high pressure voltages are so high that sparks tend to be fatal for the Micromegas. During these measurements the highest voltage ever applied to a Micromegas in our experiments was reached: $V_{mesh}=990$~V at 10~bar with argon + 10~\% isobutane. This specimen (MM6) is still operative and working fine many months later.

\begin{comment}

\begin{figure}[h]
    \centering
    \includegraphics[width=0.8\linewidth]{images/Ar1GainCurves2.png}
    \caption{Gain curves for different pressures for argon + 1\% isobutane.}
    \label{fig:Ar1GainCurves2}
\end{figure}

\end{comment}

\begin{comment}
\section{GEM tests}
\subsubsection{Objective}
\subsubsection{Methods and materials}
\subsubsection{Results}
\end{comment}

%% file: Chapters/5_DarkPhotons_Axions.tex
The lack of experimental evidence for dark matter, particularly within the WIMP paradigm, has prompted the exploration of a wide range of alternative approaches. Many models have been proposed to explain the challenge of detecting a new particle, or set of particles, that could account for the astrophysical behaviour of galaxies, galaxy clusters, and cosmological observations.

The first half of this work has been devoted to the search for light WIMPs using gaseous detectors, with particular emphasis on the TREX-DM experiment. From this point forward, the focus shifts to one of these alternative paths: the search for axions and dark photons using resonant cavities. Experimental technologies are very different but both have the same aim: to explain the hidden nature of dark matter particles.

If they only interact gravitationally with the ones from the Standard Model, laboratory searches will have little hope to see them. This possibility is seen as a world almost parallel to our own. In this hidden sector, particles may interact among them as in our Standard Model, affecting our visible Universe only at large scales due to the faint touch of gravity. 

A model like this might not be very appealing for experimental searches, but of course, ideas are not scarce: the hidden sector may have a portal to our visible Universe. The term ``portal" makes reference to the possibility of having at least one of these particles capable of interacting with Standard Model particles. The form of this interaction can be classified by the type and dimensions of the associated operator. The best motivated and most studied cases contain different types of operators depending on the spin of the mediator: \textit{vector} (spin 1), \textit{neutrino} (spin 1/2), \textit{Higgs} (scalar) and \textit{axion} (pseudo-scalar). 

In this chapter two of these cases will be briefly explained due to the common techniques employed in their experimental searches: dark photons (spin 1) and axions (pseudo-scalar). Both will eventually be shown to be coupled to electromagnetic currents and that this kind of interactions can be probed with resonant cavities.

\begin{comment}
   The vector portal is the one where the interaction takes place because of the kinetic mixing between one dark and one visible Abelian gauge boson (nonAbelian gauge bosons do not mix). This is what is called ``dark photon". This kinetic mixing takes the form of a dimension 4 operator made by the product of both field strengths. The existence of such an operator means that the two gauge bosons can go into each other as they propagate, which enables the detection of dark photons in experiments. 

   Most dark matter searches rely in the existence of a direct interaction between dark matter particles and Standard Model ones. The unfruitful efforts of past decades have motivated theoreticians to explore other models more disconnected to our own world. One of these newer paradigms is the dark sector. A completely new set of particles could coexist in our own Universe without interacting with any known particle except for the gravitational pull. This would mean a parallel Universe to our own that could explain the dark matter effects seen in the Universe. 

    This reasoning has open many lines of research. From the experimental point of view, the most promising ones are those that explore possible indirect ways of interaction of dark sector particles, what is known as ''portal". 
\end{comment}

\section{Dark photons}

\subsection{Motivation}
Dark photons, also known as para-photons or hidden photons, are the realization of the vector portal. In this model, a U(1) boson from the hidden sector mixes kinematically with ordinary photons through the dimension 4 operator of the product of both field strengths. This interaction makes possible the oscillation from one species to the other along their propagation, permitting the detection of the dark photon.

Originally, it was proposed in the context of supersymmetry but has long left this paradigm to appear  in other theories like string theory or supergravity \cite{jaeckel2010low}. Now, dark photons are considered as part of the more general category of WISPs, Weakly Interacting Slim Particles \cite{arias2012wispy}. 

This section has been elaborated following mostly two references: \cite{fabbrichesi2021physics} for general overview and experimental searches and \cite{jaeckel2013force} for details in the mathematical formulations.

\subsection{Mathematical models: massless and massive cases}

The most general Lagrangian for the kinetic terms of two U(1) bosons, $A_a$ y $A_b$, can be written as: 

\begin{equation}\label{kineticlagrangian}
    \mathcal{L}_0 = -\frac{1}{4} F^{\mu\nu}_a F_{\mu\nu}^a - \frac{1}{4} F^{\mu\nu}_b F_{\mu\nu}^b - \frac{\varepsilon}{2} F^{\mu\nu}_a F_{\mu\nu}^b
\end{equation}

Being $F^i_{\mu\nu}$ the strength tensor of each photon: $F^i_{\mu\nu} = \partial_\mu A^i_\nu - \partial_\nu A^i_\mu$, with $i\in \{a,b\}$. The last term in expression \ref{kineticlagrangian} is known as \textit{kinetic mixing} term \cite{jaeckel2013force}.

These two bosons also couple to matter of their own sector, the gauge boson $A_b$ couples to the current $J_\mu$ of ordinary Standard Model matter and the other, $A_a$, to the current $J'_\mu$, which is made of dark-sector matter:

\begin{equation}\label{currentsDP}
    \mathcal{L}_I = e J^\mu A^b_\mu + e' J'^\mu A^a_\mu
\end{equation}

\noindent where $J^\mu = \sum_f Q_f \, \bar{f} \gamma^\mu f$, with $Q_f$ the electric charge of the fermion $f$ in units of $e$, $f$ and $\bar{f}$ the fermion field and its adjoint and $\gamma^\mu$ are the gamma matrices. The total current is the sum over all fermion species. 

Diagonalizing the kinetic Lagrangian \ref{kineticlagrangian}, the measurable states can be expressed in terms of the two gauge bosons. The rotation needed for this diagonalization can be expressed as:

\begin{equation}\label{rotationDP}
    \begin{pmatrix}
    A_a^{\mu} \\
    A_b^{\mu}
    \end{pmatrix}
    =
    \begin{pmatrix}
    \dfrac{1}{\sqrt{1 - \varepsilon^2}} & 0 \\
    -\dfrac{\varepsilon}{\sqrt{1 - \varepsilon^2}} & 1
    \end{pmatrix}
    \begin{pmatrix}
    \cos\theta & -\sin\theta \\
    \sin\theta & \cos\theta
    \end{pmatrix}
    \begin{pmatrix}
    A'^{\mu} \\
    A^{\mu}
    \end{pmatrix}
\end{equation}

Now $A^{\mu}$ can be identified with the ordinary photon and $A'^{\mu}$ with the dark photon. Orthogonal rotation dependent on angle $\theta$ is an extra rotation always allowed. The parameter $\theta$ is arbitrary for the massless case and fixed by the mass in the massive case. Nothing excludes the possibility of a massive U(1) boson, so both cases will be explained further in the following paragraphs.     

\subsubsection{Massless case}

If no mass terms have to be considered in the Lagrangian, the rotation parameter $\theta$ can be freely chosen. Including the rotation \ref{rotationDP} in the Lagrangian \ref{currentsDP} the latter can be written as: 

\begin{equation}\label{rotatedlagrangianDP}
\begin{aligned}
\mathcal{L}^{\prime} & =\left[\frac{e^{\prime} \cos \theta}{\sqrt{1-\varepsilon^2}} J_\mu^{\prime}+e\left(\sin \theta-\frac{\varepsilon \cos \theta}{\sqrt{1-\varepsilon^2}}\right) J_\mu\right] A^{\prime \mu} \\
& +\left[-\frac{e^{\prime} \sin \theta}{\sqrt{1-\varepsilon^2}} J_\mu^{\prime}+e\left(\cos \theta+\frac{\varepsilon \sin \theta}{\sqrt{1-\varepsilon^2}}\right) J_\mu\right] A^\mu 
\end{aligned}
\end{equation}

And therefore, there are values of $\theta$ for which one of the two bosons only couples to its sector current.

\begin{itemize}
    \item Photons coupled only to a Standard Model current, dark photons coupled to both currents: $\sin \theta = 0$ ($\cos \theta = 1$)
    
    \begin{equation}
        \mathcal{L}^{\prime}=\left[\frac{e^{\prime}}{\sqrt{1-\varepsilon^2}} J_\mu^{\prime}-\frac{e \varepsilon}{\sqrt{1-\varepsilon^2}} J_\mu\right] A^{\prime \mu}+e J_\mu A^\mu
    \end{equation}

    This case is the limit of vanishing mass of the massive case that will be shown later.

    \item Photons coupled to both currents, dark photons coupled only to dark current: $\sin \theta = \varepsilon$ ($\cos \theta = \sqrt{1-\varepsilon^2}$)

    \begin{equation}
        \mathcal{L}^{\prime}=e^{\prime} J_\mu^{\prime} A^{\prime \mu}+\left[-\frac{e^{\prime} \varepsilon}{\sqrt{1-\varepsilon^2}} J_\mu^{\prime}+\frac{e}{\sqrt{1-\varepsilon^2}} J_\mu\right] A^\mu
    \end{equation}

    In this case, the dark sector is not directly coupled to Standard Model, the dark photon sees ordinary matter only through the effect of operators of dimension higher than 4 like the magnetic moment or the charge form factors. This is the choice defining what is commonly known as massless dark photon. The coupling between ordinary photon and dark sector matter has a coupling $e^{\prime} \varepsilon / \sqrt{1-\varepsilon^2}$, what is called ``milli-charge" and experimentally its value is constrained to be very small.
  
\end{itemize}

\subsubsection{Massive case}

If masses are considered for the U(1) bosons, new terms have to be added to the Lagrangian \ref{rotatedlagrangianDP}. One way to give mass to the photons is through the Stueckelberg Lagrangian:

\begin{equation}
    \mathcal{L}_{S t u}=-\frac{1}{2} M_a^2 A_{a \mu} A_a^\mu-\frac{1}{2} M_b^2 A_{b \mu} A_b^\mu-M_a M_b A_{a \mu} A_b^\mu
\end{equation}

These terms fix the rotation angle to diagonalize the Lagrangian: 

\begin{equation}
    \sin \theta=\frac{\delta \sqrt{1-\varepsilon^2}}{\sqrt{1-2 \delta \varepsilon+\delta^2}} \quad \cos \theta=\frac{1-\delta \varepsilon}{\sqrt{1-2 \delta \varepsilon+\delta^2}}
\end{equation}

with $\delta=M_b/M_a$. So the Lagrangian of the kinetic part ends up being: 

\begin{equation}
\begin{aligned}
\mathcal{L}^{\prime \prime} & =\frac{1}{\sqrt{1-2 \delta \varepsilon+\delta^2}}\left[\frac{e^{\prime}(1-\delta \varepsilon)}{\sqrt{1-\varepsilon^2}} J_\mu^{\prime}+\frac{e(\delta-\varepsilon)}{\sqrt{1-\varepsilon^2}} J_\mu\right] A^{\prime \mu} \, + \, \frac{1}{\sqrt{1-2 \delta \varepsilon+\delta^2}}\left[e J_\mu-\delta e^{\prime} J_\mu^{\prime}\right] A^\mu 
\end{aligned}
\end{equation}

This effect of fixing the rotation is also seen in models in which the gauge symmetry from which the dark photon originates, is spontaneously broken. In both cases, both gauge bosons are coupled to both currents.

Even in the simplest models that consider only one massive boson -we know Standard Model photons are massless so $M_b = 0$-, the mass term removes the freedom of choosing the angle $\theta$. These models are widely used in experimental searches because when $\delta = 0$ the ordinary photon couples only to ordinary matter and the massive dark photon is characterized by a direct coupling to the Standard Model current. 

Including the mass term, the kinetic ones and the interactions, assuming a single massive U(1) boson, and redefining the fields ($A^\mu \rightarrow A^\mu - \varepsilon A'^\mu$ \cite{jaeckel2013force}) to absorb the mixing term (last in \ref{kineticlagrangian}) the Lagrangian contains the following relevant terms: 

\begin{equation}
    \mathcal{L} \supset-\frac{1}{4} F_{\mu \nu} F^{\mu \nu}-\frac{1}{4} F'_{\mu \nu} F'^{\mu \nu}+\frac{m_{A'}^2}{2} A'_\mu A'^\mu + eJ_\mu A^\mu + e'J'_\mu A'^\mu - e \varepsilon J_\mu A'^\mu
\end{equation}

The last term means that the coupling of the massive dark photon to Standard Model particles is not quantized, rather, it takes the arbitrary value $e \varepsilon$. Because of this direct current-like coupling to ordinary matter, it is the model most widely considered in experimental proposals. 

The massive dark photon, when its mass tends to 0, has the same couplings as the massless case choosing $\sin \theta = 0$, therefore, this case represents the limit of vanishing mass of the massive dark photon. Other choices for $\sin \theta$ are not related with any limiting case of the massive dark photon.

\subsection{Experimental searches}

Experiments are almost always driven by the theory. Proposals particularly appealing according to current empirical observations and reachable with present technology always receive more attention. This has happened to the different dark photon models. The massive case has been much more tested than the massless one, because its direct interaction with Standard Model particles makes it more attractive for direct detection experiments.

Even so, a few dedicated dark photon experiments have been performed. Most of the experimental constraints come as secondary results from experiments devoted to related topics like axion searches, particle colliders or nuclear studies.

The indirect interaction of the massless dark photon with ordinary matter makes it a challenging theory to test. Experiments have explored two possible interactions that may hint its presence: precision measurements of the magnetic dipole operator and searches for milli-charged particles. In both, higher order operators are involved in the interaction.

Constraints over the magnetic dipole operator can be extracted from astrophysical observations. Energy loss in star evolution can consider the new dark photon channel, the most stringent constraints being those derived from bremsstrahlung process in white dwarfs and red giants. Particle production in supernovae may also point to the presence of an unseen component that may interact through magnetic dipoles. The neutrino signal registered in the supernova 1987A allows this computation, although the limit is not very stringent due to the low statistics of the few events detected. Cosmology can also be also sensitive to the presence of unseen particles. in particular, the results of nucleosynthesis may depend on the content of massless dark photons. The derivation of constraints from all these phenomena is analogous to those for pseudo-scalar hypothetical particles like axions, reviewed in the next section.

Other sources of information to constraint the effect of dark photons over the magnetic dipole operator are precision physics and colliders. Both suffer from the same limitations: due to the high order of the operator, most of the interactions produced in the laboratory are affected by lower order and more probable interactions, so evidence for higher order operators is scarce compared with other effects. In precision physics, measurements of the fine structure of the atomic levels of helium atoms and the magnetic moment of electrons and muons are used to derive upper limits for the influence of dark photons in this operator. From collider experiments, several channels in which one photon is present and some energy is missing are identified and limits have been extracted interpreting this missing energy as the dark photon that escapes.  

Milli-charged particles may be another hint of the presence of a dark sector. In the massless case of dark photons, the Standard Model photons interact with dark sector particles through a coupling with fraction of the electric charge. Searches for this interaction have been performed in observational data using the same stellar objects as for magnetic dipole measurements. Its effects have been examined in data from Lamb shifts in the hydrogen atom and many collider experiments are also sensitive to some models of milli-charged particles. SLAC and LHC, for example, have performed specific searches devoted to this hypothetical interaction \cite{ball2020search}. Also, dark matter density surveys like the one extracted with WMAP or PLANK space-borne observatories are useful to establish limits in the presence of milli-charged particles in the Universe. 

Many proposals have appeared in recent years specifically aiming for massless dark photons. In colliders, specific processes in the flavour sector, Higgs and Z sector and in the pair annihilation in some decays have been identified as reactions sensitive enough to scan new models of massless dark photons. There are also proposals to study their possible coupling to magnons in ferrimagnetic materials and their effects in neutron stars collapses through the gravitational wave patterns.

\begin{figure}[h]
    \centering
    \includegraphics[width=0.8\linewidth]{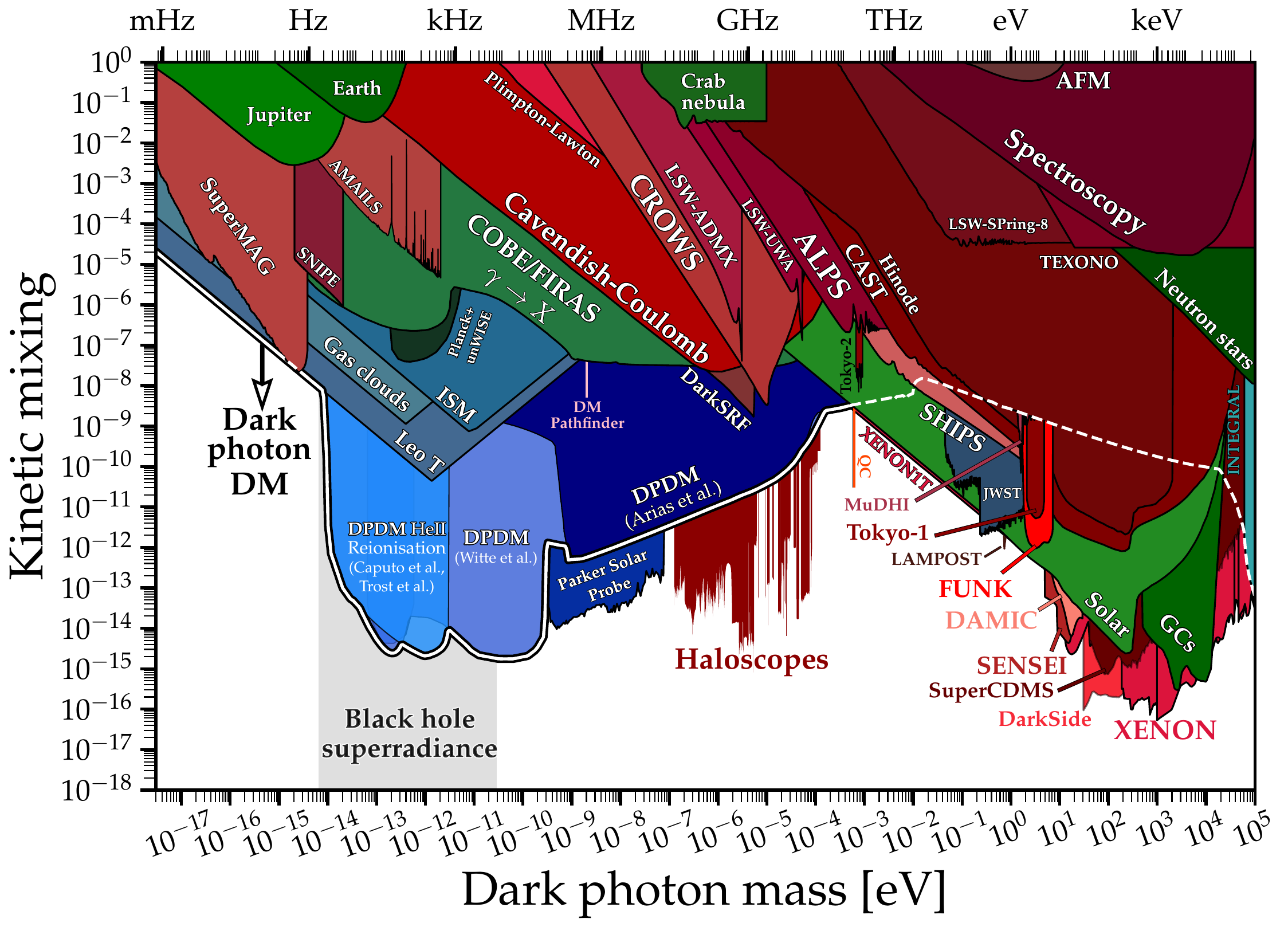}
    \caption{Current constraints on the massive dark photon mass, $m_X$ , and kinetic mixing parameter with the Standard Model photon, $\chi$. Cosmological bounds in blue, experimental bounds in red, and astrophysical bounds in green. Image extracted from \cite{caputo2021dark}.}
    \label{fig:DarkPhoton}
\end{figure}

The massive case has also been explored in astrophysics and cosmology data and in atomic and nuclear experiments like Coulomb force tests, Rydberg atoms and Lamb shifts in different elements. In addition, collider experiments can  produce constraints for massive photons, particularly in channels with missing energy, momentum or mass, and in visible final states looking for decays of dark photons produced in the collisions as long-lived particles with a displaced vertex or resonances in the spectra.

Dedicated experiments for axion and ALPs searches can set limits to massive dark photons due to the similarity of the signal produced, an excess in photon production. Therefore, light-shining-through-wall experiments, helioscopes like CAST, and haloscopes have probed wide areas of the parameter space of massive dark photons. In these experiments, the only difference between searching for axions or dark photons is usually the magnet needed for axion conversion; operating with the magnet off is enough for dark photon searches. Dark matter direct detection experiments might also be sensitive to dark photons. In figure \ref{fig:DarkPhoton} all these bounds are represented in the mass-kinetic mixing parameter space.

For a thorough review of the phenomenology, possible hints and experiments devoted to dark photon searches the review \cite{fabbrichesi2021physics} is highly recommended.

\section{Axions}

\subsection{Motivation: Strong CP problem}
The Quantum Chromodynamics (QCD) is a gauge field theory that describes the
properties of the strong interactions between quarks and gluons.  Despite its successful history of predictions across many different particle interactions, it contains a still unsolved problem known as the strong CP problem. This theory predicts a CP-symmetry violation in strong interactions that has not been observed experimentally.

The QCD Lagrangian is given by:
\begin{equation}\label{eq:QCD}
\mathcal{L}_{\text{QCD}} = -\frac{1}{4} G^a_{\mu\nu} G^{a\,\mu\nu} 
+ \sum_{f} \bar{\psi}_f (i\gamma^\mu D_\mu - m_f) \psi_f 
+ \bar{\theta} \frac{g^2}{8\pi^2} G^a_{\mu\nu} \tilde{G}^{a\,\mu\nu}
\end{equation}

where $G^a_{\mu\nu} = \partial_\mu A^a_\nu - \partial_\nu A^a_\mu + g f^{abc} A^b_\mu A^c_\nu$  is the gluon field strength tensor, \( \tilde{G}^{a\,\mu\nu} = \frac{1}{2} \epsilon^{\mu\nu\rho\sigma} G^a_{\rho\sigma} \) is its dual, \( D_\mu = \partial_\mu - i g T^a A^a_\mu \) is the covariant derivative, \( \psi_f \) is the quark field of flavour \( f \), \( m_f \) is the mass of the quark with flavour \( f \), \( \bar{\theta} \) is the CP-violating parameter and \( g \) is the QCD coupling constant.

The last term in \ref{eq:QCD} is not invariant under strong CP symmetry unless $\bar{\theta}=0$, but any value for this parameter would be completely possible. Experimentally, this parameter can be inferred from measurements of the neutron's electric dipole moment (EDMN). The most sensitive measurements of EDMN set a strong experimental bound at $|d_n| < 2.9 \times 10^{-26}$ e cm (90\% C.L.), which constrains the value of  $\bar{\theta} < 10^{-10}$ \cite{baker2006improved}. Explaining the extremely small value of this parameter is known as the strong CP problem. Although several solutions have been proposed, none have been experimentally confirmed. One such solution gives rise to the axion.

\subsection{Mathematical model: Peccei-Quinn solution}
\subsubsection{Peccei-Quinn mechanism}
An elegant answer to the strong CP problem was proposed by Peccei and Quinn in
1977~\cite{peccei1977cp, peccei16constraints}. In order to explain $\bar{\theta} = 0$ it postulates a new global and chiral symmetry $U(1)_{PQ}$ that is spontaneously broken at the energy scale of the symmetry $f_a$. It implies the existence of a new field $a$ which appears as the pseudo Nambu-Goldstone boson of the new symmetry, the axion. This is achieved through a new term $\mathcal{L}_a$ added to the QCD Lagrangian:

\begin{equation} \label{eq:Axion}
    \mathcal{L}_a = - \frac{1}{2} \partial_\mu a \, \partial^\mu a  +\frac{a}{f_a} \, \xi \, \frac{g^2}{32\pi^2} G^a_{\mu\nu} \tilde{G}_a^{\mu\nu} + \mathcal{L}(\partial_\mu a, \psi)
\end{equation}

The first term accounts for the kinetic energy of the axion field, the second term introduces the axion field $a$ and its coupling to gluons through $\xi $ a model-dependent constant, and the third gathers all possible interactions of the axion field with fermions. Considering the $\bar{\theta}$ term in equation \ref{eq:QCD} and the first term in \ref{eq:Axion} the effective potential acquired by the axion field can be expressed as:

\begin{equation}
    V_{\text{eff}} \sim \cos\left( \bar{\theta} + \frac{a}{f_a} \xi \right)
\end{equation}

and the vacuum expectation value for this potential is achieved at its minimum with respect to the axion field: 

\begin{equation}
    \left\langle \frac{\partial V_{\text{eff}}}{\partial a} \right\rangle = 0 \quad \Rightarrow \quad \langle a \rangle = -\frac{f_a}{\xi} \overline{\theta}
\end{equation}

It solves the strong CP problem by cancelling the $\bar{\theta}$ term for any value of $f_a$, providing a dynamical solution to the problem. Expanding $V_{eff}$ around its minimum the axion acquires a mass:

\begin{equation}
    m_a^2 = \left\langle \frac{\partial^2 V_{\text{eff}}}{\partial a^2} \right\rangle = -\frac{\xi}{f_a} \frac{g^2}{32\pi^2} \frac{\partial}{\partial a} \left\langle G_{\mu\nu}^a \tilde{G}_a^{\mu\nu} \right\rangle \bigg|_{\langle a \rangle}
\end{equation}

The mass is given by the Peccei-Quinn symmetry breaking scale $f_a$ so it is arbitrary. This solution of the strong CP problem predicts a new, potentially detectable particle, whose discovery would confirm the theory.

\subsubsection{Couplings}
The intensity of the interaction between axions and other particles depends on the specific structure of the axion field, determined by its Lagrangian, and the parameters that fix the precise axion model.
\\

\textbf{Coupling to gluons}

Axions couple to gluons as a consequence of the chiral anomaly of the $U(1)_{PQ}$ symmetry, so it is a direct outcome of any axion model and the most generic property of the axions. It is described by the surviving term of the axion Lagrangian after the merging with the $\bar{\theta}$ term:

\begin{equation}
    \mathcal{L}_{aG} = \frac{\alpha_s}{8\pi f_a} G_{\mu\nu}^a \tilde{G}_a^{\mu\nu} a
\end{equation}

where $\alpha_s$ is the fine-structure constant.
\\

\textbf{Coupling to photons}

The coupling to photons is done in two ways: the mixing with neutral pions and, if the fermions carry electric charge and PQ charge, through the Yukawa coupling to two photons. The axion-photon coupling is generic to all the models and most of the axion searches strategies are based on this interaction.
The axion to photons coupling can be described by the Lagrangian term:
\begin{equation}
    \mathcal{L}_{a\gamma} = -\frac{g_{a\gamma}}{4} F_{\mu\nu}^a \tilde{F}_a^{\mu\nu} \, a = g_{a\gamma} \, \vec{E} \cdot \vec{B} \, a
\end{equation}

where $F_{\mu\nu}$ is the electromagnetic field-strength tensor, $\tilde{F}^{\mu\nu}$ its dual, $\vec{E}$ and $\vec{B}$ are the electric and magnetic fields, $a$ is the axion field, and $g_a$  is the coupling constant.
\\

\textbf{Coupling to fermions}

Whether fermions carry PQ charge or not is a model dependent property, so this coupling is not as generic as the ones to gluons or photons. The possible interaction of an axion with a fermion $f$, indicated indirectly in the third term of \ref{eq:Axion},  can be expressed as:

\begin{equation}
    \mathcal{L}_{af} = \frac{g_{af}}{2\,m_f} \left( \bar{\psi}_f \gamma^\mu \gamma_5 \psi_f \right) \partial_\mu a
\end{equation}

where $\psi_f$ and $m_f$ are the fermion field and mass, and $g_{af}$ is the axion-fermion coupling constant.

\subsubsection{Axion models KSVZ - DFSF}

The original assumption from the Peccei-Quinn solution was that the $U(1)_{PQ}$ symmetry breaking scale is of the same order as the electroweak scale, so $f_a \simeq 250$ GeV. This would have been translated to an axion with a mass $\sim 100$ keV that should have been detected in reactor and accelerator experiments. 

After ruling out models with heavy and strong axions with $f_a$ ``small'', the so called \textit{visible axions},  what was left was the \textit{invisible axion} \cite{sikivie2021invisible}. This is the branding name for models with light and weakly coupled axions, so large $f_a$.

\begin{itemize}
    \item \textbf{KSVZ model}: it was proposed by Kim, and Shifman, Vainstein and Zakharov. In this model leptons and quarks do not carry PQ charge so interactions with matter only occurs via the axion-gluon coupling. These models require a new heavy quark $Q$ that is the only fermion that carries PQ charge. Here axion-electron coupling is forbidden at tree level but the coupling to nucleons is allowed. The main drawback of this model is that there is no physical motivation to introduce this new generated heavy quark.

    \item \textbf{DSFZ model}: it was proposed by Dine, Srednicki and Fischler, and Zhitnitski. In this case, Standard Model quarks and leptons carry PQ charge, so the existence of a new exotic quark is not required, but the Higgs field needs some adjustments. Axion-electron coupling at tree level is allowed. The disadvantage of this model is that fine tuning is necessary in order to obtain a breaking scale larger than the electroweak scale.
    
\end{itemize}

In both models, the axion mass and coupling to photons are proportional, so they are represented in exclusion plots as a straight band, taking into account reasonable parameters for both families of models. In figures \ref{fig:AxionPhoton_with_Projections} and \ref{fig:HaloscopesAxionLimits} they appear as yellow bands. Axions arising from these models are known as \textit{QCD axions} \cite{di2020landscape}.

\subsubsection{ALPs}
Axion models have been expanded to consider similar particles called \textit{axion-like particles}. They are predicted to arise generically, in addition to the axion, in low-energy effective field theories emerging from string theory. The main difference is that the mass and coupling to photons are entirely independent parameters, therefore any region of the parameter space explored by experiments may correspond to viable ALP candidates. Typically, interactions with matter have the same characteristics as axions, permitting the same experimental techniques to be used.
ALPs and hidden photons, as mentioned before, are two hypothesized particles that fall inside of what is known as WISPs.

\subsection{Axion searches}
Axion models do not constraint \textit{a priori} the mass, which is determined by the scale factor $f_a$ that can be arbitrary small or large. The implications of the existence of axions would shake the foundations of astrophysics, cosmology and particle physics, therefore the phenomenology has been studied in the context of all process known and some constraints have been extracted from current and past observations in these fields \cite{PDG2024review}. The most stringent limits come from astrophysics, based on the fact that axions could be produced in hot and dense environments like stars, globular clusters and white dwarfs, as will be seen in figure \ref{fig:AxionPhoton_with_Projections} of chapter 6. 
\\

\subsubsection{Astrophysics}
As mentioned when dark photons were discussed, the existence of axions may trigger a reinterpretation of some of past astrophysical and cosmological observations. From them, constraints to axion abundance may be derived, allowing to rule out some axion models. If not referenced explicitly, contents for this section have been extracted from \cite{ruiz2019ultra, garcia2015solar, PDG2024review}. \\

\textbf{Solar model}

 Photons may convert into axions in the presence of solar magnetic fields, known as Primakoff effect, or electrons may produce them through Bremsstrahlung. Because both processes mean a new way of cooling down the star, they imply an increment of the nuclear burning and of the solar temperature distribution. Constraining these effects with present solar observations leads to constraints in the axion-photon coupling of $g_{a\gamma} \leq 7 \times 10^{-10}$ GeV$^{-1}$ and in the axion-electron coupling of $g_{ae} \leq 2.5 \times 10^{-11}$ GeV$^{-1}$.\\

\textbf{Globular clusters}

These are accumulations of millions of stars densely packed formed at the same time. These objects are interesting for axion constraints because they allow to explore the ratio between two populations of stars that are affected differently by the existence of axions: the horizontal branch (HB) stars and the red giants branch (RGB). The RGB stars have a degenerate helium burning core and a hydrogen burning shell. When the RGB helium core becomes hot and dense enough, it rapidly increases the rate of fusion and becomes an HB star. Thus axion production via Primakoff effect should be larger in HB stars in comparison with RGBs, adding a new energy-loss channel which is negligible for RGB stars. So the population of RGB stars should be relatively larger than the population of HB stars inside the globular clusters.
Observations in this ratio produce a constraint in the axion-photon coupling of $g_{a\gamma} \leq \times 10^{-10}$ GeV$^{-1}$.

The brightness of RGB stars should also be affected by electron-Bremsstrahlung processes, from which constraints to $g_{ae}$ can be derived.\\

\textbf{White dwarf cooling}

White dwarfs (WD) are a remnant of low massive stars made of a degenerate carbon-oxygen core and a helium burning shell. A WD  is initially very hot but it has no energy source to maintain its temperature so it gradually cools down by the emission of neutrinos and later photons from the surface. If axion production is allowed, WD could increase its cooling speed by the emission of axions generated mainly by axion-Bremsstrahlung processes. Similar limits can be obtained from neutron star cooling. \\

\textbf{Supernova 1987 A}

A supernova is a stellar explosion that expels a big quantity of stellar material with a great force. There are different types of supernova; for axions the most interesting one is type II supernovae. When a star is massive enough to burn its carbon-oxygen core, it rapidly consumes the carbon producing heavier nuclei and increasing the core size. At some point the core exceeds the Chandrasekhar limit and the star collapses. A cataclysmic implosion takes place within seconds generating an enormous quantity of radiation and a burst of neutrinos. Axions could be emitted from the core of a type-II supernova via axion-nucleon Bremsstrahlung, shortening the neutrino burst. 
Supernova 1987 A was the first in which the neutrino flux was measured. Two experiments were sensitive to this flux: IMB and Kamiokande II. They registered a total of 20 neutrinos in a burst of 10 seconds, compatible with supernova models without axions. It should be noted that the low  statistics add uncertainty to the the limit extracted for $g_{aN}$.\\

\textbf{Superradiance of dark holes}

Gravitational wave emission of a rapidly rotating black hole may be affected by light bosonic fields such as axions. This effect may be observable through the superradiance mechanism. When the boson’s Compton wavelength is of order of the size of the black hole’s ergoregion, bosonic fields may form gravitationally bound states around the black hole that extract energy and angular momentum from the black hole, forming a coherent axion bound state emitting gravitational waves. When accretion cannot replenish the spin of the black hole, superradiance dominates the black hole spin evolution.

Stellar black hole spin measurements exploiting well-studied binaries tend to exclude the low end of axion masses. In this case, bounds do not affect the axion-photon coupling parameter but  its mass due to the gravitational nature of this effect. Limits vary depending on the analysis but a commonly accepted value is $m_a \gtrsim 10^{-11}$ eV. Gravitational wave interferometers like LIGO/Virgo are pursuing to detect effects of these axion-graviton interactions produced in black holes \cite{ng2021constraints}.

\subsubsection{Cosmology}
If axions have been present along the history of the Universe, hints may be found in cosmological probes. There are two distinct populations of cosmic relic axion: a non-thermal one behaving as cold dark matter (CDM), and a thermal one comprising a hot dark matter (HDM) component, in analogy to massive neutrinos. The hot component is restricted from cosmological observations (CMB, BBN...) to an upper bound in the axion mass $m_a \lesssim 1$ eV. The decay of such high-mass axions to photon pairs could be detected as unexpected optical or X-ray sources with narrow spectral features. 

Yet, for cosmology, the main interest in axions at present derives from their possible role as CDM. In addition to thermal processes, axions could be abundantly produced by the misalignment mechanism and therefore the axion dark matter abundance crucially depends on the cosmological history.

There are two possible scenarios: pre-inflationary PQ symmetry breaking scenario and post-inflationary PQ symmetry breaking scenario.
In the pre-inflationary scenario the PQ symmetry is broken before and during inflation and not restored afterwards. The axion field relaxes somewhere in the bottom of the ``wine-bottle-bottom'' potential (or ``mexican hat'' potential) due to topological fluctuations of the gluon fields that explicitly break the PQ symmetry. The axion field would have been homogenized by inflation, making the value of $\bar{\theta}$ unique. 
In the post-inflationary PQ symmetry breaking scenario, $\bar{\theta}$ will take different values in different patches of the present Universe because the decay is not correlated along all the Universe. Post-inflationary models can predict narrower axion mass ranges, although these ranges are still considerably large.

\subsubsection{Experimental searches}

Several techniques have been developed since axions were proposed for their detection. Although many new ideas are being explored nowadays, three are the main techniques depending on the axion source: helioscopes for solar axions, haloscopes for dark matter axions and light-shining-through-wall for axion production.\\

\textbf{Helioscopes}

\begin{figure}
    \centering
    \includegraphics[width=0.8\linewidth]{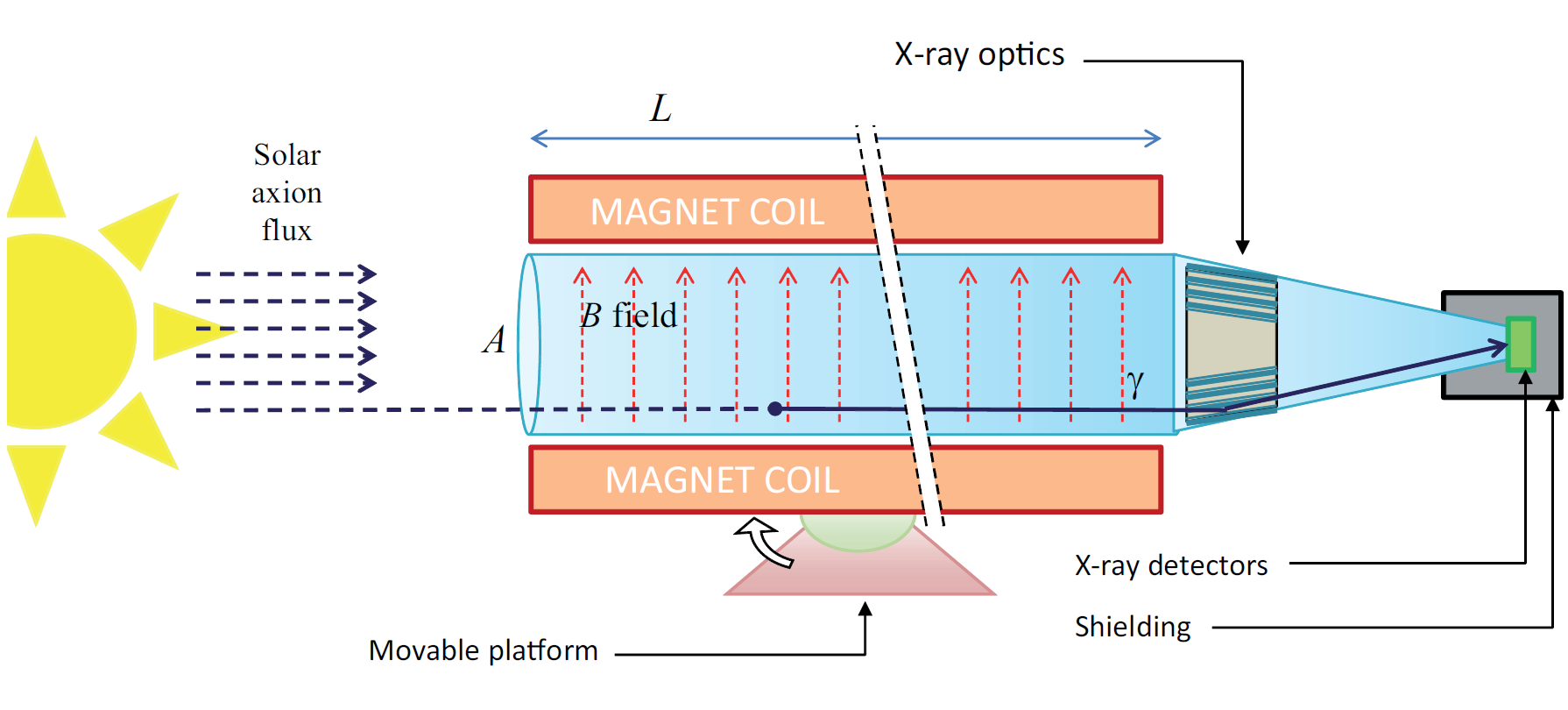}
    \caption{Schema of an helioscope from \cite{ruiz2019ultra}. Axions coming from the Sun are converted into X-ray photons by an intense transversal magnetic field. Then, they are focused towards a low-background X-ray detector where signal is registered.}
    \label{fig:Helioscope}
\end{figure}

The helioscope technique seeks axions generated at the Sun's core, it is the closest star to us so it should be the brightest source of axions. Proposed initially by P. Sikivie in 1983 \cite{sikivie1983experimental} and subsequently refined in 1989 \cite{van1989design}, this method involves directing a powerful magnet towards the Sun to induce the conversion of solar axions into photons inside the magnet. These axions arrive at the Earth with energies of the order of keV, resulting in conversion photons in the X-ray range. This is why at the other side of the magnet, X-ray
detectors are used to measure a possible excess over the radioactive background that would give the positive axion signal.
The key components of a helioscope are a movable, high-powered magnet and low-background X-ray detectors. An advanced version employs X-ray focusing optics to concentrate photons produced within the magnet onto a small area of the detector, thereby enhancing the signal-to-noise ratio. Figure \ref{fig:Helioscope} shows a conceptual scheme of the helioscope detection technique.

Helioscopes can cover axion masses in a wide range of the parameter space, up to  1 eV aprox,  because contrary to what happens with haloscopes, the signal is independent of the axion mass. There have been three generations of helioscopes since the 90s. The first helioscope was implemented in the Brookhaven National Laboratory in 1992 \cite{lazarus1992search} with a static magnet, a conversion volume of variable pressure gas and a xenon proportional chamber as X-ray detector. Later in 1998, the second generation Tokyo Axion Helioscope (SUMICO) \cite{moriyama1998direct} used a superconducting magnet on a tracking mount that allowed to follow the Sun trajectory. More recently, from 2003 until 2021 CERN Axion Solar Telescope (CAST) established the most stringent levels up to date for a helioscope \cite{altenmuller2024new}. In figure \ref{fig:AxionPhoton_with_Projections} its exclusion plot appears in dark red.

The next generation helioscope will be BabyIAXO, an intermediate stage towards IAXO, the International Axion Observatory \cite{abeln2021conceptual}, proposed to be sited at DESY, Hamburg. BabyIAXO will be a test bench for all technologies needed for the bigger version IAXO, with comparable size magnet, optics and X-ray detectors, but different number of bores. Despite this difference in active volume, BabyIAXO is by itself sensitive to unexplored areas in the parameter space of axion mass and coupling to photons, and will surpass CAST limits, the most stringent so far. All systems are currently under construction and first steps of on site installation are foreseen for 2026.
\\

\textbf{Light-shining-through-wall}

Light-shining-through-wall (LSW) is a technique based in the axion production in the laboratory. It makes use of an intense laser beam passing through a traverse magnetic field as a source of axions via Primakoff effect. Then an opaque wall is placed to block photons and behind it another magnet allows to convert the axions generated with the laser, that can easily pass through the wall, into photons again ready to be detected.
This technique has the advantage of being lees model-dependent because it does not depend on cosmological or astronomical assumptions. But on the contrary, it is penalized by the fact that the Primakoff conversion has to happen twice.

OSQAR \cite{pugnat2008results} in CERN and ALPS \cite{ehret2010new} in DESY are the most representative experiments using this technique, although their results are not yet competitive compared with other techniques. Their exclusion plots can be seen in figure \ref{fig:AxionPhoton_with_Projections}. Recently ALPS-II has resumed data taking with the improvement of a resonant cavity for the laser beam made with mirrors that increase the photon density in the conversion magnet.
\\

\textbf{Haloscopes}

The next chapter is devoted to this technique, therefore here it is enough to say that haloscopes are microwave resonant cavities immersed in magnetic fields that look for dark matter axions that may traverse the Earth. These cavities are sensitive devices tuned to certain frequency that is continuously scanned thanks to an antenna coupled to the resonant electric or magnetic field. One of the critical aspects of these experiments is the tuning of the characteristic frequency to be able to prove different axion masses. A plethora of experiments using this technique has arisen in recent years aiming for different axion masses. In the next chapter, figure \ref{fig:HaloscopesAxionLimits}, exclusion limits from many of them can be seen.

%% file: Chapters/6_QuantmSensinghaloscopes.tex
\section{Axion haloscopes}
Axion searches have been a fertile area for experimental particle physics research in the last decades. Depending on the model and the proposed origin for these axions, different detection techniques have been proposed, being most of them still in use in ongoing experiments.

\begin{figure}[h]
\centering
\includegraphics[width=0.7\textwidth]{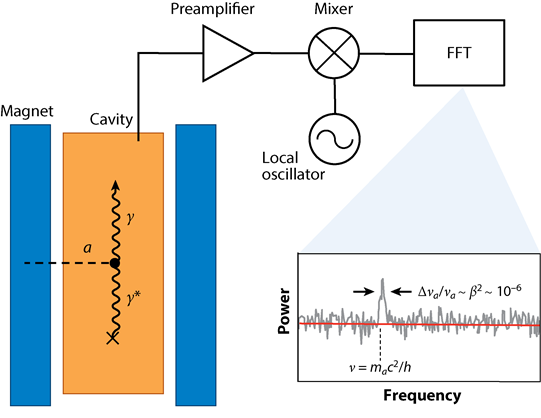}
\caption{Schematics of the axion haloscope detection principle. The axion-photon conversion is produced inside the resonant cavity, in orange, immersed in a magnetic field. The signal is extracted through a microwave antenna and the electronic chain extracts the power spectrum. If a photon appears in the cavity, a peak in the power spectrum is expected at the resonant frequency of the cavity. Image from \cite{graham2015experimental}.}
\label{fig:HaloscopeSquematics}
\end{figure}

 The axion haloscope concept was proposed by P. Sikivie \cite{sikivie1983experimental} in 1983 to look for dark matter axions, also called \textit{relic axions}, in the galactic halo. This technique makes use of the axion coupling with electromagnetic fields to convert the incoming particle into a photon in the presence of a strong magnetic field. The strategy to detect this photon consists in a resonant cavity which is tuned at the photon frequency and coupled to an antenna to measure the induced power, see figure \ref{fig:HaloscopeSquematics}. This detection technique has been in operation since its proposal, with ADMX (Axion Dark Matter eXperiment) \cite{goodman2025admx}, installed in the facilities of the University of Washington, as the leading experiment in the field. 

\subsection{Detection principle}

\begin{figure}[h]
    \centering
    \includegraphics[width=0.99\linewidth]{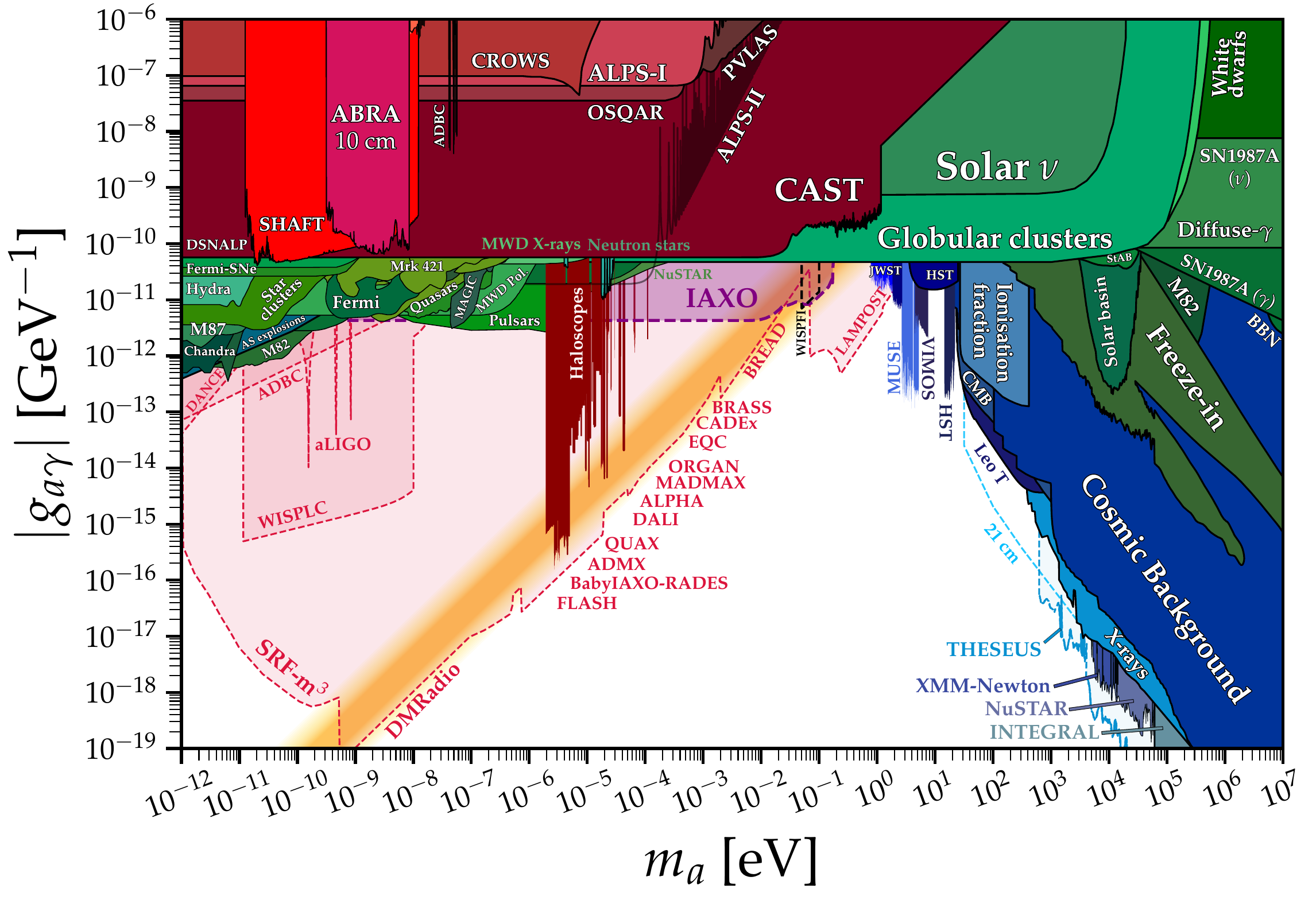}
    \caption{General landscape of axion searches. In the horizontal axis the mass of the axion model in eV, in the vertical one its coupling to photons in units of GeV$^{-1}$. In solid lines current limits, in dashed lines prospects for future experiments. Red areas are from particle physics experiments, green from astrophysical constraints and blue ones from cosmological proves. In red bars around 10$^{-5}$ eV can be seen axion haloscope experiments results. Image from \cite{AxionLimits}.}
    \label{fig:AxionPhoton_with_Projections}
\end{figure}

An axion haloscope is a microwave resonant cavity immersed in a magnetic field. When the resonant frequency of the cavity matches the one from the photon converted from the incoming axion, the power inside the cavity increases. This signal is extracted through a microwave antenna and amplified with low noise amplifiers. Typically, the frequency power spectrum is stored. This detection principle has been in use in the range of hundreds of MHz, but nowadays it is being expanded towards both higher and lower frequencies. In figure \ref{fig:AxionPhoton_with_Projections}, together with the current exclusion limits appear as a red doted line the expected sensitivity for future axion haloscope experiments.

The axion signal is expected as a peak in the power spectrum, centred at the corresponding frequency of the axion mass. If the hypothesis of the axion component of the dark matter in the galactic halo happens to be correct, the resonant cavity would be immersed in the axion bath; that is, the source would always be present and the only requirements for detection would be a cavity with the correct frequency and enough sensitivity, meaning that the extracted power from the axion-photon conversion should be higher than that coming from the noise of the system.

These detectors need to be tuned, as their resonance frequencies have a very small bandwidth. Techniques to sweep across different frequencies are continuously developed, most of which consist on moving parts that change the inner geometry of the cavity.

The power generated by the converted axion in the cavity is modelled as \cite{kim2020revisiting}:

\begin{equation}\label{eq:powerhaloscope}
 P_{g} = g_{a\gamma \gamma}^2 \left( \dfrac{m_a}{\rho_a} \right) B^2 V C \dfrac{Q_C Q_a}{Q_C + Q_a}   
\end{equation}

Some of these parameters cannot be controlled: the axion-photon coupling $g_{a\gamma \gamma}$, the axion mass $m_a$, the local density of axions in the halo $\rho_a$. Others can be optimized to increase the power of the signal: the magnetic field strength $B$, the volume of the microwave cavity $V$ and the mode-dependent form factor $C$. The last two parameters are the quality factors of the cavity $Q_C$ and the axion $Q_a$; intuitively the higher these numbers, the higher and narrower the peak of each resonance and therefore the better the sensitivity. The overall value of this term is limited by the minimum of the two, so it is very common to use instead $min(Q_C , Q_a)$.

In this ideal model, only the axion and the cavity are considered; however, the cavity is connected to the external world through some sort of antenna and it is coupled to a receiver device. This can be included in the formula adding a factor $\eta$, the fraction of power extracted through the line (magnitude of transmission coefficient $|S_{21}|$). This factor depends heavily on the precise geometry of the antenna and on temperature. In general, one aims for critical coupling in which $\eta = 0.5$, which means that half of the power generated in the cavity is extracted through the line and the other half remains in the cavity, where it will be dissipated by other processes. 

The second modification to the expression \ref{eq:powerhaloscope} is a reinterpretation of the parameter $Q_C$. The cavity is not isolated from external world, so couplings through ports affect the quality of the resonance. It will be discussed further in a while but, as a result, the internal quality factor of the cavity $Q_C$ will be replaced by $Q_L$, the loaded version of the parameter that accounts also for the external couplings. In practice, $Q_L \ll Q_a$, since $Q_a \sim 10^6$, so the expression can be written as:

$$
P_{SIG} = \eta g_{a\gamma \gamma}^2 \left( \dfrac{m_a}{\rho_a} \right) B^2 V C Q_L
$$

The ability to detect the signal coming from axion conversion depends on the noise level of the whole device, including the readout chain and the DAQ system. The signal-to-noise ratio is given by Dicke's radiometer equation \cite{diaz2021design}:

\begin{equation}\label{eq:SN_Dicke}
    \dfrac{S}{N} = \dfrac{P_{SIG}}{P_N} = \dfrac{P_{SIG}}{k_B T_{sys} \sqrt{\frac{\Delta v}{t}}}  
\end{equation}

The power of the noise is proportional to the Boltzmann constant $k_B$, the temperature of the system $T_{sys}$, the bandwidth  $\Delta v$ and the observation time $t$. In perfect conditions, $\Delta v$ should be the bandwidth of the axion resonance $\Delta v_a = \frac{m_a}{Q_a}$, to reduce the noise power outside the region of interest.  

The figure of merit used in axion haloscope experiments is the velocity of scan for a fixed signal-to-noise ratio. In other words, the time needed to achieve a certain signal-to-noise ratio per mass range scanned.
The mass range scanned around the cavity resonance is:
$$
dm_a = \frac{m_a}{Q_L}
$$
The higher the quality factor $Q_L$, the smaller the mass range probed, but also the time needed to reach the sensitivity level is shorter.

And, for fixed signal-to-noise, $S/N$, the time needed to achieve it is:
$$
dt = \Delta v_a \left(  \dfrac{\frac{S}{N} k_B T_{sys}}{P_{SIG}} \right)^2 = \dfrac{m_a}{Q_a} \left(  \dfrac{\frac{S}{N} k_B T_{sys}}{P_{SIG}} \right)^2
$$

Then the figure of merit for an axion haloscope is:
$$
\mathcal{F}  =  \dfrac{dm_a}{dt} = \dfrac{Q_a}{Q_L} \left( \dfrac{P_{SIG}}{\frac{S}{N} k_B T_{sys}} \right)^2  = Q_a Q_L \eta^2 g_{a\gamma \gamma}^4 \left( \dfrac{\rho_a}{m_a} \right)^2 B^4 C^2 V^2 \left( \frac{S}{N}k_B T_{sys} \right)^{-2}
$$

The bigger this value, the faster the haloscope will scan with the same sensitivity. Of course, there are other experimental characteristics to be taken into account for the overall scanning speed, for example the tuning mechanism or the data transmission.

\subsubsection{Cavity parameters: form\index{Form factor}, quality\index{Quality factor} and transmission\index{Transmission factor} factors}

The performance of a cavity as an axion haloscope depends on several characteristics that are quantified through their corresponding parameters. They reflect the haloscope ability to store energy in the resonance, transmit it through the ports or its coupling to external magnetic fields.

\begin{enumerate}
    \item Form factor $C$
    
    Also called geometric factor. It is a normalized parameter that quantifies the coupling between the static magnetic field $ \vec{B}$ and the dynamic electric field $ \vec{E}$ induced by the axion-photon conversion. Depending on the mode excited in the cavity this coupling could be very different. 
    $$
    C = \dfrac{|\int_V \vec{E} \cdot \vec{B} \, dV|^2}{\int_V \| \vec{B}\|^2 dV \int_V \epsilon_r \|\vec{E}\|^2 dV}
    $$
    \noindent where
    $|\cdot|$ the absolute value and $\| \cdot \|$ the usual norm in $\mathbb{R}^3$. 
    
    \vspace{1ex}
    
    \item Quality factor $Q$

    Quality factors give an idea of how good the device is at storing photons. 
    They are related to decay rates $k$: $Q = w_r/k$, and therefore,  higher decay ratios lead to smaller quality factors. These decay rates are, of course, related to finite photon lifetimes $\tau$ ($k=2\pi/\tau$), due to unwanted couplings of the resonator with electrical and magnetic surroundings \cite{zheng2021circuit}. 
    Several quality factors can be defined for different loss mechanisms, as they may have distinct behaviours when cooling down the haloscope and therefore different impact on the experiment.  
    Decay rates are additive, $k_{12} = k_1 + k_2$, and therefore quality factors are not: $\frac{1}{Q_{12}}=\frac{1}{Q_1}+\frac{1}{Q_2}$.

    Some of the most common quality factors in axion haloscope searches:
    \begin{enumerate}
        \item $Q_a$: Quality factor associated to axion resonance. It is assumed to be very big $\sim 10^6$.
        \item $Q_i$: Internal quality factor. Due to material impurities, geometrical design, etc. This value increases significantly when temperature decreases. Electromagnetic losses decrease when temperature decreases. This is what in equation \ref{eq:powerhaloscope} was called cavity quality factor $Q_C$.
        \item $Q_c$: Coupling quality factor. Losses due to coupling with the ports. The exact geometry of the antenna is crucial, and its value is roughly independent of the temperature. Typically this is the main loss channel when the cavity is cooled down.
        \item $Q_L$: Loaded quality factor. The total quality factor taking into account all decay channels. Usually, it can be expressed as $\frac{1}{Q_L}=\frac{1}{Q_i}+\frac{1}{Q_c}$
    \end{enumerate}
        
    \item Transmission factor
    
    Ratio between injected and extracted power in a resonator mode. Due to the inherent modelling of photon signals, this parameter is a complex number, determined by magnitude and phase. It is defined for each port or combination of ports and it is common to call it \textit{reflection factor} if the same port is used for injection and extraction: $S21$ denotes the transmission factor when using port 1 to inject and port 2 to extract power; $S11$ is the reflection factor for port 1. Common laboratory equipment like vector network analysers (VNA) can measure these parameters, both in magnitude and phase. Not only the cavity but also the antennas determine these parameters, which may need to be fine-tuned to improve readout speed or to increase the lifetime of photons in the cavity.
\end{enumerate}

\subsubsection{Gravitational waves\index{Gravitational waves}}

Recently, haloscopes have been proposed as gravitational wave detectors for very high frequency signals (above 100 kHz). Theory predicts a similar coupling to electromagnetic fields as for axions, and many haloscope results can be interpreted as well under the gravitational wave paradigm~ \cite{ejlli2019upper}. Sources for such high frequency oscillations remain hypothetical and many involve physics beyond the Standard Model but there are proposals up to $10^{15}$ Hz \cite{aggarwal2021challenges}, ranging from early Universe effects to mergers of exotic galactic objects. Dedicated detectors using haloscope techniques are being designed, like MAGO 2.0 \cite{berlin2023mago}, to hunt these high frequency gravitational waves.

\subsection{Present status of the technology}
In recent years, many experiments have implemented the axion haloscope technique. Most of them rely on big volumes and low temperatures to achieve the expected sensitivity. Tuning systems vary from one experiment to another but most of them have mechanical parts that can be moved, although new tuning mechanisms are being explored, like magnetic tuning with ferromagnets \cite{bourhill2016ultrahigh} or electric tuning with ferroelectric materials \cite{Garcia2023ferrTuning}.

The leading experiment in the field is ADMX \cite{khatiwada2021ADMX}. ADMX is located at the Center for Experimental Nuclear Physics and Astrophysics (CENPA) at the University of Washington, Seattle. It makes use of a cylindrical cavity of around 200 l, whose frequency can be tuned with two moving rods, inside an 8 T superconducting solenoid magnet. It is refrigerated with dilution refrigerators reaching temperatures as low as 100 mK in the cavity. Its readout chain is one of the most sensitive in the world, operating nowadays with quantum limited Josephson Parametric Amplifiers (JPA). This has allowed ADMX to become the most sensitive experiment to date in the DFSZ range of axion models between 1.9 and 3.3 $\upmu$eV. This corresponds, roughly, to a scan range from 500 MHz to 1 GHz in several data takings.

\begin{figure}[h]
    \centering
    \includegraphics[width=0.99\linewidth]{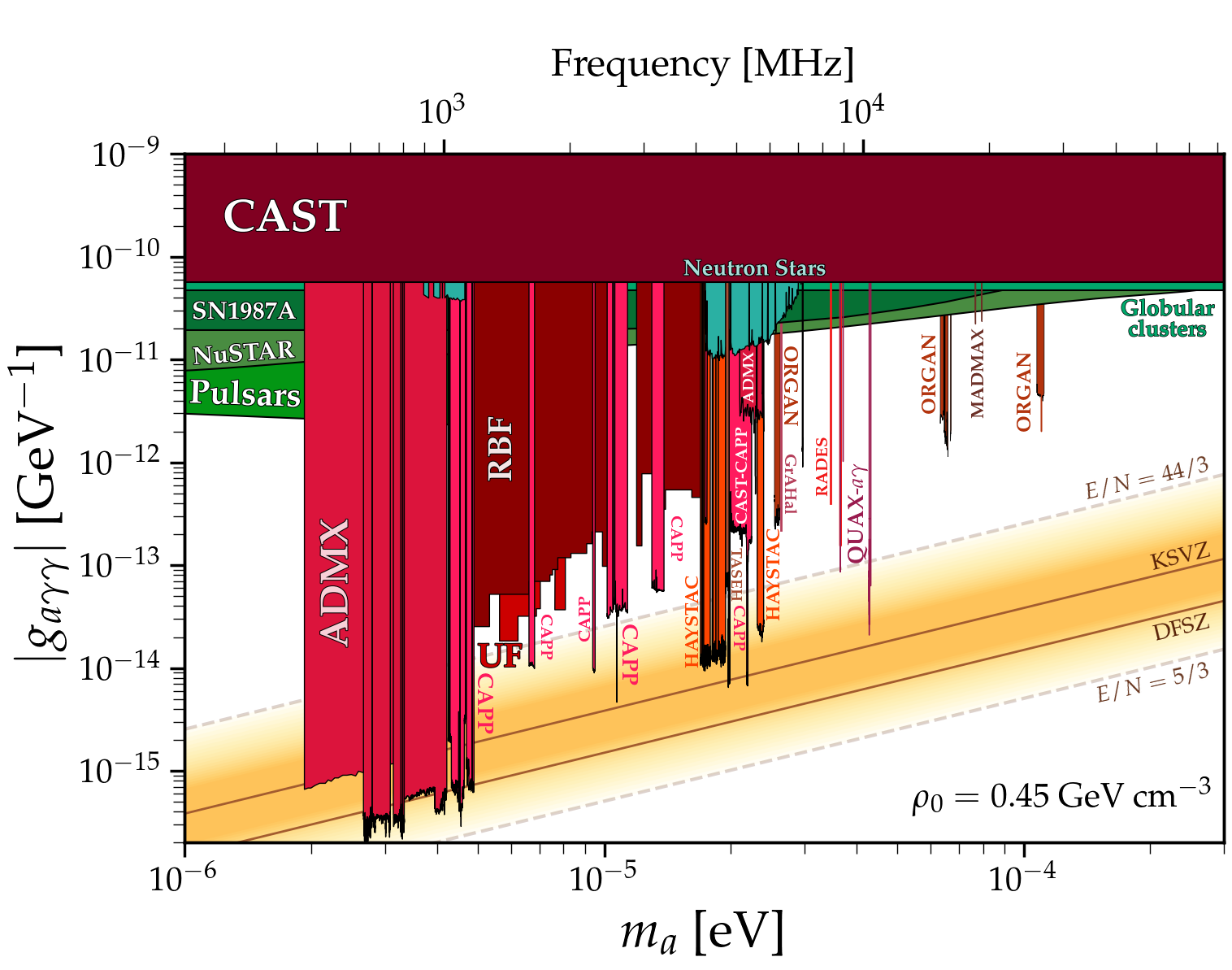}
    \caption{Exclusion limits for axion-photon coupling established by haloscopes. It is a close up of figure \ref{fig:AxionPhoton_with_Projections}. Extracted from \cite{AxionLimits}.}
    \label{fig:HaloscopesAxionLimits}
\end{figure}

In South Korea, a dedicated research centre for axion searches, the Center for Axion and Precision Physics (CAPP), has pushed axion haloscope technologies to improve several key aspects of haloscopes. They have developed new cavities to extend searches to higher frequencies \cite{yang2023extended}, have operated many different designs tuned with sapphire rods \cite{kim2023near} \cite{kwon2021capp}, in different magnets \cite{adair2022CastCapp}, techniques to improve the quality factor Q of the resonators with superconducting tapes \cite{ahn2020superconducting}, etc. Many of their results can be seen in figure~\ref{fig:HaloscopesAxionLimits}.

Haystac is another ``conventional'' resonant cavity experiment placed in Yale’s Wright Laboratory in New Haven, Connecticut. It has been the first to demonstrate near-quantum limited noise using a Josephson parametric amplifier \cite{zhong2018results}, along with the feasibility of ``state-squeezing'' to increase scan speed \cite{jewell2023new}.

Other experiments that have measured in different frequencies with sensitivities close to the axion band are QUAX in Italy \cite{rettaroli2024search}, TASEH in Taiwan \cite{chang2022taiwan}, GrAHal in France \cite{grenet2021grenoble} and ORGAN in Australia \cite{mcallister2017organ}.
There are also new concepts arising, close to haloscopes but making use of different types of resonators. ALPHA tries to develop a metamaterial made of a parallel-wire array that can be tuned adjusting wire separation. MADMAX uses dielectric layers in a magnetic field in order to resonantly enhance the photon signal. Both are aiming towards hundreds of $\upmu$eV axions. ABRACADABRA, BASE, SHAFT use different realizations of LC circuits to look for neV axions. For high frequency axions are aiming BRASS, CADEx and BREAD experiments. And there's also a proposal based in laser interferometry, DANCE. Useful references for these experiments can be found in the chapter ``Axions and Other Similar Particles'' of reference \cite{PDG2024review}.

\begin{comment}
\begin{figure}
    \centering
    \includegraphics[width=0.9\linewidth]{images/HaloscopesAxionLimits2.png}
    \caption{¿Quizá mejor esta?}
    \label{fig:enter-label}
\end{figure}    
\end{comment}

\subsection{RADES collaboration}

The works described in this thesis have been developed within the RADES (Relic Axion Detector Exploratory Setup) collaboration. It was established to develop innovative microwave resonator technologies aimed at detecting axion dark matter. The collaboration gathers the experience from several international groups in axion searches and related technologies from different institutions: CAPA institute from Universidad de Zaragoza, Max Planck Institute for Physics from Munich, Universidad Politecnica de Cartagena, IFIC from Valencia, University of Mainz; and recently, thanks to the DarkQuantum protect, several groups joined the project: Karlsruhe Institute of Technology, Aalto University form Helsinki, École Normal Supérieure from Paris, Instituto Tecnológico de Aragón, Institut de Ciència de Materials de Barcelona (ICMAB) and Institut de Física d'Altes Energies (IFAE) from Barcelona.

Traditional axion haloscope experiments rely on large, single-cavity resonators operating inside strong magnetic fields. However, the sensitivity of such systems at higher axion masses is limited by the inverse relationship between resonant frequency and cavity volume. To overcome this limitation, the RADES collaboration proposed a novel approach based on a scalable chain of smaller, coherently coupled cavities, effectively maintaining high-frequency sensitivity while increasing the total detection volume. This seminal multicavity concept was first introduced in \cite{melcon2018axion}, where the theoretical feasibility of such configurations was established.

Following the theoretical proposal, a series of multicavity prototypes were developed and tested at CERN, within the bore of the CAST (CERN Axion Solar Telescope) helioscope magnet. The first of these, known as CAST-RADES, employed a five-cavity microwave filter design, allowing for a collective resonant mode sensitive to axion conversion. The design and electromagnetic optimization of this system were detailed in \cite{melcon2020scalable}, with the objective of demonstrating the mechanical and electromagnetic viability of multicavity haloscopes in a realistic experimental environment.

Measurements carried out with these prototypes led to the first experimental limits on axion-photon coupling using this approach. The results confirmed the feasibility of the multicavity concept and provided a technological benchmark for scaling to more sensitive detectors \cite{alvarez2021first}. 

In parallel, the collaboration began exploring the use of superconducting materials to reduce losses and enhance the quality factor $Q$ of the resonators. One of the most promising developments involved lining cavity walls with high-temperature superconducting (HTS) tapes. A dedicated test of such a superconducting cavity was performed at CERN’s SM18 cryogenic facility, under strong magnetic fields. These measurements showed improved performance and opened the path to ultra-high $Q$ haloscopes operating in high magnetic field  environments\cite{ahyoune2025rades}.

Building on this technological foundation, RADES proposed the implementation of a dedicated axion haloscope within the infrastructure of BabyIAXO, the next-generation axion helioscope prototype. The BabyIAXO-RADES project envisions an advanced multicavity detector operating at cryogenic temperatures, benefitting from the large, static magnetic field of the BabyIAXO magnet. This integration would enable the exploration of axion masses in the few $\upmu$eV range with unprecedented sensitivity for compact haloscope systems \cite{ahyoune2023proposal}.

\begin{figure}[h]
    \centering
    \includegraphics[width=0.8\linewidth]{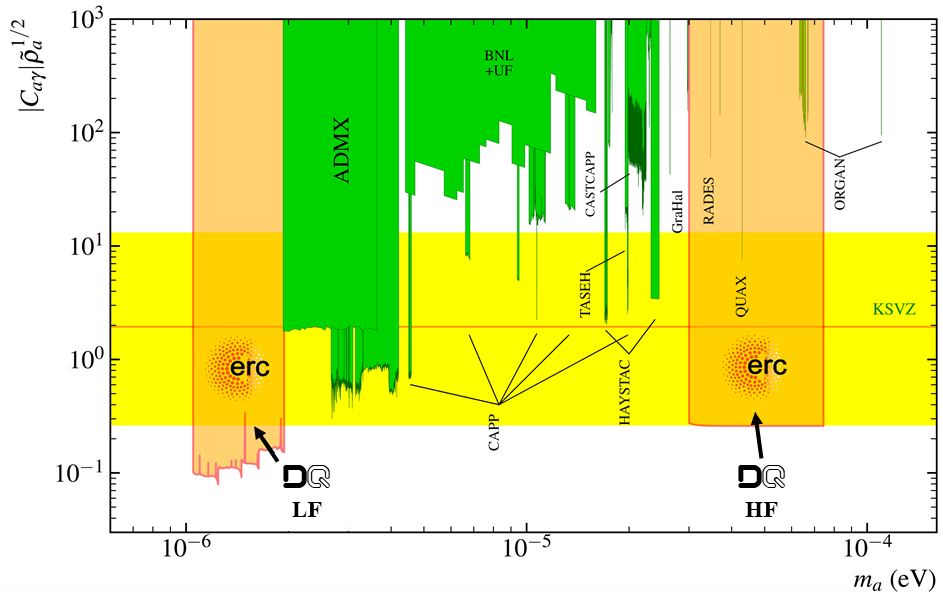}
    \caption{Exclusion limits for axion-photon coupling scaled by mass density. The two orange bands are prospects for both frequency ranges (LF and HF) foreseen in the DarkQuantum project.}
    \label{fig:DarkQuantumSensitivity}
\end{figure}

Recently, a new path opened for the collaboration with the grants awarded to DarkQuantum and Q-RADES projects. Together with the traditional axion haloscope techniques, they will foster the development of quantum sensors for axion searches with haloscopes. If successful, they will dramatically increase the sensitivity of these experiments by going beyond the Standard Quantum Limit (SQL). This and the following chapter of this work narrate the initial steps with the first single photon counter prototype developed for DarkQuantum project.

Yet, this is not the only objective of the DarkQuantum project. The ERC Synergy grant awarded for this project recognizes seven items to achieve:

\setlist{nolistsep}

\begin{itemize}[noitemsep]
    \item O1. Single-photon counter in the 8-18 GHz regime.
    \item O2. Q-limited linear amplifier (SQUID-based) at 200-500 MHz.
    \item O3. Cryogenic setup for BabyIAXO
    \item O4. HTS inner cavity coatings.
    \item O5. Tuning with ferrimagnetic crystals.
    \item O6. Full RADES HF setup underground (LSC).
    \item O7. Full RADES LF at BabyIAXO setup (DESY).
\end{itemize}

They can be grouped in two main areas of interest: low frequency (200-500 MHz) haloscope at cryogenic temperature in BabyIAXO magnet and high frequency single photon counter with quantum technologies, that will be placed in the Laboratorio Subterráneo de Canfranc. In addition, ancillary technologies like HTS coatings and tuning with ferromagnetic crystals are encouraged.
Sensitivity projections for this two regions are expected to surpass the QCD axion models band as shown in figure \ref{fig:DarkQuantumSensitivity}.

\subsection{Surpassing Standard Quantum Limit: DarkQuantum project}

One of the main limitations for the axion haloscope technology is the signal-to-noise ratio, which sets the velocity and sensitivity of the scan. The DarkQuantum project, within the RADES collaboration, proposes a new way of measuring the photon content of the cavity using a quantum sensor that overcomes the traditional Standard Quantum Limit of linear receivers, a manifestation of the uncertainty principle  \cite{lamoreaux2013analysis}. 

The concept is to count single photons rather than measuring the extracted power from the haloscope with an antenna. However, in the typical energy range explored by this technology, photons carry extremely low energies, making their individual detection highly challenging. As a result, measurements are typically performed by exploiting their wave-like properties instead. In recent years, though, single photon counters have being developed for these energies, the lowest ever achieved, in the range of few GHz. These ideas are being applied for dark matter searches with different implementations \cite{dixit2021searching} \cite{braggio2024quantum}. 
\\

Current detection techniques are fundamentally limited by the Standard Quantum Limit. For linear receivers, the noise power is modelled by 
\begin{equation}
    P_N = k_BT\sqrt{\dfrac{\Delta\nu}{t}}
\end{equation}

This was previously shown in equation \ref{eq:SN_Dicke}, the signal-to-noise expression for a haloscope. This expression gives the fluctuations in the average noise power that is detected and it is known as the Dicke radiometer equation. In the case of low temperatures, the equation must be modified to account for zero point fluctuations. This is done through the mean photon occupation number $\bar{n}$ of a cavity mode resonant at frequency $\nu$ at temperature $T$:

\begin{equation}
    \bar{n} = \dfrac{1}{e^{\frac{h\nu}{k_BT}}-1}
\end{equation}

And the modified Dicke radiometer equation is: 

\begin{equation}
    P_{LR} = h\nu(\bar{n}+1)\sqrt{\dfrac{\Delta\nu}{t}}
\end{equation}

Therefore, the Standard Quantum Limit can be derived by $P_{LR}(\bar{n}=0)$:

\begin{equation}
    P_{SQL} = h\nu\sqrt{\dfrac{\Delta\nu}{t}}
\end{equation}

Therefore, even with a mean photon occupation number of $\bar{n}=0$ in the resonance, quantum noise fluctuations stablish a minimum background noise level. The explanation of this modified version of the Dicke radiometer equation for low temperature can be found in \cite{lamoreaux2013analysis}. In addition, \cite{dixit2021thesis} includes an appendix that provides a justification for why a linear amplifier introduces a quantum of noise, derived from the uncertainty principle

For a single photon counter, the noise power expression can be derived following a similar argument. The number of photons detected in a time $t$ can be written as:

\begin{equation}
    N=\dot{n}t=\bar{n}\dfrac{1}{\tau_c}t
\end{equation}

Being $\dot{n}$ the detection rate (photons per second) and $\tau_c$ the cavity lifetime, the characteristic decay time of a photon in this resonant mode. Its fluctuation, following Poisson statistics, will be its square root. Then, the noise power for a single photon counter is the power of these fluctuations, therefore:

\begin{equation}
    P_{SPC}  = h\nu_a \dfrac{\sqrt{N}}{t} = h\nu_a\sqrt{\dfrac{2\pi\nu\bar{n}}{Q_ct}}
\end{equation}

To derive this expression, $\Delta\nu=\nu/Q_c$ and $1/\tau_c=2\pi\nu/Q_c$ were used. A perfect quantum efficiency $\eta=1$ has been assumed during this derivation for the single photon counter. This is never the case and the corresponding rescaling factor should be considered in the number of photons detected. Further clarifications can be consulted in \cite{lamoreaux2013analysis}.

\begin{figure}
    \centering
    \includegraphics[width=0.9\linewidth]{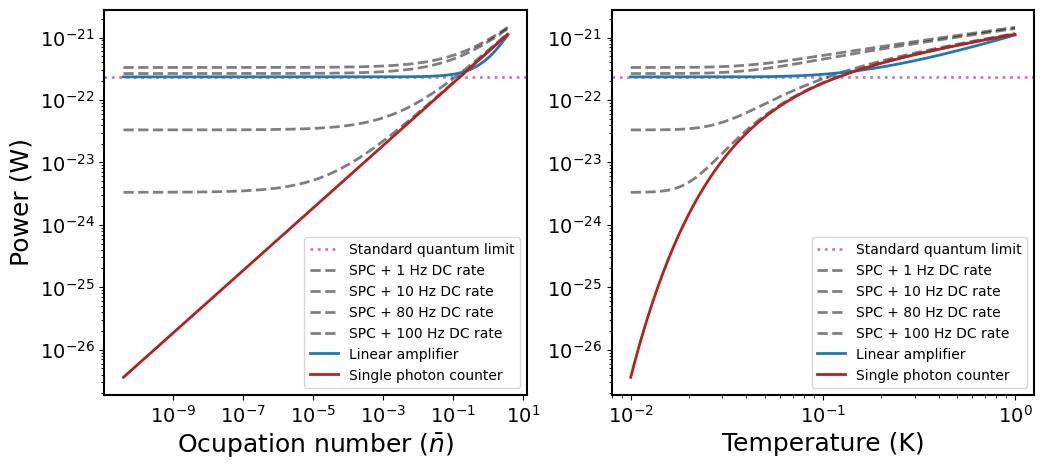}
    \caption{Noise power level for linear amplifier (blue) and single photon counter (red) in terms of mean occupation number (left) and equivalent temperature (right). The horizontal dot line marks Standard Quantum Limit. Grey lines are SPC limits with different dark count rates, lowest one for 1 Hz, upper one for 100 Hz. Plotted for target frequency 5 GHz and $Q_c = 10^6$ and $Q_a  = 10^6$.}
    \label{fig:LinearAmp_SPC_noisePower}
\end{figure}

In figure \ref{fig:LinearAmp_SPC_noisePower} there is a comparison of the two techniques, linear amplification and single photon counter. Both noise power levels have being plotted in terms of occupation number (left) and its equivalent temperature (right). These systems are operated around 8 mK, although the thermalization process is never perfect and the effective temperature tends to be slightly higher. At temperatures of few tens of mK, the noise power from the detector is no longer the limiting factor for single photon counters and usually dark count rates from non-thermal sources start to matter. In these plots, different levels of dark count rates have been represented as well. The lowest levels achieved up to now are around 1 Hz \cite{dixit2021searching}. Similar quantum readout methods developed recently showed higher dark count levels, 85 Hz \cite{braggio2024quantum}. 
The best dark count rate of the RADES prototype, as it will be discussed in chapter \ref{Ch:DarkQuantum} is around 80 Hz.

\section{Superconducting qubits as quantum sensors}

One of the limitations in axion haloscope technologies comes from the limited sensitivity when reading electromagnetic signals. Noise present in all steps in the amplification and readout chain may degrade the recorded signal and therefore weak stimuli may be hidden under the electromagnetic noise. The most crucial item for this is the first amplifying step, typically close to the cavity and at very low temperature, because it sets the noise level of the overall detector.

Recently, a different strategy has arisen that may surpass the power signal readout: the single photon counter. Instead of measuring the power carried by a wave, through the dual behaviour of photons as particles and waves, they can be counted individually. This is routine for higher energy photons like X-rays, but microwave photons behave mostly as a wave, and only under very specific conditions can their individual behaviour be perceived. This is the basic idea of quantum sensors developed for haloscopes, to be sensitive to single photons in the GHz regime.

Qubit is the name given to a mathematical model consisting in two quantum states. They are isolated from other interactions and only transitions from one state to another are possible. As a quantum system, superposition and mixed states are allowed. 

Various technologies are being developed to materialize such a mathematical model: trapped ions, lattice defects, superconducting circuits... None of the existing technologies is perfect and all of them have drawbacks. Typically, some perform better in one or another aspect interesting for certain applications so the field is split in many branches focusing in different objectives, the most widely known being quantum computation, the holy grail of quantum world. 

Fortunately, efforts are not solely focused on achieving quantum supremacy, meanwhile, other practical applications of quantum effects are being actively explored. Among the various developments, quantum sensors hold particular interest for fundamental particle physics, as they can help to improve sensitivity, discrimination capabilities, energy resolution, etc.

Related to dark matter searches, it has been proved that a qubit made of a superconducting circuit coupled to a resonant cavity can speed relic axion searches with haloscopes up. From theoretical developments in quantum optics, a mathematical model has arisen, the Jaynes-Cummings Hamiltonian, that deals with the coupling between the bosonic mode of the resonant cavity (harmonic model), the fermionic two-level system of a qubit (1/2 spin model) and their interaction.\\

\subsection{Josephson junction}

\begin{figure}[h]
\centering
\includegraphics[width=0.5\linewidth]{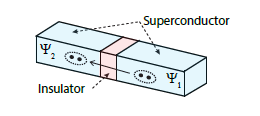}
\caption{Josephson junction schematics formed by two superconducting regions and an insulator between them. Image from \cite{naghiloo2019introduction}.}
\label{fig:JJ}
\end{figure}

The first ingredient for a superconducting circuit that behaves like a qubit is the Josephson junction. It is a weak link between two superconductors separated by a barrier that modifies the flow of electrons. Typically, it is made with two superconducting layers and an insulator or a non-superconducting metal in between, see figure \ref{fig:JJ}, but it can also be realized with geometrical patterns, for example by squeezing the point of contact between the two superconducting volumes. The mathematical relations that model the flow of electrons, current and voltage along the weak link were developed by Brian Josephson in 1962 (before my father was born, so quite a long journey).  

In a Josephson junction, a current appears between the two superconducting volumes without any voltage applied. This is due to the Cooper pairs tunnelling across the thin insulator barrier. The associated voltage is due to the kinetic energy of Cooper pairs and is often called kinetic inductance effect. This is the amazing behaviour that makes Josephson junctions so special, they are one of the main examples of macroscopic quantum behaviour, quantum effects that can be observed in the macroscopic scale.

The flow of this current $I$, and the consequent voltage $V$, can be parametrized in terms of the quantum state of each superconducting side, that can be described through a complex number: $\Psi _{1,2} = \sqrt{n_{1,2}} \ e^{i\theta_{1,2}}$. The number of carriers $n_1, n_2$ is typically similar and very big so it is the phase difference between both states what makes a bigger difference in the complex plane. Denoting by $\delta = \theta_2 - \theta_1$ this phase difference, $I_0$ the critical current above which the junction becomes a dissipative junction (not superconducting any more)  and $\Phi_0 = \frac{h}{2e}$ the flux quantum, the Josephson equations can be expressed as:

\begin{equation} \label{JJ_eqs}
\begin{split}
I &= I_0 \ \sin(\delta) \\
V &= \dfrac{\Phi_0}{2\pi} \ \dot\delta
\end{split}
\end{equation}

It can be shown that these equations correspond to a dissipation-less non-linear inductor, with  effective inductance:

\begin{equation} \label{JJ_effIncuctance}
\begin{split}
V = L  \dfrac{dI}{dt} \Rightarrow L = \dfrac{I_0 \ \cos(\delta) \ \dot \delta}{\frac{\Phi_0}{2\pi} \ \dot\delta} = \dfrac{\Phi_0}{2\pi I_0 \ \cos(\delta) } \equiv \dfrac{L_{J_0}}{\cos(\delta)}
\end{split}
\end{equation}

This effective inductance can be expressed in terms of the Josephson inductance at zero current $L_{J_0} = \frac{\Phi_0}{2\pi I_0}$. Sometimes it is useful to explicitly show its dependence with the current traversing the junction: $$L = \dfrac{L_{J_0}}{\sqrt{1-(\frac{I}{I_0})}}$$

from which it can be seen that  the closer the current to the critical one, the bigger the inductance of the Josephson junction is.

\subsection{Cooper pair box}

\begin{figure}[h]
\centering
\includegraphics[width=0.2\linewidth]{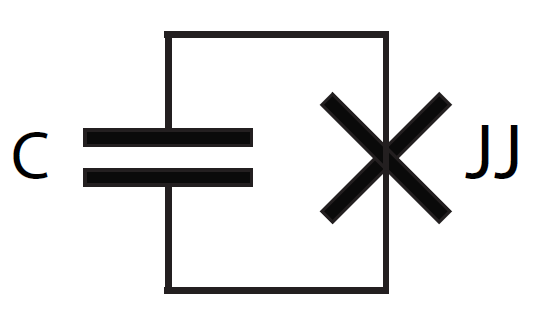}
\caption{Cooper pair box: the simplest non linear superconducting circuit.}
\label{fig:Transmon_circuit}
\end{figure}

The toolbox of electronic devices when working with superconducting circuits is limited. Of course, resistances have no sense in this regime, so the basic elements remaining are capacitors and inductors. However, a new element appears thanks to superconductivity: the non-linear inductor, also called Josephson junction. To construct a two-level system with these three elements, the most simple configuration is a loop with a Josephson junction and a capacitor. This is very similar to an LC circuit, which behaves like a harmonic oscillator. In the qubit case, the inductance is substituted by the Josephson junction because a non-linear element (anharmonic oscillator) is needed. This superconducting circuit is called a Cooper pair box and is schematically seen in figure~\ref{fig:Transmon_circuit}.

This system can be modelled adding up the energies of the two components. The energy stored in a capacitor is $E_{cap} = \frac{Q^2}{2C}$ and expressing the charge in terms of number of Cooper pairs, $m$, $Q = 2em$ so $E_{cap} = \frac{4e^2m^2}{2C} \equiv 4 \ E_C \ m^2$.
The charging energy $$E_C = \dfrac{e^2}{2C}$$ will be crucial in the design and fabrication of the qubit.

The energy of the Josephson junction can be computed as the sum of changes in the product $V \ I$, as $dU/dt = V \ I$.
$$U_{JJ} = \int_{-\infty}^{t} I(t)V(t) \ dt =  \int_{-\infty}^{t} I_0 \sin(\delta) \frac{\Phi_0}{2\pi} \dot \delta \ dt = \dfrac{I_0 \Phi_0}{2\pi} [-\cos(\delta)]_{-\infty}^{t} = \dfrac{I_0 \Phi_0}{2\pi} \ [1 - \cos(\delta)] \equiv E_J [1-\cos(\delta)]$$

Assuming that no current ($\delta = 0$) was present at time $t = - \infty$, the Josephson energy is given by: $$ E_J = \dfrac{I_0 \Phi_0}{2\pi} = \dfrac{\hbar I_0}{2e} $$

The Hamiltonian of the circuit can then be written as:

\begin{equation} \label{eq:H_CPB}
\begin{split}
H_{CPB} = 4 E_C m^2 + E_J [1-\cos(\delta)]
\end{split}
\end{equation}

Experimentally it has been shown that the number of carriers is shifted by an offset charge, $n_g$, present in the system, commonly included in the Hamiltonian: 

\begin{equation} \label{eq:H_CPB_ng}
\begin{split}
H'_{CPB} = 4 E_C (m - n_g)^2 + E_J [1-\cos(\delta)]
\end{split}
\end{equation}

This makes the Cooper pair box highly sensitive to charge fluctuations.

\begin{figure}
    \centering
    \includegraphics[width=0.7\linewidth]{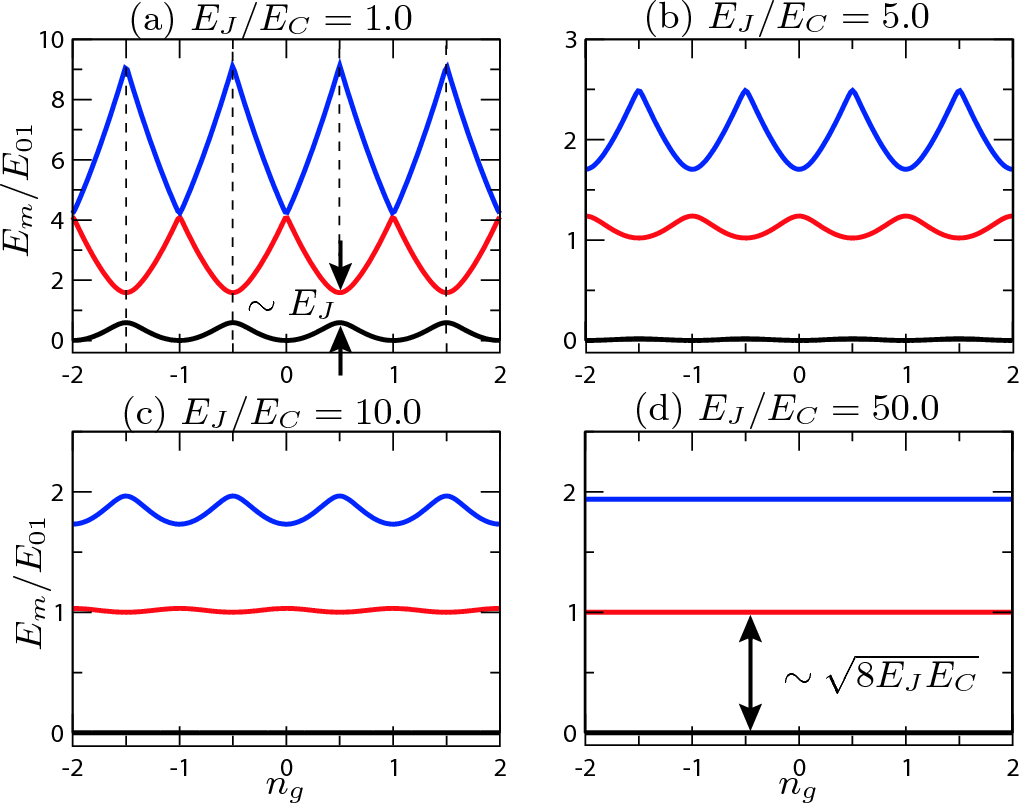}
    \caption{Eigenenergies $E_m$ for the first three levels, m = 0, 1, 2,  of the qubit Hamiltonian \ref{eq:H_CPB_ng} as a function of the effective offset charge $n_g$ for different ratios $E_J$/$E_C$. The higher the ratio, the less sensitive the system to fluctuations in the charge offset. Energies are given in units of the transition energy $E_{01}$, evaluated at the degeneracy point $n_g$ = 1/2. Image extracted from \cite{koch2007charge}.}
    \label{fig:CPBEnergyLevels}
\end{figure}

In order to avoid this limitation, transmons were developed by shunting the Josephson junction with a relatively large capacitance, so that $E_J/E_C \gg 1$, while the Cooper pair box is in the regime $E_J \sim E_C$. This reduces the sensitivity of energy levels to charge noise while maintaining a reasonable anharmonicity, as can be seen in figure \ref{fig:CPBEnergyLevels}.

\subsection{Transmon}

In order to explicitly show that a transmon can behave as a qubit, this two level system has to be derived from its Hamiltonian. The strategy for this consists in quantizing the Hamiltonian, promoting the variables $m$ and $\delta$ to operators $\hat{m}$ and $\hat{\delta}$, and solving the system in order to obtain the eigenvalues and eigenvectors. 

The Hamiltonian \ref{H_CPB} can be solved in the phase basis in terms of the Mathieu functions. But it is also very illustrative to solve it by truncation of the Taylor expansion in $\hat{\delta}$ in the charge basis. Both operators are conjugates, so their basis are the Fourier transform of each other.

In any case, the equation that has to be solved is $$ H \ \Psi_n(\delta) = E_n \ \Psi_n(\delta)$$

Denoting the basis of phase eigenvectors as $\{|\delta \rangle \}$ and the basis of charge eigenvectors as $\{|m \rangle \}$, the task here is to express the operators that form $H$, so $\hat{\delta}$ and $\hat{m}$, in a convenient way to be able to express the result when acting over vectors in the basis. 

Operators in phase basis $\{|\delta \rangle \}$: 
\begin{equation*} 
\begin{split}
\hat{\delta} | \delta \rangle &= |\delta \rangle\\
\hat{m} | \delta \rangle &= -i \dfrac{\partial}{\partial\delta } |\delta \rangle
\end{split}
\end{equation*}

Operators in charge basis $\{|m \rangle \}$: 
\begin{equation*} 
\begin{split}
e^{i\hat{\delta}} | m \rangle &= |m+1 \rangle \\
\hat{m} | m \rangle &= m |m \rangle
\end{split}
\end{equation*}

In this last case, the operator needed is $\cos(\hat{\delta}) = (e^{i\delta} +e^{-i\delta})/2$.
To prove this result, it is sufficient to express the charge state $|m\rangle$ as the Fourier transform of the respective phase states: $$ |m\rangle = \dfrac{1}{\sqrt{2\pi}} \int_0^{2\pi} e^{i \delta m} |\delta\rangle d\delta$$

resulting in  $$e^{i\hat{\delta}} | m \rangle  = \dfrac{1}{\sqrt{2\pi}} \int_0^{2\pi} e^{i \delta m} e^{i\hat{\delta}}|\delta\rangle \ d\delta = \dfrac{1}{\sqrt{2\pi}} \int_0^{2\pi} e^{i \delta m} e^{i\delta}|\delta\rangle \ d\delta = \dfrac{1}{\sqrt{2\pi}} \int_0^{2\pi} e^{i \delta (m+1)} |\delta\rangle \ d\delta = |m+1\rangle$$

With these relations, operands can be expressed as matrices in a certain basis. In this case, the charge basis is more interesting as it is discrete, and therefore
easier to implement in a computer program. 

$$\hat{m} = \sum_m m |m \rangle \langle m | = \bar{m} \mathbf{I}$$
$$\cos(\hat{\delta}) = \frac{1}{2} \ (e^{i\hat{\delta}} + e^{-i\hat{\delta}}) = \frac{1}{2} \ \left( \sum_m |m+1\rangle\langle m| + |m-1\rangle\langle m| \right) = \frac{1}{2} \ ( \mathbf{I_{-1}} + \mathbf{I_1})$$

With 

\begin{alignat*}{2}
\mathbf{I_{-1}} =
\begin{pmatrix}
0 &  & & &  &0 \\
1 & 0 & & & & \\
& 1 &\ddots & & &  \\
 & & \ddots& \ddots & & \\
 &  & & 1 & 0 & \\
0 &  &  & & 1 & 0
\end{pmatrix}
&
\qquad
&
\mathbf{I_{1}} =
\begin{pmatrix}
0 & 1 & & &  &0 \\
 & 0 & 1 & & & \\
&  &\ddots &\ddots & &  \\
 & & & \ddots & 1 & \\
 &  & &  & 0 & 1\\
0 &  &  & &  & 0
\end{pmatrix}
\end{alignat*}

At the end, $\hat{H}$ is a tridiagonal matrix.

\begin{equation} \label{H_CPB}
\begin{split}
H_{CPB} = 4 E_C \hat{m}^2 + E_J [1-\cos(\hat{\delta})] = 4 E_C \bar{m^2} \ \mathbf{I} \ + E_J [\mathbf{I}- \frac{1}{2} \ ( \mathbf{I_{-1}} + \mathbf{I_1})]
\end{split}
\end{equation}

\begin{figure}[h]
\centering
\includegraphics[width=0.8\linewidth]{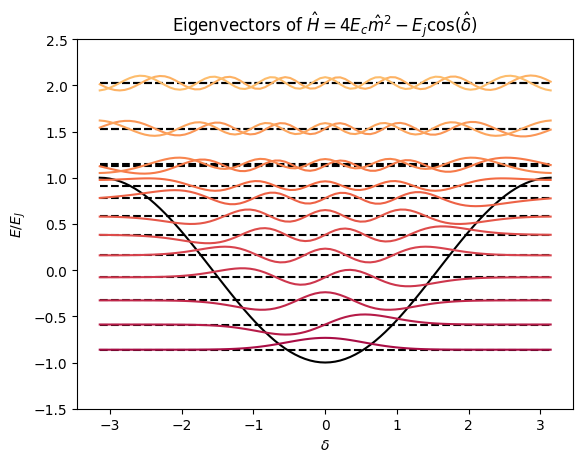}
\caption{\label{fig:HeigenVectors} Eigenvectors for a Cooper pair box Hamiltonian in the transmon regime ($E_J = 20 GHz$, $E_C = 0.2 GHz$, $n_g = 0$). In solid black, the  $ \cos(\delta)$ potential and in dashed lines $En/E_J$ eigenvalues.}
\end{figure}

$\bar{m}$ is a vector with all possible values for $m$ (and $\bar{m^2}$ with all possible values squared), in the most general framework it spans from $-\infty$ to $\infty$ and so all matrices have infinite dimension. But this model can be truncated without risk of losing information in the lowest energy levels. For example, for figure \ref{fig:HeigenVectors},  91 levels were considered, with $m$ ranging from -45 to 45, but not all of them are displayed.

\subsection{Transmon as a qubit}

The eigenvectors plotted in \ref{fig:HeigenVectors} have been computed with values for $E_J$ and $E_C$ in the regime of transmon behaviour, but the expressions are also valid for the general Cooper pair box Hamiltonian. This system has many energy levels, what in principle makes it more complex than a qubit. But thanks to the anharmonicity, transitions between energy levels are different, so individual levels can be addressed. If only the first two are considered, the transmon can operate as a qubit.

The Cooper pair box Hamiltonian in the transmon regime ($E_J/E_C \gg 1$) allows a reshaping of the expression in terms of a Taylor expansion of $\cos(\delta)$ because $\delta \ll 1$ in this regime. 

$$\cos x = \sum_{n=0}^{\infty} \frac{(-1)^n}{(2n)!} x^{2n} = 1 - \frac{x^2}{2!} + \frac{x^4}{4!} - \frac{x^6}{6!} + \cdots =  1 - \frac{x^2}{2} + \frac{x^4}{24} - \frac{x^6}{720} + \cdots
$$

Which makes the Hamiltonian: 

\begin{equation} \label{H_CPB_expanded}
\begin{split}
H_{CPB} = 4 E_C m^2 + E_J [1-\cos(\delta)] = 4 E_C m^2 + E_J [\frac{\delta^2}{2} - \frac{\delta^4}{24} + \frac{\delta^6}{720} + \cdots]
\end{split}
\end{equation}

\begin{figure}[h]
    \centering
    \includegraphics[width=0.8\linewidth]{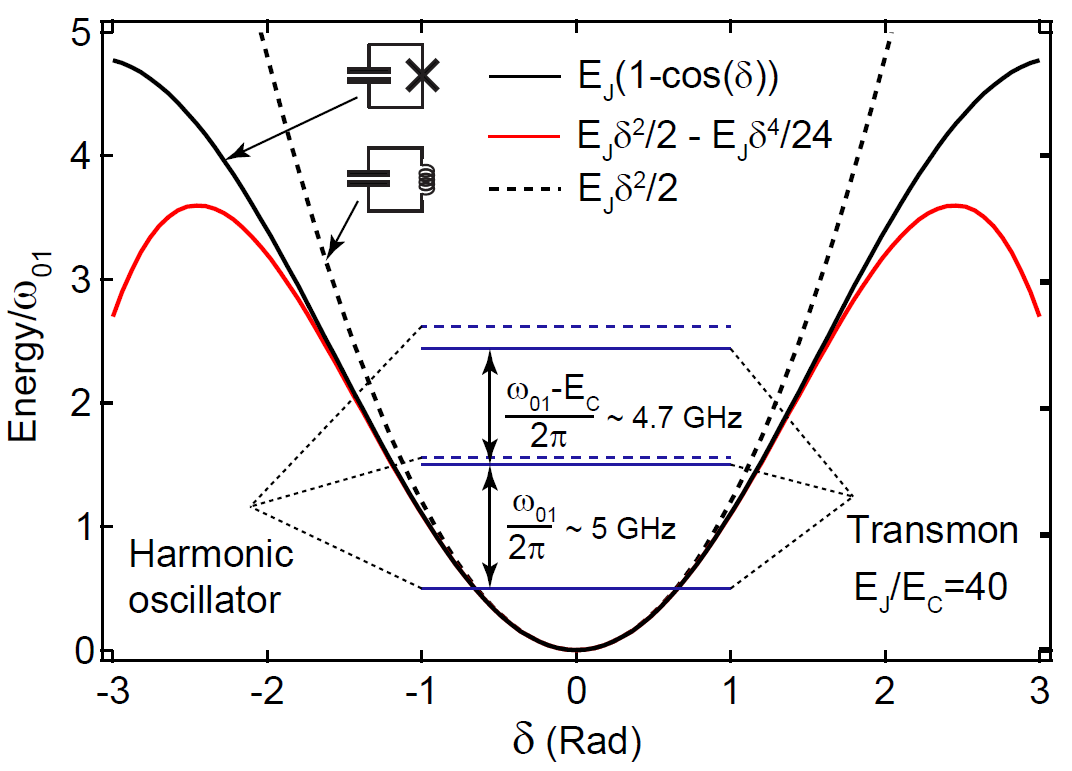}
    \caption{Transmon energy levels in solid blue compared with harmonic oscillator levels in dashed blue. The potentials for transmon (solid black), non linear oscillator (solid red) and harmonic oscillator (dashed black) are superimposed. For small oscillations around $\delta = 0$ all three are very close. Graph from \cite{naghiloo2019introduction}.  }
    \label{fig:TransmonModelAproximations}
\end{figure}

This Taylor expansion is very convenient in order to efficiently compute energy levels of a system in which a transmon is involved. 

\subsection{Qubit state representation in the Bloch sphere}

Considering only the first two energy levels, as they are distinguishable, the system is a good realization of a quantum  two-level system and therefore can be called qubit, the "quantum bit". Then, $|g\rangle = 0$ and $|e\rangle = 1$. The Hamiltonian of a qubit in a laboratory reference frame can be written as

\begin{equation} \label{QubitH}
\hat{H}_0=\hbar\left(\begin{array}{cc}
\frac{\omega_{01}}{2} & 0 \\
0 & -\frac{\omega_{01}}{2}
\end{array}\right)=\frac{\hbar \omega_{01}}{2} \hat{\sigma}_z
\end{equation}\\

where $\omega_{01}$ is the qubit transition frequency, and $\sigma_z$ is the Pauli matrix. The eigenstates of the qubit Hamiltonian, corresponding to the two energy levels, form a natural basis $\{|0\rangle$,$|1\rangle\}$ in a Hilbert space, and therefore $\hat{H}_0 |0\rangle = -\frac{\omega_{01}}{2}|0\rangle$ and $\hat{H}_0 |1\rangle = \frac{\omega_{01}}{2}|1\rangle$.\\

In a quantum system, pure states and mixed states can be considered \cite{Ufano1957description}. Pure states are linear combinations of the vectors of the basis with norm 1. In general, $\Psi = \sum_{n}c_n u_n$, with $c_n \in \mathbb{C}$, $\sum c_nc_n^* = 1$ and $u_n \in \{basis\}$, so in our two-level quantum system a pure state can be written as:

$$|\Psi\rangle = \alpha|0\rangle + \beta |1\rangle$$

with $\alpha\alpha^*+\beta\beta^*=1$. 

A mixed state is a statistical mixture of pure states:

%$$|\Phi\rangle = \sum_i p^i |\Psi\rangle^i$$
$$|\Phi\rangle = \{( p^i |\Psi\rangle^i)\}$$

with $0 \leq p^i \leq 1 \in  \mathbb{R}$ and $\sum p^i=1$. Hence, each pure state $|\Psi\rangle^i$ has a probability of occurrence of $p^i$.
In most of the cases through this work, pure states are considered but it is useful to have some insight of all possible quantum states available.

Quantum states of a two-level system can be conveniently represented as vectors on a 3D sphere. To understand how this representation arises, one must use density matrices. The general form of the density matrix for a quantum state is: 

$$\hat{\rho} = \sum_i p^i|\Psi\rangle^i\langle\Psi|^i $$

being $\{|\Psi\rangle^i\}$ pure states and $p^i$ their probabilities.

In the case of a two-level system this is a 2x2 Hermitian matrix with the following properties:

\begin{itemize}
    \item Tr($\hat{\rho}$) = 1. This follows from their meaning as probabilities, the sum of all possible outcomes has to add 1.   
    \item Tr($\hat{\rho}^2$) $\leq$ 1. The equality holds for pure states, the inequality for mixed states.
    \item $\hat{\rho}^2 = \hat{\rho} \, \iff \, $ Pure state.
\end{itemize}

In a qubit system, for a pure state like $|\psi\rangle$ the qubit density matrix looks like:

\begin{equation} \label{DensM}
\hat{\rho} = |\Psi\rangle\langle\Psi| =  \begin{pmatrix}
\alpha^2 & \alpha\beta^* \\
\alpha^*\beta & \beta^2
\end{pmatrix}
\end{equation}

In the Hilbert space of Hermitian 2×2 matrices the Pauli matrices $(\hat{\sigma}_x,\hat{\sigma}_y,\hat{\sigma}_z)$ form a basis and the qubit density matrix can be decomposed in this basis as

\begin{equation}
\rho=\frac{1}{2}\left(\hat{I}+X \sigma_x+Y \sigma_y+Z \sigma_z\right)=\frac{1}{2}\left(\begin{array}{cc}
1+Z & X-i Y \\
X+i Y & 1-Z
\end{array}\right)
\end{equation}

With X,Y, Z $\in \mathbb{R}$ and $\hat{I}$ the identity operator. Applying the second property listed for density matrices: 

\begin{equation}
\begin{aligned}
\operatorname{Tr}\left(\hat{\rho}^2\right) & =\frac{1}{4} \operatorname{Tr}\left(\begin{array}{cc}
(1+Z)^2+X^2+Y^2 & 2(X-i Y) \\
2(X+i Y) & X^2+Y^2+(1-Z)^2
\end{array}\right)= \\
& =\frac{1}{2}\left(1+X^2+Y^2+Z^2\right)=\frac{1}{2}\left(1+|\vec{V}|^2\right) \leq 1  \implies |\vec{V}|^2 \leq 1
\end{aligned}
\end{equation}

\begin{figure}[h!]
    \centering
    \includegraphics[width=0.8\linewidth]{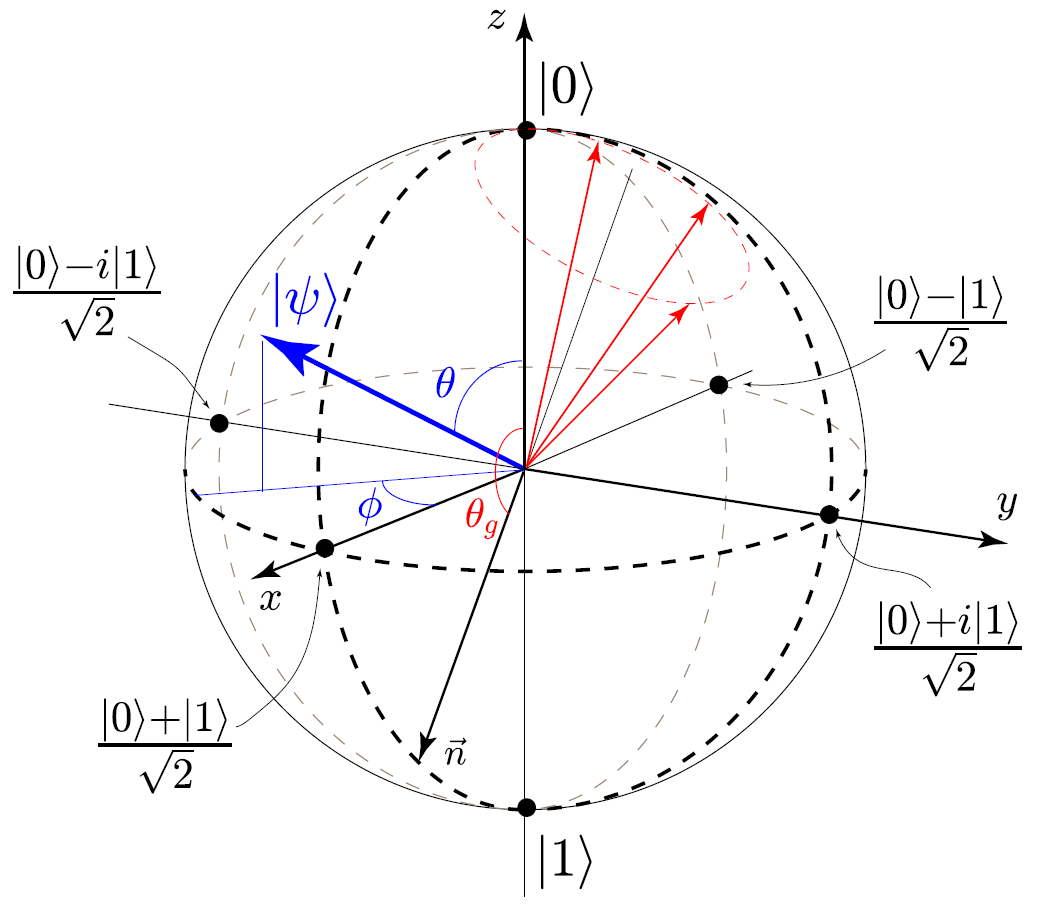}
    \caption{\textbf{Bloch sphere}. A qubit state can be represented as a vector in this sphere. For example, a pure state with the probability amplitudes $\alpha =\cos \frac{\theta}{2}$ and $\beta = e^{i\phi} \sin \frac{\theta}{2}$ corresponds to a unit vector with the polar angle $\theta$ and azimuthal angle $\phi$ shown in blue. The qubit state vectors shown in red represent the unitary evolution of the qubit  during the Rabi oscillations when the state vector rotates around the unit vector $\vec{n} = (\sin \theta_g,0,\cos \theta_g)$ given by the Rabi amplitude $\Omega$ and detuning $\Delta$. Image from \cite{danilin2018experiments}.}
    \label{fig:BlochSphere}
\end{figure}

This maps every qubit state $|\Psi\rangle$ into a 3D vector $\vec{V}$ that lives inside the unitary sphere, the so called \textit{Bloch sphere}, seen in figure \ref{fig:BlochSphere}. If it is a pure state, the equality holds and then the vector lies in the surface of the sphere.
The relation between parameters $\alpha, \,\beta$ and X, Y, Z can be easily seen expressing them in polar coordinates:

\begin{equation*}
X=\sin \theta \cos \phi, \quad Y=\sin \theta \sin \phi, \quad Z=\cos \theta
\end{equation*}

And so, the density matrix is:

\begin{equation} \label{DesMPolar}
\hat{\rho}=\frac{1}{2}\left(\begin{array}{cc}
1+\cos \theta & \sin \theta e^{-i \phi} \\
\sin \theta e^{i \phi} & 1-\cos \theta
\end{array}\right)=\left(\begin{array}{cc}
\cos ^2 \frac{\theta}{2} & \cos \frac{\theta}{2} \sin \frac{\theta}{2} e^{-i \phi} \\
\cos \frac{\theta}{2} \sin \frac{\theta}{2} e^{i \phi} & \sin ^2 \frac{\theta}{2}
\end{array}\right) 
\end{equation}

Comparing equations \ref{DensM} and \ref{DesMPolar}, pure state coefficients can be written as $\alpha = \cos{\frac{\theta}{2}}$ and $\beta = e^{i\phi}\sin{\frac{\theta}{2}}$, and therefore the state can be expressed in terms of these polar coordinates:

\begin{equation} 
|\Psi\rangle = \alpha|0\rangle + \beta |1\rangle = \cos{\frac{\theta}{2}} |0\rangle + e^{i\phi}\sin{\frac{\theta}{2}} |1\rangle
\end{equation}

The eigenstates $|0\rangle$ and $|1\rangle$ are at the poles of the sphere and all states in the equator  have exactly the same probability to measure $|0\rangle$ or $|1\rangle$.

\begin{figure}[h!]
    \centering
    \includegraphics[width=0.8\linewidth]{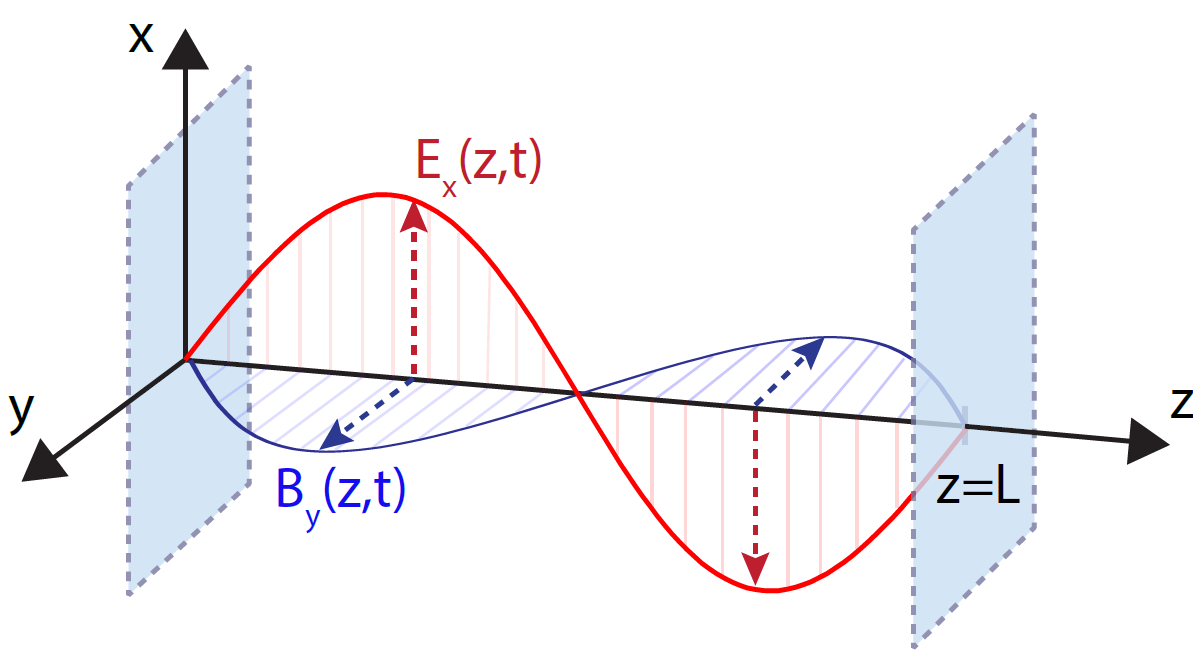}
    \caption{Electric and magnetic field standing waves in a one-dimensional cavity. The two walls -parallel and infinite- are separated by a distance L. Electromagnetic fields are only functions of Z. Here, is shown the second static mode. Image from \cite{naghiloo2019introduction}.}
    \label{fig:StandingWaveCavity}
\end{figure}

\subsection{Cavity states}

To understand the behaviour of electromagnetic waves in a 3D cavity we will briefly review the 1D model of a standing wave between two perfect conductors.

Consider a pair of infinite perfect conducting walls separated by a distance L along the $Z$ direction. For simplicity, it is assumed that the electric field is polarized along $X$ axis which implies that
the magnetic field is only along the $Y$ axis, therefore Maxwell's equations are:

\begin{subequations} \label{eq:maxwell}
\begin{align}
    \nabla \times \vec{E} = -\frac{\partial \vec{B}}{\partial t} 
    \quad &\longrightarrow \quad \frac{\partial E_x (z,t)}{\partial z} = -\frac{\partial B_y (z,t)}{\partial t} \label{eq:maxwell1} \\
    \nabla \times \vec{B} = \varepsilon_0 \mu_0 \frac{\partial \vec{E}}{\partial t} 
    \quad &\longrightarrow \quad \frac{\partial B_y (z,t)}{\partial z} = \varepsilon_0 \mu_0 \frac{\partial E_x (z,t)}{\partial t} \label{eq:maxwell2} \\
    \nabla \cdot \vec{E} = 0 
    \quad &\longrightarrow \quad \frac{\partial E_x (z,t)}{\partial x} = 0 \label{eq:maxwell3} \\
    \nabla \cdot \vec{B} = 0 
    \quad &\longrightarrow \quad \frac{\partial B_y (z,t)}{\partial y} = 0 \label{eq:maxwell4}
\end{align}
\end{subequations}

Perfect conducting walls imply that the electric field vanishes at the boundaries: $E_x(z , t) = 0$ for $z = 0, \, L$. So the solutions for electric and magnetic fields are: 

\begin{subequations} 
\begin{align}
    & E_{x}(z,t)=\mathcal{E}q(t)\sin(kz) \\
    & B_{y}(z,t)=\mathcal{E}\frac{\mu_{0}\varepsilon_{0}}{k}\dot{q}(t)\cdot\cos(kz)
\end{align}
\end{subequations}

The function $q(t)$ is the canonical position function and describes the time-evolution of the modes and has dimensions of length. The normalization constant $\mathcal{E}$ is $\mathcal{E} = \sqrt{\frac{2\omega_c^2}{V\epsilon_0}}$ with $V$ the effective volume of the cavity and $\omega_c = \frac{k}{\sqrt{\mu_0 \epsilon_0}}$. The parameter $k = m\pi/L$ with  $m = 1, 2, ...$ is the wave number. Each integer value $m$ corresponds to one mode of the cavity. Figure \ref{fig:StandingWaveCavity} shows the electric and magnetic field for the second mode of the cavity ($m = 2$).

Having canonical position $q(t)$ and momentum $p(t) = \dot{q}(t)$ functions, the energy stored in the standing wave is:

\begin{equation}
    H = \dfrac{1}{2} \left[ p^2(t) +\omega_2^2q^2(t) \right] 
\end{equation}

This Hamiltonian can be quantized by promoting the canonical variables $q(t)$ and $p(t)$ to operators $\hat{q}$, $\hat{p}$, resulting in the Hamiltonian for a quantum harmonic oscillator:

\begin{equation}
    \hat{H} = \dfrac{1}{2} \left[ \hat{p}^2(t) +\omega_2^2\hat{q}^2(t) \right] 
\end{equation}

As a quantum harmonic oscillator, the Hamiltonian can be expressed in terms of annihilation and creation operators for a photon in the corresponding mode of the cavity:

\begin{subequations} \label{eq:ladder}
\begin{align}
    \hat{a} &= \frac{1}{\sqrt{2\omega_c}} (\omega_c \hat{q} + i \hat{p}), \label{eq:a} \\
    \hat{a}^\dagger &= \frac{1}{\sqrt{2\omega_c}} (\omega_c \hat{q} - i \hat{p}). \label{eq:adag}
\end{align}
\end{subequations}

These operators satisfy the commutation relation $\left[ \hat{a}, \hat{a}^\dagger\right] = 1$.

Electric and magnetic field operators can be expressed in the same terms:

\begin{subequations} \label{eq:field_operators}
\begin{align}
    \hat{E}_x (z,t) &= \mathcal{E}_0 (\hat{a} + \hat{a}^\dagger) \sin(kz), \label{eq:Ex} \\
    \hat{B}_y (z,t) &= i \mathcal{B}_0 (\hat{a} - \hat{a}^\dagger) \cos(kz). \label{eq:By}
\end{align}
\end{subequations}

And also the Hamiltonian:

\begin{equation} \label{CavityH}
    \hat{H} = \omega_c(\hat{a}\hat{a}^\dagger +\frac{1}{2}) = \omega_c(\hat{n} +\frac{1}{2})
\end{equation}

being the operator $\hat{n} = \hat{a}\hat{a}^\dagger$ the number operator. Cavity mode states fulfill 

\begin{equation}
    \hat{H}|n\rangle = E_n |n\rangle, \quad n = 0, 1, 2, ...
\end{equation}

Eigenvectors $|n\rangle$ are associated to energy eigenstates for the single mode cavity field with the corresponding energy $E_n = \omega_c(n + 1/2 )$. The set of photon number states $\{ |n\rangle \}$, also called Fock states, form a complete basis for the space of cavity states. An arbitrary pure state of the cavity can be expressed as a linear combination of them $\sum_n c_n |n\rangle$. In the most general case, the state of the cavity is described by a mixed state, an incoherent superposition of Fock states represented by the density matrix $\rho =\sum_n P_n |n\rangle\langle n|$.

\subsection{Qubit cavity interaction}

Qubit realizations like transmons are never isolated, neither is an isolated qubit useful. Transmons interact with external electromagnetic fields like external drives or cavity modes. In general, these interactions are modeled through the following Hamiltonian:

\begin{equation}
    \hat{H}_{int} = -\hat{\vec{d}}\vec{E}
\end{equation}

where $\hat{\vec{d}}$ is the dipole moment operator of the qubit and $\vec{E}$ the electric field. 

The dipole moment operator can be expressed in general as $\hat{\vec{d}} = e\hat{\vec{r}}$. Assuming a dipole parallel to the electric field, let's say in the $X$ direction, it can be written as $\hat{\vec{d}} = d_x \hat{\sigma}_x = d_x(\hat{\sigma}_+ + \hat{\sigma}_-)$, where $\hat{\sigma}_+$ and $\hat{\sigma}_-$  are the raising and lowering operators for the qubit.

For the lowest resonant mode of a cavity, the electric field has a maximum at the centre of the cavity and minima at the walls. This means that for a quantized field in the $X$ direction $E_x(z, t) = E_0(\hat{a} + \hat{a}^\dagger) \sin(z)$, if the qubit is placed at the centre, the electric field is maximum ($\sin(z) = 1$, for $z = L/2$). 

So the Hamiltonian for the interaction with the lowest resonant mode of a cavity and the transmon in the center is:

\begin{equation} \label{IntH}
    \hat{H}_{int} = -d_x E_0 (\hat{a} + \hat{a}^\dagger)(\hat{\sigma}_+ + \hat{\sigma}_-) = g (\hat{a} + \hat{a}^\dagger)(\hat{\sigma}_+ + \hat{\sigma}_-)
\end{equation}

$g= -d_x E_0$ quantifies the interaction strength between the interaction qubit - electric field.

This reasoning can also be applied to interactions with external drives. The electric field will be different, of course, and assuming a classical representation like  $\vec{E} = \vec{E}_0(t) \cos(\omega_d t + \phi)$, the interaction term can be something like $ \hat{H}_{int} = -\dfrac{\vec{d}\vec{E}_0(t)}{2} \hat{\sigma}_x$. Usually, the angular Rabi frequency is introduced here: $\Omega(t) = -\dfrac{\vec{d}\vec{E}_0(t)}{\hbar}$ because it is useful to model qubit manipulation drives. In \cite{danilin2018experiments} there is a full derivation of this expression and a clear view of its utility to model Rabi oscillations and Ramsey fringes.

\subsection{Jaynes-Cummings model}
Once all three pieces have been examined, qubit, cavity and their interaction, the overall hybrid system can be explored. The qubit Hamiltonian \ref{QubitH} has two eigenstates $\{ |g\rangle, |e\rangle\}$ with eigenenergies $\{-\omega_q/2, +\omega_q/2 \}$. The cavity Hamiltonian \ref{CavityH} has infinite eigenvectors $\{ |n\rangle\}$ associated to eigenenergies $\{\omega_c(n+1/2) \}$. They are related to the number of photons in that state. With these two components plus the interaction term, the hybrid system can be described. In the following paragraphs this new Hamiltonian will be examined to highlight the main differences with its bare components:

\begin{equation}
    \hat{H}=\omega_c(\hat{a}^\dagger\hat{a}+\frac{1}{2})+\frac{1}{2}\omega_q\hat{\sigma}_z + g(\hat{a}+\hat{a}^\dagger)(\hat{\sigma}_- + \hat{\sigma}_+)
\end{equation}

The interaction part has four terms: 

$$\hat{H}_{int} = \hat{a}\hat{\sigma}_+ +  \hat{a}\hat{\sigma}_- +  \hat{a}^\dagger\hat{\sigma}_+ + \hat{a}^\dagger\hat{\sigma}_-$$

They describe the creation or annihilation of states in the cavity mode or qubit. For example, the term $\hat{a}\hat{\sigma}_+$ creates a qubit excitation and annihilates a cavity photon. Applying the approximation that is known as Rotating Wave Approximation, two of these terms vanish. This approximation is valid where the coupling strength $g$ is much less than both the qubit and cavity frequency, $g \ll \omega_q, \omega_c$ and also $|\omega_c - \omega_q| \ll |\omega_c + \omega_q|$.

The terms $\hat{a}\hat{\sigma}_+$ and $\hat{a}^\dagger\hat{\sigma}_-$ imply small energy variation in the system as $\pm (\omega_q -\omega_c)$ is very small compared with the total energy of the system, which is of the order of $E_{Total}\sim \omega_q +\omega_c$. For the other two terms, the situation is the opposite: they represent the excitation of the qubit and the creation of a photon, $\hat{a}^\dagger\hat{\sigma}_+$, and the deexcitation of the qubit plus the annihilation of a photon $\hat{a}\hat{\sigma}_-$, making the energy change quite relevant in terms of total energy. These two terms correspond to processes much less likely to happen in the system and they can be neglected.

Now, the Jaynes-Cummings Hamiltonian under the Rotating Wave Approximation is:

\begin{equation} \label{JCHamiltonian}
    \hat{H}_{JC}=\omega_c(\hat{a}^\dagger\hat{a}+\frac{1}{2}) +\frac{1}{2}\omega_q\hat{\sigma}_z +g(\hat{a}\hat{\sigma}_+ + \hat{a}^\dagger\hat{\sigma}_-)
\end{equation}

The matrix representation of this Hamiltonian has infinite dimensions because the number of photons in the cavity is not bound. At some point, for practical use, it has to be truncated, but in this case a general form can be obtained in the bare state basis. This basis is formed by the combination of states of the two independent systems: $\{|g,n\rangle, |e, n\rangle\}$ for $n=0,1,2 ...$. In this basis, the matrix representation of the Hamiltonian is extracted element by element applying the Hamiltonian over pairs of elements of the basis (it is a bilinear form). For example, the element $H_{11}= \langle g, 0| \hat{H} | g, 0 \rangle$ or $H_{23}= \langle g, 1| \hat{H} | e, 0 \rangle$. 

\begin{equation} \label{JCMatrix}
H_{JC}=\begin{blockarray}{l}
         \quad\quad |g,0\rangle   \quad\quad\quad |g, 1\rangle \quad\quad\quad |e, 0\rangle \quad\quad\quad \cdots \quad\quad\quad |g, n\rangle \quad\quad\quad\quad  |e, n-1\rangle \\[15pt] 
         \left(
\begin{array}
{cccccc}\frac{1}{2}\omega_{\mathrm{c}}-\frac{\omega_{\mathrm{q}}}{2} & 0 & 0 & 0 & 0 & 0 \\
0 & \frac{3}{2}\omega_{\mathrm{c}}-\frac{\omega_{\mathrm{q}}}{2} & -g & 0 & 0 & 0 \\
0 & -g & \frac{3}{2}\omega_{\mathrm{c}}+\frac{\omega_{\mathrm{q}}}{2} & 0 & 0 & 0 \\
 & &  & \ddots & \\
0 & 0 & 0 & 0 & (n+\frac{1}{2})\omega_{\mathrm{c}}-\frac{\omega_{\mathrm{q}}}{2} & -\sqrt{n+1}g \\
0 & 0 & 0 & 0 &-\sqrt{n+1}g & (n+\frac{1}{2})\omega_{\mathrm{c}}+\frac{\omega_{\mathrm{q}}}{2}
\end{array}\right)\end{blockarray}\quad\end{equation}

This matrix is diagonal by blocks of $2 \times 2$ with the following general from: 

\begin{equation}\left.M_n=\left(
\begin{array}
{cc}(n+\frac{1}{2})\omega_c-\frac{\omega_q}{2} & -\sqrt{n+1}g \\
-\sqrt{n+1}g & (n+\frac{1}{2})\omega_c+\frac{\omega_q}{2}
\end{array}\right.\right)\end{equation}

So this decomposition allows to extract eigenvalues $E_{n\pm}$ easily by diagonalizing every $M_n$. 

Eigenvalues of $H_{JC}$:

\begin{subequations} \label{eq:eigenvalues}
\begin{align}
    E_{g} &= -\frac{\Delta}{2} \label{eq:Eg} \\
    E_{\mp}& =(n+1) \, \omega_{c}\mp\frac{1}{2}\sqrt{4g^{2}(n+1)+\Delta^{2}} \label{eq:Emp}
\end{align}
\end{subequations}

with $\Delta = \omega_q - \omega_c$. 

Transitions between energy levels can be measured in experimental setups, what makes these expressions of the uttermost interest to compare theory and reality. 

\begin{figure}[h]
    \centering
    \includegraphics[width=0.6\linewidth]{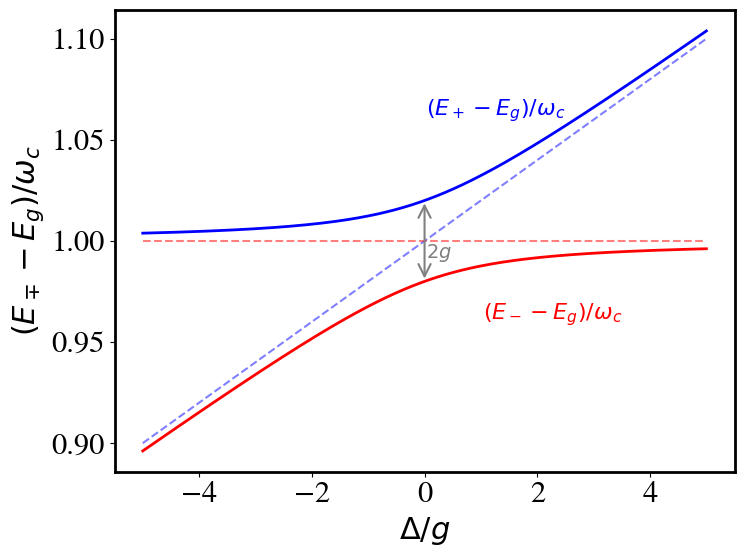}
    \caption{Avoided crossing in a hybrid qubit-cavity system. In dashed lines, the behaviour of the modes without qubit-cavity interaction. For $\Delta=0$ the distance between branches is $2g$.}
    \label{fig:AvoidedCrossing}
\end{figure}

\begin{figure}
   \begin{minipage}[t]{0.49\textwidth}
     \centering
     \includegraphics[width=0.95\linewidth]{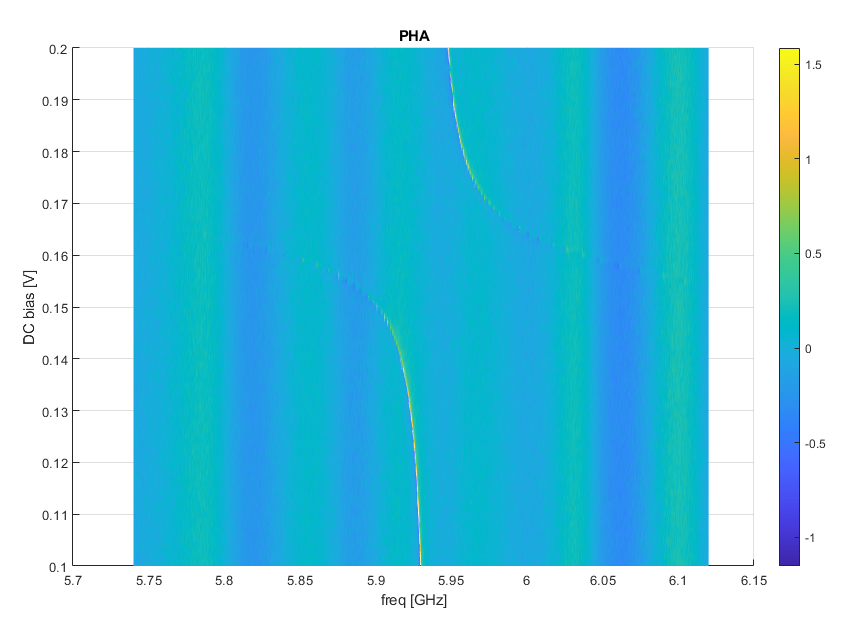}
 
   \end{minipage}\hfill
   \begin{minipage}[t]{0.49\textwidth}
     \centering
     \includegraphics[width=0.85\linewidth]{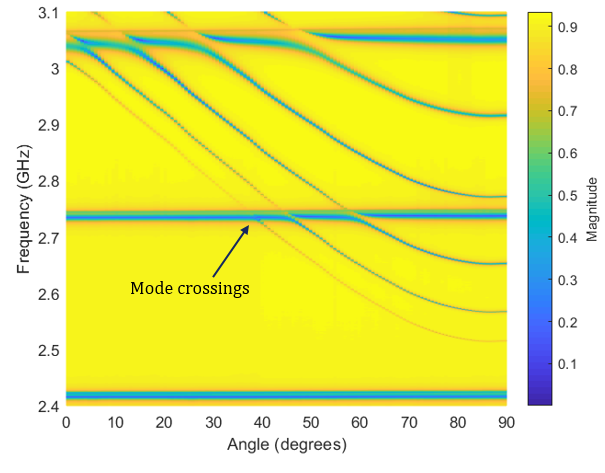}
    
   \end{minipage}
   \caption{\textit{Left}: Thin sharp lines from an avoided crossing in a flux tunable transmon in a frequency versus flux bias voltage plot. \textit{Right}: Many avoided crossings in a tunable cavity with mechanical plates. Frequency in the vertical axis, angle of the plates in the horizontal one. Different modes behave different, three modes can be seen that remain fixed. }\label{fig:AvoidedCrossings}
\end{figure}

The hybridization between the two modes makes them interchange behaviour with respect to detuning. The minimum distance between them is achieved at $\Delta = 0$ and it is $2g$, so it gives a way to measure the coupling. This is represented in figure \ref{fig:AvoidedCrossing}. In order to be able to observe this behaviour in a real set-up, one of the two elements has to change its characteristic frequency to sweep the detuning. Both options are possible, the qubit frequency varies in flux tunable transmons, in which a current is applied next to the Josephson junction and the magnetic field that it generates, modifies the qubit frequency. Cavity modes are easier to tune. Axion haloscope searches are based on this, most of them rely in movable parts that smoothly swift the frequency of the resonance of the cavity. See figures in \ref{fig:AvoidedCrossings} for both cases.

\textbf{Dressed states}

In previous paragraphs, the eigenvalues of Jaynes-Cummings Hamiltonian \ref{JCMatrix} have been discussed. Eigenvectors can be derived also from the matrix expression. As it is a block matrix, every eigenvector can be expressed in terms of the bare states basis and each one has only a maximum of 2 components. This is due to the size of the blocks, $2\times 2$. 
For each eigenvalue its eigenvector is:

\begin{subequations} \label{eq:Eigenvectors}
\begin{align}
    E_g &\longrightarrow \quad |0, -\rangle = |g\rangle |0\rangle \label{eq:Nvector0} \\
    E_+ &\longrightarrow \quad |n, +\rangle = \sin(\theta_n)|g\rangle |n+1\rangle + \cos(\theta_n)|e\rangle |n\rangle \label{eq:Nvector+} \\
    E_- &\longrightarrow \quad |n, -\rangle = \cos(\theta_n)|g\rangle |n+1\rangle - \sin(\theta_n)|e\rangle |n\rangle \label{eq:Nvector-} 
\end{align}
\end{subequations}

with $\theta_n = \frac{1}{2}\tan^{-1}(2g\sqrt{n+1}/\Delta)$ as the ``degree of hybridization". This parameter weights the contribution of each of the bare state combinations to the final hybrid state. When the cavity and qubit frequency are very close, $\Delta \rightarrow 0$, maximum hybridization is achieved: $\theta_n = \pi/4$ and the contribution of each bare state to the dressed state is $0.5$. These states are called \textit{polaritons} and have an energy difference of $2g$. In figure \ref{fig:AvoidedCrossing} these states are in the centre of the plot and the energy difference is labelled in grey. 

\subsection{Dispersive regime}

A transmon is a specific realization of a Cooper Pair Box. As mentioned before, what makes a transmon interesting is its suppression of charge noise thanks to $E_J/E_C \gg 1$. When coupling the transmon to a resonator, another parameter enters into play, $g$, the strength of this coupling. Our device will work in a regime known as ``dispersive regime", in which the coupling between both elements is weak. This condition is fulfilled when $\Delta/g \gg 1$, when the detuning is large compared with the coupling. In the dispersive regime, dressed states are very close to the independent qubit and cavity states. 

Under this condition, it has been shown \cite{koch2007charge} that in the dispersive limit the resonator-transmon interaction term in equation \ref{JCHamiltonian} can be Taylor expanded and truncated to lowest order. Then, the Hamiltonian for a two-level transmon takes the form:

\begin{equation} \label{eq:HamiltonianJC}
    \hat{H} = \omega_c' \hat{a}^\dagger\hat{a} +\dfrac{1}{2}\omega_q' \hat{\sigma_z} +\chi \hat{a}^\dagger\hat{a}\hat{\sigma_z} = \dfrac{1}{2}\omega_q' \hat{\sigma_z} + (\omega_c' + \chi\hat{\sigma_z})\hat{a}^\dagger\hat{a}
\end{equation}

Where $\chi$ is the dispersive shift and frequencies are not the bare frequencies any more, they have been displaced: $\omega_q' = \omega_q +\frac{g^2}{\Delta}$ and $\omega_c' = \omega_c -\dfrac{2g^2}{\Delta - E_C/2}$. The overall dispersive shift is 

\begin{equation} \label{eq:chi}
    \chi = \dfrac{g^2E_C/2}{\Delta(\Delta - E_C/2)}
\end{equation}

This expression shows that the state of the qubit shifts the resonance of the cavity through the term $(\omega_c' + \chi\hat{\sigma_z})$. Eigenvalues for $\hat{\sigma_z}$ are $\pm 1$, so the shift in cavity frequency depending on the state of the qubit is $2\chi$. Therefore, by measuring this frequency shift for the cavity, the state of the qubit can be identified.

Remember that anharmonicity is $\alpha = -E_C $ so the measured dispersive shift is related to anharmonicity, qubit-cavity coupling $g$ and detuning $\Delta=\omega_c - \omega_q$. These are all parameters that depend on the qubit and cavity designs.

\section{Qubit state manipulation}

Bringing all these quantum models to reality is a hard job, because reality tends to be more complex than our most elaborated models. Firstly, our ``two-level" system is never a pure two-state system. Many other levels are present in a transmon and even spurious states are possible due to impurities in the substrate or the bare material of the Josephson junction. Secondly, the temperature is not absolute zero and, despite the great performance of dilution refrigerators, any other temperature makes the states unstable. Interacting with the qubit is also a  quite ``dirty" procedure. Microwave photons have to be generated outside and far from the system, their frequency spectrum is never completely perfect and cables, attenuators and other devices along the line bring the nuisances of life very present in our experiment. All this, without mentioning the cost of all the needed microwave equipment, which is the reason behind many interesting tricks to improve the behaviour of these systems with the materials available in the lab.

In order, the topics discussed in this section will be: (i) Basic techniques for signal generation in the range of GHz: AWG, mixers, upconversion and downconversion... (ii) Resonators and couplings. (iii) Pulse sequences to characterize the transmon: frequency spectrum, Rabi oscillations, Ramsey fringes...

\subsection{Signal generation: mixers and sidebands}

\begin{figure}[h]
    \centering
    \includegraphics[width=0.9\linewidth]{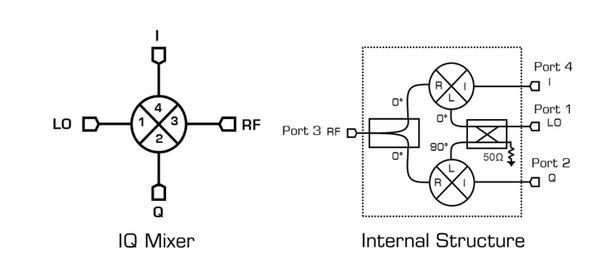}
    \caption{IQ mixer symbol and inner schema. Image from \cite{Marki}.}
    \label{fig:MixerSchema2}
\end{figure}

 The two-level system in a transmon is achieved through different energy spacing between levels, which allows to address single transitions without stimulating other energy jumps. This ability of interacting with one and not other levels relies on the quality of the signal used to stimulate the transition. Arbitrary waveform generators (AWG) are the devices used to send the signal to the system. These devices have certain properties regarding the purity of the signal they can create and, of course, they are not perfect. For our purposes, the frequency spectrum is what matters here. If only one transition should be addressed, power sent in other transition frequencies should be minimized. Spurious noise present throughout the spectrum is generally suppressed by at least 40 to 50 dB, but quantum devices are so extremely sensitive, that other strategies are set in place to further improve the purity of the signal.

\vspace{10ex}

\subsubsection{Sideband generation with mixers}

\begin{wrapfigure}[16]{R}{0.4\textwidth}
    \centering
    \includegraphics[width=0.9\linewidth]{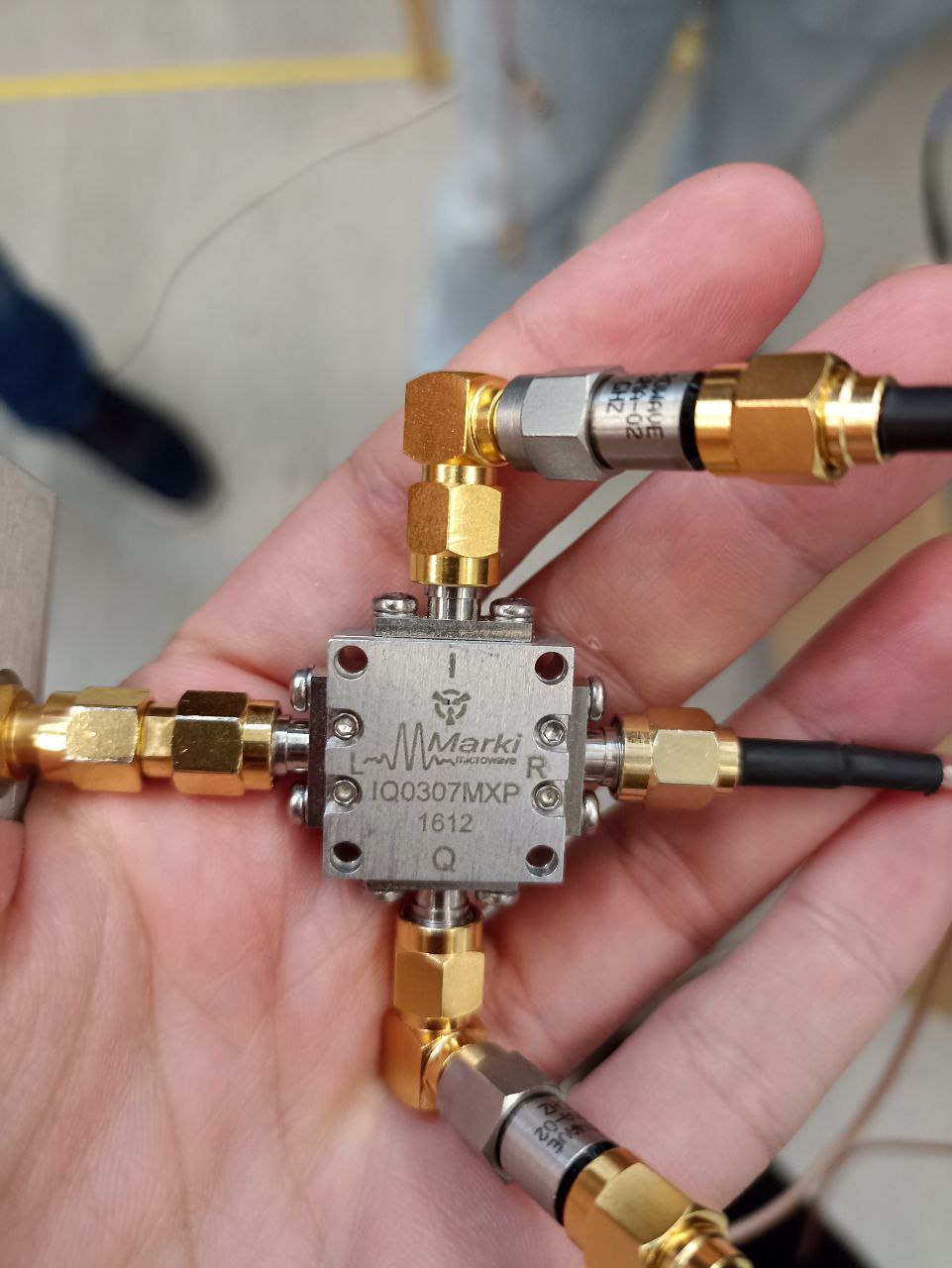}
    \caption{IQ mixer used for some of the measurements presented in this work.}
    \label{fig:MarkiMixer}
\end{wrapfigure}
\par % end paragraph

Prior to the commercial availability of digital synthesis generators, some of which became available only in the final years of this work and are capable of modulating signals with great accuracy up to 10 GHz, most arbitrary waveform generators (AWGs) could not exceed few GHz and usually don't reach the desired precision. Therefore, direct signal generation was unfeasible for qubit control. The solution was IQ mixers to add a high frequency local oscillator signal modulated with a low frequency AWG one.

Like this, with IQ mixers, signals in the range of GHz can be generated and also their quality can be improved through the sideband generation. This has been the schema used for all measurements in this thesis.

IQ mixers are 4-port devices that allow to add signals in frequency, see figure \ref{fig:MixerSchema2} for its internal schema and \ref{fig:MarkiMixer} for a picture of the device. Mixers in general are extremely useful components in telecommunication applications that can be found in radars, internet network infrastructure, radios, satellites, etc. All of the following techniques originate from these applications; quantum technologies have simply benefited from their widespread availability and well-established performance.

\begin{figure}[h!]
    \centering
    \includegraphics[width=0.9\linewidth]{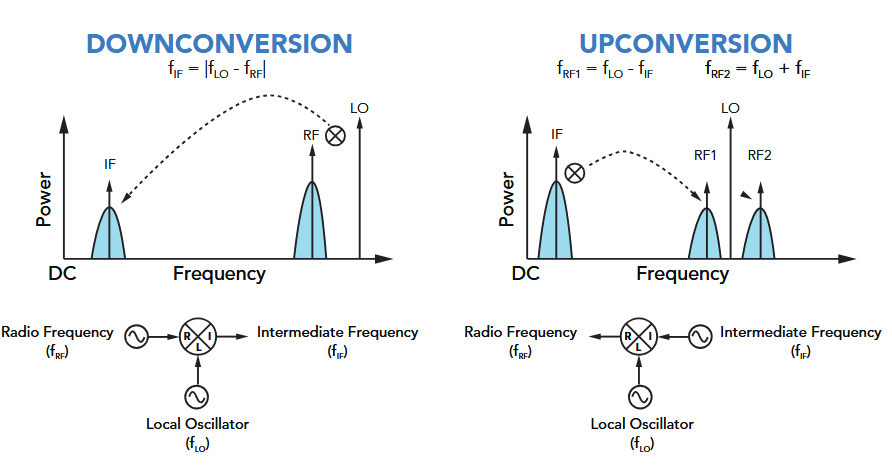}
    \caption{Downconversion and upconversion using a mixer. Image from \cite{Mixer_Basics_Primer}}
    \label{fig:UpDownConversion}
\end{figure}

These devices use the nonlinearity of diodes to add and subtract signals in frequency. Figure \ref{fig:UpDownConversion} shows the schematics of this process. There, a usual, non-IQ, mixer is used. The main difference between them is that normal mixers work with real signals, so the spectrum is always symmetric. That is not the case when using complex signals and IQ mixers. 

This upconversion technique allows to generate and modulate a high frequency signal with the help of a high frequency local oscillator (LO) and a low frequency AWG (typically below 1 GHz). 

\begin{comment}
\begin{figure}[h!]
    \centering
    \includegraphics[width=0.9\linewidth]{images/SingleSideBandUpconverter.png}
    \caption{Left: circuit diagram of an IQ mixer. Right: Upconversion selecting one of the side bands and suppressing the other. Image from \href{https://markimicrowave.com/products/connectorized/iq-mixers/mmiq-30120hm/datasheet/}{Marki}.}
    \label{fig:SingleSideBandUpconverter}
\end{figure}
\end{comment}

The other great advantage of using mixers it that they help to improve the purity of the signal through the so called ``sideband generation". When a low frequency signal (IF) is mixed with a high frequency one (LO), the output is a signal with three frequency components: $f_{LO} - f_{IF}$, $f_{LO}$, $f_{LO} + f_{IF}$. Furthermore, depending on the quality of the equipment and components in the lines, other harmonics can appear. 

\begin{figure}[h]
   \begin{minipage}[t]{0.49\textwidth}
     \centering
     \vspace{0pt}
     \includegraphics[width=0.9\linewidth]{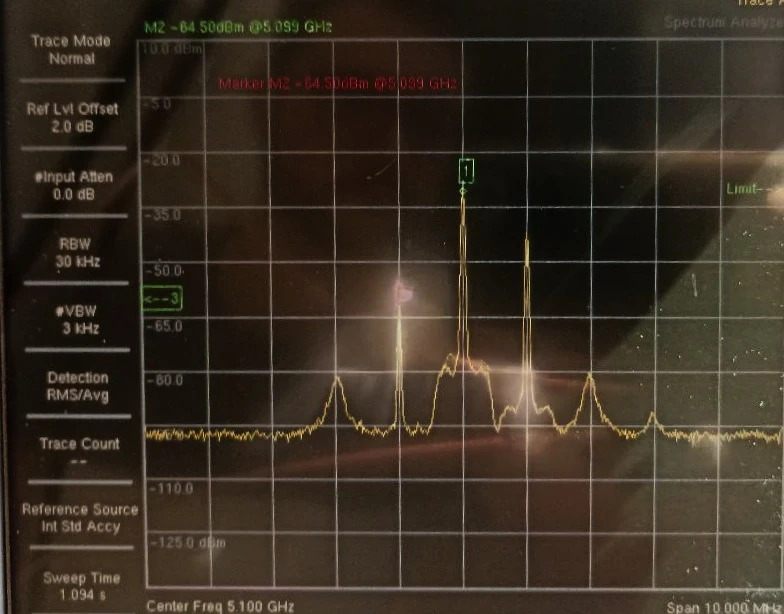}
   \end{minipage}\hfill
   \begin{minipage}[t]{0.49\textwidth}
     \centering
     \vspace{0pt}
     \includegraphics[width=0.9\linewidth]{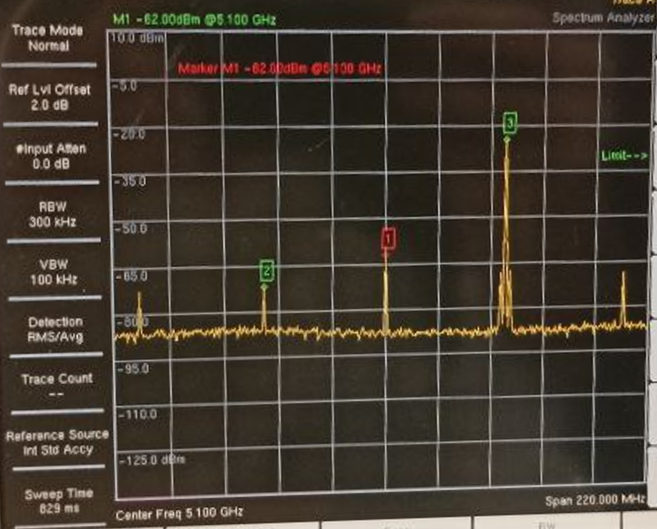}
   \end{minipage}
   \caption{\textit{Left}: Frequency spectrum after upconversion prior as mixer calibration as seen in the spectrum analyser. $f_{LO}$ = 5.1 GHz, $f_{IF}$ = 1 MHz. 
   \textit{Right}: Frequency spectrum after upconversion following mixer calibration for single side band upconversion. There is almost 40 dB difference between power in $f_{LO} + f_{IF}$ and $f_{LO}$. Frequencies are $f_{LO}$ = 5.1 GHz and $f_{IF}$ = 50 MHz.} \label{fig:MixerOptimization}
\end{figure}

In order to be able to suppress all these spurious frequency components, single side band upconversion is used. In figure \ref{fig:UpDownConversion} the schematic result of this procedure is shown. In transmon manipulation, typically the $f_{LO} + f_{IF}$ is selected, and all other frequencies are suppressed. To do so, the mixer has to be properly calibrated. What is needed is to inject signals in I and Q with certain difference in phase, amplitude and offset in order to maximize the non-linear increase in one side band and suppress it in the others. The effect of this calibration can be seen in the experimental transmission spectra shown in figure \ref{fig:MixerOptimization}.

\subsubsection{Mixer calibration}
\begin{comment}
\begin{figure}
    \centering
    \includegraphics[width=0.6\linewidth]{images/PhaseCalibration.png}
    \caption{Calibration of phase differences between I and Q signals. The power of the opposite side band is displayed in the Y axis, the phase difference in the X. The optimum for phase difference between I and Q input signals is at the minimum of this curve.}
    \label{fig:PhaseCalibration}
\end{figure}
\end{comment}

The process of optimization of phase, amplitude and offset of the signals is sequential, starting with the offset of the two channels. The characteristic we check when looking for the optimum is the suppression of LO power. As there are two offsets, several consecutive optimizations are performed, improving both each time. The resulting curve for each iteration is fitted and the optimal value for the offset is determined when the LO power is minimum.

\begin{figure}[h]
    \centering
    \includegraphics[width=0.6\linewidth]{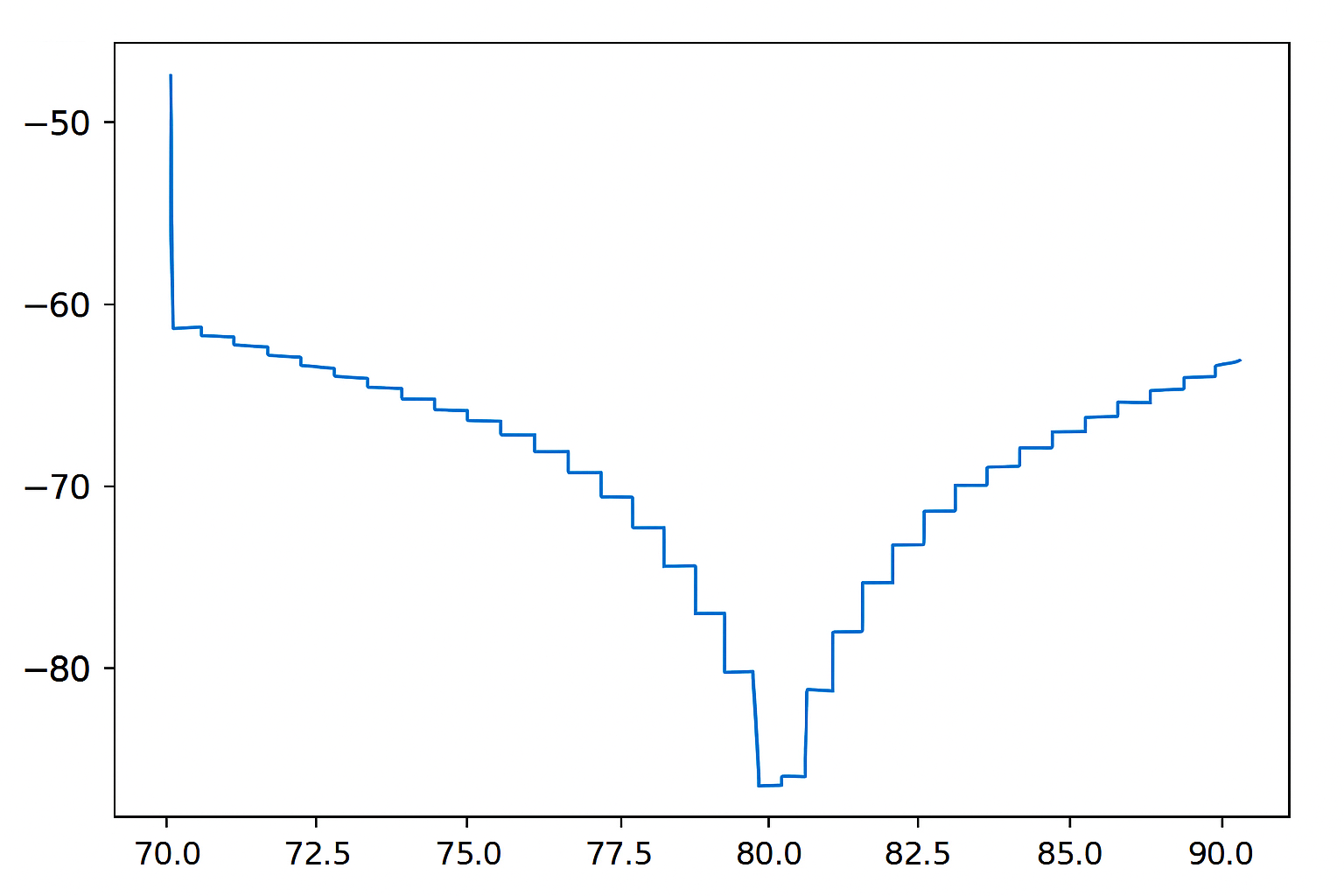}
    \caption{Calibration of phase differences between I and Q signals. The power of the opposite side band is displayed in the Y axis, the phase difference in the X. The optimum for phase difference between I and Q input signals is at the minimum of this curve.}
    \label{fig:PhaseCalibration}
\end{figure}

Then, the phase difference between I and Q signals is optimized, using the same strategy: loop over possible phases and look for the optimum. In this case, the optimum is determined suppressing the power in the other side band, $f_{LO} - f_{IF}$. Again the minimum of the resulting curve is the optimal value. In figure  \ref{fig:PhaseCalibration} an example of this calibration can be seen, where the optimum is achieved around 80º.

The last parameter to tune is the amplitude difference. This is done exactly the same away as in the phase difference: loop over possible amplitude values, check power in $f_{LO} + f_{IF}$ peak and the optimum lies where the minimum  of the curve is. What matters is the amplitude difference between signals, hence, only one needs to be modified at each step of the loop.

\begin{figure}[h!]
    \centering
    \includegraphics[width=0.8\linewidth]{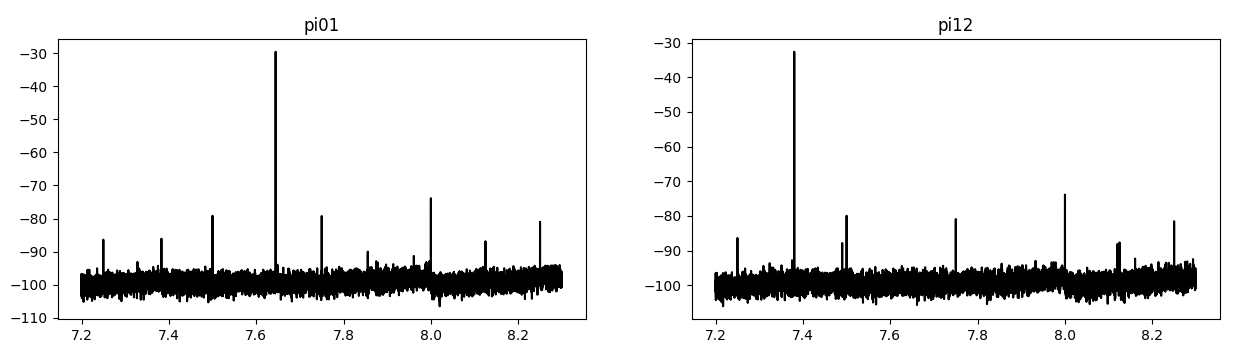}
    \caption{Automatic mixer calibration for different frequencies with the Octave device from Quantum Machines. The main selected frequency is between 40 and 50 dB higher than any other harmonic. Two different mixer calibration are shown with different selected frequencies.}
    \label{fig:MixerCalibrationOctave}
\end{figure}

The result of this calibration procedure is shown in figure \ref{fig:MixerOptimization}, and similarly in figure \ref{fig:MixerCalibrationOctave} for a commercial device, Octave from Quantum Machines, that periodically calibrates its internal mixers.

\subsection{Resonators}

Resonators are devices with one or more characteristic resonant frequencies. They can be found in many fields of physics and they make use of different types of stationary waves. Those which are interesting for quantum sensing are photon resonators. If a photon of the exact resonant frequency appears in the structure of the resonator, it will bounce for a while inside. This effect enhances the output signal that can be extracted from the resonator and it is the fundamental principle that permits resonant cavities to be used as dark matter detectors.

In quantum sensing, resonators are mainly used to read the state of the qubit. These two-level systems couple to the bosonic bath of the resonator in a way that when the state of the qubit changes, so does the resonant frequency of the resonator. There are at least two types of resonators when transmons are involved: three-dimensional cavities and plane resonators on chip. Here, several aspects of the first class of resonators will be discussed, but for other applications, having the resonator on chip is helpful because they can be created at the same time as the transmon using the same deposition techniques. Compact devices with multiple qubits coupled with different resonators and among them in almost any configuration can be fabricated in this way, which is of much interest, for example in quantum computing.

\subsubsection{Resonant modes in cavities}

Three-dimensional cavities have resonant modes according to their geometry. In general, nowadays the resonant frequencies are computed through computational methods that solve Maxwell equations to look for resonant modes of the photons. But for the simplest geometries, rectangular and cylindrical cavities, which are the most common halosocope and transmon readout cavity geometries, characteristic modes can be computed analytically.

\begin{figure}
    \centering
    \includegraphics[width=0.9\linewidth]{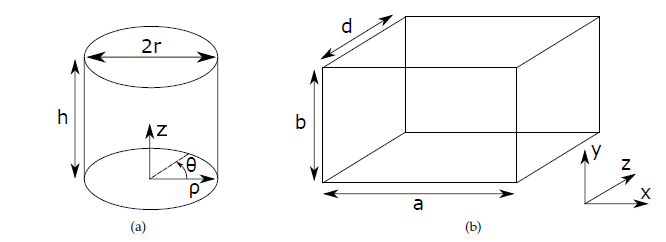}
    \caption{a) Cylindrical cavity of radius $r$ and height $h$. b) rectangular cavity of width $a$,
height $b$ and length $d$.}
    \label{fig:Cavities}
\end{figure}

Their resonant frequencies can be derived imposing boundary conditions to standing electromagnetic waves. A full derivation can be found in \cite{balanis1989advanced}. 

The modes are classified in two groups: Transversal electric, $TE$, when $E_Z = 0$ and transversal magnetic, $TM$, when $H_Z = 0$.

\begin{figure}
   \begin{minipage}[t]{0.48\textwidth}
     \centering
     \includegraphics[width=0.8\linewidth]{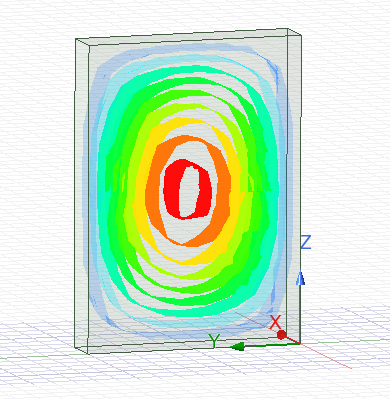}
   \end{minipage}\hfill
   \begin{minipage}[t]{0.48\textwidth}
     \centering
     \includegraphics[width=0.8\linewidth]{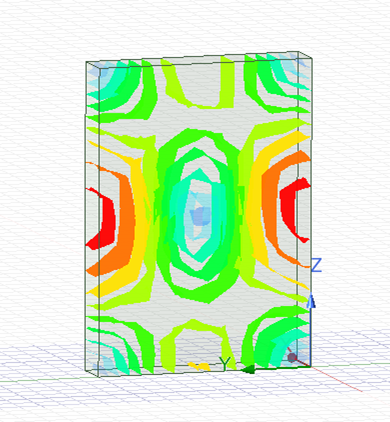}
   \end{minipage}
   \caption{\textit{Left}: Electric field pattern in $TE_{011}$ mode. \textit{Right}: Magnetic field pattern in $TE_{011}$ mode.} \label{fig:EMmodes}
\end{figure}

The resonant frequencies for these modes can be expressed as follows:

\begin{equation}\label{eq:TEcil}
f_{T E_{m n p}}^{c y l}=\frac{c}{2 \pi \sqrt{\varepsilon_r \mu_r}} \sqrt{\left(\frac{p_{m n}^{\prime}}{r}\right)^2+\left(\frac{p \pi}{h}\right)^2}, \quad \begin{aligned}
m & =0,1,2, \ldots \\
n & =1,2,3, \ldots \\
p & =1,2,3, \ldots
\end{aligned}
\end{equation}

\begin{equation}\label{eq:TMcil}
f_{T M_{m n p}}^{c y l}=\frac{c}{2 \pi \sqrt{\varepsilon_r \mu_r}} \sqrt{\left(\frac{p_{m n}}{r}\right)^2+\left(\frac{p \pi}{h}\right)^2}, \quad \begin{aligned}
m & =0,1,2, \ldots \\
n & =1,2,3, \ldots \\
p & =0,1,2, \ldots
\end{aligned}
\end{equation}

\begin{equation} \label{eq:TErec}
f_{T E_{m n p}}^{rec}=\frac{c}{2 \sqrt{\varepsilon_r \mu_r}} \sqrt{\left(\frac{m}{a}\right)^2 + \left(\frac{n}{b}\right)^2 + \left(\frac{p}{d}\right)^2 }, \quad \begin{aligned}
m & =0,1,2, \ldots \\
n & =0,1,2, \ldots \\
p & =1,2,3, \ldots \\
m & \text{ and $n$ not 0 simultaneously}
\end{aligned}
\end{equation}

\begin{equation} \label{eq:TMrec}
f_{T M_{m n p}}^{rec}=\frac{c}{2 \sqrt{\varepsilon_r \mu_r}} \sqrt{\left(\frac{m}{a}\right)^2 + \left(\frac{n}{b}\right)^2 + \left(\frac{p}{d}\right)^2 }, \quad \begin{aligned}
m & =1,2,3, \ldots \\
n & =1,2,3, \ldots \\
p & =0,1,2, \ldots
\end{aligned}
\end{equation}

where $m$, $n$, and $p$ are integers that denote the number of maxima of the electric field in the $\theta$, $\rho$, and $z$ axis for cylindrical cavities, and in the $x$, $y$, and $z$ axis for rectangular cavities, respectively; $\mu_r$ is the relative magnetic permeability ($\mu_r$ = $\varepsilon_r$ = 1 is assumed); $p_{m n}$ and $p_{m n}^{\prime}$ represent the $n-th$ zero ($n$ = 1, 2, 3, ...) of the first kind Bessel function $J_m$, and its derivative $J^{\prime}_m$ , respectively, of order $m$ ($m$ = 0, 1, 2, 3, ...); $r$ and $h$ are the radius and height of the cylindrical cavity, respectively, and $a$, $b$ and $d$ are the width, height and length of the rectangular cavity, respectively, as shown in picture \ref{fig:Cavities}. In figure \ref{fig:EMmodes} the electric and magnetic fields for the mode $TE_{011}$ of a rectangular cavity are displayed.

\begin{figure}[h]
    \centering
    \includegraphics[width=0.8\linewidth]{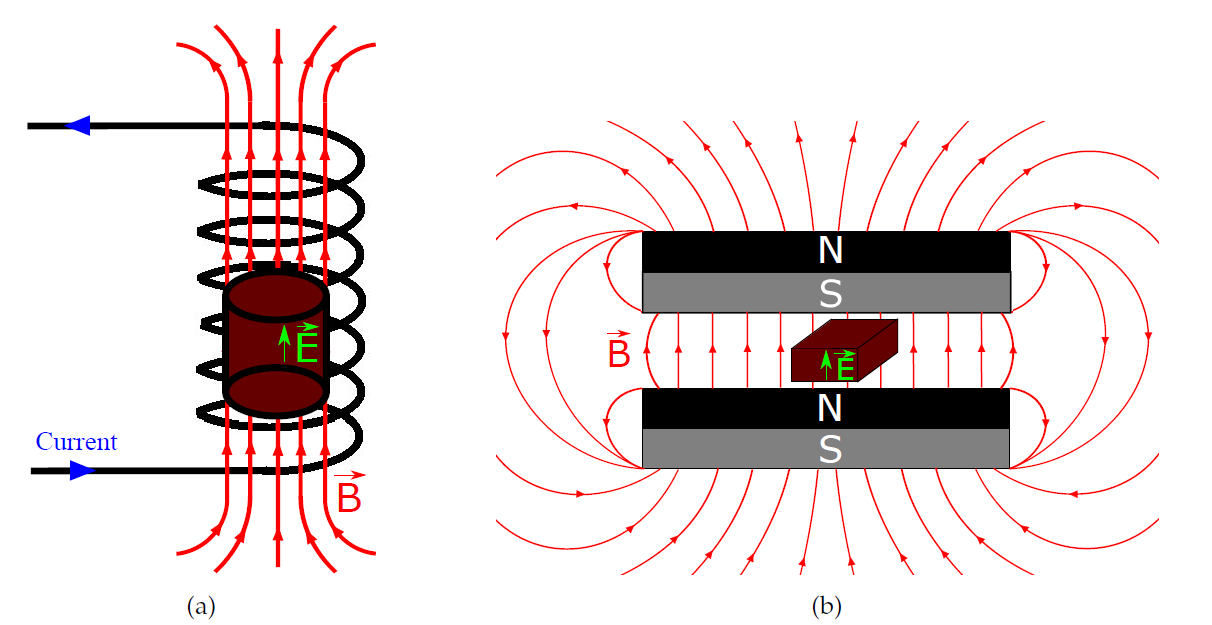}
    \caption{Types of magnets for axion haloscope searches. (a) Solenoid magnet with axial magnetic field. (b) Dipole magnet with transversal magnetic field. Image from \cite{garcia2023development}.}
    \label{fig:Magnets}
\end{figure}

Axion haloscopes are made with a resonant cavity immersed in a magnetic field. Depending on the type of magnet that generates the field, different geometries and cavity modes are used in haloscope searches. In general, with solenoid magnets, cylindrical cavities with $TM_{010}$ mode are used, and in dipole magnets, rectangular geometries with $TE_{101}$ mode are preferred, see figure \ref{fig:Magnets}. The reason for this follows from the fact that in order to maximize the scan rate in haloscope searches, the overlap between the electric field of the mode and the external magnetic field has to be maximum. As a general rule, one tries to have $\vec{E}$ parallel to $\vec{B}$. Labels of $TE$ and $TM$ modes depend on the reference system, so for the same cavity $TE_{011}$, $TE_{101}$ and $TE_{110}$ can be the same mode in different reference systems.

\begin{comment}
\begin{figure}
    \centering
    \includegraphics[width=0.7\linewidth]{images/S12.png}
    \caption{Transmission spectrum of a resonator between two ports with resonant mode at 7.192 GHz.}
    \label{fig:S12}
\end{figure}
\end{comment}

\subsubsection{Couplings to external world}

For resonators to be useful, they need a way to interact with input signals. For 3D cavities these couplings are typically antennas, coaxial antennas to couple with the internal electric field and loop antennas to couple with the internal magnetic field. The precise shape and position varies according to the geometry of the cavity and the mode of interest. 
In the case of 2D resonators, common in transmons on chip, the resonator takes the form of piece of flat wire with certain length (it is what sets the resonant frequency). In this case, a capacitance coupling is used, simply by physically placing the resonator and the readout line very close.

\begin{comment}
\begin{figure}
    \centering
    \includegraphics[width=0.5\linewidth]{images/CoplanarWaveguide.png}
    \caption{Resonator coupled to a readout line. \textcolor{red}{Cite Temperature and power characteristics of quarter‑wavelength superconducting coplanar waveguide resonator}}
    \label{fig:CoplanarWaveguide}
\end{figure}  
\end{comment}

\begin{figure}
\centering
\begin{subfigure}{.5\textwidth}
  \centering
  \includegraphics[width=.99\linewidth]{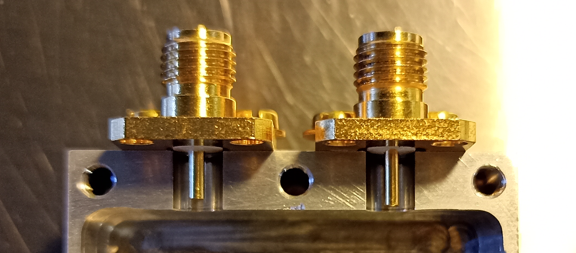}
  \caption{} \label{fig:PinPortsA}
\end{subfigure}%
\begin{subfigure}{.5\textwidth}
  \centering
  \includegraphics[width=.99\linewidth]{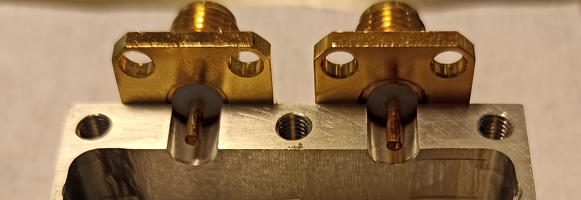}
  \caption{} \label{fig:PinPortsB}
\end{subfigure}
\caption{Different views of the pin ports. Longer pins tend to have stronger couplings, but the shape of the mode and the placement of the pin also matter.}\label{fig:PinPorts2}
\end{figure}

In coplanar 2D waveguides, the proximity and shape of the gap define the capacitance between both lines and thus the coupling. In 3D cavities with coaxial antennas, it is the position of the pin and their length what sets the coupling to the resonant modes. In general, the closer the pin is to higher values of the electric field, the higher is the coupling. For two symmetric pins, the coupling to the first mode, the one with lower frequency that has one central node, will be higher the longer the pins. 

Stronger couplings translate to a faster interaction of the resonator with the feed lines. This impacts the performance in several ways that can be tailored depending on the needs. Higher couplings mean that the photons stay less time bouncing in the cavity (the pin is very efficient extracting them); but it is also easier to inject photons (less power applied for the same number of photons injected in the cavity) and to extract them, resulting in shorter readout pulses. All these properties have advantages and disadvantages; depending on the purpose of the resonator one can be interested in maximizing the coupling, minimizing it or finding an intermediate value in which two things are balanced (inject easily but extract slow).

\subsection{Transmon characterization}

Once the qubit and the cavity have been designed, fabricated and installed in a dilution fridge, their quantum behaviour can be tested. The critical temperature of aluminium is 1.2 K. Below this temperature it is superconducting and two effects are observed: the cavity Q factor increases, as internal losses decrease a lot, and the transmon enters in the superconducting regime where its states in the phase-charge space are quantized and it can therefore be considered a qubit. 

The temperature of operation is much lower than the critical temperature of aluminium because the thermal noise decreases further and the coherence of the qubit increases. Typical base temperatures for dilution refrigerators are around 10~mK, even though the qubit itself has a residual background population of around 50 - 70~mK, as shown for example in \cite{dixit2021searching}.

\subsubsection{The \textit{punch out} method}

\begin{figure}[h]
    \centering
    \includegraphics[width=0.8\linewidth]{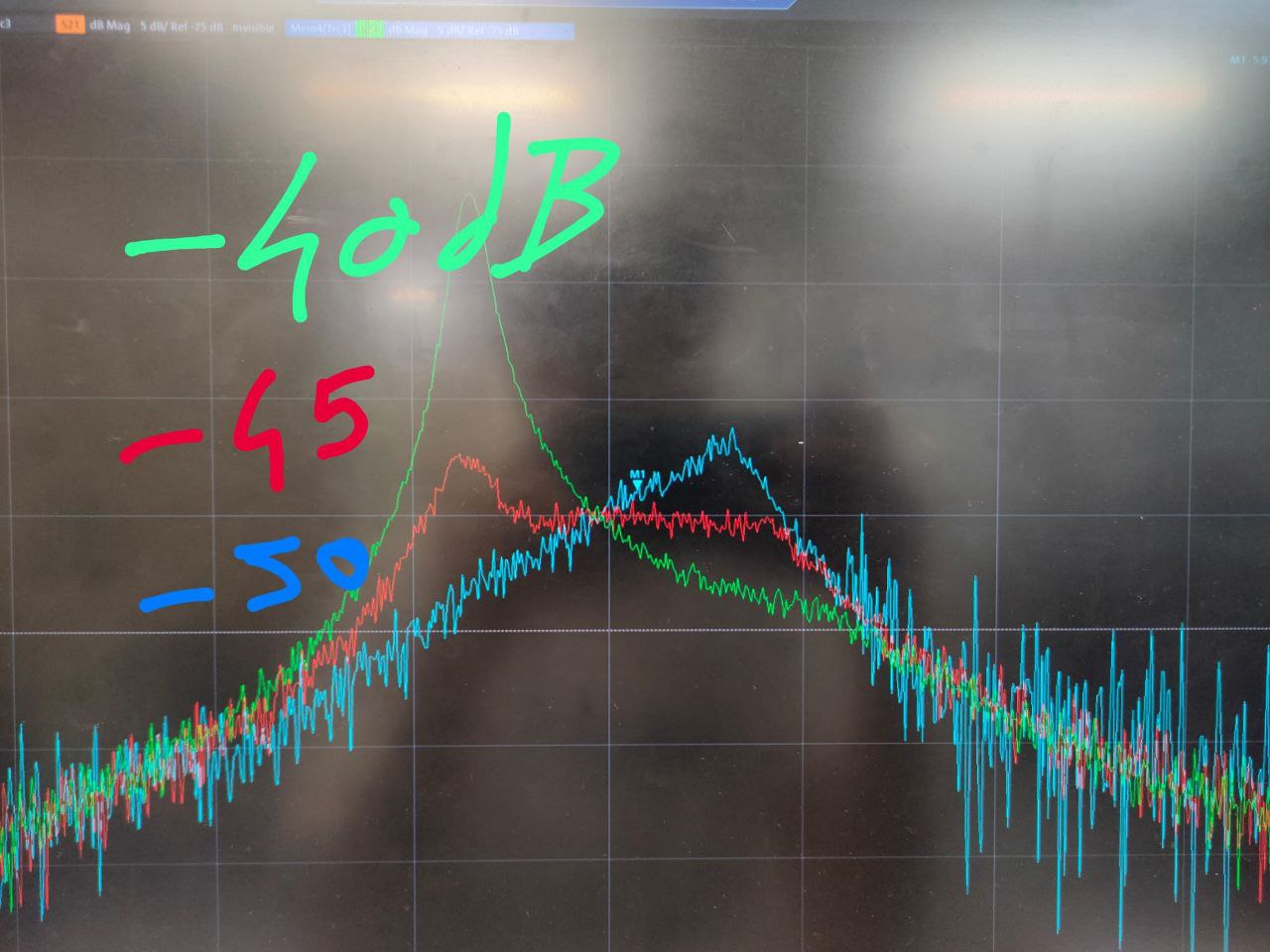}
    \caption{Transmission spectra for different signal power. Cavity and transmon coupled for low power (blue) and decouple when the power increases (green). Power levels depend on the attenuation levels included in the input channel.}
    \label{fig:PunchOut}
\end{figure}

Once everything is cold, the first check is to make sure the qubit is ``alive". This can be done by making use of the coupling between cavity and transmon. This coupling hybridizes the system and changes its characteristic resonant frequency. The way to see if this coupling is shifting the frequency is to measure the resonance with high and low power in a transmission spectrum taken from a VNA. When high power signal is injected in the cavity, the transmon decouples from the cavity so the resonance is due to the bare resonator. With low power, both components are coupled and the resonant frequency of the whole system is a combination of both, different from the bare resonator mode. In figure \ref{fig:PunchOut} this effect can be seen through three spectra with variable power. The green one, labelled by -40 dB, has the higher power, so this resonance is due to the cavity. When the power is decreased, the peak shifts to higher frequencies, an effect due to the transmon. In red, an intermediate situation in which the coupling is starting to show. 

An extra piece of information can be extracted from this measurement: if we compare modes in an avoided crossing in figure \ref{fig:AvoidedCrossing}, we see that hybridized modes (in solid lines) shift away from bare modes. This means than the hybridized mode moves from the cavity one to the opposite direction of the qubit frequency. Therefore, if the qubit frequency is lower than the cavity frequency, the peak shifts towards higher frequencies when they are coupled. And if the qubit frequency is higher than the cavity frequency, then it shifts towards lower frequencies. 
This helps to limit the range when looking for the qubit frequency, the next step in its characterization.

\subsubsection{Two-tone spectroscopy}
\begin{figure}[h!]
    \centering
    \includegraphics[width=0.8\linewidth]{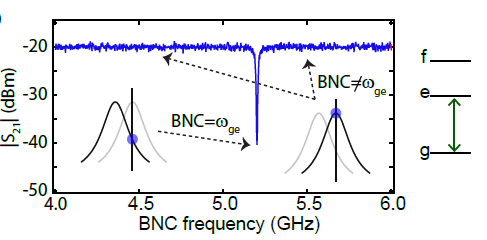}
    \caption{Two-tone spectroscopy. Magnitude of the transmission at the frequency of the resonator versus the sweeping frequency of the second tone. When the qubit transition is reached, the resonance shifts and the magnitude of the transmission decreases a lot.}
    \label{fig:TwoTone}
\end{figure}

In order to identify the qubit frequency, a two-tone measurement is performed. 

This is done injecting two different signals to the cavity, one resonating at the cavity frequency, the other one sweeping in frequency. When the second tone hits the frequency of the $|g\rangle \longrightarrow|e\rangle$ the system resonance shifts again, and the  transmission of the first tone suffers a big decrease, because at this new readout frequency, there is no maximum. This is schematically plotted in figure \ref{fig:TwoTone}.

\begin{figure}[h!]
    \centering
    \includegraphics[width=0.8\linewidth]{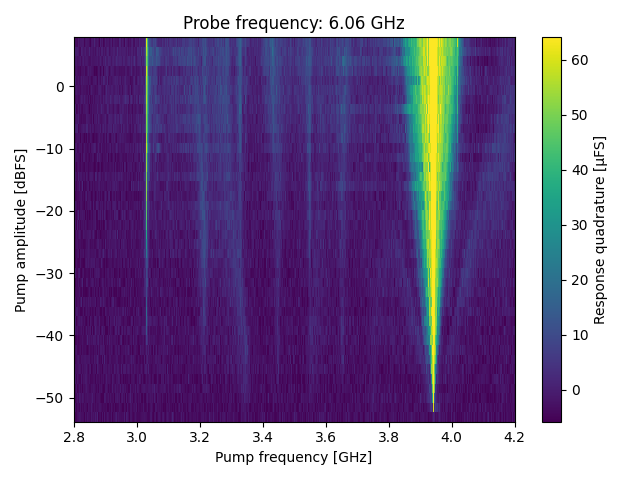}
    \caption{Two-tone measurement from a real transmon device. Here, the second tone frequency is plotted against its amplitude. Peak broadening can be seen due to the increase in power. The main peak at the right corresponds with the $|0\rangle \rightarrow|1\rangle$ qubit transition, while the second, thinner line at the left is the $|0\rangle \rightarrow|2\rangle$, a two-photon process that appears at higher power with much less probability. }
    \label{fig:TwoToneExperimental}
\end{figure}

If this transmission plot from  \ref{fig:TwoTone} is measured for different input power values in the second tone, a peak broadening can be seen, like in figure \ref{fig:TwoToneExperimental}.

\subsubsection{Readout measurement}
The qubit state can be manipulated through sending a drive tone of its frequency. This is the first qubit manipulation, however, to assess this ability of qubit manipulation, a readout technique is needed. This is achieved thanks to the coupling to the cavity. 

When the qubit state needs to be checked, a tone at cavity frequency is sent. The signal coming out of the cavity will vary according to the state of the cavity. The readout tone is a complex signal and the effect of the qubit can be seen in amplitude and phase of the output signal. It is decomposed in IQ components and plotted in the complex plane, an example will be seen in the following chapter (figure \ref{fig:OneShot}). Depending on the state of the qubit, measurements tend to be in different spots, which allows to set a region in the plane for each state. Each measurement that falls in one of the regions will be considered the corresponding state. This separation is not clean, therefore fidelity of the readout measurement is introduced to quantify this ability of measuring the correct state. 

\clearpage
\subsubsection{Rabi oscillations}

\begin{wrapfigure}[11]{R}{0.5\textwidth}
    \centering
    \includegraphics[width=0.8\linewidth]{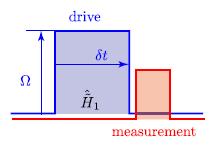}
    \caption{Schematic diagram of the pulse sequence for a Rabi measurement. From \cite{danilin2018experiments}.}
    \label{fig:SchematicRabiPulses}
\end{wrapfigure}

Once the qubit frequency has been determined with rough precision, it can be refined through Rabi oscillations. This measurement is also useful to determine the basic pulse to manipulate the state of the qubit: the $\pi$ pulse. This allows to move  the qubit from the ground state $|g\rangle$ to the first excited state $|e\rangle$. The name, $\pi$ pulse, comes from the movement in the Bloch sphere where half a turn is performed.

There are several possibilities to represent these oscillations but all of them reflect the same behaviour, the quantum oscillations between probabilities of measure $|g\rangle$ or $|e\rangle$ after a drive tone applied to the qubit. Depending on the amplitude and duration of this pulse, schematically represented in figure \ref{fig:SchematicRabiPulses}, the oscillatory pattern changes in amplitude and frequency. 

\begin{figure}[h!]
   \begin{minipage}[t]{0.48\textwidth}
     \centering
     \vspace{0pt}
     \includegraphics[width=0.99\linewidth]{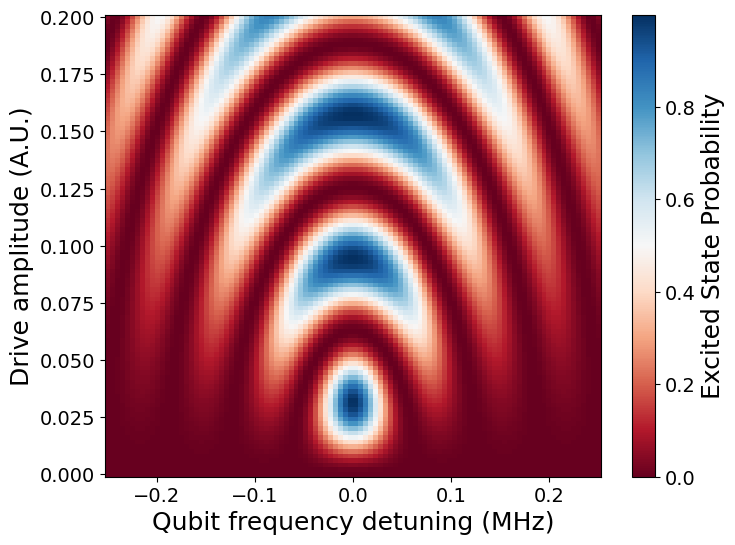}
     
   \end{minipage}\hfill
   \begin{minipage}[t]{0.48\textwidth}
     \centering
     \vspace{0pt}
     \includegraphics[width=0.99\linewidth]{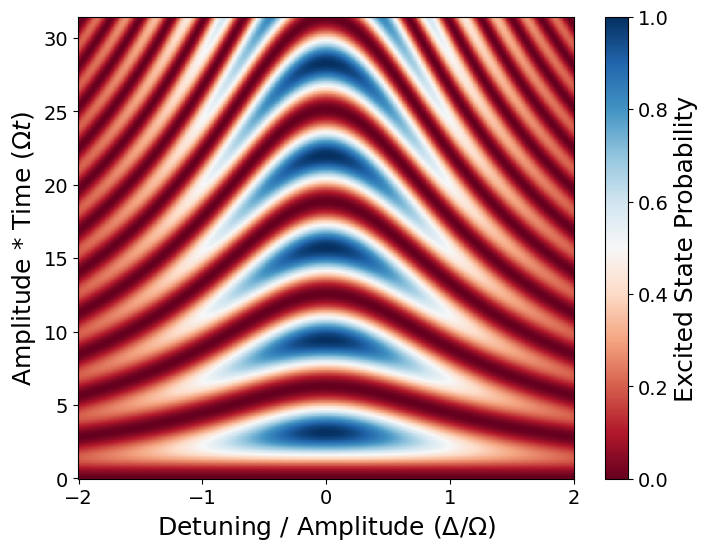}
     
   \end{minipage}
   \caption{\textit{Left}: Simulated Rabi oscillations in a drive amplitude versus qubit frequency detuning representation. \textit{Right}: Simulated Rabi oscillations in another representation. Horizontal axis is frequency detuning over amplitude, and vertical axis is amplitude times the duration of the pulse. In both, the colour gradient express the excited state probability.}\label{fig:RabiAmp}
\end{figure}

In \cite{danilin2018experiments} one can find a precise derivation of the probability of finding the qubit in excited state after applying external drive field $E_0 = \cos(\omega_d t+\phi)$. This expression \ref{eq:RabiEq} can be used to plot what is expected from a Rabi experiment, as can be seen in  \ref{fig:RabiAmp}.
 
\begin{equation}\label{eq:RabiEq}
    p_e(\Omega, \Delta, t) = |\langle e | \psi(t) \rangle|^2 = \frac{\Omega^2}{\Omega^2 + \Delta^2} \sin^2 \left( \frac{\sqrt{\Omega^2 + \Delta^2} \, t}{2} \right)
\end{equation}

\begin{figure}[h!]
   \begin{minipage}[t]{0.48\textwidth}
     \centering
     \vspace{0pt}
     \includegraphics[width=0.99\linewidth]{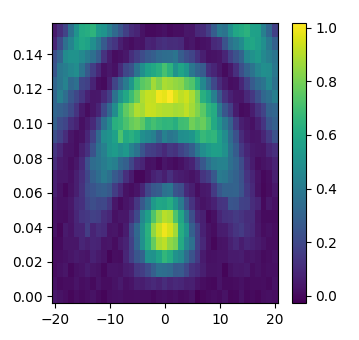}
     
   \end{minipage}\hfill
   \begin{minipage}[t]{0.48\textwidth}
     \centering
     \vspace{0pt}
     \includegraphics[width=0.99\linewidth]{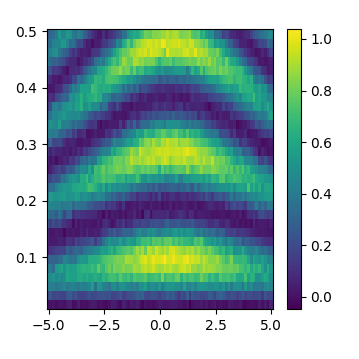}
     
   \end{minipage}
   \caption{Similar measured plots as presented in \ref{fig:RabiAmp}. \textit{Left}: Measurement of Rabi oscillations in frequency versus amplitude plot. In the X axis the detuning from qubit frequency, 40 MHz span. In Y axis the amplitude of the pulse, in volts. Time duration of the pulse is fixed. \textit{Right}: Measurement of Rabi oscillations in frequency versus duration plot. In X axis the detuning from qubit frequency, 10 MHz span. In Y axis the pulse duration in ns, from 16 ns to 512 ns in 4 ns intervals. Amplitude of the pulse is fixed. In both cases the colour gradient express the probability of being in first excited state.}\label{fig:RabiAmpMeas}
\end{figure}

For these measurements a sequence of pulses is needed. First, the drive tone is played, around the qubit frequency, with variable amplitude and duration. Right after this pulse, a readout tone is played at the cavity frequency. This schema is shown in figure \ref{fig:SchematicRabiPulses}. This tone allows to determine if the qubit is in the ground or excited state. Averaging over many measurements like this for each combination of amplitude, duration and frequency detuning for the drive pulse, the probability of ground or excited states can be computed for each point in the 2D plots, as shown in figures \ref{fig:RabiAmp} and \ref{fig:RabiAmpMeas} for simulated and measured Rabi oscillations.

\begin{comment}
    \begin{figure}[h!]
    \centering
    \includegraphics[width=0.7\linewidth]{images/RabiTimeDuration.png}
    \caption{Simulated Rabi oscillations varying duration and amplitude of the drive pulse.}
    \label{fig:RabiTimeDuration}
\end{figure}
\end{comment}

In figure \ref{fig:RabiAmp} left, the first blue dot serves to establish the parameters for a $\pi$ pulse (yellow central dot in left plot of \ref{fig:RabiAmpMeas}). The central position of this pattern can be fit to extract the frequency and amplitude of such a $\pi$ pulse.

\subsubsection{Characteristic times: T1, T2}

\begin{figure}[h]
\centering
\includegraphics[width=0.8\linewidth]{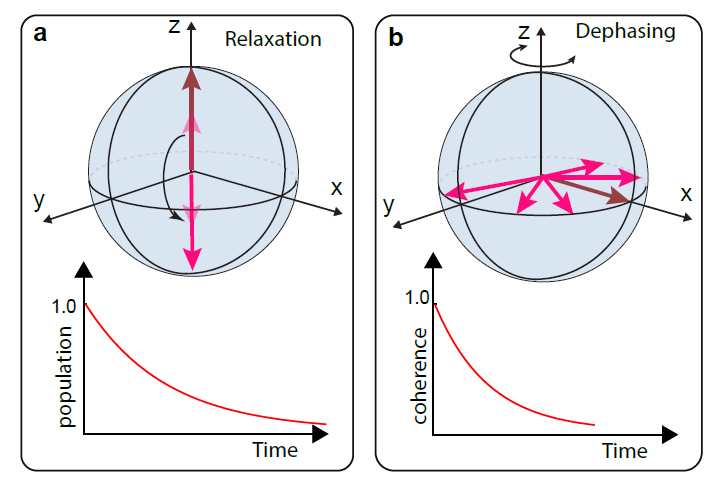}
\caption{\label{fig:RelaxationDephasing} Relaxation and dephasing in the Bloch sphere \cite{naghiloo2019introduction}. These are the two main effects characterised with T1 and T2.}
\end{figure}

Qubit states suffer from short decay times that are one of the main challenges of the technology. To interact with other elements of the system and predict the output is difficult if some of them lose their characteristic properties while measuring. This decay is characterized with two parameters: T1 and T2. T1 is the relaxation time and accounts for the time that takes the qubit to lose an excitation; T2 is the dephasing time. A superposition state is determined by the relative phase between pure states. The interaction with the environment forces this coherent states to shift, losing the coherence, the relative phase between states, which impedes recovering the original state. Both are the values that characterize the exponential decay they follow. 

\begin{figure}[h!]
\centering
\includegraphics[width=0.9\linewidth]{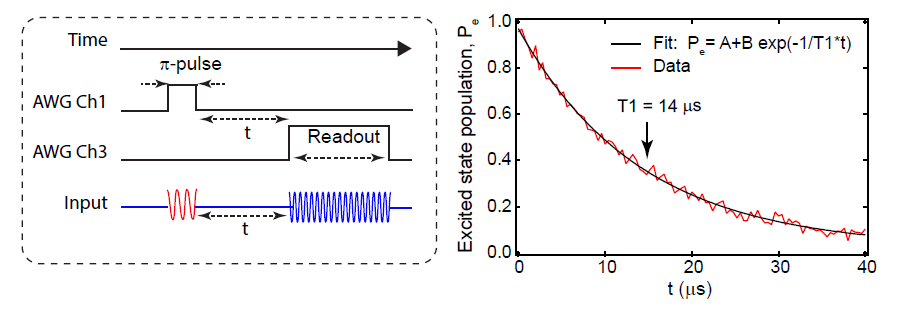}
\caption{ Schema of $T_1$ measurement \cite{naghiloo2019introduction}. \textit{Left}: Sequence of pulses for the T1 measurement. \textit{Right}: Exponential decay expected and fit to extract T1 value.}
\label{fig:T1}
\end{figure}

\begin{figure}[h!]
   \begin{minipage}[t]{0.48\textwidth}
     \centering
     \vspace{0pt}
     \includegraphics[width=0.99\linewidth]{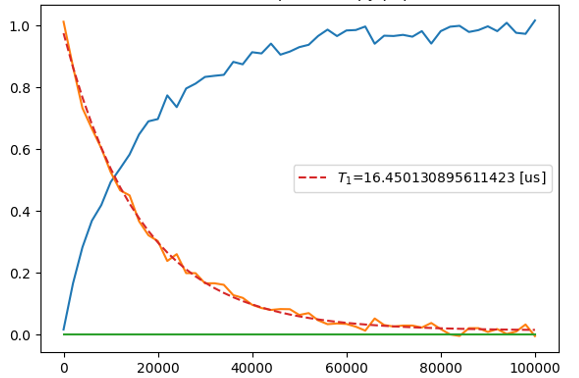}
     \caption{Relaxation exponential decay. Measured characteristic time $T_1 = 16$ $\upmu$s. In orange, the measured population in excited state, fitted to exponential decay in dashed red. In blue it is plotted the population in ground state. And in green, the population in second excited state, almost inexistent. Horizontal time scale in ns. Vertical scale excited state population. }\label{fig:T1measured}
   \end{minipage}\hfill
   \begin{minipage}[t]{0.48\textwidth}
     \centering
     \vspace{0pt}
     \includegraphics[width=0.90\linewidth]{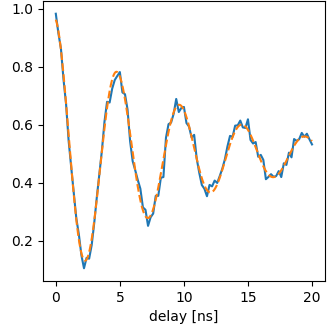}
     \caption{Oscillations in T2 measurement for T2 = 9.5 us. In the vertical axis the probability of being in the excited state, in the horizontal one the delay between $\pi/2$ pulses. } \label{fig:T2measurement}
   \end{minipage}
\end{figure}

In order to measure both characteristic times, different pulse sequences are needed. For T1, the qubit is placed in an excited state, wait some time and then measure through a readout pulse. Repeating many times this measurement varying the waiting time, an averaged population can be estimated. The longer the waiting time, the smaller the probability to find the qubit in the excited state. This pulse sequence and the exponential decay obtained is depicted in figure \ref{fig:T1}. And an experimental measurement in figure \ref{fig:T1measured}.

\begin{figure}[h]
    \centering
    \includegraphics[width=0.8\linewidth]{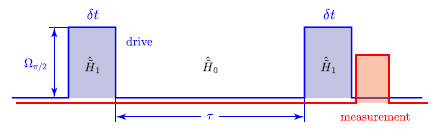}
    \caption{Sequence of pulses for a Ramsey chevrons \cite{danilin2018experiments}.}
    \label{fig:RamseySequence}
\end{figure}

To measure T2, the pulse sequence is more elaborate. It is an interferometric sequence, in which two $\pi/2$ pulses are applied with a delay between them. Depending on the dephasing during this delay time, the response changes. The idea is to rotate the state, initially the ground state, along the X axis (or Y, it is not important, just keep the same all the time), $\pi/2$ degrees so to place it in the equator of the Bloch representation. Then wait  and rotate in the same axis but towards the opposite direction $-\pi/2$. If the state remains in the position left by the first pulse, it returns to the ground state. But if some dephasing happened, then the state ends up in a superposition state. Repeating the sequence several times, the probability for each state and each delay time can be extracted. The pulse sequence and the decay pattern are shown in \ref{fig:RamseySequence} and an experimental measurement of these oscillations in \ref{fig:T2measurement}.

\begin{comment}
\begin{wrapfigure}[13]{L}{0.5\textwidth}
\centering
\includegraphics[width=0.9\linewidth]{images/T2measurement.png}
\caption{\label{fig:T2measurement} Oscillations in T2 measurement for T2 = 9.5 us. }
\end{wrapfigure}

\begin{wrapfigure}[16]{R}{0.5\textwidth}
\centering
\includegraphics[width=0.99\linewidth]{images/T1measured.png}
\caption{\label{fig:T1measured} Measured $T_1$. In orange: population in excited state, fitted to exponential decay in dashed red. In blue: population in ground state. In green population in second excited state (none). Time scale in ns. }
\end{wrapfigure}
\end{comment}

\subsubsection{Ramsey chevrons}

\begin{figure}[h]
\centering
\includegraphics[width=0.9\linewidth]{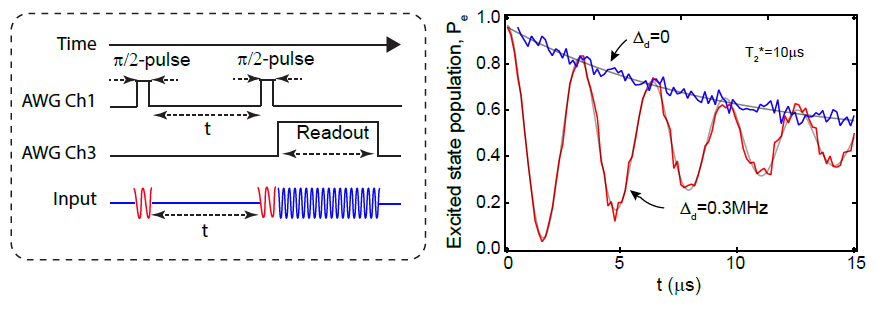}
\caption{\label{fig:RamseyNahiloo} Ramsey measurement sequence \cite{naghiloo2019introduction}.}
\end{figure}

\begin{figure}[h]
\centering
\includegraphics[width=0.8\linewidth]{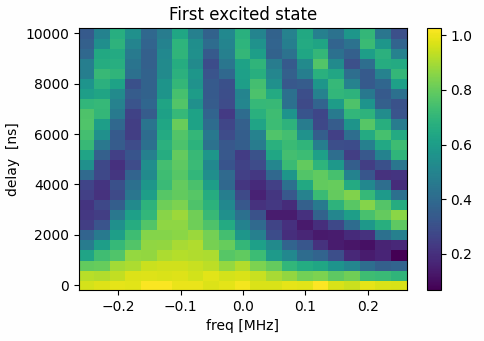}
\caption{ Population in first excited state in Ramsey measurements. X axis: frequency offset in MHz. Y axis: time in ns.}\label{fig:RamseyT2}
\end{figure}

A nice way to visualize the effects of dephasing is through Ramsey chevrons. The pulse sequence is identical to that used in a standard T2 measurement, where the delay between two $\pi/2$ pulses is varied. However, in this case, the detuning frequency of the pulses is also swept, resulting in a two-dimensional plot of detuning frequency versus delay time. If the frequency is perfectly on resonance with the qubit transition, an exponential decay is obtained. This measurement is very sensitive to the frequency of the $\pi/2$ pulses. It is because of this that Ramsey chevrons are very useful, sweeping not only time delay between $\pi/2$ pulses but also frequency of the pulses. This was measured in figure \ref{fig:RamseyT2} where typical Ramsey pattern can be seen. If one vertical slice is taken, therefore fixing the frequency detuning, oscillations appear like in \ref{fig:T2measurement} when off-resonance. Usually to measure $T_2$ an offset is applied on purpose to force that situation, where faster oscillations appear and it is easier to fit. In figure \ref{fig:RamseyNahiloo} the sequence of pulses and the oscillations when on and off resonance can be seen. 

Again, this pattern can be reproduced following the evolution of the state under the proper evolution operands. In \cite{danilin2018experiments} it is magnificently explained and simulated chevrons are displayed.

\subsubsection{Stark shift}

\begin{figure}[h!]
\centering
\includegraphics[width=0.9\linewidth]{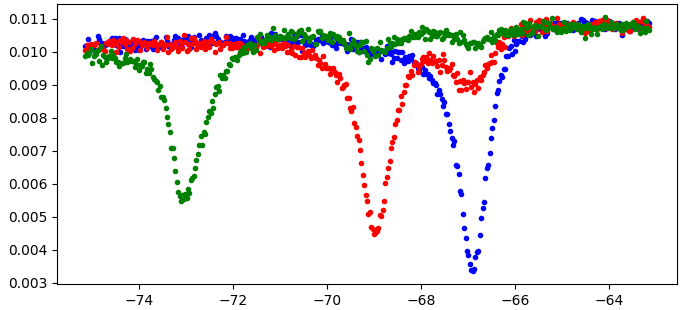}
\caption{ Cavity resonance for different states of the qubit: ground state in blue, first excited state in red and second excited state in green.}\label{fig:ChiThreeLevels}
\end{figure}

Qubit hybridization with the cavity shifts the resonance of the cavity. Depending on the state of the qubit the peak moves and this movement is characterized by $\chi$ as shown in equations \ref{eq:HamiltonianJC} and \ref{eq:chi}. The shift between peaks is $2\chi$ because possible eigenvalues of $\hat{\sigma_z}$ are $1$ and $-1$, therefore the distance is twice $\chi$. In figure \ref{fig:ChiThreeLevels} three reflection spectra are shown. There, the resonance shifts according to the state of the qubit, which is prepared just before measuring.

\begin{figure}[h]
\centering
\includegraphics[width=0.8\linewidth]{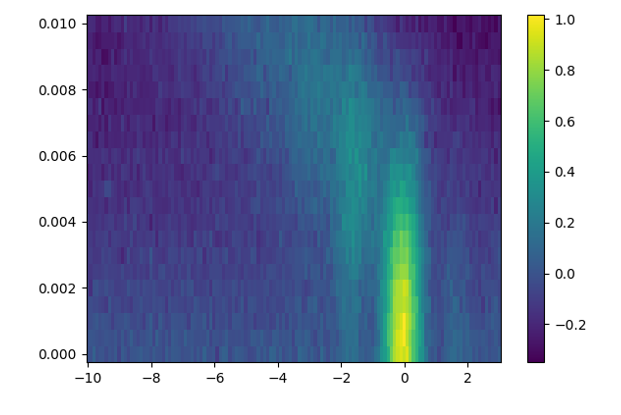}
\caption{ Stark shift visualization. X axis represents frequency detuning in MHz around qubit frequency and Y axis the amplitude of the pulse to inject photons in the cavity in arbitrary units. The colour gradient represents the probability of the first excited state. When the injected power in the cavity is low, the qubit $| g \rangle \to |e\rangle$ frequency remains centred in the calibrated one, but when power increases, this qubit transition frequency shifts showing the characteristic vertical shadows towards the left.} \label{fig:StarkShiftRefined}
\end{figure}

This peak movement can be considered from the opposite point of view: qubit resonance shifts due to the presence of photons in the cavity. This effect is called the Stark shift. The Hamiltonian in  \ref{eq:HamiltonianJC} can be rearranged to show explicitly this shift putting together all terms with $\hat{\sigma_z}$ as it was done there with $\hat{a}^\dagger\hat{a}$. This shift can be measured, as in figure \ref{fig:StarkShiftRefined}, and has interesting implications for this work as this is the principle used for photon detection in single photon counters for microwave photons. 

\begin{figure}[h!]
\centering
\includegraphics[width=0.8\linewidth]{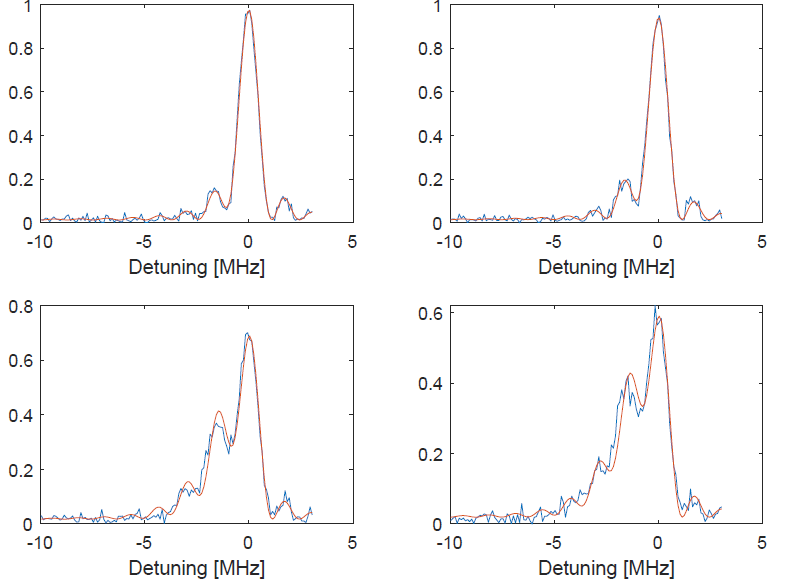}
\caption{ Experimental data from \ref{fig:StarkShiftRefined} for different fixed amplitudes (horizontal slices). In X detuning, in Y population in first excited state. In blue experimental data, in orange fitted to analytical function \ref{eq:InjectedPhotons}, product of Poisson distribution and population distribution of the qubit. From top to bottom, left to right, the mean photon content of the cavity corresponding to each spectrum is: 0.0598, 0.1256, 0.5615, 0.7049. } \label{fig:NumberOfPhotonsFit}
\end{figure}

The Stark shift allows to calibrate the amplitude and duration of the pulses to inject a variable amount of photons in the cavity. This photon injection follows a Poisson distribution, whose effects can be seen in horizontal slices of figure \ref{fig:StarkShiftRefined}, plotted as spectra with fixed injected power in plots of \ref{fig:NumberOfPhotonsFit}.  Multiple peaks appear when one, two, three... photons are injected. The mean value could be very small but this only means that most of the time there are zero photons injected and only occasionally is one photon inside. The average number of photons can be extracted from these spectra fitting them with the product of the Poisson distribution and the probability of finding an excited state of the qubit for these drive tone parameters. This last term was presented before, in the expression \ref{eq:RabiEq} for Rabi oscillations. The total function to fit is:

\begin{equation}\label{eq:InjectedPhotons}
    f(\Delta; \lambda, \Omega, \chi, t) = \sum_k \dfrac{\lambda^k e^{-\lambda}}{k!} \cdot \frac{\Omega^2}{\Omega^2 + (\Delta+k\chi)^2} \sin^2 \left( \dfrac{\sqrt{\Omega^2 + (\Delta+k\chi)^2}}{2} \, t \right)
\end{equation}

\begin{figure}[h]
\centering
\includegraphics[width=0.5\linewidth]{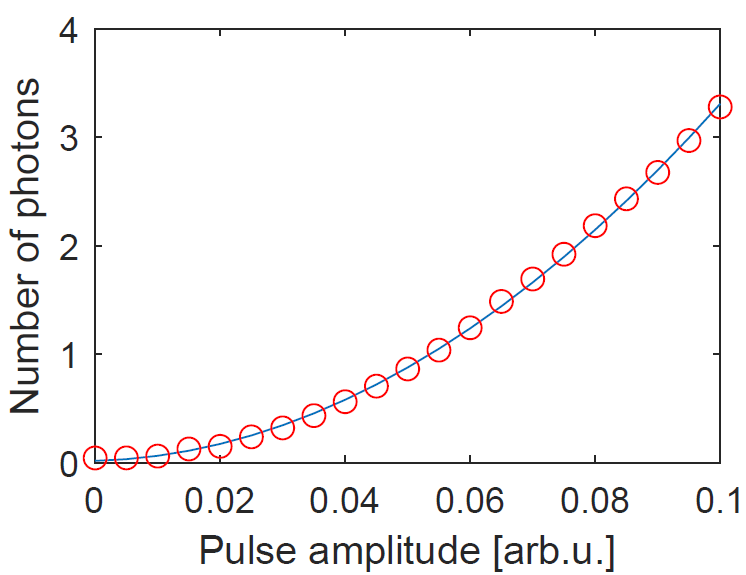}
\caption{ Average number of fitted photons for different pulse amplitudes (in red) and  fitted second degree polynomial (in blue), where the relation of the number of photons with the squared amplitude is clear.}
\label{fig:AverageNumberPhotons}
\end{figure}

$\Delta$ is the detuning with respect to the qubit frequency, $\Omega$ is the amplitude of the pulse, $t$ the time duration of the pulse, $\chi$ is the displacement due to the presence of a photon in the cavity and $\lambda$ is the average number of photons injected in the cavity. Sum over integer values $k \in \mathbb{N}$. This fit function is plotted together with the experimental data for some of the spectra in figure \ref{fig:NumberOfPhotonsFit}. The injected power in the resonator is directly proportional to the amplitude squared ($P=V^2/R$), and from the $\lambda$ values extracted for different injected power amplitudes, its quadratic relation is shown in figure \ref{fig:AverageNumberPhotons}.

\begin{comment}

\begin{figure}[h!]
   \begin{minipage}[t]{0.48\textwidth}
     \centering
     \vspace{0pt}
     \includegraphics[width=0.99\linewidth]{images/NumberOfPhotonsFit.png}
     
   \end{minipage}\hfill
   \begin{minipage}[t]{0.48\textwidth}
     \centering
     \vspace{0pt}
     \includegraphics[width=0.99\linewidth]{images/AverageNumberPhotons.png}
    
   \end{minipage}
   \caption{\textit{Left}: Experimental data from \ref{fig:StarkShiftRefined} for different fixed amplitudes (horizontal slices). In X detuning, in Y population in first excited state. In blue experimental data, in orange fitted to analytical function \ref{eq:InjectedPhotons}, product of Poisson distribution and population distribution of the qubit. \textit{Right}: Average number of fitted photons for different pulse amplitudes (in red) and  fitted second degree polynomial (in blue), where the relation of the number of photons with the squared amplitude is clear.} \label{fig:NumberOfPhotons}
\end{figure}

\end{comment}

%% file: Chapters/7_DarkQuantum.tex
Within the RADES collaboration, a project to develop a quantum sensor for axion searches has taken strength. As mentioned in the previous chapter, several grants devoted to the feasibility study and to developing a working setup with physics potential were awarded during 2024 to researchers of the collaboration. This is the case of DarkQuantum project.

In this frame, this chapter presents the efforts to design and operate the first prototype of a quantum sensor in RADES to detect single photons in the microwave regime.

\section{The double cavity}

Recently, two different strategies have achieved the detection of single microwave photons with quantum sensors. The first one to appear and the main reference for this work \cite{dixit2021searching} makes use of non-demolition measurements over the state of a qubit in order to suppress quantum errors. The other strategy \cite{braggio2024quantum}, and the first one to be applied to scan some frequency range in the parameter space of axion-photon conversion, is based on a four-wave mixing protocol to convert microwave photons in excited states of a qubit.

Here, a double cavity will be designed to host a transmon with a precise geometry to be coupled to both cavities. The idea of this strategy is that the qubit couple to two modes, one from the so called \textit{readout} cavity and the other from the \textit{storage} cavity, and its behaviour changes accordingly with the presence of photons in these modes. The storage cavity is where the photon conversion should happen for it to be detected. It is isolated except for the coupling with the transmon. The readout cavity has ports to inject and extract signals and it allows the manipulation of the qubit state.

The Hamiltonian that models this behaviour is a slightly modified Jaynes-Cummings Hamiltonian in the dispersive limit, were transmon-resonator detunings are much bigger than couplings:  

\begin{equation}
\widehat{H}_{e f f}=\frac{1}{2} \hbar \omega_{g e}^{\prime}\sigma_z +\left(\hbar \omega_r^{\prime}+\hbar \chi^{r} \sigma_z\right) \hat{a}^{\dagger} \hat{a} + \left( \hbar \omega_{s}^{\prime}+\hbar \chi^{s} \sigma_z \right) \hat{b}^{\dagger} \hat{b}
\end{equation}

$\omega^{\prime}$ are the frequencies of qubit and bare resonators modified by the coupling to the transmon: $\omega_{g e}^{\prime} = \omega_{g e} + \chi^{r}_{01} + \chi^{s}_{01}$, $\omega_{r}^{\prime} = \omega_{r} - \chi^{r}_{12}$ and $\omega_{s}^{\prime} = \omega_{s} - \chi^{s}_{12}$. Subindexes make reference to the levels between which the transition is taking place. For example, $\omega_{g e}$ is the transition angular frequency between ground and first excited levels of the qubit, and $\chi^{r}_{01}$ is the dispersive shift of the readout resonator when the qubit is in the ground state and in the first excited state. The measured Stark shifts can be expresed in terms of the first qubit transitions: $\chi^{n} = \chi^{n}_{01} - \chi^{n}_{12}/2$ with $n \in \{r,s\}$, readout and storage resonators, and they are related to the coupling between different qubit energy levels $g_{i,i+1}$ following the relations: $\chi^{n}_{i, i+1} = \dfrac{g^2_{i,i+1}}{\omega_{i,i+1} - \omega_n}$. The couplings between the first two transitions are related: $g_{12}=\sqrt{2}g_{01}$.

For further clarification, the derivation for a single resonator plus qubit is neatly described in section 2.7 of \cite{danilin2018experiments}.

Frequencies and shifts can be measured in the real experiment and some of the properties of the detector depend on these parameters. They have to be taken into account during the design step, although the precision in the simulation of these parameters is limited due to non-idealities in the final manufactured device. 

In the following sections, details on the transmon design will be given, along with how its coupling to cavity modes affects the overall performance of the detector. But first, a double cavity is needed. 

The constraints we forced on this first design were small: the prototype was going to operate at fixed frequency and it was not going to be placed inside a magnetic field. The qubit frequency would be around 4.5~GHz, meaning that resonant modes for both cavities would be a bit higher. Ultimately, a setup was designed with a 5~GHz storage cavity and a 7~GHz readout cavity. As a general rule, the closer the frequencies, the higher the coupling with the qubit. Geometry plays a significant role as well, however, as a first step, having qubit frequencies closer to the storage frequencies than to readout frequencies allows a significant coupling to storage cavity. In the next sections these couplings between cavities and qubit will be further discussed.
\\

\subsubsection{Drawings and simulations}

Once the cavity frequencies are decided, an electromagnetic simulation is performed in the geometry to avoid any unwanted resonances. First, we estimate the dimensions of the two cavities using the expressions \ref{eq:TErec} and \ref{eq:TMrec} for the modes in a rectangular cavity, then the rest of the details are added in the simulation.
In order to simplify the iteration process, it was decided to use the same length in two of the dimensions of both cavities. After several iterations, the dimensions of the cavities were as shown in table \ref{table:CavDims}.

\begin{table}[h] 
\centering
\begin{tabular}{l|ccc}
                         & \textbf{X} & \textbf{Y}& \textbf{Z} \\ \hline
Readout cavity (6.8 GHz) & 26.6 mm & 40 mm & 5 mm \\
Storage cavity (5 GHz)   & 45 mm   & 40 mm & 5 mm
\end{tabular}
\caption{Dimensions of for storage and readout cavities.}\label{table:CavDims}
\end{table}

\begin{figure}[h]
   \begin{minipage}[t]{0.48\textwidth}
     \centering
     \includegraphics[width=0.99\linewidth]{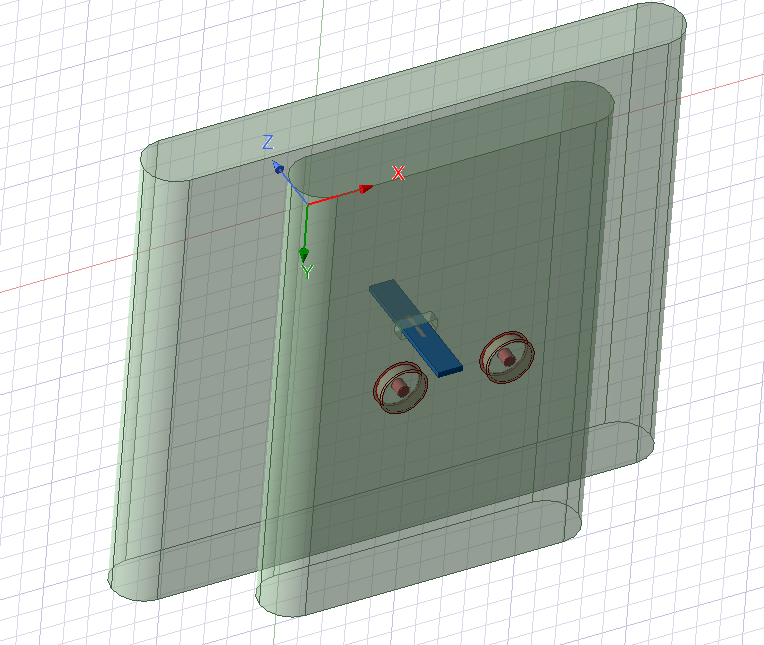}
     
   \end{minipage}\hfill
   \begin{minipage}[t]{0.48\textwidth}
     \centering
     \includegraphics[width=0.99\linewidth]{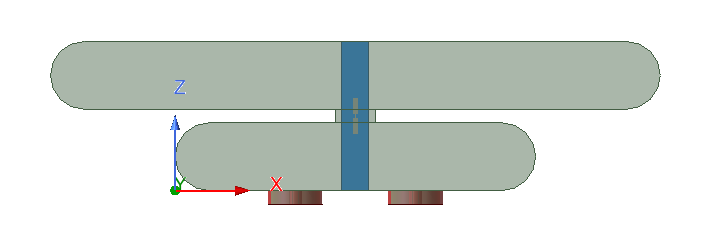}
     
   \end{minipage}
   \caption{\textit{Left}: HFSS electromagnetic model of the double cavity. \textit{Right}: Top view of the double cavity in HFSS}\label{fig:CavAnss}
\end{figure}

Foreseeing difficulties in the construction, rounded corners were added to facilitate the machining of the pieces. The final design can be seen in figure \ref{fig:CavAnss}. Finally, the target frequency of the readout cavity was lowered slightly from 7 to 6.8 GHz, forcing a slightly higher coupling with the qubit. The resonance of the readout cavity is not crucial because it has no scientific impact in the dark matter scan, where only the storage cavity frequency matters. For the readout cavity, it is essential that the external circuits can handle its main frequency. All microwave components (attenuators, amplifiers, etc.) have a range of frequencies of operation; in this case we had a cut-off frequency of 8 GHz in one amplifier, so all our frequencies are below this value. 

\begin{figure}[h]
    \centering
    \includegraphics[width=0.9\linewidth]{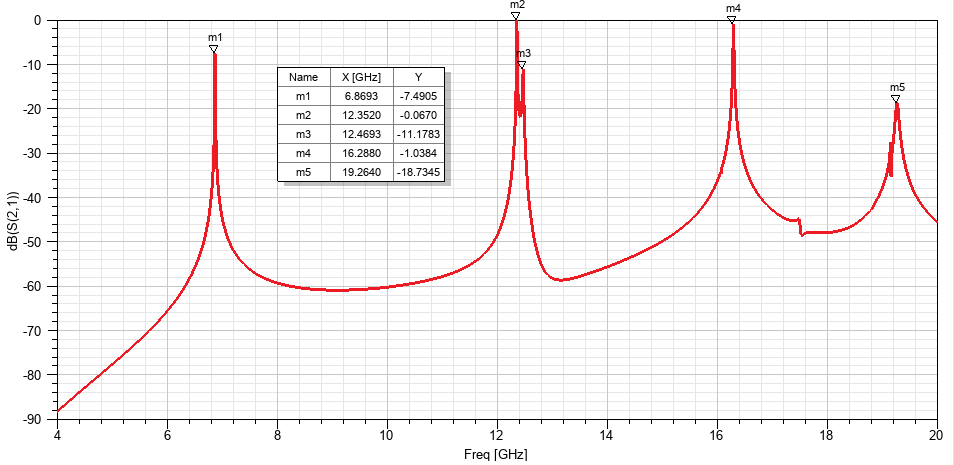}
    \caption{Simulated S12 transmission spectrum of the readout cavity. The first resonance at 6.8 GHz is the readout mode.}
    \label{fig:CavSpectrum}
\end{figure}

Resonances were simulated with the Ansys HFSS (High Frequency Structure Simulator) electromagnetic simulation software, obtaining values for the lowest frequency modes: 5.024 and 6.869 GHz for storage and readout respectively.
Although storage cavity modes are not available by design through the coaxial ports, the readout cavity can be seen in transmission measurements shown in figure \ref{fig:CavSpectrum}.

\begin{figure}
\centering
\begin{subfigure}{.5\textwidth}
  \centering
  \includegraphics[width=.95\linewidth]{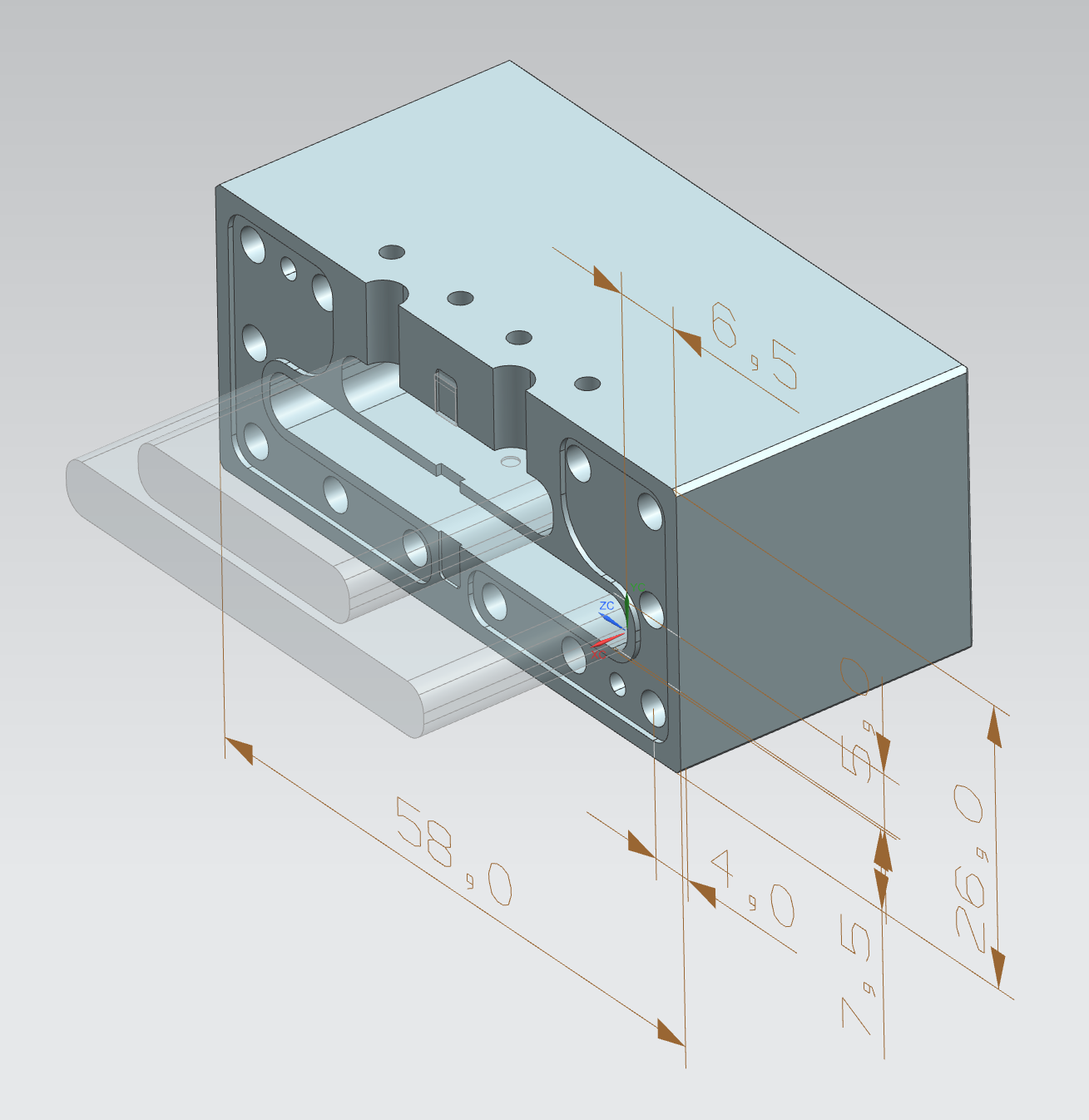}
  \label{fig:CavDesignDraw}
\end{subfigure}%
\begin{subfigure}{.5\textwidth}
  \centering
  \includegraphics[width=.95\linewidth]{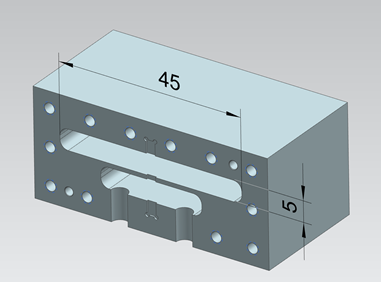}
  \label{fig:DoubleCavAcotada}
\end{subfigure}
\caption{Final design drawings for the double cavity. The entire cavity is formed by two identical pieces like the one shown here. } \label{fig:DoubleCav}
\end{figure}

From the electromagnetic simulation to the real device there is still one step: fabrication. Figure \ref{fig:DoubleCav} shows the mechanical design that was finally produced, including a small gap to hold the transmon and holes for screws and ports. 

\begin{comment}
\begin{figure}[h]
    \centering
    \includegraphics[width=0.5\linewidth]{images/Double_Cavity_drawing.png}
    \caption{Final design drawings for the double cavity. Here one have of the cavity is shown with the two empty volumes shadowed.}
    \label{fig:CavDesignDraw}
\end{figure}
\end{comment}

\subsubsection{Characterization of the machined cavity}

This double cavity was installed in a dilution refrigerator in Aalto University, in the Low Temperature Laboratory. The base temperature of the refrigerator was 10~mK, but the real temperature of the device was probably few mK higher, less than 20~mK in any case. This is due to imperfect thermal anchorages.

At these extremely low temperatures, resonances were explored giving frequency values shown in table \ref{table:CavFreqs}. The readout cavity mode was measured through the ports but the characterization of the inner cavity was performed with the help of a coupled transmon and a four-wave mixing process to inject photons into the storage cavity.

\begin{comment}
   % 1280/961
\begin{figure}
    \centering
    \includegraphics[width=0.5\linewidth, height=0.66597\linewidth]{images/CavityMachined.jpg} 
    \caption{Aluminum double cavity.}
    \label{fig:CavityMachined}
\end{figure} 
\end{comment}

\begin{figure}[h]
   \begin{minipage}[c]{0.48\textwidth}
    \centering
    \includegraphics[width=1.1\linewidth,angle=-90,origin=c]{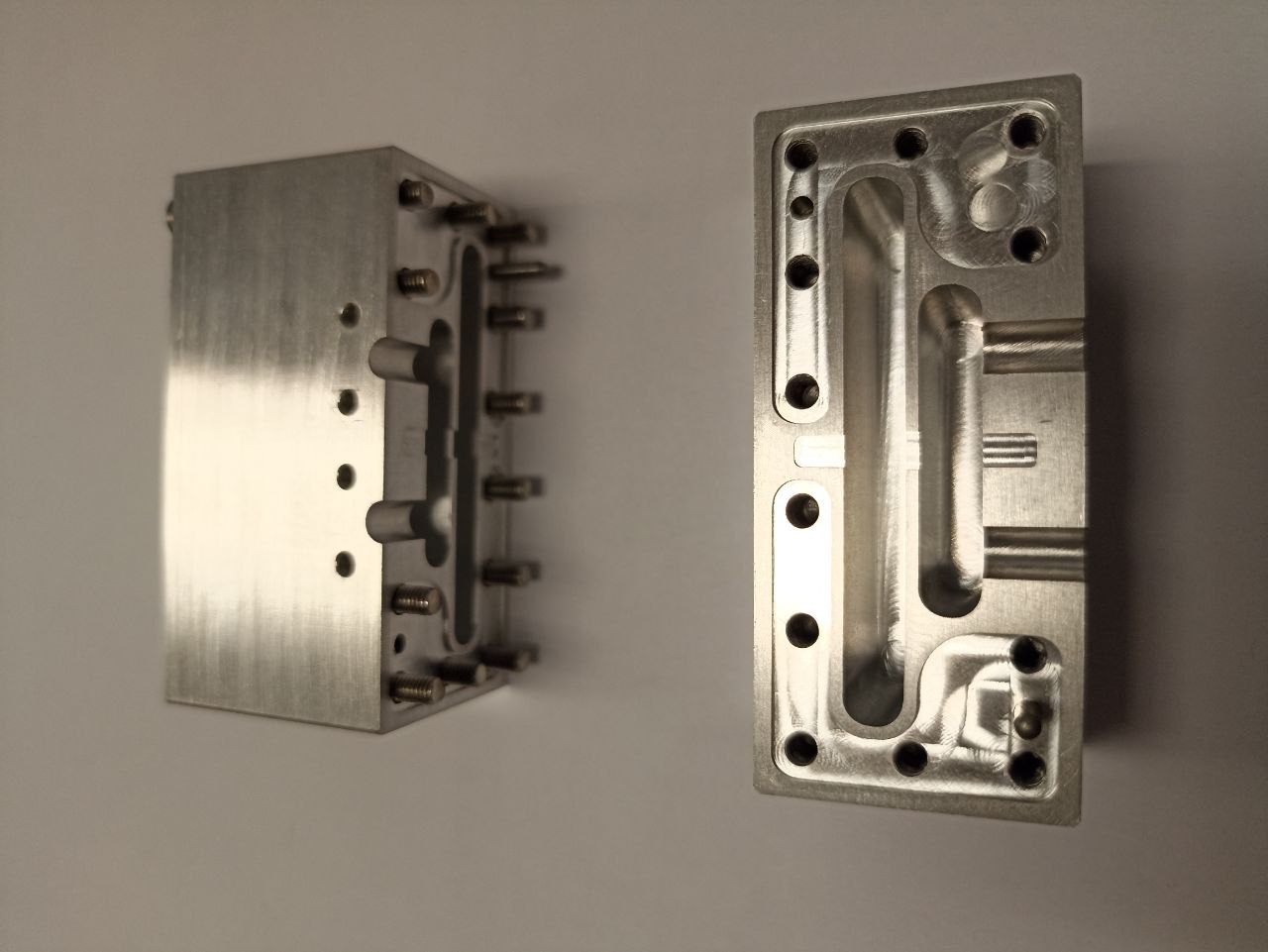} 
   
   \end{minipage}\hfill
   \begin{minipage}[c]{0.48\textwidth}
     \centering
     \includegraphics[width=0.7\linewidth]{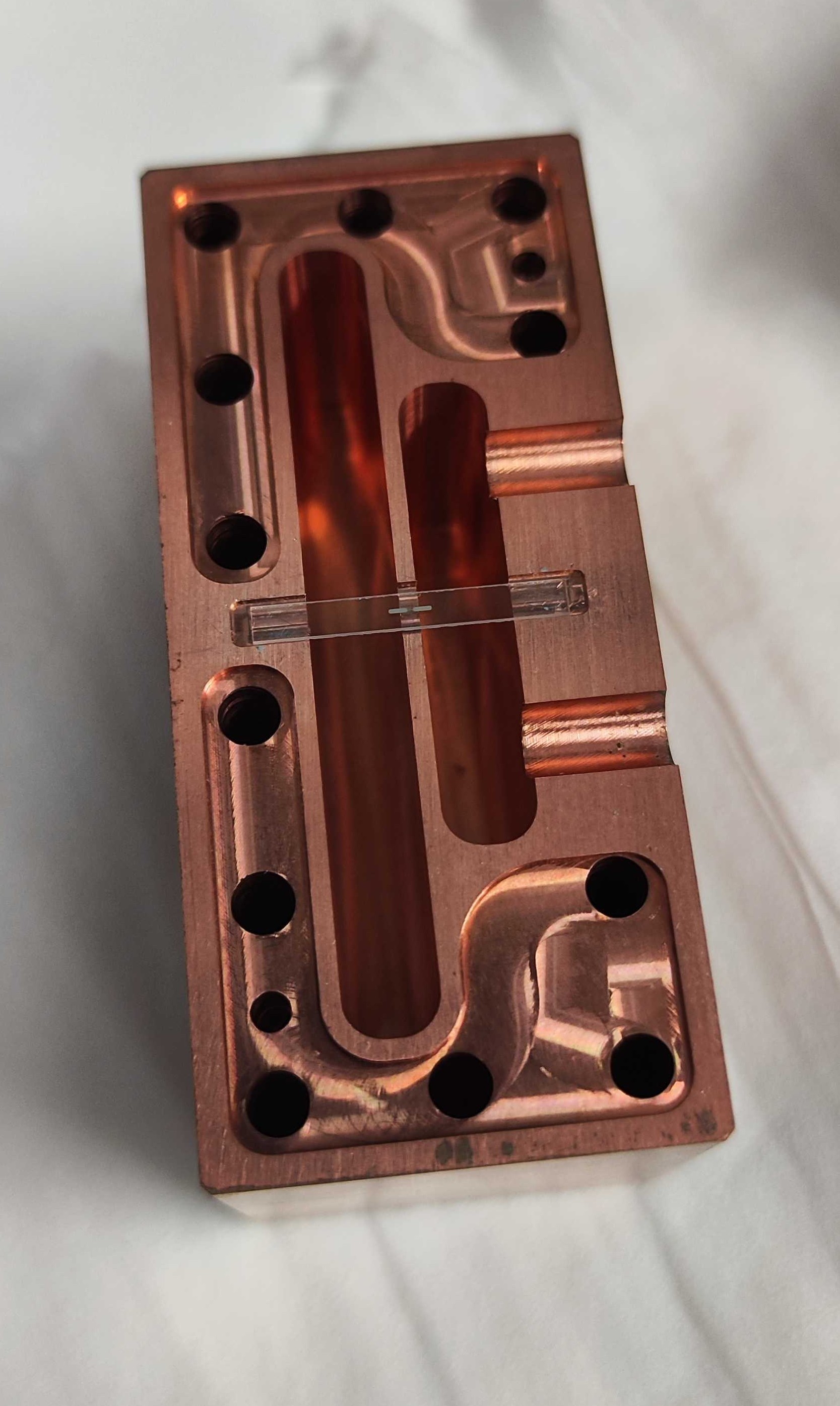}
     
   \end{minipage}
   \caption{\textit{Left}: Aluminium double cavity. \textit{Right}: Copper double cavity with transmon installed, transparent sapphire substrate can be seen in the centre. }\label{fig:CavityMachined}
\end{figure}

\begin{table}[h] 
\centering
\begin{tabular}{ccc}
&\textbf{Readout ($f_r$) }& \textbf{Storage ($f_s$)}  \\[0.05cm] \hline
\addlinespace[5pt] 
Simulated & 6.869 GHz     & 5.024 GHz  \\
Measured & 6.867 GHz     & 5.051 GHz    
\end{tabular}
\caption{Simulated and measured values for readout and storage cavities lowest modes.}\label{table:CavFreqs}
\end{table}

\section{Transmons}

\begin{figure}[h]
   \begin{minipage}[t]{0.5\textwidth}
    \centering
    \includegraphics[width=0.6\linewidth]{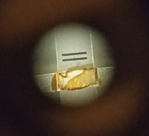}
    
   \end{minipage}\hfill
   \begin{minipage}[t]{0.5\textwidth}
     \centering
     \includegraphics[width=0.95\linewidth]{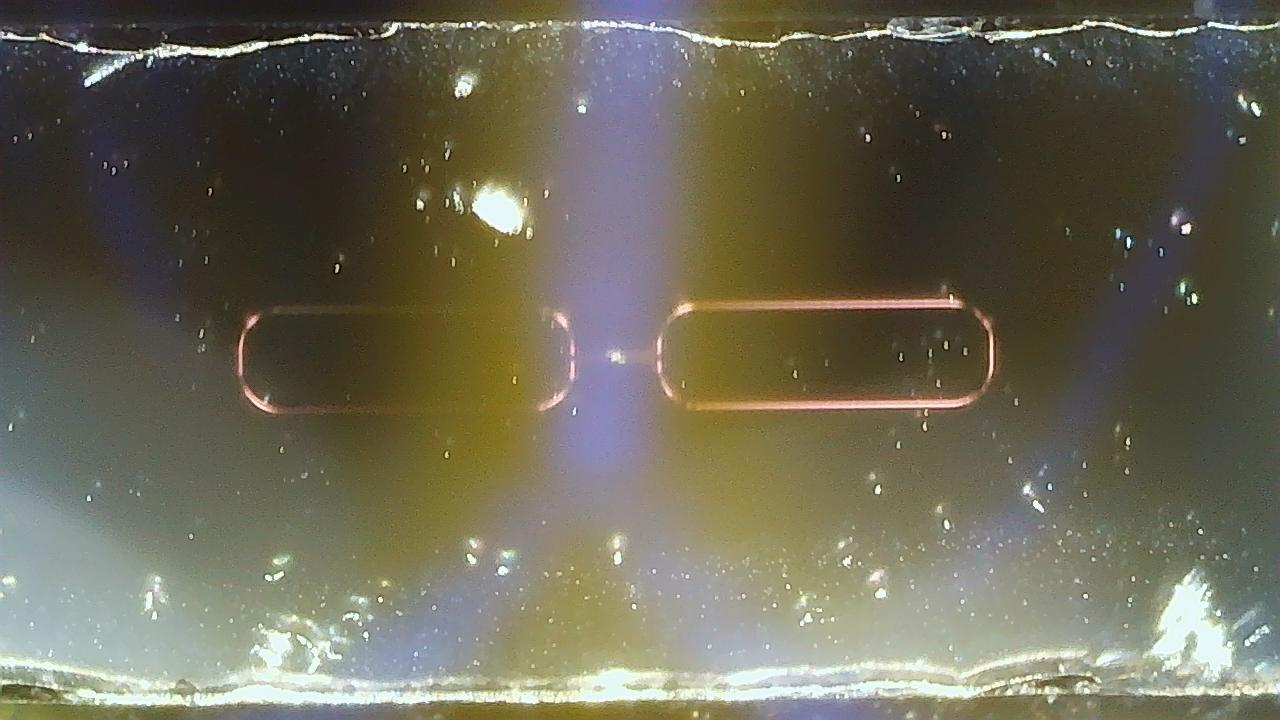}
     
   \end{minipage}
   \caption{\textit{Left}: Magnified view of an aluminium transmon over a sapphire substrate installed in a cavity. \textit{Right}: Aluminium transmon with long pads before removing the last protective layer with the design described in figure \ref{fig:TransmonDesign2}. }\label{fig:TransZoomAndChip}
\end{figure}

Previously, during my stay in Aalto University, I had had the chance to see the performance of several transmons with different geometries in single and double cavities. Two examples of these can be seen in figures \ref{fig:TransZoomAndChip}. This gave me an overview of the huge range of possibilities that superconducting devices can offer. Thanks to this, when the time came to design a full prototype, with cavity and transmon, I was aware of the behaviour I could expect and how to design accordingly.

\begin{figure}
    \centering
    \includegraphics[width=0.7\linewidth]{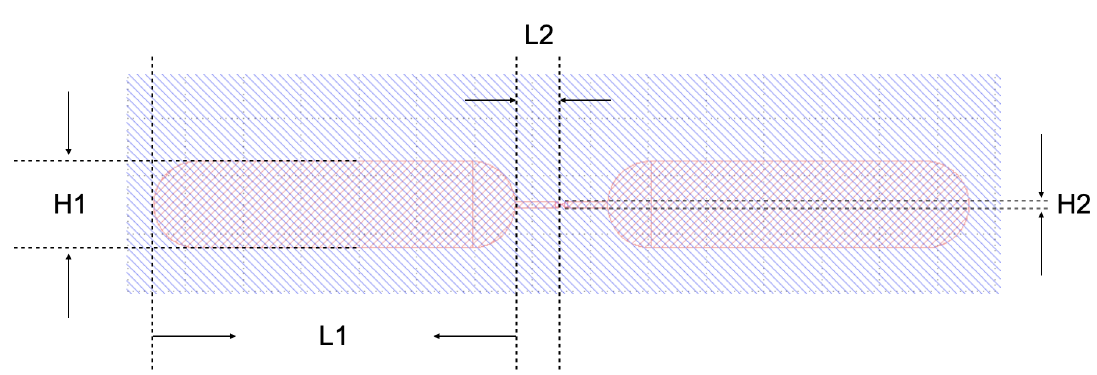}
    \caption{General transmon design and main dimensions.}
    \label{fig:TransmonDesign2}
\end{figure}

The transmon to be designed, had to be able to couple with two cavities, with a relaxation coherent time, $T_1$, of several tens of $\upmu$s. Its design was an iteration process trying different values for the dimensions showed in figure \ref{fig:TransmonDesign2}. Every configuration was simulated in the double cavity with HFSS and subsequently the data extracted from the simulation was used to estimate the parameters of the qubit (frequencies, anharmonicities, energies, couplings...) using the "Black-box" model \cite{nigg2012black}.

\begin{figure}
    \centering
    \includegraphics[width=0.8\linewidth]{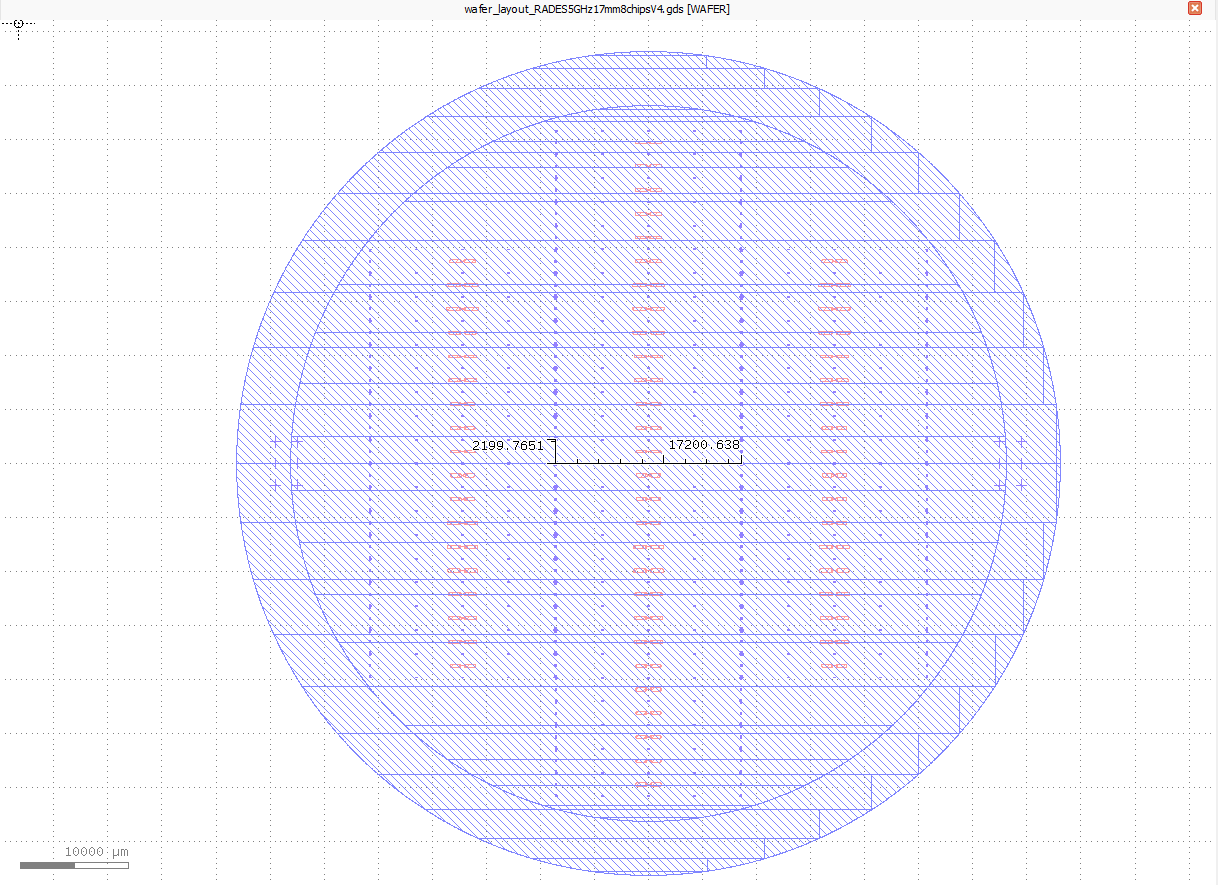}
    \caption{Wafer design with 64 transmons in 2 x 17 mm substrate. It was produced from a standard 0.43 mm thickness sapphire wafer.}
    \label{fig:WaferDesign}
\end{figure}

\begin{figure}
    \centering
    \includegraphics[width=0.8\linewidth]{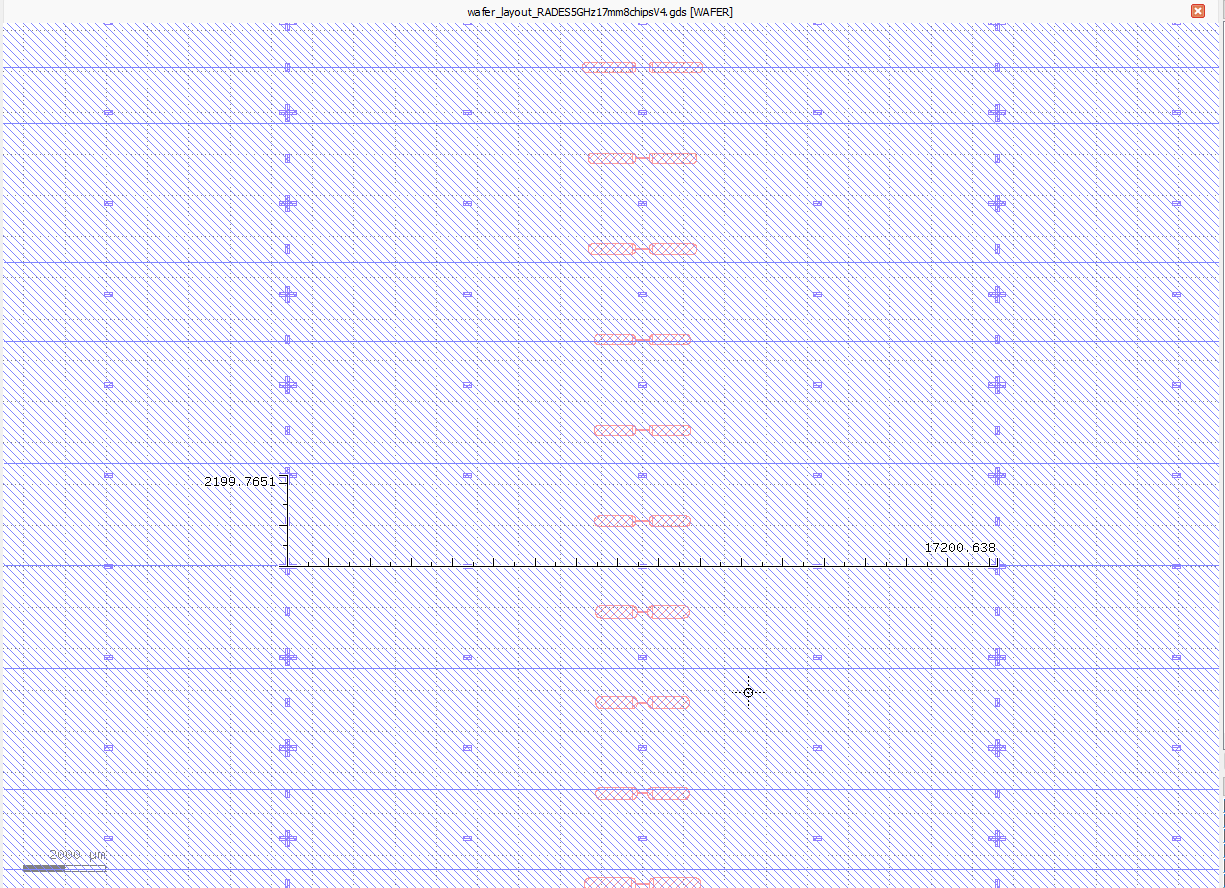}
    \caption{Close up of several transmons in the wafer.}
    \label{fig:WaferInset}
\end{figure}

The production is done in wafers, so 64 transmons were produced in the same batch. We took the opportunity of selecting multiple transmon geometries so that a wide range of possible behaviours were available. Eight were selected to have four copies of each, see table \ref{tab:TransmonDes}. In figures \ref{fig:WaferDesign} and \ref{fig:WaferInset} the final design of the wafer can be seen.

\begin{table}[] 
\centering
\begin{tabular}{c|cccccccc}
        & Q1   & Q2   & Q3   & Q4   & Q5   & Q6   & Q7   & Q8   \\ \hline
L1 (mm) & 1    & 1.15 & 1.25 & 1    & 1    & 1.15 & 1.3  & 1.1  \\
H1 (mm) & 0.3  & 0.3  & 0.3  & 0.3  & 0.25 & 0.25 & 0.25 & 0.25 \\
L2 (mm) & 0.15 & 0.12 & 0.15 & 0.12 & 0.15 & 0.15 & 0.15 & 0.12 \\
H2 (mm) & 0.02 & 0.02 & 0.02 & 0.02 & 0.02 & 0.02 & 0.02 & 0.02
\end{tabular}
\caption{Dimensions for eight different transmon configurations. See figure \ref{fig:TransmonDesign2} for dimension labeling.}
\label{tab:TransmonDes}
\end{table}

\subsubsection{Black box simulation}
The design of the single-photon counter must consider both the resonators and the qubit as an integrated system. So far, the theoretical framework for the qubit and the resonator has been presented in chapter \ref{QuantumTech}. The finite element simulation of the double-cavity structure was shown earlier in this chapter, and the qubit itself can also be simulated. 

However, the key aspect from the design point of view is the interaction between these two elements of the single photon counter. The Hamiltonian model can handle resonances from the cavity and excitations of the qubit at the same time. The mutual interaction affects the overall performance of the device. To handle this, what we call \textit{Black Box Quantization} \cite{nigg2012black} was developed, with the aim to use the information from the finite element simulation and include cavity and transmon modes in the same Hamiltonian that can be solved analytically.

It is based on a linearization procedure in which the whole system is modelled as an RLC circuit plus a non-linear contribution arising from the transmon. The Josephson junction is partitioned into its linear and non-linear components. The linear parts are grouped with the rest of the linear elements of the circuit and their effective impedance can be computed. The impedance from the linear components is extracted from the HFSS simulation. Then, the non-linear terms are added to the impedance employing Foster's theorem to calculate the total impedance as the sum of the effective lumped impedance of the modes $p$ of the circuit:

\begin{equation}
    Z(\omega) = \sum_{p=1}^{M} \left( j\omega C_p + \frac{1}{j\omega L_p} + \frac{1}{R_p} \right)^{-1}
\end{equation}

\begin{comment}
    \begin{figure}[h]
    \centering
    \includegraphics[width=0.7\linewidth]{images/CircuitFoster.png}
    \caption{Figure from \cite{nigg2012black}}
    \label{fig:CircuitFoster}
    \end{figure}
\end{comment}

As it is easier to work with zeroes instead of poles, admittance will be used $Y(\omega) = Z(\omega)^{-1}$. The resonant frequencies of the linear circuit are given by the imaginary parts of the poles of the impedance, or conversely the zeroes of the admittance, while the real parts give the width of the poles or, effectively, the decay rate.

From the finite element simulation software one can extract the $Y_c(\omega)$ seen from the qubit location. This allows to add the non-linear contribution from the Josephson junction as perturbation terms:

\begin{equation}\label{eq:AdmittanceComponents}
    Y(\omega) = j\omega C_J - \frac{j}{\omega L_J} + Y_c(\omega)
\end{equation}

\begin{figure}[h]
    \centering
    \includegraphics[width=0.8\linewidth]{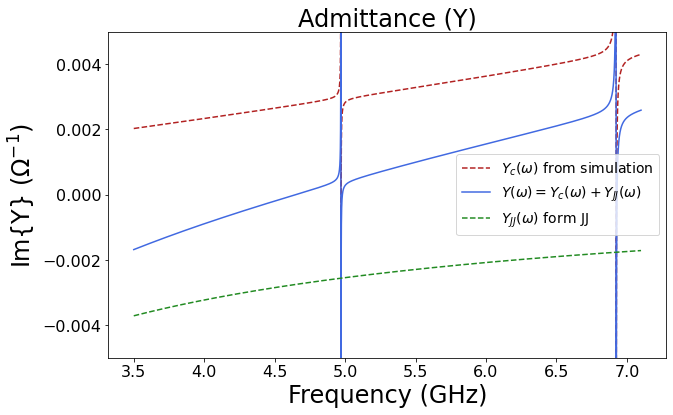}
    \caption{Imaginary part of the admittance by components, see equation \ref{eq:AdmittanceComponents}.}
    \label{fig:AdmitanceComponents}
\end{figure}

Zeroes of this admittance function set the modes of the system. And from admittance data, equivalent RLC components for each mode can be computed. 

\begin{figure}[h]
    \centering
    \includegraphics[width=0.9\linewidth]{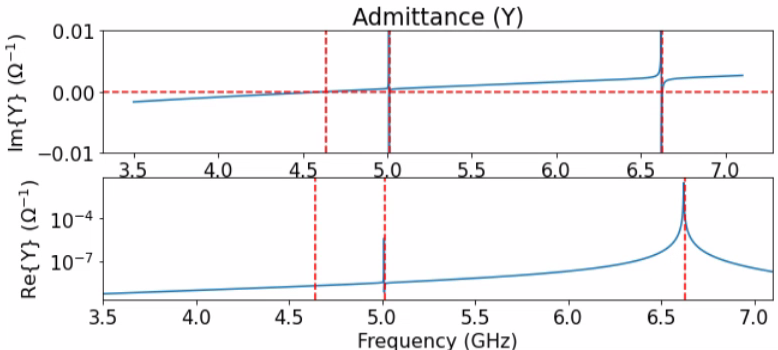}
    \caption{Imaginary and real parts of the admittance. Red lines denote the resonances of the system, from lower to higher frequencies: qubit, storage cavity, readout cavity.}
    \label{fig:AdmittanceSimulation}
\end{figure}

This RLC equivalent allows to construct the non-linear quantized Hamiltonian expanding up to certain order in the quantum flux operator. This Hamiltonian has infinite dimension so it has to be truncated to compute eigenvalues. Typically, 3 modes are considered -qubit, readout cavity, storage cavity- with 10 levels per mode. Solving the Hamiltonian means extracting their eigenvalues and eigenvectors. This is the key information in order to compute qubit transition frequencies, anharmonicity, couplings, relaxation and dephasing times, etc.

The Hamiltonian takes the form shown in expression \ref{H_CPB}. In order to extract eigenvalues, it is truncated -working at order 8 is more than enough to account for non linearities- and solved with the QuTiP \cite{lambert2024qutip5quantumtoolbox} library.

\begin{equation}\label{HQutip}
    H = H_0 - \frac{1}{24}\frac{E_J}{\hbar}H_{nl}^4 + \frac{1}{720}\frac{E_J}{\hbar}H_{nl}^6 - \frac{1}{40320}\frac{E_J}{\hbar}H_{nl}^8
\end{equation}

\begin{align*}
    H_0 &= \sum_n \omega_n \hat{a}_n\hat{a}^\dagger_n \\
    H_{nl} &= \sum_n \phi_n (\hat{a}_n + \hat{a}^\dagger_n) \\
    \phi_n &= \sqrt{\frac{2e^2}{\hbar}Z_n} = \sqrt{\frac{2e^2}{\hbar} \sqrt{\frac{L_n}{C_n}}}
\end{align*}

Equation \ref{HQutip} is the explicit form of the equations solved using QuTiP. The code used for this can be found here \cite{CodeBlackBox}. Three modes and 10 levels for each mode are considered, this makes $10^3$ eigenstates, considering all combinations of possible photon content in each mode.

It is relevant to stress that the term of order 2 in the cosine expansion is included in $H_0$. The expansion up to this term has the shape of a harmonic oscillator, $H_{CPB} \simeq 4E_C\hat{N}^2 +\frac{E_J}{2}\hat{\varphi}^2$.  Quantizing it and expressing in terms of creation and annihilation operators results in $H = \omega(\hat{a}^\dagger\hat{a}+\frac{1}{2})$, where in terms of simplicity the constant term is neglected. In chapter 2 of \cite{naghiloo2019introduction} and in \cite{dixit2021thesis} further details can be found.

In summary, from $Y(\omega)$ simulated in HFSS and values for $L_J$ and $C_J$ from the fabrication process, mode capacitances, impedances and resistances can be computed:
\begin{align*}
    C_n &= \frac{1}{2} \, \textrm{Im} \, Y'(\omega_n) \\
    R_n &= \dfrac{1}{\textrm{Re} \,Y(\omega_n) }\\
    L_n &= \dfrac{1}{2\omega_n^2 \, \textrm{Im} \, Y'(\omega_n)}
\end{align*}

From this, the truncated Hamiltonian can be solved, which gives the energy spectrum of the hybrid system. This allows to compute coupling strengths between modes, Stark shifts, frequencies, anharmonicity... even estimations for T1 coherence time. 

If $e_{qsr}$ is the eigenvalue of the Hamiltonian with $q$ excitations in qubit mode, $s$ in storage cavity mode and $r$ in readout cavity mode, then most of these parameters are obtained comparing energies between some of these levels. For example, anharmonicity of the qubit is defined as the difference between 2 times the energy of the first excited state minus the energy of the second excited state:

$$\alpha_q = 2e_{100}-e_{200}$$

Similar for the Stark shift, the difference in energy transitions for the cavity modes when the qubit is excited or not:
\begin{align*}
2\chi_{qs} = (e_{010} - e_{000}) - (e_{110} -e_{100})\\
2\chi_{qr} = (e_{001} - e_{000}) - (e_{101} -e_{100})
\end{align*}

And then the coupling strengths are given by:
\begin{equation}
    g_n = \sqrt{\chi_{qn} \Delta_{qn} \frac{\Delta_{qn} + \alpha_q}{\alpha_q}}
\end{equation}

With  $n \in \{s,r\}$ and $\Delta_{qs} = e_{100} - e_{010}$, $\Delta_{qr} = e_{100} - e_{001}$.

These parameters are the ones that need to be tuned during the transmon design stage. A proper non-linear device in this context should have an anharmonicity of few hundreds MHz (200-300 MHz), $2\chi_{qs}$ of couple of MHz (1-3 MHz) and bigger than $2\chi_{qr}$ and $E_J/E_C$ of around 100. $L_J = 12$ nH and $C_J=3.5$ pF were defined by fabrication and were not modified, they are related to the specific geometry of the Josephson junction. Target values are not perfect due to fluctuations in the manufacturing process, which may vary substantially depending on the provider. The design process from our side was related to the geometry of the cavities and the transmon pads.

\subsubsection{Transmon fabrication}

Transmons for this experiment were produced in the National Institute of Standards and Technology (NIST) in Boulder, Colorado, EEUU, by our colleague Dr. Akash Dixit. The fabrication technique is described in detail in his thesis \cite{dixit2021searching}. Also very useful are the diagrams in  \cite{naghiloo2019introduction} where an instructive explanation of the fabrication process can be found. 

\begin{figure}[h!]
    \centering
    \includegraphics[width=0.6\linewidth]{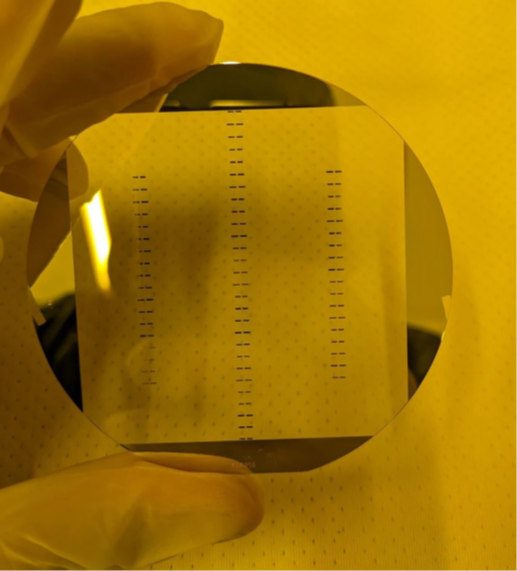}
    \caption{Wafer with transmons before slicing.}
    \label{fig:WaferTransmons}
\end{figure}

The two main features of these devices, the pads and the Josepshon junction, are produced in two steps. For the growth of the antennas optical lithography is used, while for the Josepshon junction electron-beam lithography is needed. Transmons are fabricated on 430 $\upmu$m thick C-plane (0001) sapphire wafers with a diameter of 50.8 mm. Several transmon structures can be developed in a wafer at the same time. Typically, better resolution is achieved in the centre of the wafer due to increasing non-orthogonality of the beams when you move further off the centre. 

Before starting, wafers are cleaned with organic solvents and annealed at 1200ºC for 1.5 hours. Fabrication starts by the pads that are the biggest features of the transmon. They are produced through optical lithography. First, a layer of 75~nm niobium is deposited in the wafer by electron beam evaporation. A resistor layer is developed over the niobium in which the pad's shapes are carved with a focused beam of light (wavelength 405 nm). Another resistor is developed on top of it, filling the patterns drawn in the first resistor. Features are finally etched in an inductively-coupled plasma (ICP) equipment from ``Plasma-Therm" removing all niobium under the first protective layer. Only shapes under the second resistive layer survive. The rest of the second layer photoresistor is removed by chemical means. 

Once the pads are ready, the wafer is cleaned again with organic solvents and baked in a vacuum chamber to remove any water excess and prevent outgassing. 

The Josephson junction is developed through electron-beam lithography with bi-layer resistor. The need to combine optical and electron-beam lithography arises from the size of the structure. While the pads have dimensions of the order of  hundreds of microns, the junction has structures with hundreds of nanometers: specifically, the width of the two wires that overlap to form the stack of layers, aluminium - aluminium oxide - aluminium, is of the order of 200 nm. 

This process starts by developing a bi-layer of resistors. The bottom one is thicker and softer and the outer one is thinner but harder in order to achieve sharper patterns.  Then, an argon ion mill is used to remove native oxide from the pads in the regions that will overlap with the junction. They are produced following the Manhattan pattern, in which two consecutive evaporation steps with oxidation in between generate the needed layers for the junction. The result of this process is a stack of tree layers: 35 nm aluminium, 1 nm aluminium oxide and 120 nm aluminium. Then, once the junction is ready, the remaining resistive is removed through a chemical lift-off process. After both the evaporation and lift-off, the device is exposed to an ion-producing fan to avoid electrostatic discharge of the junctions. 

\begin{figure}[h!]
    \centering
    \includegraphics[width=0.5\linewidth]{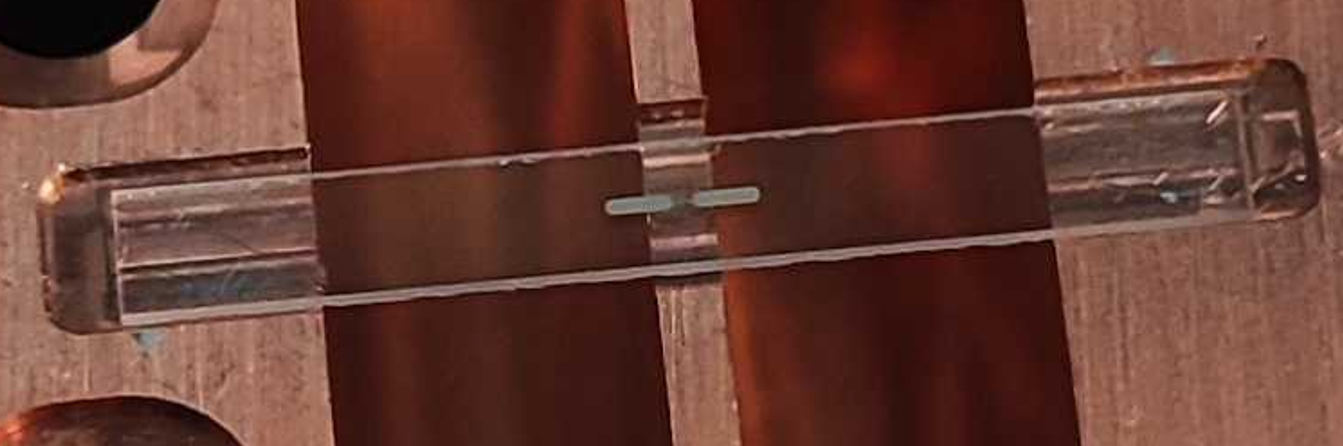}
    \caption{Final shape of a single transmon after slicing it from the sapphire wafer.}
    \label{fig:TransmonChip}
\end{figure}

\section{Measurement logic}

The purpose of these developments is to build a quantum sensor, able to detect single photons in the microwave range, with the aim to apply it to the search of relic axions. 

This quantum sensor makes use of the hybridization of quantum states in the qubit with resonator modes of the storage cavity, in a way that allows to check the photon content of the inner cavity without relying on a single quantum measurement. This is achieved through non-demolition measurements, that instead of checking only once the state of the qubit, make use of the different oscillation frequency of hybrid states depending on the photon content of the cavity. 

\begin{figure}
    \centering
    \includegraphics[width=0.9\linewidth]{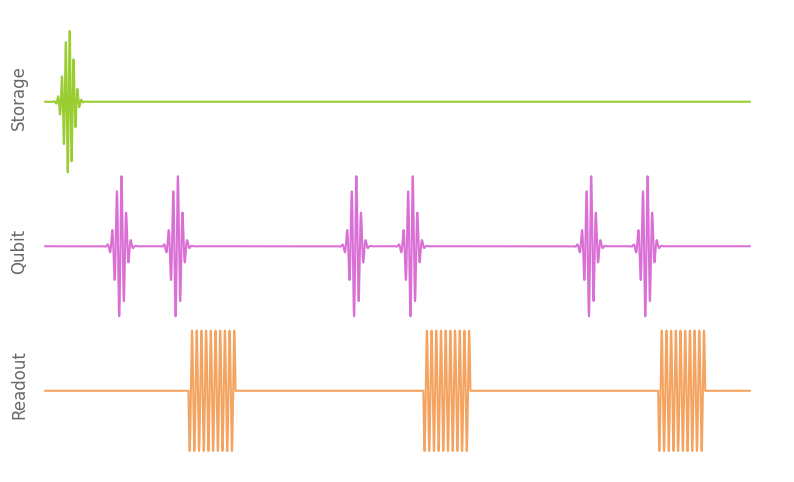}
    \caption{Representation of a sequence of parity measurements. The storage signal represents the photon coming from axion conversion: it can appear any time during the measurement but only appearances prior to the parity sequence will be efficiently detected. Each of the three channels has its frequency, pulse characteristics, etc. This plot is not to scale regarding times or pulse amplitudes. For a proper comparison between pulse durations see figure \ref{fig:PulseSequenceVisualization}.}
    \label{fig:PulseSequence}
\end{figure}

The shift in qubit frequency due to the presence of a photon, $2\chi$, has to be characterized before the measurements begin. The parity measurement, whose sequence of pulses can be seen in figure \ref{fig:PulseSequence}, follows an interferometric schema in which two $\pi/2$ pulses are performed with a certain amount of time between them. This waiting time, called parity time $t_p = \pi/|2\chi|$ is precisely determined from the measured shift. The qubit has to be initialized in a Fock state $|g\rangle$ or $|e\rangle$, so that the first $2\chi$ pulse moves it to a superposition state $1 / \sqrt{2}(|g\rangle \pm|e\rangle)$. The qubit state precesses at a rate of $|2\chi|$ when there is a photon in the cavity so waiting exactly $t_p = \pi/|2\chi|$ the state has acquired a phase of $\pi$ in the Bloch sphere representation. In that moment, the second  $\pi/2$ in the reverse direction, so $-\pi/2$, pushes it back to a Fock state. If there is a photon in the cavity, this moves the state of the qubit from  $|g\rangle$ to $|e\rangle$ or vice versa. The qubit state is measured through a dispersive drive, called "readout pulse". An alternating pattern is achieved if several measurements like this are performed. This schema only works with the presence of a photon in the cavity. If not, the interferometry is out of frequency and the same state is measured at the end of the pulse sequence. 

The advantage of this pulse sequence is that quantum false positive errors are exponentially suppressed at the cost of linearly increased time of measurement. 

The time length of the pulses and waiting times have to be tailored depending on the performance of the device. Relaxation and dephasing times of cavities and qubit, as well as coupling values, define how fast the readout and $\pi/2$ pulses can be and how long one has to wait between parity measurements in order to empty the readout cavity.

\begin{figure}
    \centering
    \includegraphics[width=0.99\linewidth]{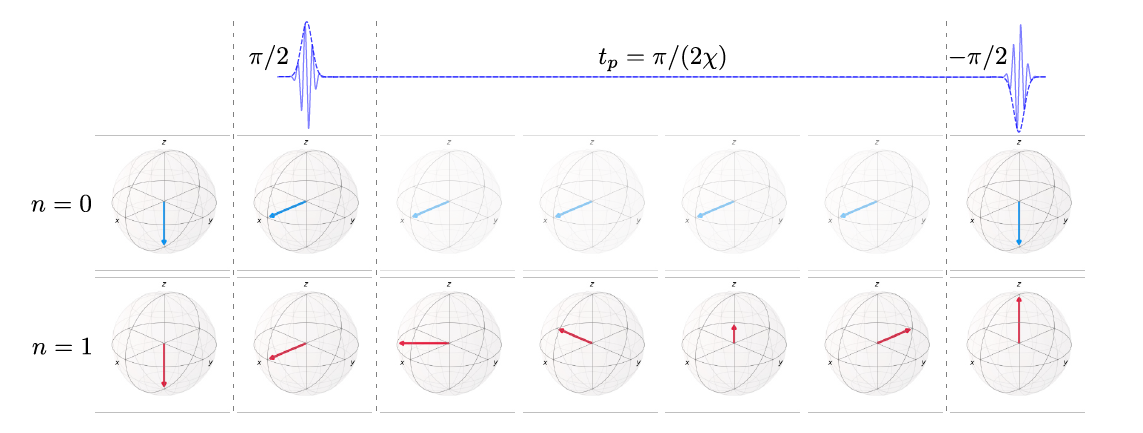}
    \caption{Effect of pulses of a parity measurement according to the photon content in the storage cavity. If the storage cavity is empty, the final state of the qubit is the same as in the beginning; when there is a photon, the state flips \cite{dixit2021thesis}.}
    \label{fig:ParityMeasurement}
\end{figure}

After every couple of parity pulses a readout tone at the frequency of the readout resonator is sent. With this tone, the state of the qubit is measured. The output of this measurement is a complex tone coming out of the system. It can be plotted in the complex plane (as well as in the amplitude and phase plane if needed, which is commonly done when measuring a resonance with a VNA). We call IQ plane representation the usual representation of a complex number in the complex plane. The name comes from "in phase" (I) and "quadrature" (Q) due to the expression

$$ x = I + Q i = r\cdot e^{\theta i}$$

with $x\in \mathbb{C}$ and $Q, I \in \mathbb{R}$, $r \in [0, +\infty)$, $\theta \in [0,2\pi)$. Of course, from Euler's formula: $I = r \cos(\theta)$ and $Q = r \sin(\theta)$.
\\

\begin{figure}[h!]
\centering
\includegraphics[width=0.9\linewidth]{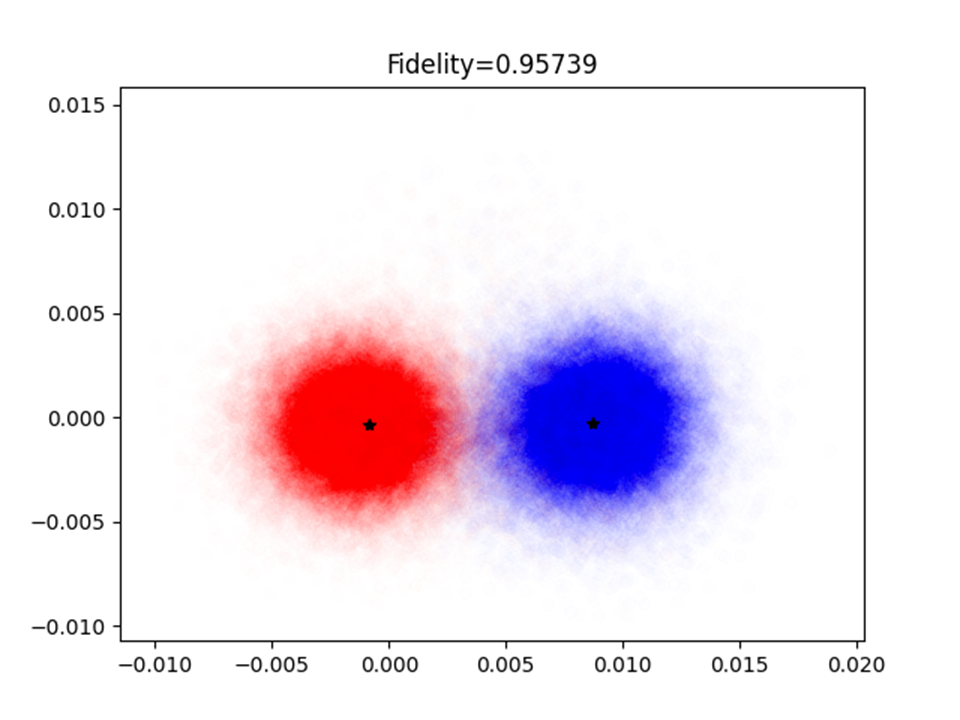}
\caption{Thousands of measurements represented in the IQ plane. Each dot is a single measurement of a state prepared in ground (blue) or first excited state (red). The readout fidelity is extracted from the percentage of times an excited state is correctly read.}
\label{fig:OneShot}
\end{figure}

\textbf{Markov chain analysis to estimate occupation}

Data is stored in form of arrays of $\{0,1\}$. If each parity sequence has $N$ measurements, a state of the qubit is measured for everyone of them. The precise IQ values that represent excited or ground states are calibrated beforehand. In figure \ref{fig:OneShot} such a calibration can be seen. States were prepared beforehand and measured right afterwards. The blue population are measurements with qubit in the ground state, red when qubit was prepared in the first excited state. If needed, measurements can be rotated so that differences between populations are maximized in the I variable, in this case I in X axis. In this way, the value of I in each measurement allows to distinguish between states. This split is not complete, as can be seen in the figure both populations slightly overlap. In this case, fidelity, the probability to measure exited state when the qubit is prepared in first excited state, is 0.95. Values above 0.9 are common in superconducting qubits.

\begin{figure}[h]
    \centering
    \includegraphics[width=0.9\linewidth]{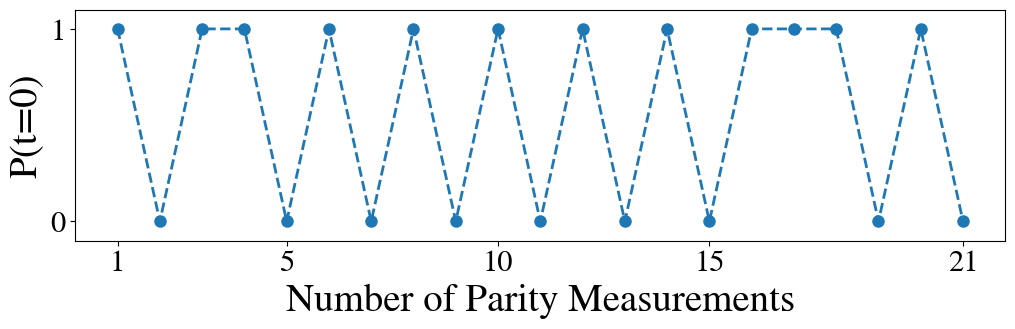}
    \caption{Parity sequence with 21 measurements. In this case, a photon is present in the storage cavity, so the sequence moves the states up and down, from ground to excited states. Several measurement errors can be seen along the sequence, where suddenly no oscillations happens.}
    \label{fig:ParitySequence}
\end{figure}

Once a parity sequence has been performed, it has to be analysed to determine if it corresponds to the presence of a photon in the storage cavity. This is done through a Bayesian analysis. A structure called "hidden Markov model" is used to determine the initial cavity state according to the measurements performed afterwards. The analysis tracks all the possible qubit and cavity states and their possible errors during measurements that would result in the observed sequence of readouts. With this technique, an exponential suppression of detector false positives is achieved by making repeated measurements of the same photon.

The idea of a "hidden Markov model" structure lies in the possible hidden states of the system that can produce certain observation. Knowing the evolution between states and the accuracy of the measurement, a probability for the initial state can be computed. This value is refined with every measurement done in the sequence. 

Hidden states are composed by states of the qubit, $q_i$ and storage  cavity $n_i$: 

\[
\begin{array}{c}
q_i \in \{g,e\} \\ 
n_i \in \{0,1\}
\end{array}
\]

Observations are the measurements over the qubit state. They do not necessarily coincide with the real state of the qubit, sometimes quantum measurements have errors, so a different space can be defined with the observed states that we call readout space: 

$$R_i \in \{\mathcal{G}, \mathcal{E}\}$$ 

The system evolves after every measurement, which is modelled through a transmission matrix $\mathbf{T}$ that depends on the characteristics of the device. Each observation may be affected by imperfect readout fidelity, which is captured by the emission matrix $\mathbf{E}$.

\begin{equation}
\mathbf{T} = 
     \begin{blockarray}{c}
         |0 g\rangle   \quad\quad |0 e\rangle \quad\quad |1 g\rangle \quad\quad |1 e\rangle  \\[2pt]       
\left[\begin{array}{cccc}
P_{00} P_{g g} & P_{00} P_{g e} & P_{01} P_{g e} & P_{01} P_{g g} \\[2pt]
P_{00} P_{e g} & P_{00} P_{e e} & P_{01} P_{e e} & P_{01} P_{e g} \\[2pt]
P_{10} P_{g g} & P_{10} P_{g e} & P_{11} P_{g e} & P_{11} P_{g g} \\[2pt]
P_{10} P_{e g} & P_{10} P_{e e} & P_{11} P_{e e} & P_{11} P_{e g}
\end{array}\right] \end{blockarray}\begin{aligned}
& |0 g\rangle \\
& |0 e\rangle \\
& |1 g\rangle \\
& |1 e\rangle
\end{aligned}
\end{equation}

\begin{equation}
\mathbf{E}=\frac{1}{2} 
     \begin{blockarray}{c}
         \mathcal{G} \quad \quad \mathcal{E} \\[2pt]
\left[\begin{array}{cc}
F_{g \mathcal{G}} & F_{g \mathcal{E}} \\[2pt]
F_{e \mathcal{G}} & F_{e \mathcal{E}} \\[2pt]
F_{g \mathcal{G}} & F_{g \mathcal{E}} \\[2pt]
F_{e \mathcal{G}} & F_{e \mathcal{E}}
\end{array}\right] \end{blockarray}\begin{aligned}
& |0 g\rangle \\
& |0 e\rangle \\
& |1 g\rangle \\
& |1 e\rangle
\end{aligned}
\end{equation}

These two matrices relate initial states, in the left column, to final state of the transmission or the measurement, in upper row. Elements of these matrices are obtained through dedicated characterization measurements in the device. In the next paragraphs, the origin of each term and its calibration is described.

\subsubsection*{Cavity transitions}
The states of the storage cavity are measured in number of photons. The probabilities of these transitions are related to the time spent in each parity measurement, $t_m$, and to the ability to store photons in the cavity, so $T_1^S$ and $T_2^S$ have to be measured. The intrinsic photon population of the cavity, $\Bar{n}_c$, is also needed, which is related to the effective temperature of the device.

\begin{itemize}
    \item Decay: $P_{10} = |1\rangle \longrightarrow |0\rangle = 1 - e^{-t_m/T_1^S}$
    \item Heating: $P_{01} = |0\rangle \longrightarrow |1\rangle = \Bar{n}_c \left(1 - e^{-t_m/T_1^S} \right) $
    \item Dephasing: The Heisenberg uncertainty relation dictates that if the number of photons of a quantum state is known, the phase is completely unknown. Therefore coherence of the cavity state is not relevant for the Hidden Markov Model.
\end{itemize}

To obtain probabilities $P_{00}$ and $P_{11}$, where the state does not change, unitary relations are applied: $P_{00} + P_{01} = 1$ and $P_{11} + P_{10} = 1$.

\subsubsection*{Qubit transitions}
The qubit states used for parity measurements are superposition states and so they can be affected by dephasing, apart from decay and heating, and all three processes can produce readout errors. The parameters to characterize these probabilities are related to transmon properties. Decay and coherence times are needed,  $T_1^q$ and $T_2^q$. And also values for the parity sequence protocol: time of the parity measurement, $t_m$, and also the time between $\pi/2$ pulses in the parity measurement, $t_p$. For the dephasing, only shifts between these two pulses may affect, as the initial phase is not relevant (initial states are not superposed). 
\begin{itemize}
    \item Decay: $ P_{eg}^{\downarrow} = |e\rangle \longrightarrow |g\rangle = 1 - e^{-t_m/T_1^q}$
    \item Heating: $P_{ge}^{\uparrow}  = |g\rangle \longrightarrow |e\rangle = \Bar{n}_q \left(1 - e^{-t_m/T_1^q} \right)$
    \item Dephasing: $ P^{\phi} = 1 - e^{-t_p/T_2^q}$
\end{itemize}

Overall transition probabilities related to qubit states are composed by dephasing and decay or heating: 
$$ P_{ge} = P_{eg}^{\downarrow} + P^{\phi}$$
$$ P_{eg} = P_{ge}^{\uparrow} + P^{\phi}$$

For qubit residual population, the probability of having the qubit in the first excited state compared to ground state has to be computed. To do so, a Rabi measurement between the first and second excited state is performed twice. First, initializing the qubit state in $|e\rangle$ and second, without initializing, so as for it to be in the ground state most of the times. The comparison of amplitudes of both Rabi oscillations give the relation between $P(e)$ and $P(g)$. In this way,  $r = \dfrac{P(e)}{P(g)}$ is measured, and $\Bar{n}_q = P(e) = \dfrac{r}{r+1}$. The unitary relation $P(g) + P(e) = 1$ is used here. 

From this measurement, the effective temperature of the qubit can be computed: 
$$ \dfrac{P(e)}{P(g)} = r = e^{\frac{-\hbar w_q}{k_B T_q}} \Rightarrow k_B T_q = \dfrac{-\hbar w_q}{ln(r)}$$

\subsubsection*{Readout fidelities}
The elements of the emission matrix are directly related to the readout fidelities. $F_{g\mathcal{G}}$ represents the probability of reading the state $\mathcal{G}$ having the qubit state $|g\rangle$. This probability is independent of the state of the storage cavity so $F_{g\mathcal{G}} = P(\mathcal{G}|g) =
P(\mathcal{G}|g0)+P(\mathcal{G}|g1) = 2 P(\mathcal{G}|g0) = 2 E_{\mathcal{G}|g0}$ because both conditional probabilities are equal. Elements in the emission matrix, like $E_{\mathcal{G}|g0}$, are split between storage cavity - qubit states so each one of them can be expressed in terms of $\frac{1}{2}F_{ij}$. These readout fidelities are measured in advance as shown in figure \ref{fig:OneShot}. Again, the unitary relation is used: $F_{g\mathcal{G}} + F_{g\mathcal{E}} = 1$ and $F_{e\mathcal{G}} + F_{e\mathcal{E}} = 1$.

\begin{table}[]
\centering
\begin{tabular}{@{}ccc@{}}
\toprule
Parameter & Label & Measurement \\ 
\midrule
\midrule
Storage cavity decay time &  $T_1^S$  & 4 wave mixing, wait, $\pi_{ge}$ sweep, readout \\
Storage cavity dephasing time&   $T_2^S$   & 4 wave mixing , $\pi/2$, wait, $\pi/2$, readout \\
Storage cavity residual population& $\Bar{n}_c$ &  Parity measurements \\
Qubit decay time &   $T_1^q$    &  $\pi$, wait, readout   \\
Qubit dephasing time &   $T_2^q$    & $\pi/2$, wait, $\pi/2$, readout \\
Qubit residual population &   $\Bar{n}_q$    & Double Rabi $|e\rangle$-$|f\rangle$            \\
$\pi/2$ pulse between $|g\rangle$ and $|e\rangle$ & $\pi_{ge}/2$ & Rabi oscillations            \\ 
Parity measurement time length&   $t_m$    & Add $\pi$, $t_p$ and waiting times. \\
Waiting time between $\pi/2$ pulses  & $t_p$ & Stark shift\\
$|g\rangle$ readout fidelity&$F_{g\mathcal{G}}$ &  Wait, readout \\
$|e\rangle$ readout fidelity&$F_{e\mathcal{E}}$& $\pi$, readout \\
\bottomrule
\end{tabular}
\caption{Parameters for the Hidden Markov Model.}
\label{tab:ParametersMarkov}
\end{table}

\subsubsection{Hidden Markov Model}

Once measurements are done and evolution matrices are generated with the parameters in table \ref{tab:ParametersMarkov} calibrated, a Bayesian analysis is performed in order to estimate the probabilities of initial cavity population $P(n_0 = 0)$ and $P(n_0=1)$. It is based on a forward-backward algorithm that makes use of previous and future measurements to compute the probabilities of the i-th measurement \cite{dixit2021searching}:

\begin{equation}\label{eq:Markov}
P\left(n_0\right)=\sum_{|g\rangle, |e\rangle} \sum_{s_0 \in\left[\left|n_0, g\right\rangle,\left|n_0, e\right\rangle\right]} \sum_{s_1} \cdots \sum_{s_N} E_{s_0, R_0} T_{s_0, s_1} E_{s_1, R_1} \ldots T_{s_{N-1}, s_N} E_{s_N, R_N}
\end{equation}

For every parity sequence of $N$ measurements, the probabilities along all possible paths compatible with the measurements are added, to account for all combinations of hidden states and readout errors that may lead to such sequence of parity measurements. The hidden state reconstructed this way involves a combination of cavity and qubit states $\{ |0g\rangle, |0e\rangle, |1g\rangle, |1e\rangle \}$, so one must add over qubit states at the end to extract the cavity state. This is the motivation for the last sum in equation\ref{eq:Markov}.

As a Bayesian method, a prior state is needed to initiate the recursive algorithm. In all analysis presented here a flat prior is used. This is a vector that keeps the same probability for each of the four possible hidden states $\{ |0g\rangle, |0e\rangle, |1g\rangle, |1e\rangle \}$, so a flat prior means a vector like this: $(0.25, \, 0.25, \, 0.25, \, 0.25)$.

\begin{figure}
    \centering
    \includegraphics[width=0.9\linewidth]{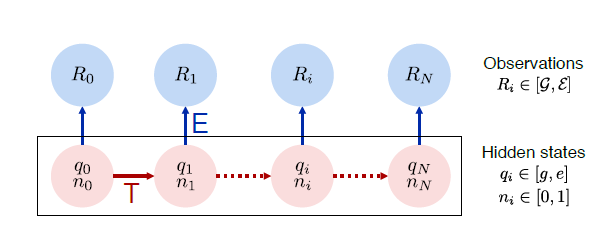}
    \caption{Hidden Markov Model schema \cite{dixit2021searching}.}
    \label{fig:MarkovModel}
\end{figure}

\begin{figure}
    \centering
    \begin{subfigure}{0.49\textwidth}
          \centering
          \includegraphics[width=.99\linewidth]{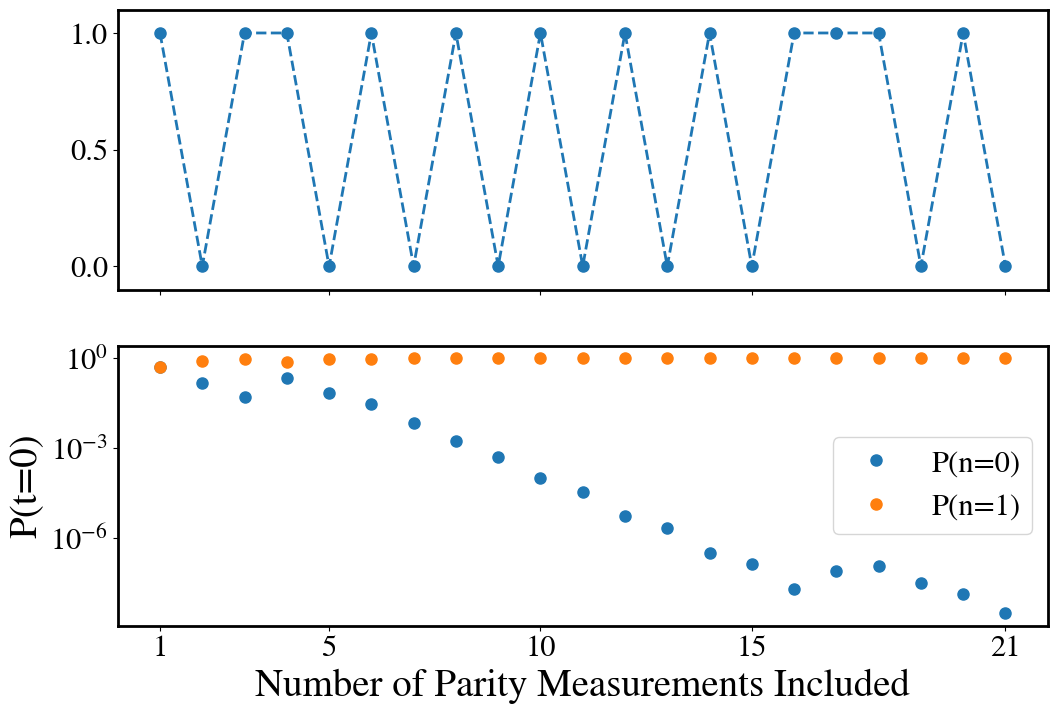}
          \caption{Photon in storage cavity.}
          \label{fig:ProbsEvent}
    \end{subfigure}\hfill
    \begin{subfigure}{0.49\textwidth}
          \centering
          \includegraphics[width=.99\linewidth]{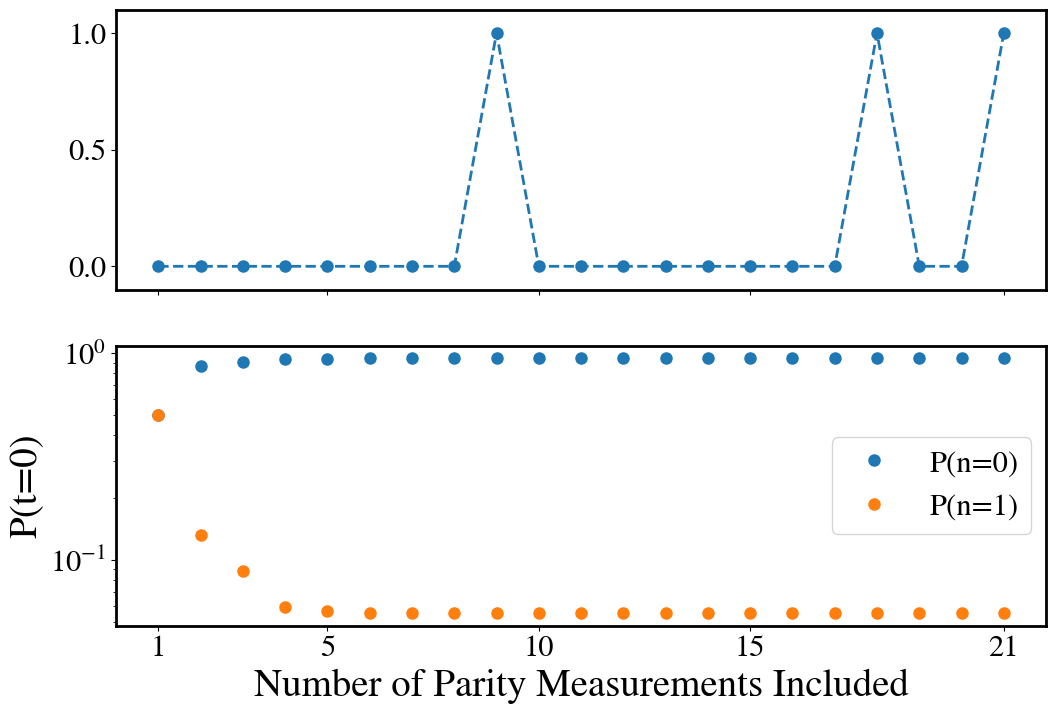}
          \caption{No photon in storage cavity.}
          \label{fig:ProbEventNoPhoton}
    \end{subfigure}
    \caption{Evolution of probabilities with $N$, number of parity measurements. \textit{Upper}: Measured state in every parity measurement. \textit{Lower}: Evolution of probabilities.}
    \label{fig:ProbEvent}
\end{figure}

For each sequence of measurements, one can get the probability of having 0 or 1 photon in the storage cavity just adding over qubit states: $P(n_0=0) = P_{|0g\rangle} +P_{|0e\rangle}$ and $P(n_0=1) = P_{|1g\rangle} +P_{|1e\rangle}$. This is shown for two events in \ref{fig:ProbsEvent} and \ref{fig:ProbEventNoPhoton}.
\begin{comment}
\begin{figure}[h]
    \centering
    \begin{minipage}{0.48\textwidth}
        \centering
        \includegraphics[width=0.9\linewidth]{images/ProbsEvent.png}
        \caption{Evolution of probabilities with the number of par for an event with a photon in storage cavity.}
        \label{fig:ProbsEvent}
    \end{minipage}%
    \begin{minipage}{0.48\textwidth}
        \centering
        \includegraphics[width=0.9\linewidth]{images/ProbEventNoPhoton.png}
        \caption{Evolution of probabilities with $N$ for an event with no photon in storage cavity.}
        \label{fig:ProbEventNoPhoton}
    \end{minipage}
\end{figure}
\end{comment}

In order to distinguish between photon contents, a likelihood test is created taking the quotient of both probabilities:

$$ \lambda = \dfrac{P(n_0=1)}{P(n_0=0)}$$

\begin{figure}[h]
    \centering
    \begin{minipage}[t]{.49\textwidth}
        \centering
        \includegraphics[width=0.9\linewidth]{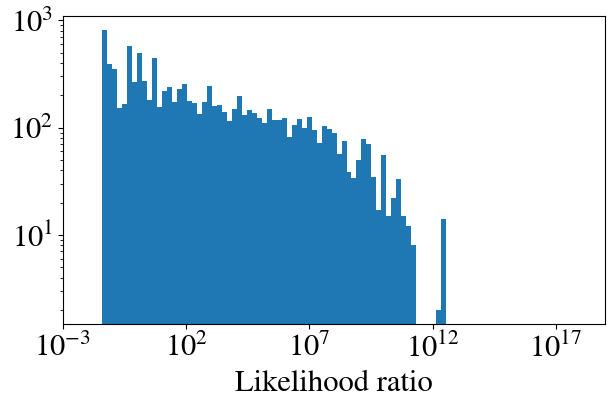}
        
    \end{minipage}\hfill
    \begin{minipage}[t]{.49\textwidth}
        \centering
        \includegraphics[width=0.9\linewidth]{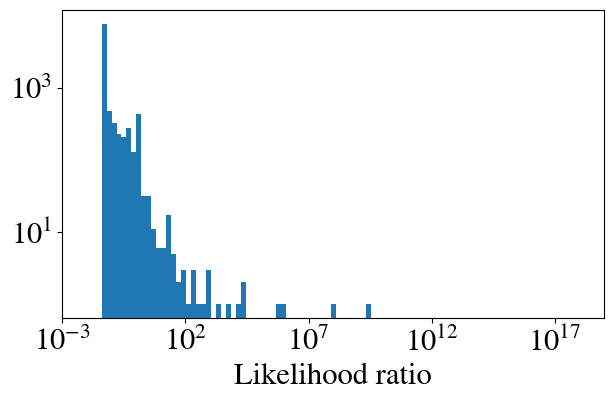}
        
    \end{minipage}
    \caption{\textit{Left}: Distribution of likelihood test for 10000 sequences injecting a photon in the storage cavity just before measuring. \textit{Right}: Distribution of likelihood test for 10000 sequences of background measurements.}\label{fig:LikelihoodDistribution}
\end{figure}

This variable has a distribution along several orders of magnitude. For a calibration run in which photons are injected before each parity measurement, the distribution of $\lambda$ is shown in figure \ref{fig:LikelihoodDistribution} left. The same distribution in a background run has a very different shape, see \ref{fig:LikelihoodDistribution} right. A criterium has to be chosen to decide from which threshold value of $\lambda$ a photon is present. This will be explained with experimental data in the next section.

\begin{comment}
\begin{figure}[h]
    \centering
    \begin{minipage}[t]{.49\textwidth}
        \centering
        \includegraphics[width=0.9\linewidth]{images/LikelihoodDistributionInjecting.png}
        \caption{Distribution of likelihood test for 1000 sequences injecting a photon in the storage cavity just before measuring.}
        \label{fig:LikelihoodDistributionInjecting}
    \end{minipage}\hfill
    \begin{minipage}[t]{.49\textwidth}
        \centering
        \includegraphics[width=0.9\linewidth]{images/LikelihoodDistributionNoPhotons.png}
        \caption{Distribution of likelihood test for 1000 sequences of background measurements.}
        \label{fig:LikelihoodDistributionNoPhotons}
    \end{minipage}
\end{figure} 
\end{comment}

\section{Performance of the single photon counter}

\begin{figure}[b!]
    \centering
    \includegraphics[width=0.8\linewidth]{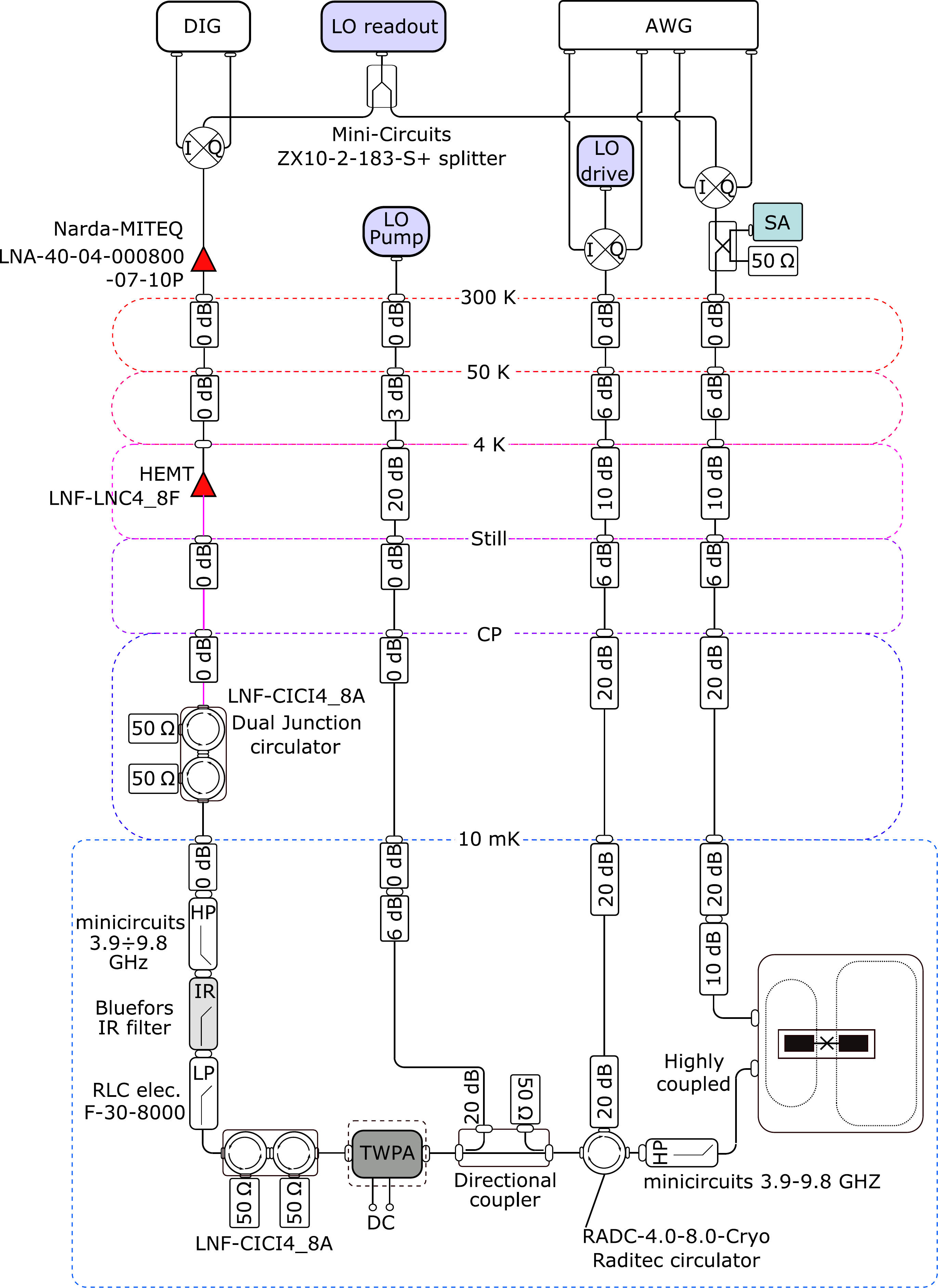}
    \caption{Layout of the input and output RF lines inside the cryostat during the parity measurements. }\label{fig:CryostatDP}
\end{figure}

The device described in this work was measured with the help of our collaborators in Aalto University, Helsinki, Finland. The aluminium cavity shown in figure \ref{fig:CavityMachined} was used with one of the transmons of figure \ref{fig:WaferTransmons} of the type Q1 identified in table \ref{tab:TransmonDes}. A dilution refrigerator Bluefors XLDsl was used to reach mK temperatures. Input and output lines, filters, circulators, amplification elements and RF equipment used during the measurements are depicted in figure \ref{fig:CryostatDP}.

Many parameters of the detector have to be characterized before parity measurements, summarized in table \ref{tab:Parameters}.  

\begin{table}[h!]
\centering
\begin{tabular}{@{}ccc@{}}
\midrule
Device parameter             & Symbol & Value \\ 
\midrule
\midrule
Transmon frequency $|g\rangle \longrightarrow |e\rangle$ & $w_q$ &   $2\pi \times 4.629$ GHz    \\
Transmon frequency $|e\rangle \longrightarrow |f\rangle$ & $w_{ef}$ &   $2\pi \times 4.357$ GHz  \\
Sideband qubit-storage frequency $|f0\rangle \longrightarrow |g1\rangle$ & $w_{f0-g1}$ &   $2\pi \times 3.935$ GHz  \\
Transmon anharmonicity       & $\alpha_q$ & $-2\pi \times 272$ MHz \\
Transmon decay time          & $T_1^q$ & 16.7 $\upmu$s \\
Transmon decoherence time    & $T_2^q$ & 6.2 $\upmu$s\\
Transmon background photon   & $\Bar{n}_q$ &   -    \\ \hline
Storage frequency            & $w_s$ &   $2\pi \times 5.051$ GHz      \\
Storage decay time           & $T_1^s$ & 20 $\upmu$s\\
Storage decoherence time     & $T_2^s$ &   -    \\
Storage-Transmon Stark shift & $2\chi$ & $-2\pi \times 4.6$ MHz \\
Storage background photon    & $\Bar{n}_s$ &    -   \\ \hline
Readout frequency            & $w_r$ &  $2\pi \times 6.867$ GHz       \\
Readout-Transmon Stark shift & $2\chi_{qr}$        &    -   \\
Readout fidelity $|g\rangle$ & $F_{g\mathcal{G}}$ & 0.95739\\
Readout fidelity $|e\rangle$ & $F_{e\mathcal{E}}$       &   -    \\ 
Readout time                 & $t_r$   &  800 ns     \\
Parity time                  & $t_p$   &  88 ns     \\
90º rotation pulse time      & $\pi /2$ & 52 ns \\
Waiting time between measurements     & $t_w$ & 100/1000 ns \\
Time between parity sequences & $t_t$ & 1 ms \\ \hline
\end{tabular}

\caption{Parameters of the single photon counter. Some of them were not measured, they are not crucial for the analysis but ideally they should be measured. }\label{tab:Parameters}
\end{table}

Four different configurations were tested in Aalto University with the same setup. The parity sequence allows a lot of fine tuning and these runs were the first attempts to explore possible variations. In each configuration two runs were recorded: calibration and background. Calibrations are runs in which, just before the parity sequence, a photon is injected to the storage cavity through the sideband $|f0\rangle \longleftrightarrow |g1\rangle$. These runs allow to quantify the efficiency of the protocol. In background runs only spontaneous photons should reach the storage cavity. 

\begin{itemize}
    \item \textbf{Run 1}: $\pi/2$ (52~ns) + $t_p$ (88~ns) + $\pi/2$ (52~ns) + Readout (800~ns) + 100~ns delay after each parity measurement. $t_m = 1.092$~$\upmu$s. Total parity sequence for $N=21$: $22.932$ $\upmu$s
    \item \textbf{Run 2}: $\pi/2$ (52 ns) + $t_p$ (88 ns)+ $\pi/2$ (52 ns)  + Readout (800 ns) + 100 ns delay after each parity measurement. $t_m = 1.092$ $\upmu$s. Total parity sequence for $N=21$: $22.932$ $\upmu$s
    \item \textbf{Run 3}: $\pi/2$ (52 ns) + $t_p$ (88 ns) + $\pi/2$ (52 ns) + Readout (800 ns) + 1000~ns delay after each parity measurement. $t_m = 1.992$ $\upmu$s. Total parity sequence for $N=21$: $41.832$ $\upmu$s
    \item \textbf{Run 4}: $\pi$ (184 ns) + Readout (800 ns) + 1000 ns delay after each measurement. $t_m = 1.984$ $\upmu$s. Total parity sequence for $N=21$: $41.664$ $\upmu$s

\end{itemize}

The first three runs are standard parity protocol with different waiting times between parity measurements. Runs 1 and 2 have a waiting time of 100~ns and Run 3 of 1000~ns.

Run 4 is different, instead of an interferometric measurement, the rotation from one state to the other is done in a single $\pi$ pulse. It is a pulse tailored in frequency, phase and amplitude that only works properly when the qubit frequency is shifted due to the Stark shift, i.e., when there's a photon in the storage cavity. In this case, the analysis has to be slightly modified due to the fact that there is no waiting time between pulses, $t_p = 0$, and therefore no dephasing is considered. 

\begin{figure}[h!]
    \centering
    \includegraphics[width=0.99\linewidth]{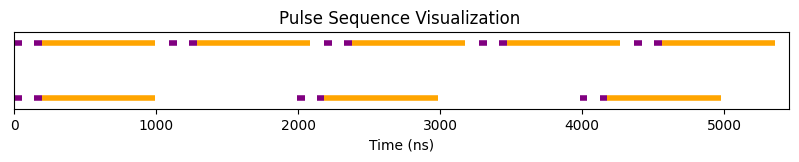}
    \caption{Different waiting times between parity measurements, upper series 100~ns (Runs 1 and 2) and lower 1000 ns (Run 3). Same $\pi/2$ (52 ns, in purple) and readout (800 ns, in yellow) pulses. Parity time delay $t_p = 88$ ns (gap between purple pulses). For $N=21$, total pulse sequence time is 22.932 $\upmu$s and 41.832 $\upmu$s.}
    \label{fig:PulseSequenceVisualization}
\end{figure}

\begin{figure}[h]
    \centering
    \includegraphics[width=0.9\linewidth]{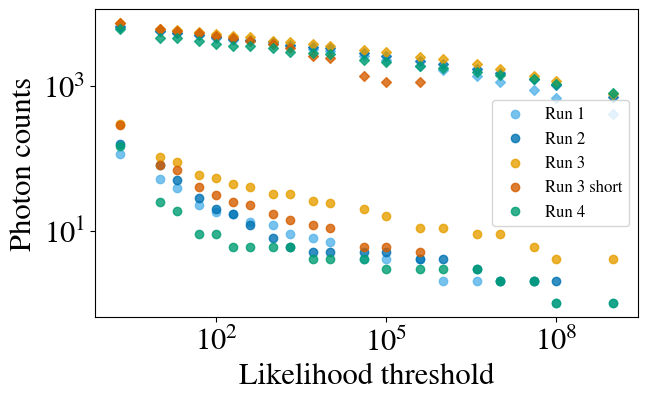}
    \caption{Photon count for all runs (10000 measurements per run). Upper trends with squares from calibrations, lower ones with dots from background. Run 3 was also analysed with shorter parity sequence, $N=11$ instead of $N=21$.}
    \label{fig:PhotonCount}
\end{figure}

For each run, the Hidden Markov Model is applied, and the study of the photon count is done depending on the threshold, see figure \ref{fig:PhotonCount}.

Calibration runs are performed together with background runs. As mentioned before, a photon is injected in the storage cavity just before the parity sequence. Doing so, the detector efficiency can be estimated, allowing for correction of the photon counts measured during background runs.

\begin{figure}[h]
    \centering
    \includegraphics[width=0.9\linewidth]{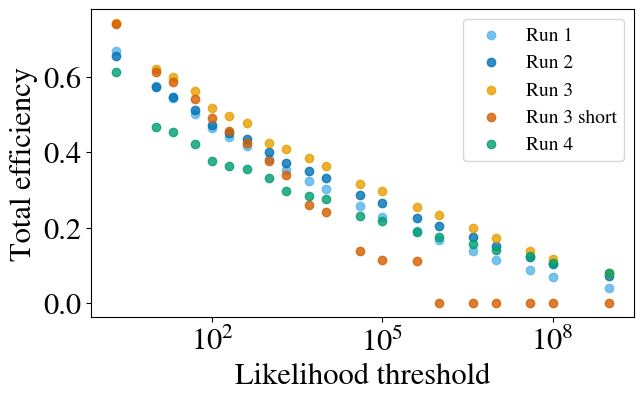}
    \caption{Efficiency as a function of the threshold.}
    \label{fig:Efficiency}
\end{figure}

\begin{figure}[h]
    \centering
    \includegraphics[width=0.9\linewidth]{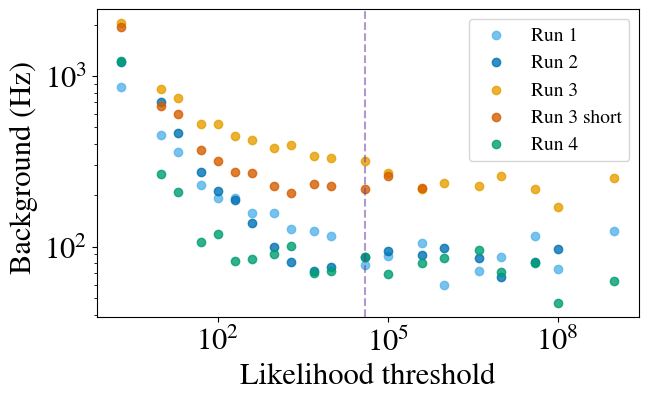}
    \caption{Evolution of measured background rate corrected by efficiency with the likelihood threshold. Rate is computed with $10000\times 0.00002$ s because the storage cavity relaxation time is $T_1^s=20$ $\upmu$s. Vertical line at proposed threshold of $\lambda_{th}=4 \times 10^4$.}
    \label{fig:BackgroundRate}
\end{figure}

To determine the background level measured with this setup implies several decisions that may not seem clear. The main one: how to set the threshold value? In \cite{dixit2021searching, zhao2025flux} a calibration is done beforehand in which a different amount of photons is injected in the storage cavity and measured. They fit the resulting relation to $\bar{n}_{meas}=\eta \, \Bar{n}_{inj}+\delta$, being $\eta$ the efficiency and $\delta$ the false positive probability. Then, the threshold can be established when the evolution of the false-positive, probability-corrected efficiency tends to stabilize. The lack of such an exhaustive calibration in our case forces us to fix the threshold from figure \ref{fig:BackgroundRate} using the criteria of flat background. Following this, the proposed threshold is $\lambda_{th}=4 \times 10^4$.

With this level, the background rate for each run is shown in table \ref{tab:BkgRate}.

\begin{table}
\centering
\begin{tabular}{ccc}
& Bkg  (Hz) & Efficiency   \\ \hline
Run 1 &  77.94  &   0.257 \\
Run 2 &  87.35  &   0.286  \\
Run 3 &  315.96 &   0.317 \\
Run 4 &  86.62  &   0.231   
\end{tabular}
\caption{Efficiency-corrected background rate for $\lambda_{th}=4 \times 10^4$ and intrinsic efficiency. }
\label{tab:BkgRate}
\end{table}

In \cite{dixit2021searching} the measured background rate was 1.1 Hz at a frequency of 6.011 GHz, and in \cite{zhao2025flux} it was 64 Hz at around 5.68 GHz. In our setup, the background level is comparable to this second result, being between 78 and 87 Hz at 5.051 GHz. However, in one of the runs, Run 3, the rate was much higher. This was unexpected and we have not found any clear explanation: the changes introduced in the readout protocol of this run regarded longer waiting times, which could have produced a reduction in efficiency and therefore readout rates.

\begin{comment}
A possible explanation would be that it is due to external photon sources: the setup was installed in a shared dilution refrigerator along with other samples that could have been the origin of the extra photons interfering with our measurement if they were measuring in this very moment. This could be a way to identify noise sources coming from different sources inside the refrigerator and the first step to mitigate them.  
\end{comment}

\section{Dark photon sensitivity}

The single photon counter provides for every measurement a binary response (yes, there is a photon, or no, there is not a photon). This means that at the end of the run, after setting the threshold and analysing the data, the result is $N$ counts over $N_{meas}$ measurements. 

In order to extract a dark photon limit from these measurements, assuming all events come from this source, an expression for the detection rate of dark photon is needed. 

\begin{equation}\label{eq:probDP}
    N_{DM} = \dfrac{dN_{DM}}{dt} T_1^s N_{meas} = \dfrac{\varepsilon^2\rho_{DM}Q_{DM}Q_sGV}{\omega_s} N_{meas}
\end{equation}

Equation \ref{eq:probDP} reflects the expected number of positive measurements ($N_{DM}$) over the number of total measurements ($N_{meas}$) due to the flux of dark photons. This expression is related with the mixing angle $\varepsilon$, $G=\dfrac{1}{3}\dfrac{2^6}{\pi^4}$ is the geometrical factor for the electromagnetic mode coupled to photons in the rectangular cavity \cite{dixit2021thesis}, $V$ is its volume, 9 cm$^3$, assumes Standard Halo Model with $\rho_{DM}=0.45$ GeV/cm$^3$ as the dark matter density, $Q_{DM}\approx 10^6$ represents the line-width of the expected dark matter signal,  $Q_s$ quality factor of the resonance and $\omega_s$ its angular frequency, with $T_1^s = Q_s / \omega_s$.

In order to compute the 95\% C.L. for $\varepsilon$, the cumulative probability of detecting $N$ positive signals or less is computed. The possible results of a single measurement is modelled as a binomial distribution with $n=N_{meas}$ and $p = \eta \dfrac{N_{DM}}{N_{meas}}$, being $\eta$ the efficiency of detection.

\begin{equation}
    P(\leq N) = 
    \sum_{k=0}^{N} \frac{N_{\text{meas}}!}{k!(N_{\text{meas}} - k)!} 
    \left( \frac{\eta \varepsilon^2 \rho_{\text{DM}} Q_{\text{DM}} Q_s G V}{\omega_s} \right)^k 
    \left( 1 - \frac{\eta \varepsilon^2 \rho_{\text{DM}} Q_{\text{DM}} Q_s G V}{\omega_s} \right)^{N_{\text{meas}} - k}
\end{equation}

\begin{figure}
    \centering
    \includegraphics[width=0.8\linewidth]{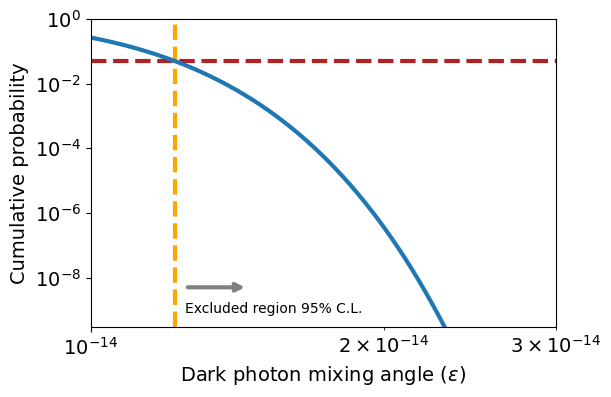}
    \caption{Cumulative probability as a function of mixing angle $\varepsilon$. When it decreases below 0.05, red dashed line, $\varepsilon$ values are excluded at 95\% confidence level. The intersection between the cumulative probability and the 0.05 level sets $\varepsilon^{95\%} = 1.4027 \cdot 10^{-14}$. Values used for this computation are shown in table \ref{tab:ExclusionDP}.}
    \label{fig:CumulativeProb}
\end{figure}

When this cumulative probability falls below 0.05 the corresponding $\varepsilon$ value sets the 95\% confidence level. In plot \ref{fig:CumulativeProb} this reasoning is visualized. 

This $\varepsilon^{95\%}$ can be translated into photon occupation number if needed thanks to equation \ref{eq:probDP} because $$n^{95\%} = \dfrac{N_{DM}(\varepsilon^{95\%})}{N_{meas}}$$

\begin{figure}
    \centering
    \includegraphics[width=1\linewidth]{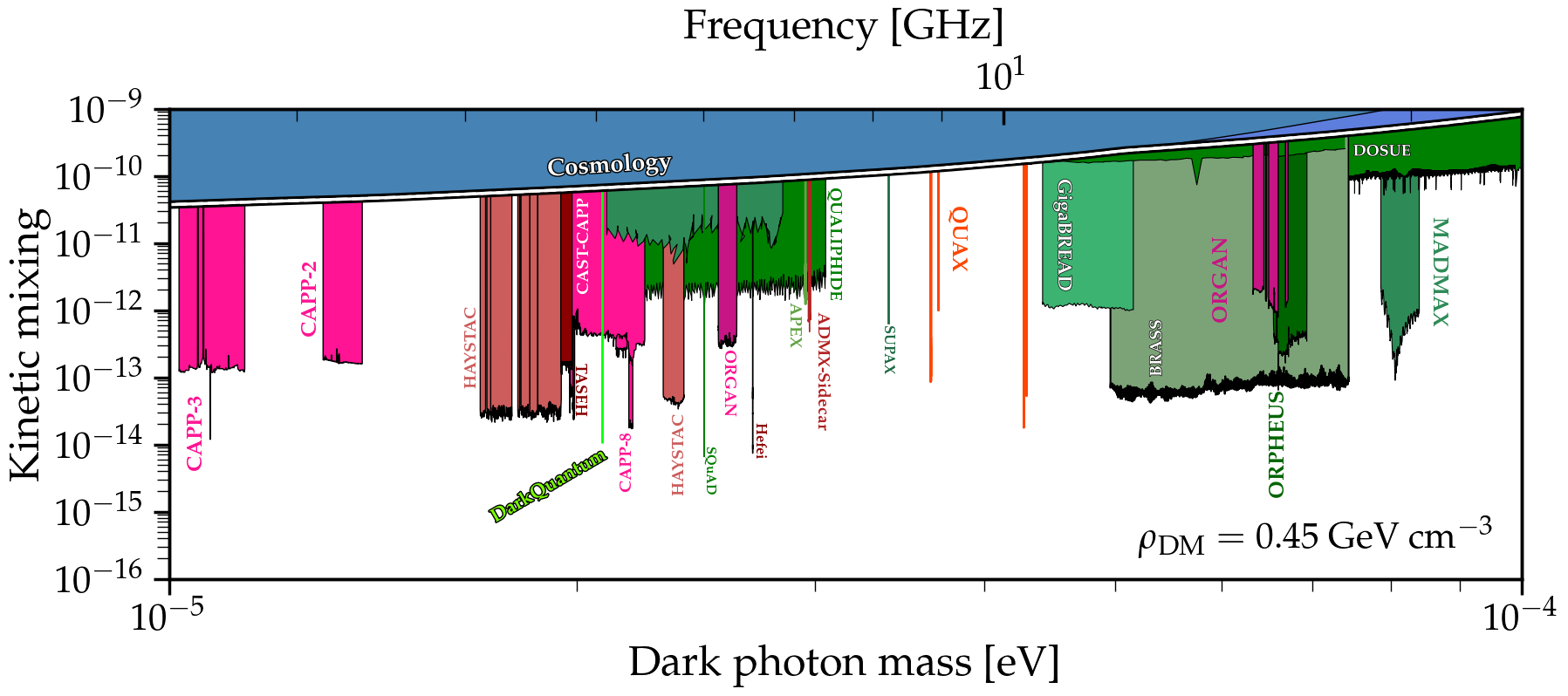}
    \caption{Exclusion limit for dark photons from the quantum single photon counter at 95\% C.L. shown in light green under the label DarkQuantum compared with previous experiments. Runs 1 and 2 presented in the text are used to compute this limit. Plot adapted from \cite{AxionLimits}.}
    \label{fig:DarkPhotonSensitivity}
\end{figure}

Table \ref{tab:ExclusionDP} shows the values used in this work to compute the exclusion limit from the measurements presented in the previous section, Run 1, together with values used for exclusion limit in \cite{dixit2021searching}. In that article, a 90\% confidence level was used but in order to compare with the usual 95\% confidence level of other experiments displayed in figure \ref{fig:DarkPhotonSensitivity} limits with both confidence levels have been computed. The result from \cite{dixit2021searching} appears in the figure under the label SQuAD.

\begin{table}[]
\renewcommand{\arraystretch}{1.3}
\begin{tabular}{l|c|c|c}

\hline
\multirow{2}{*}{\centering\textbf{Parameter}} & \textbf{A. Dixit} \cite{dixit2021searching} & \textbf{This work} & \textbf{This work} \\
 & \small{($\rho_{DM}=0.4$, 90\% CL)} & \small{($\rho_{DM}=0.4$, 90\% CL)} & \small{($\rho_{DM}=0.45$, 95\% CL)} \\
\hline
\hline
$f_s$ (GHz) & 6.011 & 5.051 & 5.051 \\
\hline
$Q_s$ & 20621451 (20.6M) & 634727 (0.6M) & 634727 (0.6M) \\
\hline
$N_{DM}$ & 9 & 4 & 4 \\
\hline
$N_{meas}$ & 15141 & 10000 & 10000 \\
\hline
Efficiency $\eta$ & 0.409 & 0.257 & 0.257 \\
\hline
$n^{CL}$ & 0.00242 & 0.00311 & 0.00358 \\
\hline
$\varepsilon^{CL}$ & $1.6799 \cdot 10^{-15}$ & $1.1402 \cdot 10^{-14}$ & $1.2217 \cdot 10^{-14}$ \\
\hline
\end{tabular}
\caption{Table with exclusion limit levels for the mixing angle from measurements in this work, for $CL = 90\%$ and $95\%$ confidence levels, compared with numbers in \cite{dixit2021searching}.}
\label{tab:ExclusionDP}

\end{table}

For the results explained above, only Run 1 was taken into account but it is possible to combine the results of several runs to extract a more stringent limit. This prototype is a non tunable device, so given any arbitrary sensitivity, the width is restricted to a very narrow line, therefore small area can be proved anyway. With this limitations in mind, the combined analysis has been restricted to Runs 1 and 2, that have exactly the same readout protocol. 

Two experiments following independent binomial distributions are considered, $B_1(n_1, p_1)$ and $B_2(n_2, p_2)$, with parameters $n_1 = N_{meas}^1$, $n_2 = N_{meas}^2$, $p_1 = \eta_1\dfrac{N_{Dm}^1}{N_{meas}^1}$ and $p_2 = \eta_2\dfrac{N_{Dm}^2}{N_{meas}^2}$. For each $\varepsilon$ tested, the cumulative distribution probability (CDF) is computed for each distribution and $N_{meas}^1$ and $N_{meas}^2$: 
\begin{align*}
P_1 &= \text{CDF}[B_1(N_{meas}^1 | n_1, p_1)] \\
P_2 &= \text{CDF}[B_2(N_{meas}^2 | n_2, p_2)]
\end{align*}

As both distributions are independent, the combined CDF is the product of both probabilities: $P_{Combined}= P_1P_2$ and therefore $\varepsilon$ values are ruled out with 95\% C.L. if $P_{Combined}(\varepsilon) \leq 0.05$, the same argument showed in figure \ref{fig:CumulativeProb}. The results of Runs 1 and 2 and the combined limit are shown in table \ref{tab:CombinedDPresults}. Figure \ref{fig:DarkPhotonSensitivity} includes this combined exclusion limit under the label DarkQuantum. In \cite{ThesisCodes} most of the software developed for this analysis can be found.

\begin{table}[]
\renewcommand{\arraystretch}{1.3}
\begin{tabular}{l|c|c|c}

\hline
\multirow{2}{*}{\centering\textbf{Parameter}} & \textbf{Run 1} & \textbf{Run 2} & \textbf{Runs 1 + 2} \\
 & \small{($\rho_{DM}=0.45$, 95\% CL)} & \small{($\rho_{DM}=0.45$, 95\% CL)} & \small{($\rho_{DM}=0.45$, 95\% CL)} \\
\hline
\hline
$f_s$ (GHz) & 5.051 & 5.051 & 5.051 \\
\hline
$Q_s$ & 634727 (0.6M) & 634727 (0.6M) & 634727 (0.6M) \\
\hline
$N_{DM}$ & 4 & 5 & 4 - 5 \\
\hline
$N_{meas}$ & 10000 & 10000 & 20000 \\
\hline
Efficiency $\eta$ & 0.257 & 0.286 & 0.257 - 0.286 \\
\hline
$n^{CL}$ & 0.00358 & 0.00368 & 0.00261 \\
\hline
$\varepsilon^{CL}$ & $ 1.2217 \cdot 10^{-14}$ & $ 1.2387 \cdot 10^{-14}$ & $ 1.0446 \cdot 10^{-14}$ \\
\hline
\end{tabular}
\caption{Table with exclusion limits for the mixing angle of dark photons at 95\% C.L. from Runs 1, 2 and the combined analysis (explained in the text) using a quantum sensor as single photon counter.}
\label{tab:CombinedDPresults}

\end{table}

This prototype does not have a frequency tuning system. In the future, once such a system is implemented, background counts will be measured also in adjacent frequency bins. This will allow background subtraction from these adjacent bins, highlighting possible signal contributions as increases over global background level, which enables more stringent exclusion limits.

The experiment described in this chapter represents the most sensitive measurement to date of dark photons at this frequency. However, the most significant achievement of this work lies in the successful implementation of a single-photon counter based on a superconducting qubit. This accomplishment required a coordinated effort across multiple disciplines in physics and engineering: transmon fabrication, design of the double cavity, its operation at ultra-low temperature, the RF signal generation for the parity measurement and finally the analysis of the results to produce this exclusion limit.

This work marks the first step within a collaboration that aims to develop a new approach to axion haloscopes. This challenge will require the integration of a frequency-tuning mechanism, a high-Q cavity, and most critically, a quantum device capable of operating in the presence of magnetic fields. Ongoing research efforts include the exploration of novel materials, advanced shielding techniques, faster readout architectures, optimized quantum state manipulation, and many other innovations that will enable the development of a detector capable of probing deeper and faster than ever before.

%% file: Chapters/8_Conclusiones.tex
\renewcommand{\thefigure}{SC.\arabic{figure}}
\setcounter{figure}{0}

Dark matter is still one of the biggest challenges to our understanding of the Universe. This thesis has presented a comprehensive exploration of experimental and theoretical approaches towards its comprehension and detection. Beginning with the foundational motivations for dark matter searches and culminating in state-of-the-art quantum sensor implementations, the work spans from traditional gaseous detector development to superconducting qubit-based single-photon counters. 
In this final chapter, the content of each of the previous ones is summarized with emphasis on the experimental results and their implications for the broader field of particle and astroparticle physics.
\\

The thesis opens in Chapter 1 by reviewing the compelling astronomical and cosmological evidence pointing toward the existence of dark matter. Observational data from galaxy rotation curves, gravitational lensing in galaxy clusters and anisotropies in the cosmic microwave background all support the presence of a non-luminous, non-baryonic form of matter that constitutes approximately 27\% of the Universe's energy content. The limitations of the Standard Model to explain this behaviour support the hypothesis of dark matter as a new particle beyond known physics. Several theoretical models are introduced to explain this new component, including Weakly Interacting Massive Particles (WIMPs), axions and dark photons, among others.
Various detection strategies, direct, indirect, and collider-based, are discussed, setting the stage for the subsequent chapters that detail specific experimental efforts.
\\

The first technique explored in this work is based on gaseous detectors. Chapter 2 provides a foundational overview of their detection principles, a brief history of their development and a description of the newest member of the family, the Micromegas detector. This is relevant for the understanding of the TREX-DM experiment, developed at the University of Zaragoza and explained in the following chapter. Key physical processes such as photon and charged particle interactions, electron drift, diffusion, and avalanche multiplication are reviewed. The discussion is supported by quantitative modelling, including calculations of mean free paths, diffusion coefficients, and amplification factors under different gas mixtures and pressures.

The use of argon and neon mixtures, target gases for TREX-DM, with varying isobutane concentrations is analysed for their impact on detector performance. Empirical and simulated data are used to extract parameters critical to optimizing gaseous detectors for low-energy rare event searches, such as ionization energy, drift velocity or electron diffusion. 
\\

Chapter 3 is devoted to the TREX-DM (TPC for Rare Event eXperiments-Dark Matter) experiment, a low-background gaseous time projection chamber for low-mass WIMP detection. The chapter details the TPC design and installation, including Micromegas readouts, the electronics and data acquisition system, and shielding strategy. A key focus is placed on the detector’s ability to operate at high pressures with ultra-low background levels.

A robust data handling infrastructure, based on the REST-for-Physics framework, has been developed for gaseous detectors with Micromegas planes. Automated data processing pipelines produce analysis-ready outputs, including hit maps, energy spectra, and temporal evolution plots.

The experimental characterization of Micromegas detectors reveals consistent gains and energy resolutions across different pressures and configurations. Two different Micromegas designs have been tested in TREX-DM, and a comparison between them in terms of threshold and energy resolution is presented. Particle track reconstruction relies on the correct channel identification in the readout planes. The process of developing the mapping between electronic channels and physical ones is reviewed with attention to the signs that can hint incongruences in the mapping. 

An unexpected low-energy background source was identified coming from the $^{222}$Rn. Alpha particles coming from its decays were monitored and reduced as much as possible to suppress the secondary emissions affecting the energy region of interest.  

\begin{figure}[h!]
    \centering
    \includegraphics[width=0.8\linewidth]{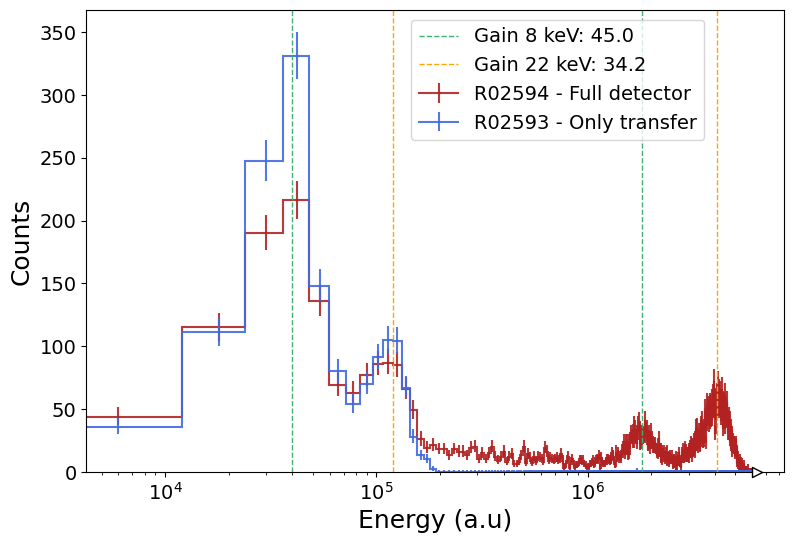}
    \caption{Micromegas and GEM $^{109}$Cd calibration spectra. In red, full detector spectrum, in blue GEM with no voltage difference so that only events in the transfer region are collected. This shows that in the red spectrum, both types of events are present, these from the drift volume that are preamplified by the GEM and those form the transfer region with no amplification. This allows to measure the intrinsic preamplification factor obtained by the GEM, which is computed from the 8 and 22 keV peaks identified by vertical lines. The preamplification value is 45 for the 8 keV peak and 34 for the 22 keV peak.}
    \label{fig:MM_GEM_Cd109spectraConclusionesEng}
\end{figure}

A major advancement presented in this chapter is the integration of GEM foils atop Micromegas planes. Experimental tests of both small-scale and full-size prototypes demonstrate gain increases of up to a factor of 80, significantly enhancing detection sensitivity. However, challenges related to electrical discharges and long-term stability are noted, particularly after GEM implementation in TREX-DM’s operational environment. There, measured gain increase of factor 45, as can be seen in figure \ref{fig:MM_GEM_Cd109spectraConclusionesEng}, and threshold of around 20 eV were tested thanks to $^{109}$Cd and $^{37}$Ar calibration sources.
\\

Related with the improvements developed to enhance the sensitivity of TREX-DM, several ancillary tests were performed in the installations of the University of Zaragoza with the aim of future upgrades of the experiment. Chapter 4 showcases two of these innovations in gaseous detector technology: the implementation of UV LED-based internal calibration systems and a detailed characterization of gain behaviour with argon + 10\% isobutane gas mixture.

One of the thesis’ novel contributions is the first demonstration of the use of a UV LED to extract photoelectrons from a metallic photocathode within a gas chamber for detector calibration. Previously, expensive lasers and noble gas lamps were used. Two photocathode materials, aluminium and copper, were tested. Aluminium yielded a higher electron extraction efficiency due to its lower work function. Although copper exhibited reduced performance, its ubiquity in radiopure experiments justifies further investigation. The LED-based calibration system offers advantages in simplicity, compactness and compatibility with low-background environments compared to traditional radioactive or laser sources. It opens the possibility for in-situ and routine calibrations in sealed chambers.

Extensive measurements of gain curves under varying pressures (1–10 bar) for argon plus 1\% and 10\% isobutane gas mixtures were performed. These studies reveal optimal operating regimes and inform design decisions for future gas-based rare-event detectors. One key result is that argon with 1\% isobutane performs better at high pressures, while argon with 10\% isobutane offers better gains at lower pressures.
\\

In the second part of the thesis, the scope of the thesis shifts from WIMP detection towards axion searches. Chapter 5 is devoted to the theoretical motivations and experimental strategies towards the detection of new particles from the hidden sector: dark photons and axions. These particles, though not part of the Standard Model, are well-motivated extensions that could explain dark matter’s elusive nature.
For dark photons, both massless and massive models are reviewed. The thesis discusses how direct interactions of massive dark photons with Standard Model particles allow more accessible experimental searches. The constraints derived from astrophysical observations, precision measurements and collider experiments are outlined. Similarly, axion models and axion-like particles (ALPs) are introduced. The Peccei-Quinn solution to the strong CP problem is discussed, along with the DSFZ and KSVZ axion models. A broad overview of experimental searches is provided, from helioscopes and haloscopes to astrophysical probes. This theoretical background sets the stage for understanding the relevance and impact of later experimental chapters involving resonant cavities and quantum sensors.
\\

Axions are one of the most appealing theoretical solutions for the strong-CP problem. They present a set of properties that make them ideal candidates for dark matter particles and can be detected through the Primakoff conversion into photons. Chapter 6 introduces haloscopes, resonant microwave cavities immersed in magnetic fields, as one of the fundamental experimental setups for detecting axion-induced photons. Historical and ongoing efforts, including the leading ADMX experiment, are reviewed. The focus then shifts to quantum-limited detection techniques. The Standard Quantum Limit (SQL), which bounds the sensitivity of linear amplifiers, is introduced, motivating the use of quantum non-demolition detectors and single-photon counters. The DarkQuantum proposal is presented as an innovative approach to surpass the SQL using superconducting qubits and circuit QED techniques.

This chapter also discusses the physical principles of superconducting qubits: Josephson junctions, Cooper pair boxes, and its physical realization: transmons, and their integration with resonators and cavities. The Jaynes-Cummings model is used to describe the interaction between a qubit and a microwave cavity. Experimental techniques such as pulse generation, qubit manipulation, and readout schemes are introduced. Characterization results of transmon qubits are reported, including their transition frequencies, coherence properties, and dispersive shifts, all of which are crucial for single-photon detection as discussed in the following chapter.
\\

\begin{figure}
    \centering
    \includegraphics[width=1\linewidth]{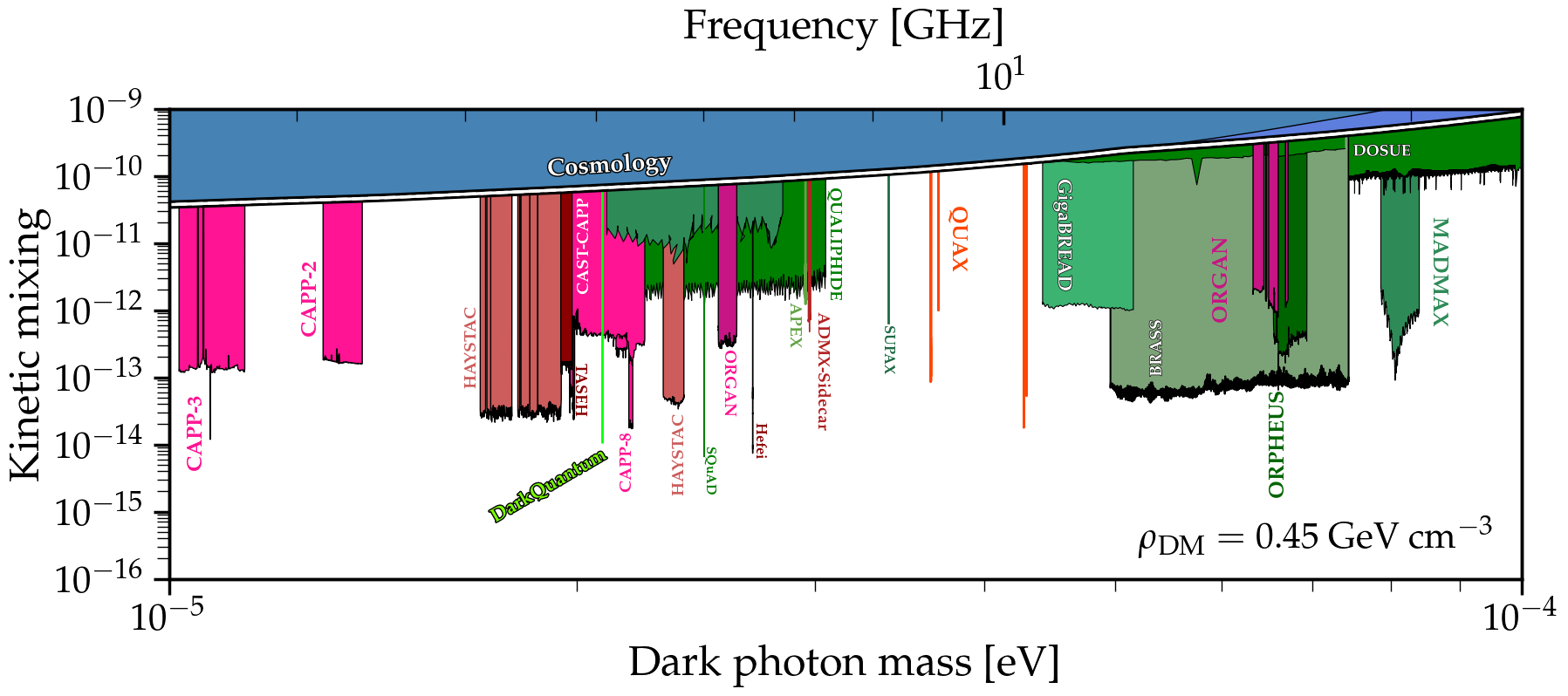}
    \caption{Exclusion limit for dark photons from the quantum single photon counter at 95\% confidence level shown in light green under the label DarkQuantum compared with previous experiments. }
    \label{fig:DarkPhotonSensitivityConclusionsEng}
\end{figure}

Chapter 7 describes in detail the development and experimental evaluation of the DarkQuantum microwave photon counter. The system is based on a double-cavity structure: a readout cavity coupled to a transmon qubit, and a storage cavity designed to capture and hold incoming microwave photons. The experimental work reported includes fabrication, setup, and calibration of the qubit-cavity system, as well as pulse sequences to manipulate and read out the qubit state. Qubit spectroscopy, Rabi oscillations, and $T_1$, $T_2$ coherence measurements are used to characterize device performance. A major outcome of this work is the demonstration of single-photon sensitivity in the microwave regime. This is validated using a sequence of control pulses and qubit state readout schemes tailored to reveal the presence of a photon in the storage cavity. To assess the system’s potential for dark matter searches, a first experimental exclusion limit for dark photons was computed. This result is competitive with other, much larger-scale experiments. Table \ref{tab:ExclusionDP} compares the exclusion levels achieved with those from a previous study, showing that this work reaches the most sensitive measurement to date at its specific frequency, 5.051 GHz. Figure \ref{fig:DarkPhotonSensitivityConclusionsEng} shows the achieved exclusion limit compared with previous experimental bounds. The sensitivity achieved confirms the viability of the prototype as a quantum sensor for rare event detection and opens the path toward scalable dark matter detection using circuit QED platforms.
\\

This thesis embodies a significant multidisciplinary effort bridging experimental particle physics, quantum optics and advanced instrumentation. Several main contributions can be highlighted:
\\

    \textbf{Advances in gaseous detector technologies}
    
    The integration of UV LEDs with photocathodes for calibration, combined with the implementation of GEM-Micromegas hybrid readouts, constitutes a meaningful advance for rare-event searches in low-background environments.
    Detailed characterization of a new gas mixture behaviour under pressure provides a valuable reference for the design of next-generation gas-based detectors, particularly in experiments like TREX-DM.
\\

    \textbf{Haloscopes: Transition from classical to quantum sensing}
    
    A cornerstone of this thesis is the transition from traditional detection techniques to quantum-enhanced sensors. The DarkQuantum prototype represents an innovative and successful realization of a single-microwave-photon detector based on transmon qubits, achieving performance levels competitive with larger-scale experiments. The experimental results present a meaningful exclusion limit for dark photon interactions using a quantum sensor. This milestone proves the feasibility of deploying compact, cryogenic, quantum-enhanced haloscopes as powerful tools for dark matter exploration. The performance under magnetic fields will set the feasibility of this technology for axion searches.
\\

In summary, the work presented here illustrates the scientific and technical richness of dark matter searches. From the deployment of high-pressure gaseous TPCs to the operation of superconducting quantum circuits, this thesis traverses a wide range of technologies, all contributing to the same ambitious goal: uncovering the nature of dark matter.

As experimental capabilities continue to evolve, particularly in quantum technologies, the approaches developed and demonstrated here will likely form the basis for future detectors that explore uncharted territories in the dark sector with unprecedented sensitivity.

%% file: Chapters/8_Conclusiones_ES.tex
\begin{otherlanguage}{spanish}
\renewcommand{\thefigure}{RC.\arabic{figure}}
\setcounter{figure}{0}

La materia oscura sigue siendo uno de los mayores desafíos para nuestra comprensión del Universo. Esta tesis presenta una revisión integral de los esfuerzos experimentales y teóricos que se han llevado a cabo para su comprensión y detección. Comenzando con las motivaciones fundamentales para la búsqueda de materia oscura y culminando con la implementación de sensores cuánticos de última generación, el trabajo abarca desde el desarrollo de detectores gaseosos tradicionales hasta contadores de fotones individuales basados en qubits superconductores.

En este capítulo final, se resume el contenido de los anteriores poniendo énfasis en los resultados experimentales alcanzados y sus implicaciones en el ámbito de la física de partículas y astropartículas.
\\

La tesis comienza con el capítulo 1 revisando las evidencias astronómicas y cosmológicas que apuntan a la existencia de materia oscura. Los datos observacionales de las curvas de rotación galácticas, los efectos de lente gravitacional en cúmulos de galaxias y las anisotropías en el fondo cósmico de microondas, respaldan la presencia de una forma de materia no luminosa y no bariónica que constituye aproximadamente el 27\% del contenido energético del Universo. Las limitaciones del Modelo Estándar para explicar este comportamiento refuerzan la hipótesis de que la materia oscura sea una nueva partícula más allá de la física conocida. A continuación, se introducen varios modelos teóricos que podrían explicar esta nueva componente, incluyendo partículas masivas débilmente interactuantes (WIMPs por sus iniciales en inglés), axiones y fotones oscuros, entre otros.

Se discuten diversas estrategias de detección, directas, indirectas y en colisionadores, estableciendo el contexto para los capítulos posteriores que detallan esfuerzos experimentales específicos.
\\

La primera técnica explorada en este trabajo se basa en detectores gaseosos. El capítulo 2 ofrece una visión general de sus principios de detección, esenciales para comprender el experimento TREX-DM, desarrollado en la Universidad de Zaragoza, y sus posteriores mejoras. Se revisan los procesos físicos clave como las interacciones de fotones y partículas cargadas, la recolección de electrones, su difusión y su multiplicación por avalancha. La discusión se apoya en modelos cuantitativos, incluyendo cálculos de recorridos libres medios, coeficientes de difusión y factores de amplificación bajo diferentes mezclas y presiones de gas.

Se analiza el uso de mezclas de argón y neón, gases utilizados en TREX-DM, con diferentes concentraciones de isobutano por su impacto en el rendimiento del detector. Datos empíricos y simulaciones se emplean para extraer parámetros críticos en la optimización de detectores gaseosos para la búsqueda de eventos raros de baja energía, como la energía de ionización, la velocidad de deriva o la difusión electrónica.
\\

El capítulo 3 está dedicado al experimento TREX-DM (TPC for Rare Event eXperiments-Dark Matter), una cámara de proyección temporal gaseosa de bajo fondo optimizada para la detección de WIMPs de baja masa. El capítulo detalla el diseño e implementación de la TPC, incluyendo los planos de lectura Micromegas, el sistema de adquisición de datos y la electrónica, y la estrategia de blindaje ante la radiación externa. Se hace especial énfasis en la capacidad del detector para operar a altas presiones con niveles de fondo ultra bajos.

Se ha desarrollado una infraestructura robusta de manejo de datos basada en el software REST-for-Physics para detectores gaseosos con planos Micromegas. Procesos automáticos analizan los datos y generan informes preliminares, que incluyen mapas de impactos, espectros de energía y evolución temporal, y que permiten evaluar el desempeño del detector.

La caracterización experimental de los detectores Micromegas ha mostrado ganancias y resoluciones energéticas consistentes en diferentes presiones y configuraciones. Se han probado dos diseños distintos de Micromegas en TREX-DM, presentándose aquí una comparación en términos de umbral y resolución energética. La reconstrucción de trayectorias de partículas depende de la correcta identificación de canales en los planos de lectura. Se  repasa también el proceso de mapeo entre canales electrónicos y físicos, prestando atención a los indicios de posibles incongruencias.

Además, se identificó una fuente inesperada de fondo a baja energía proveniente del $^{222}$Rn. Así que se monitorizaron las partículas alfa provenientes de sus decaimientos y se mitigaron en la medida de lo posible, mediante filtros y operación en ciclo abierto, para suprimir las emisiones secundarias que afectan la región de interés energético.

Un avance significativo presentado en este capítulo es la integración de láminas GEM sobre planos Micromegas. Pruebas experimentales de prototipos a pequeña y gran escala han demostrado incrementos de ganancia de hasta un factor 80, mejorando notablemente el umbral de detección. Sin embargo, se han encontrado desafíos relacionados con las descargas eléctricas y su estabilidad a largo plazo, particularmente tras la implementación de la GEM en el entorno operativo de TREX-DM, donde se consiguieron medir ganancias con un factor de 45, como se puede ver en la figura \ref{fig:MM_GEM_Cd109spectraConclusiones}, y umbrales en torno a 20 eV gracias al uso de fuentes de calibración de $^{109}$Cd y $^{37}$Ar.
\\

\begin{figure}[h!]
    \centering
    \includegraphics[width=0.8\linewidth]{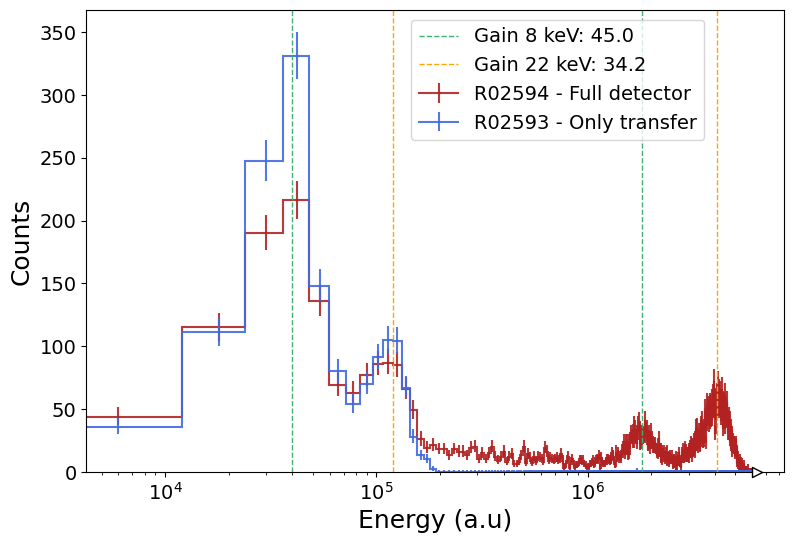}
    \caption{Espectro de calibración con $^{109}$Cd del sistema conjunto Micromegas con GEM. En rojo el espectro con ambos planos, en azul el espectro con GEM sin voltaje, en el que solo se recogen los eventos de la región de transferencia. En el espectro rojo ambos tipos de eventos están presentes, los del volumen de conversión amplificados por la GEM y los de la región de transferencia. Esto permite medir el factor de preamplificación de la GEM, calculado a partir de los picos de 8 y 22 keV, señalados con lineas verticales. Estos factores son 45 para el de 8 keV y 34 para el de 22 keV.}
    \label{fig:MM_GEM_Cd109spectraConclusiones}
\end{figure}

Relacionado con las mejoras desarrolladas para aumentar la sensibilidad de TREX-DM, se realizaron varias pruebas auxiliares en las instalaciones de la Universidad de Zaragoza de cara a futuras actualizaciones del experimento. El capítulo 4 muestra dos de estas innovaciones para detectores gaseosos: la implementación de sistemas de calibración basados en LEDs ultravioleta (UV) y la caracterización detallada del comportamiento de una mezcla de argón + 10\% de isobutano a diferentes presiones.

Una de las contribuciones novedosas de la tesis es la primera demostración del uso de un LED UV para extraer fotoelectrones de un fotocátodo metálico dentro de una cámara de gas para calibración del detector. Anteriormente, se utilizaban láseres costosos y lámparas de gases nobles. Se probaron dos materiales para los fotocátodos: aluminio y cobre. El aluminio presentó una mayor eficiencia de extracción electrónica debido a su menor función de trabajo. Aunque el cobre mostró un rendimiento inferior, su abundancia en experimentos radiopuros justifica su estudio. El sistema de calibración basado en LED ofrece ventajas en simplicidad, es compacto y compatible con entornos de bajo fondo, en especial en comparación con fuentes radiactivas o láseres tradicionales. Esto abre la posibilidad de realizar calibraciones in situ y rutinarias en cámaras selladas.

También se realizaron extensas mediciones de curvas de ganancia bajo diferentes presiones (1–10 bar) para mezclas de argón con 1\% y 10\% de isobutano. Estos estudios revelan regímenes operativos óptimos y orientan decisiones de diseño para futuros detectores gaseosos de eventos raros. Un resultado clave es que el argón con 1\% de isobutano funciona mejor a altas presiones, mientras que el argón con 10\% de isobutano ofrece mejores ganancias a bajas presiones.
\\

En la segunda parte de esta tesis el enfoque se desplaza desde la detección de WIMPs hacia la búsqueda de axiones. El capítulo 5 está dedicado a las motivaciones teóricas y estrategias experimentales para detectar nuevas partículas del sector oculto: fotones oscuros y axiones. Estas partículas, aunque no pertenecen al Modelo Estándar, son extensiones bien motivadas que podrían explicar la naturaleza escurridiza de la materia oscura.

Para los fotones oscuros, se revisan tanto modelos con masa como sin masa. Se discute cómo las interacciones directas de fotones oscuros masivos con partículas del Modelo Estándar permiten búsquedas experimentales más accesibles. Se describen las restricciones derivadas de observaciones astrofísicas, medidas de precisión y experimentos en colisionadores. Igualmente, se introducen los modelos de axiones y partículas tipo axión (ALPs). Se discute la solución de Peccei-Quinn al problema CP fuerte, junto con los modelos de axión DSFZ y KSVZ. Se ofrece una visión amplia de las búsquedas experimentales, desde helioscopios y haloscopios hasta sondas astrofísicas. Este marco teórico prepara el terreno para comprender la relevancia e impacto de los capítulos experimentales posteriores que involucran cavidades resonantes y sensores cuánticos.
\\

Los axiones son una de las soluciones teóricas más atractivas al problema CP fuerte. Presentan un conjunto de propiedades que los convierten en candidatos ideales a partículas de materia oscura, que pueden detectarse mediante su conversión a través del efecto Primakoff en fotones. El capítulo 6 introduce los haloscopios, cavidades resonantes de microondas inmersas en campos magnéticos, como uno de los esquemas experimentales fundamentales para detectar fotones inducidos por axiones. Se revisan esfuerzos históricos y actuales, incluido el experimento líder ADMX. Luego, el foco se desplaza hacia técnicas de detección cuánticas. Se introduce el Límite Cuántico Estándar, que impone un límite a la sensibilidad de los amplificadores lineales, motivando el uso de detectores cuánticos sin demolición y contadores de fotones individuales. Se presenta la propuesta DarkQuantum como un enfoque innovador para superar el SQL mediante qubits superconductores y técnicas de circuitos QED.

Este capítulo también discute los principios físicos de los qubits superconductores: uniones Josephson, cajas de pares de Cooper, y su forma final: los transmones, así como su integración con resonadores y cavidades. Se utiliza el modelo de Jaynes-Cummings para describir la interacción entre un qubit y una cavidad de microondas. Se introducen técnicas experimentales como la generación de pulsos, manipulación de qubits y protocolos de lectura. Se presentan resultados de caracterización de qubits, incluyendo sus frecuencias de transición, propiedades de coherencia y desplazamientos dispersivos, todos ellos cruciales para la detección de fotones individuales que se presenta en el siguiente capítulo.
\\

El capítulo 7 describe en detalle el desarrollo y la evaluación experimental del contador cuántico de fotones de microondas dentro del proyecto DarkQuantum. El sistema se basa en una estructura de doble cavidad: una cavidad de lectura acoplada a un qubit transmon que a su vez se acopla también a una cavidad de almacenamiento diseñada para capturar y retener fotones de microondas entrantes. El trabajo experimental incluye la fabricación, montaje y calibración del sistema qubit-cavidad, así como las secuencias de pulsos para manipular y leer el estado del qubit. Se utilizan espectroscopía de qubits, medidas de oscilaciones de Rabi y se caracterizan los tiempos de coherencia $T_1$, $T_2$ para evaluar el rendimiento del dispositivo.

Uno de los logros principales de este trabajo es la demostración de sensibilidad a fotones individuales en el régimen de microondas. Esto se valida mediante una secuencia de pulsos de control y protocolos de lectura del estado del qubit diseñados para revelar la presencia de un fotón en la cavidad de almacenamiento. Para evaluar el potencial del sistema en búsquedas de materia oscura, se calculó un primer límite experimental de exclusión para fotones oscuros. Este resultado es competitivo en comparación a otros experimentos de mayor envergadura. La Tabla \ref{tab:ExclusionDP} compara los niveles de exclusión alcanzados con los del estudio pionero en esta técnica, mostrando que este trabajo logra la medición más sensible hasta la fecha en su frecuencia específica, 5.051 GHz. La Figura \ref{fig:DarkPhotonSensitivityConclusions} muestra el límite de exclusión obtenido en comparación con los límites experimentales anteriores. La sensibilidad lograda confirma la viabilidad de este prototipo como sensor cuántico para la detección de sucesos raros y abre el camino hacia la detección de materia oscura utilizando sensores con circuitos QED.

\begin{figure}
    \centering
    \includegraphics[width=1\linewidth]{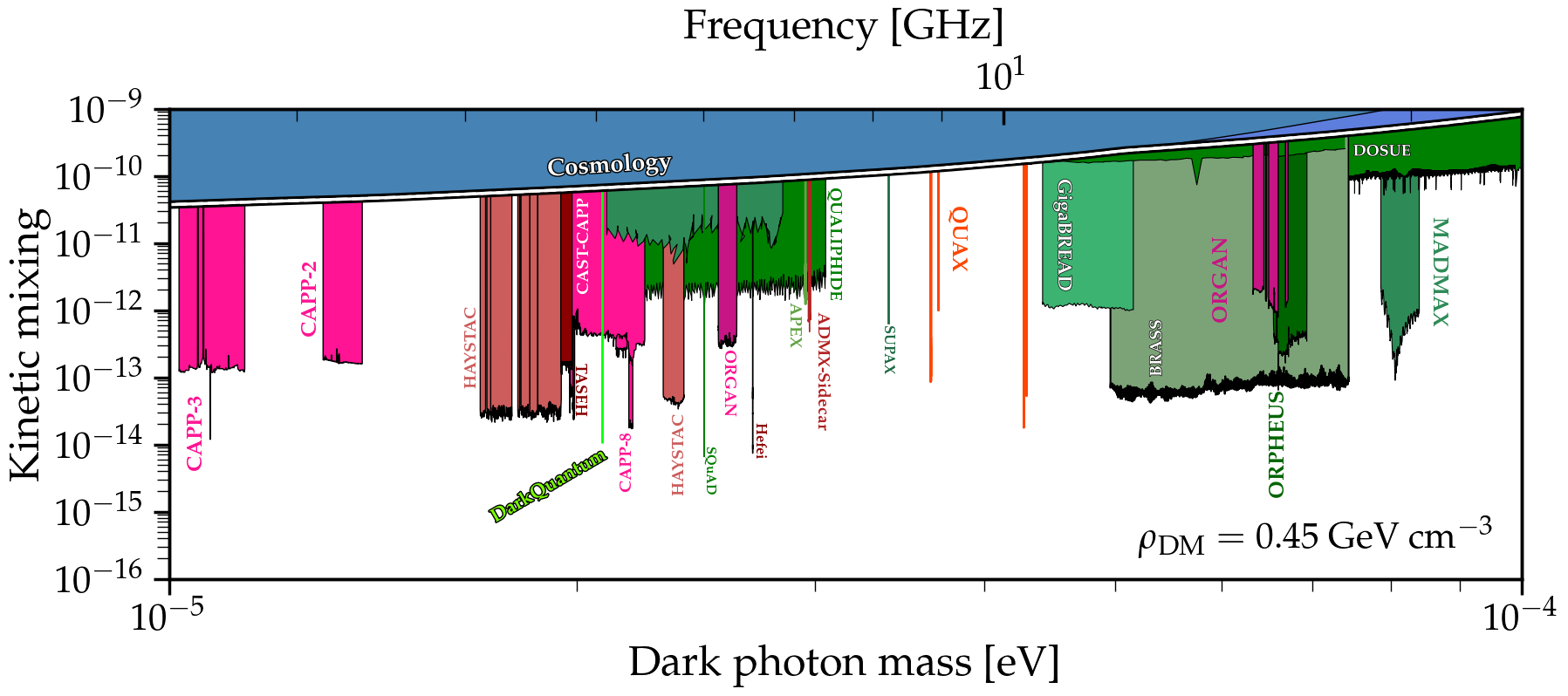}
    \caption{Límite de exclusión para fotones oscuros con el contador de fotones individuales desarrollado en este trabajo al 95\% de nivel de confianza. Se muestra en verde claro con la leyenda DarkQuantum junto a las curvas de sensibilidad de experimentos previos. }
    \label{fig:DarkPhotonSensitivityConclusions}
\end{figure}

Esta tesis representa un esfuerzo multidisciplinario que conecta la física experimental de partículas, la óptica cuántica y la instrumentación avanzada. Se pueden destacar varias contribuciones principales:
\\

    \textbf{Avances en tecnologías de detectores gaseosos}
    
    La integración de LEDs UV con fotocátodos para calibración, combinada con la implementación de lecturas híbridas GEM-Micromegas, constituye un avance importante para la búsqueda de eventos raros en entornos de bajo fondo. La caracterización detallada del comportamiento de nuevas mezclas gaseosas bajo presión proporciona una referencia valiosa para el diseño de la próxima generación de detectores gaseosos, en particular para el futuro de TREX-DM.
\\

    \textbf{Haloscopios: Transición de la detección clásica a la cuántica}
    
    Un pilar fundamental de esta tesis es la transición de técnicas de detección tradicionales hacia sensores mejorados cuánticamente. El sistema presentado bajo el paraguas del proyecto DarkQuantum representa una realización innovadora y exitosa de un detector de fotones individuales de microondas basado en transmones, alcanzando niveles de sensibilidad competitivos. Los resultados experimentales permiten extraer un límite de exclusión significativo para interacciones de fotones oscuros utilizando un sensor cuántico a una cierta frecuencia. Este hito demuestra la viabilidad de implementar haloscopios compactos, criogénicos y cuánticos para la exploración de la materia oscura. El rendimiento bajo campos magnéticos definirá la viabilidad de esta tecnología para la búsqueda de axiones.
\\

En resumen, el trabajo presentado aquí ilustra la riqueza científica y técnica de las búsquedas de materia oscura. Desde el despliegue de cámaras TPC de gas a alta presión hasta la operación de circuitos cuánticos superconductores, esta tesis recorre una amplia gama de tecnologías, todas contribuyendo a un mismo objetivo: desvelar la naturaleza de la materia oscura.

A medida que las capacidades experimentales continúan evolucionando, particularmente en tecnologías cuánticas, los enfoques desarrollados y demostrados en este trabajo serán la base de futuros detectores que exploren territorios inexplorados en el conocido como \textit{dark sector} alcanzando una sensibilidad sin precedentes.

\end{otherlanguage}

%% file: biblio.bib
@article{sikivie1983experimental,
  title={\href{https://journals.aps.org/prl/abstract/10.1103/PhysRevLett.51.1415}{Experimental tests of the" invisible" axion}},
  author={Sikivie, Pierre},
  journal={Physical Review Letters},
  volume={51},
  number={16},
  pages={1415},
  year={1983},
  publisher={APS},
  howpublished={\url{https://journals.aps.org/prl/abstract/10.1103/PhysRevLett.51.1415}}
}

@article{graham2015experimental,
  title={\href{https://www.annualreviews.org/doi/pdf/10.1146/annurev-nucl-102014-022120}{Experimental searches for the axion and axion-like particles}},
  author={Graham, Peter W and Irastorza, Igor G and Lamoreaux, Steven K and Lindner, Axel and van Bibber, Karl A},
  journal={Annual Review of Nuclear and Particle Science},
  volume={65},
  pages={485--514},
  year={2015},
  publisher={Annual Reviews},
  howpublished={\url{https://www.annualreviews.org/doi/pdf/10.1146/annurev-nucl-102014-022120}}
}

@article{diaz2021design,
  title={\href{https://www.mdpi.com/2218-1997/8/1/5}{Design of new resonant haloscopes in the search for the dark matter axion: A review of the first steps in the RADES collaboration}},
  author={D{\'\i}az-Morcillo, Alejandro and Garc{\'\i}a Barcel{\'o}, Jos{\'e} Mar{\'\i}a and Lozano Guerrero, Antonio Jos{\'e} and Navarro, Pablo and Gimeno, Benito and Arguedas Cuendis, Sergio and {\'A}lvarez Melc{\'o}n, Alejandro and Cogollos, Cristian and Calatroni, Sergio and D{\"o}brich, Babette and others},
  journal={Universe},
  volume={8},
  number={1},
  pages={5},
  year={2021},
  publisher={MDPI},
  howpublished={\url{https://www.mdpi.com/2218-1997/8/1/5}}
}

@article{kim2020revisiting,
  title={\href{https://iopscience.iop.org/article/10.1088/1475-7516/2020/03/066/meta}{Revisiting the detection rate for axion haloscopes}},
  author={Kim, Dongok and Jeong, Junu and Youn, SungWoo and Kim, Younggeun and Semertzidis, Yannis K},
  journal={Journal of Cosmology and Astroparticle Physics},
  volume={2020},
  number={03},
  pages={066--066},
  year={2020},
  howpublished={\url{https://iopscience.iop.org/article/10.1088/1475-7516/2020/03/066/meta}}
}

@phdthesis{zheng2021circuit,
    author = {Zheng, Guoji},
    title = {\href{https://repository.tudelft.nl/islandora/object/uuid:ee9e2137-630b-454b-8f37-228f068bcc89}{Circuit quantum electrodynamics with single electron spins in silicon}},
    school = {Delft University of Technology},
    year = {2021},
    howpublished={\url{https://repository.tudelft.nl/islandora/object/uuid:ee9e2137-630b-454b-8f37-228f068bcc89}}
}

@article{aggarwal2021challenges,
  title={\href{https://link.springer.com/article/10.1007/s41114-021-00032-5}{Challenges and opportunities of gravitational-wave searches at MHz to GHz frequencies}},
  author={Aggarwal, Nancy and Aguiar, Odylio D and Bauswein, Andreas and Cella, Giancarlo and Clesse, Sebastian and Cruise, Adrian Michael and Domcke, Valerie and Figueroa, Daniel G and Geraci, Andrew and Goryachev, Maxim and others},
  journal={Living reviews in relativity},
  volume={24},
  pages={1--74},
  year={2021},
  publisher={Springer},
  howpublished={\url{https://link.springer.com/article/10.1007/s41114-021-00032-5}}
}

@article{ejlli2019upper,
  title={\href{https://link.springer.com/article/10.1140/epjc/s10052-019-7542-5}{Upper limits on the amplitude of ultra-high-frequency gravitational waves from graviton to photon conversion}},
  author={Ejlli, Aldo and Ejlli, Damian and Cruise, Adrian Mike and Pisano, Giampaolo and Grote, Hartmut},
  journal={The European Physical Journal C},
  volume={79},
  number={12},
  pages={1032},
  year={2019},
  publisher={Springer},
  howpublished={\url{https://link.springer.com/article/10.1140/epjc/s10052-019-7542-5}}
}

@article{berlin2023mago,
  title={\href{https://arxiv.org/abs/2303.01518}{MAGO 2.0: Electromagnetic cavities as mechanical bars for gravitational waves}},
  author={Berlin, Asher and Blas, Diego and d'Agnolo, Raffaele Tito and Ellis, Sebastian AR and Harnik, Roni and Kahn, Yonatan and Sch{\"u}tte-Engel, Jan and Wentzel, Michael},
  journal={arXiv},
  note = {preprint arXiv: 2303.01518},
  year={2023},
  howpublished={\url{https://arxiv.org/abs/2303.01518}}
}

@misc{AxionLimits,
  author       = {Ciaran O'Hare},
  title        = {\href{https://cajohare.github.io/AxionLimits/}{AxionLimits}},
  howpublished = {GitHub repository: \url{https://cajohare.github.io/AxionLimits/}},
  year         = {Accessed: 2025},
  publisher    = {Zenodo}
}

@article{bourhill2016ultrahigh,
  title={\href{https://journals.aps.org/prb/abstract/10.1103/PhysRevB.93.144420}{Ultrahigh cooperativity interactions between magnons and resonant photons in a YIG sphere}},
  author={Bourhill, Jeremy and Kostylev, Nikita and Goryachev, Maxim and Creedon, DL and Tobar, ME},
  journal={Physical Review B},
  volume={93},
  number={14},
  pages={144420},
  year={2016},
  publisher={APS},
  howpublished = {\url{https://journals.aps.org/prb/abstract/10.1103/PhysRevB.93.144420}}
}

@article{Garcia2023ferrTuning,
  author={García-Barceló, J. M. and Melcón, A. Álvarez and Cuendis, S. Arguedas and Díaz-Morcillo, A. and Gimeno, B. and Kanareykin, A. and Lozano-Guerrero, A. J. and Navarro, P. and Wuensch, W.},
  journal={IEEE Access}, 
  title={\href{https://ieeexplore.ieee.org/abstract/document/10078243}{On the Development of New Tuning and Inter-Coupling Techniques Using Ferroelectric Materials in the Detection of Dark Matter Axions}}, 
  year={2023},
  volume={11},
  pages={30360-30372},
  %doi={10.1109/ACCESS.2023.3260783},
  howpublished = {\url{https://ieeexplore.ieee.org/abstract/document/10078243}}
}

@book{balanis1989advanced,
  title={\href{https://archive.org/details/WileyAdvancedEngineeringElectromagnetics2ndEd.2012_201508}{Advanced engineering electromagnetics}},
  author={Balanis, Constantine A},
  year={1989},
  publisher={John Wiley \& Sons}
}

@article{braggio2024quantum,
  title={\href{https://arxiv.org/pdf/2403.02321}{Quantum-enhanced sensing of axion dark matter with a transmon-based single microwave photon counter}},
  author={Braggio, C and Balembois, L and Di Vora, R and Wang, Z and Travesedo, J and Pallegoix, L and Carugno, G and Ortolan, A and Ruoso, G and Gambardella, U and others},
  journal={arXiv},
  note = {preprint arXiv: 2403.02321},
  year={2024},
  howpublished={\url{https://arxiv.org/pdf/2403.02321}}
}

@article{dixit2021searching,
  title={\href{https://journals.aps.org/prl/pdf/10.1103/PhysRevLett.126.141302}{Searching for dark matter with a superconducting qubit}},
  author={Dixit, Akash V and Chakram, Srivatsan and He, Kevin and Agrawal, Ankur and Naik, Ravi K and Schuster, David I and Chou, Aaron},
  journal={Physical review letters},
  volume={126},
  number={14},
  pages={141302},
  year={2021},
  publisher={APS},
  howpublished={\url{https://journals.aps.org/prl/pdf/10.1103/PhysRevLett.126.141302}}
}

@article{zhao2025flux,
  title={\href{https://arxiv.org/pdf/2501.06882}{A Flux-Tunable cavity for Dark matter detection}},
  author={Zhao, Fang and Li, Ziqian and Dixit, Akash V and Roy, Tanay and Vrajitoarea, Andrei and Banerjee, Riju and Anferov, Alexander and Lee, Kan-Heng and Schuster, David I and Chou, Aaron},
  journal={arXiv},
  note = {preprint arXiv: 2501.06882},
  year={2025}, 
  howpublished={\url{https://arxiv.org/pdf/2501.06882}}
}

@phdthesis{danilin2018experiments,
  title={\href{https://aaltodoc.aalto.fi/server/api/core/bitstreams/82bbdf7d-a5ab-4ab5-9eda-b755701ea82a/content}{Experiments with a transmon artificial atom-state manipulation and detection of magnetic fields}},
  author={Danilin, Sergey},
  year={2018},
  school={Aalto University},
  howpublished={\url{https://aaltodoc.aalto.fi/server/api/core/bitstreams/82bbdf7d-a5ab-4ab5-9eda-b755701ea82a/content}}
}

@article{nigg2012black,
  title={\href{https://arxiv.org/pdf/1204.0587}{Black-box superconducting circuit quantization}},
  author={Nigg, Simon E and Paik, Hanhee and Vlastakis, Brian and Kirchmair, Gerhard and Shankar, Shyam and Frunzio, Luigi and Devoret, MH and Schoelkopf, RJ and Girvin, SM},
  journal={Physical review letters},
  volume={108},
  number={24},
  pages={240502},
  year={2012},
  publisher={APS},
  howpublished={\url{https://arxiv.org/pdf/1204.0587}}
}

@article{naghiloo2019introduction,
  title={\href{https://arxiv.org/pdf/1904.09291}{Introduction to experimental quantum measurement with superconducting qubits}},
  author={Naghiloo, Mahdi},
  journal={arXiv},
  note = {preprint arXiv: 1904.09291},
  year={2019},
  howpublished={\url{https://arxiv.org/pdf/1904.09291}}
}

@article{Ufano1957description,
  title={\href{https://journals.aps.org/rmp/pdf/10.1103/RevModPhys.29.74}{Description of states in quantum mechanics by density matrix and operator techniques}},
  author={Fano, Ugo},
  journal={Reviews of modern physics},
  volume={29},
  number={1},
  pages={74},
  year={1957},
  publisher={APS},
  howpublished={\url{https://journals.aps.org/rmp/pdf/10.1103/RevModPhys.29.74}}
}

@article{koch2007charge,
  title={\href{https://journals.aps.org/pra/pdf/10.1103/PhysRevA.76.042319}{Charge-insensitive qubit design derived from the Cooper pair box}},
  author={Koch, Jens and Yu, Terri M and Gambetta, Jay and Houck, Andrew A and Schuster, David I and Majer, Johannes and Blais, Alexandre and Devoret, Michel H and Girvin, Steven M and Schoelkopf, Robert J},
  journal={Physical Review A—Atomic, Molecular, and Optical Physics},
  volume={76},
  number={4},
  pages={042319},
  year={2007},
  publisher={APS}, 
  howpublished={\url{https://journals.aps.org/pra/pdf/10.1103/PhysRevA.76.042319}}
}

@book{fabbrichesi2021physics,
  title={\href{https://arxiv.org/pdf/2005.01515}{The physics of the dark photon: a primer}},
  author={Fabbrichesi, Marco and Gabrielli, Emidio and Lanfranchi, Gaia},
  year={2021},
  publisher={Springer},
  howpublished={\url{https://arxiv.org/pdf/2005.01515}}
}

@phdthesis{ruiz2019ultra,
  title={\href{https://zaguan.unizar.es/record/87032/files/TESIS-2020-013.pdf}{Ultra-low background micromegas {X-ray} detectors for axion searches in {IAXO} and {BabyIAXO}}},
  author={Ruiz Ch{\'o}liz, Elisa },
  year={2019},
  school={Universidad de Zaragoza}, 
  howpublished={\url{https://zaguan.unizar.es/record/87032/files/TESIS-2020-013.pdf}}
}

@phdthesis{garcia2015solar,
  title={\href{https://zaguan.unizar.es/record/31618/files/TESIS-2015-059.pdf}{Solar Axion search with Micromegas detectors in the CAST Experiment with 3He as buffer gas}},
  author={Garc{\'\i}a, Juan Antonio},
  year={2015},
  school={Universidad de Zaragoza}, 
  howpublished={\url{https://zaguan.unizar.es/record/31618/files/TESIS-2015-059.pdf}}
}

@phdthesis{gracia2016micromegas,
  title={\href{https://zaguan.unizar.es/record/47876/files/TESIS-2016-060.pdf}{Micromegas for the search of solar axions in CAST and low-mass {WIMPs} in {TREX-DM}}},
  author={Gracia Garza, Javier},
  year={2016},
  school={Universidad de Zaragoza},
  howpublished={\url{https://zaguan.unizar.es/record/47876/files/TESIS-2016-060.pdf}}
}

@phdthesis{mirallas2024planos,
  title={\href{https://zaguan.unizar.es/record/145132/files/TESIS-2024-402.pdf}{Desarrollo de grandes planos de lectura Micromegas para experimentos de búsqueda de sucesos poco probables}},
  author={Mirallas Sánchez, Héctor},
  year={2024},
  school={Universidad de Zaragoza}
}

@book{blum2008particle,
  title={\href{https://link.springer.com/book/10.1007/978-3-540-76684-1}{Particle detection with drift chambers}},
  author={Blum, Walter and Riegler, Werner and Rolandi, Luigi},
  year={2008},
  publisher={Springer Science \& Business Media}
}

@article{biagi1999monte,
  title={\href{https://magboltz.web.cern.ch/magboltz/}{{Monte Carlo} simulation of electron drift and diffusion in counting gases under the influence of electric and magnetic fields}},
  author={Biagi, SF},
  journal={Nuclear Instruments and Methods in Physics Research Section A: Accelerators, Spectrometers, Detectors and Associated Equipment},
  volume={421},
  number={1-2},
  pages={234--240},
  year={1999},
  publisher={Elsevier},
  howpublished={\url{https://magboltz.web.cern.ch/magboltz/}}
}

@article{Derre2000,
  author    = {J. Derre and others},
  title     = {\href{https://inspirehep.net/literature/513370}{Fast signals and single electron detection with a MICROMEGAS photodetector}},
  journal   = {Nuclear Instruments and Methods in Physics Research Section A},
  volume    = {449},
  year      = {2000},
  pages     = {314-318},
  howpublished={\url{https://inspirehep.net/literature/513370}}
}

@article{Cebrian_2010,
   title={\href{http://dx.doi.org/10.1088/1475-7516/2010/10/010}{Micromegas readouts for double beta decay searches}},
   volume={2010},
   number={10},
   journal={Journal of Cosmology and Astroparticle Physics},
   publisher={IOP Publishing},
   author={Cebrián, S and Dafni, T and Ferrer-Ribas, E and Galán, J and García, J.A and Giomataris, I and Gómez, H and Herrera, D.C and Iguaz, F.J and Irastorza, I.G and Luzón, G and Rodríguez, A and Seguí, L and Tomás, A},
   year={2010},
   month=oct, 
   pages={010–010},
   %ISSN={1475-7516},
   %url={http://dx.doi.org/10.1088/1475-7516/2010/10/010},
   %DOI={10.1088/1475-7516/2010/10/010}
}

@article{IGUAZ2012448,
    title = {\href{https://www.sciencedirect.com/science/article/pii/S1875389212017233}{New developments in {Micromegas Microbulk} detectors}},
    journal = {Physics Procedia},
    volume = {37},
    pages = {448-455},
    year = {2012},
    note = {Proceedings of the 2nd International Conference on Technology and Instrumentation in Particle Physics (TIPP 2011)},
    author = {F.J. Iguaz and S. Andriamonje and F. Belloni and E. Berthoumieux and M. Calviani and T. Dafni and De Oliveira and E. Ferrer-Ribas and J. Galáan and J.A. Garcáıa and I. Giomataris and C. Guerrero and  Gunsing and D.C. Herrera and I.G. Irastorza and T. Papaevangelou and A. Rodráıguez and A. Tomáas},
    %issn = {1875-3892},
    %doi = {https://doi.org/10.1016/j.phpro.2012.02.392},
    %url = {https://www.sciencedirect.com/science/article/pii/S1875389212017233},
    howpublished={\url{https://www.sciencedirect.com/science/article/pii/S1875389212017233}}
}

@misc{ObisGas,
  author = {Obis, L.},
  title = {\href{https://github.com/lobis/gas-files}{gas-files}},
  howpublished = {GitHub repository: \url{https://github.com/lobis/gas-files}},
  year = {Accessed: 2025},
  commit = {522a5fcb00e9301ba60b2a16c44f96f7a3c838a9}
}

@article{berger2010xcom,
  title={\href{https://physics.nist.gov/PhysRefData/Xcom/html/xcom1.html}{{XCOM} Photon Cross Sections Database, {NIST}}},
  author={Berger, MJ and Hubbell, JH and Seltzer, SM and Chang, J and Coursey, JS and Sukumar, R and Zucker, DS and Olsen, K},
  journal={PML, Radiation Physics Division},
  howpublished = {\url{https://physics.nist.gov/PhysRefData/Xcom/html/xcom1.html}},
  year={2010}
}

@article{EstarPestarAstar,
  author = {Martin Berger and J Coursey and M Zucker},
  title = {\href{https://physics.nist.gov/PhysRefData/Star/Text/intro.html}{{ESTAR, PSTAR, and ASTAR}: Computer programs for calculating stopping-power and range tables for electrons, protons and helium ions (version 1.21)}},
  year = {1999},
  month = {01},
  publisher = {http://physics.nist.gov/Star},
  howpublished = {\url{https://physics.nist.gov/PhysRefData/Star/Text/intro.html}},
  language = {en},
}

@misc{CodeBlackBox,
  author = {Díez-Ibáñez, D.},
  title = {\href{https://github.com/DavidDiezIb/BlackBoxTransmon}{{BlackBoxTransmon}}},
  year = {Accessed: 2025},
  howpublished = {GitHub repository: \url{https://github.com/DavidDiezIb/BlackBoxTransmon}},
  commit = {83dfc6640eae7e599d4ee751b658ff6e9a4a0488}
}

@misc{ThesisCodes,
  author = {Díez-Ibáñez, D.},
  title = {\href{https://github.com/DavidDiezIb/ThesisCodes}{{ThesisCodes}}},
  year = {Accessed: 2025},
  howpublished = {GitHub repository: \url{https://github.com/DavidDiezIb/ThesisCodes}},
}

@phdthesis{dixit2021thesis,
  title={\href{https://search.proquest.com/openview/9912a02db5521501cbe5fbb1618cb652/1?pq-origsite=gscholar&cbl=18750&diss=y}{Searching for dark matter with superconducting qubits}},
  author={Dixit, Akash V},
  year={2021},
  school={University of Chicago},
  howpublished = {\url{https://search.proquest.com/openview/9912a02db5521501cbe5fbb1618cb652/1?pq-origsite=gscholar&cbl=18750&diss=y}}
}

@article{rubin1970rotation,
  title={\href{https://adsabs.harvard.edu/full/1970ApJ...159..379R/0000383.000.html}{Rotation of the {Andromeda} nebula from a spectroscopic survey of emission regions}},
  author={Rubin, Vera C and Ford Jr, W Kent},
  journal={Astrophysical Journal, vol. 159, p. 379},
  volume={159},
  pages={379},
  year={1970},
  howpublished = {\url{https://adsabs.harvard.edu/full/1970ApJ...159..379R/0000383.000.html}}
}

@article{zwicky1933rotverschiebung,
  title={\href{https://adsabs.harvard.edu/full/1933AcHPh...6..110Z/0000119.000.html}{Die rotverschiebung von extragalaktischen nebeln}},
  author={Zwicky, Fritz},
  journal={Helvetica Physica Acta, Vol. 6, p. 110-127},
  volume={6},
  pages={110--127},
  year={1933},
  howpublished = {\url{https://adsabs.harvard.edu/full/1933AcHPh...6..110Z/0000119.000.html}}
}

@article{rubin1980rotational,
  title={\href{https://adsabs.harvard.edu/pdf/1980apj...238..471r}{Rotational properties of 21 SC galaxies with a large range of luminosities and radii, from {NGC 4605/R= 4kpc/to UGC 2885/R= 122 kpc}}},
  author={Rubin, Vera C and Ford Jr, W Kent and Thonnard, Norbert},
  journal={Astrophysical Journal},
  volume={238},
  pages={471--487},
  year={1980},
  howpublished = {\url{https://adsabs.harvard.edu/pdf/1980apj...238..471r}}
}

@article{clowe2006direct,
  title={\href{https://iopscience.iop.org/article/10.1086/508162/pdf}{A direct empirical proof of the existence of dark matter}},
  author={Clowe, Douglas and Brada{\v{c}}, Maru{\v{s}}a and Gonzalez, Anthony H and Markevitch, Maxim and Randall, Scott W and Jones, Christine and Zaritsky, Dennis},
  journal={The Astrophysical Journal},
  volume={648},
  number={2},
  pages={L109},
  year={2006},
  publisher={IOP Publishing},
  howpublished = {\url{https://iopscience.iop.org/article/10.1086/508162/pdf}}
}

@article{PDG2024review,
  title={\href{https://pdg.lbl.gov/}{Review of particle physics}},
  author={Navas, S and Amsler, C and Gutsche, T and Hanhart, C and Hern{\'a}ndez-Rey, JJ and Louren{\c{c}}o, C and Masoni, A and Mikhasenko, M and Mitchell, RE and Patrignani, C and others},
  journal={Physical Review D},
  volume={110},
  number={3},
  pages={030001},
  year={2024},
  howpublished = {\url{https://pdg.lbl.gov/}}
}

@article{aghanim2020planck,
  title={\href{https://arxiv.org/pdf/1807.06209}{Planck 2018 results-{VI}. Cosmological parameters}},
  author={Aghanim, Nabila and Akrami, Yashar and Ashdown, Mark and Aumont, Jonathan and Baccigalupi, Carlo and Ballardini, Mario and Banday, Anthony J and Barreiro, RB and Bartolo, Nicola and Basak, S and others},
  journal={Astronomy \& Astrophysics},
  volume={641},
  pages={A6},
  year={2020},
  publisher={EDP sciences},
  howpublished = {\url{https://arxiv.org/pdf/1807.06209}}
}

@misc{AstroWiki,
  author = {Yelland, A.},
  title = {\href{https://github.com/AYelland/AstroWiki_2.0}{AstroWiki 2.0}},
  year = {Accessed: 2025},
  howpublished = {GitHub repository: \url{https://github.com/AYelland/AstroWiki_2.0}},
  commit = {3f93dcfa88865026861517f981dda1b6f3295f6b}
}

@phdthesis{Coarasa2021anais,
  author={Coarasa Casas, Iv{\'a}n },
  title={\href{https://zaguan.unizar.es/record/110884} {Anais 112. Searching for the annual modulation of dark matter with a 112.5 kg {NaI (Tl)} detector at the {Canfranc Underground Laboratory}}},
  year={2021},
  school={Universidad de Zaragoza},
  howpublished = {\url{https://zaguan.unizar.es/record/110884}}               
}

@article{bertone2005particle,
  title={\href{https://arxiv.org/abs/hep-ph/0404175}{Particle dark matter: Evidence, candidates and constraints}},
  author={Bertone, Gianfranco and Hooper, Dan and Silk, Joseph},
  journal={Physics reports},
  volume={405},
  number={5-6},
  pages={279--390},
  year={2005},
  publisher={Elsevier},
  howpublished = {\url{https://arxiv.org/abs/hep-ph/0404175}}   
}

@article{bertone2017dark,
  title={\href{https://arxiv.org/abs/1703.00013}{How dark matter came to matter}},
  author={de Swart, Jaco G and Bertone, Gianfranco and van Dongen, Jeroen},
  journal={Nature Astronomy},
  volume={1},
  number={3},
  pages={0059},
  year={2017},
  publisher={Nature Publishing Group UK London},
  howpublished = {\url{https://arxiv.org/abs/1703.00013}}
}

@article{alcock2000macho,
  title={\href{https://iopscience.iop.org/article/10.1086/309512/pdf}{The {MACHO} Project: Microlensing Results from 5.7 Years of {Large Magellanic Cloud} Observations}},
  author={Alcock, Charles and Allsman, RA and Alves, David R and Axelrod, TS and Becker, Andrew C and Bennett, DP and Cook, Kem H and Dalal, N and Drake, Andrew John and Freeman, KC and others},
  journal={The Astrophysical Journal},
  volume={542},
  number={1},
  pages={281},
  year={2000},
  publisher={IOP Publishing},
  howpublished = {\url{https://iopscience.iop.org/article/10.1086/309512/pdf}}
}

@article{bertone2018new,
  title={\href{https://www.nature.com/articles/s41586-018-0542-z.pdf}{A new era in the search for dark matter}},
  author={Bertone, Gianfranco and Tait, Tim MP},
  journal={Nature},
  volume={562},
  number={7725},
  pages={51--56},
  year={2018},
  publisher={Nature Publishing Group UK London},
  howpublished = {\url{https://www.nature.com/articles/s41586-018-0542-z.pdf}}
}

@article{steigman1985cosmological,
    title = {\href{https://www.sciencedirect.com/science/article/pii/0550321385905371}{Cosmological constraints on the properties of weakly interacting massive particles}},
    journal = {Nuclear Physics B},
    volume = {253},
    pages = {375-386},
    year = {1985},
    author = {Gary Steigman and Michael S. Turner}
    %issn = {0550-3213},
    %doi = {https://doi.org/10.1016/0550-3213(85)90537-1},
    %url = {https://www.sciencedirect.com/science/article/pii/0550321385905371},
    
}

@article{milgrom1983modification,
  title={\href{https://adsabs.harvard.edu/pdf/1983apj...270..365m}{A modification of the Newtonian dynamics as a possible alternative to the hidden mass hypothesis}},
  author={Milgrom, Mordehai},
  journal={Astrophysical Journal},
  volume={270},
  pages={365--370},
  year={1983},
  howpublished = {\url{https://adsabs.harvard.edu/pdf/1983apj...270..365m}}
}

@article{amendola2018fate,
  title={\href{https://arxiv.org/abs/1711.04825}{Fate of large-scale structure in modified gravity after {GW170817} and {GRB170817A}}},
  author={Amendola, Luca and Kunz, Martin and Saltas, Ippocratis D and Sawicki, Ignacy},
  journal={Physical review letters},
  volume={120},
  number={13},
  pages={131101},
  year={2018},
  publisher={APS},
  howpublished = {\url{https://arxiv.org/abs/1711.04825}}
}

@article{abbott2017gw170817,
  title={\href{https://journals.aps.org/prl/pdf/10.1103/PhysRevLett.119.161101}{{GW170817}: observation of gravitational waves from a binary neutron star inspiral}},
  author={Abbott, Benjamin P and Abbott, Rich and Abbott, TDea and Acernese, Fausto and Ackley, Kendall and Adams, Carl and Adams, Thomas and Addesso, Paolo and Adhikari, Rana X and Adya, Vaishali B and others},
  journal={Physical review letters},
  volume={119},
  number={16},
  pages={161101},
  year={2017},
  publisher={APS},
  howpublished = {\url{https://journals.aps.org/prl/pdf/10.1103/PhysRevLett.119.161101}}
}

@article{universe4110131,
    AUTHOR = {Trevisani, Nicolò},
    TITLE = {\href{https://www.mdpi.com/2218-1997/4/11/131}{Collider Searches for Dark Matter {(ATLAS + CMS)}}},
    JOURNAL = {Universe},
    VOLUME = {4},
    YEAR = {2018},
    NUMBER = {11},
    ARTICLE-NUMBER = {131},
    %URL = {https://www.mdpi.com/2218-1997/4/11/131},
    %ISSN = {2218-1997},
    %DOI = {10.3390/universe4110131}
}

@article{baudis2016dark,
  title={\href{https://iopscience.iop.org/article/10.1088/0954-3899/43/4/044001/pdf?casa_token=TK-znc8ZBc4AAAAA:jxlu2iXHB7cvNVx8aTiln0alViG-2uKtIZy1aj3k9XKMFmXgQn82WZoAnUxBL8jyLIbINZ9zFz4NtmHSLOJM7nuGJN_C}{Dark matter detection}},
  author={Baudis, Laura},
  journal={Journal of Physics G: Nuclear and Particle Physics},
  volume={43},
  number={4},
  pages={044001},
  year={2016},
  publisher={IOP Publishing},
  howpublished = {\url{https://iopscience.iop.org/article/10.1088/0954-3899/43/4/044001/pdf?casa_token=TK-znc8ZBc4AAAAA:jxlu2iXHB7cvNVx8aTiln0alViG-2uKtIZy1aj3k9XKMFmXgQn82WZoAnUxBL8jyLIbINZ9zFz4NtmHSLOJM7nuGJN_C}}
}

@article{hooper2011dark,
  title={\href{https://www.sciencedirect.com/science/article/pii/S0370269311001742}{Dark matter annihilation in the Galactic Center as seen by the {Fermi Gamma Ray Space Telescope}}},
  author={Hooper, Dan and Goodenough, Lisa},
  journal={Physics Letters B},
  volume={697},
  number={5},
  pages={412--428},
  year={2011},
  publisher={Elsevier},
  howpublished = {\url{https://www.sciencedirect.com/science/article/pii/S0370269311001742}}
}

@article{aartsen2018search,
  title={\href{https://link.springer.com/article/10.1140/epjc/s10052-018-6273-3}{Search for neutrinos from decaying dark matter with {IceCube}}},
  author={Aartsen, MG and Ackermann, M and Adams, J and Aguilar, JA and Ahlers, M and Ahrens, M and Samarai, I Al and Altmann, D and Andeen, K and Anderson, T and others},
  journal={The European Physical Journal C},
  volume={78},
  number={10},
  pages={1--9},
  year={2018},
  publisher={Springer},
  howpublished = {\url{https://link.springer.com/article/10.1140/epjc/s10052-018-6273-3}}
}

@article{albert2020search,
  title={\href{https://www.sciencedirect.com/science/article/pii/S0370269320302434?}{Search for dark matter towards the Galactic Centre with 11 years of {ANTARES} data}},
  author={Albert, A and Andr{\'e}, Michel and Anghinolfi, Marco and Anton, Gisela and Ardid, M and Aubert, J-J and Aublin, J and Baret, B and Basa, St{\'e}phane and Belhorma, B and others},
  journal={Physics Letters B},
  volume={805},
  pages={135439},
  year={2020},
  publisher={Elsevier},
  howpublished = {\url{https://www.sciencedirect.com/science/article/pii/S0370269320302434?}}
}

@article{abe2020indirect,
  title={\href{https://journals.aps.org/prd/abstract/10.1103/PhysRevD.102.072002}{Indirect search for dark matter from the {Galactic Center} and halo with the {Super-Kamiokande} detector}},
  author={Abe, K and Bronner, C and Haga, Y and Hayato, Y and Ikeda, M and Imaizumi, S and Ito, H and Iyogi, K and Kameda, J and Kataoka, Y and others},
  journal={Physical Review D},
  volume={102},
  number={7},
  pages={072002},
  year={2020},
  publisher={APS},
  howpublished = {\url{https://journals.aps.org/prd/abstract/10.1103/PhysRevD.102.072002}}
}

@article{lewin1996review,
  title={\href{https://cds.cern.ch/record/298578/files/SCAN-9603159.pdf}{Review of mathematics, numerical factors, and corrections for dark matter experiments based on elastic nuclear recoil}},
  author={Lewin, J David and Smith, Peter F},
  journal={Astroparticle Physics},
  volume={6},
  number={1},
  pages={87--112},
  year={1996},
  publisher={Elsevier},
  howpublished = {\url{https://cds.cern.ch/record/298578/files/SCAN-9603159.pdf}}
}

@article{DiezIbanez2019deteccion,
  title={\href{https://zaguan.unizar.es/record/87438}{Detecci{\'o}n de {WIMPs} con el detector {TREX-DM}}},
  author={D{\'\i}ez-Ib{\'a}{\~n}ez, David },
  year={2019},
  journal={Trabajo de fin de Master, Universidad de Zaragoza},
  howpublished = {\url{https://zaguan.unizar.es/record/87438}} 
}

@article{schumann2019direct,
  title={\href{https://iopscience.iop.org/article/10.1088/1361-6471/ab2ea5/pdf}{Direct detection of {WIMP} dark matter: concepts and status}},
  author={Schumann, Marc},
  journal={Journal of Physics G: Nuclear and Particle Physics},
  volume={46},
  number={10},
  pages={103003},
  year={2019},
  publisher={IOP Publishing},
  howpublished = {\url{https://iopscience.iop.org/article/10.1088/1361-6471/ab2ea5/pdf}}
}

@article{amare2025towards,
  title={\href{https://arxiv.org/abs/2502.01542}{Towards a robust model-independent test of the {DAMA/LIBRA} dark matter signal: {ANAIS-112} results with six years of data}},
  author={Amar{\'e}, Julio and Apilluelo, Jaime and Cebri{\'a}n, Susana and Cintas, David and Coarasa, Iv{\'a}n and Garc{\'\i}a, Eduardo and Mart{\'\i}nez, Mar{\'\i}a and Ortigoza, Ysrael and de Sol{\'o}rzano, Alfonso Ortiz and Pardo, Tamara and others},
  journal={arXiv preprint},
note = {preprint arXiv: 2502.01542},
  year={2025},
  howpublished = {\url{https://arxiv.org/abs/2502.01542}}
}

@article{adhikari2022three,
  title={\href{https://arxiv.org/abs/2111.08863}{Three-year annual modulation search with {COSINE-100}}},
  author={Adhikari, G and Barbosa de Souza, E and Carlin, N and Choi, JJ and Choi, S and Ezeribe, AC and Fran{\c{c}}a, LE and Ha, C and Hahn, IS and Hollick, SJ and others},
  journal={Physical Review D},
  volume={106},
  number={5},
  pages={052005},
  year={2022},
  publisher={APS},
  howpublished = {\url{https://arxiv.org/abs/2111.08863}}
}

@article{aguilar2020results,
  title={\href{https://journals.aps.org/prl/pdf/10.1103/PhysRevLett.125.241803}{Results on low-mass weakly interacting massive particles from an 11 kg d target exposure of damic at {SNOLAB}}},
  author={Aguilar-Arevalo, Alexis and Amidei, D and Baxter, Daniel and Cancelo, G and Vergara, BA Cervantes and Chavarria, AE and D’Olivo, JC and Estrada, Juan and Favela-Perez, F and Gaior, Romain and others},
  journal={Physical Review Letters},
  volume={125},
  number={24},
  pages={241803},
  year={2020},
  publisher={APS},
  howpublished = {\url{https://journals.aps.org/prl/pdf/10.1103/PhysRevLett.125.241803}}
}

@article{arnquist2023damic,
  title={\href{https://www.scipost.org/SciPostPhysProc.12.014/pdf}{The {DAMIC-M} experiment: Status and first results}},
  author={Arnquist, Isaac J and Avalos, Nicolas and Bailly, Philippe and Baxter, David and Bertou, Xavier and Bogdan, Mircea and Bourgeois, Christian and Brandt, J and Cadiou, Arnaud and Castell{\'o}-Mor, Nuria and others},
  journal={SciPost Physics Proceedings},
  number={12},
  pages={014},
  year={2023},
  howpublished = {\url{https://www.scipost.org/SciPostPhysProc.12.014/pdf}}
}

@inproceedings{bernabei2023annual,
  title={\href{https://www.worldscientific.com/doi/epdf/10.1142/S2010194523610086}{Annual modulation results from {DAMA/LIBRA}}},
  author={Bernabei, R and Belli, P and Caracciolo, V and Cerulli, R and Di Marco, A and Leoncini, A and Merlo, V and Montecchia, F and Cappella, F and d’Angelo, A and others},
  booktitle={International Journal of Modern Physics: Conference Series},
  volume={51},
  pages={2361008},
  year={2023},
  organization={World Scientific},
  howpublished = {\url{https://www.worldscientific.com/doi/epdf/10.1142/S2010194523610086}}
}

@article{jiang2018limits,
  title={\href{https://journals.aps.org/prl/abstract/10.1103/PhysRevLett.120.241301}{Limits on light weakly interacting massive particles from the first 102.8 kg$\times$ day data of the {CDEX-10} experiment}},
  author={Jiang, H and Jia, LP and Yue, Q and Kang, KJ and Cheng, JP and Li, YJ and Wong, HT and Agartioglu, M and An, HP and Chang, JP and others},
  journal={Physical review letters},
  volume={120},
  number={24},
  pages={241301},
  year={2018},
  publisher={APS},
  howpublished = {\url{https://journals.aps.org/prl/abstract/10.1103/PhysRevLett.120.241301}}
}

@article{agnese2018results,
  title={\href{https://journals.aps.org/prl/pdf/10.1103/PhysRevLett.120.061802}{Results from the super cryogenic dark matter search experiment at {Soudan}}},
  author={Agnese, R and Aramaki, T and Arnquist, IJ and Baker, W and Balakishiyeva, D and Banik, S and Barker, D and Basu Thakur, R and Bauer, DA and Binder, T and others},
  journal={Physical review letters},
  volume={120},
  number={6},
  pages={061802},
  year={2018},
  publisher={APS},
  howpublished = {\url{https://journals.aps.org/prl/pdf/10.1103/PhysRevLett.120.061802}}
}

@article{arnaud2020first,
  title={\href{https://journals.aps.org/prl/abstract/10.1103/PhysRevLett.125.141301}{First germanium-based constraints on {sub-MeV dark matter with the EDELWEISS} experiment}},
  author={Arnaud, Quentin and Armengaud, E and Augier, C and Benoit, A and Berg{\'e}, L and Billard, J and Broniatowski, A and Camus, P and Cazes, A and Chapellier, M and others},
  journal={Physical Review Letters},
  volume={125},
  number={14},
  pages={141301},
  year={2020},
  publisher={APS},
  howpublished = {\url{https://journals.aps.org/prl/abstract/10.1103/PhysRevLett.125.141301}}
}

@article{abdelhameed2019first,
  title={\href{https://journals.aps.org/prd/abstract/10.1103/PhysRevD.100.102002}{First results from the {CRESST-III} low-mass dark matter program}},
  author={Abdelhameed, Ahmed H and Angloher, G and Bauer, P and Bento, A and Bertoldo, E and Bucci, C and Canonica, L and D’Addabbo, Antonio and Defay, X and Di Lorenzo, S and others},
  journal={Physical Review D},
  volume={100},
  number={10},
  pages={102002},
  year={2019},
  publisher={APS},
  howpublished = {\url{https://journals.aps.org/prd/abstract/10.1103/PhysRevD.100.102002}}
}

@article{balogh2023news,
  title={\href{https://iopscience.iop.org/article/10.1088/1748-0221/18/02/T02005/pdf}{The {NEWS-G} detector at {SNOLAB}}},
  author={Balogh, L and Beaufort, C and Brossard, A and Caron, J-F and Chapellier, M and Coquillat, J-M and Corcoran, EC and Crawford, S and Dastgheibi-Fard, A and Deng, Y and others},
  journal={Journal of Instrumentation},
  volume={18},
  number={02},
  pages={T02005},
  year={2023},
  publisher={IOP Publishing},
  howpublished = {\url{https://iopscience.iop.org/article/10.1088/1748-0221/18/02/T02005/pdf}}
}

@article{arnaud2018first,
  title={\href{https://www.sciencedirect.com/science/article/pii/S0927650517301871}{First results from the {NEWS-G} direct dark matter search experiment at the {LSM}}},
  author={Arnaud, Quentin and Asner, D and Bard, J-P and Brossard, A and Cai, B and Chapellier, M and Clark, M and Corcoran, EC and Dandl, T and Dastgheibi-Fard, A and others},
  journal={Astroparticle Physics},
  volume={97},
  pages={54--62},
  year={2018},
  publisher={Elsevier},
  howpublished = {\url{https://www.sciencedirect.com/science/article/pii/S0927650517301871}}
}

@article{ajaj2019search,
  title={\href{https://journals.aps.org/prd/abstract/10.1103/PhysRevD.100.022004}{Search for dark matter with a 231-day exposure of liquid argon using {DEAP-3600 at SNOLAB}}},
  author={Ajaj, R and Amaudruz, P-A and Araujo, GR and Baldwin, M and Batygov, M and Beltran, B and Bina, CE and Bonatt, J and Boulay, MG and Broerman, B and others},
  journal={Physical Review D},
  volume={100},
  number={2},
  pages={022004},
  year={2019},
  publisher={APS},
  howpublished = {\url{https://journals.aps.org/prd/abstract/10.1103/PhysRevD.100.022004}}
}

@article{aprile2017xenon1t,
  title={\href{https://link.springer.com/article/10.1140/epjc/s10052-017-5326-3}{The {XENON1T} dark matter experiment}},
  author={Aprile, Elena and Aalbers, J and Agostini, F and Alfonsi, M and Amaro, Fernando D and Anthony, M and Antunes, B and Arneodo, F and Balata, M and Barrow, P and others},
  journal={The European Physical Journal C},
  volume={77},
  number={12},
  pages={1--23},
  year={2017},
  publisher={Springer},
  howpublished = {\url{https://link.springer.com/article/10.1140/epjc/s10052-017-5326-3}}
}

@article{cui2017dark,
  title={\href{https://journals.aps.org/prl/pdf/10.1103/PhysRevLett.119.181302}{Dark matter results from 54-ton-day exposure of {PandaX-II} experiment}},
  author={Cui, Xiangyi and Abdukerim, Abdusalam and Chen, Wei and Chen, Xun and Chen, Yunhua and Dong, Binbin and Fang, Deqing and Fu, Changbo and Giboni, Karl and Giuliani, Franco and others},
  journal={Physical review letters},
  volume={119},
  number={18},
  pages={181302},
  year={2017},
  publisher={APS},
  howpublished = {\url{https://journals.aps.org/prl/pdf/10.1103/PhysRevLett.119.181302}}
}

@article{akerib2017results,
  title={\href{https://journals.aps.org/prl/abstract/10.1103/PhysRevLett.118.021303}{Results from a search for dark matter in the complete {LUX} exposure}},
  author={Akerib, DS and Alsum, S and Ara{\'u}jo, HM and Bai, X and Bailey, AJ and Balajthy, J and Beltrame, P and Bernard, EP and Bernstein, A and Biesiadzinski, TP and others},
  journal={Physical review letters},
  volume={118},
  number={2},
  pages={021303},
  year={2017},
  publisher={APS},
  howpublished = {\url{https://journals.aps.org/prl/abstract/10.1103/PhysRevLett.118.021303}}
}

@article{agnes2018low,
  title={\href{https://journals.aps.org/prl/abstract/10.1103/PhysRevLett.121.081307}{Low-mass dark matter search with the {DarkSide-50} experiment}},
  author={Agnes, P and Albuquerque, Ivone Freire da Mota and Alexander, T and Alton, AK and Araujo, GR and Asner, David M and Ave, M and Back, Henning O and Baldin, B and Batignani, G and others},
  journal={Physical review letters},
  volume={121},
  number={8},
  pages={081307},
  year={2018},
  publisher={APS},
  howpublished = {\url{https://journals.aps.org/prl/abstract/10.1103/PhysRevLett.121.081307}}
}

@article{aalseth2018darkside,
  title={\href{https://link.springer.com/article/10.1140/epjp/i2018-11973-4}{{DarkSide-20k: A 20 tonne two-phase LAr TPC for direct dark matter detection at LNGS}}},
  author={Aalseth, Craig E and Acerbi, F and Agnes, P and Albuquerque, IFM and Alexander, T and Alici, A and Alton, AK and Antonioli, P and Arcelli, S and Ardito, R and others},
  journal={The European Physical Journal Plus},
  volume={133},
  pages={1--129},
  year={2018},
  publisher={Springer},
  howpublished = {\url{https://link.springer.com/article/10.1140/epjp/i2018-11973-4}}
}

@article{amole2019dark,
  title={\href{https://journals.aps.org/prd/pdf/10.1103/PhysRevD.100.022001}{Dark matter search results from the complete exposure of the {PICO-60} $C_3 F_8$ bubble chamber}},
  author={Amole, Chanpreet and Ardid, M and Arnquist, IJ and Asner, DM and Baxter, D and Behnke, E and Bressler, M and Broerman, B and Cao, G and Chen, CJ and others},
  journal={Physical Review D},
  volume={100},
  number={2},
  pages={022001},
  year={2019},
  publisher={APS},
  howpublished = {\url{https://journals.aps.org/prd/pdf/10.1103/PhysRevD.100.022001}}
}

@article{battat2017low,
  title={\href{https://www.sciencedirect.com/science/article/abs/pii/S0927650517300075}{Low threshold results and limits from the DRIFT directional dark matter detector}},
  author={Battat, James BR and Ezeribe, AC and Gauvreau, J-L and Harton, JL and Lafler, R and Law, E and Lee, ER and Loomba, D and Lumnah, A and Miller, EH and others},
  journal={Astroparticle Physics},
  volume={91},
  pages={65--74},
  year={2017},
  publisher={Elsevier},
  howpublished = {\url{https://www.sciencedirect.com/science/article/abs/pii/S0927650517300075}}
}

@article{vahsen2020cygnus,
  title={\href{https://arxiv.org/pdf/2008.125875}{{CYGNUS: Feasibility} of a nuclear recoil observatory with directional sensitivity to dark matter and neutrinos}},
  author={Vahsen, Sven E and O'Hare, CAJ and Lynch, WA and Spooner, NJC and Baracchini, E and Barbeau, P and Battat, JBR and Crow, B and Deaconu, C and Eldridge, C and others},
  journal={arXiv},
  note = {preprint arXiv: 2008.12587},
  year={2020},
  howpublished = {\url{https://arxiv.org/pdf/2008.125875}}
}

@article{alexandrov2021directionality,
  title={\href{https://iopscience.iop.org/article/10.1088/1475-7516/2021/04/047/pdf}{Directionality preservation of nuclear recoils in an emulsion detector for directional dark matter search}},
  author={Alexandrov, A and De Lellis, G and Di Crescenzo, A and Golovatiuk, A and Tioukov, V},
  journal={Journal of Cosmology and Astroparticle Physics},
  volume={2021},
  number={04},
  pages={047},
  year={2021},
  publisher={IOP Publishing},
  howpublished = {\url{https://iopscience.iop.org/article/10.1088/1475-7516/2021/04/047/pdf}}
}

@article{Littlejohn,
  author        = {Littlejohn, Robert },
  title         = {\href{https://bohr.physics.berkeley.edu/classes/221/1112/notes/photelec.pdf}{Lecture notes in {Quantum Mechanics, course Physics 221AB, Notes} 34}},
  year          = {2020},
  journal     = {University of California, Berkeley},
  howpublished = {\url{https://bohr.physics.berkeley.edu/classes/221/1112/notes/photelec.pdf}}
}

@article{billard2022direct,
  title={\href{https://iopscience.iop.org/article/10.1088/1361-6633/ac5754/pdf}{Direct detection of dark matter—APPEC committee report}},
  author={Billard, Julien and Boulay, Mark and Cebri{\'a}n, Susana and Covi, Laura and Fiorillo, Giuliana and Green, Anne and Kopp, Joachim and Majorovits, B{\'e}la and Palladino, Kimberly and Petricca, Federica and others},
  journal={Reports on Progress in Physics},
  volume={85},
  number={5},
  pages={056201},
  year={2022},
  publisher={IOP Publishing}
}

@article{castel2019background,
  title={\href{https://link.springer.com/content/pdf/10.1140/epjc/s10052-019-7282-6.pdf}{Background assessment for the TREX dark matter experiment}},
  author={Castel, J and Cebri{\'a}n, S and Coarasa, I and Dafni, T and Gal{\'a}n, J and Iguaz, FJ and Irastorza, IG and Luz{\'o}n, G and Mirallas, H and Ortiz de Sol{\'o}rzano, A and others},
  journal={The European Physical Journal C},
  volume={79},
  number={9},
  pages={782},
  year={2019},
  publisher={Springer}
}

@article{cast2017new,
  title={\href{https://www.nature.com/articles/nphys4109}{New CAST limit on the axion-photon interaction}},
author={Anastassopoulos, V and Aune, S and Barth, K and Belov, A and Cantatore, G and Carmona, JM and Castel, JF and Cetin, SA and Christensen, F and Collar, JI and others},
  journal={Nature Physics},
  volume={13},
  number={6},
  pages={584--590},
  year={2017},
  publisher={Nature Publishing Group UK London}
}

@article{altenmuller2024new,
  title={\href{https://journals.aps.org/prl/pdf/10.1103/PhysRevLett.133.221005}{New upper limit on the axion-photon coupling with an extended CAST run with a Xe-based Micromegas detector}},
  author={Altenm{\"u}ller, K and Anastassopoulos, V and Arguedas-Cuendis, S and Aune, S and Baier, J and Barth, K and Br{\"a}uninger, H and Cantatore, G and Caspers, F and Castel, JF and others},
  journal={Physical Review Letters},
  volume={133},
  number={22},
  pages={221005},
  year={2024},
  publisher={APS}
}

@article{abeln2021conceptual,
  title={\href{https://link.springer.com/article/10.1007/JHEP05(2021)137}{Conceptual design of BabyIAXO, the intermediate stage towards the International Axion Observatory}},
  author={Abeln, A and Altenm{\"u}ller, K and Arguedas Cuendis, S and Armengaud, E and Atti{\'e}, D and Aune, S and Basso, S and Berg{\'e}, L and Biasuzzi, B and Borges De Sousa, PTC and others},
  journal={Journal of High Energy Physics},
  volume={2021},
  number={5},
  pages={1--80},
  year={2021},
  publisher={Springer}
}

@article{altenmuller2024background,
  title={\href{https://www.frontiersin.org/journals/physics/articles/10.3389/fphy.2024.1384415/full}{Background discrimination with a Micromegas detector prototype and veto system for BabyIAXO}},
  author={Altenm{\"u}ller, Konrad and Castel, Juan Francisco and Cebri{\'a}n, S and Dafni, Theopisti and D{\'\i}ez-Iba{\~n}ez, David and Ezquerro, A and Ferrer-Ribas, Esther and Galan, Javier and Galindo, Javier and Garc{\'\i}a, Juan Antonio and others},
  journal={Frontiers in Physics},
  volume={12},
  pages={1384415},
  year={2024},
  publisher={Frontiers Media SA}
}

@article{chen2017pandax,
  title={\href{https://link.springer.com/article/10.1007/s11433-017-9028-0}{PandaX-III: Searching for neutrinoless double beta decay with high pressure 136 Xe gas time projection chambers}},
  author={Chen, Xun and Fu, ChangBo and Galan, Javier and Giboni, Karl and Giuliani, Franco and Gu, LingHui and Han, Ke and Ji, XiangDong and Lin, Heng and Liu, JiangLai and others},
  journal={Science China Physics, Mechanics \& Astronomy},
  volume={60},
  pages={1--40},
  year={2017},
  publisher={Springer}
}

@article{altenmuller2022alphacamm,
  title={\href{https://iopscience.iop.org/article/10.1088/1748-0221/17/08/P08035}{AlphaCAMM, a Micromegas-based camera for high-sensitivity screening of alpha surface contamination}},
  author={Altenm{\"u}ller, Konrad and Castel, Juan F and Cebri{\'a}n, Susana and Dafni, Theopisti and D{\'\i}ez-Ib{\'a}{\~n}ez, David and Gal{\'a}n, Javier and Galindo, Javier and Garc{\'\i}a, Juan Antonio and Irastorza, Igor G and Luz{\'o}n, Gloria and others},
  journal={Journal of Instrumentation},
  volume={17},
  number={08},
  pages={P08035},
  year={2022},
  publisher={IOP Publishing}
}

@article{iguaz2016trex,
  title={\href{https://link.springer.com/content/pdf/10.1140/epjc/s10052-016-4372-6.pdf}{TREX-DM: a low-background Micromegas-based TPC for low-mass WIMP detection}},
  author={Iguaz, FJ and Garza, JG and Aznar, F and Castel, JF and Cebri{\'a}n, S and Dafni, T and Garc{\'\i}a, JA and Irastorza, IG and Lagraba, A and Luz{\'o}n, G and others},
  journal={The European Physical Journal C},
  volume={76},
  pages={1--28},
  year={2016},
  publisher={Springer}
}

@article{iguaz2022microbulk,
  title={\href{https://iopscience.iop.org/article/10.1088/1748-0221/17/07/P07032/pdf}{Microbulk Micromegas in non-flammable mixtures of argon and neon at high pressure}},
  author={Iguaz, FJ and Dafni, T and Canellas, C and Castel, JF and Cebri{\'a}n, S and Garza, JG and Irastorza, IG and Luz{\'o}n, G and Mirallas, H and Ruiz-Ch{\'o}liz, E},
  journal={Journal of Instrumentation},
  volume={17},
  number={07},
  pages={P07032},
  year={2022},
  publisher={IOP Publishing}
}

@article{sauli1997gem,
  title={\href{https://wiki.iac.isu.edu/images/d/d7/Sauli_NIMA386_1997_531.pdf}{GEM: A new concept for electron amplification in gas detectors}},
  author={Sauli, Fabio},
  journal={Nuclear Instruments and Methods in Physics Research Section A: Accelerators, Spectrometers, Detectors and Associated Equipment},
  volume={386},
  number={2-3},
  pages={531--534},
  year={1997},
  publisher={Elsevier}
}

@article{giomataris1996micromegas,
  title={\href{https://cds.cern.ch/record/299159/files/SCAN-9603270.pdf}{MICROMEGAS: a high-granularity position-sensitive gaseous detector for high particle-flux environments}},
  author={Giomataris, Yannis and Rebourgeard, Ph and Robert, Jean Pierre and Charpak, Georges},
  journal={Nuclear Instruments and Methods in Physics Research Section A: Accelerators, Spectrometers, Detectors and Associated Equipment},
  volume={376},
  number={1},
  pages={29--35},
  year={1996},
  publisher={Elsevier}
}

@article{giomataris2006micromegas,
  title={\href{https://www.sciencedirect.com/science/article/abs/pii/S0168900205026501}{Micromegas in a bulk}},
  author={Giomataris, I and De Oliveira, R and Andriamonje, S and Aune, S and Charpak, G and Colas, P and Fanourakis, G and Ferrer, E and Giganon, A and Rebourgeard, Ph and others},
  journal={Nuclear Instruments and Methods in Physics Research Section A: Accelerators, Spectrometers, Detectors and Associated Equipment},
  volume={560},
  number={2},
  pages={405--408},
  year={2006},
  publisher={Elsevier}
}

@article{andriamonje2010development,
  title={\href{https://iopscience.iop.org/article/10.1088/1748-0221/5/02/P02001/pdf}{Development and performance of Microbulk Micromegas detectors}},
  author={Andriamonje, S and Attie, D and Berthoumieux, E and Calviani, M and Colas, P and Dafni, T and Fanourakis, G and Ferrer-Ribas, E and Galan, J and Geralis, T and others},
  journal={Journal of Instrumentation},
  volume={5},
  number={02},
  pages={P02001},
  year={2010},
  publisher={IOP Publishing}
}

@article{trzaska2019cosmic,
  title={\href{https://link.springer.com/article/10.1140/epjc/s10052-019-7239-9}{Cosmic-ray muon flux at Canfranc Underground Laboratory}},
  author={Trzaska, Wladyslaw Henryk and Slupecki, Maciej and Bandac, Iulian and Bayo, Alberto and Bettini, Alessandro and Bezrukov, Leonid and Enqvist, Timo and Fazliakhmetov, Almaz and Ianni, Aldo and Inzhechik, Lev and others},
  journal={The European Physical Journal C},
  volume={79},
  number={8},
  pages={1--5},
  year={2019},
  publisher={Springer}
}

@article{altenmuller2022rest,
  title={\href{https://www.sciencedirect.com/science/article/pii/S0010465521003933}{REST-for-Physics, a ROOT-based framework for event oriented data analysis and combined Monte Carlo response}},
  author={Altenm{\"u}ller, Konrad and Cebri{\'a}n, Susana and Dafni, Theopisti and D{\'\i}ez-Ib{\'a}{\~n}ez, David and Gal{\'a}n, Javier and Galindo, Javier and Garc{\'\i}a, Juan Antonio and Irastorza, Igor G and Luz{\'o}n, Gloria and Margalejo, Cristina and others},
  journal={Computer Physics Communications},
  volume={273},
  pages={108281},
  year={2022},
  publisher={Elsevier}
}

@misc{REST,
  author = {Galán, Javier and Obis, Luis and Ni, Kaixian and Margalejo, Cristina and Díez-Ibáñez, David and García, Juan Antonio and von Oy, Johanna and others},
  title = {\href{https://github.com/rest-for-physics}{REST-for-Physics}},
  year = {Accessed: 2025},
  howpublished = {GitHub repository: \url{https://github.com/rest-for-physics}},
  commit = {0dde7aed59971df3141b3f7cb283362016877034}
}

@article{pollacco2018get,
  title={\href{https://www.sciencedirect.com/science/article/pii/S0168900218300342}{GET: A generic electronics system for TPCs and nuclear physics instrumentation}},
  author={Pollacco, EC and Grinyer, GF and Abu-Nimeh, F and Ahn, T and Anvar, S and Arokiaraj, Alex and Ayyad, Y and Baba, H and Babo, Mathieu and Baron, P and others},
  journal={Nuclear Instruments and Methods in Physics Research Section A: Accelerators, Spectrometers, Detectors and Associated Equipment},
  volume={887},
  pages={81--93},
  year={2018},
  publisher={Elsevier}
}

@misc{EzquerroSlowControl,
  author = {Ezquerro, Álvaro},
  title = {\href{https://github.com/AlvaroEzq/trex-slowcontrolHV}{trex-slowcontrolHV}},
  year = {Accessed: 2025},
  howpublished = {GitHub repository: \url{https://github.com/AlvaroEzq/trex-slowcontrolHV}},
  commit = {59a6d22b924fb4e1262415d81ee2bd46e18dfa23}
}

@article{goodman2025admx,
  title={\href{https://journals.aps.org/prl/pdf/10.1103/PhysRevLett.134.111002}{ADMX axion dark matter bounds around 3.3 $\mu$ eV with Dine-Fischler-Srednicki-Zhitnitsky discovery ability}},
  author={Goodman, C and Guzzetti, M and Hanretty, C and Rosenberg, LJ and Rybka, G and Sinnis, J and Zhang, D and Clarke, John and Siddiqi, I and Chou, AS and others},
  journal={Physical Review Letters},
  volume={134},
  number={11},
  pages={111002},
  year={2025},
  publisher={APS}
}

@phdthesis{garcia2023development,
  title={\href{https://portalinvestigacion.upct.es/documentos/654e7fc72f15e346a1178d4f}{Development of resonant cavity-based microwave filters for axion detection}},
  author={Garc{\'i}a Barcel{\'o}, Jos{\'e} Mar{\'i}a},
  year={2023},
  school={Universidad Polit{\'e}cnica de Cartagena}
}

@article{khatiwada2021ADMX,
  title={\href{https://arxiv.org/pdf/2010.00169}{Axion dark matter experiment: Detailed design and operations}},
  author={Khatiwada, R and Bowring, D and Chou, AS and Sonnenschein, Andrew and Wester, W and Mitchell, Don V and Braine, T and Bartram, C and Cervantes, Raphael and Crisosto, N and others},
  journal={Review of Scientific Instruments},
  volume={92},
  number={12},
  year={2021},
  publisher={AIP Publishing}
}

@article{adair2022CastCapp,
  title={\href{https://www.nature.com/articles/s41467-022-33913-6}{Search for dark matter axions with CAST-CAPP}},
  author={Adair, CM and Altenm{\"u}ller, K and Anastassopoulos, V and Arguedas Cuendis, S and Baier, J and Barth, K and Belov, A and Bozicevic, D and Br{\"a}uninger, H and Cantatore, G and others},
  journal={Nature Communications},
  volume={13},
  number={1},
  pages={6180},
  year={2022},
  publisher={Nature Publishing Group UK London}
}

@article{kwon2021capp,
  title={\href{https://journals.aps.org/prl/pdf/10.1103/PhysRevLett.126.191802}{First Results from an Axion Haloscope at CAPP around 10.7 $\mu$ eV}},
  author={Kwon, Ohjoon and Lee, Doyu and Chung, Woohyun and Ahn, Danho and Byun, HeeSu and Caspers, Fritz and Choi, Hyoungsoon and Choi, Jihoon and Chong, Yonuk and Jeong, Hoyong and others},
  journal={Physical Review Letters},
  volume={126},
  number={19},
  pages={191802},
  year={2021},
  publisher={APS}
}

@article{kim2023near,
  title={\href{https://journals.aps.org/prl/pdf/10.1103/PhysRevLett.130.091602}{Near-quantum-noise axion dark matter search at CAPP around 9.5 $\mu$ eV}},
  author={Kim, Jinsu and Kwon, Ohjoon and Kutlu, {\c{C}}a{\u{g}}lar and Chung, Woohyun and Matlashov, Andrei and Uchaikin, Sergey and Van Loo, Arjan Ferdinand and Nakamura, Yasunobu and Oh, Seonjeong and Byun, HeeSu and others},
  journal={Physical review letters},
  volume={130},
  number={9},
  pages={091602},
  year={2023},
  publisher={APS}
}

@article{yang2023extended,
  title={\href{https://journals.aps.org/prl/pdf/10.1103/PhysRevLett.131.081801}{Extended axion dark matter search using the CAPP18T haloscope}},
  author={Yang, Byeongsu and Yoon, Hojin and Ahn, Moohyun and Lee, Youngjae and Yoo, Jonghee},
  journal={Physical Review Letters},
  volume={131},
  number={8},
  pages={081801},
  year={2023},
  publisher={APS}
}

@article{ahn2020superconducting,
  title={\href{https://arxiv.org/pdf/2002.08769}{Superconducting cavity in a high magnetic field}},
  author={Ahn, Danho and Kwon, Ohjoon and Chung, Woohyun and Jang, Wonjun and Lee, Doyu and Lee, Jhinhwan and Youn, Sung Woo and Youm, Dojun and Semertzidis, Yannis K},
  journal={arXiv},
  note = {preprint arXiv: 2002.08769},
  year={2020}
}

@article{jewell2023new,
  title={\href{https://journals.aps.org/prd/pdf/10.1103/PhysRevD.107.072007}{New results from HAYSTAC’s phase II operation with a squeezed state receiver}},
  author={Jewell, MJ and Leder, AF and Backes, KM and Bai, Xiran and van Bibber, K and Brubaker, BM and Cahn, SB and Droster, A and Esmat, Maryam H and Ghosh, Sumita and others},
  journal={Physical Review D},
  volume={107},
  number={7},
  pages={072007},
  year={2023},
  publisher={APS}
}

@article{zhong2018results,
  title={\href{https://journals.aps.org/prd/pdf/10.1103/PhysRevD.97.092001}{Results from phase 1 of the HAYSTAC microwave cavity axion experiment}},
  author={Zhong, L and Al Kenany, S and Backes, KM and Brubaker, BM and Cahn, SB and Carosi, G and Gurevich, YV and Kindel, WF and Lamoreaux, SK and Lehnert, KW and others},
  journal={Physical Review D},
  volume={97},
  number={9},
  pages={092001},
  year={2018},
  publisher={APS}
}

@article{chang2022taiwan,
  title={\href{https://arxiv.org/pdf/2204.14265}{Taiwan axion search experiment with haloscope: CD102 analysis details}},
  author={Chang, Hsin and Chang, Jing-Yang and Chang, Yi-Chieh and Chang, Yu-Han and Chang, Yuan-Hann and Chen, Chien-Han and Chen, Ching-Fang and Chen, Kuan-Yu and Chen, Yung-Fu and Chiang, Wei-Yuan and others},
  journal={Physical Review D},
  volume={106},
  number={5},
  pages={052002},
  year={2022},
  publisher={APS}
}

@article{mcallister2017organ,
  title={\href{https://arxiv.org/pdf/1706.00209}{The ORGAN experiment: An axion haloscope above 15 GHz}},
  author={McAllister, Ben T and Flower, Graeme and Ivanov, Eugene N and Goryachev, Maxim and Bourhill, Jeremy and Tobar, Michael E},
  journal={Physics of the dark universe},
  volume={18},
  pages={67--72},
  year={2017},
  publisher={Elsevier}
}

@article{rettaroli2024search,
  title={\href{https://journals.aps.org/prd/pdf/10.1103/PhysRevD.110.022008}{Search for axion dark matter with the QUAX--LNF tunable haloscope}},
  author={Rettaroli, A and Alesini, D and Babusci, D and Braggio, C and Carugno, G and D’Agostino, D and D’Elia, A and Di Gioacchino, D and Di Vora, R and Falferi, P and others},
  journal={Physical Review D},
  volume={110},
  number={2},
  pages={022008},
  year={2024},
  publisher={APS}
}

@article{grenet2021grenoble,
  title={\href{https://arxiv.org/pdf/2110.14406}{The Grenoble Axion Haloscope platform (GrAHal): development plan and first results}},
  author={Grenet, Thierry and Ballou, Rafik and Basto, Quentin and Martineau, Killian and Perrier, Pierre and Pugnat, Pierre and Quevillon, J{\'e}r{\'e}mie and Roch, Nicolas and Smith, Christopher},
  journal={arXiv},
  note = {preprint arXiv: 2110.14406},
  year={2021}
}

@article{lamoreaux2013analysis,
  title={\href{https://journals.aps.org/prd/pdf/10.1103/PhysRevD.88.035020}{Analysis of single-photon and linear amplifier detectors for microwave cavity dark matter axion searches}},
  author={Lamoreaux, SK and Van Bibber, KA and Lehnert, KW and Carosi, G},
  journal={Physical Review D—Particles, Fields, Gravitation, and Cosmology},
  volume={88},
  number={3},
  pages={035020},
  year={2013},
  publisher={APS}
}

@article{yazaki2017klein,
  title={\href{https://www.jstage.jst.go.jp/article/pjab/93/6/93_PJA9306B-04/_pdf}{How the Klein-Nishina formula was derived: Based on the Sangokan Nishina source materials}},
  author={Yazaki, Yuji},
  journal={Proceedings of the Japan Academy, Series B},
  volume={93},
  number={6},
  pages={399--421},
  year={2017},
  publisher={The Japan Academy}
}

@article{baker2006improved,
  title={\href{https://journals.aps.org/prl/abstract/10.1103/PhysRevLett.97.131801}{Improved experimental limit on the electric dipole moment of the neutron}},
  author={Baker, CA and Doyle, DD and Geltenbort, P and Green, K and Van der Grinten, MGD and Harris, PG and Iaydjiev, P and Ivanov, SN and May, DJR and Pendlebury, JM and others},
  journal={Physical review letters},
  volume={97},
  number={13},
  pages={131801},
  year={2006},
  publisher={APS}
}

@article{peccei1977cp,
  title={\href{https://journals.aps.org/prl/abstract/10.1103/PhysRevLett.38.1440}{CP conservation in the presence of pseudoparticles}},
  author={Peccei, Roberto D and Quinn, Helen R},
  journal={Physical Review Letters},
  volume={38},
  number={25},
  pages={1440},
  year={1977},
  publisher={APS}
}

@article{peccei16constraints,
  title={Constraints imposed by CP conservation in the presence of instantons.},
  year={1977},
  author={Peccei, RD and Quinn, HR},
  journal={Phys. Rev. D},
  volume={16},
  number={1791},
  pages={36}
}

@article{arias2012wispy,
  title={\href{https://iopscience.iop.org/article/10.1088/1475-7516/2012/06/013/pdf}{WISPy cold dark matter}},
  author={Arias, Paola and Cadamuro, Davide and Goodsell, Mark and Jaeckel, Joerg and Redondo, Javier and Ringwald, Andreas},
  journal={Journal of Cosmology and Astroparticle Physics},
  volume={2012},
  number={06},
  pages={013},
  year={2012},
  publisher={IOP Publishing}
}

@article{ng2021constraints,
  title={\href{https://journals.aps.org/prl/abstract/10.1103/PhysRevLett.126.151102}{Constraints on ultralight scalar bosons within black hole spin measurements from the LIGO-Virgo GWTC-2}},
  author={Ng, Ken KY and Vitale, Salvatore and Hannuksela, Otto A and Li, Tjonnie GF},
  journal={Physical Review Letters},
  volume={126},
  number={15},
  pages={151102},
  year={2021},
  publisher={APS}
}

@article{van1989design,
  title={\href{https://core.ac.uk/download/pdf/231869211.pdf}{Design for a practical laboratory detector for solar axions}},
  author={Van Bibber, K and McIntyre, PM and Morris, DE and Raffelt, GG},
  journal={Physical Review D},
  volume={39},
  number={8},
  pages={2089},
  year={1989},
  publisher={APS}
}

@article{lazarus1992search,
  title={\href{https://journals.aps.org/prl/abstract/10.1103/PhysRevLett.69.2333}{Search for solar axions}},
  author={Lazarus, DM and Smith, GC and Cameron, R and Melissinos, AC and Ruoso, G and Semertzidis, Yannis K and Nezrick, FA},
  journal={Physical Review Letters},
  volume={69},
  number={16},
  pages={2333},
  year={1992},
  publisher={APS}
}

@article{moriyama1998direct,
  title={\href{https://www.sciencedirect.com/science/article/abs/pii/S0370269398007667}{Direct search for solar axions by using strong magnetic field and X-ray detectors}},
  author={Moriyama, Shigetaka and Minowa, Makoto and Namba, Toshio and Inoue, Yoshizumi and Takasu, Yuko and Yamamoto, Akira},
  journal={Physics Letters B},
  volume={434},
  number={1-2},
  pages={147--152},
  year={1998},
  publisher={Elsevier}
}

@article{jaeckel2010low,
  title={\href{https://arxiv.org/pdf/1002.0329}{The low-energy frontier of particle physics}},
  author={Jaeckel, Joerg and Ringwald, Andreas},
  journal={Annual Review of Nuclear and Particle Science},
  volume={60},
  number={1},
  pages={405--437},
  year={2010},
  publisher={Annual Reviews}
}

@article{jaeckel2013force,
  title={\href{https://arxiv.org/pdf/1303.1821}{A force beyond the Standard Model-Status of the quest for hidden photons}},
  author={Jaeckel, Joerg},
  journal={arXiv},
  note = {preprint arXiv: 1303.1821},
  year={2013}
}

@article{caputo2021dark,
  title={\href{https://journals.aps.org/prd/pdf/10.1103/PhysRevD.104.095029}{Dark photon limits: A handbook}},
  author={Caputo, Andrea and Millar, Alexander J and O’Hare, Ciaran AJ and Vitagliano, Edoardo},
  journal={Physical Review D},
  volume={104},
  number={9},
  pages={095029},
  year={2021},
  publisher={APS}
}

@article{pugnat2008results,
  title={\href{https://journals.aps.org/prd/abstract/10.1103/PhysRevD.78.092003}{Results from the OSQAR photon-regeneration experiment: No light shining through a wall}},
  author={Pugnat, Pierre and Duvillaret, Lionel and Jost, Remy and Vitrant, Guy and Romanini, Daniele and Siemko, Andrzej and Ballou, Rafik and Barbara, Bernard and Finger, Michael and Finger, Miroslav and others},
  journal={Physical Review D—Particles, Fields, Gravitation, and Cosmology},
  volume={78},
  number={9},
  pages={092003},
  year={2008},
  publisher={APS}
}

@article{ehret2010new,
  title={\href{https://www.sciencedirect.com/science/article/pii/S0370269310005526}{New ALPS results on hidden-sector lightweights}},
  author={Ehret, Klaus and Frede, Maik and Ghazaryan, Samvel and Hildebrandt, Matthias and Knabbe, Ernst-Axel and Kracht, Dietmar and Lindner, Axel and List, Jenny and Meier, Tobias and Meyer, Niels and others},
  journal={Physics Letters B},
  volume={689},
  number={4-5},
  pages={149--155},
  year={2010},
  publisher={Elsevier}
}

@article{nikezic2005radon,
  title={\href{https://www.sciencedirect.com/science/article/pii/S0168583X05008347}{Radon progeny behavior in diffusion chamber}},
  author={Nikezi{\'c}, D and Stevanovi{\'c}, N},
  journal={Nuclear Instruments and Methods in Physics Research Section B: Beam Interactions with Materials and Atoms},
  volume={239},
  number={4},
  pages={399--406},
  year={2005},
  publisher={Elsevier}
}

@article{pablo2025micromegas,
  title={\href{https://pmc.ncbi.nlm.nih.gov/articles/PMC11926529/pdf/openreseurope-5-21540.pdf}{Micromegas with GEM preamplification for enhanced energy threshold in low-background gaseous time projection chambers}},
  author={Castel, Juan Francisco and Cebri{\'a}n, Susana  and Dafni, Theopisti and D{\'\i}ez-Ib{\'a}{\~n}ez, David  and Gal{\'a}n, Javier and Garc{\'\i}a, Juan Antonio  and Ezquerro, {\'A}lvaro  and G. Irastorza, Igor and Luz{\'o}n, Gloria and Margalejo, Cristina  and others},
  journal={Open Research Europe},
  volume={5},
  pages={53},
  year={2025}
}

@misc{be2013table,
  title={\href{https://cea.hal.science/cea-02476417/file/Monographie_BIPM-5_Tables_Vol7.pdf}{Table of radionuclides (Vol. 7-A= 14 to 245)}},
  author={B{\'e}, Marie-Martine and Chist{\'e}, Vanessa and Dulieu, Christophe and Mougeot, Xavier and Chechev, Valery and Kondev, Filip and Nichols, Alan and Huang, Xiaolong and Wang, Baosong},
  year={2013}
}

@article{gerbier2014news,
  title={\href{https://arxiv.org/pdf/1401.7902}{NEWS: a new spherical gas detector for very low mass WIMP detection}},
  author={Gerbier, G and Giomataris, I and Magnier, P and Dastgheibi, A and Gros, M and Jourde, D and Bougamont, E and Navick, XF and Papaevangelou, T and Galan, J and others},
  journal={arXiv},
  note = {preprint arXiv: 1401.7902},
  year={2014}
}

@article{colas2002electron,
  title={\href{https://doi.org/10.1016/S0168-9002(01)01760-0}{Electron drift velocity measurements at high electric fields}},
  author={Colas, P and Delbart, A and Derre, J and Giomataris, I and Jeanneau, F and Lepeltier, V and Papadopoulos, I and Rebourgeard, Ph},
  journal={Nuclear Instruments and Methods in Physics Research Section A: Accelerators, Spectrometers, Detectors and Associated Equipment},
  volume={478},
  number={1-2},
  pages={215--219},
  year={2002},
  publisher={Elsevier}
}

@article{mcdonald2019electron,
  title={\href{https://iopscience.iop.org/article/10.1088/1748-0221/14/08/P08009}{Electron drift and longitudinal diffusion in high pressure xenon-helium gas mixtures}},
  author={McDonald, AD and Woodruff, K and Al Atoum, B and Gonz{\'a}lez-D{\'\i}az, D and Jones, BJP and Adams, C and {\'A}lvarez, V and Arazi, L and Arnquist, IJ and Azevedo, CDR and others},
  journal={Journal of Instrumentation},
  volume={14},
  number={08},
  pages={P08009},
  year={2019},
  publisher={IOP Publishing}
}

@article{pellecchia2020uv,
  title={\href{https://iopscience.iop.org/article/10.1088/1748-0221/15/04/C04011/pdf}{A UV laser test bench for micro-pattern gaseous detectors}},
  author={Pellecchia, Antonello and Ranieri, Antonio and Verwilligen, Piet},
  journal={Journal of Instrumentation},
  volume={15},
  number={04},
  pages={C04011},
  year={2020},
  publisher={IOP Publishing}
}

@article{lang1971theory,
  title={\href{https://journals.aps.org/prb/pdf/10.1103/PhysRevB.3.1215?casa_token=2psIjGIKOCwAAAAA\%3AIqY69glI-hxdyC_jCPIrxAFjSe7qDvg_3ribi-8psyzcemZc4_VjoaceHivpspIeIOqUG7bNSClwUZQ}{Theory of metal surfaces: work function}},
  author={Lang, ND and Kohn, WJPRB},
  journal={Physical Review B},
  volume={3},
  number={4},
  pages={1215},
  year={1971},
  publisher={APS}
}

@article{polyanskiy2024,
  title={\href{https://refractiveindex.info/}{Refractiveindex. info database of optical constants}},
  author={Polyanskiy, Mikhail N},
  journal={Scientific Data},
  volume={11},
  number={1},
  pages={94},
  year={2024},
  publisher={Nature Publishing Group UK London}
}

@article{iguaz2012characterization,
  title={\href{https://iopscience.iop.org/article/10.1088/1748-0221/7/04/P04007/pdf}{Characterization of microbulk detectors in argon-and neon-based mixtures}},
  author={Iguaz, FJ and Ferrer-Ribas, E and Giganon, A and Giomataris, I},
  journal={Journal of Instrumentation},
  volume={7},
  number={04},
  pages={P04007},
  year={2012},
  publisher={IOP Publishing}
}

@book{lide1995crc,
  title={CRC handbook of chemistry and physics: a ready-reference book of chemical and physical data},
  author={Lide, David R},
  year={1995},
  publisher={CRC press}
}

@article{sikivie2021invisible,
  title={\href{https://journals.aps.org/rmp/pdf/10.1103/RevModPhys.93.015004}{Invisible axion search methods}},
  author={Sikivie, Pierre},
  journal={Reviews of Modern Physics},
  volume={93},
  number={1},
  pages={015004},
  year={2021},
  publisher={APS}
}

@misc{Marki,
  title = {\href{https://markimicrowave.com/products/connectorized/iq-mixers/mmiq-30120hm/datasheet/}{MMIQ-30120HM - Datasheet }},
  howpublished = {\url{https://markimicrowave.com/products/connectorized/iq-mixers/mmiq-30120hm/datasheet/}},
  author = {Marki},
  note = {accessed: 03-06-2025}
}

@article{Mixer_Basics_Primer,
  title = {\href{https://markimicrowave.com/assets/c2c4688b-15c7-4421-a703-254cb238f9fb/Mixer_Basics_Primer.pdf}{Mixer basics primer. A tutorial for RF \& Microwave Mixers}},
  author = {Marki, Ferenc and Marki, Christopher},
  journal={Marki microwave},
  year={2010}
}

@article{di2020landscape,
  title={\href{https://www.sciencedirect.com/science/article/pii/S0370157320302477}{The landscape of QCD axion models}},
  author={Di Luzio, Luca and Giannotti, Maurizio and Nardi, Enrico and Visinelli, Luca},
  journal={Physics Reports},
  volume={870},
  pages={1--117},
  year={2020},
  publisher={Elsevier}
}

@article{durrer2015cosmic,
  title={\href{https://iopscience.iop.org/article/10.1088/0264-9381/32/12/124007/pdf}{The cosmic microwave background: the history of its experimental investigation and its significance for cosmology}},
  author={Durrer, Ruth},
  journal={Classical and Quantum Gravity},
  volume={32},
  number={12},
  pages={124007},
  year={2015},
  publisher={IOP Publishing}
}

@article{diez2025efficient,
  title={\href{https://www.sciencedirect.com/science/article/pii/S0010465525003078}{Efficient cosmic ray generator for particle detector simulations}},
  author={D{\'\i}ez Ib{\'a}{\~n}ez, David and Obis Aparicio, Luis},
  journal={Computer Physics Communications},
  pages={109805},
  year={2025},
  publisher={Elsevier}
}

@article{alvarez2021first,
  title={\href{https://link.springer.com/content/pdf/10.1007/jhep10(2021)075.pdf}{First results of the CAST-RADES haloscope search for axions at 34.67 $\mu$eV}},
  author={{\'A}lvarez Melc{\'o}n, A and Arguedas Cuendis, Sergio and Baier, Justin and Barth, K and Br{\"a}uninger, H and Calatroni, S and Cantatore, G and Caspers, F and Castel, JF and Cetin, SA and others},
  journal={Journal of High Energy Physics},
  volume={2021},
  number={10},
  pages={1--16},
  year={2021},
  publisher={Springer}
}

@article{melcon2020scalable,
  title={\href{https://link.springer.com/content/pdf/10.1007/JHEP07(2020)084.pdf}{Scalable haloscopes for axion dark matter detection in the 30 $\mu$eV range with RADES}},
  author={Melc{\'o}n, A {\'A}lvarez and Cuendis, S Arguedas and Cogollos, Cristian and D{\'\i}az-Morcillo, Alejandro and D{\"o}brich, Babette and Gallego, Juan Daniel and Barcel{\'o}, JM and Gimeno, Benito and Golm, Jessica and Irastorza, Igor G and others},
  journal={Journal of High Energy Physics},
  volume={2020},
  number={7},
  pages={1--28},
  year={2020},
  publisher={Springer}
}

@article{ahyoune2023proposal,
  title={\href{https://onlinelibrary.wiley.com/doi/pdf/10.1002/andp.202300326}{A Proposal for a Low-Frequency Axion Search in the 1-2 $\mu$eV Range and Below with the BabyIAXO Magnet}},
  author={Ahyoune, Saiyd and {\'A}lvarez Melc{\'o}n, Alejandro and Arguedas Cuendis, Sergio and Calatroni, Sergio and Cogollos, Cristian and Devlin, Jack and D{\'\i}az-Morcillo, Alejandro and D{\'\i}ez-Ib{\'a}{\~n}ez, David and D{\"o}brich, Babette and Galindo, Javier and others},
  journal={Annalen der Physik},
  volume={535},
  number={12},
  pages={2300326},
  year={2023},
  publisher={Wiley Online Library}
}

@article{ahyoune2025rades,
  title={\href{https://arxiv.org/pdf/2403.07790}{RADES axion search results with a high-temperature superconducting cavity in an 11.7 T magnet}},
  author={Ahyoune, S and Melc{\'o}n, A {\'A}lvarez and Cuendis, S Arguedas and Calatroni, S and Cogollos, C and D{\'\i}az-Morcillo, A and D{\"o}brich, B and Gallego, JD and Garc{\'\i}a-Barcel{\'o}, JM and Gimeno, B and others},
  journal={Journal of High Energy Physics},
  volume={2025},
  number={4},
  pages={1--23},
  year={2025},
  publisher={Springer}
}

@article{melcon2018axion,
  title={\href{https://arxiv.org/pdf/1803.01243}{Axion searches with microwave filters: the RADES project}},
  author={Melc{\'o}n, Alejandro {\'A}lvarez and Cuendis, Sergio Arguedas and Cogollos, Cristian and D{\'\i}az-Morcillo, Alejandro and D{\"o}brich, Babette and Gallego, Juan Daniel and Gimeno, Benito and Irastorza, Igor G and Lozano-Guerrero, Antonio Jos{\'e} and Malbrunot, Chlo{\'e} and others},
  journal={Journal of Cosmology and Astroparticle Physics},
  volume={2018},
  number={05},
  pages={040},
  year={2018},
  publisher={IOP Publishing}
}

@article{cebrian2010micromegas,
  title={\href{https://arxiv.org/pdf/1009.1827}{Micromegas readouts for double beta decay searches}},
  author={Cebri{\'a}n, S and Dafni, T and Ferrer-Ribas, E and Gal{\'a}n, J and Garcia, JA and Giomataris, I and G{\'o}mez, H and Herrera, DC and Iguaz, FJ and Irastorza, IG and others},
  journal={Journal of Cosmology and Astroparticle Physics},
  volume={2010},
  number={10},
  pages={010},
  year={2010},
  publisher={IOP Publishing}
}

@article{calvet2014versatile,
  title={\href{https://ieeexplore.ieee.org/document/6729073}{A versatile readout system for small to medium scale gaseous and silicon detectors}},
  author={Calvet, Denis},
  journal={IEEE Transactions on Nuclear Science},
  volume={61},
  number={1},
  pages={675--682},
  year={2014},
  publisher={IEEE}
}

@article{brun1997root,
  title={\href{https://cds.cern.ch/record/491486/files/p11.pdf}{ROOT—An object oriented data analysis framework}},
  author={Brun, Rene and Rademakers, Fons},
  journal={Nuclear instruments and methods in physics research section A: accelerators, spectrometers, detectors and associated equipment},
  url={https://root.cern/},
  volume={389},
  number={1-2},
  pages={81--86},
  year={1997},
  publisher={Elsevier}
}

@article{agostinelli2003geant4,
  title={\href{https://www.sciencedirect.com/science/article/pii/S0168900203013688}{GEANT4—a simulation toolkit}},
  author={Agostinelli, Sea and Allison, John and Amako, K al and Apostolakis, John and Araujo, Henrique and Arce, Pedro and Asai, Makoto and Axen, D and Banerjee, Swagato and Barrand, GJNI and others},
  journal={Nuclear instruments and methods in physics research section A: Accelerators, Spectrometers, Detectors and Associated Equipment},
  url={https://geant4.web.cern.ch/},
  volume={506},
  number={3},
  pages={250--303},
  year={2003},
  publisher={Elsevier}
}

@article{ball2020search,
  title={\href{https://journals.aps.org/prd/pdf/10.1103/PhysRevD.102.032002}{Search for millicharged particles in proton-proton collisions at s= 13 TeV}},
  author={Ball, A and Beauregard, G and Brooke, J and Campagnari, C and Carrigan, M and Citron, M and De La Haye, J and De Roeck, A and Elskens, Y and Franco, R Escobar and others},
  journal={Physical Review D},
  volume={102},
  number={3},
  pages={032002},
  year={2020},
  publisher={APS}
}

@phdthesis{Oscar2025development,
  author={P{\'e}rez L{\'a}zaro, {\'O}scar },
  title={\href{https://arxiv.org/pdf/2507.02172} {Development of new ultra-low-background particle detectors based on Micromegas technology for the search of Dark Matter at the low-mass frontier}},
  year={2025},
  school={Universidad de Zaragoza},
  howpublished = {\url{https://arxiv.org/pdf/2507.02172}}               
}

@article{cebrian2013micromegas,
  title={\href{https://iopscience.iop.org/article/10.1088/1748-0221/8/01/P01012/pdf}{Micromegas-TPC operation at high pressure in xenon-trimethylamine mixtures}},
  author={Cebrian, S and Dafni, T and Ferrer-Ribas, E and Giomataris, I and Gonzalez-Diaz, D and G{\'o}mez, H and Herrera, DC and Iguaz, FJ and Irastorza, IG and Luzon, G and others},
  journal={Journal of Instrumentation},
  volume={8},
  number={01},
  pages={P01012},
  year={2013},
  publisher={IOP Publishing}
}

@article{lambert2024qutip5quantumtoolbox,
      title={\href{https://arxiv.org/abs/2412.04705}{QuTiP 5: The Quantum Toolbox in Python}}, 
      author={Neill Lambert and Eric Giguère and Paul Menczel and Boxi Li and Patrick Hopf and Gerardo Suárez and Marc Gali and Jake Lishman and Rushiraj Gadhvi and Rochisha Agarwal and Asier Galicia and Nathan Shammah and Paul Nation and J. R. Johansson and Shahnawaz Ahmed and Simon Cross and Alexander Pitchford and Franco Nori},
      year={2024},
      eprint={2412.04705},
      archivePrefix={arXiv},
      url={https://qutip.org/}, 
}
